%% file: master-wp-v1120-final.tex
\documentclass[11pt]{report}
\usepackage{graphicx}
\usepackage{amsmath}
\usepackage{amssymb}
\usepackage{wrapfig}
\usepackage{color}
\usepackage[usenames,dvipsnames]{xcolor}
\usepackage{makeidx}
\usepackage{multicap,color}
\usepackage{multicol}
\usepackage{multirow}

\usepackage{url}
\usepackage[dvips]{epsfig,graphicx}
\usepackage{epstopdf} 

\usepackage[linktocpage]{hyperref}

\usepackage[textwidth=6.0in,textheight=9.0in]{geometry}
\allowdisplaybreaks[2]

\newcommand{\gev}{\operatorname{GeV}}
\newcommand{\mev}{\operatorname{MeV}}
\newcommand{\fm}{\operatorname{fm}}

\newcommand{\fb}{\operatorname{fb}}
\newcommand{\pb}{\operatorname{pb}}

\newcommand{\jpsi}{J\mskip -2mu/\mskip -0.5mu\Psi}

\newcommand{\lsim}{\raisebox{-4pt}{$\,\stackrel{\textstyle <}{\sim}\,$}}

\newcommand{\beq}{\begin{equation}}
\newcommand{\eeq}{\end{equation}}

\newcommand{\open}{{<\kern -0.3em{\scriptscriptstyle )}}}

\newcommand{\sqrts}{\mbox{$\sqrt{s}$}}

\newcommand{\ep}{$e$$p$}

\newcommand{\AuAu}{AuAu}
\newcommand{\PbPb}{PbPb}

\newcommand{\pT}{\mbox{$p_T$}}

\newcommand{\gevc}{\mbox{${\mathrm{GeV/}}c$}}

\newcommand{\lumi}{\mbox{$\mathrm{cm}^{-2}\mathrm{sec}^{-1}$}}

\newcommand{\fmc}{\mbox{$\mathrm{fm}/c$}}
\newcommand{\Qss}{\mbox{$Q_s^2$}}

\newcommand{\as}{\alpha_s}
\def\eq#1{{Eq.~(\ref{#1})}}
\def\fig#1{{Fig.~\ref{#1}}}

\definecolor{BlueA}     {hsb}{1.00,0.0,1.0}
\definecolor{BlueB}     {hsb}{0.50,0.5,0.5}
\definecolor{ColorA}    {hsb}{0.18,0.5,1.0}
\definecolor{ColorB}    {hsb}{0.18,0.5,0.7}
\definecolor{gray50}    {gray}{0.7}
\definecolor{Blue}           {named}{Blue}
\definecolor{SkyBlue}        {named}{SkyBlue}
\definecolor{CornflowerBlue} {named}{CornflowerBlue}
\definecolor{RoyalBlue}      {named}{RoyalBlue}
\definecolor{BlueGreen}      {named}{BlueGreen}
\definecolor{MBlue}          {named}{MidnightBlue}
\definecolor{Cyan}           {named}{Cyan}
\definecolor{darkblue}       {rgb}{0,0,0.5}
\definecolor{Lavender}       {named}{Lavender}
\definecolor{Magenta}        {named}{Magenta}
\definecolor{Thistle}        {named}{Thistle}
\definecolor{VioletRed}      {named}{VioletRed}
\definecolor{Rhodamine}      {named}{Rhodamine}
\definecolor{Orchid}         {named}{Orchid}
\definecolor{Plum}           {named}{Plum}
\definecolor{pink}           {rgb}{1,0.5,0.5}
\definecolor{lila}           {rgb}{0.6,0.0,1.0}
\definecolor{Magenta1}       {cmyk}{0.0,0.2,0.0,0.0}
\definecolor{Red}            {named}{Red}
\definecolor{darkred}        {rgb}{0.5,0,0}
\definecolor{lightred}       {rgb}{1.0,0,0}
\definecolor{Red1}           {cmyk}{0.0,0.4,0.2,0.0}
\definecolor{Green}          {named}{Green}
\definecolor{SGreen}         {named}{SpringGreen}
\definecolor{FGreen}         {named}{ForestGreen}
\definecolor{LGreen}         {named}{LimeGreen}
\definecolor{FGreen}         {named}{YellowGreen}
\definecolor{darkgreen}      {rgb}{0,0.5,0}
\definecolor{Yellow3}        {rgb}{0.0,0.0,1.0}
\definecolor{Yellow4}        {rgb}{0.5,0.5,0}
\definecolor{Yellow1}        {cmyk}{0.0,0.0,0.6,0.0}
\definecolor{Yellow2}        {cmyk}{0.0,0.0,0.2,0.0}
\definecolor{Yellow}         {named}{Yellow}
\definecolor{GYellow}        {named}{GreenYellow}
\definecolor{YOrange}        {named}{YellowOrange}
\definecolor{Orange}         {named}{Orange}
\definecolor{OrangeRed}      {named}{OrangeRed}
\definecolor{White}          {named}{White}
\definecolor{LightYellow}{cmyk}{0.0,0.0,0.1,0.0}

\makeatletter

\newenvironment{figurehere}
  {\def\@captype{figure}}
  {}
\makeatother

\begin{document}

%
%
\newgeometry{textwidth=8.5in,textheight=11.0in}
\hspace*{-0.53in}
\includegraphics[width=8.5in,height=10.99in]{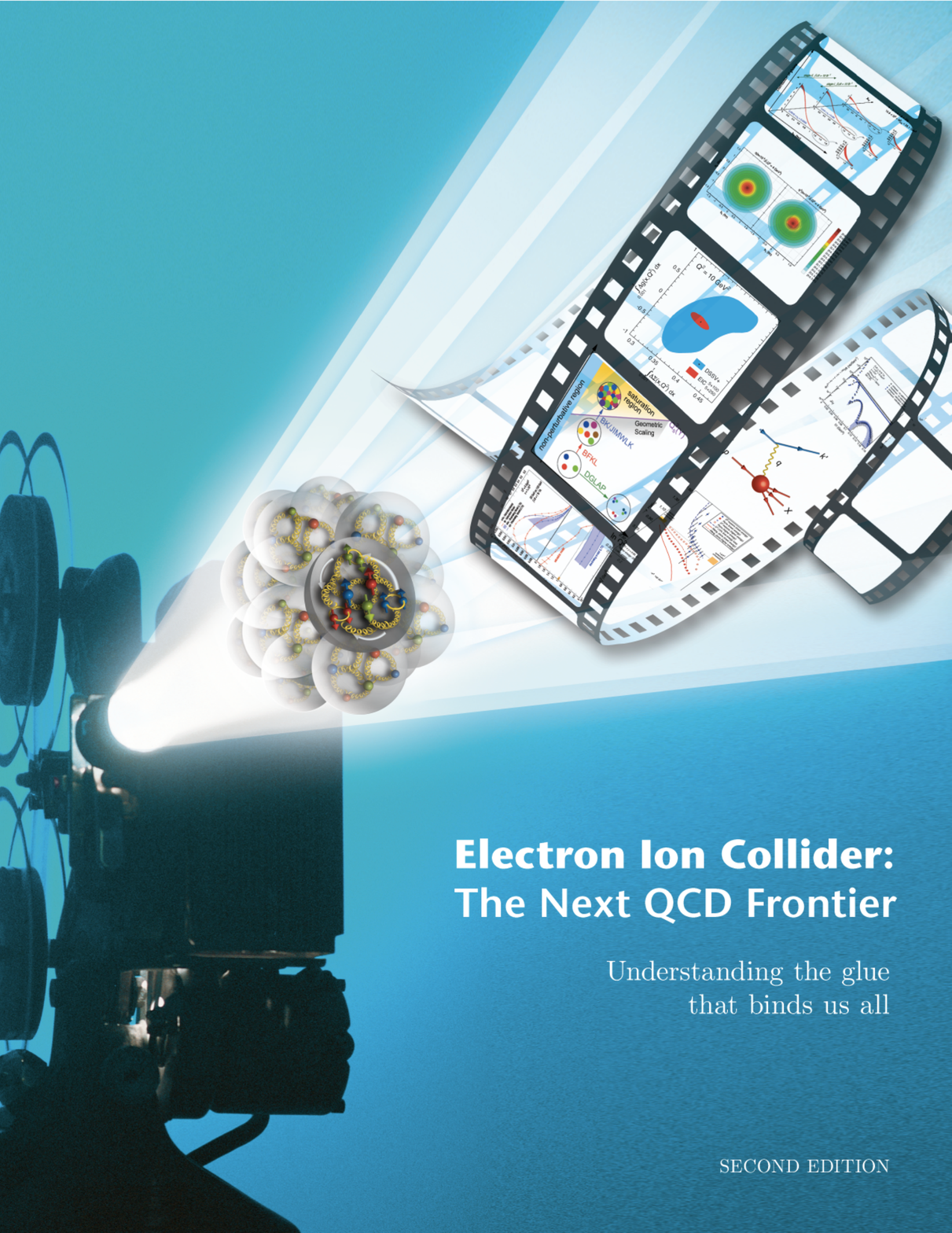}
\restoregeometry

\newpage
\thispagestyle{empty}
\mbox{}

%
%
\newpage
\thispagestyle{empty}
\hfill \begin{minipage}[b]{0.23\textwidth}
\rightline{BNL-98815-2012-JA}
\rightline{JLAB-PHY-12-1652}
\rightline{arXiv:1212.1701}
\rightline{December, 2014}
\end{minipage}

\vspace*{5cm}
\centerline{\bf\Huge
Electron Ion Collider:}
\vspace*{0.5cm}
\centerline{\bf\huge The Next QCD Frontier}
\vskip 0.25cm
\centerline{\LARGE
Understanding the glue that binds us all}

\vskip 12cm
\centerline{\Large
Second Edition}

\newpage
\thispagestyle{empty}
\mbox{}

%
%
\input{./files_tex/authors}

%
%
\input{./files_tex/abstract}

\newpage
\thispagestyle{empty}
\mbox{}

%
%

\setcounter{tocdepth}{4}
\tableofcontents

\newpage
\thispagestyle{empty}
\mbox{}

%
\chapter*{Executive Summary:\ Exploring the Glue that Binds Us All}
\addcontentsline{toc}{chapter}{Executive Summary:\ Exploring the Glue that Binds Us All}

\input{./files_tex/executive}

%
\chapter{Overview:\ Science, Machine and Deliverables of the EIC}
\setcounter{page}{1}
\pagenumbering{arabic}

\input{./files_tex/introduction}

%
%
\chapter{Spin and Three-Dimensional Structure of the Nucleon}
	\input{./files_tex/ep-intro}
	\include{./files_tex/sidebarDIS}

	\include{./files_tex/sidebarCrossSection}
	\input{./files_tex/helicity}

	\include{./files_tex/sidebarTMD} 
	\input{./files_tex/tmd}

	\include{./files_tex/sidebarGPD} 
        \input{./files_tex/exclusive}

%
%
\chapter{The Nucleus: A Laboratory for QCD}	
         \input{./files_tex/eA-intro}
	\include{./files_tex/sidebarDiffraction}

	\input{./files_tex/eA-dense} 
	\input{./files_tex/eA-cold}

	\input{./files_tex/eA-connection}

%
%
\chapter{Possibilities at the Luminosity Frontier: Physics Beyond the Standard Model}
	\input{./files_tex/electroweak}

%
%
\chapter{The Accelerator Designs and Challenges}
	\input{./files_tex/accelerator}

%
%
\chapter{The EIC Detector Requirements and Design Ideas}
	\input{./files_tex/detector}

%
\onecolumn
\input{./files_tex/acknowledgment}
\addcontentsline{toc}{chapter}{Acknowledgments}

%
%
\bibliographystyle{h-physrev5}

\clearpage
\phantomsection
\addcontentsline{toc}{chapter}{References}

\small{
\bibliography{./bibliography/eic_white_paper.with.archive}{}
}

\newpage
\thispagestyle{empty}
\mbox{}

%
%
\newpage
\newgeometry{textwidth=8.5in,textheight=11.0in}
\hspace*{-0.53in}
\includegraphics[width=8.5in,height=10.99in]{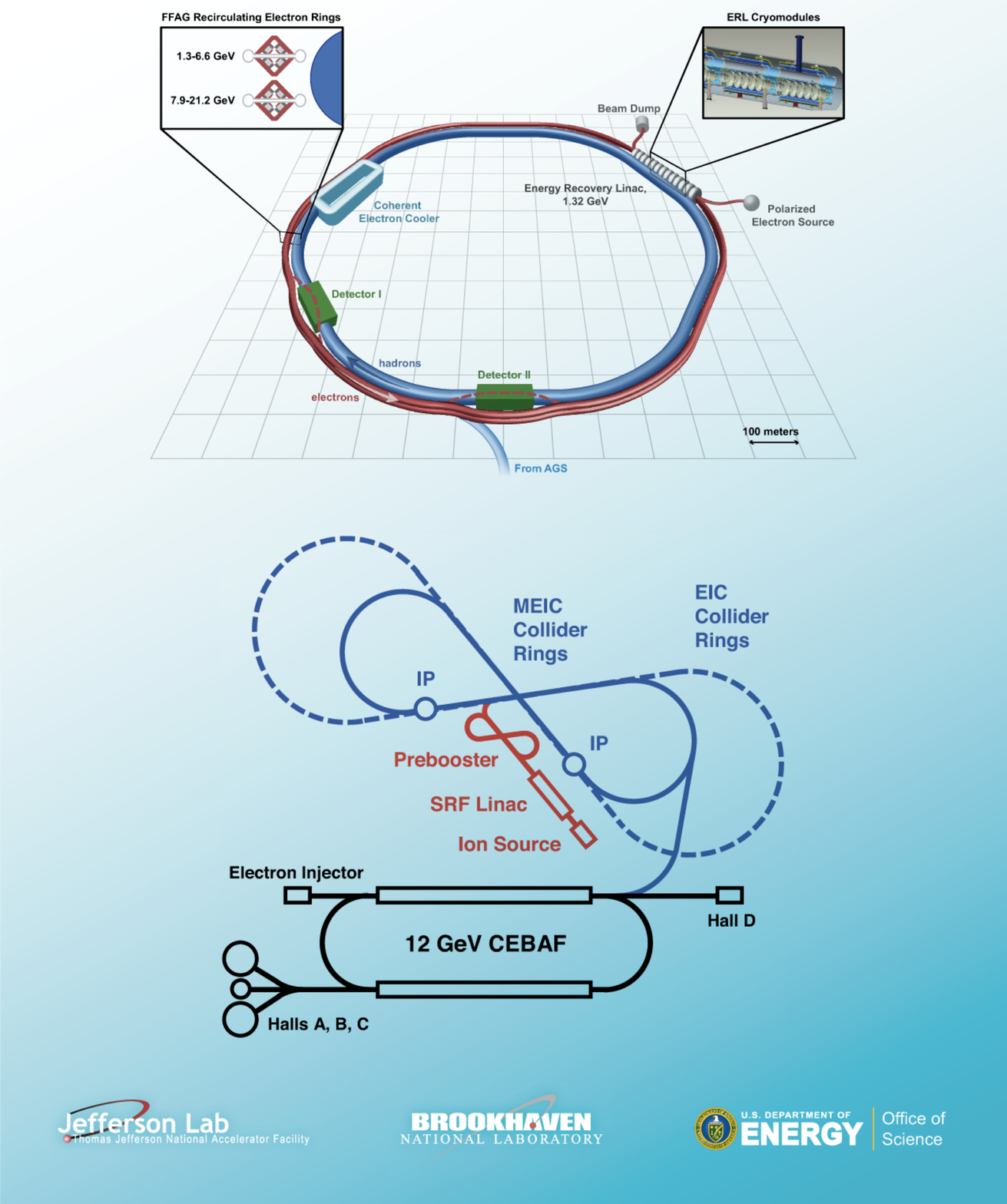}
\restoregeometry
\end{document}

%% file: files_tex/authors.tex
%
%
\renewcommand{\thefootnote}{\fnsymbol{footnote}}
\newpage
\setcounter{page}{1}
\pagenumbering{roman}

\centerline{\Large\bf Authors}
\vskip 0.3in

\begin{center}
A.~Accardi$^{14,28}$,
J.~L.~Albacete$^{16}$,
M.~Anselmino$^{29}$,
N.~Armesto$^{37}$,
E.~C. Aschenauer$^{3,\dagger}$,
A.~Bacchetta$^{36}$,
D.~Boer$^{33}$,
W.K.~Brooks$^{38,\dagger}$,
T.~Burton$^3$,
N.-B.~Chang$^{23}$,
W.-T.~Deng$^{13,23}$,
A.~Deshpande$^{25,}$\footnote{Editors}$^{,}$\footnote{Writing Committee Members},
M.~Diehl$^{11,\dagger}$,
A.~Dumitru$^{2}$,
R.~Dupr\'{e$^{7}$},
R.~Ent$^{28,}$\footnote{Laboratory Management Representatives},
S.~Fazio$^3$,
H.~Gao$^{12,\dagger}$,
V.~Guzey$^{28}$,
H.~Hakobyan$^{38}$,
Y.~Hao$^{3}$,
D.~Hasch$^{15}$,
R.~Holt$^{1,\dagger}$,
T.~Horn$^{5,\dagger}$,
M.~Huang$^{23}$,
A.~Hutton$^{28,\dagger}$,
C.~Hyde$^{20}$,
J.~Jalilian-Marian$^{2}$,
S.~ Klein$^{17}$,
B.~Kopeliovich$^{38}$,
Y.~Kovchegov$^{19,\dagger}$,
K.~Kumar$^{25,\dagger}$,
K.~Kumeri\v{c}ki$^{40}$,
M.~A.~C.~Lamont$^{3}$,
T.~Lappi$^{34}$,
J.-H.~Lee$^{3}$,
Y.~Lee$^{3}$,
E.~M.~Levin$^{26,38}$,
F.-L.~Lin$^{28}$,
V.~Litvinenko$^{3}$,
T.~W.~Ludlam$^{3,\ddagger}$,
C.~Marquet$^{8}$,
Z.-E.~Meziani$^{27,*,\dagger}$,
R.~McKeown$^{28,\ddagger}$,
A.~Metz$^{27}$,
R. Milner$^{18}$,
V.~S.~Morozov$^{28}$,
A.~H.~Mueller$^{9,\dagger}$,
B.~M\"{u}ller$^{3,12,\ddagger}$,
D.~M\"{u}ller$^{22}$,
P.~Nadel-Turonski$^{28}$,
H.~Paukkunen$^{34}$,
A.~Prokudin$^{28}$,
V.~Ptitsyn,$^{3}$,
X.~Qian$^{4}$,
J.-W.~Qiu$^{3,*,\dagger}$,
M.~Ramsey-Musolf$^{35,\dagger}$,
T.~Roser$^{3,\dagger}$,
F.~Sabati\'e$^{7,\dagger}$,
R.~Sassot$^{31}$,
G.~Schnell$^{30}$,
P.~Schweitzer$^{32}$,
E.~Sichtermann$^{17,\dagger}$,
M.~Stratmann$^{39}$,
M.~Strikman$^{21}$,
M.~Sullivan$^{24}$,
S.~Taneja$^{10,25}$,
T.~Toll$^{3}$,
D.~Trbojevic$^{3}$,
T.~Ullrich$^{3,\dagger}$,
R.~Venugopalan$^{3}$,
S.~Vigdor$^{3,\ddagger}$,
W.~Vogelsang$^{39,\dagger}$,
C.~Weiss$^{28}$,
B.-W.~Xiao$^{6}$,
F.~Yuan$^{17,\dagger}$,
Y.-H.~Zhang$^{28}$,
L.~Zheng$^{3,6}$\\[1cm]

\noindent
$^{1}${\it Argonne National Laboratory, USA}

$^{2}${\it Baruch College, CUNY, USA}

$^3${\it  Brookhaven National Laboratory, USA}

$^{4}${\it California Institute of Technology, USA}

$^{5}${\it  The Catholic University of America, USA}

$^{6}${\it Central China Normal University, China}

$^{7}${\it CEA, Centre de Saclay, France}

$^{8}${\it CERN, Switzerland}

$^{9}${\it Columbia University, USA}

$^{10}${\it Dalhousie University, Canada}

$^{11}${\it DESY, Germany}

$^{12}${\it Duke University, USA}

$^{13}${\it Frankfurt University, FIAS, Germany}

$^{14}${\it Hampton University, USA}

$^{15}${\it INFN, Frascati, Italy}

$^{16}${\it IPNO, Universit«e Paris-Sud 11, CNRS/IN2P3, France}

$^{17}${\it Lawrence Berkeley National Laboratory, USA}

$^{18}${\it Massachusetts Institute of Technology, USA}

$^{19}${\it The Ohio State University, USA}

$^{20}${\it Old Dominion University, USA}

$^{21}${\it Pennsylvania State University, USA}

$^{22}${\it Ruhr-University Bochum, Germany}

$^{23}${\it Shandong University, China}

$^{24}${\it Stanford Linear Accelerator Center, USA}

$^{25}${\it Stony Brook University, USA}

$^{26}${\it Tel Aviv University, Israel}

$^{27}${\it Temple University, USA}

$^{28}${\it Thomas Jefferson National Accelerator Facility, USA}

$^{29}${\it Torino University \& INFN, Italy}

$^{30}${\it University of Basque Country, Bilbao, Spain}

$^{31}${\it University de Buenos Aires, Argentina}

$^{32}${\it University of Connecticut, USA}

$^{33}${\it University of Groningen, The Netherlands}

$^{34}${\it University of Jyvaskyla, Finland}

$^{35}${\it University of Massachusetts at Amherst, USA}

$^{36}${\it University of Pavia, Italy}

$^{37}${\it University of Santiago de Campostella, Spain}

$^{38}${\it Universidad T\'ecnica Federico Santa Maria, Chile}

$^{39}${\it University of  T\"ubingen, Germany}


$^{40}${\it University of Zagreb, Croatia}
\end{center}

%% file: files_tex/abstract.tex
%
%
\newpage

\centerline{\Large\bf {Abstract}}

\vskip 0.3in

This White Paper presents the science case of an Electron-Ion Collider (EIC), focused
on the structure and interactions of gluon-dominated matter, 
with the intent to articulate it to the broader nuclear science community.
It was commissioned by the
managements of Brookhaven National Laboratory (BNL) and Thomas Jefferson National
Accelerator Facility (JLab) with the objective of presenting a summary of scientific
opportunities and goals of the EIC as a follow-up to the 2007 NSAC Long Range plan.
This document is a culmination of a community-wide effort in nuclear science following
a series of workshops on EIC physics over the past decades and, in particular, 
the focused ten-week program
on "Gluons and quark sea at high energies" at the Institute for Nuclear Theory
in Fall 2010. It contains a brief description of a few golden physics measurements
along with accelerator and detector concepts required to achieve them. It has been
benefited profoundly from inputs by the users' communities of BNL and JLab. 
This White Paper offers the promise to propel the QCD science program in the U.S., 
established with the CEBAF accelerator at JLab and the RHIC collider at BNL, 
to the next QCD frontier.
\renewcommand{\thefootnote}{\arabic{footnote}}

\vspace{10cm}
\centerline{\large\bf Editors' Note for the Second Edition}
\vspace{0.25cm}

The first edition of this White Paper was released in 2012. 
In the current (second) edition, the science case for the EIC is further sharpened 
in view of 
the recent data from BNL, CERN and JLab experiments and the lessons learnt from them.  
Additional improvements were made by taking into account suggestions from the
larger nuclear physics community including those made at the EIC Users Group meeting 
at Stony Brook University in July  2014, and the QCD Town Meeting 
at Temple University in September 2014.
\vspace{0.25cm}

\rightline{Abhay Deshpande, Zein-Eddine Meziani \& Jian-Wei Qiu}
\rightline{November 2014}

%% file: files_tex/executive.tex

\vspace{-0.2in}

\begin{multicols}{2}
Nuclear science is concerned with the origin and structure of the core
of the atom, the nucleus and the nucleons (protons and neutrons)
within it, which account for essentially all of the mass of the
visible universe. Half a century of investigations have revealed that
nucleons are themselves composed of more basic constituents called
quarks, bound together by the exchange of gluons, and have led to the
development of the fundamental theory of strong interactions known as
Quantum Chromo-Dynamics (QCD). Understanding these constituent
interactions and the emergence of nucleons and nuclei from the
properties and dynamics of quarks and gluons in QCD is a fundamental
and compelling goal of nuclear science.

QCD attributes the forces among quarks and gluons to their ``color
charge''. In contrast to the quantum electromagnetism, where the force
carrying photons are electrically neutral, gluons carry color charge.
This causes the gluons to interact with each other, generating a
significant fraction of the nucleon mass and leading to a
little-explored regime of matter, where abundant gluons dominate its
behavior.  Hints of this regime become manifest when nucleons or
nuclei collide at nearly the speed of light, as they do in colliders
such as HERA, RHIC and the LHC. The quantitative study of matter in this
new regime requires a new experimental facility: an Electron Ion
Collider (EIC).

In the last decade, nuclear physicists have developed new
phenomenological tools to enable remarkable tomographic images of the
quarks and gluons inside unpolarized as well as polarized protons and neutrons. 
These tools will be further developed and utilized to study predominantly the valence quarks 
in the nucleon at the upgraded 12 GeV CEBAF at JLab and COMPASS
at CERN. Applying these new tools to study the matter dominated by
gluons and sea quarks originating from gluons will require the higher
energy and beam polarization of an EIC.

As one increases the energy of the electron-nucleon collision, the
process probes regions of progressively higher gluon density. However,
the density of gluons inside a nucleon must eventually saturate to
avoid untamed growth in the strength of the nucleon-nucleon
interaction, which would violate the fundamental principle of
unitarity. To date this saturated gluon density regime has not been
clearly observed, but an EIC could enable detailed study of this
remarkable aspect of matter.  This pursuit will be facilitated by
electron collisions with heavy nuclei, where coherent contributions
from many nucleons effectively amplify the gluon density being probed.

The EIC was designated in the 2007 Nuclear Physics Long Range Plan as
``embodying the vision for reaching the next QCD frontier''
\cite{NSAC:lrp}. It would extend the QCD science programs in the
U.S. established at both the CEBAF accelerator at JLab and RHIC at BNL
in dramatic and fundamentally important ways.  The most intellectually
pressing questions that an EIC will address that relate to our
detailed and fundamental understanding of QCD in this {\it frontier}
environment are:
\end{multicols}

\begin{itemize}
\item {\bf How are the sea quarks and gluons, and their spins,
distributed in space and momentum inside the nucleon?}  How are these
quark and gluon distributions correlated with overall nucleon
properties, such as spin direction?  What is the role of the orbital
motion of sea quarks and gluons in building the nucleon spin?
\item {\bf Where does the saturation of gluon densities set in?}  Is
there a simple boundary that separates this region from that of more
dilute quark-gluon matter? If so, how do the distributions of quarks
and gluons change as one crosses the boundary? Does this saturation
produce matter of universal properties in the nucleon and all nuclei
viewed at nearly the speed of light?
\item {\bf How does the nuclear environment affect the distribution of
quarks and gluons and their interactions in nuclei?}  How does the
transverse spatial distribution of gluons compare to that in the
nucleon? How does nuclear matter respond to a fast moving color charge
passing through it? Is this response different for light and heavy
quarks?
\end{itemize}

Answers to these questions are essential for understanding the nature
of visible matter. An EIC is the ultimate machine to provide answers
to these questions for the following reasons:
\begin{itemize}
\item A collider is needed to provide kinematic reach well into the
gluon-dominated regime;
\item Electron beams are needed to bring to bear the unmatched
precision of the electromagnetic interaction as a probe;
\item Polarized nucleon beams are needed to determine the correlations
of sea quark and gluon distributions with the nucleon spin;
\item Heavy ion beams are needed to provide precocious access to the
regime of saturated gluon densities and offer a precise dial in the
study of propagation-length for color charges in nuclear matter.
\end{itemize}

\begin{multicols}{2}
The EIC would be distinguished from all past, current, and
contemplated facilities around the world by being at the intensity
frontier with a versatile range of kinematics and beam polarizations,
as well as beam species, allowing the above questions to be tackled at
one facility. In particular, the EIC design exceeds the capabilities
of HERA, the only electron-proton collider to date, by adding a)
polarized proton and light-ion beams; b) a wide variety of heavy-ion
beams; c) two to three orders of magnitude increase in luminosity to
facilitate tomographic imaging; and d) wide energy variability to
enhance the sensitivity to gluon distributions.  Achieving these
challenging technical improvements in a single facility will extend
U.S. leadership in accelerator science and in nuclear science.

The scientific goals and the machine parameters of the EIC were
delineated in deliberations at a community-wide program held at the
Institute for Nuclear Theory (INT) \cite{Boer:2011fh}. The physics
goals were set by identifying critical questions in QCD that remain
unanswered despite the significant experimental and theoretical
progress made over the past decade. This White Paper is prepared for
the broader nuclear science community, and presents a summary of those
scientific goals with a brief description of the golden measurements
and accelerator and detector technology advances required to achieve
them.
\end{multicols}

%% file: files_tex/introduction.tex

\section{Scientific Highlights}
\label{sec:execsum}

\subsection{Nucleon Spin and its 3D Structure and Tomography}

\begin{figure*}[htb!]
\begin{center}
\includegraphics[width=0.76\textwidth,height=2.0in]{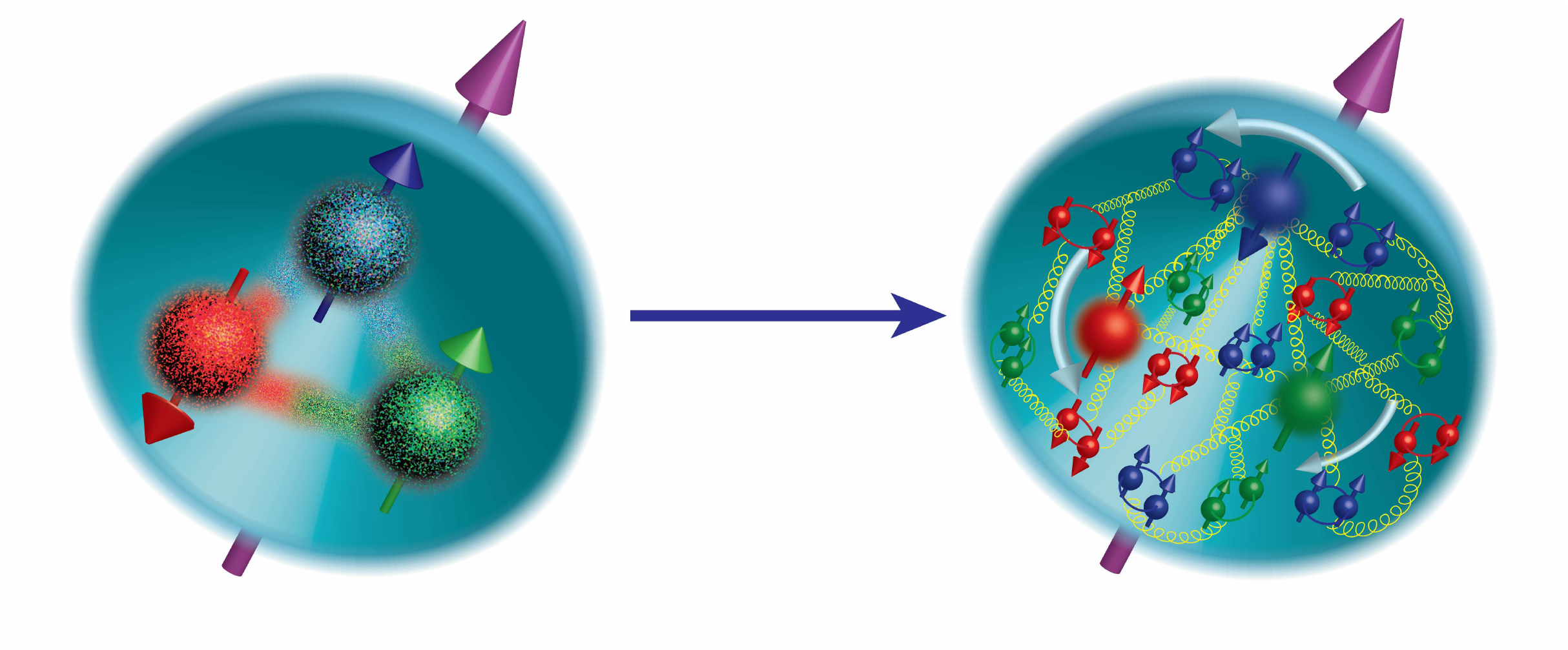}
\end{center} 
\caption{\label{fig:spin_evol}{ Evolution of our understanding of
nucleon spin structure.  {\bf Left:} In the 1980s, a nucleon's spin
was naively explained by the alignment of the spins of its constituent
quarks.  {\bf Right:} In the current picture, valence quarks, sea
quarks and gluons, and their possible orbital motion are expected to
contribute to overall nucleon spin.  }  }
\end{figure*}

Several decades of experiments on deep inelastic scattering (DIS) of
electron or muon beams off nucleons have taught us about how quarks
and gluons (collectively called partons) share the momentum of a
fast-moving nucleon.  They have not, however, resolved the question of
how partons share the nucleon's spin and build up other nucleon
intrinsic properties, such as its mass and magnetic moment. The
earlier studies were limited to providing the longitudinal momentum
distribution of quarks and gluons, a one-dimensional view of nucleon
structure. The EIC is designed to yield much greater insight into the
nucleon structure (Fig.~\ref{fig:spin_evol}, from left to right), by
facilitating multi-dimensional maps of the distributions of partons in
space, momentum (including momentum components transverse to the
nucleon momentum), spin, and flavor.

\vskip 0.1in

\newpage
The 12 GeV upgrade of CEBAF at JLab and the COMPASS at CERN
will initiate such studies in predominantly valence quark region.
However, these programs will be
dramatically extended at the EIC to explore the role of the gluons and
sea quarks in determining the hadron structure and properties.  This
will resolve crucial questions, such as whether a substantial
``missing'' portion of nucleon spin resides in the gluons. By
providing high-energy probes of partons' transverse momenta, the EIC
should also illuminate the role of their orbital motion contributing
to nucleon spin.

\bigskip
\noindent{\bf The Spin and Flavor Structure of the Nucleon}

\begin{multicols}{2}
An intensive and worldwide experimental program over the past two decades
has shown that the spin of quarks and antiquarks is only responsible
for $\sim 30$\% of the proton spin.  Recent RHIC results indicate that
the gluons' spin contribution in the currently explored kinematic
region is non-zero, but not yet sufficient to account for the missing
70\%.  The partons' total helicity contribution to the proton spin is
very sensitive to their minimum momentum fraction $x$ accessible by
the experiments. With the unique capability to reach two orders of
magnitude lower in $x$ and to span a wider range of momentum transfer
$Q$ than previously achieved, the EIC would offer the most powerful
tool to precisely quantify how the spin of gluons and that of quarks
of various flavors contribute to the protonÕs spin. The EIC would
realize this by colliding longitudinally polarized electrons and
nucleons, with both inclusive and semi-inclusive DIS measurements. In
the former, only the scattered electron is detected, while in the
latter, an additional hadron created in the collisions is to be
detected and identified.  
\end{multicols}
\begin{figurehere}
\begin{center}
\begin{minipage}[b]{0.55\textwidth}
\centering
\includegraphics[width=0.98\textwidth,height=2.4in]{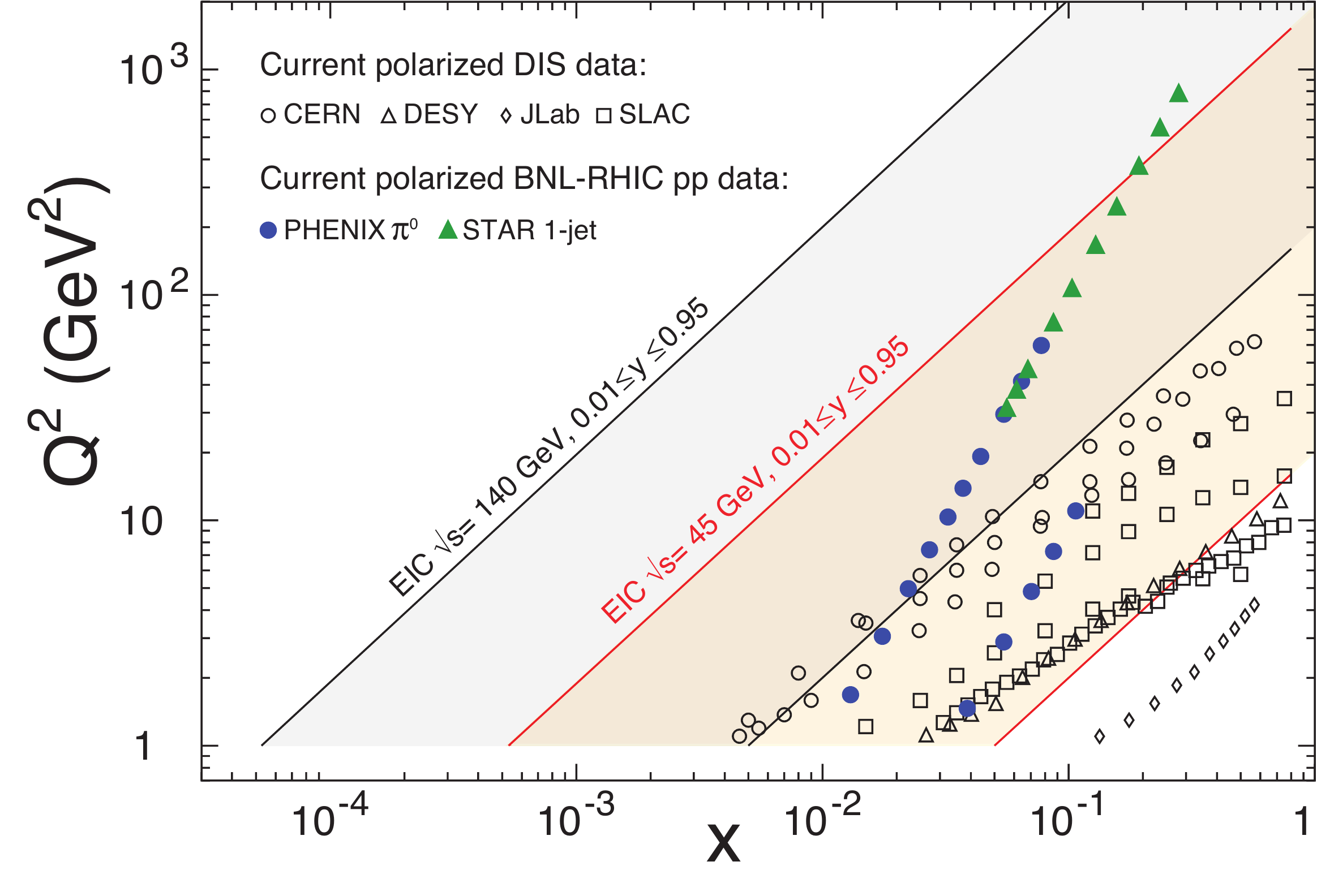}
\end{minipage}
\begin{minipage}[b]{0.44\textwidth}
\centering
\includegraphics[width=0.94\textwidth]{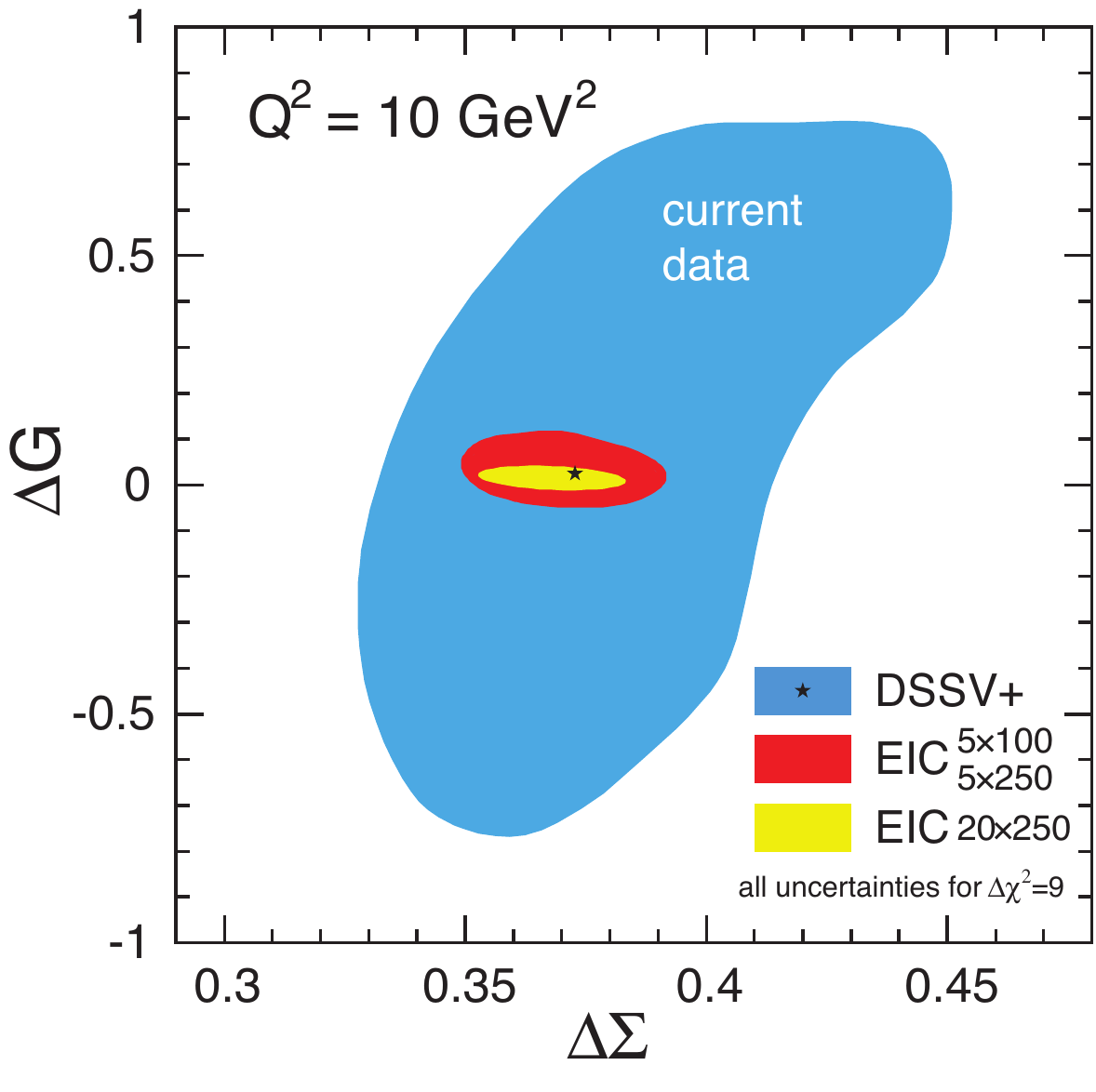}
\end{minipage}
\end{center}
\vskip -0.3in
\caption{\label{fig:kin_helicity}{ {\bf Left:} The range in parton
momentum fraction $x$ vs. the square of the momentum transferred by
the electron to the proton $Q^{2}$ accessible with the EIC in $e$+$p$
collisions at two different center-of-mass energies, compared to
existing data.  {\bf Right:} The projected reduction in the
uncertainties of the gluon's helicity contribution $\Delta $G vs. the
quark helicity contribution $\Delta \Sigma/2$ to the proton spin from
the region of parton momentum fractions $x > 0.001$ that would be
achieved by the EIC for different center-of-mass energies.  }  }
\end{figurehere} 
\begin{multicols}{2}
Figure~\ref{fig:kin_helicity} (Right) shows the reduction in
uncertainties of the contributions to the nucleon spin from the spin
of the gluons, quarks and antiquarks, evaluated in the $x$ range from
0.001 to 1.0.  This would be achieved by the EIC in its early operations.
In future, the kinematic range could be further
extended down to $x \sim 0.0001$ reducing significantly the
uncertainty on the contributions from the unmeasured small-$x$
region. While the central values of the helicity contributions in
Fig.~\ref{fig:kin_helicity} are derived from existing data, they could
change as new data become available in the low- $x$ region. The
uncertainties calculated here are based on the state-of-the art
theoretical treatment of all available data related to the
nucleon spin puzzle. Clearly, the EIC will make a huge impact on our
knowledge of these quantities, unmatched by any other existing or
anticipated facility.  The reduced uncertainties would definitively
resolve the question of whether parton spin preferences alone can
account for the overall proton spin, or whether additional
contributions are needed from the orbital angular momentum of partons
in the nucleon.
\end{multicols}

\noindent {\bf The Confined Motion of Partons Inside the Nucleon}

\begin{multicols}{2}
Semi-inclusive DIS (SIDIS) measurements have two natural momentum
scales: the large momentum transfer from the electron beam needed to
achieve the desired spatial resolution, and the momentum of the
produced hadrons perpendicular to the direction of the momentum
transfer, which prefers a small value sensitive to the motion of
confined partons. Remarkable theoretical advances over the past decade
have led to a rigorous framework where information on the confined
motion of the partons inside a fast-moving nucleon is matched to
transverse-momentum dependent parton distributions (TMDs). In
particular, TMDs are sensitive to correlations between the motion of
partons and their spin, as well as the spin of the parent nucleon.
These correlations can arise from spin-orbit coupling among the
partons, about which very little is known to date. TMDs thus allow us
to investigate the full three-dimensional dynamics of the proton,
going well beyond the information about longitudional momentum
contained in conventional parton distributions.  With both electron
and nucleon beams polarized at collider energies, the EIC will
dramatically advance our knowledge of the motion of confined gluons
and sea quarks in ways not achievable at any existing or proposed
facility.
\end{multicols}

\vspace*{-0.25cm}
\begin{figurehere}
\begin{center}
\begin{minipage}[b]{0.38\textwidth} \centering
{\hskip -0.1in}
\includegraphics[width=0.96\textwidth]{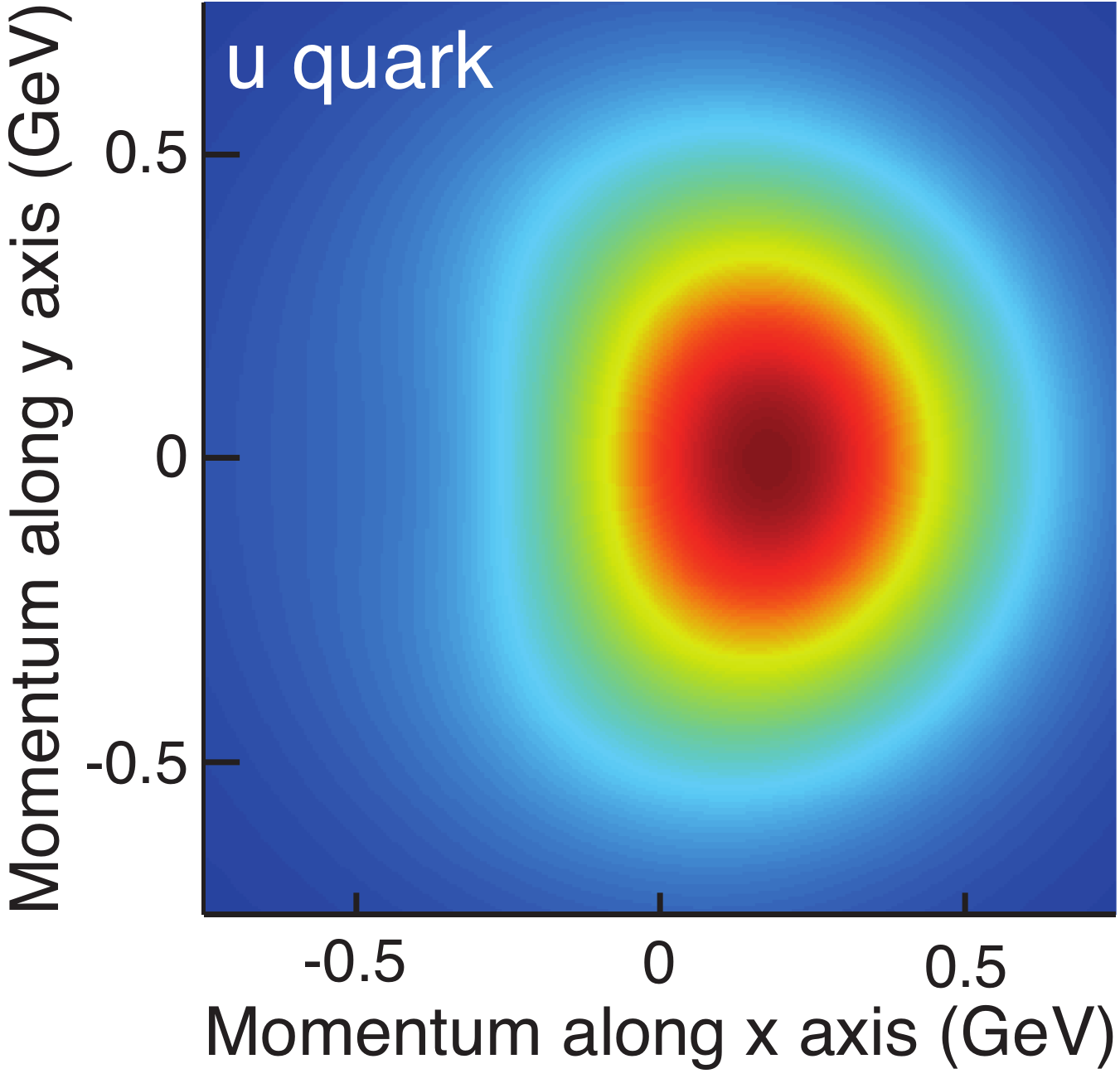}
\end{minipage} 
\hskip 0.05in
\begin{minipage}[b]{0.60\textwidth} \centering
\includegraphics[width=0.98\textwidth]{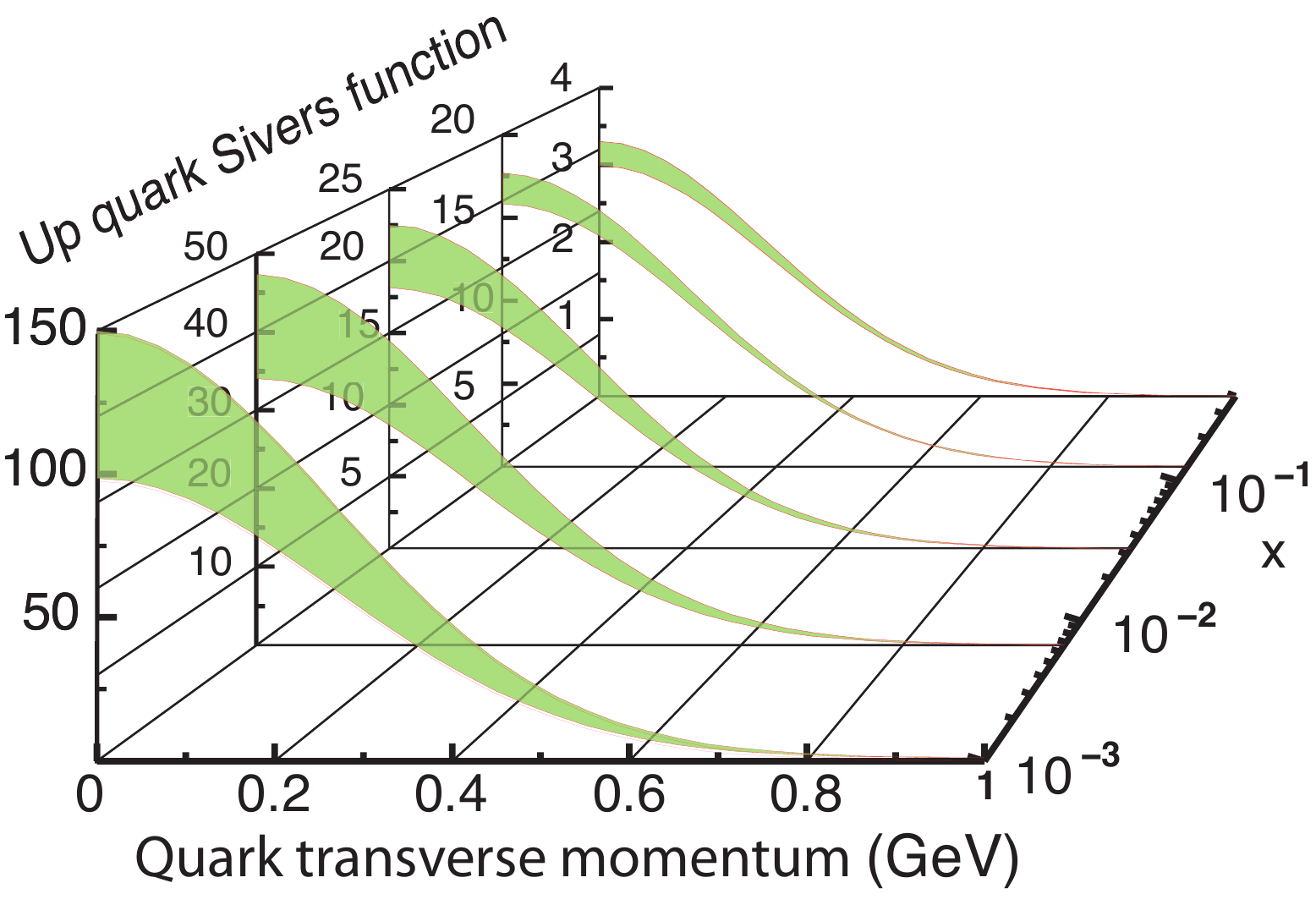}
\end{minipage}
\end{center} 
\vskip -0.05in
\caption{\label{fig:ptomography}{ {\bf Left:} The transverse-momentum
distribution of an up quark with longitudinal momentum fraction $x=0.1$
in a transversely polarized proton moving in the z-direction, while
polarized in the y-direction. The color code indicates the probability
of finding the up quarks.  {\bf Right:} The transverse-momentum
profile of the up quark Sivers function at five $x$ values accessible
to the EIC, and corresponding statistical uncertainties.  }  }
\end{figurehere} 
\begin{multicols}{2}
Figure~\ref{fig:ptomography} (Left) shows the transverse-momentum
distribution of up quarks inside a proton moving in the z direction
(out of the page) with its spin polarized in the y direction. The
color code indicates the probability of finding the up quarks.  The
anisotropy in transverse momentum is described by the Sivers
distribution function, which is induced by the correlation between the
proton's spin direction and the motion of its quarks and gluons.
While the figure is based on a preliminary extraction of this
distribution from current experimental data, nothing is known about
the spin and momentum correlations of the gluons and sea quarks. The
achievable statistical precision of the quark Sivers function from 
EIC kinematics is also shown in Fig.~\ref{fig:ptomography}
(Right). Currently no data exist for extracting such a picture in the
gluon-dominated region in the proton. The EIC will be crucial to
initiate and realize such a program.
\end{multicols}

\noindent{\bf The Tomography of the Nucleon - Spatial Imaging of
Gluons and Sea Quarks}

\begin{multicols}{2}
By choosing particular final states in
electron+proton scattering, the EIC will probe the transverse spatial
distribution of sea quarks and gluons in the fast-moving proton as a
function of the parton's longitudinal momentum fraction, $x$. This
spatial distribution yields a picture of the proton that is
complementary to the one obtained from the transverse-momentum
distribution of quarks and gluons, revealing aspects of proton
structure that are intimately connected with the dynamics of QCD at
large distances. 
\end{multicols}
\vspace*{-0.15cm}
\begin{figurehere}
\begin{center}
\includegraphics[width=0.92\textwidth]{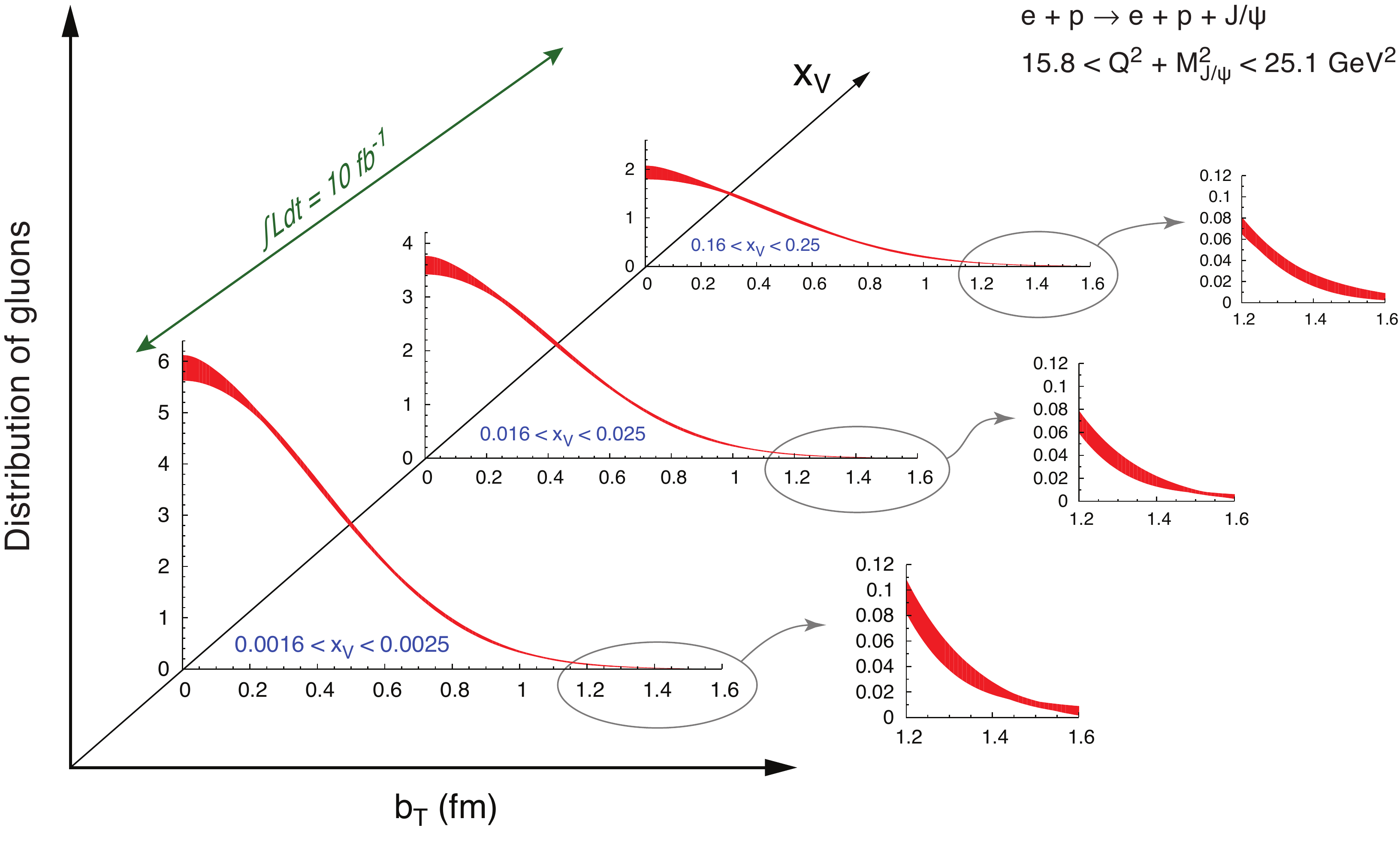}
\end{center} 
\vskip -0.1in
\caption{\label{fig:gimaging}{ The projected precision of the transverse
spatial distribution of gluons as obtained from the cross-section of
exclusive $\jpsi$ production. It includes statistical and
systematic uncertainties due to extrapolation into the unmeasured
region of momentum transfer to the scattered proton. The distance of
the gluon from the center of the proton is $b_{T}$ in femtometers, and
the kinematic quantity $x_{V}=x_B\, (1+M_{\jpsi}^2/Q^2)$ determines
the gluon's momentum fraction. The collision energies assumed for
the top large $x_V$ plot and the lower $x_V$ plots 
are $E_{e} =5, 20$ GeV and $E_{p} = 100, 250$
GeV, respectively.  }  }
\end{figurehere}
\begin{multicols}{2}
With its broad range of collision energies, its high
luminosity and nearly hermetic detectors, the EIC could image the
proton with unprecedented detail and precision from small to large
transverse distances.  The accessible parton momentum fractions $x$
extend from a region dominated by sea quarks and gluons to one where
valence quarks become important, allowing a connection to the precise
images expected from the 12 GeV upgrade at JLab and COMPASS at CERN.
This is illustrated in Fig.~\ref{fig:gimaging}, which shows the
precision expected for the spatial distribution of gluons as measured
in the exclusive process: electron + proton $\rightarrow$ electron + proton + $\jpsi$.

The tomographic images obtained from cross-sections and polarization
asymmetries for exclusive processes are encoded in generalized parton
distributions (GPDs) that unify the concepts of parton densities and
of elastic form factors. They contain detailed information about
spin-orbit correlations and the angular momentum carried by partons,
including their spin and their orbital motion.  The combined kinematic
coverage of the EIC and of the upgraded CEBAF as well as COMPASS is
essential for extracting quark and gluon angular momentum
contributions to the proton's spin.
\end{multicols}

\vskip 0.1in
\subsection{The Nucleus, a QCD Laboratory}

\begin{figure}[h]
\begin{center}
\includegraphics[width=0.98\textwidth]{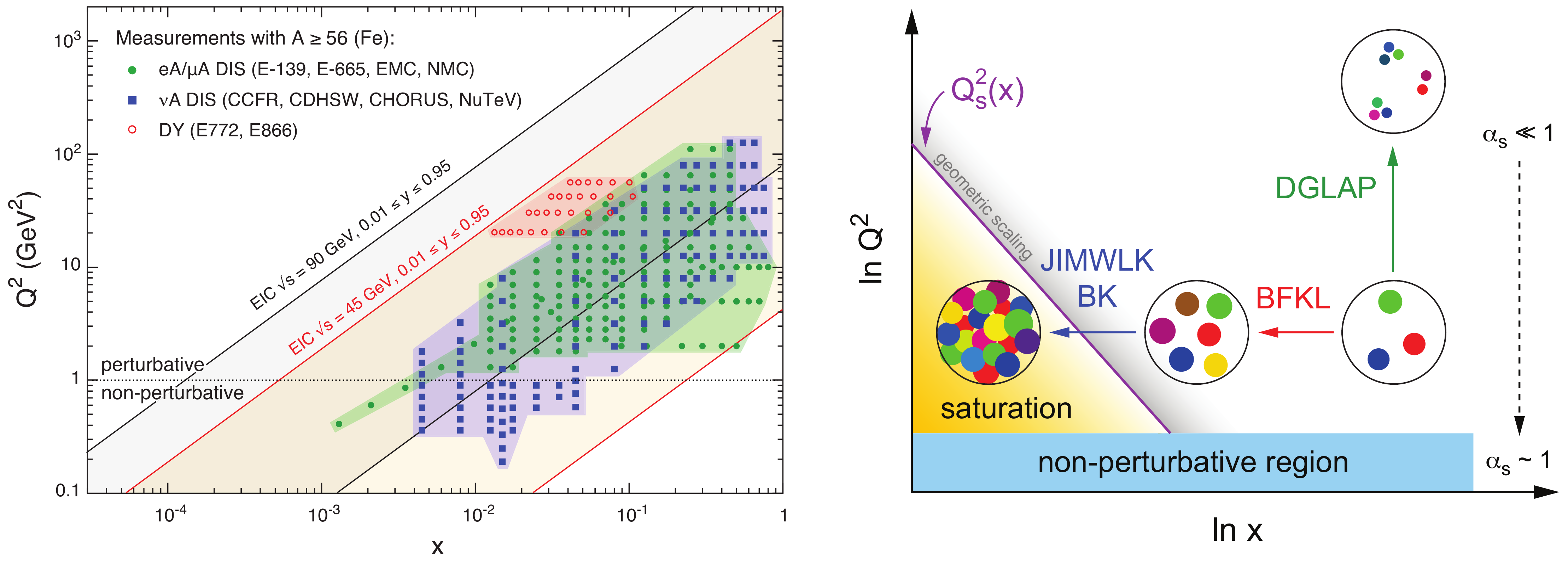}
\end{center} 
\vskip -0.1in
\caption{\label{fig:kin-wf}{ {\bf Left:} The range in the square of the transferred 
momentum by the electron to the nucleus, $Q^2$, versus the parton momentum fraction 
$x$ accessible to the EIC in $e$-A collisions at two different center-of-mass energies, 
compared with the existing data.
{\bf Right:} The schematic probe resolution vs. energy landscape,
indicating regions of non-perturbative and perturbative QCD, including
in the latter, low to high saturated parton density, and the
transition region between them.  }  }
\end{figure}

The nucleus is a QCD ``molecule'', with a complex structure
corresponding to bound states of nucleons.  Understanding the
formation of nuclei in QCD is an ultimate long-term goal of nuclear
physics.  With its wide kinematic reach, as shown in
Fig.~\ref{fig:kin-wf} (Left), the capability to probe a variety of
nuclei in both inclusive and semi-inclusive DIS measurements, the EIC
will be the first experimental facility capable of exploring the
internal 3-dimensional sea quark and gluon structure of a fast-moving
nucleus. Furthermore, the nucleus itself is an unprecedented QCD
laboratory for discovering the collective behavior of gluonic matter
at an unprecedented occupation number of gluons, and for studying the
propagation of fast-moving color charges in a nuclear medium.

\vskip 0.1in

\noindent {\bf QCD at Extreme Parton Densities}

\begin{multicols}{2}
In QCD, the large
soft-gluon density enables the non-linear process of gluon-gluon
recombination to limit the density growth. Such a QCD self-regulation
mechanism necessarily generates a dynamic scale from the interaction
of high density massless gluons, known as the saturation scale, $Q_s$,
at which gluon splitting and recombination reach a balance.  At this
scale, the density of gluons is expected to saturate, producing new and
universal properties of hadronic matter. The saturation scale $Q_s$
separates the condensed and saturated soft gluonic matter from the
dilute, but confined, quarks and gluons in a hadron, as shown in
Fig.~\ref{fig:kin-wf} (Right).
\end{multicols}
\vspace{-0.05in}
\begin{figurehere}
\begin{center}
\includegraphics[width=0.98\textwidth,height=2.9in]{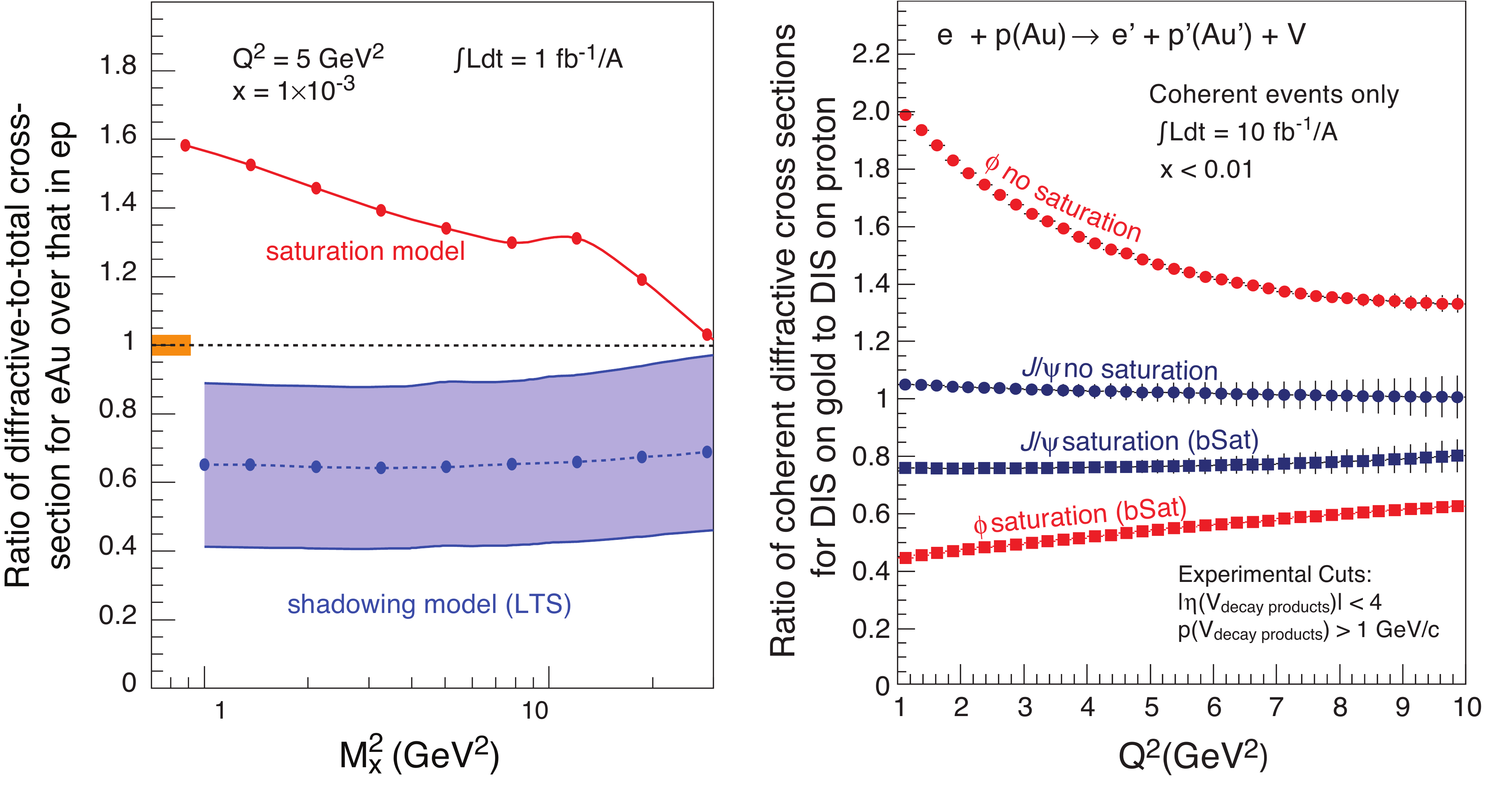}
\end{center} \vskip -0.2in
\caption{\label{fig:sat_diff}{ {\bf Left:} The ratio of diffractive
over total cross-section for DIS on gold normalized to DIS on proton
plotted for different values of M$_{\rm X}^2$, the mass squared of
hadrons produced in the collisions for models assuming saturation and
non-saturation.  
The statistical error bars are too small to depict and
the projected systematic uncertainty for the measurements is shown by the orange bar.
The theoretical uncertainty for the predictions of the LTS model is shown by
the grey band. 
{\bf Right:} The ratio of the coherent diffractive cross-section in $e$+Au
to $e$+$p$ collisions normalized by $A^{4/3}$ and plotted as a function of
$Q^2$ for both saturation and non-saturation models.  The $1/Q$ is
effectively the initial size of the quark-antiquark systems ($\phi$
and $\jpsi$) produced in the medium.  }  }
\end{figurehere} 

\begin{multicols}{2}
The existence of such a state of saturated, soft gluon matter, often
referred to as the Color Glass Condensate (CGC), is a direct consequence
of gluon self-interactions in QCD. It has been conjectured that the
CGC of QCD has universal properties common to nucleons and all nuclei,
which could be systematically computed if the dynamic saturation scale
$Q_s$ is sufficiently large.  However, such a semi-hard $Q_s$ is
difficult to reach unambiguously in electron-proton scattering without
a multi-TeV proton beam.  Heavy ion beams at the EIC could provide
precocious access to the saturation regime and the properties of the
CGC because the virtual photon in forward lepton scattering probes
matter coherently over a characteristic length proportional to 1/$x$,
which can exceed the diameter of a Lorentz-contracted nucleus.  Then,
all gluons at the same impact parameter of the nucleus, enhanced by
the nuclear diameter proportional to A$^{1/3}$ with the atomic weight
A, contribute to the probed density, reaching saturation at far lower
energies than would be needed in electron+proton collisions. While
HERA, RHIC and the LHC have only found hints of saturated gluonic
matter, the EIC would be in a position to seal the case, completing
the process started at those facilities.  

Figure~\ref{fig:sat_diff} illustrates some of the dramatic predicted
effects of gluon density saturation in electron+nucleus
vs. electron+proton collisions at an EIC.  The left frame considers
coherent diffractive processes, defined to include all events in which
the beam nucleus remains intact and there is a rapidity gap containing
no produced particles. 
As shown in the figure, the fraction of such diffractive events are 
greatly enhanced by gluon saturation (the red points) in comparison
with the predictions of shadowing model (the blue points).
In all gluon saturation models, the coherent destructive multiple interaction 
among colored gluons suppresses both the coherent diffractive and total 
DIS cross-sections on nuclei compared to those on the proton, but, 
the suppression on the coherent diffractive events with the nucleus 
remained intact is much weaker than that of the total cross section
leading to a dramatic enhancement in the double ratio as shown 
in Fig.~\ref{fig:sat_diff} (Left).
An early measurement of coherent
diffraction in $e+A$ collisions at the EIC would provide the first
unambiguous evidence for gluon saturation.

Figure~\ref{fig:sat_diff} (Right) shows that gluon saturation is
predicted to suppress vector meson production in $e+A$ relative to $e+p$
collisions at the EIC.  The vector mesons result from quark-antiquark
pair fluctuations of the virtual photon, which hadronize upon the exchange
of gluons with the beam proton or nucleus. The magnitude of the
suppression depends on the size (or color dipole moment) of the
quark-antiquark pair, being significantly larger for produced $\phi$
(red points) than for $\jpsi$ (blue) mesons. An EIC measurement of the
processes in Fig.~\ref{fig:sat_diff} (Right) will provide a powerful
probe to explore the properties of saturated gluon matter.
\end{multicols}

\medskip
\noindent {\bf The Tomography of the Nucleus}

\begin{multicols}{2}
With its capability to
measure the diffractive and exclusive processes with a variety of ion
beams, the EIC will also provide the first 3-dimensional images of
sea quarks and gluons in a fast-moving nucleus with sub-femtometer
resolution.  For example, the EIC could obtain the spatial
distribution of gluons in a nucleus by measuring the coherent
diffractive production of $\jpsi$ in electron-nucleus scattering,
similar to the case of electron-proton scattering shown in
figure~\ref{fig:gimaging}.
\end{multicols}

\medskip
\noindent {\bf Propagation of a Color Charge in QCD Matter}

\begin{multicols}{2}
One of
the key pieces of evidence for the discovery of the quark gluon plasma
(QGP) at RHIC is Òjet quenchingÓ, manifested as a strong suppression
of fast-moving hadrons produced in the very hot matter created in
relativistic heavy-ion collisions. The suppression is believed to be
due to the energy loss of colored partons traversing the QGP. It has been
puzzling that the production is nearly as much suppressed for heavy as
for light mesons, even though a heavy quark is much less likely to
lose its energy via medium-induced radiation of gluons. Some of the
remaining mysteries surrounding heavy vs. light quark interactions in
hot matter can be illuminated by EIC studies of related phenomena in
a better known cold nuclear matter.  For example, the variety of ion beams available
for electron-nucleus collisions at the EIC would provide a femtometer
filter to test and to help determine the correct mechanism by which
quarks and gluons lose energy and hadronize in nuclear matter (see
schematic in Fig.~\ref{fig:attenuation} (Left)).  
\end{multicols}

\vspace{0.1in}
\begin{figurehere}
\begin{center}
\begin{minipage}[c]{0.45\textwidth} \centering
\vspace{-0.1in}
\includegraphics[width=0.96\textwidth]{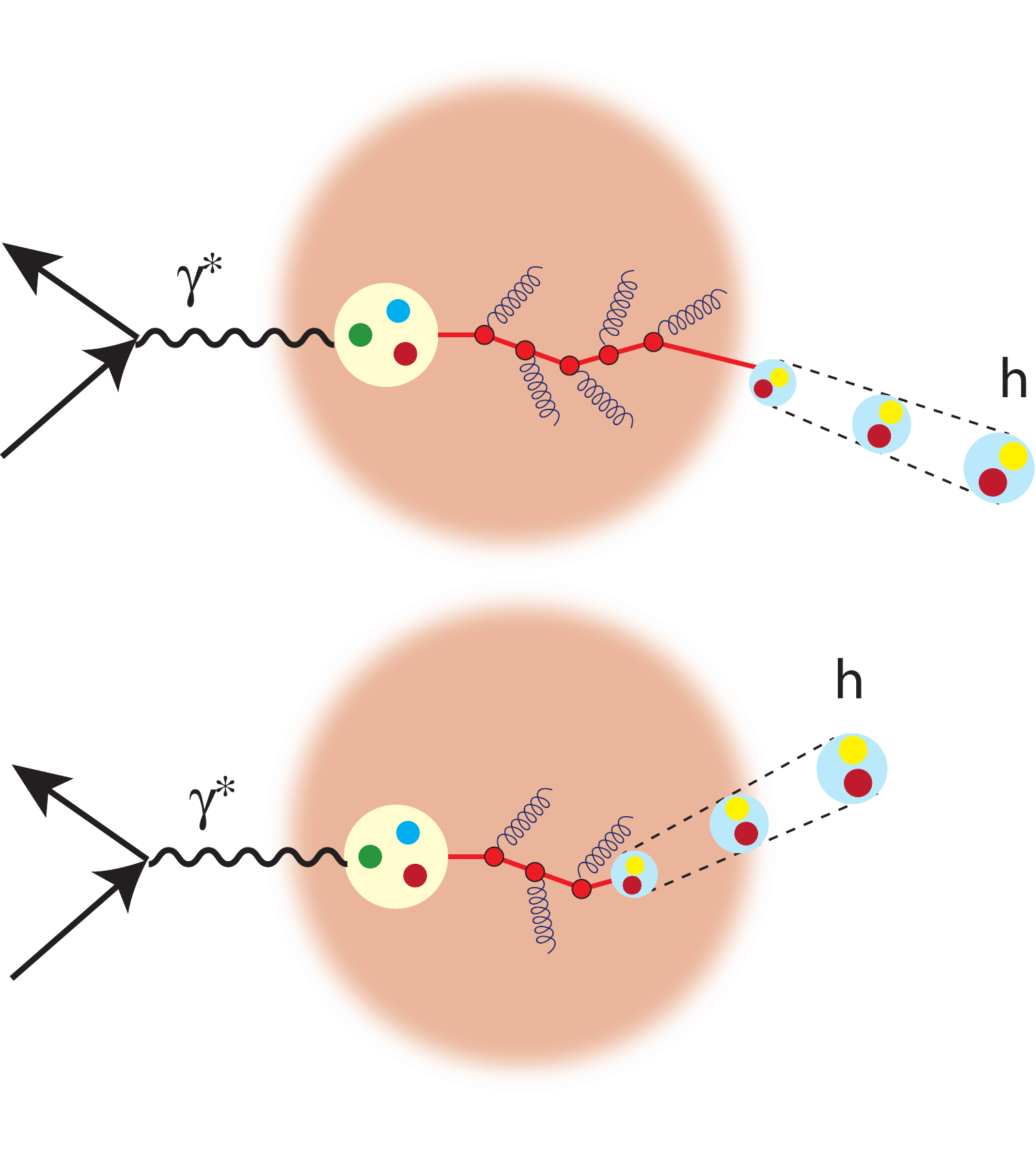}
\end{minipage} \hskip 0.1in
\begin{minipage}[c]{0.5\textwidth} \centering
\includegraphics[width=0.99\textwidth]{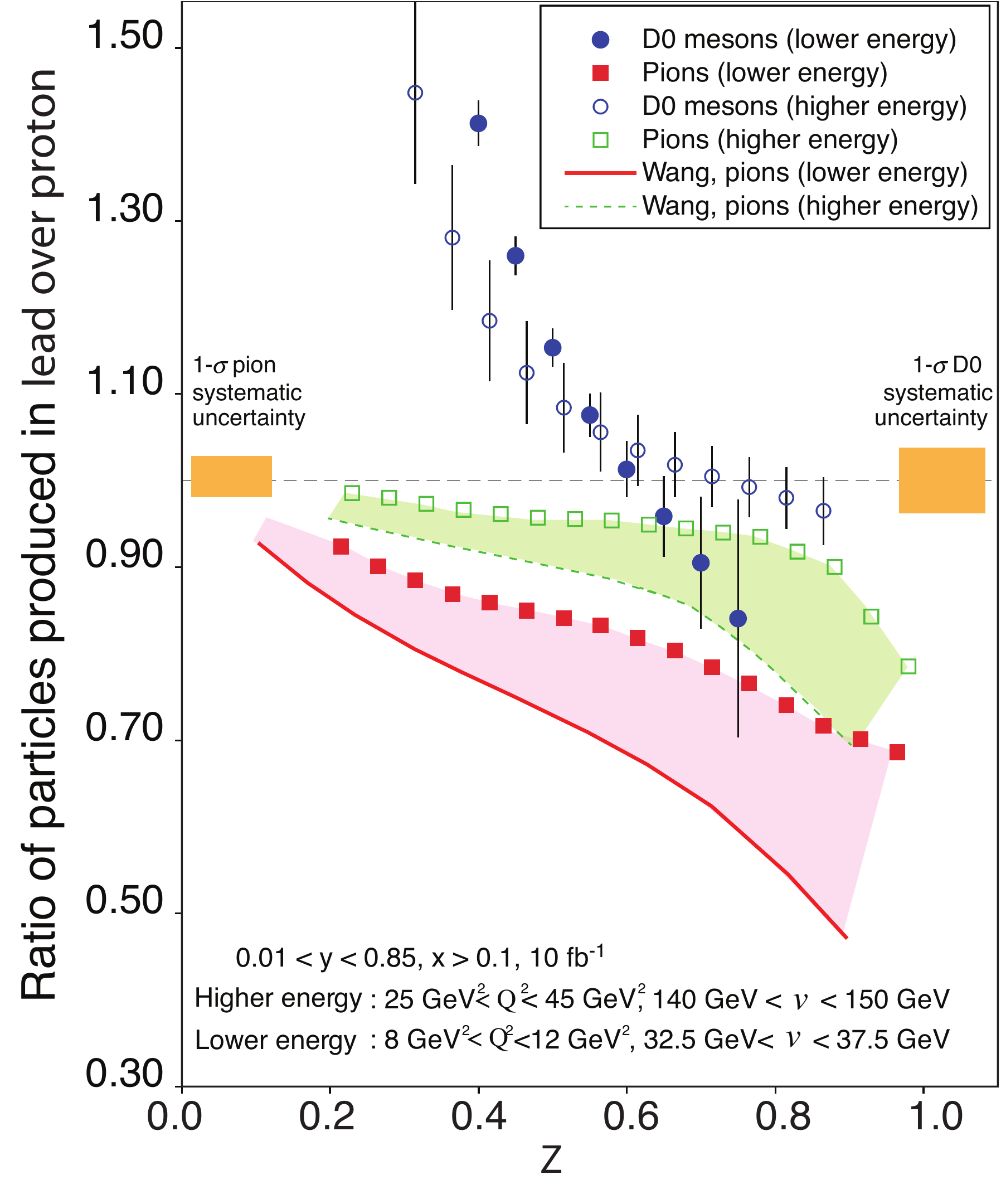}
\end{minipage}
\end{center} \vskip -0.1in
\caption{\label{fig:attenuation}{ {\bf Left:} A schematic illustrating
the interaction of a parton moving through cold nuclear matter: the
hadron is formed outside (top) or inside (bottom) the nucleus.  {\bf
Right:} The ratio of the semi-inclusive cross-section for producing a pion
(red) composed of light quarks, and a $D^{0}$ meson (blue) composed of
heavy quarks in $e$+lead collisions to $e$+deuteron collisions, plotted as a 
function of $z$, the ratio of the momentum carried by the produced
hadron to that of the virtual photon ($\gamma^{*}$), as shown in the
plots on the left.  }  }
\end{figurehere} 

\vspace{0.2in}
\begin{multicols}{2}
Figure~\ref{fig:attenuation} (Right) shows the ratio of the number of
produced mesons in electron+nucleus and electron+deuteron collisions
for pions (light mesons) and D$^0$-mesons (heavy mesons) at both low
and high virtual photon energy $\nu$, as a function of $z$ -- that is,
the momentum fraction of the virtual photon taken by the observed
meson. The calculation of the lines and blue circle symbols assumes that 
the mesons are formed outside of the nucleus, as shown in the top
sketch of Fig.~\ref{fig:attenuation} (Left), while the square symbols
are simulated according to a model where a color neutral pre-hadron
was formed inside the nucleus, like in the bottom sketch of
Fig.~\ref{fig:attenuation} (Left).  The location of measurements
within the shaded area would provide the first direct information on
when the mesons are formed. Unlike the suppression expected for pion
production at all $z$, the ratio of heavy meson production could be
larger than unity due to very different hadronization properties of
heavy mesons. The discovery of such a dramatic difference in
multiplicity ratios between light and heavy mesons at the EIC will 
shed light on the hadronization process and on what governs the
transition from quarks to hadrons.
\end{multicols}

\newpage
\noindent{{\bf The Distribution of Quarks and Gluons in the
Nucleus}}

\begin{multicols}{2}
The EMC experiment at CERN and experiments in the
following two decades clearly revealed that the distribution of quarks
in a fast-moving nucleus is not a simple superposition of their
distributions within nucleons.  Instead, the ratio of nuclear over
nucleon structure functions follows a non-trivial function of Bjorken
$x$, deviating significantly from unity, with a suppression as $x$ 
decreases (often referred to as nuclear shadowing). Amazingly, there
is as of yet no knowledge whether the same holds true for gluons. With
its much wider kinematic reach in both $x$ and $Q$, the EIC could
measure the suppression of the structure functions to a much lower
value of $x$, approaching the region of gluon saturation.  In
addition, the EIC could for the first time reliably quantify the
nuclear gluon distribution over a wide range of momentum fraction $x$.
\end{multicols}

\subsection{{Physics Possibilities at the Intensity Frontier}}

\begin{multicols}{2}
The subfield of Fundamental Symmetries in nuclear physics has an
established history of key discoveries, enabled by either the
introduction of new technologies or the increase in energy and
luminosity of accelerator facilities. While the EIC is primarily being
proposed for exploring new frontiers in QCD, it offers a unique new
combination of experimental probes potentially interesting to the
investigations in Fundamental Symmetries. For example, the
availability of polarized beams at high energy and high luminosity,
combined with a state-of-the-art hermetic detector, could extend
Standard Model tests of the running of the weak-coupling constant far
beyond the reach of the JLab12 parity violation program, namely toward
the Z-pole scale previously probed at LEP and SLC.
\end{multicols}

\section{{The EIC and its Realization}}

Two independent designs for a future EIC have evolved in the United
States. Both use the existing infrastructure and facilities available
to the US nuclear science community. At Brookhaven National Laboratory
(BNL), the eRHIC design (Figure~\ref{fig:machines}, top) utilizes a new
electron beam facility based on an Energy Recovery LINAC (ERL) to be
built inside the RHIC tunnel to collide with RHICÕs existing
high-energy polarized proton and nuclear beams. At Jefferson
Laboratory (JLab), the Medium energy Electron Ion Collider (MEIC) design
(Figure~\ref{fig:machines}, bottom) employs a new electron and ion
collider ring complex together with the 12 GeV upgraded CEBAF, now
under construction, to achieve similar collision parameters.

\medskip
The EIC machine designs are aimed at achieving
\begin{itemize}
\item Highly polarized ($\sim$ 70\%) electron and nucleon beams
\item Ion beams from deuteron to the heaviest nuclei (uranium or lead)
\item Variable center of mass energies from $\sim 20\, - \sim$100 GeV,
upgradable to $\sim$140 GeV
\item High collision luminosity $\sim$10$^{33-34}$ cm$^{-2}$s$^{-1}$
\item Possibilities of having more than one interaction region
\end{itemize}

\newpage
\noindent
\begin{figurehere}
\begin{center} \hskip 0.2in
\includegraphics[width=0.94\textwidth
                          ]{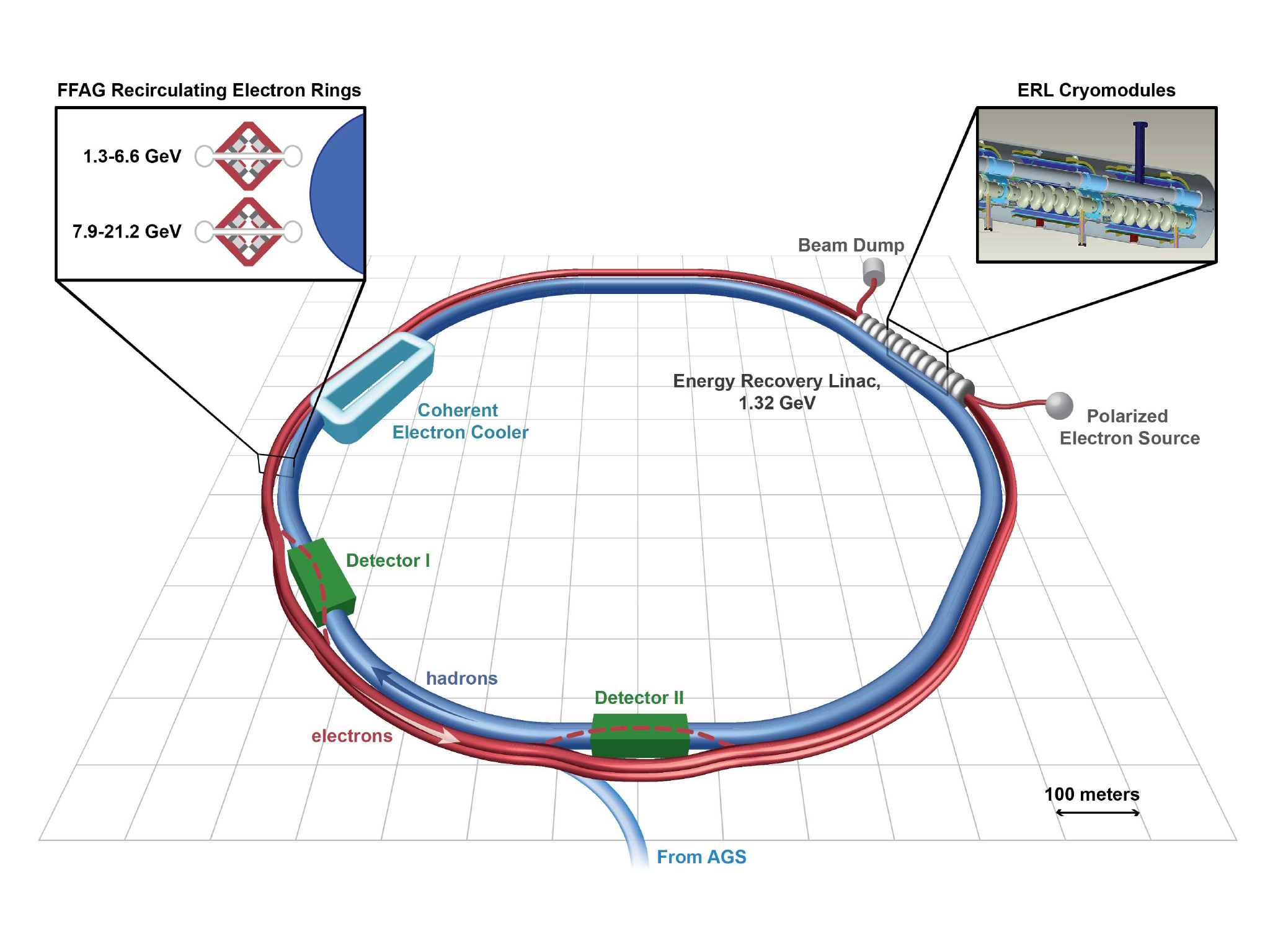}
\vskip -0.28in
\includegraphics[width=0.73\textwidth
                          ]{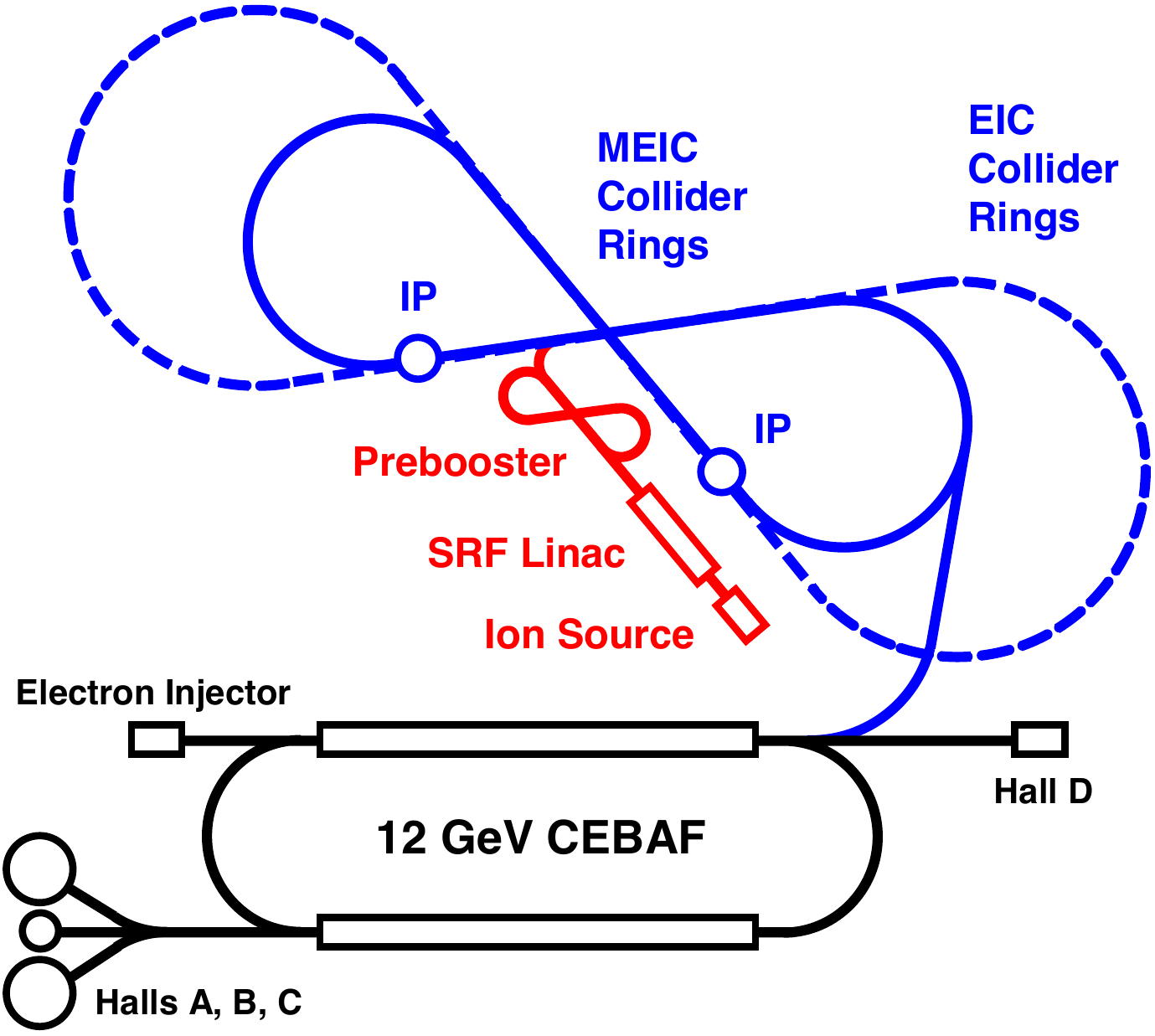}
\end{center} \vskip -0.1in
\caption{\label{fig:machines}{ {\bf Top:} The schematic of eRHIC at
BNL, which would require construction of an electron beam facility
(red) to collide with the RHIC blue beam at up to three interaction
points.  {\bf Botton:} The schematic of MEIC at JLab, which would 
require construction of an ion linac (red), and an electron-ion collider 
ring (blue) with at least two interaction points, around the 12 GeV 
CEBAF.
} }
\end{figurehere}  

\begin{multicols}{2}
The EIC requirements will push accelerator designs to the limits of
current technology, and will therefore need significant R\&D.  Cooling
of the hadron beam is essential to attain the luminosities demanded by
the science. The development of coherent electron cooling is now underway
at BNL, while the JLab design is based on conventional electron
cooling techniques, but proposes to extend them to significantly
higher energy and to use bunched electron beams for the first time.

An energy recovery linac at the highest possible energy and intensity
are key to the realization of eRHIC at BNL, and this technology is
also important for electron cooling in MEIC at JLab.  The eRHIC design
at BNL also requires a high intensity polarized electron source that
would be an order of magnitude higher in intensity than the current
state of the art, while the MEIC design at JLab will utilize a novel
figure-8 storage ring design for both electrons and ions.

The physics-driven requirements on the EIC accelerator parameters and
extreme demands on the kinematic coverage for measurements makes
integration of the detector into the accelerator a particularly
challenging feature of the design.  Lessons learned from past
experience at HERA have been considered while designing the EIC
interaction region. Driven by the demand for high precision on
particle detection and identification of final state particles in both
$e$+$p$ and $e$+A programs, modern particle detector systems will be at the
heart of the EIC. In order to keep the detector costs manageable, R\&D
efforts are under way on various novel ideas for: compact (fiber
sampling and crystal) calorimetry, tracking (NaI coated GEMs, GEM size
and geometries), particle identification (compact DIRC, dual radiator
RICH and novel TPC) and high radiation tolerance for
electronics. Meeting these R\&D challenges will keep the U.S. nuclear
science community at the cutting edge in both accelerator and detector
technology.
\end{multicols}

\bigskip
\section{Physics Deliverables of the EIC}

Both realizations of the EIC, the eRHIC and the MEIC, are expected to evolve
over time from $\sim 20-100$~GeV in center-of-mass-energy to $\sim 140$~GeV
with polarized nucleon and electron beams, a wide range of heavy ion beams for nuclear DIS,
and a luminosity for electron+proton collisions approaching 10$^{34}$ cm$^{-2}$s$^{-1}$.
With such a facility, the EIC physics program would
have an excellent start toward addressing the following fundamental
questions with key measurements:
\begin{itemize}
\item {\bf Proton spin:} Within just a few months of operation, the
EIC would be able to deliver decisive measurements, which no other
facility in the world could achieve, on how much the intrinsic spin of
quarks and gluons contribute to the proton spin as shown in
Fig.~\ref{fig:kin_helicity} (Right).
\item {\bf The motion of quarks and gluons in the proton:}
Semi-inclusive measurements with polarized beams would enable us to
selectively probe with precision the correlation between the spin of a
fast moving proton and the confined transverse motion of both quarks
and gluons within. Images in momentum space as shown in
Fig.~\ref{fig:ptomography} are simply unattainable without the
polarized electron and proton beams of the proposed EIC.
\item {\bf The tomographic images of the proton:} By measuring
exclusive processes, the EIC, with its unprecedented luminosity and
detector coverage, would create detailed images of the proton gluonic
matter distribution, as shown in Fig.~\ref{fig:gimaging}, as well as
images of sea quarks. Such measurements would reveal aspects of proton
structure that are intimately connected with QCD dynamics at large
distances.
\item {\bf QCD matter at an extreme gluon density:} By measuring the
diffractive cross-sections together with the total DIS cross-sections
in electron+proton and electron+ nucleus collisions as shown in
Fig.~\ref{fig:sat_diff}, the EIC would provide the first unambiguous
evidence for the novel QCD matter of saturated gluons.  The EIC is
poised to explore with precision the new field of the collective
dynamics of saturated gluons at high energies.
\item {\bf Quark hadronization:} By measuring pion and D$^0$ meson
production in both electron+proton and electron+nucleus collisions,
the EIC would provide the first measurement of the quark mass
dependence of the hadronization along with the response of nuclear
matter to a fast moving quark.
\end{itemize}

The Relativistic Heavy Ion Collider (RHIC) at BNL has revolutionized
our understanding of hot and dense QCD matter through its discovery of
the strongly-coupled quark gluon plasma that existed a few
microseconds after the birth of the universe. Unprecedented studies of
the nucleon and nuclear structure -- including the nucleon spin, and
the nucleon's tomographic images in the valence quark region -- have
been and will be possible with the high luminosity fixed target
experiments at Jefferson Laboratory using the 6 and 12 GeV CEBAF,
respectively. The EIC promises to propel both programs to the next QCD
frontier, by unraveling the three-dimensional sea quark and gluon
structure of visible matter. Furthermore, the EIC will probe the existence
of a universal state of saturated gluon matter and has the capability
to explore it in detail.  The EIC will thus enable the US to continue
its leadership role in nuclear science research through the quest for
understanding the unique gluon-dominated nature of visible matter in
the universe.

%% file: files_tex/ep-intro.tex
\section{Introduction}
\label{sec:parton-intro}

Among the most intriguing aspects of quantum chromodynamics (QCD) is
the relation between its basic degrees of freedom, quarks and gluons,
and the observable physical states, i.e.\ hadrons such as the proton.
Parton distributions are the most prominent quantities that describe
this relationship.  They are relevant in connection with several key
issues of the strong interaction:

\begin{itemize}
\item \textbf{What is the dynamical origin of sea quarks and gluons inside the proton?}

  Parton distributions describe the proton as a system of many quarks,
  anti-quarks and gluons.  At high resolution, the presence of partons with
  small momentum fraction $x$ can largely be understood as the result of
  parton radiation, similar to the appearance of electrons, positrons and
  photons generated from a single electron in an electromagnetic cascade.
  This parton radiation can be computed using perturbation theory in the
  small coupling ($\alpha_s$) limit.  However, comparison with experimental data
  shows that even at low
  resolution, the proton does not \emph{only} consist of quarks carrying
  about a third of the proton momentum, as one might naively expect from
  the familiar constituent quark picture, where the proton is made up of
  two $u$ quarks and one $d$ quark.  Instead, even at low resolution, the
  proton contains both gluons and low-momentum quarks and anti-quarks
  (termed sea quarks)~\cite{Gluck:1998xa,JimenezDelgado:2008hf} .  These must be generated by dynamics beyond the
  reach of perturbation theory, and their origin remains to be understood.
  Note that calculations in lattice QCD tell us that even the proton mass
  is largely due to the binding energy of the gluons that keep the quarks
  together.
\item \textbf{How does the proton spin originate at the microscopic level?}

  The fact that quarks have spin $1/2$ and gluons spin $1$ plays an
  essential role in their interactions among themselves.  An outstanding
  question is how the total spin of the proton is built up from the
  polarization and the orbital angular momentum of quarks, anti-quarks and
  gluons.  Starting with the seminal results of the EMC experiment
  \cite{Ashman:1987hv}, a series of increasingly precise measurements in
  the last decades revealed that the polarization of the quarks and
  anti-quarks combined, only provides about 30\% of the nucleon
  spin. Present lattice calculations \cite{Hagler:2009ni} suggest that the
  missing 70\% is \emph{not} provided by the orbital angular momentum of
  quarks alone, and recent results from RHIC point
  towards a significant contribution from the polarization of gluons~\cite{RHICwp:2012}.
  This highlights again the importance of gluons for the basic properties
  of the nucleon.
\item \textbf{How is hadron structure influenced by chiral symmetry and
    its breaking?}

  QCD has an approximate chiral symmetry, which is dynamically broken.  As
  a consequence the strong interaction generates Goldstone bosons --- the
  pions --- whose mass is remarkably small compared with that of other
  hadrons.  These almost massless bound-states propagate over distances
  significantly larger 
  than the typical hadronic scale.  They are critical in
  generating the force that binds neutrons and protons within nuclei, but
  also appear to greatly influence the properties of isolated nucleons.
  No understanding of matter is complete without a detailed explanation of
  the role of pions.  It is thus crucial to expose the role played by
  pions in nucleon structure.
\item \textbf{How does confinement manifest itself in the structure of
    hadrons?}

  At distances around 1 femtometer ($\fm$) the strong force becomes so
  strong that
  quarks and gluons are confined in hadrons and cannot exist as free
  particles.  As a consequence, the structure of the proton differs
  profoundly from that of weakly bound systems such as atoms (whose
  overall size is proportional to the inverse electron mass).  The spatial
  distribution of partons in the proton and their distribution in
  transverse momentum is characterized by scales of the order of a $\fm$
  or a few hundred $\mev$, which are similar to the confinement scale and
  very different from the $u$ and $d$ quark masses.  Experimental mapping
  and theoretical computation of these distributions should 
  further our understanding of confinement.
\end{itemize}
The EIC will be unique in mapping out the quark-gluon structure of the
proton in several ways that will take our knowledge to a new level.
Specifically, the EIC will enable us to investigate:
\begin{itemize}
\item the distribution of sea quarks and gluons in momentum and in
  position space, in order to better understand their dynamical interplay,
\item their polarization and their orbital angular momentum, the latter
  being closely connected with their transverse position and transverse
  motion since it is a cross product ($\vec{L} = \vec{r} \times \vec{p}$),
\item correlations between polarization and the distribution of partons in
  momentum or position space, which may be regarded as the QCD analog of
  spin-orbit correlations in atomic or nuclear physics,
\item the change of distributions when going from small to large $x$, to
  compare the characteristics of sea and valence quarks and to understand
  their relation to each other,
\item the dependence of the above characteristics on the quark flavor.
  This is of particular interest when comparing distributions, i.e.\
  $\bar{u}$ with $\bar{d}$, $\bar{s}$ with $(\bar{u} + \bar{d})/2$ or $s$
  with $\bar{s}$.  Significant differences between those distributions are
  a direct imprint of non-perturbative dynamics because perturbative
  parton radiation is not able to generate them.  This imparts special
  interest to the polarization carried by sea quarks of different flavors,
  above and beyond its contribution to the overall spin of the proton.
\end{itemize}
To quantify these properties and to connect them with experimental data,
we have a powerful formalism at our disposal, which has seen significant
progress in the last one and a half decades.  Parton distributions come in
different varieties, with an increasing level of complexity:
\begin{itemize}
\item The familiar parton
  distribution functions (PDFs) $\color{Magenta}{f(x)}$ give the number
  density of partons with longitudinal momentum fraction $x$ in a fast-moving
  proton, where the longitudinal direction is given by the proton
  momentum.  They are measured in inclusive or semi-inclusive processes,
  the first and foremost being inclusive deep inelastic lepton-proton
  scattering (DIS).  PDFs form the backbone of our knowledge about hadron
  structure, and for most cases their determination is an enterprise at
  the precision frontier.

  A powerful tool for disentangling the distributions for different quark
  and anti-quark flavors is semi-inclusive DIS (SIDIS), where a specified
  hadron is detected in the final state.  SIDIS involves fragmentation
  functions, which describe another key phenomenon of the strong
  interaction, namely the fragmentation of a parton into a hadron.
  Fragmentation functions and parton distributions provide two different
  settings to investigate the consequences of confinement.  The
  possibilities to study the fragmentation process in nuclei will be
  discussed in Sec.~\ref{sec:nuclei}.
\item Transverse-momentum dependent parton densities (TMDs)
  $\color{darkgreen}{f(x,\boldsymbol{k}_T)}$ describe the joint distribution of partons in
  their longitudinal momentum fraction $x$ and their momentum
  $\boldsymbol{k}_T$ transverse to the proton direction.  To measure TMDs
  requires more detailed information about the kinematics of a scattering
  process.  In the  appropriate kinematics of 
  SIDIS, the transverse momentum of the detected final-state
  hadron can be computed from a $k_T$ dependent parton density
  and from a $k_T$ dependent fragmentation function, which
  describes the transverse momentum transferred during the hadronization
  process.
\item Generalized parton distributions (GPDs) $\color{SkyBlue}{H(x,\xi,t)}$ 
  appear in
  exclusive scattering processes such as deeply virtual Compton scattering
  (DVCS: $\gamma^*p\to \gamma p$), in which all final-state particles are
  detected.  They depend on two longitudinal momentum fractions $x$ and
  $\xi$ (see the Sidebar on page~\pageref{sdbar:GPD})  
  and on the squared momentum
  transfer $t$ to the proton or equivalently, on its transverse component
  $\boldsymbol{\Delta}_T$.  Setting $\xi=0$ and performing a Fourier
  transform with respect to $\boldsymbol{\Delta}_T$ one obtains an impact
  parameter distribution $\color{Blue}{f(x,\boldsymbol{b}_T)}$, which 
  describes the joint
  distribution of partons in their longitudinal momentum and their
  transverse position $\boldsymbol{b}_T$ inside the proton, as sketched in
  figure~\ref{fig:bspace-density}.

  Integrating the generalized quark distribution $\color{RoyalBlue}{H(x,0,t)}$ 
  over $x$ and taking an appropriate sum over quark flavors, one obtains the
  electromagnetic Dirac form factor $F_1(t)$ of the proton.  This provides
  a connection between parton distributions and form factors, which have
  played a major role in exploring the proton structure ever since the
  seminal experiment of Hofstadter.  More generally, the integral $\int
  dx\, x^{n-1} \color{SkyBlue}{H(x,\xi,t)}$ gives generalized form factors 
  for a large set
  of local operators that cannot be directly measured but can be
  computed on the lattice.  This provides a connection with one of the
  main tools for calculations in the non-perturbative sector of QCD.
\end{itemize}

\begin{multicols}{2}
Indeed, measurements at the EIC and lattice calculations will have a high
degree of complementarity.  For some quantities, notably the $x$ moments
of unpolarized and polarized quark distributions, a precise determination
will be possible both in experiment and on the lattice.  Using this to
validate the methods used in lattice calculations, one will gain
confidence in computing quantities whose experimental determination is
very hard, such as generalized form factors.  Furthermore, one can gain
insight into the underlying dynamics by computing the same
quantities with values of the quark masses that are not realized in
nature, so as to reveal the importance of these masses for specific
properties of the nucleon.  On the other hand, there are many aspects of
hadron structure beyond the reach of lattice computations, in particular,
the distribution and polarization of quarks and gluons at small $x$, for
which collider measurements are our only source of information.

\begin{figurehere}
\begin{center}
\includegraphics[width=0.40\textwidth]{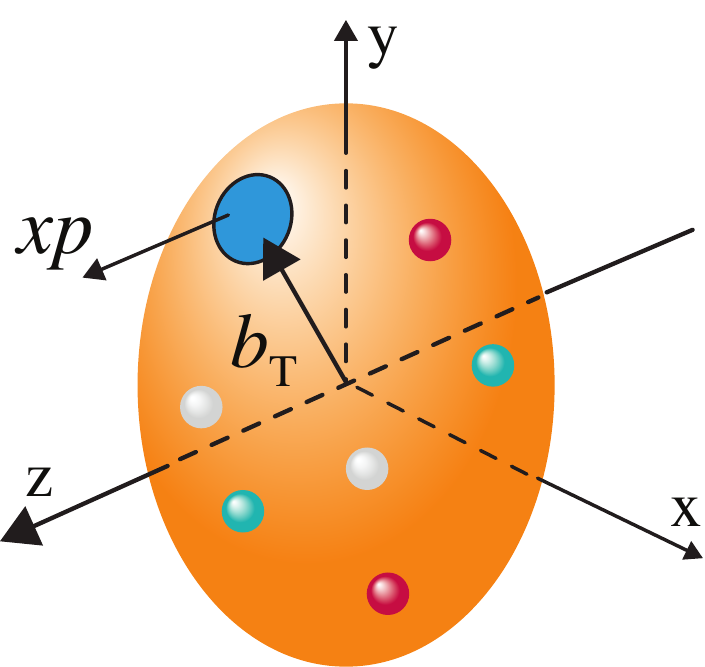}
\end{center}
\caption{\label{fig:bspace-density} Schematic view of a parton with longitudinal 
momentum fraction $x$ and transverse position $\boldsymbol{b}_T$ in the proton.}
\end{figurehere}
\vspace*{0.25cm}

Both impact parameter distributions $\color{Blue}{f(x,\boldsymbol{b}_T)}$
and transverse-momentum distributions
$\color{darkgreen}{f(x,\boldsymbol{k}_T)}$ describe proton structure in
three dimensions, or more accurately in $2+1$ dimensions (two transverse
dimensions in either configuration or momentum space, along with one
longitudinal dimension in momentum space).  Note that in a fast-moving
proton, the transverse variables play very different roles than the
longitudinal momentum.

\begin{figure}
\begin{center}
\includegraphics[width=\textwidth,clip=true]{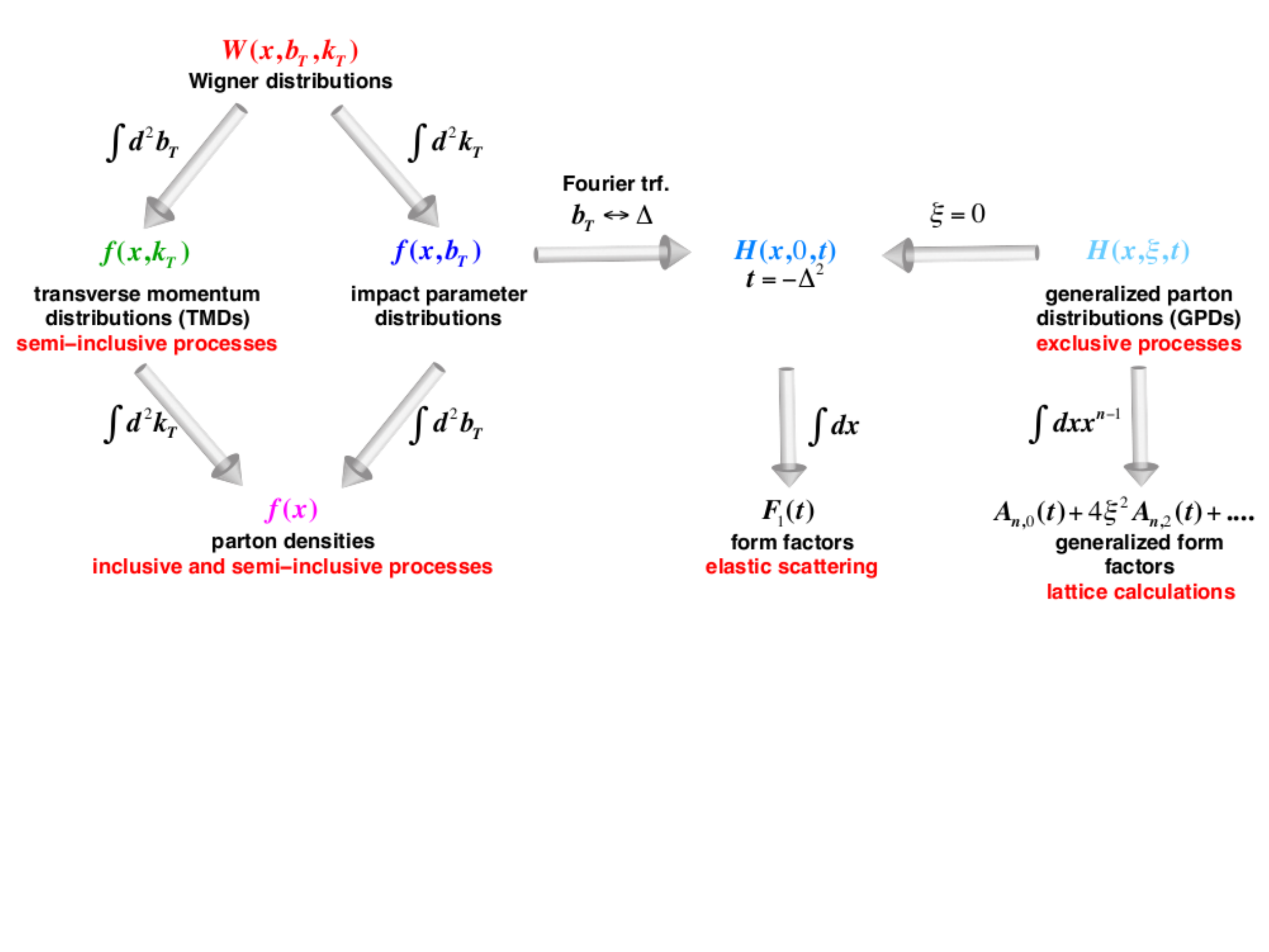}
\end{center}
\vskip -1.8in
\caption{\label{fig:Wigner-connection} Connections between different
  quantities describing the distribution of partons inside the proton.
  The functions given here are for unpolarized partons in an unpolarized
  proton; analogous relations hold for polarized quantities.}
\end{figure}

It is important to realize that $\color{Blue}{f(x,\boldsymbol{b}_T)}$ and
$\color{darkgreen}{f(x,\boldsymbol{k}_T)}$ are \emph{not} related to each
other by a Fourier
transform (nevertheless it is common to denote both functions by the same
symbol $f$).  Instead, $\color{Blue}{f(x, \boldsymbol{b}_T)}$ and 
$\color{darkgreen}{f(x, \boldsymbol{k}_T)}$
give complementary information about partons, and both types of quantities
can be thought of as descendants of Wigner distributions
$\color{red}{W(x, \boldsymbol{b}_T, \boldsymbol{k}_T)}$ 
\cite{Belitsky:2003nz}, which are used extensively in
other branches of physics \cite{Hillery:1983ms}.  Although there is no
known way to measure Wigner distributions for quarks and gluons, they provide
a unifying theoretical framework for the different aspects of hadron
structure we have discussed.  Figure~\ref{fig:Wigner-connection} shows the
connection between these different aspects and the experimental
possibilities to explore them.

All parton distributions depend on a scale which specifies the resolution
at which partons are resolved, and which in a given scattering process is
provided by a large momentum transfer.  For many processes in $e$+$p$
collisions, the relevant hard scale is $Q^2$ 
(see the Sidebar on page~\pageref{sdbar:DIS}).
The evolution equations that describe the scale dependence of parton
distributions provide an essential tool, both for the validation of the
theory and for the extraction of parton distributions from cross section
data.  They also allow one to convert the distributions seen at high
resolution to lower resolution scales, where contact can be made with
non-perturbative descriptions of the proton.

An essential property of any particle is its spin, and parton
distributions can depend on the polarization of both the parton and the
parent proton.  The spin structure is particularly rich for TMDs and GPDs
because they single out a direction in the transverse plane, thus opening
the way for studying correlations between spin and $\boldsymbol{k}_T$ or
$\boldsymbol{b}_T$.  Information about transverse degrees of freedom is
essential to access orbital angular momentum, and specific TMDs and GPDs
quantify the orbital angular momentum carried by partons in different
ways.

The theoretical framework we have sketched is valid over a wide range of
momentum fractions $x$, connecting in particular the region of valence
quarks with the one of gluons and the quark sea.  While the present
chapter is focused on the nucleon, the concept of parton distributions is
well adapted to study the dynamics of partons in nuclei, as we will see in
Sec.~\ref{sec:nuclei}.  For the regime of small $x$, which is probed in
collisions at the highest energies, a different theoretical description is
at our disposal.  Rather than parton distributions, a basic quantity in
this approach is the amplitude for the scattering of a color dipole on a
proton or a nucleus.  The joint distribution of gluons in $x$ and in
$\boldsymbol{k}_T$ or $\boldsymbol{b}_T$ can be derived from this dipole
amplitude.  This high-energy approach is essential for addressing the
physics of high parton densities and of parton saturation, as discussed in
Sec. \ref{HGDN}.  On the other hand, in a regime of moderate
$x$, around $10^{-3}$ for the proton and higher for heavy nuclei, the
theoretical descriptions based on either parton distributions or color
dipoles are both applicable and can be related to each other.  This will
provide us with valuable flexibility for interpreting data in a wide
kinematic regime.

The following sections highlight the physics opportunities in measuring
PDFs, TMDs and GPDs to map out the quark-gluon structure of the proton at
the EIC.  An essential feature throughout will be the broad reach of the EIC in
the kinematic plane of the Bjorken variable $x$ 
(see the Sidebar on page~\pageref{sdbar:DIS})
and the invariant momentum transfer $Q^2$ to the electron.
While $x$ determines the momentum fraction of the partons probed,
$Q^2$ specifies the scale at which the partons are resolved.  Wide
coverage in $x$ is hence essential for going from the valence quark
regime deep into the region of gluons and sea quarks, whereas a large
lever arm in $Q^2$ is the key for unraveling the information contained in
the scale evolution of parton distributions.
\end{multicols}  

%% file: files_tex/sidebarDIS.tex

\newpage
\pagecolor{LightYellow}

\phantomsection
\label{sdbar:DIS}

\vspace*{-1.00cm}
\begin{center}
{\textbf {\textit {\textcolor{blue}{\Large Deep Inelastic Scattering: Kinematics}}}}
\line(1,0){435}
\end{center}

\begin{multicols}{2}
\vspace*{-1.50cm}
\begin{figurehere}
\begin{center}
\includegraphics[width=0.50\textwidth]{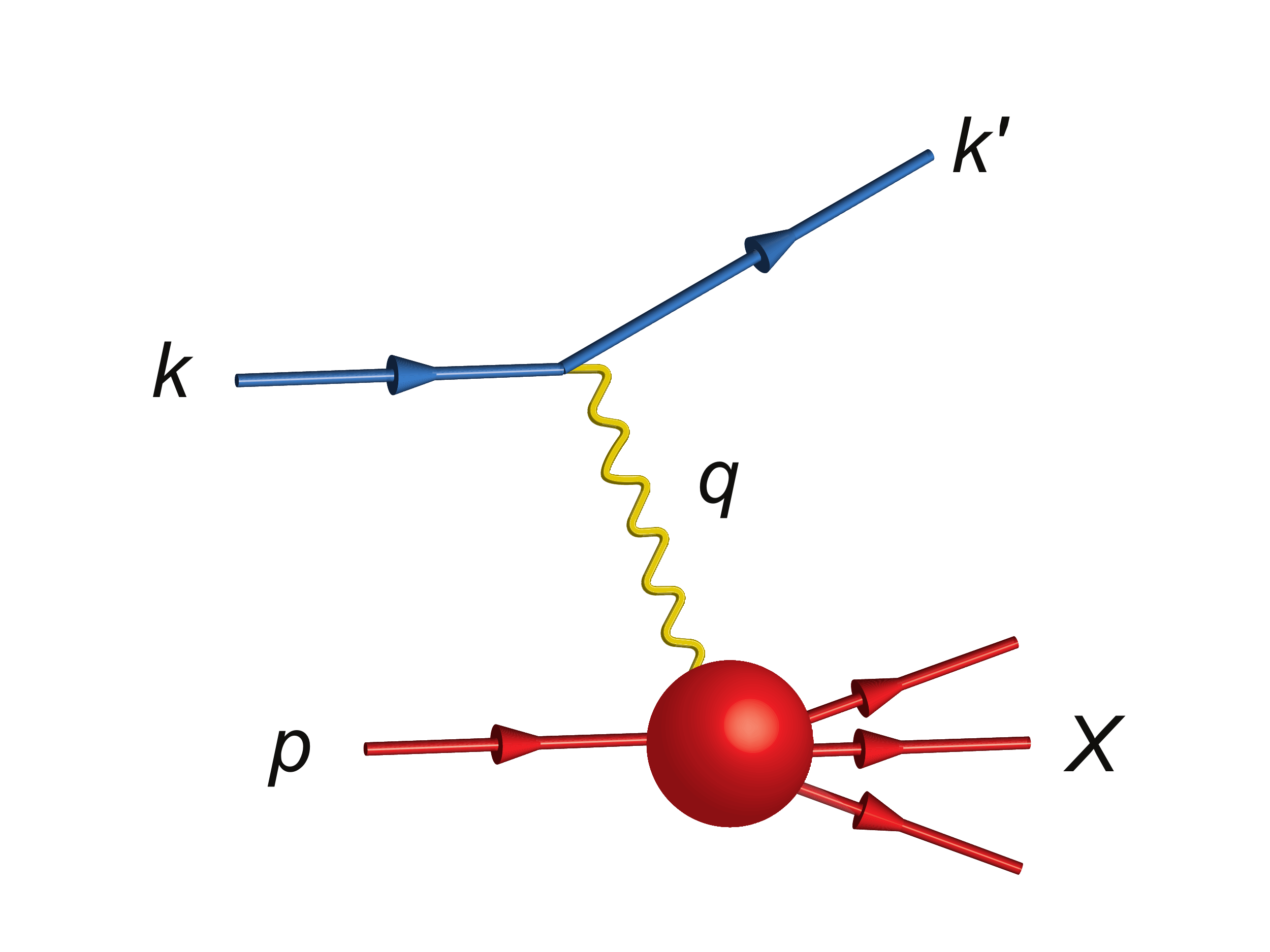}
\end{center}
\vspace*{-0.90cm}
\caption{\label{sidebar:fig:sidebarDISGraph}
A schematic diagram of the Deep Inelastic Scattering (DIS) process. }
\end{figurehere}
          
\vskip 0.15in
          
\noindent
{\bf Deep Inelastic Scattering:}\\ 
$\mathbf{e + p \longrightarrow e + X}$,
proceeds through the exchange of a virtual photon between the electron and the proton.
The kinematic description remains the same for the exchange of a $Z$ or $W$ boson, 
which becomes important at high momentum transfer.\\[0.1cm]
Depending on the physics situation, the process is discussed in different reference 
frames:
     \begin{itemize}
        \item the {\bf collider frame}, 
              where a proton with energy $E_p$ and an electron with energy 
              $E_e$ collide head-on
        \item the {\bf rest frame} of the hadronic system $X$, i.e. the 
              center-of-mass of the $\gamma^* p$ collision
        \item the {\bf rest frame} of the proton
     \end{itemize}
\noindent {\bf {Kinematic Variables:}} \\
In the following, we neglect the proton mass, $M$, where appropriate and the electron 
mass throughout. \\[0.1cm]
$\large\boldsymbol{k, k'}$ are the four-momenta of the incoming and outgoing lepton \\
$\large\boldsymbol{p}$ is the four-momentum of a nucleon\\[0.25cm]
{\bf Lorentz invariants:}
\begin{itemize}
\item the squared $e$+$p$ collision energy $\boldsymbol{s} = (p + k)^2 = 4 E_p E_e$
\item the squared momentum transfer to the lepton 
      $\boldsymbol{Q^2} = - q^2 = -(k - k^{\prime})^2$,
      equal to the virtuality of the exchanged photon. Large values of $Q^2$ 
      provide a hard scale to the process, which allows one to resolve quarks 
      and gluons in the proton.
\item the Bjorken variable $\boldsymbol{x_B} = Q^2/(2 p \cdot q)$, often simply denoted 
      by $\boldsymbol{x}$. It determines the momentum fraction of the parton on which the 
      photon scatters. Note that $0 < x < 1$ for $e$+$p$-collisions.
\item the inelasticity $\boldsymbol{y} = (q \cdot p) /(k \cdot p)$ is limited to 
      values $0 < y < 1$
      and determines in particular the polarization of the virtual photon.
      In the collider frame, the energy of the scattered electron is 
      $E'_e = E_e (1 - y) + Q^2 /(4 E_e)$;
      detection of the scattered electron thus typically requires a cut on $y < y_{max}$.
\end{itemize}
These invariants are related by $Q^2 = x y s$. The available phase space is often 
represented in the plane of $x$ and $Q^2$. For a given $e$+$p$ collision energy, lines 
of constant $y$ are then lines with a slope of 45 degrees in a double logarithmic 
$x-Q^2$-plot.\\[0.1cm]
{\bf Two more important variables: }\\[0.1cm]
$\large\boldsymbol{W^2}$ $= (p + q)^2 = Q^2 (1 - 1/x)$ is the squared invariant mass 
           of the produced hadronic system $X$.\\
           DIS is characterized by the Bjorken limit, 
           where $Q^2$ and $W^2$ become large at a fixed value of $x$.
           Note: for a given $Q^2$, small $x$ corresponds to a high $\gamma^* p$ 
           collision energy.\\[0.25cm]
$\large\boldsymbol{\nu}$ $= q \cdot p /M = y s /(2 M)$ is the energy lost by the lepton 
            (i.e. the energy carried away by the virtual photon) in the proton 
            rest frame.\\[0.25cm]
For scattering on a nucleus of atomic number $A$, replace the proton momentum 
$p$ by $P/A$ in the definitions, where $P$ is the momentum of the nucleus.  
Note that for the Bjorken variable one then has $0 < x <A$.

\end{multicols}

%% file: files_tex/sidebarCrossSection.tex

\newpage
\pagecolor{LightYellow}

\phantomsection
\label{sdbar:crosssection}

\vspace*{-1.00cm}
\begin{center}
{\textbf {\textit {\textcolor{blue}{\Large Deep Inelastic Scattering: Structure Functions}}}}
  \line(1,0){435}
\end{center}

\noindent The cross-sections for neutral-current deep inelastic scattering 
($e+N \longrightarrow e^{\prime}+X$) on unpolarized 
nucleons and nuclei can be written in the one photon exchange approximation 
(neglecting electroweak effects) in terms of two structure functions $F_2$ and $F_L$:
\begin{equation}
    \frac{d^2 \sigma}{dx\, dQ^2} = 
    \frac{4 \pi \alpha^2}{x Q^4} 
    \left[  \left( 1 - y + \frac{y^2}{2} \right) F_2(x, Q^2) - \frac{y^2}{2} F_L(x, Q^2) \right]\, .
   \label{eq:csDIS}
\end{equation}
For practical purposes, often the reduced cross-section, $\sigma_r$, is used:
\begin{equation}
    \sigma_r =  \left( \frac{d^2 \sigma}{dx\, dQ^2}\right)\ \frac{x Q^4}{2 \pi \alpha^2 [1 + (1-y)^2]}
    = F_2(x, Q^2) - \frac{y^2}{1+(1-y)^2} F_L(x, Q^2)\, .
   \label{eq:redcsDIS}
\end{equation}

For longitudinally polarized proton and electron beams, the neutral current cross-section 
for deep inelastic scattering can be written in terms of one structure function $g_1$:
\begin{equation}
  \frac{1}{2}\left[
  \frac{d^2\sigma^{\,\begin{subarray}{c}
       \rightarrow \\[-1.3mm] 
       \leftarrow
             \end{subarray}}}{dx\,dQ^2}-
              \frac{d^2\sigma^{\,\begin{subarray}{c}
       \rightarrow \\[-1.3mm] 
       \rightarrow
             \end{subarray}}}{dx\,dQ^2}\right]
               \ \simeq \
    \frac{4\pi\,\alpha^2}{Q^4}
       y \left( 2 - y \right) g_1(x,Q^2)\, ,
   \label{eq:g1sb}
\end{equation}
where the superscript arrows represent electron and proton longitudinal spin directions 
and the terms suppressed by $x^2 M^2/Q^2$ have been neglected.

     Experimentally $F_2$, $F_L$ and $g_1$ can be measured in inclusive scattering,
     i.e., the final hadronic state, $X$, does not need to be analyzed.
     The relevant kinematic variables $x$, $Q^2$, and $y$, can be reconstructed from
     the measured scattered lepton alone.

    $F_2$, $F_L$ and $g_1$ are proportional to the 
    cross-section for the hadronic subprocess $\gamma^*+p \rightarrow X$, which gets
    contributions from the different polarization states of the virtual photon.
    $F_2$ corresponds to the sum over transverse and longitudinal polarizations and the 
    structure function $F_L$ to longitudinal polarization 
    of the virtual photon (i.e., helicity =0). 
    The $g_1$ structure function is sensitive to the transverse polarization of the virtual photon 
    (i.e., helicity =$\pm$1).

    Equation \ref{eq:redcsDIS} shows that the longitudinal structure function 
    $F_L$ starts to contribute to
    the cross-section at larger values of $y$ but is negligible at
    very small values of $y$. To separate the structure functions 
    $F_L$ and $F_2$ for a given $x$ and 
    $Q^2$, one needs to measure the cross-section for different values of 
    $y$ and hence different $e$+$p$ collision energies.

    At large $Q^2$ and to leading order (LO) in the strong coupling $\alpha_s$,
    the structure functions $F_2$ and $g_1$ are respectively sensitive to the sum over 
    unpolarized and longitudinally polarized quark and anti-quark distributions in
    the nucleon, 
    \begin{equation}
     F_2(x,Q^2) = x \sum e_q^2 \left. \left[ q(x,Q^2) + \bar{q}(x,Q^2) \right] \right.,
     \label{eq:f2qpm}
    \end{equation}

    \begin{equation}
      g_1(x,Q^2) = \frac{1}{2} \sum e_q^2 \left. \left[ \Delta q(x,Q^2) + \Delta \bar{q}(x,Q^2) \right] \right.,
      \label{eq:g1qpmsb}
    \end{equation}
    where $e_q$ denotes a quark's electric charge.
    
    At large $Q^2$, one has $F_L = 0$ at LO, i.e., this structure function receives its 
    first contributions at order $\alpha_s$.  It 
    is thus particularly sensitive to gluons, especially at low 
    $x$ where the gluon densitiy is much larger than the densities 
    for quarks and anti-quarks.

\begin{figure}[h!]
\begin{minipage}[b]{0.49\textwidth}
\centering
\includegraphics[width=\textwidth,height=3.52in]{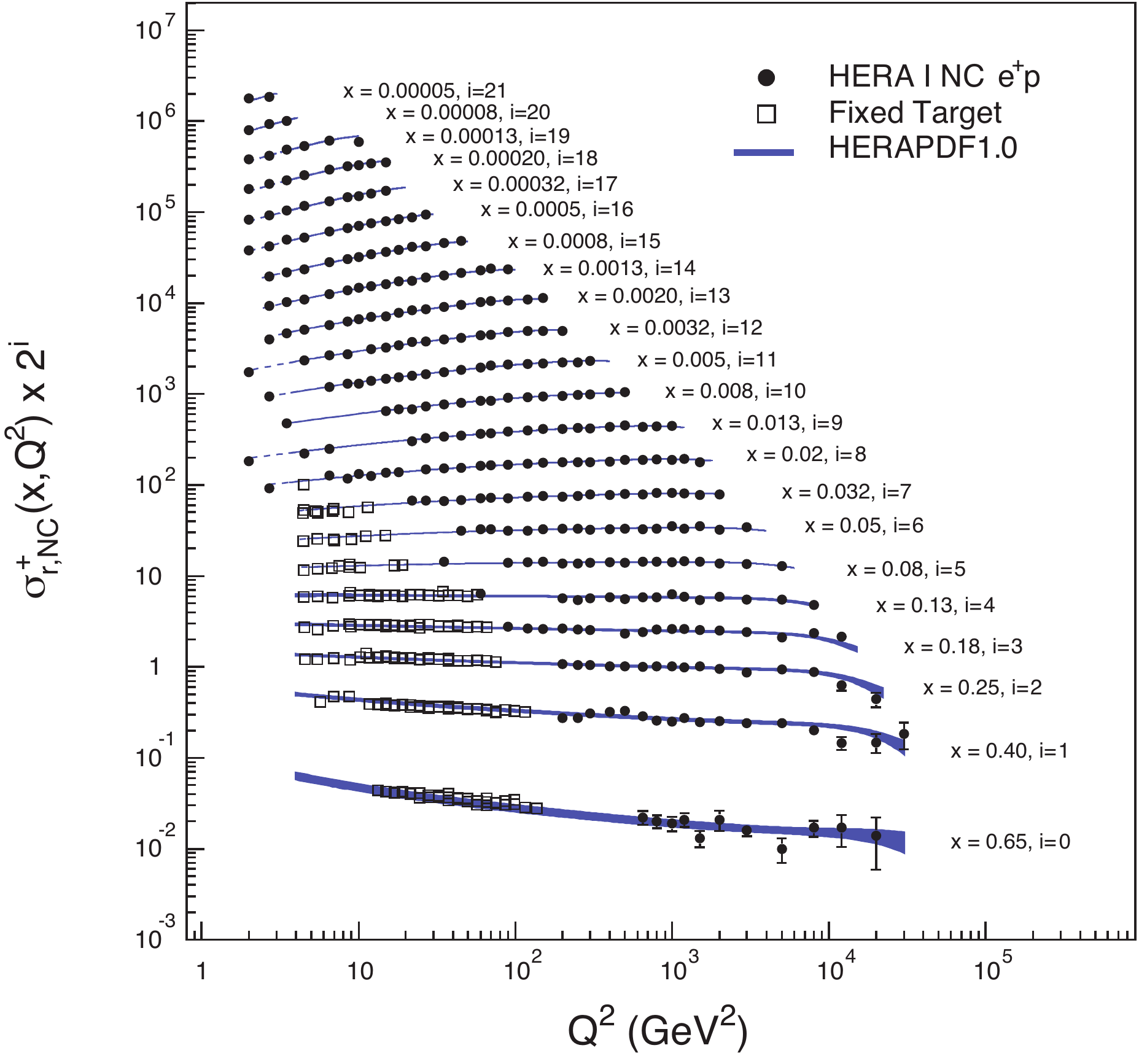}
\end{minipage}
\hskip 0.2in
\begin{minipage}[b]{0.46\textwidth}
\centering
\includegraphics[width=\textwidth,height=3.5in]{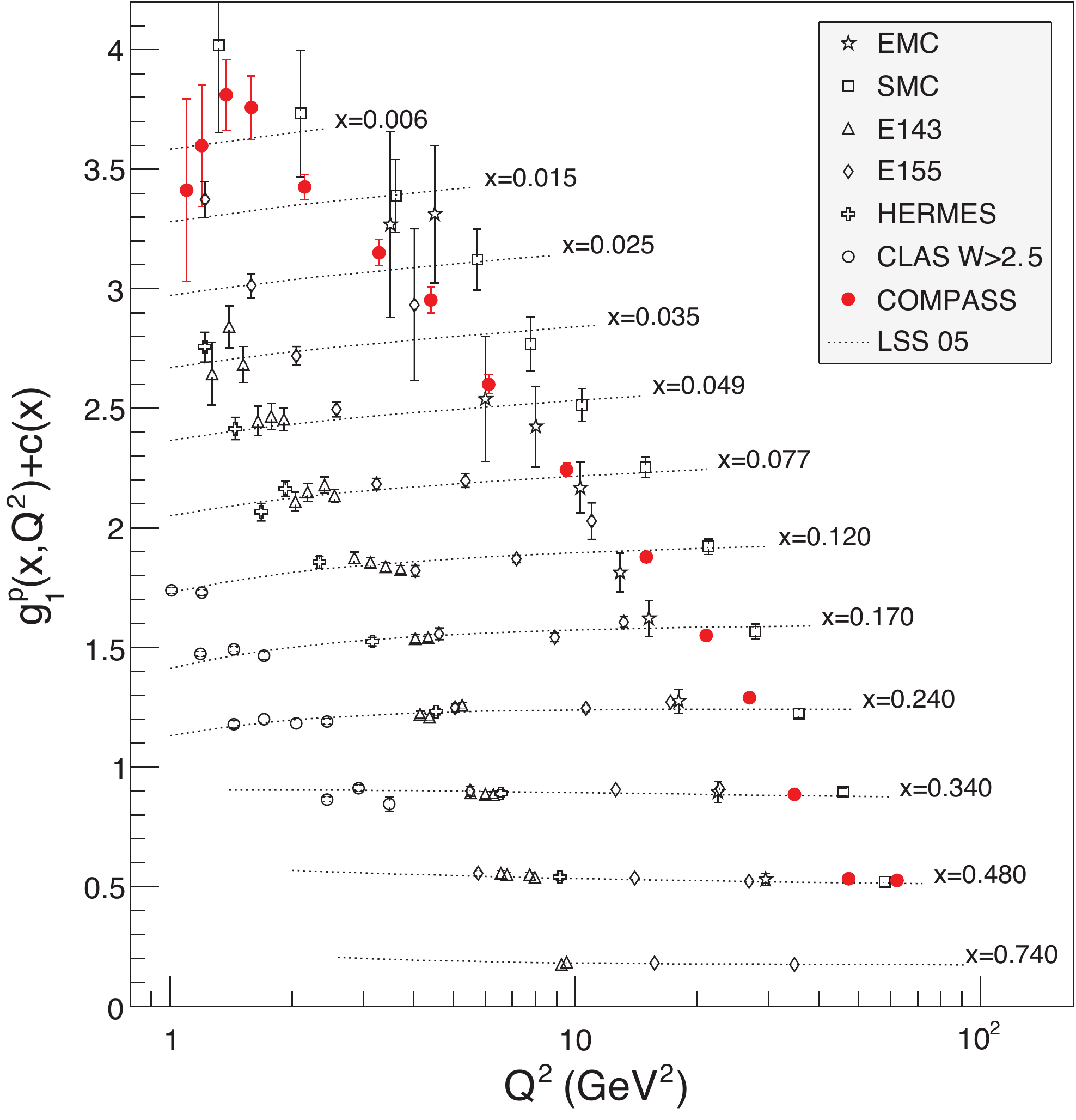}
\end{minipage}
\caption{ \label{sidebar:fig:sidebarReducedCrossSectionPlot}{
{\bf Left:} The $ep$ reduced cross-section as measured at HERA and from  
        fixed-target experiments as a function of $Q^2$ for fixed values of
       $x$. The data are compared to a pQCD fit.
{\bf Right:} The spin-dependent structure function $g_1(x,Q^2)$
       as a function of $x$ and $Q^2$. The world data are compared to a pQCD fit.
}}
\end{figure}

    Figure ~\ref{sidebar:fig:sidebarReducedCrossSectionPlot} (Left) 
    shows the world data of the reduced cross-section, $\sigma_r \propto F_2$, as a
    function of $Q^2$ for a wide range of fixed values of $x$ for scattering 
    on a proton.
    The apparent scaling of the data with $Q^2$ at large $x$ in early
    DIS data from SLAC was termed ``Bjorken scaling" and motivated the
    parton model. Violations of this scaling are predicted by the QCD evolution
    equations for parton densities.  They are especially strong at small $x$.
    We note that our experimental knowledge of $F_L$ is considerably less
    precise than that of $F_2$.

    Figure ~\ref{sidebar:fig:sidebarReducedCrossSectionPlot} (Right) 
    shows the world data of the
    polarised structure function $g_1$ as a function of $Q^2$ for fixed values of $x$ 
    for scattering on a proton. The covered $x-Q^2$ range is significantly smaller than 
    that for the unpolarized measurements, which is due to the fact that there 
    has been no collider with both polarized lepton- and hadron-beams.

\newpage 
\pagecolor{white}

%% file: files_tex/helicity.tex
\section{The Longitudinal Spin of the Nucleon}
\label{sec:helicity}

{\large\it Conveners:\ Ernst Sichtermann and Werner Vogelsang}

\subsection{Introduction}

Deep-inelastic processes, when carried out with longitudinally
polarized nucleons, probe the {\it helicity} parton distribution
functions of the nucleon. For each flavor
$f=u,d,s,\bar{u},\bar{d},\bar{s},g$ these are defined by
\begin{equation} \label{qdef} \Delta f(x,Q^2) \equiv f^+(x,Q^2) \; -
\; f^-(x,Q^2) \; ,
\end{equation} with $f^+$ ($f^-$) denoting the number density of
partons with the same (opposite) helicity as the nucleons, as a
function of the momentum fraction $x$ and the resolution scale $Q$. Similar to
the unpolarized quark and gluon densities, the $Q^2$-dependences of
$\Delta q(x,Q^2)$, $\Delta \bar{q}(x,Q^2)$ and the gluon helicity
distribution $\Delta g(x,Q^2)$ are related by QCD radiative processes
that are
calculable~\cite{Dokshitzer:1977sg,Gribov:1972ri,Altarelli:1977zs,Zijlstra:1993sh,Mertig:1995ny,Vogelsang:1995vh,Vogt:2008yw}.

When integrated over all momentum fractions and appropriately summed
over flavors, the $\Delta f$ distributions give the quark and gluon
spin contributions $S_q,S_g$ to the proton spin which appear in the
fundamental proton helicity sum
rule~\cite{Jaffe:1989jz,Ji:1996ek,Wakamatsu:2010qj,Ji:2012sj}
(see~\cite{Wakamatsu:2012zz} for a brief review and additional
references):
\begin{equation} \label{ssr} \frac{1}{2}=S_q + L_q+S_g+ L_g \;.
\end{equation} 
Here, we have
\begin{eqnarray} 
\label{sigma} 
S_q(Q^2) 
\hskip -0.1in &=& \hskip -0.1in 
\frac{1}{2} \int_0^1
\Delta\Sigma(x,Q^2)dx\,
\equiv\, \frac{1}{2} \int_0^1 \left( \Delta
u+\Delta\bar{u}+ \Delta d+\Delta\bar{d}
+ \Delta s+\Delta\bar{s}\right)(x,Q^2) dx\;,
\nonumber \\[2mm]
S_g(Q^2)
\hskip -0.1in &=& \hskip -0.1in 
\int_0^1 \Delta g(x,Q^2)dx\;,
\end{eqnarray} 
\noindent
where the factor $1/2$ in the first equation is the
spin of each quark and anti-quark. The $\Delta f$ distributions are
thus key ingredients to solving the proton spin problem.

As discussed in the Sidebar on page~\pageref{sdbar:crosssection}, 
experimental access to the $\Delta f$ in lepton-scattering is obtained through the
spin-depen\-dent structure function $g_1(x,Q^2)$, which appears in the
polarization difference of cross sections when the lepton and the
nucleon collide with their spins anti-aligned or aligned:
\begin{equation} \frac{1}{2}\left[
\frac{\mathrm{d}^2\sigma^{\,\begin{subarray}{c} \rightarrow \\[-1.3mm]
\leftarrow
             \end{subarray}}}{\mathrm{d} x\,\mathrm{d} Q^2}-
\frac{\mathrm{d}^2\sigma^{\,\begin{subarray}{c} \rightarrow \\[-1.3mm]
\rightarrow
             \end{subarray}}}{\mathrm{d} x\,\mathrm{d} Q^2}\right] \
\simeq \ \frac{4\pi\,\alpha^2}{Q^4} y \left( 2 - y \right)
g_1(x,Q^2)\, .
   \label{eq:g1}
\end{equation} 
The expression above assumes photon exchange between
the lepton and the nucleon. At high energies, also $W$ or $Z$ exchange
contribute and lead to additional structure functions. These have thus
far not been accessible in polarized deep-inelastic scattering
experiments and would be a unique opportunity at an EIC. We will
briefly address them below.

In leading order in the strong coupling constant, the structure
function $g_1(x,Q^2)$ of the proton can be written as 
(see the Sidebar on page~\pageref{sdbar:crosssection})
\begin{equation} g_1(x,Q^2) = \frac{1}{2} \sum e_q^2 \left. \left[
\Delta q(x,Q^2) + \Delta \bar{q}(x,Q^2) \right] \right.,
  \label{eq:g1qpm}
\end{equation} where $e_q$ denotes a quark's electric charge.  Higher
order expansions contain calculable QCD coefficient
functions~\cite{Dokshitzer:1977sg,Gribov:1972ri,Altarelli:1977zs}.
The structure function $g_1(x,Q^2)$ is thus directly sensitive to the
nucleon spin structure in terms of the combined quark and anti-quark
spin degrees of freedom. The gluon distribution $\Delta g$ enters the
expression for $g_1$ only at higher order in perturbation theory;
however, it drives the scaling violations (i.e. the $Q^2$-dependence)
of $g_1(x,Q^2)$. Deep-inelastic measurements hence can also give
insight into gluon polarization, {\it provided} a large lever arm in
$Q^2$ is available at fixed $x$.

\subsection{Status and Near Term Prospects}
\begin{multicols}{2}
The EMC experiment~\cite{Ashman:1987hv,Ashman:1989ig}, using a longitudinally
polarized muon beam and a stationary target that contained polarized
protons, was the first experiment to explore $g_1(x,Q^2)$ down to
momentum fractions $x$ as low as $0.01$. When extrapolated over
unmeasured $x < 0.01$ and combined with the couplings in leptonic
hyperon decays and the assumption of $\mathrm{SU(3)}$ flavor
symmetry~\cite{Close:1988de,Close:1993mv}, this led to the famous
conclusion that the quark and anti-quark spins constitute only a small
fraction of the proton spin.  In addition, with these assumptions, the
polarization of the strange quark sea in the polarized proton is found
to be negative.  Significant progress has been made since the EMC
observations on the proton's spin
composition.  One main focus has been on measurements with
longitudinally polarized lepton beams scattering off longitudinally
polarized nucleons in stationary targets.  Inclusive data have been
obtained in experiments at
CERN~\cite{Adeva:1998vv,Alexakhin:2006vx,Alekseev:2010hc},
DESY~\cite{Ackerstaff:1997ws,Airapetian:2006vy}, Jefferson
Laboratory~\cite{Zheng:2004ce,Dharmawardane:2006zd}, and
SLAC~\cite{Anthony:1996mw,Abe:1998wq,Abe:1997cx,Anthony:1999rm,Anthony:2000fn}
in scattering off targets with polarized protons and neutrons.  The
kinematic reach and precision of the data on $g_1(x,Q^2)$ so far is
similar to that of the unpolarized structure function $F_2(x,Q^2)$
just prior to the experimental program at the HERA electron-proton
collider (cf. Sidebar on page~\pageref{sdbar:crosssection}).

\begin{figure}[th!]
\begin{center}
\includegraphics[width=0.80\textwidth]{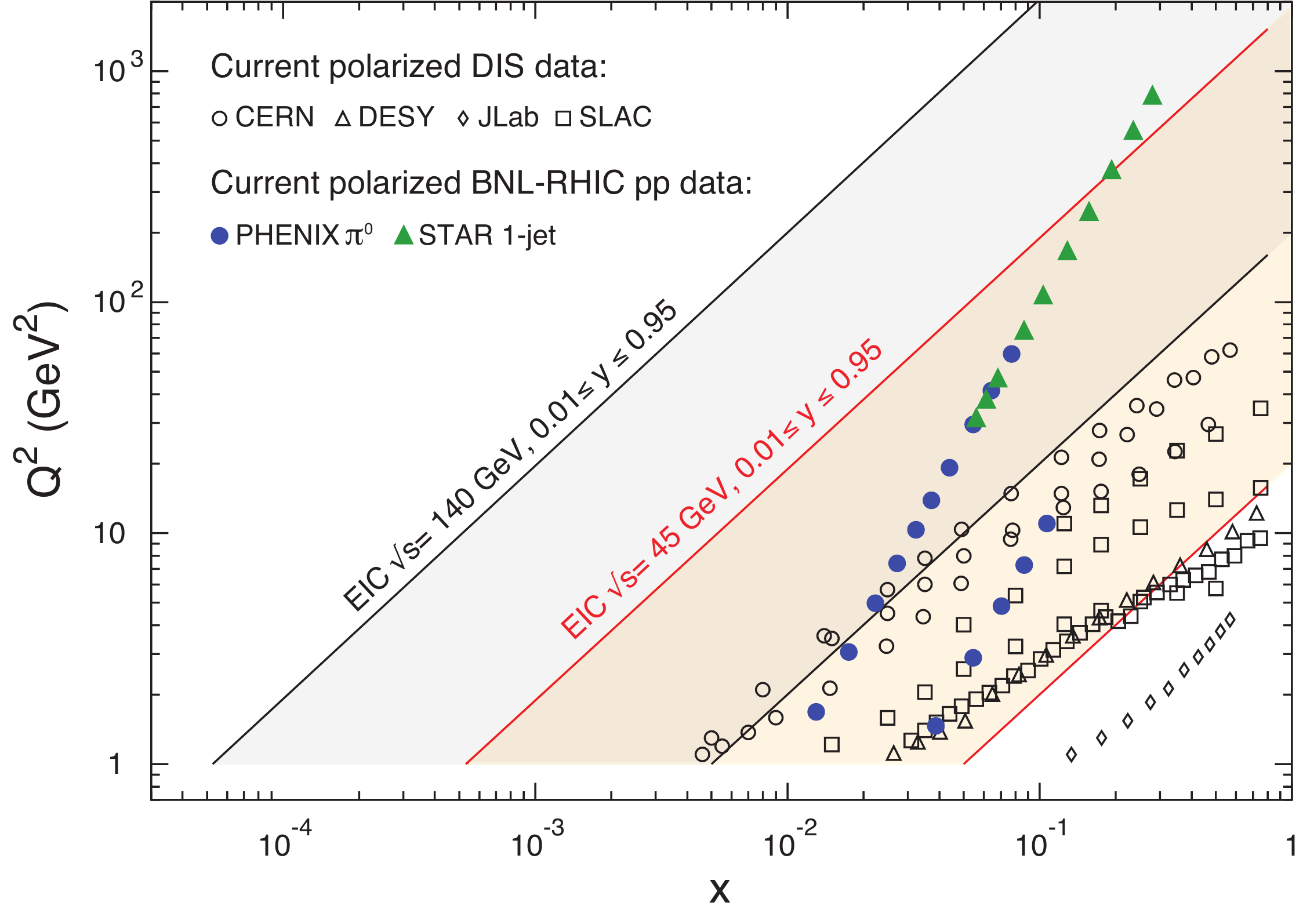}
\end{center}
\caption{\label{fig:world}{Regions in $x$, $Q^2$ covered by previous
spin experiments and anticipated to be accessible at an EIC. The
values for the existing fixed-target DIS experiments are shown as data
points.  The RHIC data are shown at a scale $Q^2 = p_T^2$, where $p_T$
is the observed jet (pion) transverse momentum, and an $x$ value that
is representative for the measurement at that scale. The $x$-ranges
probed at different scales are wide and have considerable overlap.
The shaded regions show the $x$, $Q^2$ reach of an EIC for
center-of-mass energy $\sqrt{s}=45$~GeV and $\sqrt{s}=140$~GeV,
respectively.  }}
\end{figure}

Figure~\ref{fig:world} provides a survey of the regions in $x$ and $Q^2$ 
covered by the world
polarized-DIS data, which is roughly $0.004 < x < 0.8$ for $Q^2 >
1\,\mathrm{GeV}^2$.  For a representative value of $x \simeq 0.03$,
the $g_1(x,Q^2)$ data are in the range $1\,\mathrm{GeV}^2 < Q^2 <
10\,\mathrm{GeV}^2$.  This is to be compared to $1\,\mathrm{GeV}^2 <
Q^2 < 2000\,\mathrm{GeV}^2$ for the unpolarized data on $F_2(x,Q^2)$
at the same $x$. The figure also shows the vast expansion in $x,Q^2$
reach that an EIC would provide, as will be discussed below.  
Over the past 15 years, an additional powerful line of experimental
study of nucleon spin structure has emerged: {\it semi-inclusive}
deep-inelastic scattering.  In these measurements, a charged or
identified hadron $h$ is observed in addition to the scattered
lepton. The relevant spin-dependent structure function,
\end{multicols}
\begin{center}
\begin{equation} 
g_1^h(x,Q^2,z) = \frac{1}{2} \sum_q \left. e_q^2
\left[ \Delta q(x,Q^2) D^h_q(z,Q^2) + \Delta \bar{q}(x,Q^2)
D^h_{\bar{q}}(z,Q^2) \right] \right.  ,
  \label{eq:g1h}
\end{equation} 
\end{center}
\begin{multicols}{2}
\noindent
depends on fragmentation functions
$D^h_{q,\bar{q}}(z,Q^2)$, where $z$ is the momentum fraction that is
transferred from the outgoing quark or anti-quark to the observed
hadron $h$. The non-perturbative fragmentation functions are at
present determined primarily from precision data on hadron production
in $e^+e^-$ annihilation through perturbative QCD
analysis~\cite{Hirai:2007cx,deFlorian:2007hc,deFlorian:2007aj,Albino:2008fy,deFlorian:2014xna}.
Data from the $B$-factories and the LHC are helping to
further improve their determination~\cite{deFlorian:2014xna}. Also measurements of hadron
multiplicities at an EIC would contribute to a better knowledge of
fragmentation functions.  Insights from the semi-inclusive
measurements are complementary to those from the inclusive
measurements. Specifically, they make it possible to delineate the
quark and anti-quark spin contributions by flavor, since $\Delta q$
and $\Delta \bar{q}$ appear with different weights in
Eq.~(\ref{eq:g1h}). A large body of semi-inclusive data sensitive to
nucleon helicity structure has been collected by the experiments at
CERN~\cite{Adeva:1997qz,Alekseev:2007vi,Alekseev:2010ub} and
DESY~\cite{Airapetian:2004zf}.
\end{multicols}

\begin{multicols}{2}
A further milestone in the study of the nucleon was the advent of
RHIC, the world's first polarized proton+proton collider. In the
context of the exploration of nucleon spin structure, the RHIC spin
program is a logical continuation. Very much in the spirit of the
unpolarized hadron colliders in the 1980's, RHIC entered the scene to
provide complementary information on the nucleon that is not readily
available in fixed-target lepton scattering.  The measurement of the
spin-dependent gluon distribution $\Delta g(x,Q^2)$ in the proton is a
major focus and strength of RHIC. Here the main tools are spin
asymmetries in the production of inclusive
pions~\cite{Adler:2004ps,Adare:2007dg,:2008px,Abelev:2009pb,Adare:2014hsq} and
jets~\cite{Abelev:2006uq,Abelev:2007vt,Sarsour:2009zd,Djawotho:2011zz,Adamczyk:2014ozi}
at large transverse momentum perpendicular to the beam axis, which
sets the hard scale $Q$ in these reactions. Their reach in $x$ and $Q^2$ is also indicated in
Fig.~\ref{fig:world}. Unlike DIS, the processes used at RHIC do not
probe the partons locally in $x$, but rather sample over a region in
$x$. RHIC also provides complementary information on
$\Delta u,\Delta \bar{u}, \Delta d,\Delta \bar{d}$ for $0.05 < x <
0.5$~\cite{Adare:2010xa,Aggarwal:2010vc,Adamczyk:2014xyw,Gal:2014fha}, 
with a beautiful technique that exploits the violation of parity
(mirror symmetry) in nature and does not rely on knowledge of
fragmentation.  The carriers of the charged-current weak interactions,
the $W$ bosons, naturally select left-handed quarks and right-handed
anti-quarks, and their production in $p$+$p$ collisions at RHIC and
calculable leptonic decay hence provide an elegant probe of nucleon
helicity structure.

Combined next-to-leading order QCD
analyses~\cite{deFlorian:2008mr,deFlorian:2009vb,deFlorian:2014yva,Nocera:2014gqa} 
of the published data from
inclusive and semi-inclusive deep-inelastic scattering and from $p$+$p$
scattering at RHIC have been performed, which provide the best
presently available information of the nucleon's helicity structure.
The main results of the first such analysis~\cite{deFlorian:2008mr,deFlorian:2009vb} 
are displayed in Fig.~\ref{fig:deltag-impact}. Here we describe the main qualitative 
features found in the latest studies:
\end{multicols}

\begin{itemize}
\item The combination of the large body of inclusive deep-inelastic
scattering data off targets containing polarized protons and neutrons
has established that the up quarks and anti-quarks combine to have net
polarization along the proton spin, whereas the down quarks and
anti-quarks combine to carry negative polarization.  The ``total''
$\Delta u+\Delta\bar{u}$ and $\Delta d+\Delta\bar{d}$ helicity
distributions are very well constrained by now at medium to large $x$.
\item The light sea quark and anti-quark distributions still carry
large uncertainties, even though there are some constraints by the
semi-inclusive data and, most recently,  from measurements of spin-dependence 
in leptonic $W$ decay in $\sqrt{s} = 500\,\mathrm{GeV}$ polarized proton+proton
collisions at RHIC~\cite{Adamczyk:2014xyw,Gal:2014fha}. RHIC probes the 
$\Delta u$, $\Delta d$, $\Delta \bar{u}$ and $\Delta \bar{d}$ densities
for $0.05 < x < 0.5$ at a scale set by the $W$-mass~\cite{Bunce:2000uv}.  
The sea shows hints of not being SU(2)-flavor symmetric: the $\Delta \bar{u}$ 
distribution has a tendency to be mainly positive, while the $\Delta \bar{d}$ 
anti-quarks carry opposite polarization. This pattern has been predicted at least 
qualitatively by a number of models of the nucleon (for a review,
see~\cite{Kumano:1997cy}).  More sensitive constraints on $\Delta u$,
$\Delta d$, $\Delta \bar{u}$ and $\Delta \bar{d}$ are 
are anticipated~\cite{deFlorian:2010aa} from additional RHIC measurements with 
higher integrated luminosity. The large luminosities and high
resolution available at the Jefferson Laboratory after an upgrade to
$12\,\mathrm{GeV}$ electron beam energy will extend the kinematic
reach of the existing Jefferson Laboratory inclusive and semi-inclusive
deep-inelastic scattering data to twice smaller $x$ as well as to
larger $x$ than have thus far been measured.
\item Strange quarks appear to be deeply involved in nucleon spin
structure. As we mentioned earlier, from the inclusive deep-inelastic
data, along with SU(3) flavor symmetry considerations, one derives a
strong negative value of the integrated strange helicity distribution.
Strange quarks and anti-quarks would thus be polarized opposite to the
nucleon.  This would need to be viewed as part of the reason why the
total quark and anti-quark spin contribution $S_q$ is so much smaller
than expected in simple models. On the other hand, significant SU(3)
flavor symmetry breaking effects have been discussed in the
literature~\cite{Cabibbo:2003cu,Savage:1996zd,Zhu:2002tn,Ratcliffe:2004jt,Sasaki:2008ha}. 
The semi-inclusive measurements with identified
kaons~\cite{Alekseev:2010ub,Airapetian:2004zf} are hence of particular
interest since they yield the most direct and best sensitivity thus
far to the polarization of strange quarks and anti-quarks, albeit with
considerable dependence on the kaon fragmentation
functions~\cite{Leader:2011tm}.  No evidence for sizable $\Delta
s(x,Q^2)$ or $\Delta \bar{s}(x,Q^2)$ has been found in polarized
semi-inclusive measurements with fixed targets. As a consequence,
$\Delta s$ would need to obtain its negative integral purely from the
contribution from the thus far unmeasured small-$x$ region. This exemplifies the
need for simultaneous measurements of the kaon production
cross-sections and their spin-dependence in semi-inclusive
deep-inelastic scattering at smaller $x$.
\item Constraints on the spin-dependent gluon distribution $\Delta g$
predominantly come from RHIC, with some information also entering from
scaling violations of the deep-inelastic structure function
$g_1(x,Q^2)$.  The production cross sections for inclusive hadrons and
jets at RHIC receive contributions from gluon-gluon and quark-gluon
scattering and probe $\Delta g(x,Q^2)$ over the range $0.02 < x <
0.4$. Note that the $x$ is not explicitly resolved
in measurements of inclusive pion and jet probes. Initial results
from RHIC saw small double-spin asymmetries for inclusive jets
and hadrons. As a result, the global analysis~\cite{deFlorian:2008mr,deFlorian:2009vb} 
concluded that there were no indications of a sizable contribution of gluon spins to the
proton spin. This has changed recently: The latest much more precise STAR results
for the double-spin asymmetry in jet production~\cite{Adamczyk:2014ozi}
provide, for the first time, evidence of a non-vanishing polarization of
gluons in the nucleon in the RHIC kinematic regime~\cite{deFlorian:2014yva}.
This is a major breakthrough for this field. The limited
$x$-range and unresolved $x$-dependence preclude definitive
conclusions on the total gluon spin contribution to the proton spin,
$S_g$, although it appears likely now that gluons are an important player
for the proton spin. Continued measurements at $\sqrt{s} = 200\,\mathrm{GeV}$ will
enhance the sensitivity primarily at large $x$, and measurements of
correlated probes are anticipated to yield insights in $x$-dependence.
Forthcoming measurements at $\sqrt{s} = 500\,\mathrm{GeV}$ are
expected to extend the small-$x$ reach to $2\div3$ times smaller
values and modest further gains may be possible with new instruments
at larger pseudorapidity.  Extrapolation over the unmeasured $x \lsim
0.01$ region is precarious, and definitive resolution of the gluon spin
contribution to the nucleon spin thus relies on a new generation of
experiments.
\end{itemize}

\subsection{Open Questions and the Role of an EIC} 

The overarching
scientific question --- {\it How is the spin of the proton distributed
among its quark and gluon constituents?} --- will remain only
partially answered even after the completion of the present programs
and their upgrades.  Concerning the helicity parton distributions, the
remaining key open issues will be:
\begin{itemize}
\item {\it What is the gluon spin contribution to the proton spin?} As
we saw, there is now initial knowledge about $\Delta g$ in a
relatively narrow region of $x$. Clearly, more extended
coverage is required to determine this intrinsic property of the
proton and constrain the integral of the distribution.
\item {\it What polarization is carried by the proton's light sea?}
Previous and present experiments give a hint at interesting flavor
structure of sea quark polarization.  Still, even after the completion of
the RHIC program with $W$ bosons, we will likely have little precision
on, for example $\Delta\bar{u}-\Delta \bar{d}$, a quantity that
features prominently in virtually all models of the nucleon in ways
that are complementary to the unpolarized light sea. Exploring in
detail the proton's sea quark ``landscape'' would provide
unprecedented insight into non-perturbative QCD.
\item {\it What role do strange quarks play in nucleon spin
structure?}  Strange quarks play a special role for understanding QCD
as their mass is of the order of $\Lambda_\mathrm{QCD}$ and they are
hence to be considered neither light (as the up and down quarks), nor
heavy (as the charm and heavier quarks). Present experimental
information on their role in nucleon spin structure is quite puzzling,
as we described above.  There is clearly a strong need to determine
$\Delta s$ and $\Delta \bar{s}$ over a wide range in 
$x$. This will also probe important aspects of SU(3) flavor symmetry
and its breaking in QCD.
\end{itemize} 
\begin{multicols}{2}
In order to fully solve the proton spin problem one
evidently also needs to obtain information on quark and gluon {\it
orbital} angular momenta in the nucleon. This requires a new suite of
measurements, using exclusive processes such as deeply-virtual Compton
scattering and transverse-spin asymmetries. The associated physics and
the prospects of measurements at an EIC will be described in
Sections~\ref{sec:tmd} and~\ref{sec:exclusive}.

The envisioned polarized electron ion collider brings unique
capabilities to the study of nucleon spin.  Its high center-of-mass
energy of up to $\sqrt{s} = 173\,\mathrm{GeV}$ affords access to a
vast region in $x$ and $Q^2$ that will probe $1 \div 2$ orders of
magnitude smaller values in $x$ than the body of existing and
forthcoming data and comparably harder scales $Q^2$, as is clearly
visible from Fig.~\ref{fig:world}.  The high luminosity and
polarization will allow one to do so with precision using a suite of
probes (see~Table~\ref{tbl:helicity}). In this way,
the EIC aims to provide answers to the questions raised above.
\end{multicols}
\begin{table*}[h] \centering
\noindent\makebox[\textwidth]{%
\footnotesize
\begin{tabular}{|c|c|c|c|c|} \hline Deliverables & Observables & What
we learn & Requirements \\ \hline \hline 
polarized gluon & scaling violations & gluon contribution & coverage down
 to $x\simeq 10^{-4}$;\\ 
distribution $\Delta g$ & in inclusive DIS & to proton spin &
${\cal{L}}$ of about $10\;\mathrm{fb}^{-1}$\\ \hline 
polarized quark and & semi-incl.\ DIS for & quark contr.\ to proton spin;
 & similar to DIS; \\ 
antiquark densities & pions and kaons & asym.\ like $\Delta
\bar{u}-\Delta\bar{d}$; $\Delta s$ & good particle ID \\ \hline 
novel electroweak & inclusive DIS & flavor separation & $\sqrt{s}\geq
100\,\mathrm{GeV}$; ${\cal{L}}\geq 10\;\mathrm{fb}^{-1}$ \\ 
spin structure functions & at high $Q^2$ & at medium $x$ and large $Q^2$ &
positrons; polarized d or $^{3}$He beam\\ \hline
\end{tabular}}
\caption{\label{tbl:helicity} Key measurements to determine the quark
and gluon helicity distributions in the polarized nucleon.}
\end{table*} 
\begin{multicols}{2}
We will now discuss the scientific highlights of an EIC, insofar as
they pertain to nucleon helicity structure.

Arguably {\it the} golden ``flagship'' measurement of nucleon spin
structure at the EIC will be a precision study of the proton's spin
structure function $g_1(x,Q^2)$ and its scaling violations, over wide
ranges in $x$ and $Q^2$. The methods to measure $g_1(x,Q^2)$ are well
known experimentally and $g_1(x,Q^2)$ is also understood very well
theoretically.  The small $x$ region is key to determining and
understanding the role of sea quarks and gluons in the spin
decomposition of the nucleon.  The structure function $g_1(x,Q^2)$
presently is {\it terra incognita} for $x < 0.004$ and $Q^2 >
1\,\mathrm{GeV}^2$ (see Fig.~\ref{fig:world}). Low-$x$ measurements of
$g_1$ reduce the present uncertainty associated with the required
extrapolation when computing the quark and anti-quark spin
contribution $S_q$ to the proton spin. The $Q^2$-dependence of
$g_1(x,Q^2)$ will give unprecedented insight into gluon polarization.
The EIC will also vastly expand our knowledge of the quark flavor structure,
a key element in mapping out the proton ``landscape''.
A powerful measurement available to achieve this is {\it
semi-inclusive deep-inleastic scattering} which at the EIC would
extend to much higher $Q^2$ than in fixed-target scattering, where the
interpretation becomes significantly cleaner, less afflicted by
corrections suppressed by $1/Q^2$, and better tractable theoretically.
The kinematic coverage in $x$ and $Q^2$ will be similar to what can
be achieved in inclusive scattering.  
\end{multicols}
\begin{figure}[th!]
\begin{center}
\includegraphics[scale=0.65]{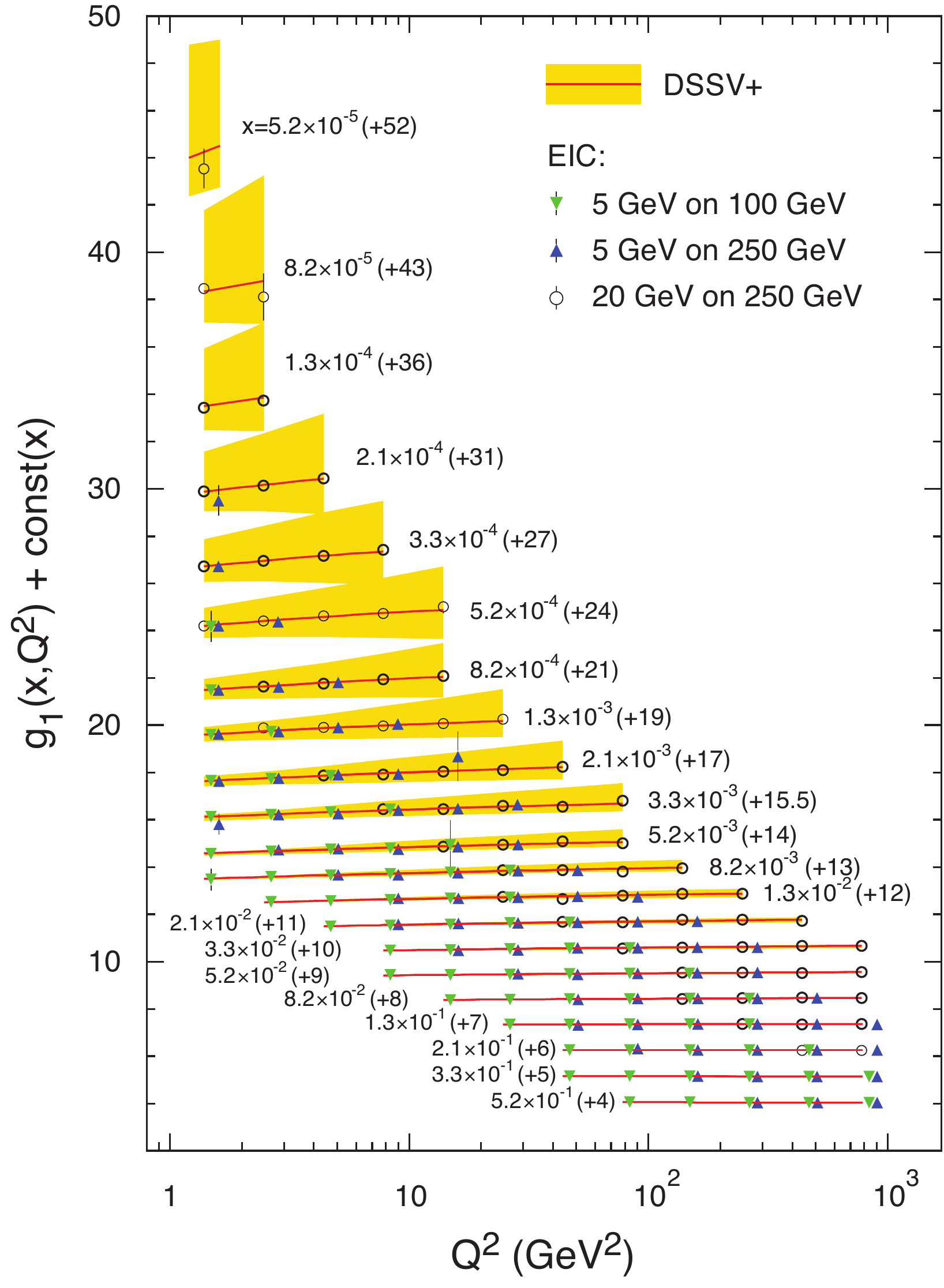}
\end{center}
\vskip -0.2in
\caption{\label{fig:g1-pseudo}{EIC pseudo-data on the inclusive spin
structure function $g_1(x,Q^2)$ versus $Q^2$ at fixed $x$ for 5 GeV
and 20 GeV electron beams colliding with 100 GeV and 250 GeV proton
beam energies at an EIC, as indicated. The error bars indicate the
size of the statistical uncertainties. The data set for each $x$ is
offset by a constant $c(x)$ for better visibility. The bands indicate
the current uncertainty as estimated in the ``DSSV+'' analysis (see text).}  }
\end{figure} 

\begin{multicols}{2}
To illustrate the tremendous impact of EIC measurements of inclusive
and semi-inclusive polarized deep-inelastic scattering on our
knowledge of helicity parton distributions, a series of perturbative
QCD analyses were performed~\cite{Aschenauer:2012ve} with realistic
pseudo-data for various center-of-mass energies.  The data simulations
were based on the PEPSI Monte Carlo
generator~\cite{Mankiewicz:1991dp}.  The precision of the data sets
corresponds to an accumulated integrated luminosity of
$10\,\mathrm{fb}^{-1}$ (or one to two months of running for most
energies at the anticipated luminosities) and an assumed operations
efficiency of $50\%$.  A minimum $Q^2$ of $1\,\mathrm{GeV}^2$ was
imposed, as well as $W^2>10\,\mathrm{GeV}^2$, a depolarization factor
of the virtual photon of $D(y)>0.1$, and $0.01\le y\le 0.95$.
Figure~\ref{fig:g1-pseudo} shows the pseudo-data for the inclusive
structure function $g_1(x,Q^2)$ of the proton versus $Q^2$ at fixed
$x$.

Collisions at $\sqrt{s} \simeq 70\,\mathrm{GeV}$ at an EIC 
are seen to provide access to $x$ values down to about $2\times 10^{-4}$.
The anticipated size of the asymmetry $A_1(x,Q^2) \simeq
g_1(x,Q^2)/F_1(x,Q^2)$ at these $x$ values is $\mathcal{O}(10^{-3})$, which
sets the scale for the required data samples and control of experiment
systematics.  These and other aspects are discussed further in
Section~\ref{sec:detectors}.  Data from a higher-energy 
EIC, shown for electron beam energies up to 20\,GeV, is seen to provide access to
significantly smaller $x$ and larger $Q^2$.  As demonstrated in
Fig.~\ref{fig:g1-pseudo1}, the combination of measurements
with a wide range of center-of-mass energies at an EIC will make it
possible to directly determine
$\mathrm{d}g_1(x,Q^2)/\mathrm{d}\log(Q^2)$ with good sensitivity,
which directly probes the gluon distribution $\Delta g$.  
\end{multicols}
\begin{figurehere}
\begin{center}
\includegraphics[scale=0.65]{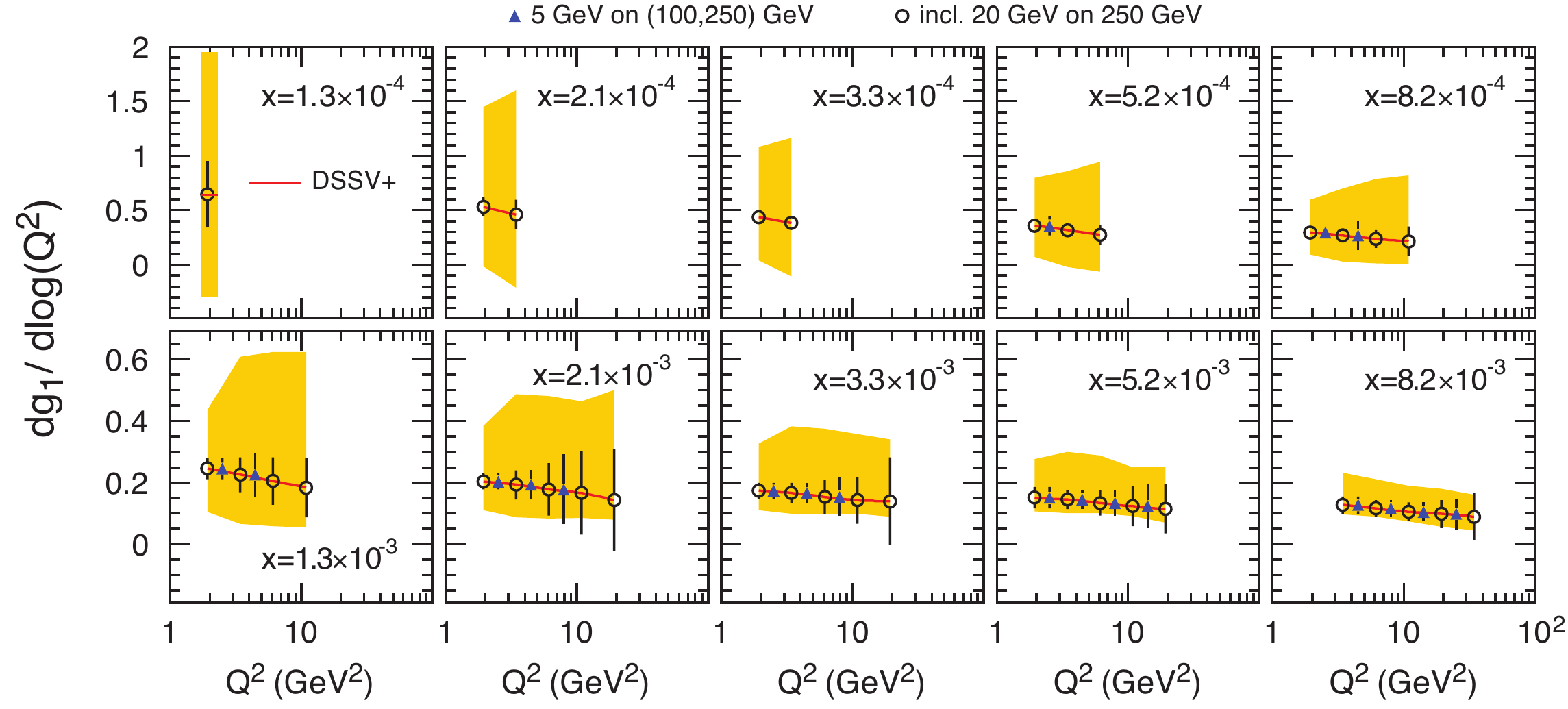}
\end{center}
\vskip -0.2in
\caption{\label{fig:g1-pseudo1}{The derivative of $g_1(x,Q^2)$ with
$\log(Q^2)$ for different $x$ values for the same combination of
electron and proton beam energies as used in Fig.~\ref{fig:g1-pseudo},
together with the ``DSSV+'' uncertainty bands.}}
\end{figurehere} 

\begin{figure}[th!]
\begin{center}
\includegraphics[width=0.46\textwidth]{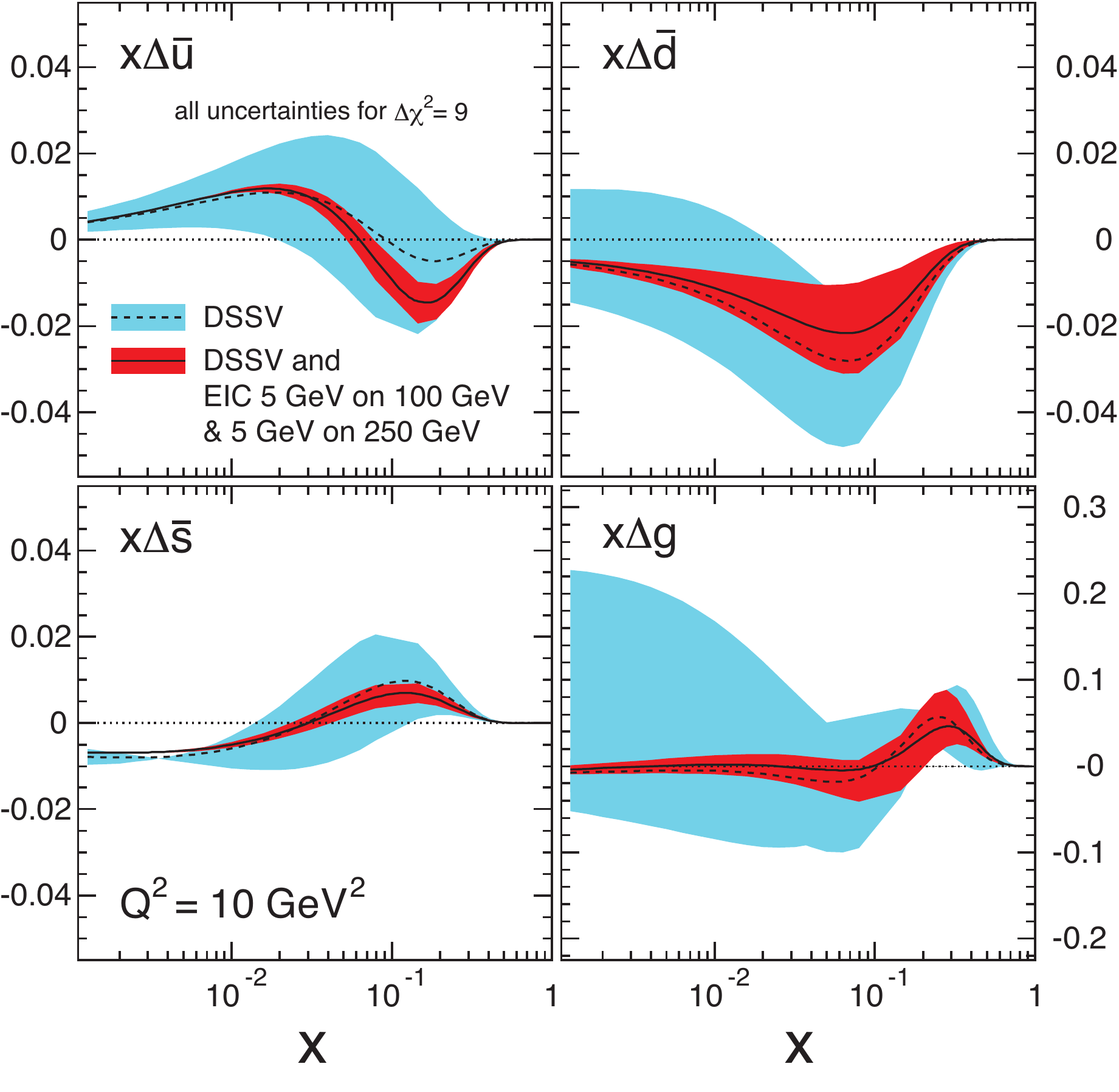}
\includegraphics[width=0.47\textwidth]{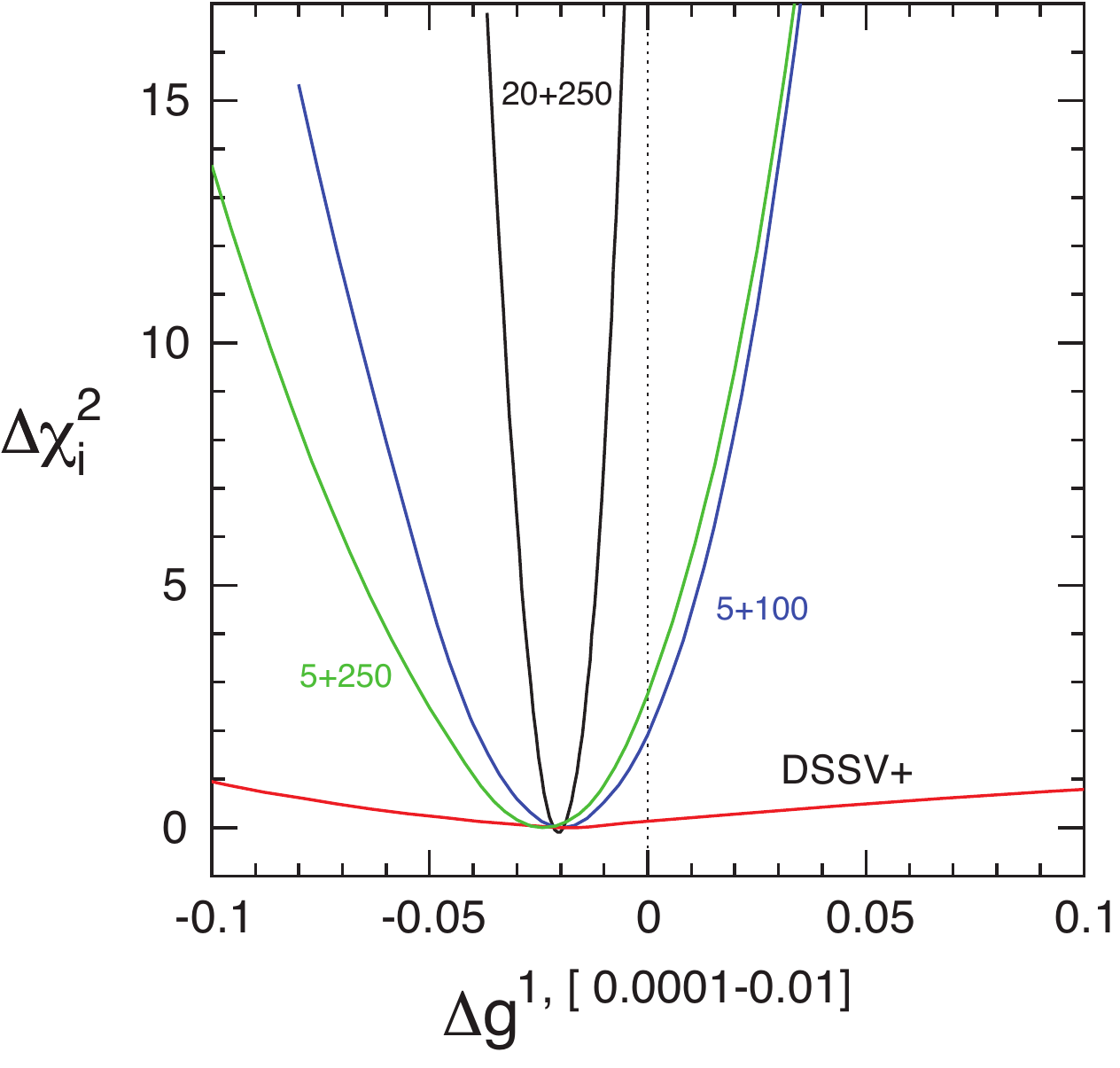}
\end{center} \vspace*{-0.4cm}
\caption{\label{fig:deltag-impact} {Left: Uncertainty bands on
helicity parton distributions, in the first DSSV analysis~\cite{deFlorian:2008mr,deFlorian:2009vb} 
(light bands) and with EIC
data (darker bands), using projected inclusive and semi-inclusive EIC
data sets (see text). Note that for this analysis only data with
$x\geq 10^{-3}$ were used, for which $Q^2\geq 2.5$~GeV$^2$.  Right:
$\chi^2$ profiles for the truncated $x$ integral of $\Delta g$ over
the region $10^{-4}\leq x\leq 10^{-2}$ with and without including the
generated EIC pseudo-data in the fit. Results are shown for three
different EIC center-of-mass energies.}}
\end{figure}

\begin{multicols}{2}
The pseudo-data for $g_1$ and for semi-inclusive spin asymmetries were
included~\cite{Aschenauer:2012ve} in the global analysis of
helicity-dependent parton distribution functions based on the DSSV
framework
\cite{deFlorian:2008mr,deFlorian:2009vb}.~\footnote{As described earlier, these first 
DSSV papers do not yet contain the latest information from RHIC on $\Delta g$,
which were not yet available at the time of~\cite{Aschenauer:2012ve}. However, 
this is not an issue here as the figures below are merely meant to demonstrate the
{\it improvements} an EIC would provide on the knowledge of the helicity distributions. 
We note that for the studies presented here the analysis of~\cite{deFlorian:2008mr,deFlorian:2009vb} 
has been complemented with recent lepton scattering data~\cite{Alekseev:2010hc,Alekseev:2010ub} 
from CERN. It will henceforth be referred to as ``DSSV+'' analysis.}
Figure~\ref{fig:deltag-impact}
(left) shows the results of this analysis in terms of the sea quark
and gluon helicity distributions. For comparison, the present
uncertainty bands are also displayed. As one can see, an impressive
reduction in the width of the bands would be expected from EIC data,
in particular, towards lower values of $x$.
Evidently, extractions of $\Delta g$ from scaling violations, and of
the light-flavor helicity distributions $\Delta u$, $\Delta d$ and
their anti-quark distributions from semi-inclusive scattering will be
possible with exquisite precision. With dedicated studies of kaon
production, the strange and anti-strange distributions will also be
accessible.  All this is anticipated to yield new insights into the
question of why it is that the combined quark and anti-quark spin
contribution to the proton spin turns out to be so small.  
\end{multicols}

\vspace*{-0.2cm}
\begin{multicols}{2}
The right part of the figure shows the $\chi^2$ profile of the
truncated first moment of the gluon helicity distribution,
$\int_{0.0001}^{0.01} dx \Delta g(x,Q^2)$, at $Q^2=10$~GeV$^2$, again
compared to the ``DSSV+'' estimate.
Also here, the impact of EIC data is evident. One also observes the
importance of high energies. For instance, running at the highest
energy clearly constrains the small-$x$ region much better. Overall,
the EIC data greatly improves the $\chi^2$ profile, even more so when
all data 
in Fig.~\ref{fig:g1-pseudo} are included. 

\begin{figure*}[t]
\begin{center} \vspace*{-0.5cm}
\includegraphics[width=0.7\textwidth]{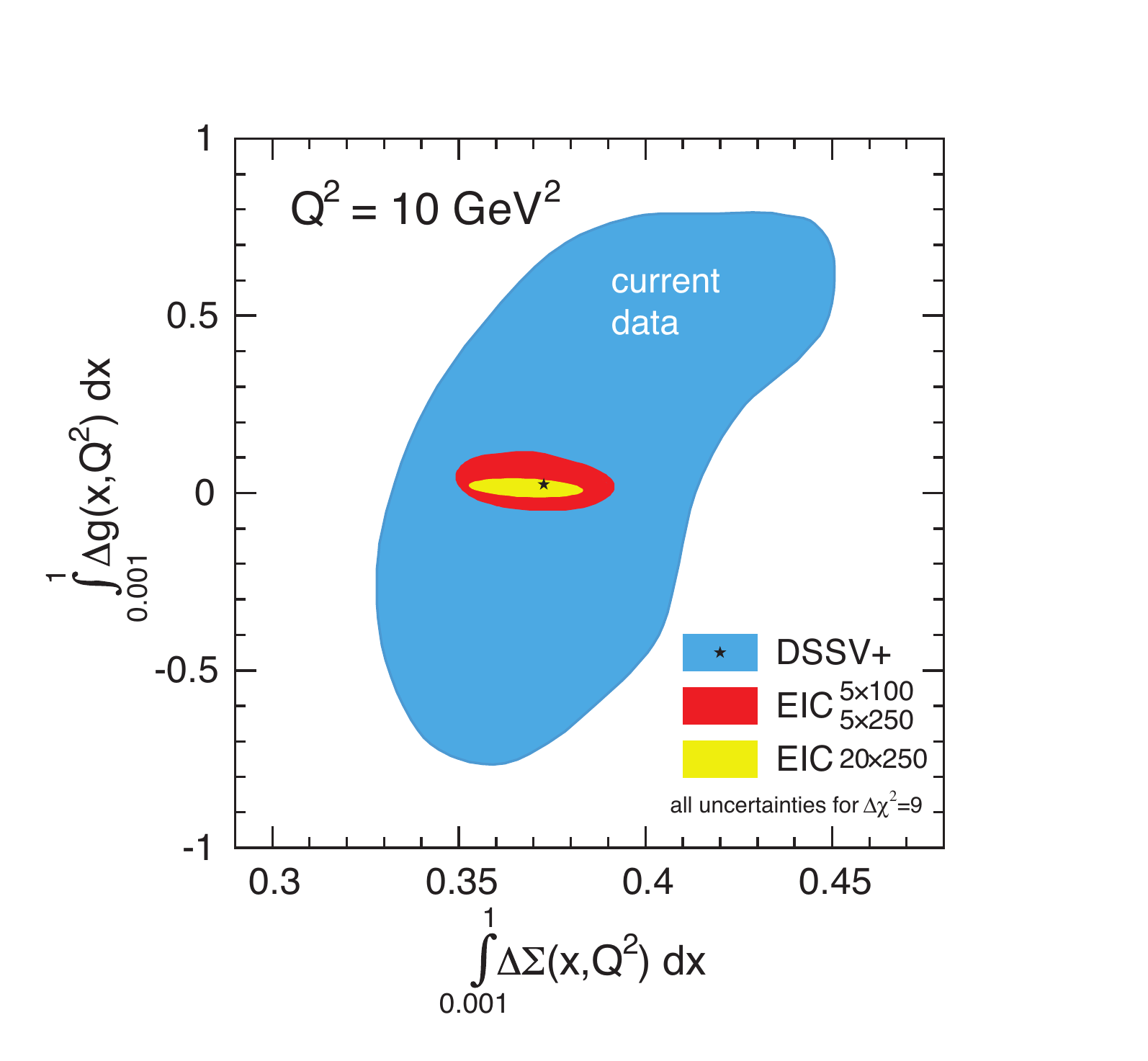}
\end{center} \vspace*{-0.5cm}
\caption{\label{fig:dg-dsigma} {Accuracies for the correlated truncated integrals of
$\Delta\Sigma$ and $\Delta g$ over $0.001\leq x\leq 1$, on the basis of the ``DSSV+'' analysis
(outer area) and projected for an EIC (inner areas)~\cite{Aschenauer:2012ve}.}}
\end{figure*}

The light shaded area in Fig.~\ref{fig:dg-dsigma} displays the present
accuracies of the integrals of $\Delta \Sigma$ and $\Delta g$ over
$0.001\leq x\leq 1$, along with their correlations. The inner areas
represent the improvement to be obtained from the EIC, based on the
global analysis studies with pseudo-data described above. We stress
that similar relative improvements would occur for any other benchmark
set of polarized parton distribution functions, such as the latest 
DSSV~\cite{deFlorian:2014yva} set. The results shown in the figure
clearly highlight the power of an EIC in mapping out nucleon helicity
structure.  The anticipated kinematic range and precision of EIC data
will give unprecedented insight into the spin contributions $S_q$ and
$S_g$.  Their measurements, by subtracting from the total proton spin
$1/2$, will provide stringent and independent constraints on the total
contribution of quark and gluon orbital momenta, $L_q+L_g$.  

Besides polarized proton beams, the EIC design envisions beams of
polarized deuterons or helium-3. 
The neutron's $g_1(x,Q^2)$ can thus
be determined, potentially with a precision that is comparable to the
data on $g_1(x,Q^2)$ of the proton.  The difference of the moments of
proton and neutron $g_1(x,Q^2)$ allows a test of the fundamental sum
rule by Bjorken~\cite{Bjorken:1968dy}.  The data from polarized fixed
target experiments have verified the sum rule to a precision of about
$10\%$ of its value.  The extended kinematic range and improved
precision of EIC data allow for more stringent tests of this sum rule,
as well as its corrections, to an accuracy that is currently
anticipated to be driven mostly by advances in hadron beam polarimetry
(cf.~Section~\ref{sec:polarimetry}).  

An additional, and unique, avenue for delineating the flavor structure
of the quark and anti-quark spin contribution to the proton spin at
the EIC is electroweak deep-inelastic scattering. At high $Q^2$, the
deep-inelastic process also proceeds significantly via exchange of $Z$
and $W^\pm$ bosons. This gives rise to novel structure functions that
are sensitive to different combinations of the proton's helicity
distributions. For instance, in the case of charged-current interactions
through $W^-$, the inclusive structure functions contribute, 
\end{multicols}
\vspace*{-0.5cm}
\begin{eqnarray}\label{g5lo} g_1^{W^-} (x,Q^2) &=& \left[\Delta u +
\Delta\bar{d}+\Delta c +\Delta\bar{s} \right] (x,Q^2) \; ,
\nonumber\\[2mm] g_5^{W^-}(x,Q^2) &=& \left[-\Delta u +
\Delta\bar{d}-\Delta c +\Delta\bar{s} \right](x,Q^2) \;,
\end{eqnarray} 

\vspace*{-0.5cm}
\begin{multicols}{2}
\noindent
where $\Delta c$ denotes the proton's
polarized charm quark distribution.  The analysis of these structure
functions does not rely on knowledge of fragmentation.  Studies show
that both neutral-current and charged-current interactions would be
observable at the EIC, even with relatively modest integrated
luminosities. To fully exploit the potential of the EIC for such
measurements, positron beams are required, albeit not necessarily
polarized. Besides the new insights into nucleon structure this would
provide, studies of spin-dependent electroweak scattering at short
distances with an EIC would be beautiful physics in itself, much in
the line of past and ongoing electroweak measurements at HERA,
Jefferson Laboratory, RHIC, and the LHC.  
\begin{figure*}[t]
\begin{center}
\includegraphics[scale=0.55]{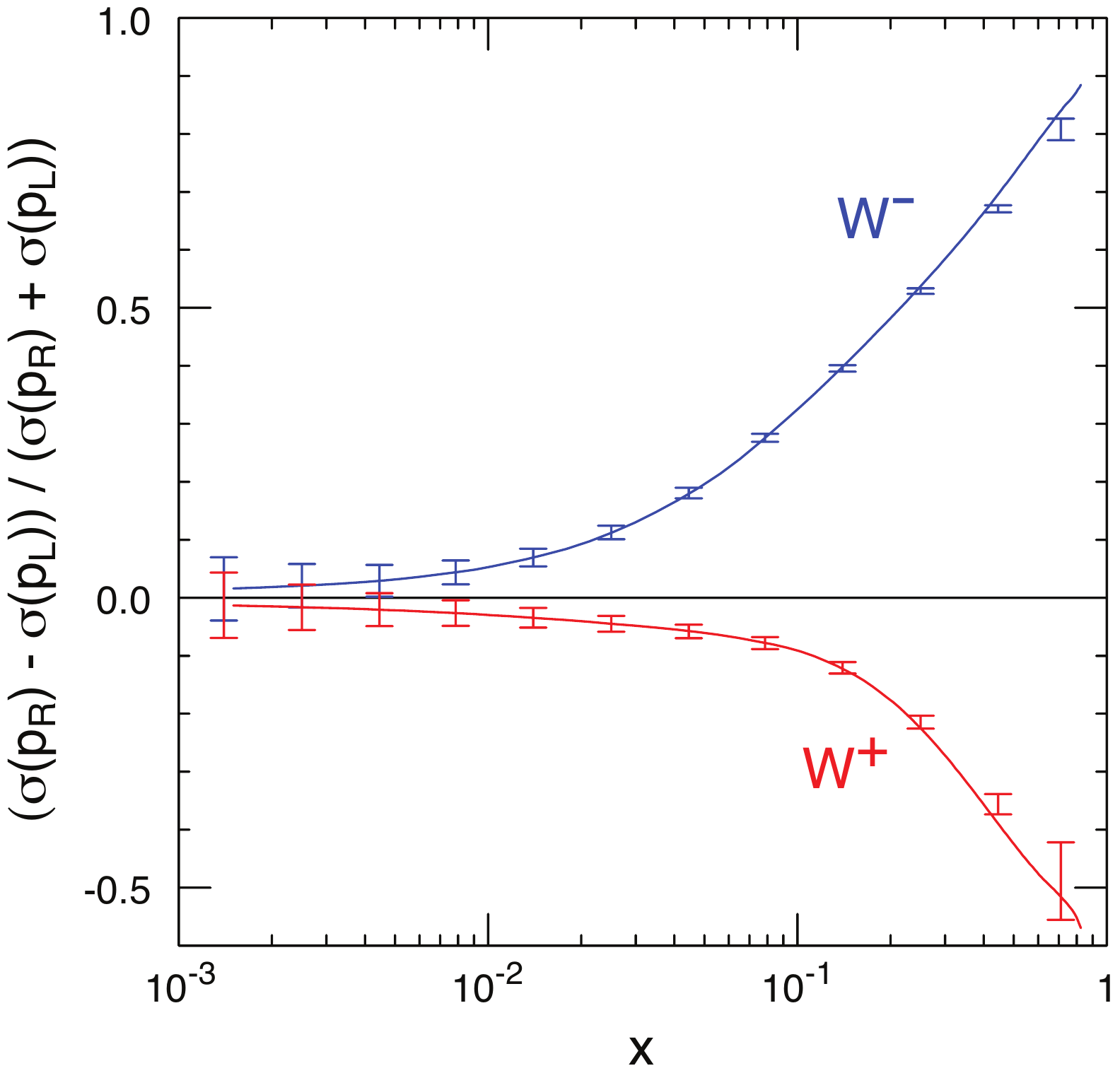}
\end{center}
\caption{\label{fig:ew}{Single-longitudinal spin asymmetries for $W^-$
and $W^+$ exchange at an EIC, using polarized protons.  A collision
energy of $\sqrt{s}=141$~GeV was assumed and cuts $Q^2>1$~GeV$^2$ and
$0.1<y<0.9$ were applied. The uncertainties shown are statistical, for
10~fb$^{-1}$ integrated luminosity.}}
\end{figure*}
\noindent
As an illustration of the EIC's
potential in this area, Fig.~\ref{fig:ew} shows production-level
estimates for charged-current interactions through $W^-$ and $W^+$
exchange at collision energy $\sqrt{s}=141$~GeV. Cuts of
$Q^2>1$~GeV$^2$ and $0.1<y<0.9$ have been applied. The figure shows
the parity-violating single-longitudinal spin asymmetry
$(\sigma(p_R)-\sigma(p_L))/(\sigma(p_R)+\sigma(p_L))$ obtained from
the cross sections for positive ($p_R$) or negative ($p_L$) proton
helicity. The figure also shows production-level statistical
uncertainties for measurements at an EIC with 10~fb$^{-1}$ integrated
luminosity. As one can see, large asymmetries are expected in the
region of moderate to large $x$, where the energies of the observed
jet are typically large.  Their measurement provides unique insights
into the flavor composition of the proton spin.
A more detailed study has recently been published~\cite{Aschenauer:2013iia}.
\end{multicols}

%% file: files_tex/sidebarTMD.tex

\newpage
\pagecolor{LightYellow}

\phantomsection
\label{sdbar:tmd}

\vspace*{-1.00cm}

\begin{center}
{\textbf {\textit {\textcolor{blue}{\Large Semi-inclusive Deep Inelastic Scattering}}}}
\line(1,0){435}
\end{center}

Semi-inclusive hadron production in deep inelastic
scattering (SIDIS) provides a powerful probe of the transverse momentum dependent (TMD) 
quark distributions of nucleons.
Common kinematic variables have been described in the DIS section 
(see the Sidebar on page~\pageref{sdbar:DIS}). 
In SIDIS, the kinematics of the final state hadrons can be specified as follows
\begin{multicols}{2}
\includegraphics[width=0.40\textwidth]{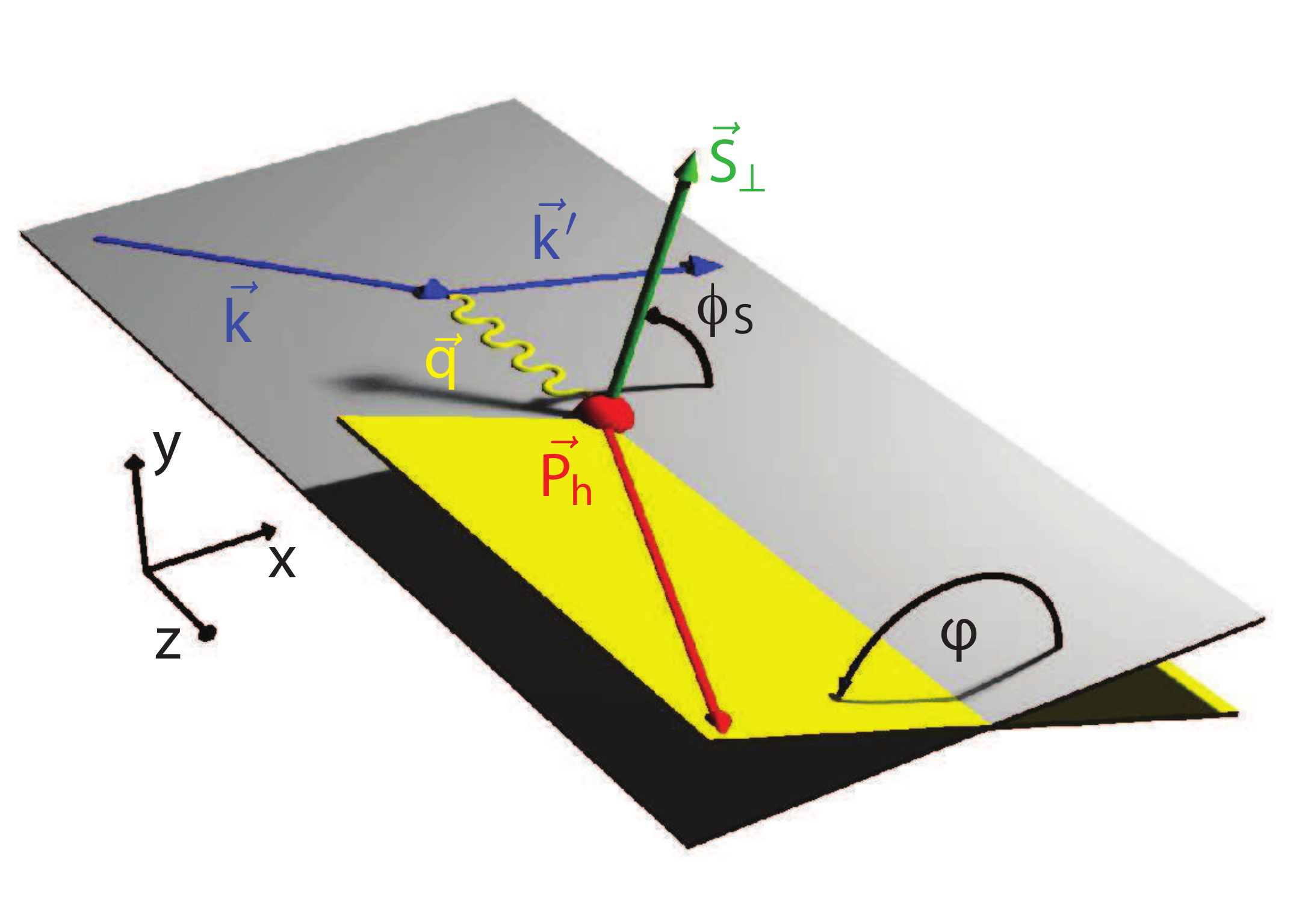}
\mfcaption{\label{f:anglestrento} Semi-inclusive hadron
production in DIS processes: $e+N\rightarrow e^\prime+h+X$,
in the target rest frame. $\mbox{\boldmath $P$}_{hT}$ and
$\mbox{\boldmath $S$}_{\perp}$ are the transverse components of
$\mbox{\boldmath $P$}_h$ and $\mbox{\boldmath $S$}$ with respect
to the virtual photon momentum $\mbox{\boldmath $q$}=\mbox{\boldmath
$k$}-\mbox{\boldmath $k^\prime$}$.}
\begin{description}
      \item[$\boldsymbol{\phi_h}$, $\boldsymbol{\phi_s}$] 
      Azimuthal angles of the final state
      hadron and the transverse polarization vector of the nucleon with
      respect to the lepton plane.

      \item[$\boldsymbol{P_{hT}}$] Transverse momentum of the final state hadron 
      with respect to
      the virtual photon in the center-of-mass of the virtual photon and the nucleon.

      \item [$\boldsymbol{z}=P_h\cdot P/q\cdot P$] gives the momentum fraction of the 
      final state hadron with respect to the virtual photon.
\end{description}
\end{multicols}

\begin{figure}[h]
  \begin{center}
    \begin{minipage}[t]{0.84\linewidth}
      \includegraphics[width=0.8\linewidth]{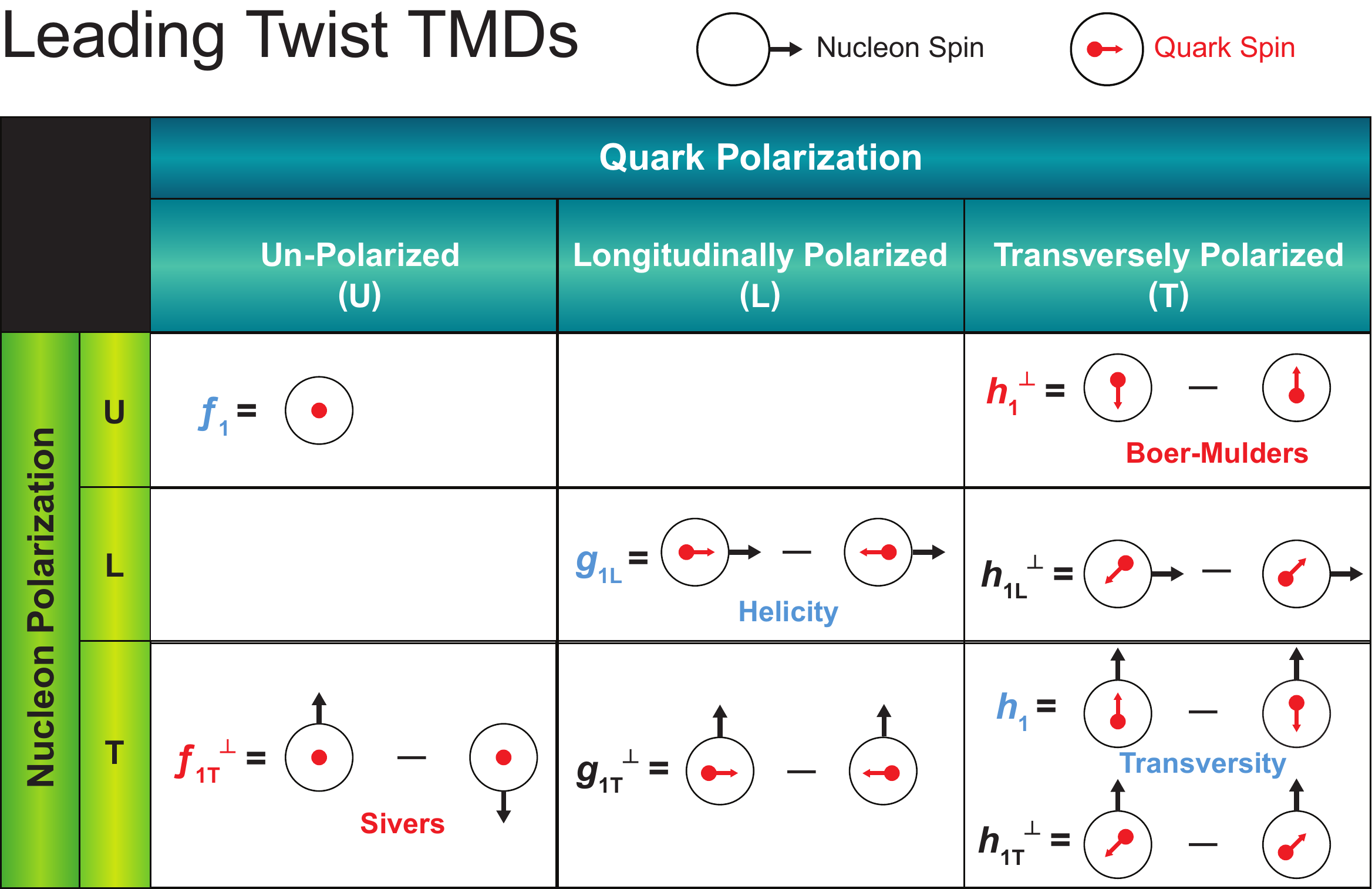}
    \end{minipage}\hfill
    \vspace*{-5.75cm}
    \hspace*{10.3cm}
    \begin{minipage}[t]{0.3\textwidth}
\caption{\label{f:TMDtable} Leading twist TMDs classified according
to the polarizations of the quark (f, g, h) and nucleon (U, L, T).
The distributions $f_{1T}^{\perp,q}$ and $h_1^{\perp,q}$ are called naive-time-reversal-odd
TMDs. For gluons a similar classification of TMDs exists.}
    \end{minipage}
  \end{center}
\end{figure}

The differential SIDIS cross section can be written as a convolution of the transverse 
momentum dependent quark distributions $f(x,k_T)$, fragmentation functions 
$D(z,p_T)$, and a factor for 
a quark or antiquark to scatter off the photon.
At the leading power of $1/Q$, we can probe eight different TMD quark distributions 
as listed in Fig.~\ref{f:TMDtable}. These distributions represent various correlations 
between the transverse momentum of the quark $\mbox{\boldmath $k$}_{T}$, the 
nucleon momentum \mbox{\boldmath $P$}, the nucleon spin \mbox{\boldmath $S$}, and 
the quark spin $\mbox{\boldmath $s$}_q$. 

\newpage \pagecolor{white}

%% file: files_tex/tmd.tex
\section{Confined Motion of Partons in Nucleons: TMDs}
\label{sec:tmd} 

{\large\it Conveners:\ Haiyan Gao and Feng Yuan}

\subsection{Introduction}
\label{subsec:intro-tmd}
\begin{multicols}{2}
DIS is a powerful way to probe the internal structure of nucleons.
After four decades of experiments scattering high energy leptons off
nucleons, our knowledge of the nucleon structure has made impressive
progress.  However, our understanding of the nucleon structure from
inclusive DIS experiments is basically one-dimensional.
From inclusive DIS we ``only" learn about the longitudinal motion of
partons in a fast moving nucleon, whose transverse momenta are not
resolved.  Meanwhile, the past decade has witnessed tremendous
experimental achievements which led to fascinating new insights into
the structure of the nucleon through semi-inclusive hadron production
in DIS (SIDIS) and hard exclusive processes in DIS. These less
inclusive methods enable us to investigate the partonic structure of
the nucleon beyond one-dimensional space. As discussed at the
beginning of this chapter, these developments have stimulated
theoretical advances from a simple parton model description of nucleon
structure to multi-dimensional distributions of partons, including the
generalized parton distributions (GPDs), the transverse momentum
dependent parton distributions (TMDs), and the quantum phase space
Wigner distributions.  The focus of this section is on the TMDs, their
theoretical properties and phenomenological implications, and the
experimental access to them.  TMDs open a new window to understand
some of the most fundamental aspects of QCD.  Several fascinating
topics are related to the study of TMDs:
\end{multicols}
\begin{itemize}

\item {\it 3D-imaging.} The TMDs represent the intrinsic motion of
partons inside the nucleon (confined motion!) and allow reconstruction
of the nucleon structure in momentum space. Such information, when
combined with the analogous information on the parton spatial
distribution from GPDs, leads to a 3-dimensional imaging of the
nucleon.

\item {\it Orbital motion.} Most TMDs would vanish in the absence of
parton orbital angular momentum, and thus enable us to quantify the
amount of orbital motion.

\item {\it Spin-orbit correlations.} Most TMDs and related observables
are due to couplings of the transverse momentum of quarks with the
spin of the nucleon (or the quark).  Spin-orbit correlations in QCD,
akin to those in hydrogen atoms and topological insulators, can
therefore be studied.

\item {\it Gauge invariance and universality.} The origin of some TMDs
and related spin asymmetries, at the partonic level, depend on
fundamental properties of QCD, such as its color gauge invariance.
This leads to clear differences between TMDs in different processes,
which can be experimentally tested.

\end{itemize}

\begin{multicols}{2}
The ``simplest" TMD is the unpolarized function $f_1^q(x,k_T)$, which
describes, in a fast moving nucleon, the probability of finding a
quark carrying the longitudinal momentum fraction $x$ of the nucleon
momentum, and a transverse momentum $k_T=|\mbox{\boldmath $k$}_T|$. It
is related to the collinear (`integrated') PDF by $\int{\rm d}^2
\mbox{\boldmath $k$}_T \, f_1^q(x,k_T)=f_1^q(x)$.  In addition to
$f_1^q(x,k_T)$, there are two other TMDs: $g_{1L}^q(x,k_T)$ and
$h_1^q(x,k_T)$, whose integrals correspond to the collinear PDFs: the
longitudinal polarized structure function discussed in the previous
section and the quark transversity distribution. The latter is related
to the tensor charge of the nucleon.  These three distributions can be
regarded as a simple transverse momentum extension of the associated
integrated quark distributions.  More importantly, the power and rich
possibilities of the TMD approach arise from the simple fact that
$\mbox{\boldmath $k$}_{T}$ is a vector, which allows for various
correlations with the other vectors involved: the nucleon momentum
\mbox{\boldmath $P$}, the nucleon spin \mbox{\boldmath $S$}, and
the parton spin (say a quark, $\mbox{\boldmath $s$}_q$).  Accordingly,
there are eight independent TMD quark distributions as shown in
Fig.~\ref{f:TMDtable}.  Apart from the straightforward extension
of the normal PDFs to the TMDs, there are five TMD quark
distributions, which are sensitive to the direction of ${k_{T}}$, and
will vanish with a simple $k_T$ integral.

Because of the correlations between the quark transverse momentum and
the nucleon spin, the TMDs naturally provide important information on
the dynamics of partons in the transverse plane in {\sl momentum
space}, as compared to the GPDs which describe the dynamics of partons
in the transverse plane in {\sl position space}.  Measurements of the
TMD quark distributions provide information about the correlation
between the quark orbital angular momentum and the nucleon/quark spin
because they require wave function components with nonzero orbital
angular momentum.  Combining the wealth of information from all of
these functions could thus be invaluable for disentangling spin-orbit
correlations in the nucleon wave function, and providing important
information about the quark orbital angular momentum.
\end{multicols}
\vskip -0.4cm
\begin{figure*}[h]
\begin{center} 
\includegraphics[width=0.85\textwidth]{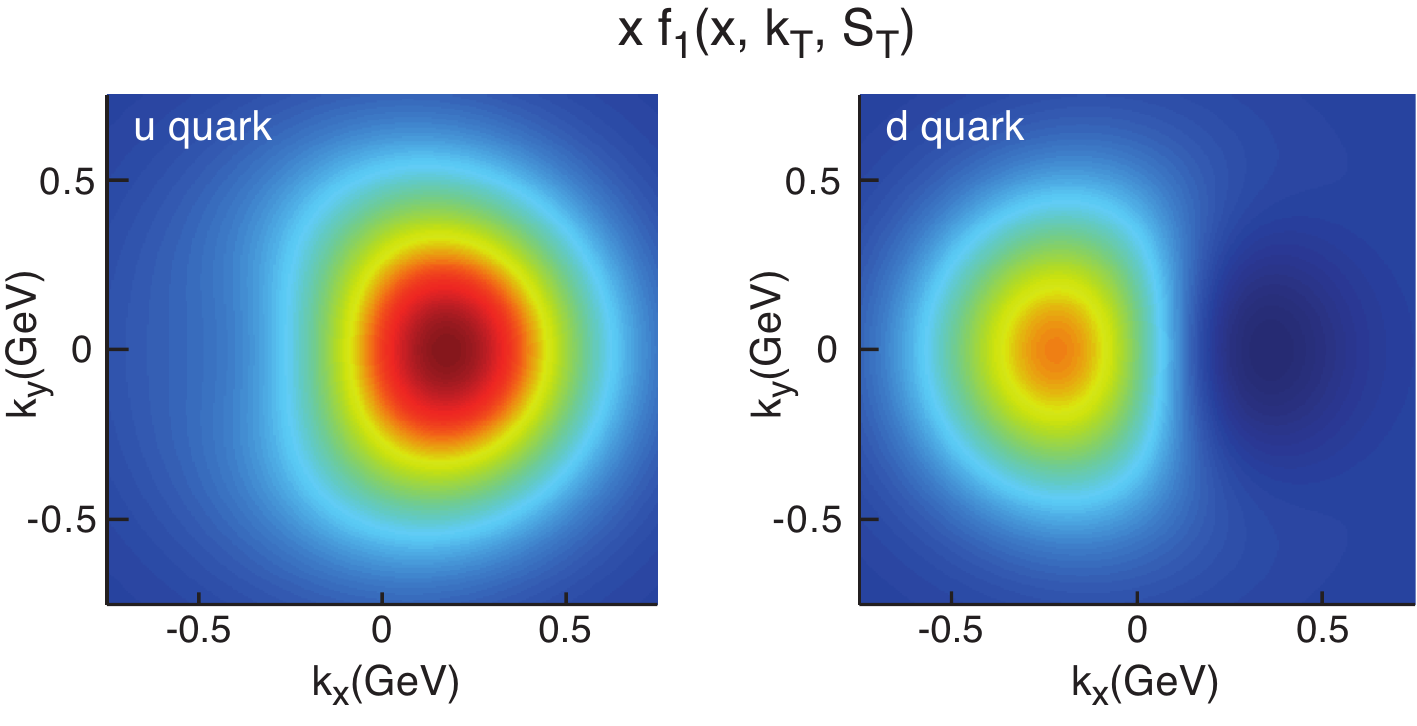}
\vskip -0.4cm  
\caption{ The density in the transverse-momentum plane for unpolarized
quarks with $x = 0.1$ in a nucleon polarized along the $\hat{y}$
direction.  The anisotropy due to the proton polarization is described
by the Sivers function, for which the model of
~\cite{Anselmino:2010bs} is used. The deep red (blue) indicates large
negative (positive) values for the Sivers function.}
\label{tmd-2}
\end{center}
\end{figure*} 
\vskip -0.4cm
\begin{multicols}{2}
One particular example is the quark Sivers function $f_{1T}^{\perp q}$
which describes the transverse momentum distribution correlated with
the transverse polarization vector of the nucleon. As a result, the
quark distribution will be azimuthally asymmetric in the transverse
momentum space in a transversely polarized nucleon.
Figure~\ref{tmd-2} demonstrates the deformations of the up and down
quark distributions.  There is strong evidence of the Sivers effect in
the DIS experiments observed by the HERMES, COMPASS, and JLab Hall A
collaborations~\cite{Airapetian:2004tw,Alekseev:2008aa,Qian:2011py}.
An important aspect of the Sivers functions that has been revealed
theoretically in last few years is the process dependence and the
color gauge
invariance~\cite{Brodsky:2002cx,Collins:2002kn,Belitsky:2002sm,Boer:2003cm}.
Together with the Boer-Mulders function, they are denoted as naive
time-reversal odd (T-odd) functions.  In SIDIS, where a leading hadron is 
detected in coincidence with the
scattered lepton, the quark Sivers function arises due to the exchange
of (infinitely many) gluons between the active struck quark and the
remnants of the target, which is referred to as final state
interaction effects in DIS.  On the other hand, for the Drell-Yan
lepton pair production process, it is due to the initial state
interaction effects.  As a consequence, the quark Sivers and
Boer-Mulders functions differ by a sign in these two processes.  This
non-universality is a fundamental prediction from the gauge invariance
of QCD~\cite{Collins:2002kn}.  The experimental check of this sign
change is currently one of the outstanding topics in hadronic physics,
and Sivers functions from the Drell-Yan process can be measured at RHIC.
\end{multicols}

\subsection{Opportunities for Measurements of TMDs at the EIC}
\label{secI:info-about-TMDs}

\begin{multicols}{2}
To study the transverse momentum dependent parton distributions in
high-energy hadronic processes, an additional hard momentum scale is
essential, besides the transverse momentum, for proper interpretation
of results.  This hard momentum scale needs to be much larger than the
transverse momentum. At the EIC, DIS processes
naturally provide a hard momentum scale: $Q$, the virtuality of the
photon. More importantly, the wide range of $Q^2$ values presents a
unique opportunity to systematically investigate the strong
interaction dynamics associated with the TMDs. Although there has been
tremendous progress in understanding TMDs, without a new lepton-hadron
collider, many aspects of TMDs will remain unexplored --- or at best
be explored only on a qualitative level. Existing facilities either
suffer from a much too restricted kinematic coverage or from low
luminosity or from both.

The SIDIS measurement discussed below is the necessary method to access
TMDs. We define two planes in SIDIS: the lepton plane and the hadron
plane, as shown in Fig.~\ref{f:anglestrento}, which allows us to study
different angular dependences in the hadron production cross
sections. These angular distributions are important to extract the
TMDs since each of them has a unique angular dependence.  Precision
measurements of the various angular modulations are only possible with
a comprehensive and hermetic detector.  With such a detector and the
EIC's ability to provide a wide kinematic range and high luminosity,
we see the following opportunities for measurements at an EIC that
would be impossible in current experiments:
\end{multicols}

\begin{itemize}

\item High precision quantitative measurements of all the quark TMDs
in the valence region, with the ability to go to sufficiently large
values of $Q^2$ in order to suppress potential higher twist
contaminations;

\item First-ever measurements of the TMDs for anti-quarks and gluons;

\item Multi-dimensional representations of the observables leading to
TMDs;

\item Systematic studies of perturbative QCD techniques (for
polarization observables) and studies of QCD evolution properties of
TMDs;
	        
\item The transition between the non-perturbative low transverse momentum
region and perturbative high transverse momentum region for both
polarized and unpolarized collisions due to a wide range of kinematic
coverage.
\end{itemize}

\begin{multicols}{2}
The above discussions apply to all of the eight TMD quark
distributions listed in Fig.~(\ref{f:TMDtable}). The rich physics
covered by the TMD quark and gluon distribution functions can be
thoroughly investigated at the EIC with a dedicated detector.  In the
following subsections, we will take semi-inclusive DIS as an example
for the quark Sivers function and di-hadron production for the gluon
Sivers function and highlight the impact the EIC could have on these
measurements.
\end{multicols}

\subsubsection{Semi-inclusive Deep Inelastic Scattering}

\begin{multicols}{2}
The TMDs are measured using SIDIS processes.  In such
reactions, the hadron, which results from the fragmentation of a
scattered quark, ``remembers" the original motion of the quark,
including its transverse momentum.  SIDIS depends on six kinematic
variables. In addition to the variables for inclusive DIS, $x$, $y =
(P \cdot q) / (P \cdot l)$, and the azimuthal angle $\phi_S$
describing the orientation of the target spin vector for transverse
polarization, one has three variables for the final state hadron,
which we denote by $z = (P \cdot P_h) / (P \cdot q)$ (longitudinal
hadron momentum fraction), $P_{hT}$ (magnitude of transverse hadron
momentum), and the angle $\phi_h$ for the orientation of
$\mbox{\boldmath $P$}_{hT}$ (see Fig.~\ref{f:anglestrento}). In the
one-photon exchange approximation, the SIDIS cross-section can be
decomposed in terms of structure functions. Each of them is
characterized by the unique azimuthal angular modulation in the
differential cross-sections. The extraction of these structure functions
will give access to all of the leading TMD quark distributions listed
in Fig.~(\ref{f:TMDtable}).
\end{multicols}

For example, for the spin-average and single-spin dependent
contributions, we have
\begin{equation} \frac{d\sigma}{dx\, dy\, d\phi_S \,dz\, d\phi_h\,
d P_{hT}^2} \propto F_{UU ,T} + |\mbox{\boldmath $S$}_\perp|
\sin(\phi_h-\phi_S) F_{UT,T}^{\sin\left(\phi_h -\phi_S\right)}\, +...
\label{e:crossmaster}
\end{equation} 

\begin{multicols}{2}
\noindent
where $F_{UU}$ represents the spin-average structure
function depending on the unpolarized quark distribution
$f_1^q(x,k_T)$, and $F_{UT}$ depends on the quark Sivers function
$f_{1T}^{\perp q}(x,k_T)$.  For TMD studies, one is interested in the
kinematic region defined by $P_{hT} \ll Q $, for which the structure
functions can be written as certain convolutions of TMDs.  To extract
the quark Sivers function, we measure the $\sin(\phi_h-\phi_s)$
modulation of the single transverse spin asymmetry (SSA), which is
defined by the ratio of the two cross-section terms in
Eq.~(\ref{e:crossmaster}).  This asymmetry depends on four kinematics:
$Q^2$, $x_B$, $z_h$, $P_{hT}$.  
\begin{figure*}[t]
\centering \includegraphics[width=0.75\textwidth]{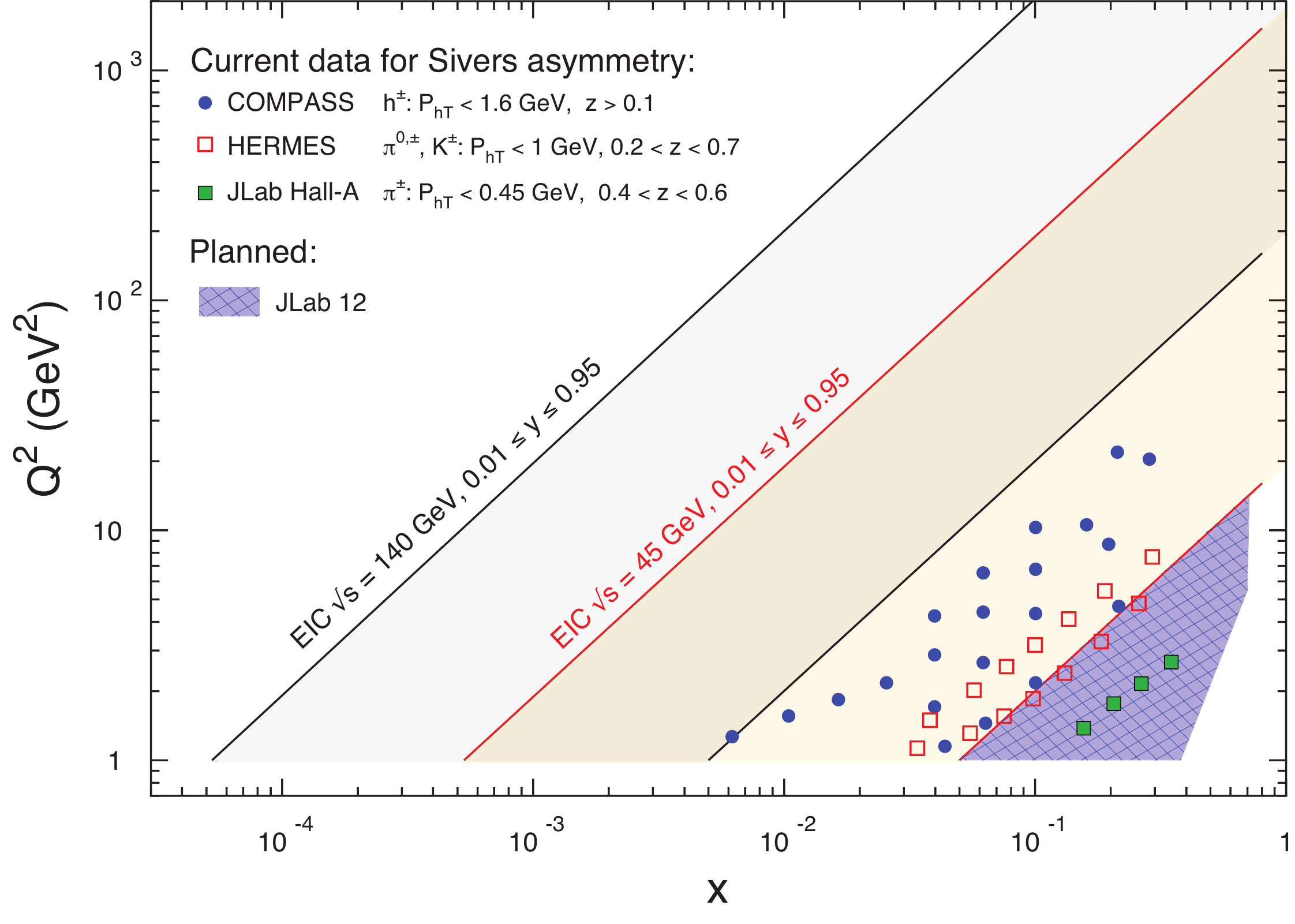}
\caption{\label{fig:TMD-EIC-kine-ranges} Kinematic coverage in $x$ and
$Q^2$ for the EIC compared to the coverage of the planned JLab12
experiment. The kinematics of the existing experimental measurements
are also shown for comparison.}
\end{figure*} 
%
A systematic and detailed study of the Sivers function, and TMDs in
general, can only be performed on the basis of precise spin- and
azimuthal-asymmetry amplitude measurements in SIDIS over a wide
kinematic range. In Fig.~(\ref{fig:TMD-EIC-kine-ranges}), we compare
the $x$-$Q^2$ coverage of the HERMES, COMPASS, and JLab 12 GeV upgrade
with the coverage of an EIC.  The wide kinematic coverage puts the EIC
in the unique position of accessing the valence region at much larger
$Q^2$ than current and near-future experiments while also accessing
low-$x$ down to values of about $10^{-5}$, where sea quarks and gluons
could be studied in detail.  The expected high luminosity will also
allow for a fully differential analysis over almost the entire
kinematic range of $x$, $Q^2$, $z$ and $P_{hT}$, which is vital
for phenomenological analyses.

\begin{figure*}[t]
\begin{center}
\includegraphics[width=0.95\textwidth]{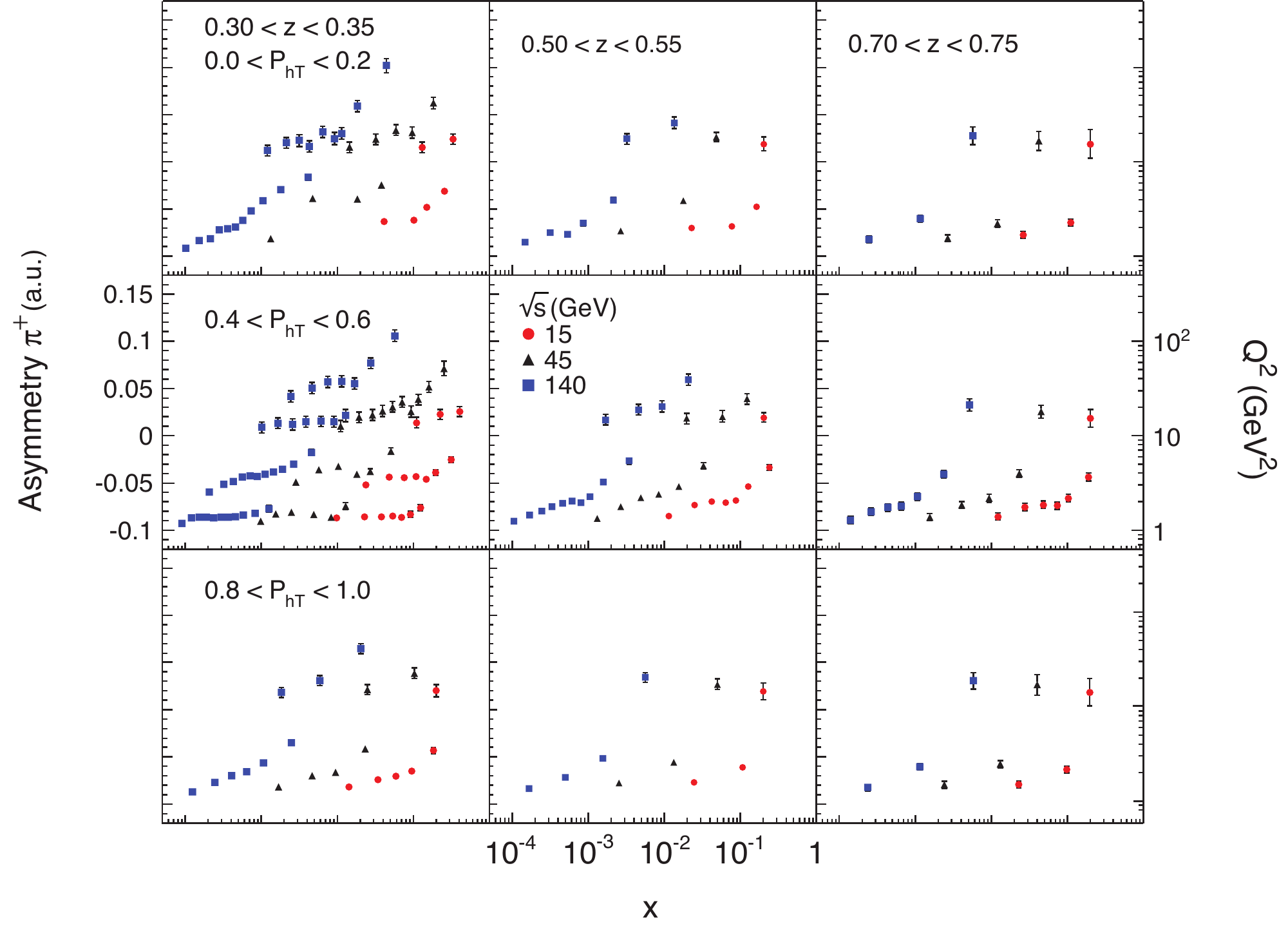}
\caption{\label{fig:TMD-asymm-simulation-pip} Four-dimensional
representation of the projected accuracy for $\pi^+$ production in
semi-inclusive DIS off the proton.  Each panel corresponds to a
specific $z$ bin with increasing value from left to right and a
specific $P_{hT}$ bin with increasing value from top to bottom, with
values given in the figure.  The position of each point is according
to its $Q^2$ and $x$ value, within the range $0.05 < y < 0.9$.  The
projected event rate, represented by the error bar, is scaled to the
(arbitrarily chosen) asymmetry value at the right axis.  Blue squares,
black triangles and red dots represent the $\sqrt{s} = 140 $ GeV,
$\sqrt{s} = 45 $ GeV and $\sqrt{s} = 15 $ GeV EIC configurations,
respectively.  Event counts correspond to an integrated luminosity of
10 fb$^{-1}$ for each of the three configurations.}
\label{fig:proton_pip}
\end{center}
\end{figure*}

In the following, we illustrate the expected impact of data
from the EIC using the parameterization from
Ref.~\cite{Anselmino:2010bs} as an arbitrarily chosen model of the
Sivers function.  This parameterization, denoted $theor_i = F(x_i,
z_i, P_{hT}^i, Q^2_i; {\bf a_0})$ with the $M$ parameters ${\bf a_0} =
\{a_1^0, ..., a_M^0\}$ fitted to existing data, serves to generate a
set of pseudo-data in each kinematic bin $i$.  In each $x_i$, $Q^2_i$,
$z_i$ and $P^i_{hT}$ bin, the obtained values, $value_i$, for the
Sivers function are distributed using a Gaussian smearing with a width
$\sigma_i$ corresponding to the simulated event rate at the
center-of-mass energy of $\sqrt{s} = 45$ GeV obtained with an
integrated luminosity of 10 fb$^{-1}$.  To illustrate the achievable
statistical precision, the event rate for the production of $\pi^{\pm}$
in semi-inclusive DIS was used, see, for example,
Fig.~\ref{fig:proton_pip}.

\begin{figure*}[t!]
\begin{center} \hspace*{-0.5cm} \mbox{
\includegraphics[width=0.5\textwidth,angle=0]{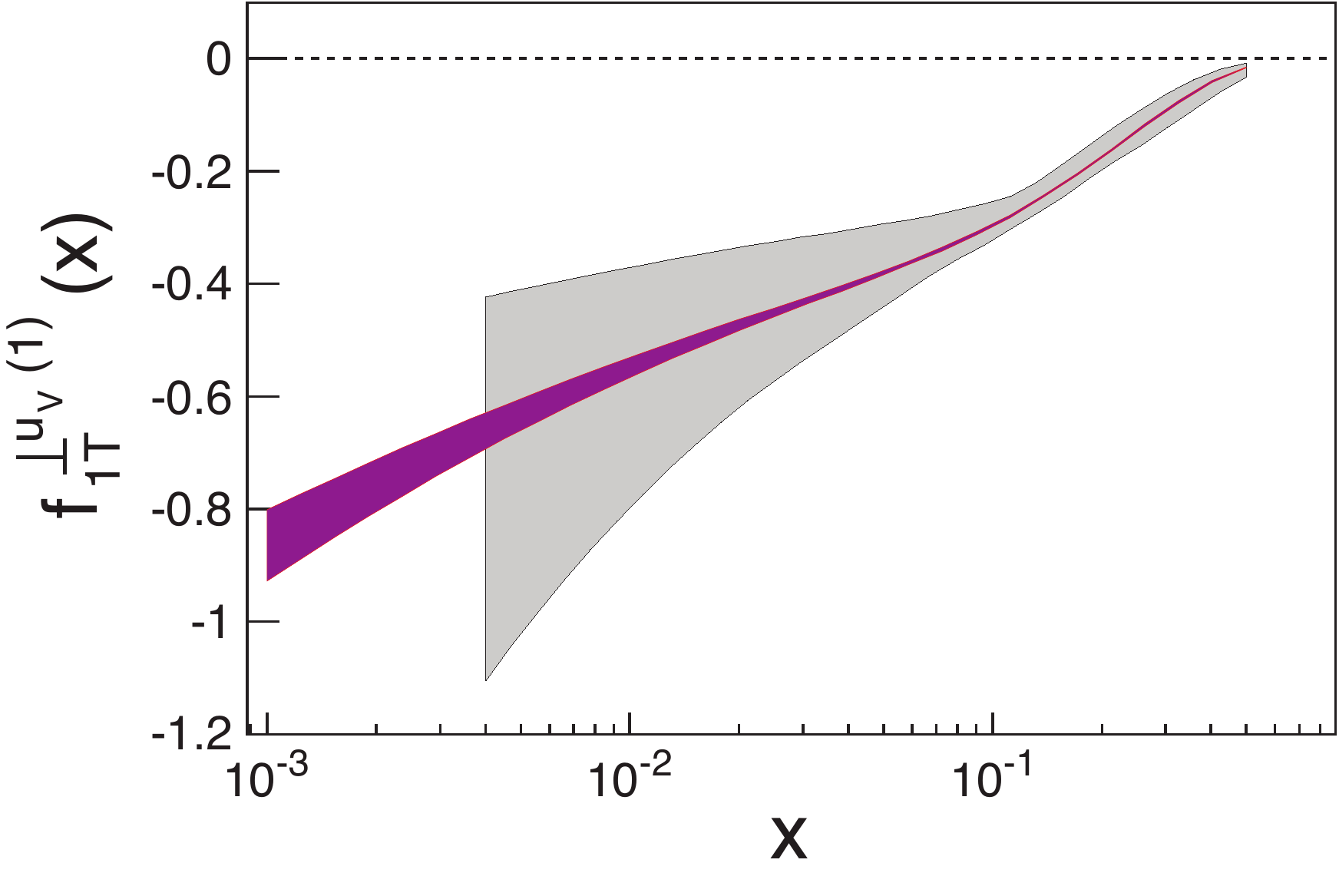}
\includegraphics[width=0.5\textwidth,angle=0]{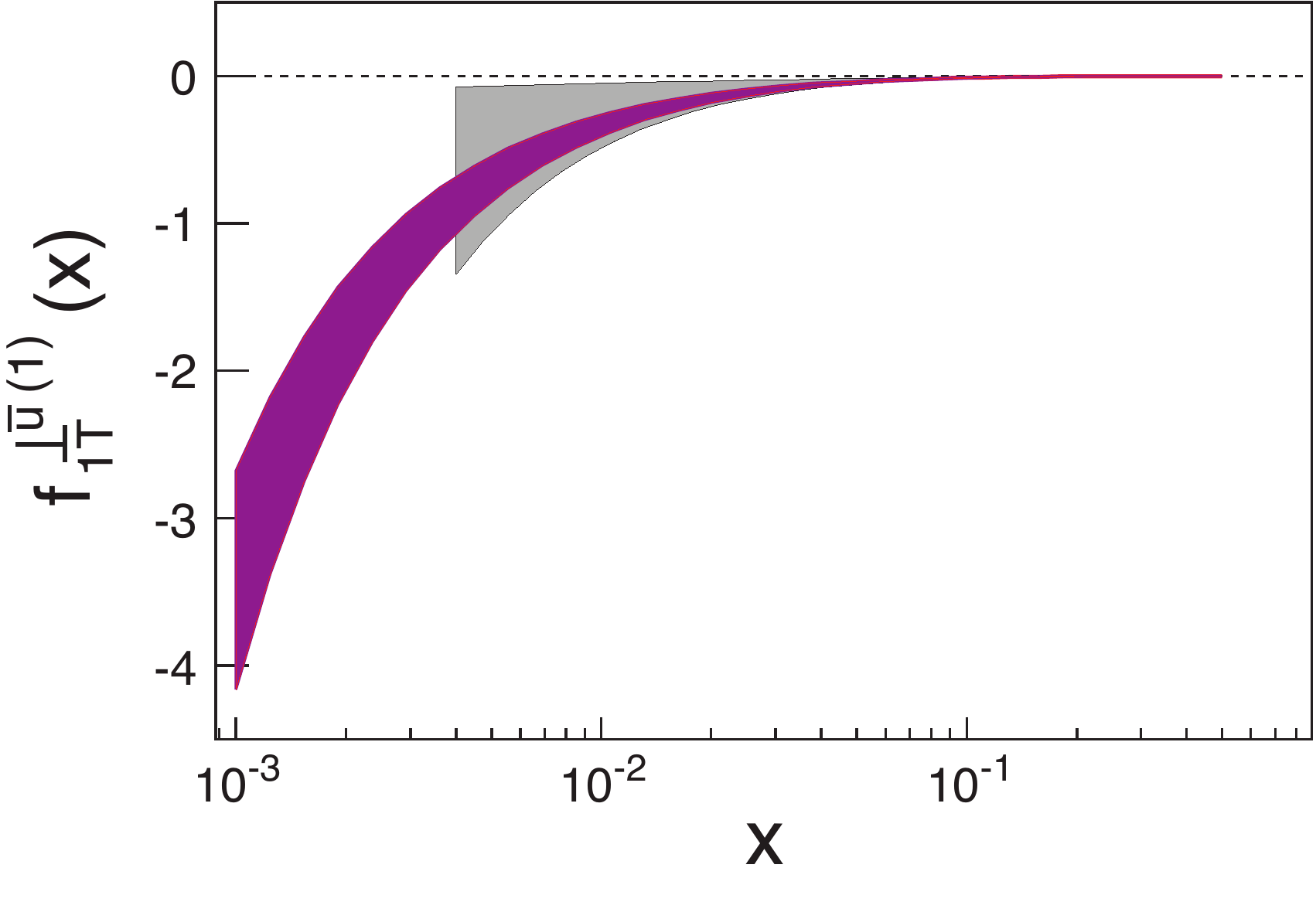}
}
\caption{\label{fig:prokudin_sivers_eic_low_energy} Comparison of the
precision (2-$\sigma$ uncertainty) of extractions of the Sivers
function for the valence (left) $u_v = u-\bar u$ and sea (right) $\bar
u$ quarks from currently available data~\cite{Anselmino:2010bs} (grey
band) and from pseudo-data generated for the EIC with energy setting
of $\sqrt{s} = 45$ GeV and an integrated luminosity of 10 fb$^{-1}$
(purple band with a red contour).  The uncertainty estimates are for
the specifically chosen underlying functional form.  }
\end{center}
\end{figure*}

This new set of pseudo-data was then analysed like the real data in
Ref.~\cite{Anselmino:2010bs}. Fig.~\ref{fig:prokudin_sivers_eic_low_energy}
shows the result for the extraction of the Sivers function for the
valence and sea up quarks.  Similar results are obtained for the down
quarks as well.  The central value of $f_{1T}^{\perp u}$, represented
by the red line, follows by construction the underlying model.  The
2-sigma uncertainty of this extraction, valid for the specifically
chosen functional form, is indicated by the purple band.  This
precision, obtainable with an integrated luminosity of 10 fb$^{-1}$,
is compared with the uncertainty of the extraction from existing data,
represented by the light grey band.  It should be emphasized that our
current knowledge is restricted to only a qualitative picture of the
Sivers function 
and the above analysis did not take into account the model dependence and the
associated theoretical uncertainties.  With the anticipated large
amount of data (see Fig.~\ref{fig:TMD-asymm-simulation-pip} for a
modest integrated luminosity 10~fb$^{-1}$), we can
clearly see that the EIC will be a powerful facility enabling access
to TMDs with unprecedented precision, and particularly in the
currently unexplored sea quark region.  
This precision is not only crucial for the
fundamental QCD test of the sign change between the Sivers asymmetries
in the DIS and Drell-Yan processes, but also important to investigate
the QCD dynamics in the hard processes in SIDIS, such as the QCD
evolution and resummation, matching between the TMD factorization and
collinear factorization approaches, etc.  Meanwhile, an exploration of
the sea quark Sivers function will provide, for the first time,
unique information on the spin-orbital correlation in the small-$x$
region.  
The right panel of Fig.~\ref{fig:ptomography} in the Introduction (Section 1.2) showed
the kinematic reach of the EIC which would enable a measurement of the
transverse momentum profile of the quark Sivers function over a wide
range in $x$, e.g. from the valence to the sea quark region. Note that
Fig.~\ref{fig:ptomography} showed the total up quark Sivers function,
while Fig.~\ref{fig:prokudin_sivers_eic_low_energy} shows the valence
and the sea quarks separately.

Here, we emphasize the importance of the high $Q^2$ reach of the EIC
for SIDIS measurements. Most of the existing experiments focus
on the $Q^2$ range of a few $GeV^2$. The EIC will, for
the first time, reach $Q^2$ values up to hundreds and more $GeV^2$.
This will provide an unique opportunity to investigate the 
scale evolution of the Sivers asymmetries, which has attracted
strong theoretical interests in the last few years 
\cite{Aybat:2011ge,Aybat:2011ta,Sun:2013dya,Sun:2013hua,Echevarria:2014xaa,Su:2014wpa}.
As a leading
power contribution in the spin asymmetries, the associated
energy evolution unveils the underlying strong interaction
dynamics in the hard scattering processes. The embedded 
universality and factorization property of the TMDs can only be 
fully investigated at the EIC with the planned kinematic coverage in $Q^2$.
In particular, the theory calculations including evolution effects
agree with the current constraints on the quark Sivers
function presented in Fig.~2.16, while they do differ at higher values of $Q^2$ 
\cite{Aybat:2011ge,Aybat:2011ta,Sun:2013dya,Sun:2013hua,Echevarria:2014xaa,Su:2014wpa}.
Moreover, a recent study has shown that at the kinematics
of HERMES and COMPASS, the leading order SIDIS suffers
significant power corrections, which however will diminish 
at higher $Q^2$ \cite{Su:2014wpa}. This makes the EIC the only machine to be
able to establish the leading partonic picture of the TMDs
in SIDIS. 

The kinematic reach of the EIC also allows the measurement of
physical observables over a wide transverse momentum range.  This is
particularly important to understand the underlying mechanism that
results in single spin asymmetries.  Recent theoretical developments
have revealed that both the transverse-momentum-dependent Sivers
mechanism and the quark-gluon-quark correlation collinear mechanism
describe the same physics in the kinematic regions where both
approaches apply~\cite{Ji:2006ub,Bacchetta:2008xw}.  The only way to
distinguish between the two and understand the underlying physics is
to measure them over wide $p_{T}$ ranges.  The high luminosities at
the EIC machine could provide a golden opportunity to explore and
understand the mechanism of the transverse spin asymmetries.  In
addition, with precision data in a large range of transverse momentum,
we shall be able to study the strong interaction dynamics in the
description of large transverse momentum observables and investigate
the transition between the non-perturbative low transverse momentum
region and the perturbative high transverse momentum region.
\end{multicols}

\newpage
\subsubsection{Access to the Gluon TMDs}

\begin{multicols}{2}
Beyond the gluon helicity measurements described in Sec.~\ref{sec:helicity},
the gluonic orbital angular momentum contribution
would be studied in hard exclusive meson production processes at the
EIC.  The transverse momentum dependent gluon distribution can provide
complementary information on the spin-orbital correlation for the
gluons in the nucleon.  Just as there are eight TMDs for quarks, there
exist eight TMDs for gluons~\cite{Meissner:2007rx}.  Experimentally,
the gluon TMDs --- in particular, the gluon Sivers function --- are
completely unexplored so far and will likely not be probed at existing
facilities.  In addition, toward the small-$x$ region, the TMD gluon
distributions have intimate connections to the saturation phenomena
discussed in Sec.~\ref{HGDN}, where the gluon distributions are
fundamental objects as well. Explorations of the TMD gluon
distributions (experimentally and theoretically) shall offer deep
insight into the QCD dynamics evolving from the valence region to the
sea region.

\begin{figure}[h]
  \begin{center}
    \begin{minipage}[t]{0.84\linewidth}
      \includegraphics[width=0.7\linewidth]{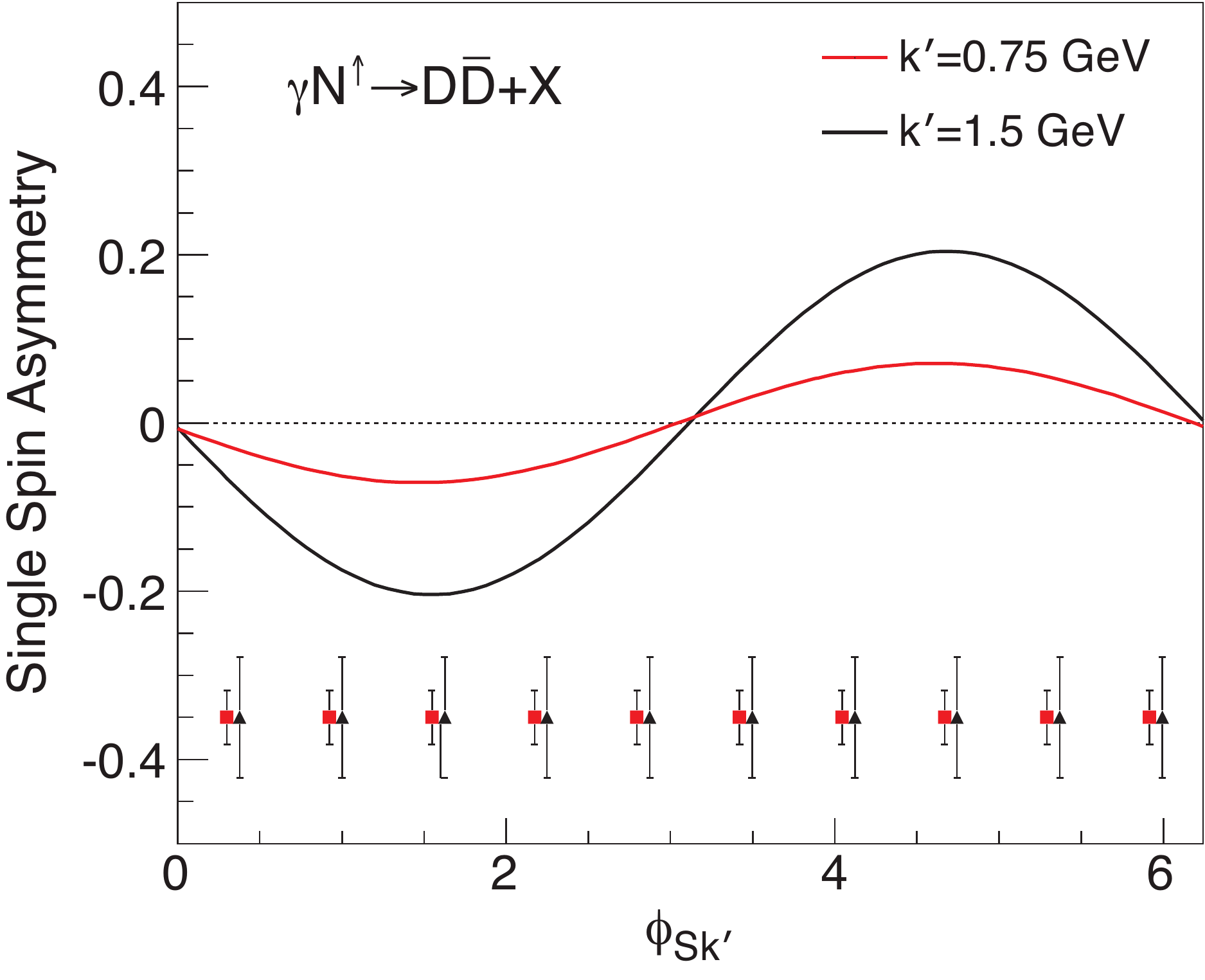}
    \end{minipage}\hfill
    \vspace*{-8.00cm}
    \hspace*{9.6cm}
    \begin{minipage}[t]{0.35\textwidth}
\caption{\label{gsivers} The single transverse spin asymmetry for
$\gamma^{\ast} N^{\uparrow}\to D^0 \bar{D}^0 + X$, where $\phi$ is the
azimuthal angle between the total transverse momentum $k_\perp'$ of
the $D$-$\bar D$ pair and the transverse polarization vector $S_\perp$
of the nucleon.  The asymmetries and the experimental projections are
calculated for two different $k_\perp'=0.75, 1.5 {\rm GeV}$ as
examples.  The kinematics are specified by $\langle W \rangle= 60
\gev$, $\langle Q^2\rangle = 4 \gev^2$.}
    \end{minipage}
  \end{center}
\end{figure}

Many processes in DIS can be used to probe the transverse momentum
dependent gluon distributions, for example, di-jet/di-hadron
production, heavy quark, and quarkonium production.  We take one
particular example: heavy meson pair ($D$-$\bar D$) production in DIS.
In this process, $D$ and $\bar D$ are produced in the current
fragmentation region: $\gamma^{\ast }N^{\uparrow} \rightarrow D(k_{1})
+ \bar D(k_{2})+X $, where $N$ represents the transversely polarized
nucleon, $D$ and $\bar D$ are the two mesons with momenta $k_1$ and
$k_2$, respectively.  Similar to the Sivers effect in semi-inclusive
hadron production in DIS discussed above, the gluon Sivers function
will introduce an azimuthal asymmetry correlating the total transverse
momentum $k_\perp'=k_{1\perp}+k_{2\perp}$ of the $D$-$\bar D$ pair
with the transverse polarization vector $S_\perp$ of the nucleon. In
experiment, this results in a single spin azimuthal asymmetry
depending on the azimuthal angle between $k'_\perp$ and $S_\perp$. In
Fig.~\ref{gsivers}, we show the sensitivity of the measurement of the
asymmetry in a typical kinematic configuration of the EIC machine \cite{Burton:2012ug}.
The two theory curves are based on a model calculation from
Ref.~\cite{Boer:2011fh}. The estimate of the projected error bars
comes from a simulation of the integrated luminosity of $100$ fb$^{-1}$.
Since the gluon Sivers effects has never been measured,
this will be the first measurement of such an effect. Beside the
$D$-$\bar D$ correlation, the di-hadron/di-jet correlations in DIS can
also give us an independent handle on the study of the gluon Sivers
function.
\end{multicols}

\newpage
\subsection{Summary}

The EIC will be a unique facility to systematically investigate the
transverse momentum dependent parton distributions
comprehensively. While the measurements of quark TMDs have begun in
fixed target experiments, the gluon TMDs can only be studied at an
EIC, and such studies would be unprecedented.  The QCD dynamics
associated with the transverse momentum dependence in hard processes
can be rigorously studied at the EIC because of its wide kinematic
coverage. The comparison of the Sivers single spin asymmetry and
Boer-Mulders asymmetry between DIS and Drell-Yan processes can provide
an important test of the fundamental prediction of QCD. In summary, we
list these important science questions to be addressed at the EIC in
Table \ref{tab:TMD-sciencematrix}.

\begin{table}[htb]
\begin{tabular}{|c|c|c|c|c|} \hline
Deliverables & Observables & What we learn \\
\hline\hline
Sivers $\&$ &  SIDIS with  & Quantum Interference \& Spin-Orbital correlations \\
unpolarized &Transverse& 3D Imaging of quark's motion: valence + sea \\
TMD quarks  & polarization; & 3D Imaging of gluon's motion \\
and gluon& di-hadron (di-jet)  &   QCD dynamics in a unprecedented $Q^2$ ($P_{hT}$) range\\
\hline
\hline
Chiral-odd &  SIDIS with & 3${}^{\rm rd}$ basic quark PDF: valence + sea, tensor charge \\
functions: &Transverse& Novel spin-dependent hadronization effect  \\
Transversity; & polarization & QCD dynamics in a chiral-odd sector \\
Boer-Mulders &  & with a wide $Q^2$ ($P_{hT}$) coverage \\
\hline
\end{tabular}
\caption{\label{tab:TMD-sciencematrix} 
Science Matrix for TMD: 3D structure in transverse momentum space: (upper) the golden
measurements; (lower) the silver measurements.
}
\end{table}

%% file: files_tex/sidebarGPD.tex
\newpage
\setlength{\oddsidemargin}{0em}
\setlength{\textwidth}{6in}
\setlength{\topmargin}{-0.55in}
\setlength{\textheight}{9in}
\definecolor{lightgray}{rgb}{0.85,0.85,0.85}
\definecolor{LightYellow}{cmyk}{0.0,0.0,0.1,0.0}

\pagecolor{LightYellow}

\phantomsection
\label{sdbar:GPD}

\vspace*{-1.00cm}
\begin{center}
{\textbf {\textit {\textcolor{blue}{\Large Exclusive Processes and Generalized Parton Distributions}}}}
 \line(1,0){435}
\end{center}

\begin{figure}[b]
\begin{center}
\includegraphics[width=0.74\textwidth]{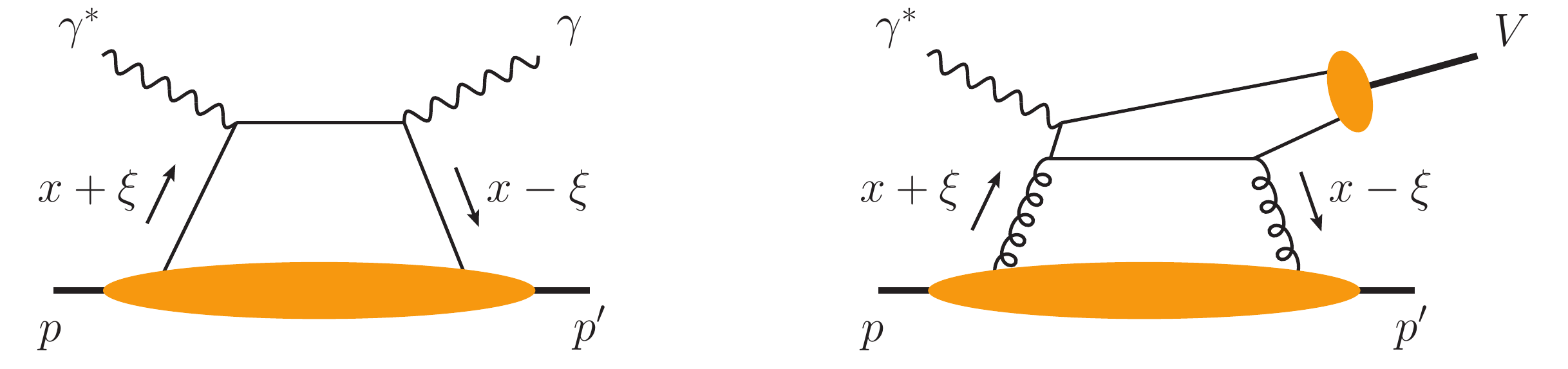}
\end{center}
\vspace*{-0.5cm}
\caption{\label{fig:gpd-graphs} Graphs for deeply virtual Compton scattering (left) 
and for exclusive vector meson production (right) in terms of generalized parton 
distributions, which are represented by the lower blobs. The upper filled oval in the 
right figure represents the meson wave function.}
\end{figure}

Generalized parton distributions (GPDs) can be extracted from suitable
exclusive scattering processes in $e$+$p$ collisions.  Examples are deeply
virtual Compton scattering (DVCS: $\gamma^* + p\to \gamma + p$) and the production
of a vector meson ($\gamma^* + p\to V + p$).  The virtual photon is provided
by the electron beam, as usual in deep inelastic scattering processes
(see the Sidebar on page~\pageref{sdbar:DIS}). 
GDPs depend on three kinematical variables and a resolution scale:

\begin{multicols}{2}
\begin{itemize}
\item $x+\xi$ and $x-\xi$ are longitudinal parton momentum fractions with
  respect to the average proton momentum $(p+p')/2$ before and after the
  scattering, as shown in Figure~\ref{fig:gpd-graphs}.

  Whereas $x$ is integrated over in the scattering amplitude, $\xi$ is
  fixed by the process kinematics. For DVCS one has $\xi = x_B/(2-x_B)$
  in terms of the usual Bjorken variable $x_B = Q^2 /(2p\cdot q)$.  For
  the production of a meson with mass $M_V$ one finds instead $\xi =
  x_V/(2-x_V)$ with $x_V = (Q^2 + M_V^2) /(2p\cdot q)$.
\item The crucial kinematic variable for parton imaging is the transverse
  momentum transfer $\boldsymbol{\Delta}_T = \boldsymbol{p}'_T -
  \boldsymbol{p}_T$ to the proton.  It is related to the invariant
  square $t=(p'-p)^2$ of the momentum transfer by
         $t = -(\boldsymbol{\Delta}_T^2 + 4\xi^2 M^2) /(1-\xi^2)$,
  where $M$ is the proton mass.
\item The resolution scale is given by $Q^2$ in DVCS and light meson
  production, whereas for the production of a heavy meson such as the
  $\jpsi$ it is $M_{\smash{\jpsi}}^2 + Q^2$.
\end{itemize}
\end{multicols}

Even for unpolarized partons, one has a nontrivial spin structure,
parameterized by two functions for each parton type.  $H(x,\xi,t)$ is
relevant for the case where the helicity of the proton is the same
before and after the scattering, whereas $E(x,\xi,t)$ describes a proton
helicity flip.  For equal proton four-momenta, $p=p'$, the distributions
$H(x,0,0)$ reduce to the familiar quark, anti-quark and gluon densities
measured in inclusive processes, whereas the forward limit $E(x,0,0)$ is
unknown.

Weighting with the fractional quark charges $e_q$ and integrating over
$x$, one obtains a relation with the electromagnetic Dirac and Pauli form
factors of the proton:
\begin{align}
\sum_q e_q \int dx\, H^q(x,\xi,t) &= F^p_1(t) \,,
&
\sum_q e_q \int dx\, E^q(x,\xi,t) &= F^p_2(t)
\end{align}
and an analogous relation to the neutron form factors.  At small $t$ the
Pauli form factors of the proton and the neutron are both large, so that
the distributions $E$ for up and down quarks cannot be small everywhere.

\vspace*{-0.75cm}

\newpage \pagecolor{white}

%% file: files_tex/exclusive.tex
\section{Spatial Imaging of Quarks and Gluons}
\label{sec:exclusive}

{\large\it Conveners:\ Markus Diehl and Franck Sabati\'e}

\subsection{Physics Motivations and Measurement Principle}
\label{sec:excl-intro}

\begin{multicols}{2}
\noindent{\bf Spatial imaging} \
Elastic electron-nucleon scattering has played a major role in our
understanding of strong interactions ever since the Hofstadter experiment
showed that protons and neutrons are not point-like particles.
Measurements of the electromagnetic nucleon form factors have become ever
more precise \cite{Arrington:2011kb} and give detailed information about
the spatial distribution of electric charge and magnetization in the
nucleon.  Further information (albeit with less accuracy) can be obtained
from neutral and charged weak currents.  However, elastic scattering does
not reveal the distribution of gluons, which carry only color charge, and
it is not selectively sensitive to sea quarks.

Hard exclusive scattering processes bring the idea of imaging to a new
qualitative level by probing the transverse distribution of quarks,
anti-quarks and gluons as a function of their longitudinal momentum in the
nucleon.  One may regard this as a tomography of the nucleon, with
two-dimensional spatial images being taken for different ``slices'' of
the parton momentum fraction, $x$.  In different terms, one maps out in
this way the $2+1$ dimensional structure of the nucleon, with two
dimensions in space and one in momentum.

Such spatial images of partons can provide insight into the fundamental
questions about QCD dynamics inside hadrons spelled out in
Sec.~\ref{sec:parton-intro}.  In particular, quantifying the difference
in the distributions of quarks and gluons will shed light on their
dynamical interplay, and the dependence of the transverse distribution of
quarks on $x$ will reveal to what extent sea and
valence quarks have different or similar characteristics.  As the size of
effects that can be expected is not huge, measurements with high precision
are crucial to uncover them.

We will show that with a suitable setup of detectors and the interaction
region, the EIC will be able to probe partons at transverse distances $b_T$ up
to about $1.5 \fm$ or even higher.  In this region, there are definite
predictions \cite{Strikman:2003gz,Strikman:2009bd} for the impact
parameter distribution $f(x,\boldsymbol{b}_T)$ of partons, namely an
exponential falloff in $b_T$ (akin to the one produced by a Yukawa
potential) with a characteristic length that depends on $x$ and is of
order $1/(2 m_\pi) \approx 0.7 \fm$.  This behavior results from quantum
fluctuations with virtual pions at large $b_T$, sometimes referred to as the
``pion cloud'' of the nucleon.  The characteristics of these fluctuations
are a direct consequence of the breakdown of chiral symmetry in QCD and
can be computed using effective field theory methods.  From a different
point of view, one may hope that the structure of the proton of distances
on the femtometer scale will eventually help us to better understand the
mechanism of confinement.

Although the spatial imaging of partons puts highest demands on
experiment, the underlying physical principle is quite simple.  In
suitable exclusive processes one can measure the difference
$\boldsymbol{\Delta}_T$ between the transverse momentum of the proton in the
initial and the final state.  A two-dimensional Fourier transform converts
the distribution of $\boldsymbol{\Delta}_T$ into the spatial distribution of
partons in the
transverse plane \cite{Burkardt:2002hr,Diehl:2002he}.  This bears some
similarity with X-ray diffraction, where a spatial image of a crystal is
obtained by a Fourier transform from the deflection of X-rays.

To reconstruct the longitudinal momentum information in nucleon tomography
is less easy.  In exclusive processes suitable for parton imaging, the
longitudinal momentum of the parton before and after the scattering is in
fact not the same.  The generalized parton distributions that describe the
nucleon structure in these processes thus depend on two momentum fractions,
$x+\xi$ and $x-\xi$ as shown in the Sidebar on page~\pageref{sdbar:GPD}.
Whereas $\xi$ 
can be directly measured via the longitudinal momentum transferred to the
proton, $x$ is integrated over in the expression of the scattering
amplitude.  However, one finds that the typical values of $x$ in this
integral are of order $\xi$.  In the first instance, exclusive
measurements thus yield integrals over GPDs that can be turned into the
distribution of partons with a transverse position $\boldsymbol{b}_T$ in
the proton and with momentum fractions smeared around $\xi$.

Information about the separate dependence on $x$ and $\xi$ is contained in
the dependence of GPDs on the resolution scale $Q^2$, given that a change in
resolution scale changes their $x$ dependence in a calculable way while
leaving $\xi$ and $\boldsymbol{\Delta}_T$ untouched.  To reconstruct the $x$
dependence of GPDs by measuring the $Q^2$ dependence of exclusive
processes at given $\xi$ is challenging because the relevant variation in
$Q^2$ is only logarithmic.  To be successful, such a program requires
precise data in as wide a range of $Q^2$ and $\xi$ as possible.
\end{multicols}


\noindent{\bf Orbital Motion and Angular Momentum} 
\begin{multicols}{2}
Exclusive processes with polarized beams open up unique possibilities to
study spin-orbit correlations of quarks and gluons in the nucleon.  A
correlation of particular interest is the shift in the transverse
distribution of partons induced by transverse polarization
$\boldsymbol{S}_T$ of the proton, which has the form \cite{Burkardt:2002hr}
\end{multicols}
\vspace*{-0.25cm}
\begin{align}
  \label{eq:shift-from-E}
f^{\Uparrow}(x, \boldsymbol{b}_T) = f(x, \boldsymbol{b}{}_T^2)
   + \frac{(\boldsymbol{S}_T \times \boldsymbol{b}_T)^z}{M}\,
     \frac{\partial}{\partial \boldsymbol{b}{}_T^2}\;
     e(x, \boldsymbol{b}{}_T^2) \,,
\end{align}
\begin{multicols}{2}
\noindent
where $M$ is the proton mass.  The distributions $f(x,
\boldsymbol{b}{}_T^2)$ and $e(x, \boldsymbol{b}{}_T^2)$, which give the impact
parameter distribution of unpolarized partons and its polarization induced
shift, are respectively obtained by a two-dimensional Fourier transform
from the generalized parton distributions $H(x,\xi,t)$ and $E(x,\xi,t)$ at
$\xi=0$ (see the Sidebar on page~\pageref{sdbar:GPD}).  
This shift is the position space 
analog of the Sivers effect discussed in Sec.~\ref{sec:tmd},
where transverse proton polarization induces an anisotropy in the
transverse momentum of a parton.  The shifts in transverse position and in
transverse momentum give independent information about spin-orbit
correlations at the parton level. 

A dynamical connection between the two phenomena, called
\emph{chromodynamic lensing}, has been formulated in
\cite{Burkardt:2003uw}.  As explained in Sec.~\ref{sec:tmd},
the Sivers effect arises from the interaction of the scattered parton with
the proton remnant.  The shift in the spatial distribution of the parton
described by Eq.~\eqref{eq:shift-from-E} goes along with a shift in the
spatial distribution of the remnant, which leads to an anisotropy in the
transverse momentum of the scattered parton.  This connection is
explicitly seen in simple model calculations  where
the proton is represented as a bound state of a quark and a diquark,
with their interaction via gluon exchange being treated in perturbation
theory \cite{Meissner:2007rx,Gamberg:2010xi}.  At the EIC, it will be possible
to measure both the Sivers effect and the GPDs $H$ and $E$ that enter in
Eq.~\eqref{eq:shift-from-E}.  The comparison of their size, sign and $x$
dependence will yield information about the non-perturbative interactions
between active and spectator partons in the nucleon.

The spin-orbit correlation described by Eq.~\eqref{eq:shift-from-E} is
intimately connected with the orbital angular momentum carried by partons
in the nucleon and thus with the proton spin puzzle, i.e., with the
question of how the spin of the proton is distributed at the microscopic
level.  Writing the densities in Eq.~\eqref{eq:shift-from-E} or the associated
GPDs in terms of nucleon wave functions, one indeed finds that $E$
originates from the interference of wave functions whose orbital angular
momentum differs by one unit \cite{Burkardt:2005km}.  A different way to
quantify this connection is Ji's sum rule \cite{Ji:1996ek,Ji:1998pc}
\end{multicols}
\vspace*{-0.25cm}
\begin{align}
  \label{eq:ji-sum-rule}
J^q &= \frac{1}{2} \int dx\, x\,
   \bigl[ H^q(x,\xi,t=0) + E^q(x,\xi,t=0) \bigr] \,,
\end{align}
\begin{multicols}{2}
\noindent
which represents the \emph{total} angular momentum $J^q$ (including both
helicity and orbital contributions) carried by quarks and anti-quarks of
flavor $q$ as an integral over GPDs.  An analogous relation holds for
gluons.  There is a close connection between Ji's sum rule and the shift
in ${b}$-space (see Eq.~\eqref{eq:shift-from-E})
\cite{Burkardt:2005hp}.  Let us
mention that the very definition of angular momentum for quarks and gluons
is non-trivial and involves several conceptual aspects at the core of
non-abelian gauge theories, see e.g.\ \cite{Wakamatsu:2010qj,Ji:2012sj}
and references therein.

$J^q$ is a generalized form factor at $t=0$ that can be computed in
lattice QCD \cite{Hagler:2009ni}, and we foresee that such computations
will have reached maturity by the time the EIC is operational.  In turn, a
precise determination of Eq.~\eqref{eq:ji-sum-rule} with GPDs extracted from
exclusive scattering processes is extremely challenging, especially
because it requires knowledge of $H$ and $E$ for all $x$ at fixed $\xi$.
A reliable estimate of the associated theoretical uncertainties will only
be possible when high-precision data enable us to gain a better
understanding of the dependence of GPDs on their different kinematic
arguments.  On the other hand, exclusive scattering experiments can
investigate the dependence of $H$ and $E$ on the longitudinal momentum
fractions in a wide kinematic range.  Measurements at the EIC will in
particular probe the region of sea quarks, whose contribution to the
angular momentum sum rule is suppressed compared to valence quarks because
of the factor $x$ in the integral given in Eq.~\eqref{eq:ji-sum-rule}.  In this sense,
computations in lattice QCD and measurements of exclusive reactions are
highly complementary.
\end{multicols}


\subsection{Processes and Observables}
\label{sec:gpd-processes}

\begin{multicols}{2}
A large number of exclusive channels can be experimentally investigated at
the EIC, and each of them will give specific physics information.  An overview
of key measurements is given in Table~\ref{tab:gpd-matrix}.

For most processes, we have formal proofs of factorization
\cite{Collins:1996fb,Collins:1998be}, which provide a solid ground for
their interpretation in terms of GPDs (akin to the factorization proofs
that enable us to extract conventional parton densities from inclusive
processes, see Sec. \ref{sec:helicity}).  For these proofs to apply, the
photon virtuality $Q^2$ must be large, in particular much larger than the
invariant momentum transfer $t$ to the hadron.  In terms of imaging, the
precision $\sim 1/Q$ with which partons are resolved is then much finer
than the precision $\sim 1 \big/\!\sqrt{|t|}$ with which their position in
the hadron is determined \cite{Diehl:2002he}.  This permits a clean
separation between the object that is being imaged and the probe used to
obtain the image.
\end{multicols}
\newpage
Deeply virtual Compton scattering (DVCS) is measured in the reaction $ep
\to ep\gamma$ and plays a privileged role in several respects:
\begin{itemize}
\item Its theoretical description is most advanced, with radiative
  corrections being available up to order $\alpha_s^2$
  \cite{Kumericki:2007sa,Pire:2011st,Moutarde:2013qs} and corrections of order $1/Q$ to the limit of
  large $Q^2$ being well understood in their structure
  \cite{Belitsky:2005qn}.  Recently, results have even been obtained for
  corrections of order $1/Q^2$ due to the finite target mass and to
  nonzero $t$ \cite{Braun:2011dg,Braun:2012hq,Braun:2014sta}.
\item It has a large number of angular and polarization observables that can
  be calculated using the factorization theorem and thus constrain GPDs
  \cite{Belitsky:2001ns,Belitsky:2012ch}.  With longitudinal electron polarization and
  both longitudinal and transverse polarization of the proton, one has
  enough observables to disentangle the distributions $H$ and $E$
  discussed above, as well as their counterparts $\tilde{H}$ and
  $\tilde{E}$ for longitudinally polarized partons.
\item Several contributions that are suppressed by $1/Q$ can be extracted
  from suitable observables and be calculated in terms of twist-three
  distributions, which are closely connected to those accessible in
  semi-inclusive processes at high transverse momentum (see Sec.~
  \ref{secI:info-about-TMDs}).
\item Compton scattering interferes with the Bethe-Heitler process, which
  is calculable in QED.  This allows one to extract the complex phase of
  the Compton scattering amplitude, which in turn gives more detailed
  information about GPDs.
\item Further information about the phase of the Compton amplitude can be
  extracted if both $e^-$ and $e^+$ beams are available (even if the
  latter are unpolarized).  In the absence of a positron beam, some of
  this information may be obtained by running at different beam energies
  (using a Rosenbluth-type separation of different contributions to the
  cross-section).
\end{itemize}

\begin{table}[h]
\small
\label{matrix}
\begin{center}
\begin{tabular}{|c|c|c|c|}
\hline
Deliverables   & Observables                  & What we learn              & Requirements \\
\hline\hline
 GPDs of       & DVCS and $\jpsi,\rho^0,\phi$ & transverse spatial distrib. & $\int dt\, L \sim 10 ~\text{to}~ 100 \fb^{-1}$; \\
 sea quarks    & production cross-section     & of sea quarks and gluons;
 & leading proton detection; \\
 and gluons    & and polarization             & total angular momentum     & polarized $e^-$ and $p$ beams; \\
               & asymmetries                  & and spin-orbit correlations & wide range of $x$ and $Q^2$; \\ \cline{1-3} 
 GPDs of       & electro-production of         & dependence on              & range of beam energies; \\
 valence and   & $\pi^+, K$ and $\rho^+, K^*$ & quark flavor and           & $e^+$ beam \\
 sea quarks    &                              & polarization               & valuable for DVCS \\ \hline
\hline
\end{tabular}
\caption{\label{tab:gpd-matrix} Key measurements for imaging partons in
  the transverse plane.  With an EIC running at lower energies, one can 
  investigate the transition from the valence to the sea quark regime and
  measure the processes in the lower block, while an EIC with higher 
  energies provides access to a wide region dominated by sea quarks and gluons.}
\end{center}
\end{table}

\vspace{-0.25in}
\begin{multicols}{2}
Closely related to DVCS is time-like Compton scattering, $\gamma p\to
\ell^+\ell^- p$, i.e.\ photo-production of a lepton pair with large
invariant mass \cite{Pire:2011st,Moutarde:2013qs,Goritschnig:2014eba}. 
An advantage of this process is that
the analog of the DVCS beam charge asymmetry is an asymmetry in the
angular distribution of the produced lepton pair, which can be measured
without positron beams.

Compton scattering thus has the potential to yield detailed and precise
information about GPDs for different polarizations of the partons and the
proton. A limitation it shares with inclusive DIS is that it is sensitive
only to the sum of quark and anti-quark distributions in a particular
flavor combination and that it involves gluon distributions only via a
logarithmic dependence on $Q^2$.  Exclusive meson production offers
substantial help in the separation of different quark and anti-quark
flavors and of gluons, which is of special interest as discussed in
Sec.~\ref{sec:parton-intro}.  The extraction of the flavor dependence of GPDs
will only be possible if GPDs are truly universal. Hints of this universality 
have been unveiled recently by a common analysis of all DVCS and exclusive meson
production data with a common GPD set \cite{Kroll:2012sm}. The theoretical description of these
processes is more involved: it requires knowledge of the relevant meson
wave functions, and theoretical progress is still needed to achieve
control over radiative corrections \cite{Ivanov:2004vd,Diehl:2007hd} and
over corrections to the large $Q^2$ limit \cite{Goloskokov:2006hr}.
Measuring at $Q^2$ well above $10 \gev^2$ can substantially decrease the
theoretical uncertainties.  This holds in particular for parton imaging,
given that at lower $Q^2$ the measured $t$ dependence receives
contributions from the finite meson size as well as from the structure of
the proton target.  Let us highlight specific features of different
production channels.
\end{multicols}
\begin{itemize}
\item $\jpsi$ production provides selective access to unpolarized gluons.
  In this case, the hard scale of the process is $Q^2 +
  M_{\smash{\jpsi}}^{2}$ rather than $Q^2$, so that both photo- and
  electro-production can be used to probe GPDs.  Electro-production has
  smaller rates but reduced theoretical uncertainties.  Furthermore, the
  cross-sections $\sigma_L$ and $\sigma_T$ for longitudinal and transverse
  photon polarization, which can be separated experimentally from the
  angular distribution in the decay $\jpsi \to \ell^+\ell^-$, provide two
  independent observables to validate the theory description.
\item The production of the neutral vector mesons $\rho^0$, $\phi$,
  $\omega$ involves unpolarized gluons and sea quarks in particular flavor
  combinations.  $\rho^+$ production provides direct information about the
  difference of $u$ and $d$ distributions, whereas the production of
  $K^*(829)$ is sensitive to strange quarks in the proton
  \cite{Diehl:2005gn}.

  The factorization theorem allows us to compute the cross-section
  $\sigma_L$ for longitudinal photon polarization and the associated
  transverse proton spin asymmetry, whereas other observables require a
  model for effects suppressed by $1/Q$ \cite{Goloskokov:2007nt}.  An
  experimental separation of $\sigma_L$ and $\sigma_T$ can be performed
  using the vector meson decay, i.e., a Rosenbluth separation with
  different beam energies is not required.
\item Production of the pseudoscalar mesons $\pi$, $K$, $\eta$ and $\eta'$
  provides information about different flavor combinations for
  longitudinally polarized quarks and anti-quarks, encoded in distributions
  $\tilde{H}$ and $\tilde{E}$.  Again, only $\sigma_L$ can be computed
  from the factorization theorem.  To separate $\sigma_L$ and $\sigma_T$
  one has to apply the Rosenbluth method and hence needs data for
  different beam energies.

  The calculations of $1/Q$ suppressed terms in
  \cite{Ahmad:2008hp,Goloskokov:2011rd} found that $\sigma_T$ can be of
  substantial size due to contributions from GPDs for transversely
  polarized quarks, which are closely related to the transversity
  distribution $h_1(x)$ introduced in the Sidebar on page~\pageref{sdbar:tmd}. 
\item The production of $\pi^+\pi^-$ pairs in the continuum or on the
  $f_2(1270)$ resonance is one of the very few processes sensitive to the
  \emph{difference} of quark and anti-quark distributions
  \cite{Warkentin:2007su}, thus providing access to the $x$ dependence of
  the distributions whose integrals over $x$ give the electromagnetic
  nucleon form factors.
\item The production of two mesons with a large rapidity gap between them
  is again sensitive to GPDs for transversely polarized quarks
  \cite{Ivanov:2002jj}.
\end{itemize}
\begin{multicols}{2}
Finally there is the possibility to study the generalized parton
distributions in the pion using DVCS or meson production on a virtual pion
emitted from the proton beam
\cite{Strikman:2003gz,Strikman:2009bd,Amrath:2008vx}.  The experimental
signature is a recoil neutron as well as a recoil $\pi^+$ in the final
state.  For a clear theoretical interpretation of such a measurement, the
emitted pion must have only a small virtuality, i.e., it must be almost
real.  As shown in \cite{Amrath:2008vx}, this requires both high energy
and high luminosity, which will be available at the EIC for the first time.
\end{multicols}


\subsection{Parton Imaging Now and in the Next Decade}

\begin{multicols}{2}
Pioneering measurements for imaging low-$x$ partons have been performed in
the last decade at the HERA collider, where the experiments H1 and ZEUS
measured DVCS and exclusive vector meson production with up to $28 \gev$
electrons or positrons scattering on $920 \gev$ unpolarized protons.  Most
precise information about the spatial gluon distribution comes from
$\jpsi$ photo-production (with the smallest statistical errors among all
relevant final states), and DVCS has provided us with first information
about sea quarks at momentum fractions $x$ around $10^{-3}$.  These
measurements provide evidence for differences between the spatial
distribution of small-$x$ gluons, small-$x$ quarks and the distribution of
valence quarks one can infer from the electromagnetic nucleon form
factors.  For gluons they also show a weak dependence of the average
impact parameter on $x$.  With an integrated luminosity of $500
\operatorname{pb}^{-1}$ many of the HERA results on imaging are however
limited by statistical errors and leave open many important questions, in
particular regarding sea quarks and the dependence of impact parameter
distributions on the resolution scale $Q^2$.

Possibilities to extend the HERA measurements of $\gamma p\to \jpsi\, p$
and $\gamma p \to \Upsilon p$ to higher energies are offered by
ultraperipheral proton-proton or proton-nucleus collisions at the LHC.
The quasi-real incident photon is radiated off a beam proton or nucleus in
this case, the beam particle being scattered with a very low momentum
transfer.

Groundbreaking measurements in the region of moderate- to large-$x$ have
been made by fixed-target experiments with $28 \gev$ electrons and
positrons at HERMES and with up to $6 \gev$ electrons at JLab, proving in
particular that angular and polarization asymmetries can be measured in
DVCS and interpreted in terms of GPDs.  However, most of these
measurements are at rather small $Q^2$ or have sizeable statistical
uncertainties, which puts serious limitations on the precision of
extracted GPDs and precludes the use of $Q^2$ evolution as a tool.

The precise measurements of electromagnetic nucleon form factors, as well
as the calculation of generalized form factors in lattice QCD
\cite{Hagler:2009ni}, are already providing valuable information about
the spatial distribution of partons in regions of $x$ typically 
above 0.1 or so.  Both research areas are anticipated to make significant
progress in the future and will constitute an important complement of
imaging through exclusive processes, as discussed in
Sec.~\ref{sec:parton-intro}.

First measurements for imaging partons with $x$ between $10^{-1}$ and
$10^{-2}$, i.e., in the transition region between valence and sea quarks,
will be possible with the COMPASS experiment at CERN, which will have the
benefit of both $\mu^+$ and $\mu^-$ beams to measure the charge asymmetry
in DVCS.  The anticipated integrated luminosity around $100 \pb^{-1}$
will, however, limit the accuracy of measurements at $Q^2$ above $5\gev^2$
and the possibilities to explore simultaneously the dependence on $x$,
$Q^2$ and $t$.  At present it is not clear whether polarized protons will
be available.

\begin{figure}[h!]
\begin{center}
\includegraphics[width=0.96\textwidth]{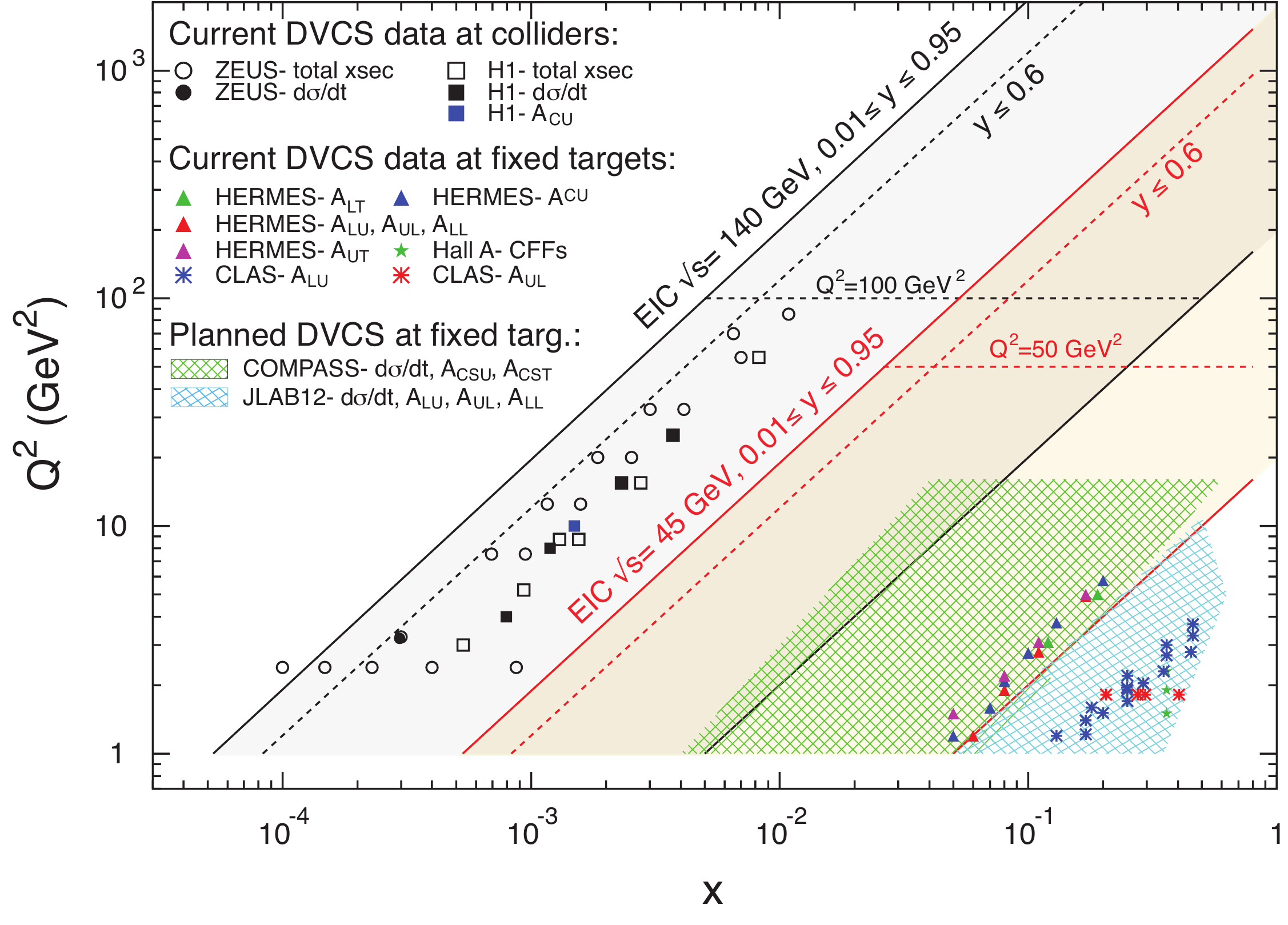}
\end{center}
\vskip -0.25in
\caption{\label{fig:x-q2-dvcs} An overview of existing and planned
  measurements of DVCS in the $x,Q^2$ plane.}
\end{figure}

A first era of precise parton imaging will begin with the 12 GeV upgrade
at JLab, with very high statistics and sufficiently high $Q^2$ to probe
partons at high-$x$, including the effects of
polarization.  Figure~\ref{fig:x-q2-dvcs} gives an overview of existing
and anticipated measurements of DVCS in the $x,Q^2$ plane.

To realize the full physics potential of parton imaging that we have
discussed in the previous section will require the EIC.  Such a machine
will, for the first time, make it possible to image partons with high
statistics and with
polarization in a wide range of small- to moderate-$x$.  At high-$x$ it
will complement the JLab 12 program with measurements at large-$Q^2$, thus
opening up the possibility to extract physics from scaling violations
for high-momentum partons.  

Let us finally mention that it is very difficult to obtain information on
GPDs from exclusive processes in $p$+$p$ collisions.  This is due to the
effect of soft gluon exchange between spectator partons in the two
protons, which precludes a simple theoretical interpretation of such
reactions.  Lepton-proton scattering thus provides a privileged way to
quantify the spatial structure of the proton via GPDs.  On the other hand,
the information gained in lepton-proton scattering can help to better
understand important features of proton-proton collisions, in particular
the dynamics of multi-parton interactions
\cite{Frankfurt:2003td,Diehl:2011yj}.
\end{multicols}

\newpage
\subsection{Accelerator and Detector Requirements}

\begin{multicols}{2}
The experimental study of DVCS and meson electro-production requires
high luminosity: cross-sections are at best a few percent of
the inclusive DIS cross-section, and the data need to be kinematically
binned in up to five variables ($x, Q^2, t, \phi, \phi_S$), where $\phi$
($\phi_S$) is the angle between the hadron production (proton beam
polarization) plane and the electron scattering plane.  Luminosities as
high as $10^{34} \operatorname{cm}^{-2} \operatorname{s}^{-1}$ are crucial
for the measurement of DVCS spin asymmetries and for the exploration of
the high-$t$ region, as well as for certain meson production channels,
especially at low-$x$.  A large lever arm in $Q^2$ at fixed $x$ is
required for testing the power behavior predicted by factorization
theorems, and beyond this for the use of evolution effects to disentangle
gluons from quarks in Compton scattering. If several collision energies
and hence several beam configurations are needed to achieve this, one
needs accurate measurements of integrated luminosities in order to
cross-normalize data sets. A significant lever arm in $y$
at fixed $x$ and $Q^2$ is mandatory for
the separation of $\sigma_L$ and $\sigma_T$, which is essential for
pseudoscalar mesons and helpful for DVCS in case a positron beam is not
available, as explained in Sec.~\ref{sec:gpd-processes}.

To measure truly exclusive processes, it is essential to detect all final
state particles. Hermeticity of the EIC detector is therefore a crucial
requirement. The most critical aspect is the ability to detect the recoil
baryon, which in the region of interest has a transverse momentum up to a
few $\gev$. This corresponds to very small scattering angles with respect
to the proton beam. At large proton beam energies, the detection of the
recoil proton may require Roman Pots integrated in the machine lattice,
whereas at lower proton beam energies, or high proton transverse
momenta, it should be possible to detect the proton in the main
EIC detector. Note that the transverse momentum acceptance is directly
related to the region in $b_T$ space where reliable images can be obtained.
The emittance of the proton beam at the location of the detectors needs to
be kept reasonably low so that the detectors can be placed as close to the
proton beam as possible.  Near perfect hermeticity is also essential in
the case of low-$y$ events, which are needed to explore high $x$ at a
given $Q^2$.  Indeed, in this case, $y$ is measured using a hadronic method
and depends on the sum over the energy minus the longitudinal momentum of
all the hadronic final-state particles.

Specifically for DVCS, but also for $\pi^0$ production, the photon
detection coverage is particularly important over the full rapidity range.
Note that for DVCS, both the photon and the electron tend to be emitted
backward in the same hemisphere when the electron energy increases.

As far as particle identification is concerned, the situation varies
depending on the beam energies.  In the most general case, the separation
of electrons and pions requires particular care in the momentum range
between about $4$ and $10 \gev$.  For the identification of light mesons,
mostly in the barrel section, the same care will be necessary in the same
momentum range. A ring imaging Cherenkov counter (RICH) or a DIRC
complementing a time-of-flight system will likely be needed in the barrel
section of the detector (see Sec.~\ref{sec:layout}).  Note that in
addition to standard particle identification, the missing mass method
might be used at low collision energies
to discriminate between particle types, depending on the
kinematics and the resolution that can be achieved.

To measure $\jpsi$ production, one would use ideally both the decays into
$\mu^+\mu^-$ and $e^+e^-$. In both cases, the momentum resolution needs to
be sufficiently good to avoid contamination from the non-resonant
background as well as from the exclusive and semi-inclusive
$\psi(\mathrm{2S})$ production channels, which have the same decay modes.

As pointed out in Sec.~\ref{sec:gpd-processes}, polarization is
critically important in order to disentangle the different GPDs entering
DVCS and other processes. Specifically, transverse proton polarization is
essential to access the information about orbital angular momentum encoded
in the distribution $E$. High values of electron and especially proton
polarization are ideal for precise measurements.  The electron and proton
polarizations should be measured with sufficient accuracy, so as not to
become significant sources of systematic error.
\end{multicols}

\subsection{Parton Imaging with the EIC}

\begin{figure*}[h!]
\begin{center}
\includegraphics[width=0.49\textwidth]{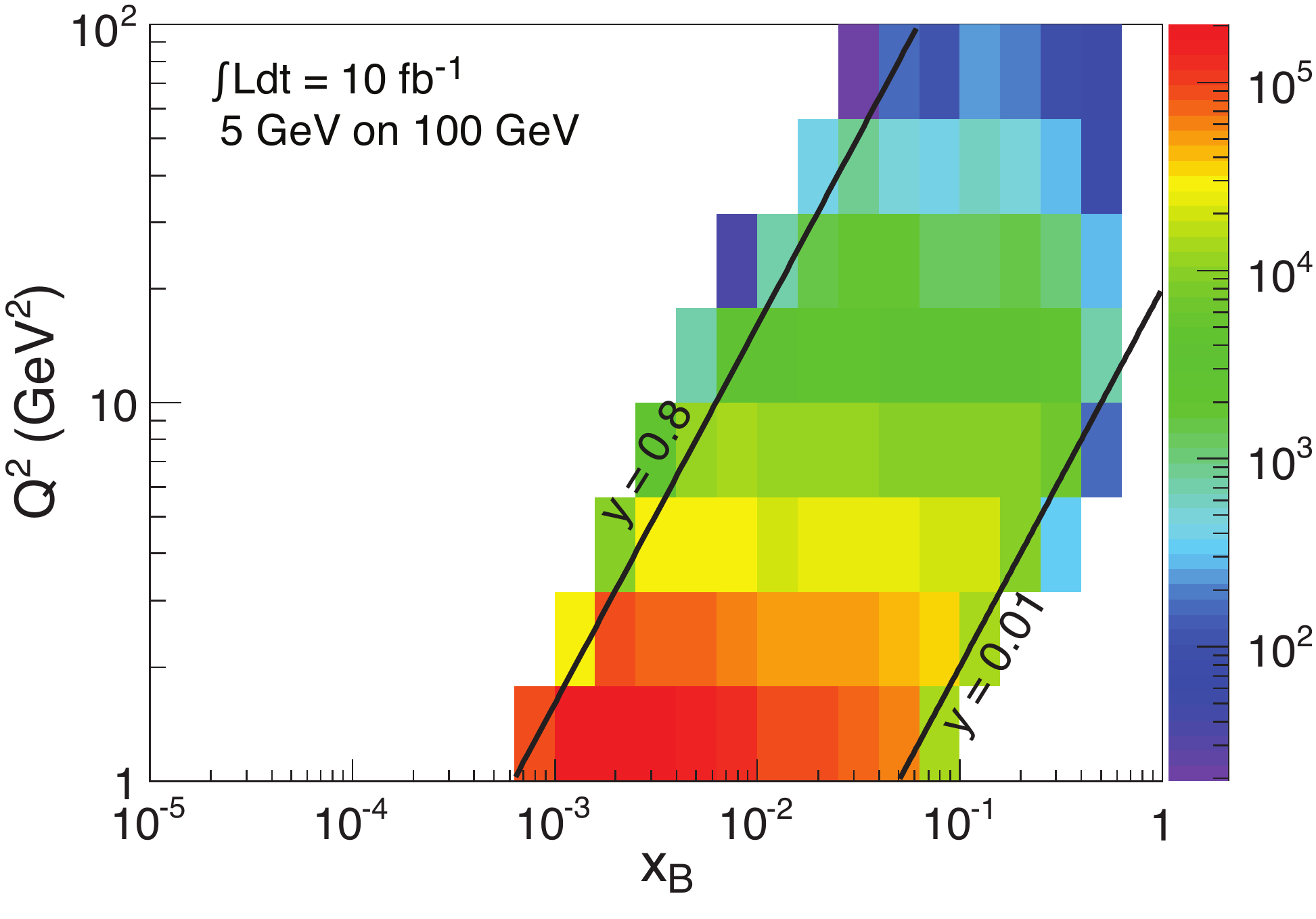}
\includegraphics[width=0.49\textwidth]{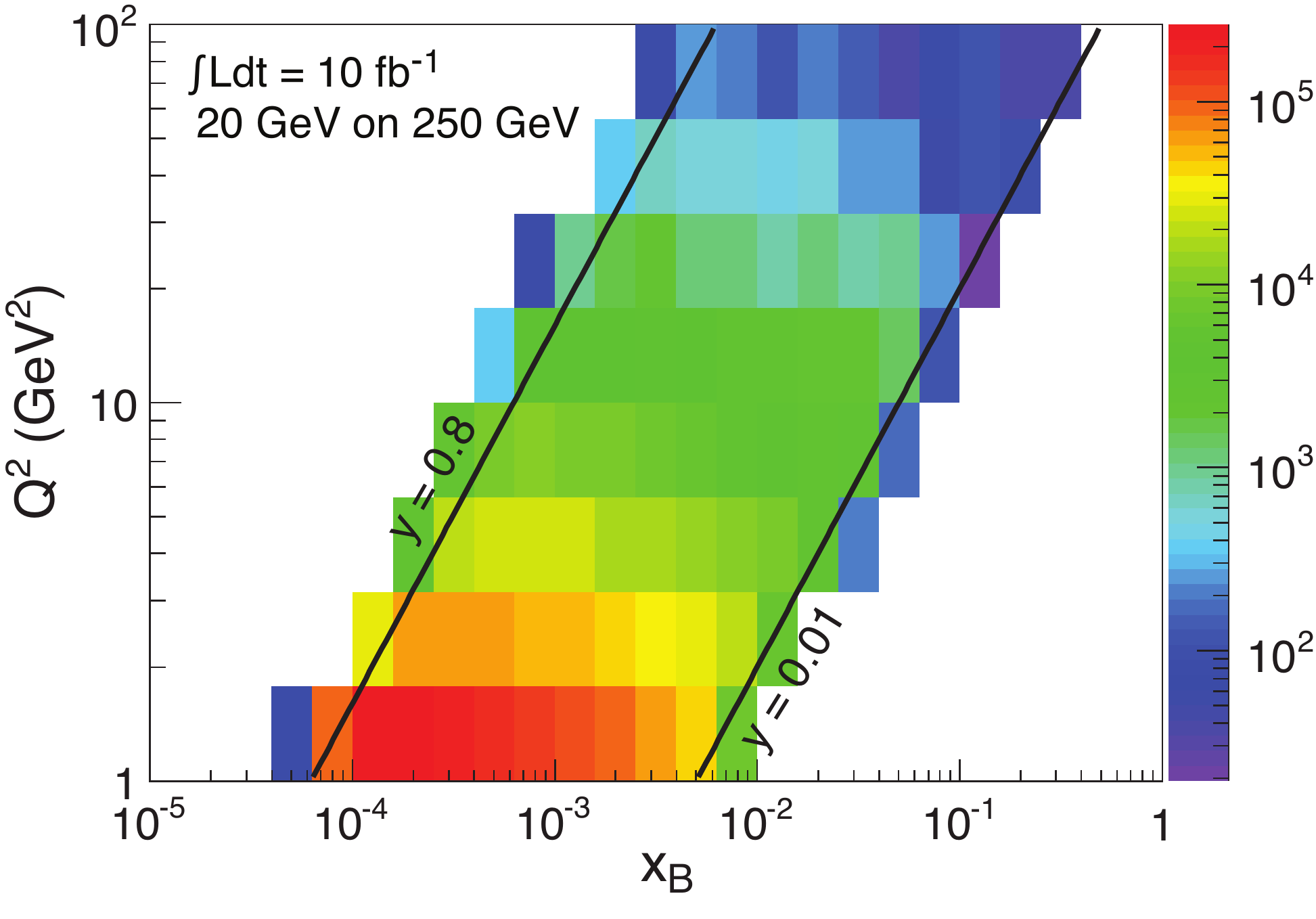}
\end{center}
\vskip -0.25in
\caption{\label{fig:dvcs-x-Q} Expected distribution of DVCS events in bins
  of $x$ and $Q^2$.  Event numbers correspond to Compton scattering,
  i.e.\ the contribution of the Bethe-Heitler process to the process
  $ep\to ep\gamma$ has been subtracted.}
\end{figure*}

\begin{multicols}{2}
Let us show the potential of an EIC for imaging partons using the DVCS
process, which plays a privileged role as we discussed in
Sec.~\ref{sec:gpd-processes}.  The following projections are based on
events simulated according to GPD models that give a good description of
the existing DVCS data \cite{Kumericki:2007sa,Kumericki:2009uq}.
Acceptance cuts for the detected electron, photon and proton corresponding
to the detector layout in Sec.~\ref{sec:kinmat} and Sec.~\ref{sec:layout}
have been applied.  Figure~\ref{fig:dvcs-x-Q} shows that a fine binning of
DVCS events in both $x$ and $Q^2$ is possible in a wide kinematic range.
\end{multicols}

\begin{figure*}[t!]
\begin{center}
\includegraphics[height=0.98\textwidth,angle=-90]{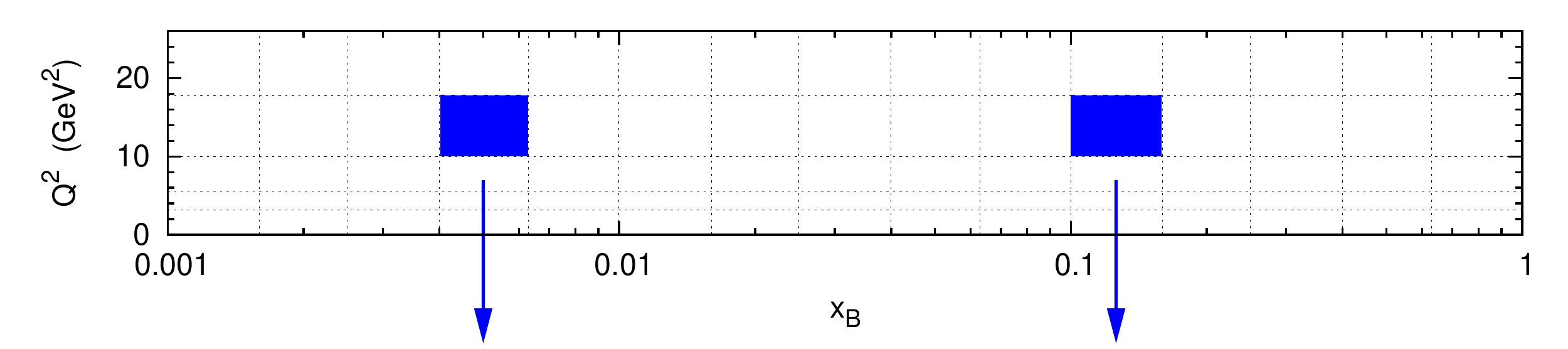}\\[0.5em]
\includegraphics[width=0.96\textwidth]{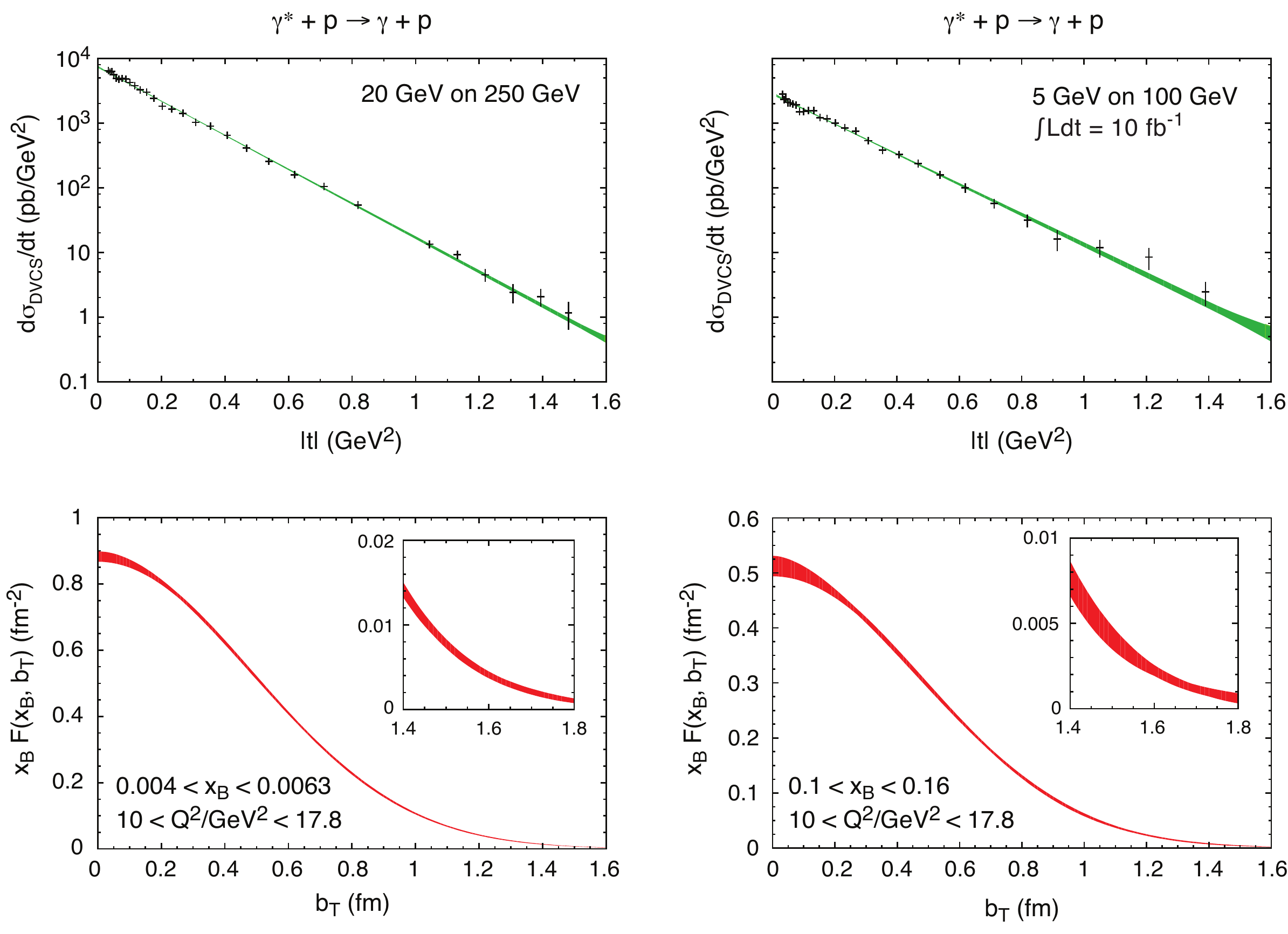}
\end{center}
\vskip -0.2in
\caption{\label{fig:dvcs-t-b} Top: The DVCS cross-section in two bins of
  $x$ and $Q^2$.
  The error bars reflect statistical and assumed systematic uncertainties,
  but not the overall normalization uncertainty from the luminosity measurement.
  For the left
  panels the assumed luminosity is $10 \fb^{-1}$ for $|t| < 1 \gev^2$ and
  $100 \fb^{-1}$ for $|t| > 1 \gev^2$.  Bottom: The distribution of partons in
  impact parameter $b_T$ obtained from the DVCS cross-section.  The bands
  represent the parametric errors in the fit of $d\sigma_{\text DVCS}/dt$
  and the uncertainty from different extrapolations to the regions of
  unmeasured (very low and very high) $t$, as specified in Sec. 3.6 of
  \protect\cite{Boer:2011fh}.}
\end{figure*}
\begin{multicols}{2}
With the lower set of beam energies, 
one finds ample statistics in bins with large $Q^2$ for $x$ as high as
$0.2$.  The combination of such data with fixed-target results will give a
substantial lever arm in $Q^2$ and permit the study of evolution effects
in the kinematic regime where valence quarks are important.

The top panels of Figure~\ref{fig:dvcs-t-b} show the $t$ dependence of the
DVCS cross-section in two bins of $x$ and $Q^2$, accessible with $E_e = 5
\gev$, $E_p = 100 \gev$ and with $E_e = 20 \gev$ $E_p = 250 \gev$,
respectively.  The simulated data have been smeared for resolution, and
the error bars include both statistics and an estimate of systematic
uncertainties.  The scattered proton is assumed to be detected in Roman
pots for $|t|$ above $(175 \mev)^2$, see Sec.~\ref{sec:recoil} and
chapter 7.3 of \cite{Boer:2011fh}.  More detail on the simulation is given
elsewhere~\cite{Diehl:DIS2012,Fazio:2012ue,Aschenauer:2013hhw}.  From the DVCS cross-section, one
can reconstruct the scattering amplitude, which can then be Fourier transformed
into $b_T$ space.  The resulting images correspond to the particular
combination of quarks, anti-quarks and gluons ``observed'' in Compton
scattering.  We explained earlier that the momentum fraction of those
partons is ``smeared'' around the measured value of $\xi = x/(2-x)$,
whereas the variable $\boldsymbol{b}_T$ is legitimately interpreted as a
transverse parton position \cite{Diehl:2002he}.  The bottom panels of
Figure~\ref{fig:dvcs-t-b} show that precise images are obtained in a wide
range of $b_T$, including the large $b_T$ region where a characteristic
dependence on $b_T$ and $x$ due to virtual pion fluctuations is
predicted as discussed in Sec.~\ref{sec:excl-intro}.  We emphasize that
a broad acceptance in $t$ is essential to achieve this accuracy.  If, for
instance, the measured region of $|t|$ starts at $(300 \mev)^2$ instead of
$(175 \mev)^2$, the associated extrapolation uncertainty exceeds $50\%$
for $b_T > 1.5 \fm$ with the model used here.
\end{multicols}

\begin{multicols}{2}
The simulations presented here assume an exponential $t$ dependence of the
GPDs and hence of the DVCS cross-section.  As shown in Sec. 3.6 of
\cite{Boer:2011fh}, GPDs that have a dipole form in $t$ lead to larger
uncertainty bands in $b_T$ space, with uncertainties becoming significant
below $0.2 \fm$.  This reflects a larger uncertainty from the
extrapolation of the cross-section to the unmeasured large-$t$ region,
where a dipole form decreases much less quickly than an exponential law.
In such a 
scenario, measurement up to the largest possible $t$ values is crucial for
the accuracy of imaging at small impact parameters.
\begin{figurehere}
\begin{center}
\includegraphics[width=0.46\textwidth]{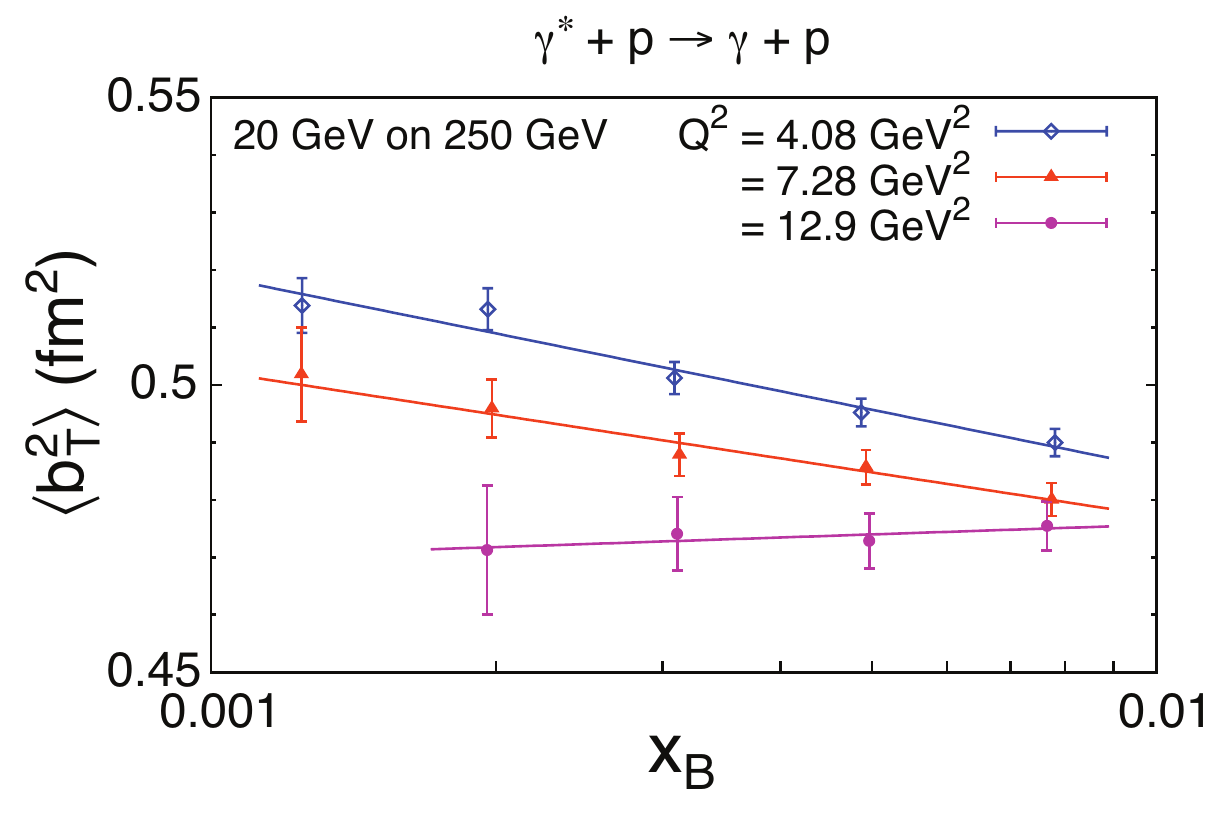}
\end{center}
\vskip -0.3in
\caption{\label{fig:dvcs-b2-fit} Average values of $\boldsymbol{b}_T^2$
  obtained from the DVCS cross-section in different bins of $x$ and
  $Q^2$.  The assumed luminosity is as for the left panels of
  figure~\protect\ref{fig:dvcs-t-b}.  The lines indicate linear fits of
  $\langle \boldsymbol{b}_T^2 \rangle$ vs.\ $\log x$ at fixed $Q^2$.
  Within errors, the fit for $Q^2 = 12.9 \gev^2$ is consistent with a
  vanishing or a small negative slope.}
\end{figurehere}

\vspace*{0.5cm}
Figure~\ref{fig:dvcs-b2-fit} shows that the quality of EIC measurements
allows one to resolve the correlation of the average impact
parameter $\langle \boldsymbol{b}{}_T^2 \rangle$ with $x$ and with $Q^2$.
The change of $\langle \boldsymbol{b}{}_T^2 \rangle$ with $Q^2$ reflects the
dynamics of perturbative parton radiation embodied in evolution equations.
By contrast, the logarithmic broadening of $\langle \boldsymbol{b}{}_T^2
\rangle$ with decreasing $x$ (taken as an input in the GPD model used
for the simulation) reflects non-perturbative dynamics, which has been
linked to the physics of confinement \cite{Bartels:2000ze}.  To exhibit
and separate these effects requires simultaneous binning in $Q^2$, $x$
and $t$ and high precision, which will only be possible at the~EIC.

The unpolarized DVCS cross-section is mainly sensitive to the distribution
$H$, i.e.\ to unpolarized partons in an unpolarized proton.  Information
about the phase of the corresponding amplitude can be extracted from the
longitudinal spin asymmetry of the electron beam (not shown here).
Sizeable values of this asymmetry are expected for $y$ not too small and
not too large (say between $0.2$ and $0.8$).  This method can in
particular give good constraints in regions where
$d\sigma_{\text{DVCS}}/dt$ has large uncertainties due to the subtraction
of the Bethe-Heitler cross-section.

\begin{figure*}[t!]
\begin{center}
\includegraphics[width=1.00\textwidth]{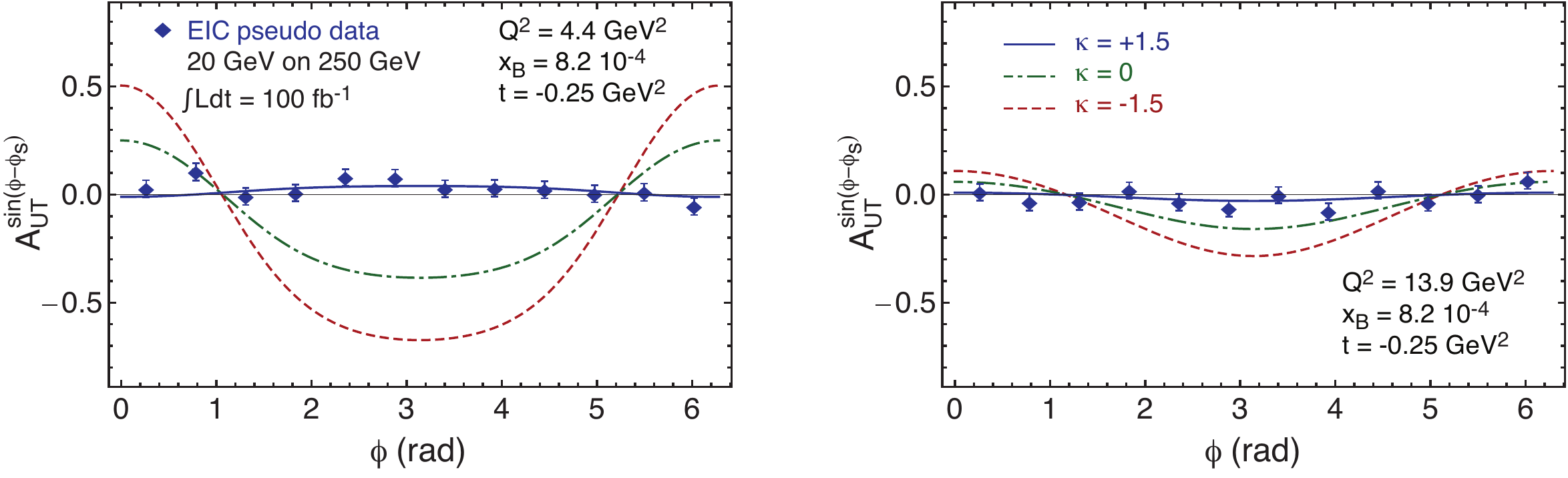}\\[1.5em]
\includegraphics[width=1.00\textwidth]{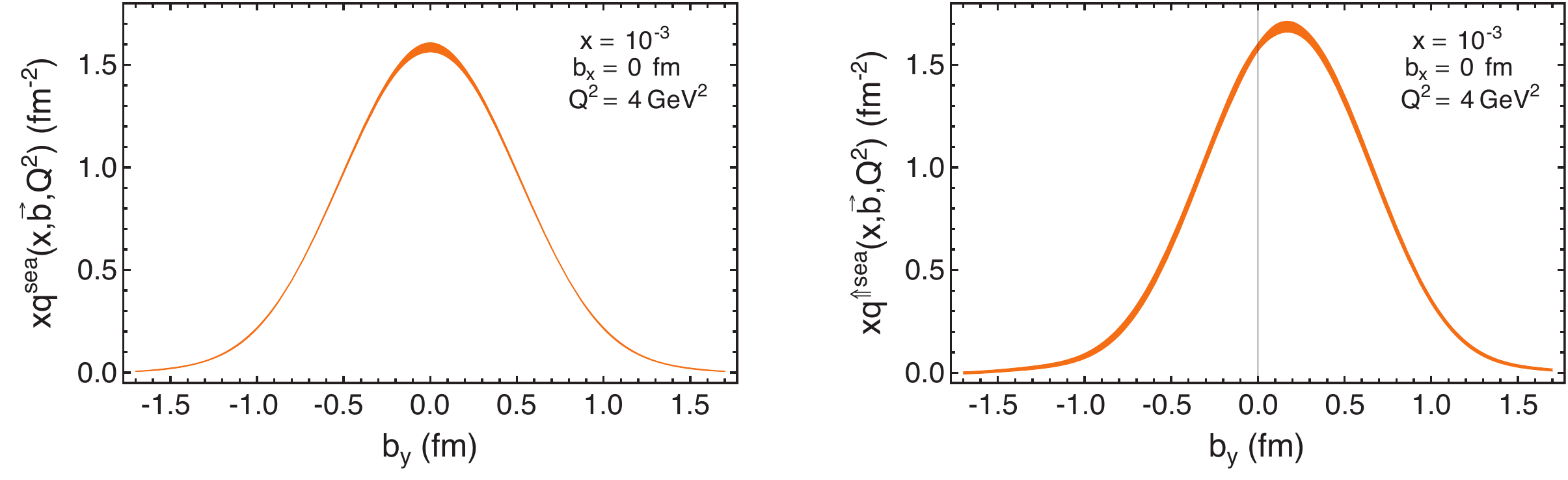}\\[1.5em]
\includegraphics[width=1\textwidth]{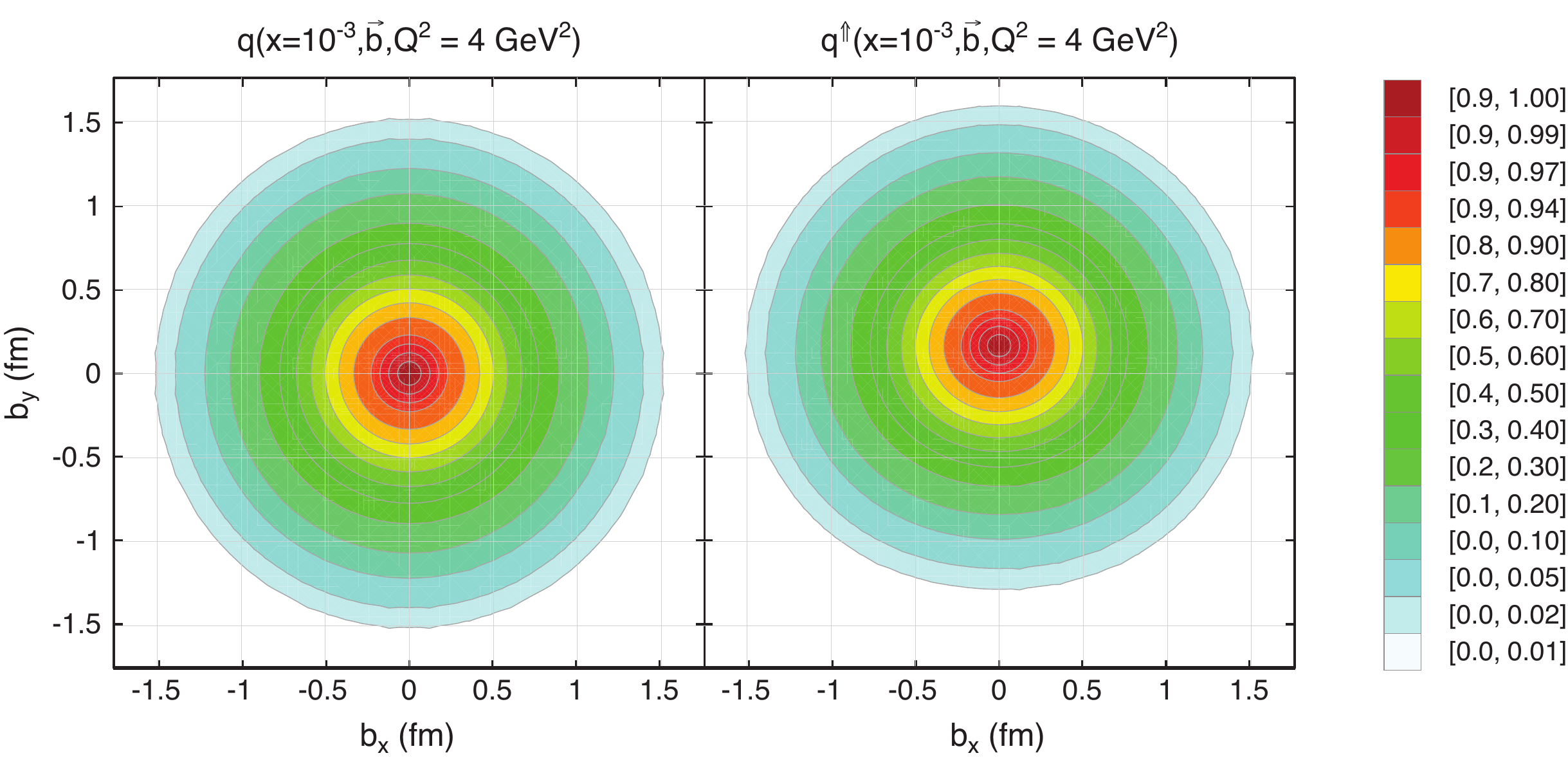}
\end{center}
\vskip -0.1in
\caption{\label{fig:TSA-gpd-b} Top: The DVCS polarization asymmetry
  $A_{UT}^{\smash{\,\sin(\phi-\phi_S)}}$ for a transversely polarized
  proton (see \protect\cite{Bacchetta:2004jz} for a precise definition).
  Middle: The spatial distribution of sea quarks in an unpolarized proton
  (left) and in a proton polarized along the positive $x$ axis (right)
  obtained from a GPD fit to simulated data for $d\sigma_{\text{DVCS}}/dt$
  and $A_{UT}^{\smash{\,\sin(\phi-\phi_S)}}$.  The bands represent the
  parametric errors of the fit and the uncertainty from extrapolating the
  $t$ spectrum outside the measured region.  Bottom: The corresponding
  density of partons in the transverse plane.}
\end{figure*}

Information about the other distributions, $E$, $\tilde{H}$ and
$\tilde{E}$, can be extracted from a number of polarization asymmetries.
For the sake of simplicity, we focus in the following on the region of
small $x$, where $\tilde{H}$ and $\tilde{E}$ are expected to be small
and can be neglected in a first approximation.  Access to $E$, and thus to
orbital angular momentum, can then be obtained from a particular angular
asymmetry measurable with transverse proton polarization.  The top panel
of Figure~\ref{fig:TSA-gpd-b} shows simulated data for this asymmetry
calculated with a specific model of $E$ and $H$.  The curves
have been obtained for different values of $\kappa = E(x,\xi,0)
/H(x,\xi,0)$, which determines the size of the transverse shift in the
density (see Eq.~\eqref{eq:shift-from-E}), and the data points correspond to $\kappa
= +1.5$ for sea quarks.  Since the asymmetry receives contributions from
both $H$ and $E$ it would be nonzero even for vanishing $E$.  The
projected errors are for a polarization of $80\%$ and include estimated
systematic uncertainties.  We see that the EIC could clearly distinguish
between \mbox{different scenarios}.

Assuming a functional form of the GPDs, one can extract both $H$ and
$E$ in a fit to the DVCS cross-section and the transverse proton spin
asymmetry.  The middle and lower panels of Figure~\ref{fig:TSA-gpd-b}
show the $b_T$ space densities obtained from a fit to simulated data for $20
\gev$ electrons scattering on $250 \gev$ protons in the kinematic region
with $3.2 \gev^2 < Q^2 < 17.8 \gev^2$ and $10^{-4} < x <
10^{-2}$.  Details of this study are given in
\cite{Diehl:DIS2012,Muller:DIS2012}.  We see that the parametric 
uncertainty of the results is very small and\break
\end{multicols}

\begin{multicols}{2}
\noindent
allows one to resolve the
transverse shift of the distribution in a polarized proton (about
$0.15\fm$ in the example).  Given its lever arm in $Q^2$, the fit also
permits a determination of the distribution $H$ for gluons from evolution
effects, with the resulting density profile shown in
Figure~\ref{fig:gpd-gluon-b}.
\begin{figurehere}
\begin{center}
\includegraphics[width=0.48\textwidth]{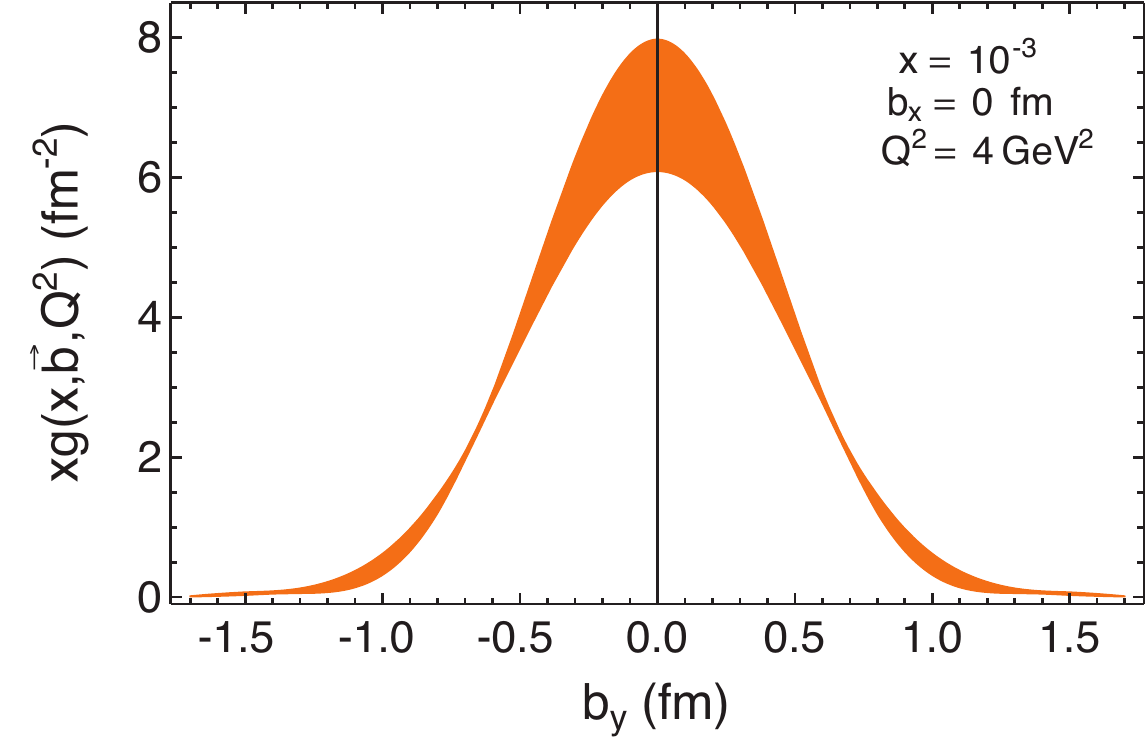}
\end{center}
\vskip -0.2in
\caption{\label{fig:gpd-gluon-b} The $b_T$ space density for gluons obtained
  in the same fit as the densities in Figure~\protect\ref{fig:TSA-gpd-b}.}
\end{figurehere}
\vspace*{0.5cm}

\begin{figure*}[h!]
\begin{center}
\includegraphics[width=0.49\textwidth]{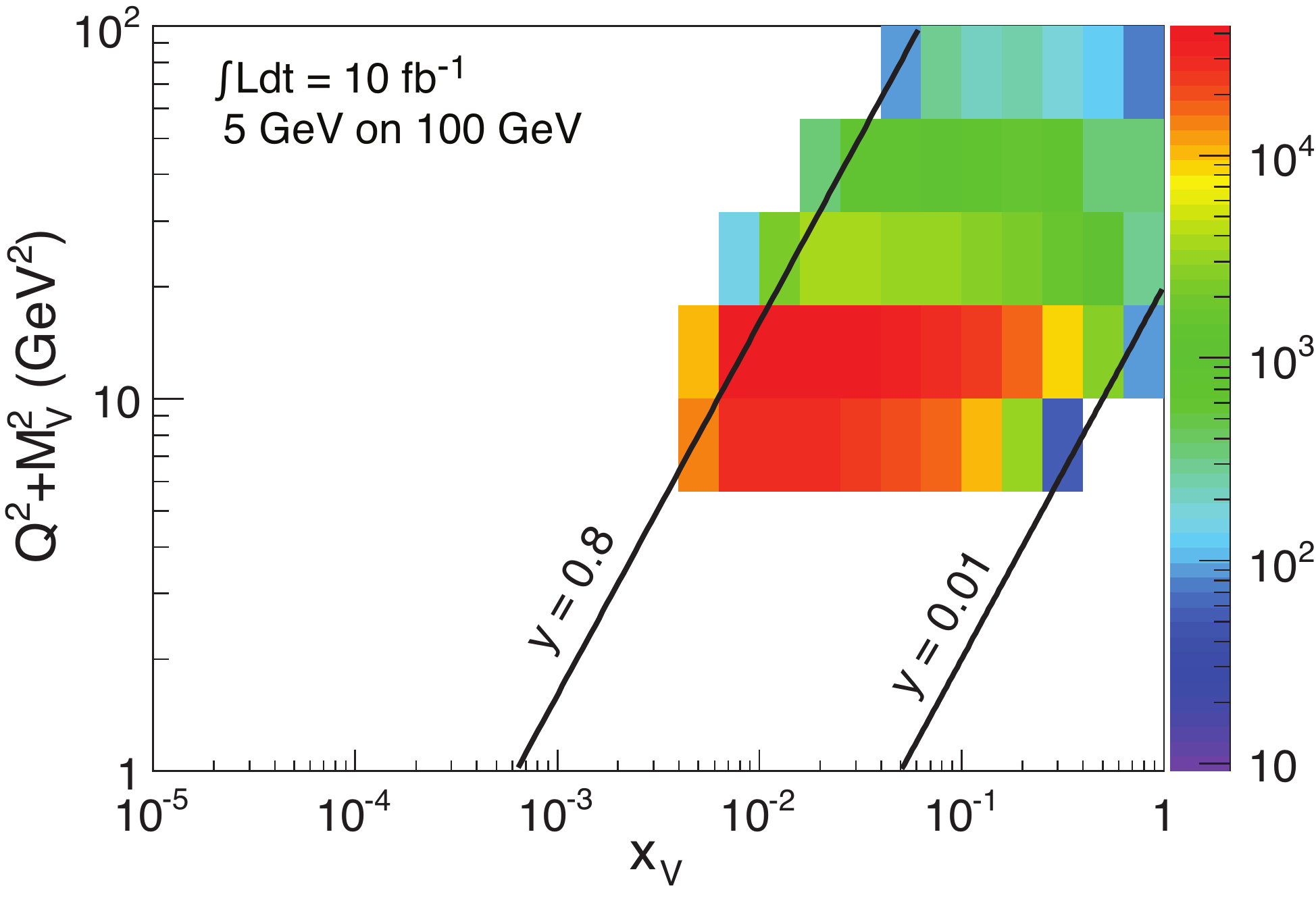}
\includegraphics[width=0.49\textwidth]{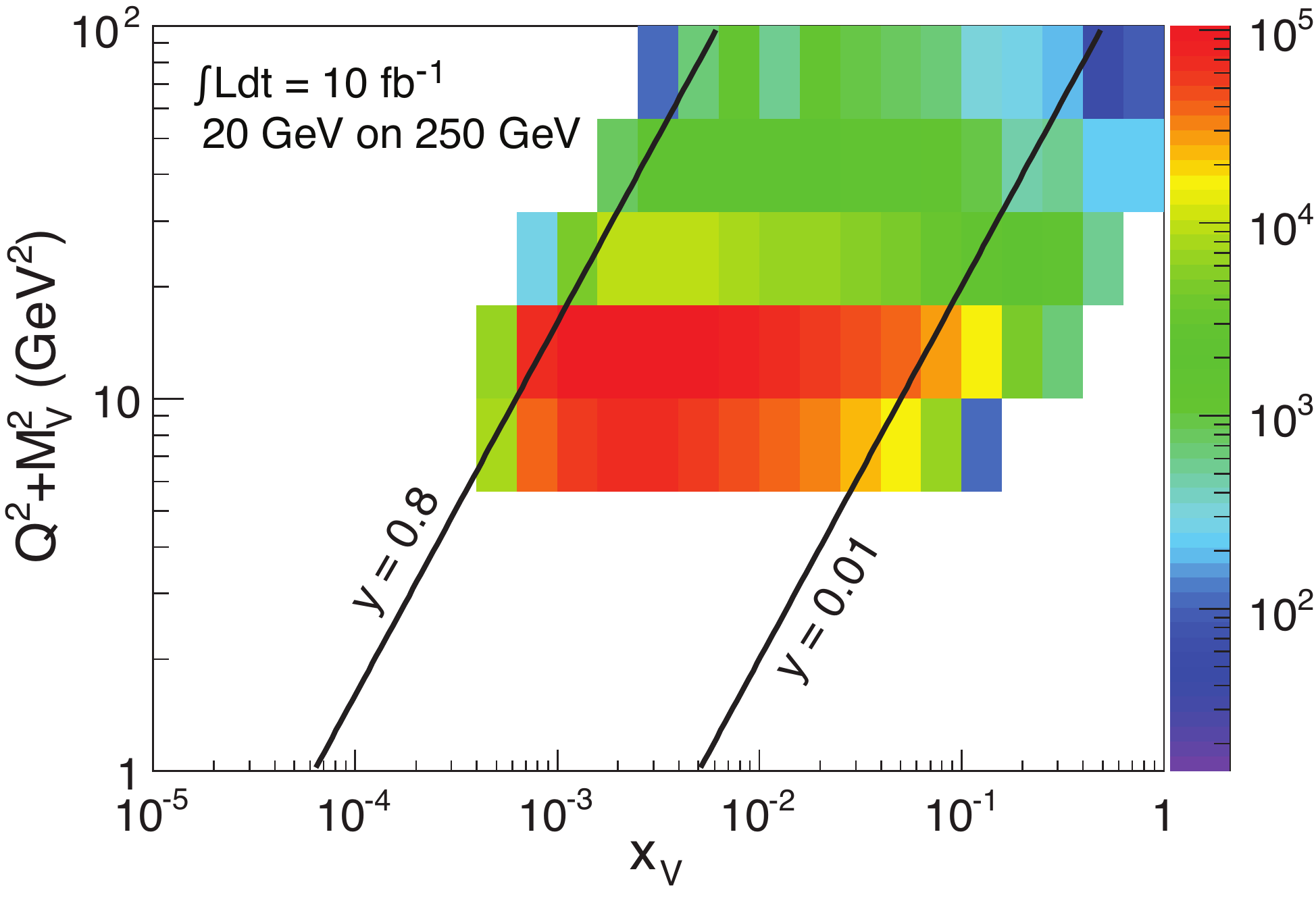}
\end{center}
\vspace{-0.3in}
\caption{\label{fig:jpsi-x-Q} Expected number of events for exclusive
  $\jpsi$ production in bins of $x_{V}$ and~$Q^2$.}
\end{figure*}

As discussed in Sec.~\ref{sec:gpd-processes}, exclusive $\jpsi$
production offers direct access to the distribution of unpolarized gluons.
The scaling variable for this process is $x_V$ 
and the hard scale is $Q^2 + M_{\smash{\jpsi}}^2$ (see the Sidebar on 
page~\pageref{sdbar:GPD}).
The expected distribution of events in $x_{\text{V}}$
and $Q^2$ in Figure~\ref{fig:jpsi-x-Q} shows that high-statistics studies
will be possible not only for photo- but also for electro-production, with
the additional benefits mentioned earlier.

\begin{figure*}[t!]
\begin{center}
\vspace*{-1.5in}
\includegraphics[height=0.98\textwidth,angle=-90]{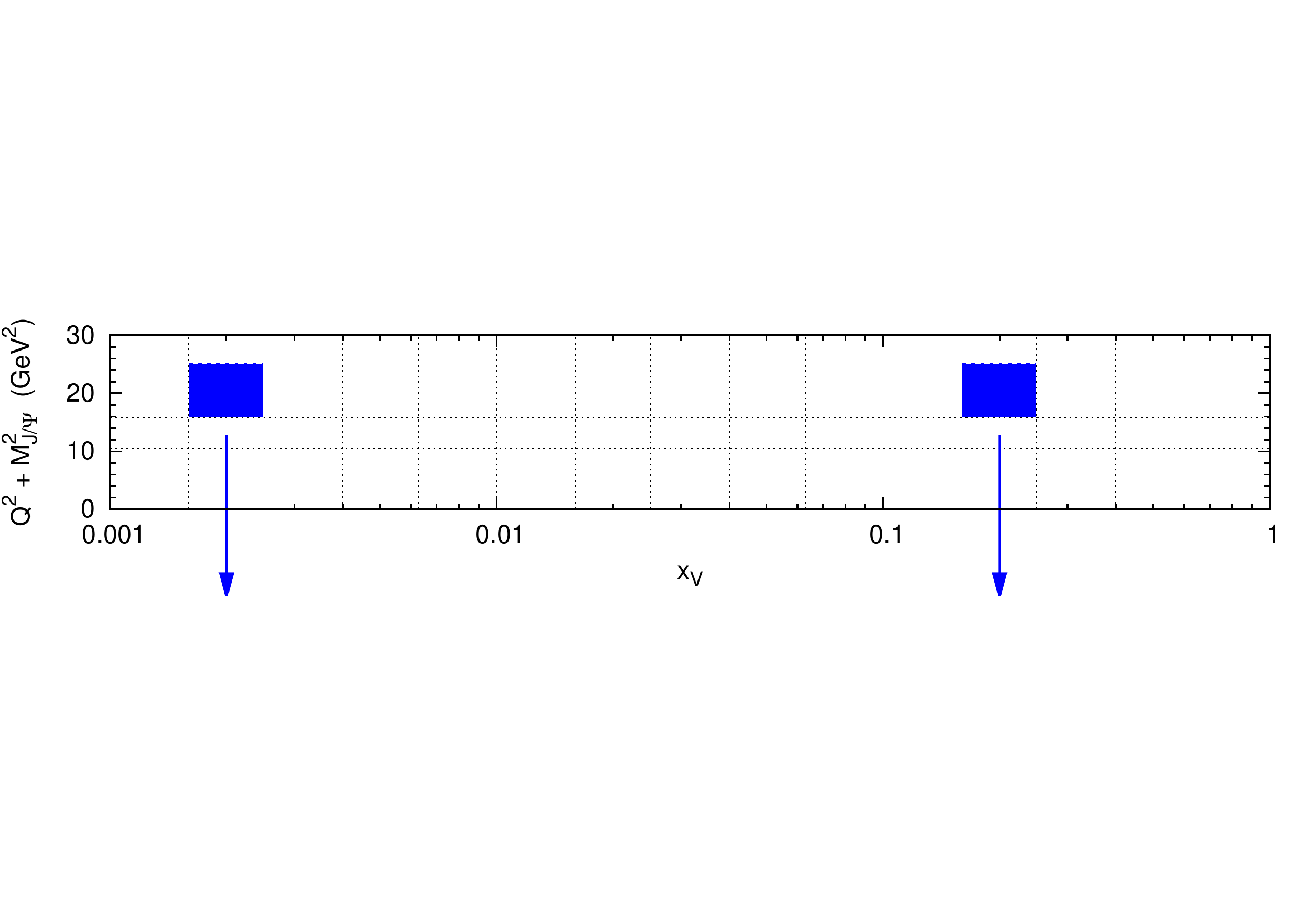}\\
[-1.45in]
\includegraphics[width=0.96\textwidth]{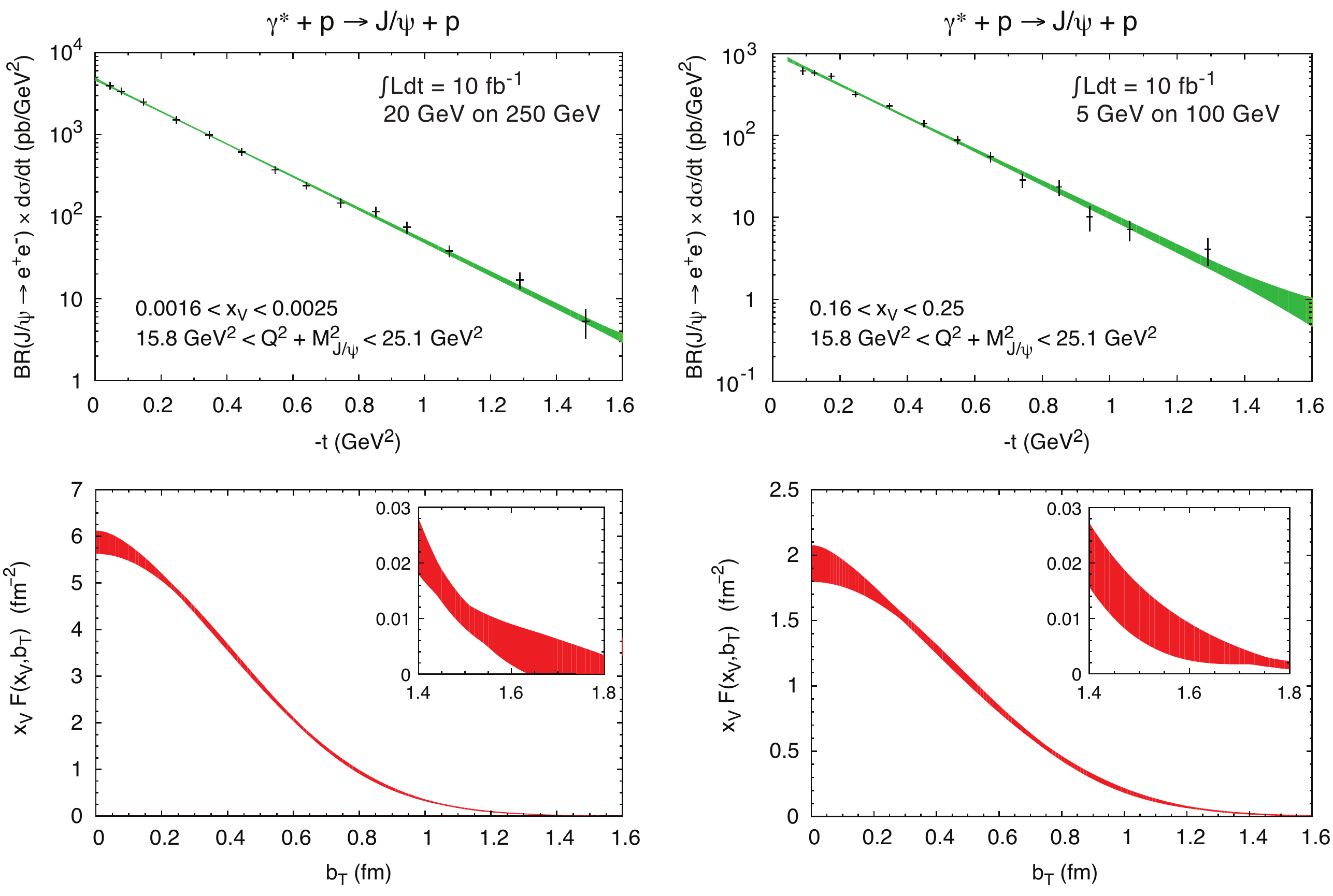}
\end{center}
\vspace{-0.25in}
\caption{\label{fig:jpsi-t-b} Top: cross-section for $\gamma^* p\to \jpsi
  p$ in two bins of $x_{V}$ and $Q^2$.  Bottom: the distribution of gluons
  in impact parameter $b_T$ obtained from the $\jpsi$ production cross
  section.  The bands have the same meaning as in
  Figure~\protect\ref{fig:dvcs-t-b}.}
\end{figure*}

Examples for the expected spectrum in $t$ are shown in
Fig.~\ref{fig:jpsi-t-b}, with details given in \cite{Diehl:DIS2012}.
Also shown are the $b_T$ space images obtained from the $\gamma^* p\to
\jpsi p$ scattering amplitude by a Fourier transform. The distributions
thus contain a contribution from the small but finite size of the
$\jpsi$ meson, which needs to be disentangled in a full GPD analysis.
We see from the Figure that with data from the low and high energy coverage of an EIC,
this process will enable us to accurately probe the spatial distribution
of gluons over two orders of magnitude in $x$, up to the
region where the dominant partons are valence quarks.  The transverse
proton spin asymmetry \cite{Koempel:2011rc} will in addition give
constraints on the distribution $E$ for gluons and thus strongly
complement what can be achieved with DVCS.
\end{multicols}

\vspace*{-0.2in}

\vspace*{-0.2in}
\subsection{Opportunities with Nuclei}
\label{sec:excl-nuclei}
\vspace{-0.2in}

\begin{multicols}{2}
Although the focus of this section is on imaging the proton, let us
briefly point out that exclusive reactions with nuclear beams offer a
variety of physics
opportunities.  Light nuclei such as ${}^{3}$He or the deuteron can
provide an effective neutron target, which can be used for
disentangling $u$ and $d$ distributions, just as for the usual parton
densities measured in inclusive processes.  Such measurements are even
more powerful if the nuclei can be polarized.

Coherent exclusive processes, in which the nucleus stays intact, give new
handles for the understanding of collective dynamics such as shadowing,
anti-shadowing or the EMC effect.  An overview and references can be found
in Sec.~5.9.1 of \cite{Boer:2011fh}.  Coherent exclusive reactions such
as $\jpsi$ production on heavy nuclear targets have the potential to map
out the geometry of the nucleus in high-energy processes and thus to
quantify the initial conditions of heavy-ion collisions.  As discussed in
Sec.~\ref{keyM}, they may offer detailed information about parton
saturation by exhibiting the
$\boldsymbol{b}_T$ dependence of the amplitude
$N(x,\boldsymbol{r}_T,\boldsymbol{b}_T)$ for scattering a color dipole of
size $\boldsymbol{r}_T$ at a transverse distance $\boldsymbol{b}_T$ from
the center of the nucleus.

Scattering processes at high $Q^2$ in which two or more
nucleons are simultaneously knocked out of a nucleus provide an
opportunity to study short-range correlations between nucleons in a
nucleus.  Fixed-target experiments \cite{Malki:2000gh,Subedi:2008zz} have
obtained intriguing results, which not only provide detailed insight into
the nucleon-nucleon interaction at short distances but also have
astrophysical implications \cite{Frankfurt:2008zv}.  At the EIC, one will have
the unique opportunity to study the role of gluon degrees of freedom in
these short-range correlations.  For instance, in exclusive $\jpsi$
production off light nuclei accompanied by knockout nucleons, see
Sec.~5.12 of \cite{Boer:2011fh}.  Such
studies have the potential to greatly increase our understanding of
nuclear forces in the transition region between hadronic and partonic
degrees of freedom.
\end{multicols}

%% file: files_tex/eA-intro.tex
\section{Introduction}
\label{sec:intro-nuclei}

\begin{multicols}{2}
\setlength{\parskip}{0mm}
QCD, the accepted theory of strong interactions, is in general very
successful in describing a broad range of hadronic and nuclear
phenomena. One of the main achievements in our understanding of QCD is
the variation of the strong coupling constant and 
asymptotic freedom, which is the name for the theoretically predicted
and experimentally established fact that quarks and gluons are almost
free at very short (asymptotic) distances inside the hadrons
\cite{Gross:1973id,Politzer:1973fx}. QCD is often studied in deep
inelastic scattering (DIS) experiments, in which one probes the inner
structure of the proton or nucleus by scattering a small probe (a
lepton) on it. The lepton probes the quark distribution in the proton
or nucleus by exchanging a photon with it. 
Past DIS experiments were very successful in determining the quark structure of the proton
and of some light and intermediate-size nuclei.

Despite the many successes in our understanding of QCD, some profound
mysteries remain. One of them is quark confinement: quarks can not be
free (for a long time) in nature and are always confined inside bound
states -- the hadrons. Another one is the mass of the proton (and
other hadrons), which, at $938$~MeV, is much larger than the sum of the
valence quark masses (about $10$~MeV). Both of these problems at the
moment can only be tackled by numerical QCD simulations on the
lattice. The current consensus is that the gluons are responsible for
both the quark confinement and much of the hadronic mass.  The gluons,
which bind quarks together into mesons (bound states of a quark and an
anti-quark) and baryons (bound states of three quarks), significantly
contribute to the masses of hadrons. At the same time, gluons are
significantly less well-understood than quarks. Unlike photons, the
carriers of the electromagnetic force, gluons interact with each
other. The underlying non-linear dynamics of this self-interaction is
hard to put under theoretical control. Gluons are quite
little-studied for particles providing over $98 \%$ of the proton and
neutron masses, generating much of the visible matter mass in the
Universe.\footnote{One may compare the gluons to the Higgs boson, the
  search for which received a lot of attention in recent decades.
  While the recently discovered Standard Model Higgs accounts for the
  masses of all the known quarks along with the $W^\pm$ and $Z$
  bosons, this would still add up to only about $5 \%$ of the mass in
  the visible Universe.} In addition, \mbox{it is} known that gluons play a
dominant role in high energy DIS, hadronic and nuclear collisions,
being responsible for much of the particle production and total 
cross-sections in these processes. In high-energy heavy-ion collisions it is
the gluons that are likely to be responsible for production and
thermalization of the medium made out of deconfined quarks and gluons,
known as the quark-gluon plasma (QGP). Clearly any progress in our
understanding of gluon dynamics would profoundly improve our knowledge
of the strong force, allowing us to better control and more deeply
understand this fundamental interaction.

In this chapter, we illustrate that DIS experiments on large nuclei
(heavy ions) at high energies are the best way to study gluon
dynamics.  We show that a large number of nucleons in a heavy ion
likely results in strong gluon fields in its wave function probed at
high energy, possibly leading to the phenomenon of {\sl parton (gluon)
  saturation}, also known as the Color Glass Condensate (CGC). The
transition to this non-linear regime is characterized by the {\sl
  saturation momentum} $Q_s$, which can be large for heavy ions. Our
current theoretical understanding suggests that this strong gluon
field combines complex non-linear QCD dynamics with a perturbatively
large momentum scale $Q_s$, allowing one to perform small-coupling
theoretical calculations due to the asymptotic freedom property of
QCD. An electron-ion collider (EIC) would allow us to probe the wave
functions of high energy nuclei with an energetic electron: by
studying these interactions one may probe the strong gluon fields of
the CGC. While experiments at HERA, RHIC, and LHC found evidence
consistent with saturation, an EIC would have the potential to seal
the case, completing the discovery process started at those
accelerators.

Nuclei are made out of nucleons, which in turn, are bound states of
the fundamental constituents probed in high energy scattering or at
short distance, namely quarks and gluons.  The binding of nucleons
into a nucleus must be sensitive to how these quarks and gluons are
confined into nucleons, and must influence how they distribute inside
the bound nucleons.  The European Muon Collaboration (EMC) discovery
at CERN that revealed a peculiar pattern of nuclear modification of
the DIS cross-section as a function of Bjorken $x$, confirmed by
measurements at several facilities in the following two decades, shows
clear evidence that the momentum distributions of quarks in a
fast-moving nucleus are strongly affected by the binding and the
nuclear environment.  With much wider kinematic reach in both $x$ and
$Q$, and unprecedented high luminosity, the EIC not only can explore
the influence of the binding on the momentum distribution of sea
quarks and gluons, but also, for the first time, determine the spatial
distribution of quarks and gluons in a nucleus by diffractive or
exclusive processes.  

The EIC is capable of exploring the emergence of hadrons from almost
massless quarks and gluons, or heavy quarks.  This is a necessary and
critical process in the formation of our visible universe shortly
after its birth.  Color neutralization is key to the formation of
hadrons, and is still not understood within QCD.  In electron-nucleus
($e$+A) collisions at the EIC, the nucleus could serve as an effective
femtometer size detector to probe the color neutralization of a fast
moving color charge.
With the span of available collision energies, the wealth of
semi-inclusive probes and the control of kinematics, the EIC is able
to explore the response of nuclear medium to the motion of the color
charge, and to probe the strength and spatial distributions of quarks
and gluons inside the colliding nucleus.

The EIC would be the world's first dedicated electron-nucleus ($e$+A)
collider. It would be an excellent laboratory for exploring QCD
dynamics. The experimental program of the machine is targeted to
answer the following fundamental questions concerning the dynamics of
quarks and gluons in a nuclear environment:
\end{multicols}
\vspace{2mm} 
\begin{itemize}

\item {\bf Can we experimentally find evidence of a novel universal
    regime of\break 
\noindent non-linear QCD dynamics in nuclei?}  The large number of
  partons in a nucleus may result in strong gluon fields leading to
  the phenomenon of gluon saturation, known as the Color Glass
  Condensate.  This universal regime of high-energy QCD is described
  by non-linear evolution equations.  Discovery of the saturation
  regime would not be complete without unambiguous experimental
  evidence in favor of this non-linear behavior that stands in strong
  contrast to the linear DGLAP evolution, which describes QCD at
  large-$x$ and $Q^2$ so successfully. An EIC can complete the
  discovery of the gluon saturation/CGC regime, tantalizing hints of
  which may have been seen at HERA, RHIC, and the LHC.  Accomplishing the
  discovery of a new regime of QCD would have a profound impact on our
  understanding of strong interactions.

\item {\bf What is the role of saturated strong gluon fields, and what
    are the degrees of freedom in this high gluon density regime?}  An
  EIC will allow us to probe the wave functions of high-energy nuclei.
  By studying these interactions, one may probe the strong gluon fields
  of the CGC, possibly the strongest fields in nature. In this regime,
  multi-parton correlations dominate and the picture of hadronic
  matter described by individual parton distributions loses its
  validity.  If quarks and gluons are not the relevant degrees of
  freedom any more, than what are the correct degrees of freedom?  
  With its broad kinematic range, an EIC will allow us to explore this
  small-$x$ regime and gain insight into the dynamic of saturation
  expanding our understanding of QCD.

\item {\bf What is the fundamental quark-gluon structure of light and
    heavy nuclei?}  The measurement of momentum and spatial (impact
  parameter) distributions of gluons and sea quarks in nuclei over an
  unprecedented kinematic range in $x$ and $Q^2$ would provide
  groundbreaking insight into the new regime of saturation and the
  fundamental structure of nuclei.  These measured distributions at
  the EIC, together with the understanding of quark and gluon
  correlations, could expand our knowledge of nuclear structure into
  the realm of fundamental interaction described by QCD.

\item {\bf Can the nucleus, serving as a color filter, provide novel
    insight into the propagation, attenuation and hadronization of
    colored quarks and gluons?}  The emergence of colorless hadrons
  from colored quarks and gluons is a rich and still mysterious
  process in QCD.  Multiple interactions between a moving color charge
  and the color field of a nucleus it is colliding with, could alter the
  color evolution of this charge and its hadronization.  Hence, it is a
  valuable probe of color neutralization.  By using the nucleus as a
  space-time analyzer the EIC will shed light on answers to the
  questions such as the following: How does the nucleus respond to the propagation
  of a color charge through it?  What are the fluctuations in the
  spatial distributions of quarks and gluons inside the nucleus?  
  What governs the transition from quarks and gluons to hadrons?

\end{itemize}
\begin{figure}[h!]
\begin{center}
\includegraphics[width=0.85\textwidth]{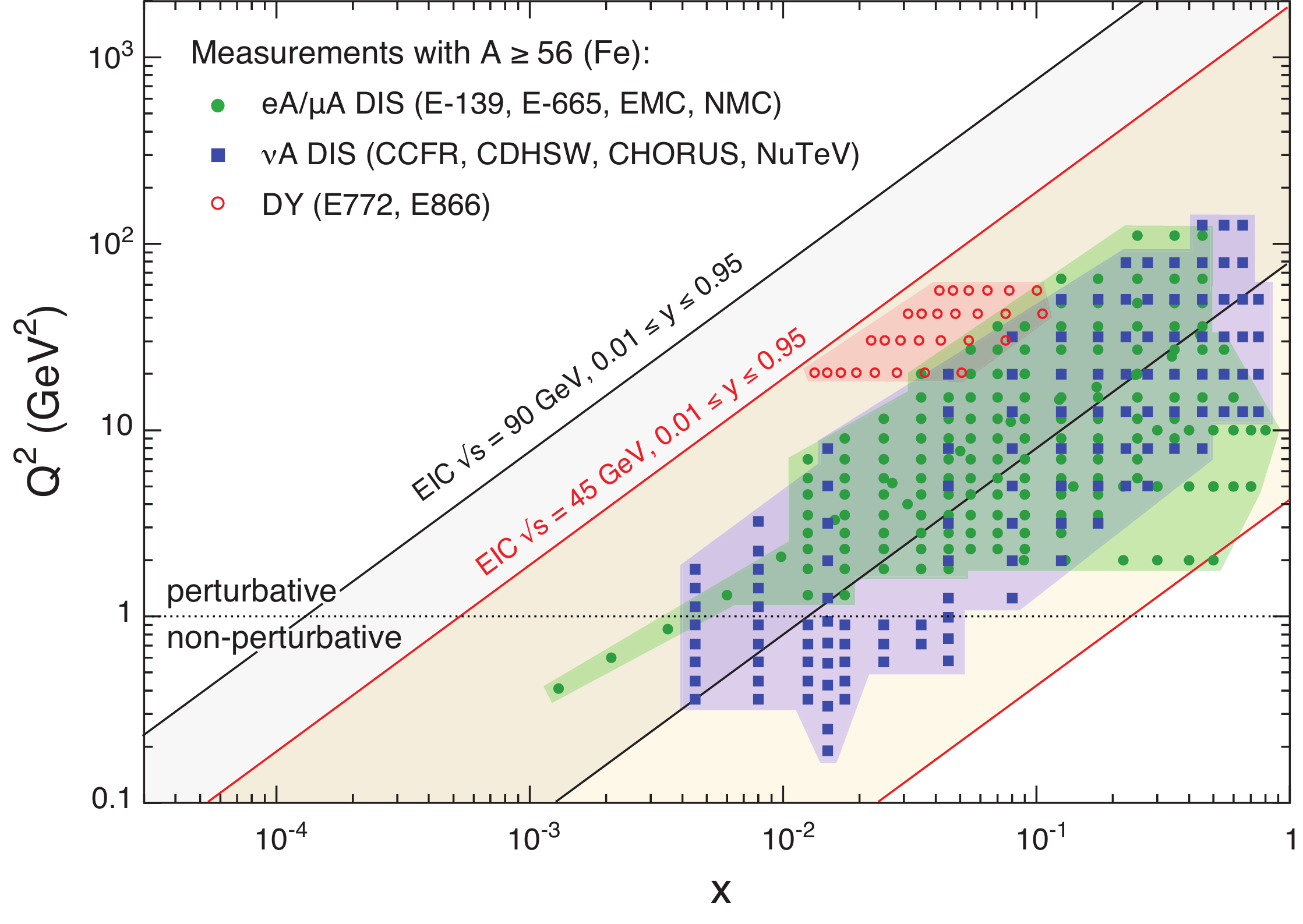}
\end{center}
\vspace*{-7mm}
\caption{The kinematic acceptance in $x$ and $Q^2$ of completed
  lepton-nucleus (DIS) and Drell-Yan (DY) experiments (all fixed
  target) compared to two EIC energy options. The acceptance bands for the EIC
  are defined by $Q^2 = x\ y\ s$ with $0.01 \le y \le 0.95$ and values
  of $s$ shown.}
\label{fig:lA-experiments}
\end{figure}
\begin{multicols}{2}
  The big questions listed above can be answered by performing a set
  of measurements using DIS on heavy ions at the EIC. The measurements
  relevant for the small-$x$ $e$+A physics are described in
  Sec.~\ref{HGDN}, while those pertaining to the large-$x$ $e$+A
  physics are discussed in Sec.~\ref{sec:nuclei}. Some of these
  measurements have analogs in $e$+$p$ collisions but have never been
  performed in nuclei; for these, $e$+$p$ collisions will allow us to
  understand universal features of the physics of the nucleon and the
  physics of nuclei. Other measurements have no analog in $e$+$p$
  collisions and nuclei provide a completely unique environment to
  explore these.  The EIC would have a capability of colliding many
  ion species at a wide range of collision energies.  With its high
  luminosity and detector coverage, as well as its high collision
  energies, the EIC could probe the confined motion as well as spatial
  distributions of quarks and gluons inside a nucleus at unprecedented
  resolution --- one tenth of a femtometer or better --- and could
  detect soft gluons whose energy in the rest frame of the nucleus is
  less than one tenth of the averaged binding energy needed to hold
  the nucleons together to form the nucleus. With large nuclei, the
  EIC could reach the saturation regime that may only be reached by
  electron-proton collisions with a multi-TeV proton beam.  The
  kinematic acceptance of an EIC compared to all other data collected
  in DIS on nuclei and in Drell-Yan (DY) experiments is shown in
  \fig{fig:lA-experiments}. Clearly an EIC would greatly extend our
  knowledge of strong interactions in a nuclear environment.
\end{multicols}

%% file: files_tex/sidebarDiffraction.tex
\newpage
\setlength{\oddsidemargin}{0em}
\setlength{\textwidth}{6in}
\setlength{\topmargin}{-0.55in}
\setlength{\textheight}{9in}
\definecolor{lightgray}{rgb}{0.85,0.85,0.85}
\definecolor{LightYellow}{cmyk}{0.0,0.0,0.1,0.0}

\pagecolor{LightYellow}

\phantomsection
\label{sdbar:diffraction}

\vspace*{-1.00cm}
\begin{center}
{\textbf {\textit {\textcolor{blue}{\Large Diffractive Scattering}}}}
\line(1,0){435}
\end{center}

    \noindent Diffractive scattering has made a spectacular comeback with the
    observation of an unexpectedly large cross-section for diffractive
    events at the HERA $e$+$p$ collider.  At HERA, hard diffractive
    events, $e(k) + N(p) \rightarrow e^{\prime}(k^{\prime}) +
    N(p^{\prime}) + X$, were observed where the proton remained intact
    and the highly virtual photon fragmented into a final state $X$
    that was separated from the scattered proton by a large rapidity gap 
    without any particles.
    These events are indicative of a color neutral
    exchange in the $t$-channel between the virtual photon and
    the proton over several units in rapidity. This color singlet
    exchange has historically been called the pomeron, which had a
    specific interpretation in Regge theory. An illustration of a
    hard diffractive event is shown in
    Fig.~\ref{fig:sidebarDiffractiveGraph}.
    
\begin{multicols}{2}
    \includegraphics[width=0.45\textwidth]{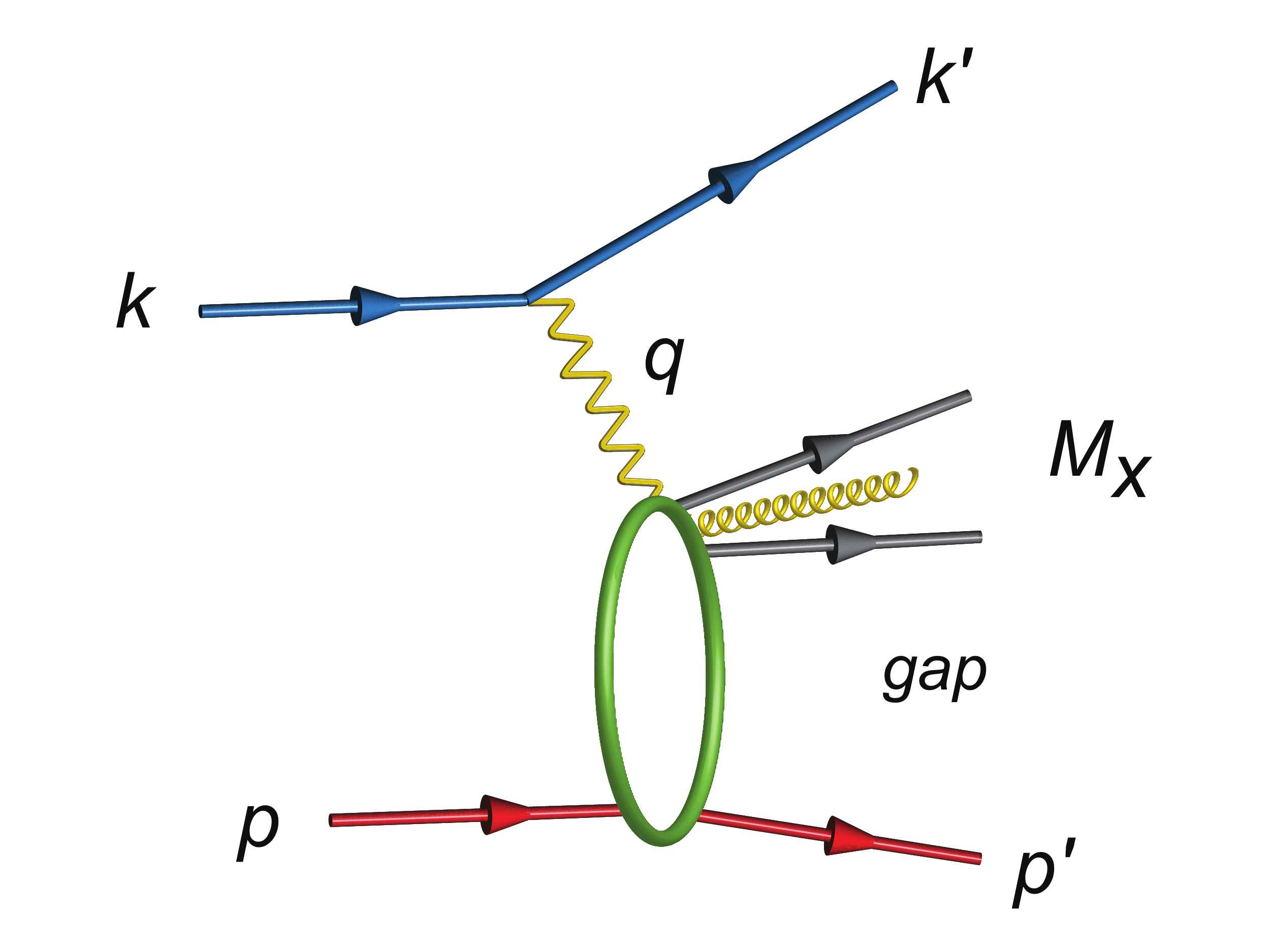}
    \mfcaption{\it Kinematic quantities for the description of a diffractive event.}
    \label{fig:sidebarDiffractiveGraph}

    The kinematic variables are similar to those for DIS with the
    following additions:
    \begin{description}
    \item[$\boldsymbol{t} = (p - p^{\prime})^2$] is the square of the momentum
        transfer at the hadronic vertex. The variable $t$ here is identical to the 
        one used in exclusive processes and generalised parton distributions 
        (see the Sidebar on page~\pageref{sdbar:GPD}).
        
    \item[$\boldsymbol{M_X^2} = (p-p^{\prime} + k-k^{\prime})^2$] 
        is the squared mass of the diffractive final state.
      
    \item[$\boldsymbol{\eta} = \ln(\tan(\theta/2))$] is the pseudorapidity of a 
        particle whose momentum has a relative angle $\theta$ to the proton beam axis. 
     For ultra-relativistic particles the pseudorapidity is equal to the rapidity,  
     $\eta \sim y = 1/2 \ln ((E+p_L)/(E-p_L))$.
      
    \end{description}
  \end{multicols}

    At HERA, gaps of several units in rapidity have been observed.  
    One finds that roughly 15\%
    of the deep inelastic cross-section corresponds to hard diffractive events with
    invariant masses $M_X > 3$\,GeV. The remarkable nature of this
    result is transparent in the proton rest frame: a 50\,TeV electron
    slams into the proton and $\approx$ 15\% of the time, the proton
    is unaffected, even though the virtual photon imparts a high momentum 
    transfer on a quark or antiquark in the target.
    A crucial question in diffraction is the nature of the color neutral
    exchange between the proton and the virtual photon.
    This interaction probes, in a novel fashion, the
    nature of confining interactions within hadrons.

    The cross-section can be formulated analogously to inclusive DIS
    by defining the \textit{diffractive} structure functions $F_2^D$
    and $F_L^D$ as
    \begin{eqnarray*}
        \frac{d^4 \sigma}{dx_B\, dQ^2\, dM_X^2\, dt} = 
        \frac{4 \pi \alpha^2}{Q^6} 
        \left[  \left( 1 - y + \frac{y^2}{2} \right) F_2^{D, 4}(x, Q^2, M_X^2, t) 
            - \frac{y^2}{2} F_L^{D, 4}(x, Q^2, M_X^2, t) \right].
    \end{eqnarray*}

    In practice, detector specifics may limit the measurements of
    diffractive events to those where the outgoing proton (nucleus) is not
    tagged, requiring instead a large rapidity gap $\Delta \eta$ in the detector. 
    $t$ can then only be measured for particular final states $X$, e.g.
    for $J/\Psi$ mesons, whose momentum can be reconstructed very precisely.
    
\newpage \pagecolor{white}

%% file: files_tex/eA-dense.tex
\section{Physics of High Gluon Densities in Nuclei}
\label{HGDN}

\setlength{\parskip}{0mm} 

{\large\it Conveners:\ Yuri Kovchegov and Thomas Ullrich}
\vskip 0.15in

In this section we present a description of the physics one would like
to access with the small-$x$ EIC program, along with the measurements
needed to answer the related fundamental questions from the beginning
of this chapter.  One needs to measure the nuclear structure functions
$F_2$ and $F_L$ (see the Sidebar on page~\pageref{sdbar:crosssection}) 
as functions of the Bjorken-$x$ variable and photon virtuality $Q^2$ (see
the Sidebar on page~\pageref{sdbar:DIS}), 
which allows us to extract quark and
gluon distribution functions of the nuclei, along with the
experimental evidence for the non-linear QCD effects. One needs to
determine the saturation scale $Q_s$ characterizing the CGC wave
function by measuring two-particle correlations.  The distribution of
gluons, both in position and momentum spaces, can be pinpointed by the
measurement of the cross-section of elastic vector meson production.
The cross-sections for diffractive (quasi-elastic) events are most
sensitive to the onset of the non-linear QCD dynamics.


\subsection{Gluon Saturation: a New Regime of QCD}

\begin{multicols}{2}
\subsubsection{Non-linear Evolution}
\label{nonlin_ev}

The proton is a bound state of three ``valence'' quarks: two up quarks
and one down quark. The simplest view of a proton reveals three quarks
interacting via the exchanges of gluons, which ``glue'' the quarks
together. But experiments probing proton structure at the HERA
collider at Germany's DESY laboratory, and the increasing body of
evidence from RHIC and the LHC, suggest that this picture is far too
simple. Countless other gluons and a ``sea'' of quarks and anti-quarks
pop in and out of existence within each hadron. These fluctuations can
be probed in high-energy scattering experiments.  Due to Lorentz time
dilation, the more we accelerate a proton and the closer it gets to
the speed of light, the longer are the lifetimes of the gluons that
arise from the quantum fluctuations. An outside ``observer'' viewing a
fast moving proton would see the cascading of gluons last longer and
longer, the larger the velocity of the proton.  So, in effect, by
speeding the proton up, one can slow down the gluon fluctuations
enough to ``take snapshots'' of them with a probe particle sent to
interact with the high-energy proton.

In DIS experiments, one probes the proton wave-function with a lepton,
which interacts with the proton by exchanging a (virtual) photon with
it (see the Sidebar on page~\pageref{sdbar:DIS}). 
The virtuality of the photon, $Q^2$,
determines the size of the region in the plane transverse to the beam
axis probed by the photon.  By the uncertainty principle, the region's width
is $\Delta r_T \sim 1/Q$. Another relevant variable is Bjorken $x$,
which is the fraction of the proton momentum carried by the struck
quark. At high energy, $x \approx Q^2/W^2$ is small ($W^2$ is the
center-of-mass energy squared of the photon-proton system).  Therefore,
small $x$ corresponds to high-energy scattering.

\begin{figure*}[htb]
\begin{center}
\includegraphics[width=0.6\textwidth]{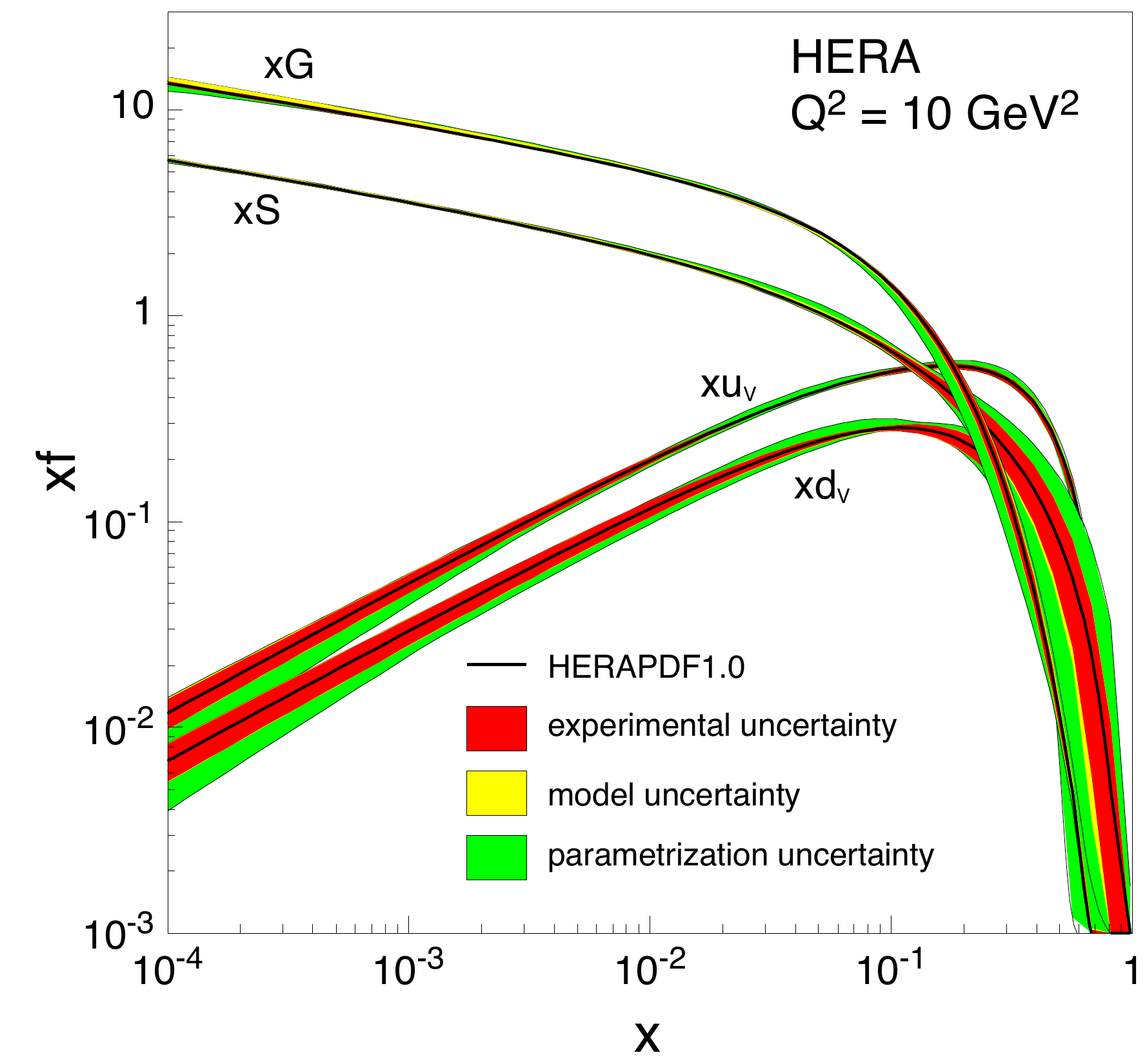}
\end{center}
\vskip -0.2in
\caption{Proton parton distribution functions plotted as functions of
  Bjorken $x$. 
  Clearly gluons dominate at
  small-$x$.}
\label{fig:hera_10GeV2}
\end{figure*}

The proton wave-function depends on both $x$ and $Q^2$. An example of
such a dependence is shown in \fig{fig:hera_10GeV2}, extracted from
the data measured at HERA for DIS on a proton.  Here we plot the
$x$-dependence of the parton (quark or gluon) distribution functions
(PDFs). At the leading order PDFs can be interpreted as providing the
number of quarks and gluons with a certain fraction $x$ of the
proton's momentum. In \fig{fig:hera_10GeV2}, one can see the PDFs of
the valence quarks in the proton, $x u_v$ and $x d_v$ which decrease
with decreasing $x$. The PDFs of the ``sea'' quarks and gluons,
denoted by $xG$ and $xS$ in \fig{fig:hera_10GeV2}, appear to grow very
strongly towards the low $x$. (Please note the logarithmic scale of
the vertical axis.) 
One can also observe that the gluon distribution dominates over those
of the valence and ``sea'' quarks at a moderate $x$ below $x
=0.1$. Remembering that low-$x$ means high energy, we conclude that
the part of the proton wave-function responsible for the interactions
in high energy scattering consists mainly of gluons.

\begin{figure*}[htb]
\begin{center}
\includegraphics[width=0.7\textwidth]{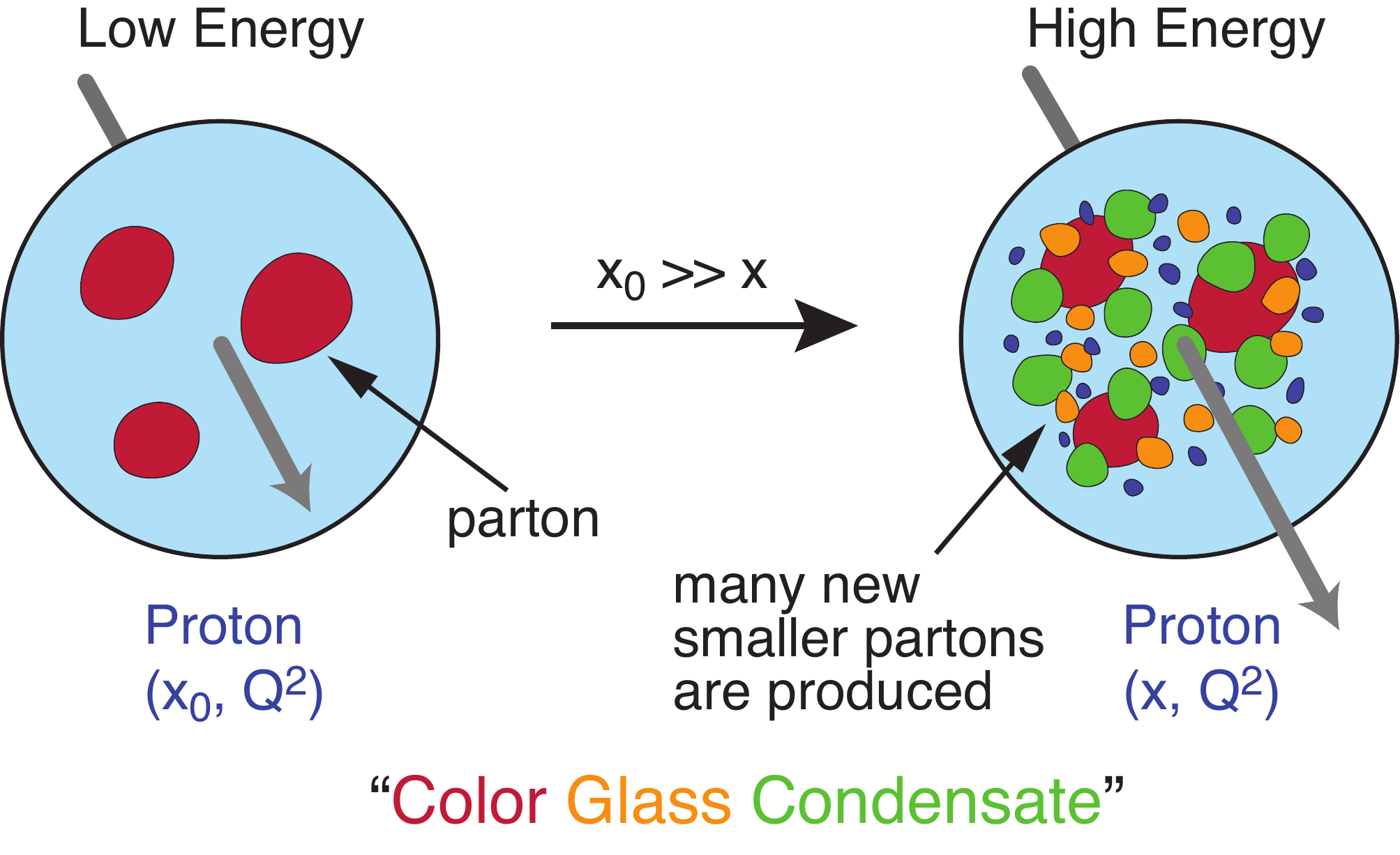}
\end{center}
\vskip -0.2in
\caption{The proton wave-function at small-$x$ (shown on the right)
  contains a large number of gluons (and quarks) as compared to the
  same wave-function at a larger $x = x_0$ (shown on the left). The
  figure is a projection on the plane transverse to the beam axis (the
  latter is shown by arrows coming ``out of the page,'' with the
  length of the arrows reflecting the momentum of the proton).}
\label{partons}
\end{figure*}

The small-$x$ proton wave-function is dominated by gluons, which are
likely to populate the transverse area of the proton, creating a high
density of gluons. This is shown in \fig{partons}, which illustrates
how at lower $x$ (right panel), the partons (mainly gluons)
are much more numerous inside the proton than at larger-$x$ (left
panel), in agreement with \fig{fig:hera_10GeV2}. This dense small-$x$
wave-function of an ultra-relativistic proton or nucleus is referred to
as the Color Glass Condensate (CGC) \cite{Iancu:2001ad}. 

\begin{figure*}[ht]
\begin{center}
\includegraphics[width=0.9\textwidth]{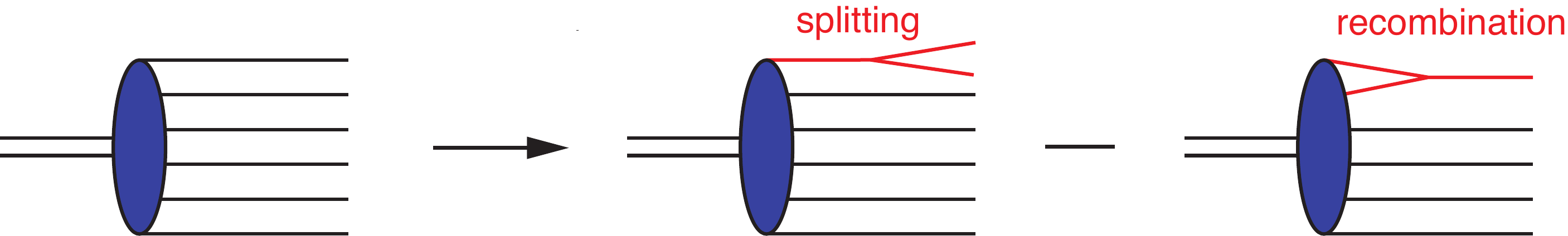}
\end{center}
\vskip -0.1in
\caption{The non-linear small-$x$ evolution of a hadronic or nuclear wave 
  functions. All partons (quarks and gluons) are denoted by straight
  solid lines for simplicity.}
\label{BKfig}
\end{figure*}

To understand the onset of the dense regime, one usually employs {\sl
  QCD evolution equations}. The main principle is as follows: While
the current state of the QCD theory does not allow for a
first-principles calculation of the quark and gluon distributions, the
evolution equations, loosely-speaking, allow one to determine these
distributions at some values of $(x, Q^2)$ if they are initially known
at some other $(x_0, Q_0^2)$. The most widely used evolution equation
is the Dokshitzer-Gribov-Lipatov-Altarelli-Parisi (DGLAP) equation
\cite{Gribov:1972ri,Altarelli:1977zs,Dokshitzer:1977sg}.  If the PDFs
are specified at some initial virtuality $Q_0^2$, the DGLAP equation
allows one to find the parton distributions at $Q^2 > Q_0^2$ at all
$x$ where DGLAP evolution is applicable. The evolution equation that
allows one to construct the parton distributions at low-$x$, given the
value of it at some $x_0 > x$ and all $Q^2$, is the
Balitsky-Fadin-Kuraev-Lipatov (BFKL) evolution equation
\cite{Balitsky:1978ic,Kuraev:1977fs}. This is a linear evolution
equation, which is illustrated by the first term on the right hand
side of \fig{BKfig}. The wave-function of a high-energy proton or
nucleus containing many small-$x$ partons is shown on the left of
\fig{BKfig}. As we make one step of evolution by boosting the
nucleus/proton to higher energy in order to probe its smaller-$x$ wave
function, either one of the partons can split into two partons,
leading to an increase in the number of partons proportional to the
number of partons $N$ at the previous step,
\begin{align}\label{BFKL}
  \frac{\partial \, N (x, r_T)}{\partial \ln (1/x)} = \as \,
  K_{\mathrm{BFKL}} \, \otimes \, N (x, r_T),
\end{align}
with $K_{\mathrm{BFKL}}$ an integral kernel and $\as$ the strong
coupling constant. In DIS at high energy, the virtual photon splits
into a quark-antiquark dipole which interacts with the proton.  The
dipole scattering amplitude $N (x, r_T)$ probes the gluon distribution
in the proton at the transverse distance $r_T \sim 1/Q$.\footnote{In
  general, the dipole amplitude also depends on the impact parameter
  $b_T$ of the dipole (cf. Sec.~\ref{sec:excl-nuclei}): for simplicity
  we suppress this dependence in $N (x, r_T)$.} Note that a Fourier
transform of $N (x, r_T)$ is related to the gluon transverse momentum
distribution (TMD) $f (x, k_T)$ from Chap. 2. The BFKL evolution
leads to the power-law growth of the parton distributions with
decreasing $x$, such that $N \sim (1/x)^\lambda$ with $\lambda$ a
positive number \cite{Balitsky:1978ic}. This behavior may account for
the increase of the gluon density at small-$x$ in the HERA data of
\fig{fig:hera_10GeV2}.

The question arises whether the gluon and quark densities can grow
without limit at small-$x$. While there is no strict bound on the
number density of gluons in QCD, there is a bound on the scattering
cross-sections stemming from unitarity. Indeed, a proton (or nucleus)
with a lot of ``sea'' gluons is more likely to interact in high
energy scattering, which leads to larger scattering cross-sections.
Therefore, the bound on cross-sections should have implications for the
gluon density. The cross-section bound arises due to the black disk
limit known from quantum mechanics.  The high-energy total scattering cross
section of a particle on a sphere of radius $R$ is bounded by
\begin{align}\label{bd}
  \sigma_{\mathrm{tot}} \, \le \, 2 \, \pi \, R^2.
\end{align}
In QCD, the black disk limit translates into the Froissart--Martin
unitarity bound, which states that the total hadronic cross-section
can not grow faster than $\ln^2 s$ at very high energies with $s$ the
center-of-mass energy squared \cite{Froissart:1961ux}. The cross
section resulting from the BFKL growth of the gluon density in the
proton or nucleus wave-function grows as a power of energy,
$\sigma_{\mathrm{tot}} \sim s^{\lambda}$, and clearly violates both the black
disk limit and the Froissart--Martin bound at very high energy.

We see that something has to modify the BFKL evolution at high energy
to prevent it from becoming unphysically large. The modification is
illustrated on the far right of \fig{BKfig}.  At very high energies
(leading to high gluon densities), partons may start to recombine with
each other on top of the splitting. The recombination of two partons
into one is proportional to the number of pairs of partons, which in
turn scales as $N^2$. We end up with the following non-linear
evolution equation:
\end{multicols}
\vspace{-8mm}
\begin{align}\label{BK}
  \frac{\partial \, N (x, r_T)}{\partial \ln (1/x)} \, = \, \as \,
  K_{\mathrm{BFKL}} \, \otimes \, N (x, r_T) - \as \, [N (x, r_T)]^2. 
\end{align}
\vskip 1mm 
\begin{multicols}{2}
\noindent
This is the Balitsky-Kovchegov (BK) evolution equation
\cite{Balitsky:1995ub,Kovchegov:1999yj,Kovchegov:1999ua},
which is valid for QCD in the limit of the large number of colors
$N_c$.\footnote{An equation of this type was originally suggested by
Gribov, Levin and Ryskin in \cite{Gribov:1984tu} and by Mueller and
Qiu in \cite{Mueller:1986wy}, though at the time it was assumed that
the quadratic term was only the first non-linear correction with
higher order terms expected to be present as well.  In
\cite{Balitsky:1995ub,Kovchegov:1999yj}, the exact form of the
equation was found, and it was shown that in the large-$N_c$ limit
\eq{BK} does not have any higher-order terms in $N$.} A generalization
of \eq{BK} beyond the large-$N_c$ limit is accomplished by the
Jalilian-Marian--Iancu--McLerran--Weigert--Leonidov--Kovner (JIMWLK)
\cite{Iancu:2001ad,Jalilian-Marian:1997gr,JalilianMarian:1998cb,JalilianMarian:1997dw,
Iancu:2000hn} evolution equation, which is a functional differential equation.

The physical impact of the quadratic term on the right of \eq{BK} is
clear: it slows down the small-$x$ evolution, leading to {\sl parton
  saturation}, when the number density of partons stops growing with
decreasing $x$. The corresponding total cross-sections satisfy the
black disk limit of \eq{bd}. The effect of gluon mergers becomes
important when the quadratic term in \eq{BK} becomes comparable to the
linear term on the right-hand-side. This gives rise to the {\sl
  saturation scale} $Q_s$, which grows as $Q_s^2 \sim (1/x)^\lambda$
with decreasing $x$ \cite{Gribov:1984tu,Iancu:2002tr,Mueller:2002zm}.
\end{multicols}

\subsubsection{Classical Gluon Fields and the Nuclear ``Oomph'' Factor}
\label{class_sec}

\begin{multicols}{2}
We have argued above that parton saturation is a universal phenomenon,
valid both for scattering on a proton or a nucleus. Here we
demonstrate that nuclei provide an extra enhancement of the saturation
phenomenon, making it easier to observe and study experimentally.

\begin{figure*}[ht]
\begin{center}
\includegraphics[width=0.65\textwidth]{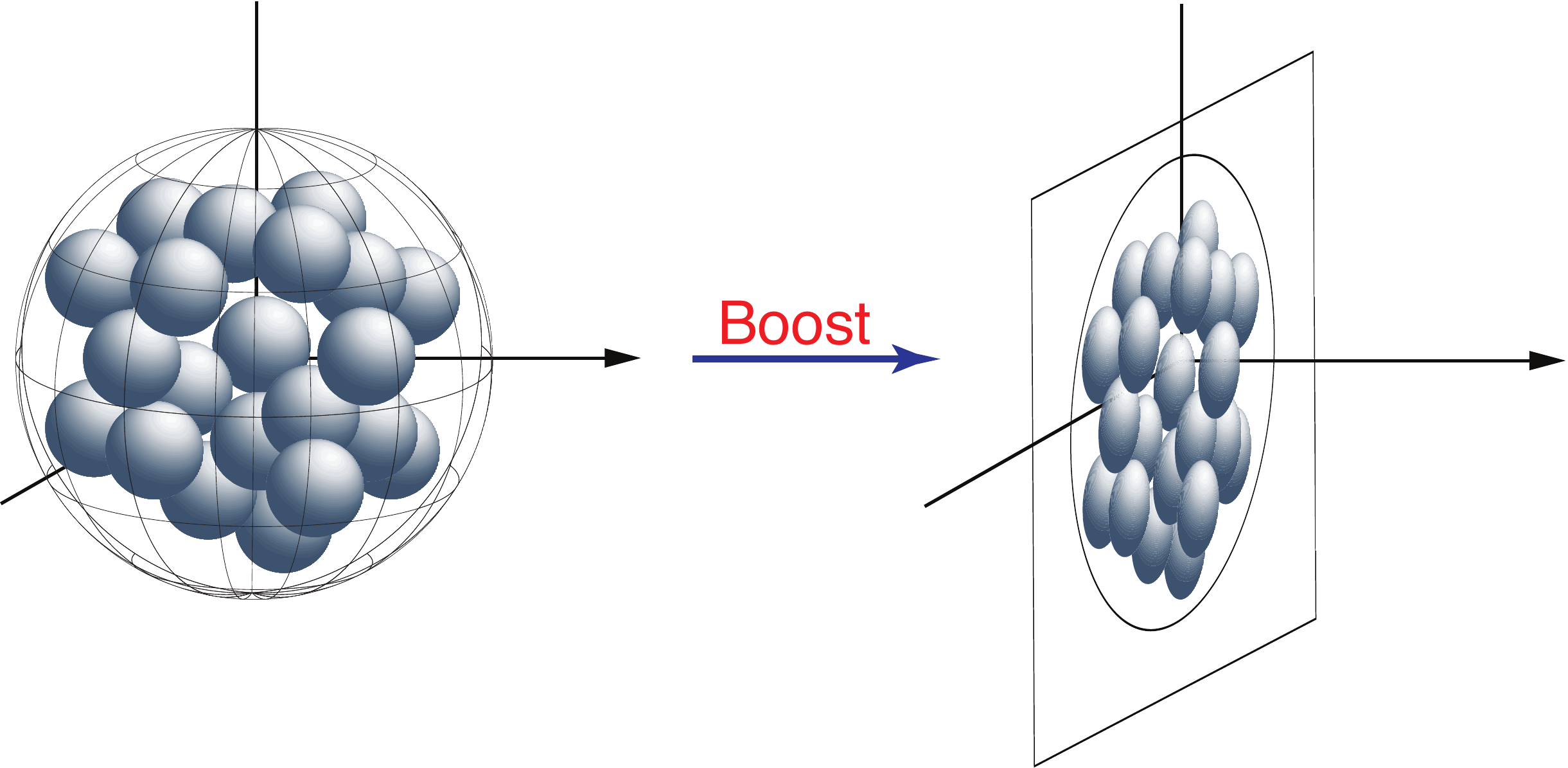}
\end{center}
\vskip -0.5cm
\caption{A large nucleus before and after an ultra-relativistic boost.}
\label{nucl_boost}
\end{figure*}

Imagine a large nucleus (a heavy ion), which was boosted to some
ultra-relativistic velocity, as shown in \fig{nucl_boost}. We are
interested in the dynamics of small-$x$ gluons in the wave-function of
this relativistic nucleus. One can show that due to the Heisenberg
uncertainty principle, the small-$x$ gluons interact with the whole
nucleus coherently in the longitudinal (beam) direction, Therefore,
only the transverse plane distribution of nucleons is important for
the small-$x$ wave-function. As one can see from \fig{nucl_boost},
after the boost, the nucleons, as ``seen'' by the small-$x$ gluons
with large longitudinal wavelength, appear to overlap with each other
in the transverse plane, leading to high parton density. A large
occupation number of color charges (partons) leads to a classical
gluon field dominating the small-$x$ wave-function of the
nucleus. This is the essence of the McLerran-Venugopalan (MV) model
\cite{McLerran:1993ni}. According to the MV model, the dominant gluon
field is given by the solution of the classical Yang-Mills equations,
which are the QCD analogue of Maxwell equations of electrodynamics.

The Yang-Mills equations were solved for a single nucleus exactly
\cite{Kovchegov:1996ty,Jalilian-Marian:1997xn}; their solution was
used to construct an unintegrated gluon distribution (gluon TMD) $\phi
(x, k_T^2)$ shown in \fig{mv2} (multiplied by the phase space factor
of the gluon's transverse momentum $k_T$) as a function of
$k_T$.\footnote{Note that in the MV model $\phi (x, k_T^2)$ is
  independent of Bjorken-$x$.  Its $x$-dependence comes in though the
  BK/JIMWLK evolution equations described above.}
\fig{mv2} demonstrates the emergence of the saturation scale $Q_s$.
The majority of gluons in this classical distribution have transverse
momentum $k_T \approx Q_s$. Note that the gluon distribution slows
down its growth with decreasing $k_T$ for $k_T < Q_s$ (from a
power-law of $k_T$ to a logarithm, as can be shown by explicit
calculations).  The distribution {\sl saturates}, justifying the name
of the saturation scale.

The gluon field arises from all the nucleons in the nucleus at a given
location in the transverse plane (impact parameter).  Away from the
edges, the nucleon density in the nucleus is approximately constant.
Therefore, the number of nucleons at a fixed impact parameter is
simply proportional to the thickness of the nucleus in the
longitudinal (beam) direction.
\begin{figurehere}
\begin{center}
\includegraphics[width=0.48\textwidth]{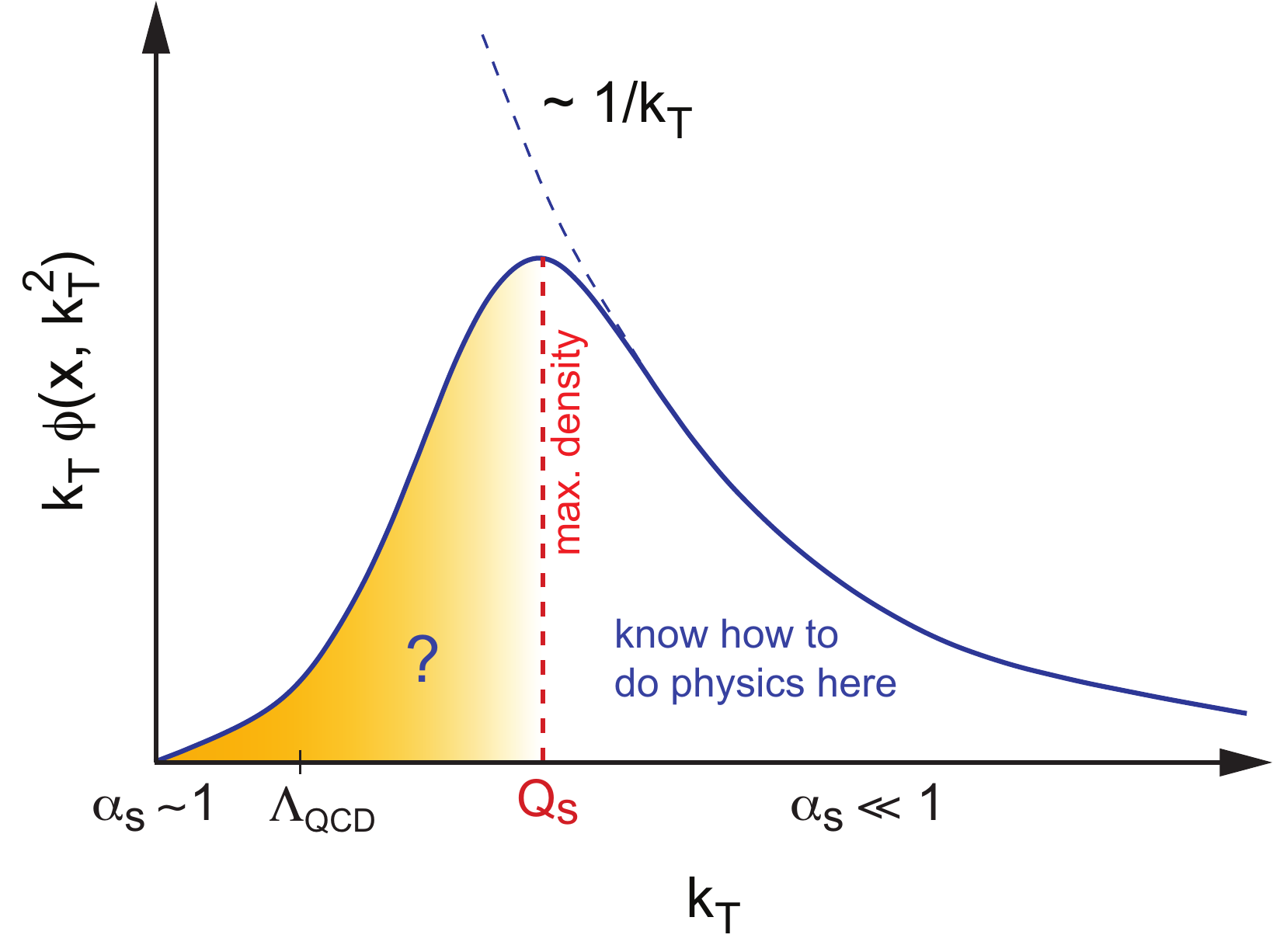}
\end{center}
\vskip -0.5cm
\caption{The unintegrated gluon distribution (gluon TMD) $\phi (x, k_T^2)$
  of a large nucleus due to classical gluon fields (solid line). The
  dashed curve denotes the lowest-order perturbative result.}
\label{mv2}
\end{figurehere}
\vspace{0.5cm} For a large nucleus, that thickness, in turn, is
proportional to the nuclear radius $R\sim A^{1/3}$ with the nuclear
mass number $A$. The transverse momentum of the gluon can be thought
of as arising from many transverse momentum ``kicks'' acquired from
interactions with the partons in all the nucleons at a given impact
parameter. Neglecting the correlations between nucleons, which is
justified for a large nucleus in the leading power of $A$
approximation, once can think of the ``kicks'' as being random.  Just
like in the random walk problem, after $A^{1/3}$ random kicks the
typical transverse momentum --- and hence the saturation scale ---
becomes $Q_s \sim \sqrt{A^{1/3}}$, such that $Q_s^2, \sim A^{1/3}$. We
see that the saturation scale for heavy ions, $Q_s^A$ is much larger
than the saturation scale of the proton, $Q_s^p$, (at the same $x$),
since $(Q_s^A)^2 \approx A^{1/3} \, (Q_s^p)^2$
\cite{Gribov:1984tu,Mueller:1986wy,McLerran:1993ni,Mueller:1989st}. This
enhancement factor $A^{1/3}$ of the saturation scale squared is often
referred to as the {\sl nuclear ``oomph'' factor}, since it reflects
the enhancement of saturation effects in the nucleus as compared to
the proton. For the gold nucleus with $A=197$, the nuclear ``oomph''
factor is $A^{1/3} \approx 6$.
\end{multicols}

\subsubsection{Map of High Energy QCD and the Saturation Scale}
\label{sec_map}

\begin{figure*}[ht]
\begin{center}
\includegraphics[width=0.5\textwidth]{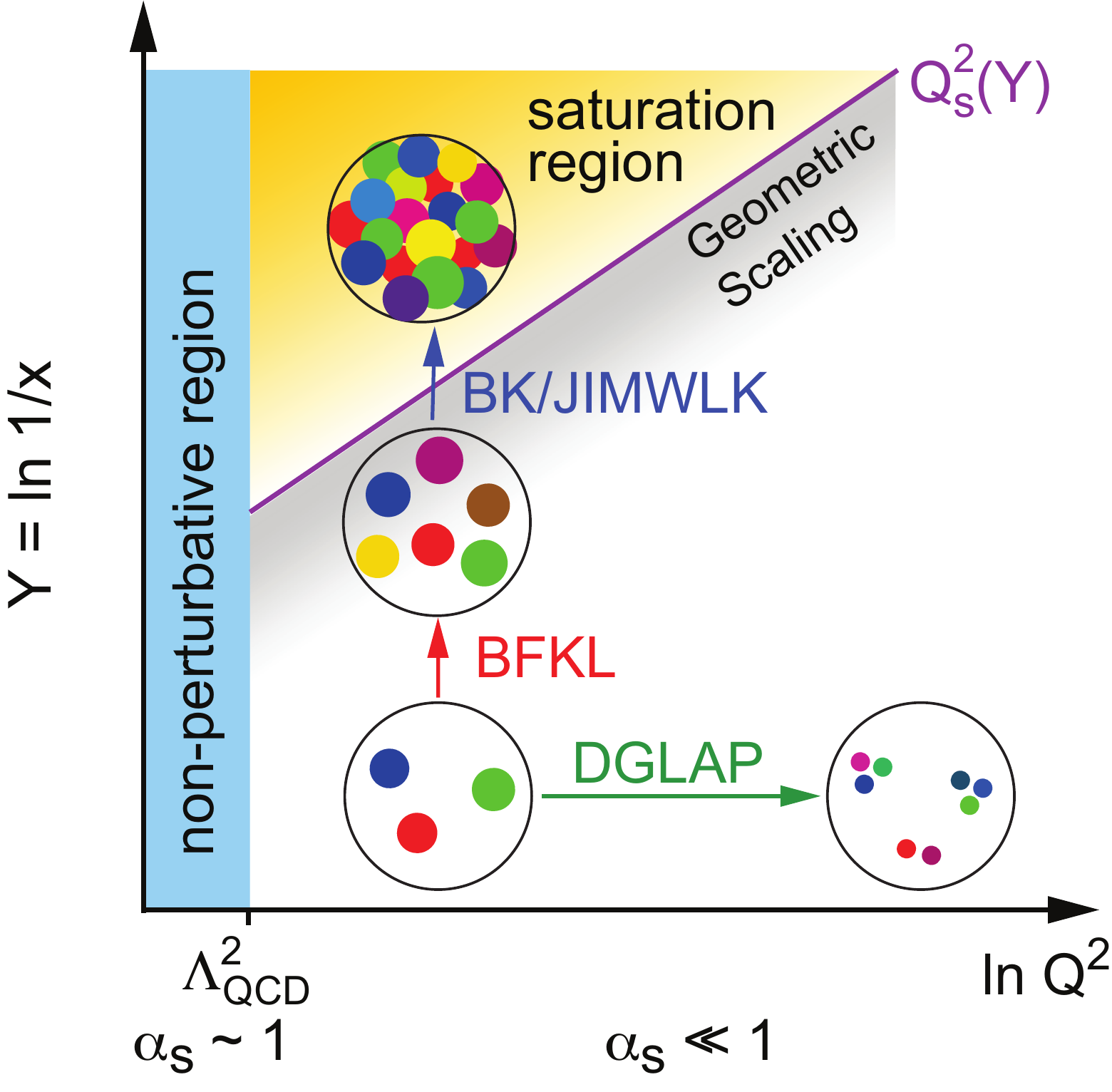}
\end{center}
\vskip -0.1in
\caption{The map of high energy QCD in the $(Q^2, Y=\ln 1/x)$ plane.}
\label{satbk}
\end{figure*}

\begin{multicols}{2}
We summarize our theoretical knowledge of high energy QCD discussed
above in \fig{satbk}, in which different regimes are plotted in the
$(Q^2, Y=\ln 1/x)$ plane.  On the left of \fig{satbk} we see the
region with $Q^2 \le \Lambda_{QCD}^2$ in which the strong coupling is
large, $\as \sim 1$, and small-coupling approaches do not work
($\Lambda_{QCD}$ is the QCD confinement scale).  In the perturbative
region, $Q^2 \gg \Lambda_{QCD}^2$, where the coupling is small, $\as
\ll 1$, we see the standard DGLAP evolution and the linear small-$x$
BFKL evolution, denoted by the horizontal and vertical arrows
correspondingly. The BFKL equation evolves the gluon distribution
towards small-$x$, where the parton density becomes large and parton
saturation sets in. The transition to saturation is described by the
non-linear BK and JIMWLK evolution equations. Most importantly, this
transition happens at $Q_s^2 \gg \Lambda_{QCD}^2$ where the
small-coupling approach is valid.

Saturation/CGC physics provides a new way of tackling the problem of
calculating hadronic and nuclear scattering cross-sections. It is
based on the theoretical observation that small-$x$ hadronic and
nuclear wave-functions --- and, therefore, the scattering
cross-sections --- are described by an internal momentum scale, the
{\sl saturation scale} $Q_s$ \cite{Gribov:1984tu}. As we argued above,
the saturation scale grows with decreasing $x$ (and, conversely, with
the increasing center-of-mass energy $\sqrt{s}$) and with the
increasing mass number of a nucleus $A$ (in the case of a nuclear wave
function) approximately as
\begin{align}\label{Qs}
  Q_s^2 (x) \, \sim \, A^{1/3} \, \left( \frac{1}{x} \right)^\lambda
\end{align}
where the best current theoretical estimates of $\lambda$ give
$\lambda$ = 0.2 -- 0.3 \cite{Albacete:2007sm}, in agreement with the
experimental data collected at HERA
\cite{Albacete:2009fh,Albacete:2010sy,gbw,GolecBiernat:1999qd} and at
RHIC \cite{Albacete:2007sm}. Therefore, for hadronic collisions at
high energy and/or for collisions of large ultra-relativistic nuclei,
the saturation scale becomes large, $Q_s^2 \gg \Lambda^2_{QCD}$. For
the total (and particle production) cross-sections, $Q_s$ is usually
the largest momentum scale in the problem.  We therefore expect it to
be the scale determining the value of the running QCD coupling
constant, making it small,
\begin{align}
  \as (Q_s^2) \, \ll \, 1,
\end{align}
and allowing for first-principles calculations of total hadronic and
nuclear cross-sections, along with extending our ability to calculate
particle production and to describe diffraction in a small-coupling
framework. For detailed descriptions of the physics of parton
saturation and the CGC, we refer the reader to the review articles
\cite{JalilianMarian:2005jf,Weigert:2005us,Iancu:2003xm,Gelis:2010nm}
and to an upcoming book \cite{KovchegovLevin}.

\eq{Qs} can be written in the following simple pocket formula if one
puts $\lambda = 1/3$, which is close to the range of $\lambda$ quoted
above. One has
\begin{align}\label{oomph}
  Q_s^2 (x) \, \sim \, \left( \frac{A}{x} \right)^{1/3}.
\end{align}
\noindent
From the pocket formula \eqref{oomph}, we see that the saturation
scale of the gold nucleus ($A=197$) is as large as that for a proton
at the 197-times smaller value of $x$! Since lower
values of $x$ can only be achieved by increasing the center-of-mass
energy, which could be prohibitively expensive, we conclude that at
the energies available at the modern-day colliders one is more likely
to complete the discovery of saturation/CGC physics started at HERA,
RHIC, and the LHC by performing DIS experiments on {\sl nuclei}.

\begin{figure*}[htb]
\begin{center}
\includegraphics[width=0.85\textwidth]{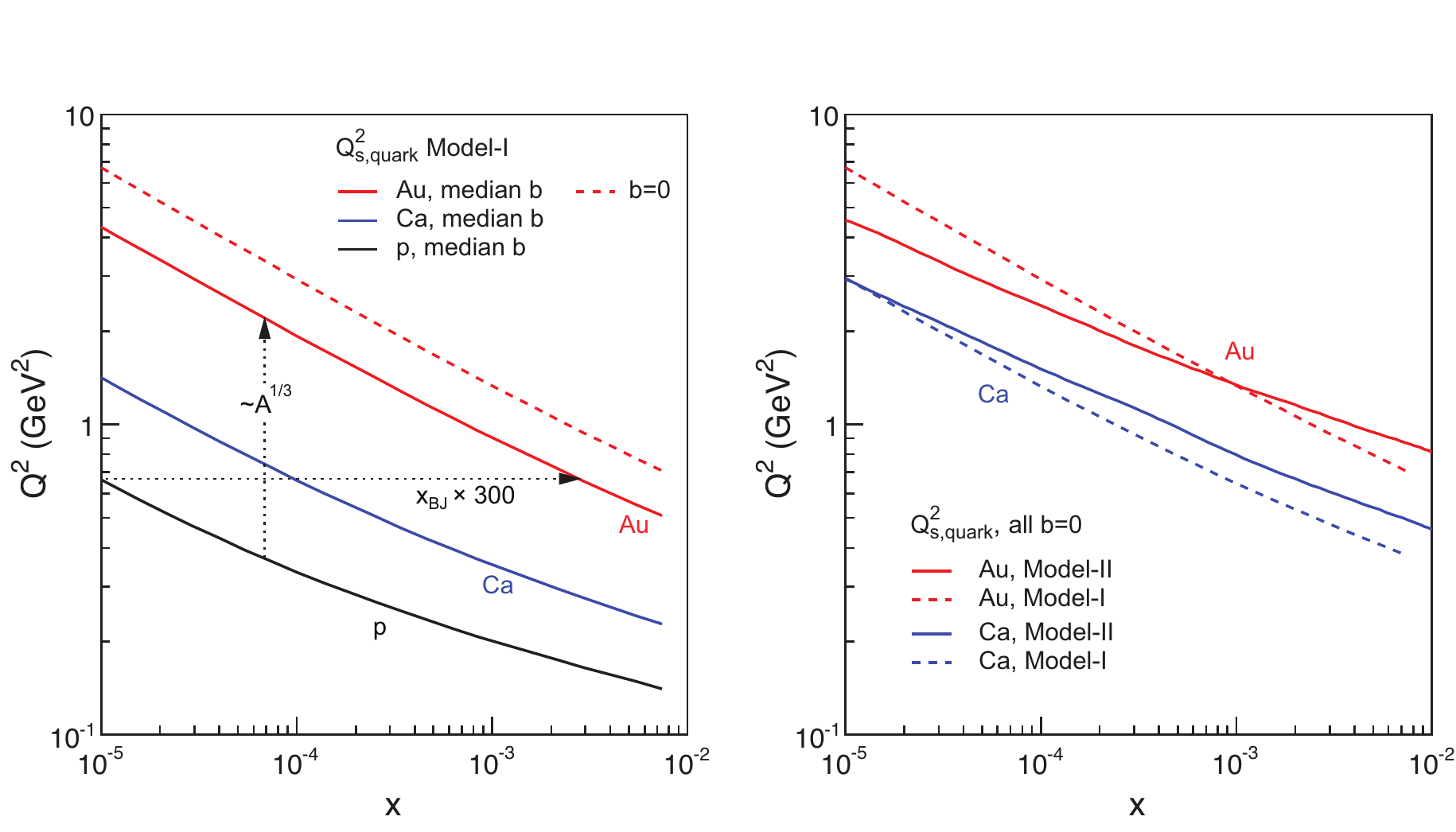}
\end{center}
\vskip -0.2in
\caption{Theoretical expectations for the saturation scale as a
  function of Bjorken $x$ for the proton along with Ca and Au
  nuclei.}
\label{fig:QsModels}
\end{figure*}

This point is further illustrated in \fig{fig:QsModels}, which shows
our expectations for the saturation scale as a function of $x$ coming
from the saturation-inspired Model-I \cite{Kowalski:2003hm} and from
the prediction of the BK evolution equation (with higher order
perturbative corrections included in its kernel) dubbed Model-II
\cite{Albacete:2009fh,Albacete:2010sy}. One can clearly see from the
left panel that the saturation scale for Au is larger than the
saturation scale for Ca, which, in turn, is much larger than the
saturation scale for the proton: the ``oomph'' factor of large nuclei
is seen to be quite significant.

As we argued above, the saturation scale squared is proportional to
the thickness of the nucleus at a given impact parameter
$b$. Therefore, the saturation scale depends on the impact parameter,
becoming larger for small $b \approx 0$ (for scattering through the
center of the nucleus, see \fig{nucl_boost}) and smaller for large $b
\approx R$ (for scattering on the nuclear periphery, see
\fig{nucl_boost}). This can be seen in the left panel of
\fig{fig:QsModels} where most values of $Q_s$ are plotted for median
$b$ by solid lines, while, for comparison, the $Q_s$ of gold is also
plotted for $b=0$ by the dashed line: one can see that the saturation
scale at $b=0$ is larger than at median $b$. The curves in the right
panel of \fig{fig:QsModels} are plotted for $b=0$: this is why they
give higher values of $Q_s$ than the median-$b$ curves shown in the
left panel for the same nuclei.

This $A$-dependence of the saturation scale, including a realistic
impact parameter dependence, is the {\em raison d'${\hat e}$tre} for
an electron-{\em ion} collider.  Collisions with nuclei probe the same
universal physics as seen with protons at values of $x$ at least two
orders of magnitude lower (or equivalently an order of magnitude
larger $\sqrt{s}$).  Thus, the nucleus is an efficient {\em amplifier}
of the universal physics of high gluon densities allowing us to study
the saturation regime in $e$+A at significantly lower energy than
would be possible in $e$+$p$. For example, as can be seen from
Fig.~\ref{fig:QsModels}, $Q_s^2 \approx 7$~GeV$^2$ is reached at $x =
10^{-5}$ in $e$+$p$ collisions requiring a collider providing a
center-of-mass energy of almost $\sqrt{s} \approx \sqrt{Q_s^2/x}
\approx 1$~TeV, while in $e$+Au collisions, only $\sqrt{s} \approx
60$~GeV is required to achieve comparable gluon density and the same
saturation scale.

\begin{figure*}[htb]
\begin{center}
\includegraphics[width=0.55\textwidth]{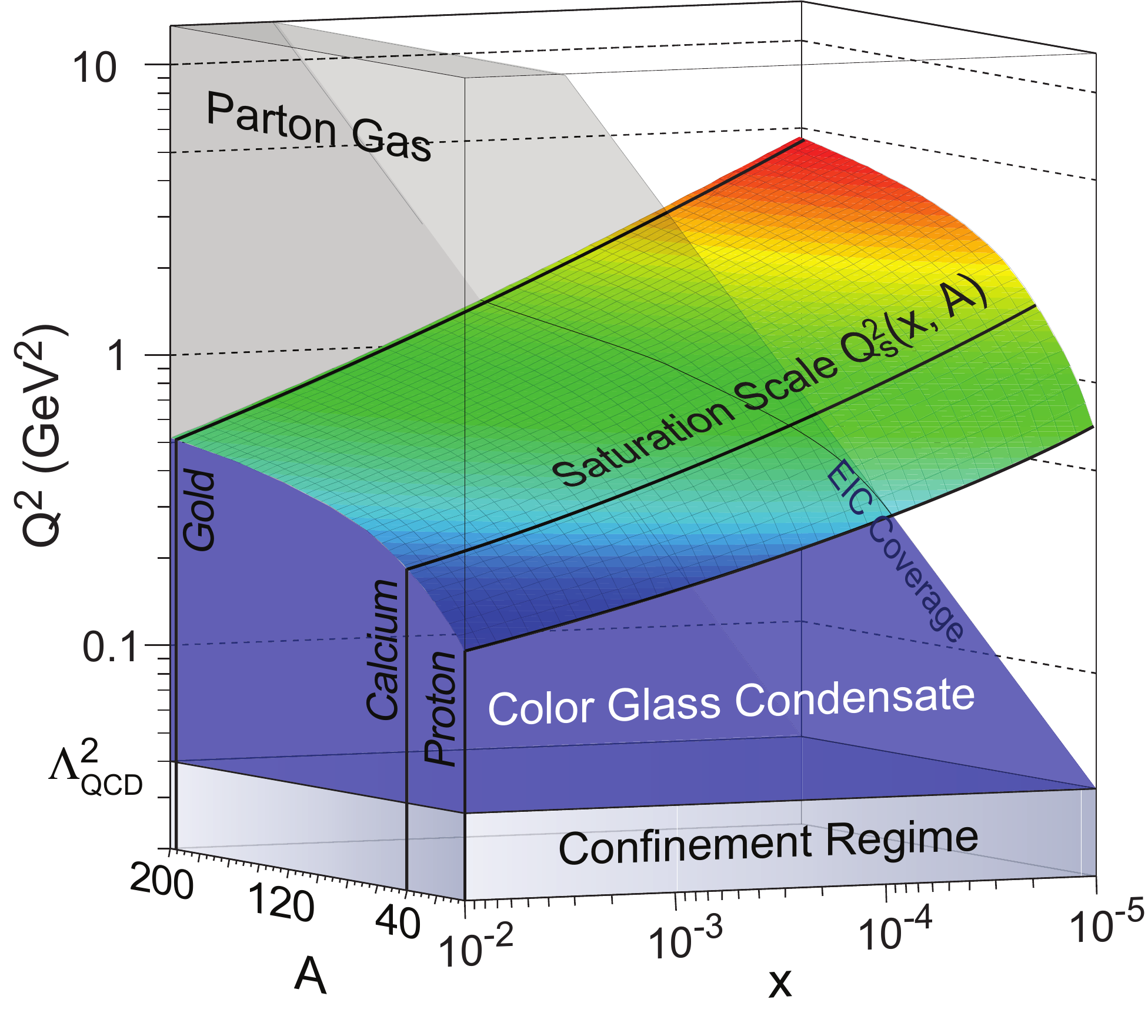}
\end{center}
\vskip -0.1in
\caption{The theoretical expectations for the saturation scale at
  medium impact parameter from Model-I as a function of Bjorken-$x$ and the nuclear mass
  number $A$.}
\label{fig:Qs3D}
\end{figure*}

To illustrate the conclusion that $Q_s$ is an increasing function of
both $A$ and $1/x$, we show a plot of its dependence on both variables
in \fig{fig:Qs3D} using Model-I of \fig{fig:QsModels}. One can see
again from \fig{fig:Qs3D} that larger $Q_s$ can be obtained by
increasing the energy or by increasing mass number $A$.

Measurements extracting the $x$, $b$ and $A$ dependence of the
saturation scale provide very useful information on the momentum
distribution and space-time structure of strong color fields in QCD at
high energies. The saturation scale defines the transverse momentum of
the majority of gluons in the small-$x$ wave-function, as shown in
\fig{mv2}, thus being instrumental to our understanding of the
momentum distributions of gluons. The impact parameter dependence of
the saturation scale tells us how the gluons are distributed in the
transverse coordinate plane, clarifying the spatial distribution of
the small-$x$ gluons in the proton or nucleus.


\subsubsection{Nuclear Structure Functions}
\label{sec_nsf}

The plots in Figs.~\ref{satbk}, \ref{fig:QsModels} and \ref{fig:Qs3D}
suggest a straightforward way of finding saturation/CGC physics: if we
perform the DIS experiment on a proton, or, better yet, on a nucleus,
and measure the DIS scattering cross-section as a function of $x$ and
$Q^2$, then, at sufficiently low $x$ and $Q^2$, one may be able to see
the effects of saturation. As explained in the Sidebar on
page~\pageref{sdbar:crosssection}, the total DIS cross-section is
related to the structure functions $F_2 (x, Q^2)$ and $F_L (x, Q^2)$
by a linear relation. One finds that the structure function $F_2$ is
more sensitive to the quark distribution $x q (x, Q^2)$ of the proton
or nucleus, while the structure function $F_L$ measures the gluon
distribution $x G (x, Q^2)$
\cite{Dokshitzer:1977sg,Nikolaev:1994ce}. Saturation effects can thus
be seen in both $F_2$ and $F_L$ at low $x$ and $Q^2$, although, since
saturation is gluon-driven, one would expect $F_L$ to manifest them
stronger.

The nuclear effects on the structure functions can be quantified by
the ratios
\vskip -0.2in
\begin{align}
  R_2 (x, Q^2) \equiv \frac{F_2^A (x, Q^2)}{A \, F_2^p (x, Q^2)},\nonumber\\ 
  R_L (x, Q^2) \equiv \frac{F_L^A (x, Q^2)}{A \, F_L^p (x, Q^2)}
 \label{eq:shadowing}
\end{align}
for the two structure functions, where the superscripts $p$ and $A$
label the structure functions for the protons and nuclei
correspondingly. Ratios like those in \eq{eq:shadowing} can be
constructed for the quark and gluon nuclear PDFs too. The ratio for
the gluon distribution compares the number of gluons per nucleon in
the nucleus to the number of gluons in a single free proton. Since the
structure function $F_L$ measures the gluon distribution $x G (x,
Q^2)$ \cite{Dokshitzer:1977sg,Nikolaev:1994ce}, the ratio $R_L (x,
Q^2)$ is close to the ratio $R_G (x, Q^2)$ of the gluon PDFs in the
nucleus and the proton normalized the same way,
\begin{align}
  \label{eq:xG_ratio}
  R_G (x, Q^2) \equiv \frac{x G_A (x, Q^2)}{A \, x G_p (x, Q^2)}.
\end{align}

\begin{figure*}[htb]
\begin{center}
\includegraphics[width=0.65\textwidth]{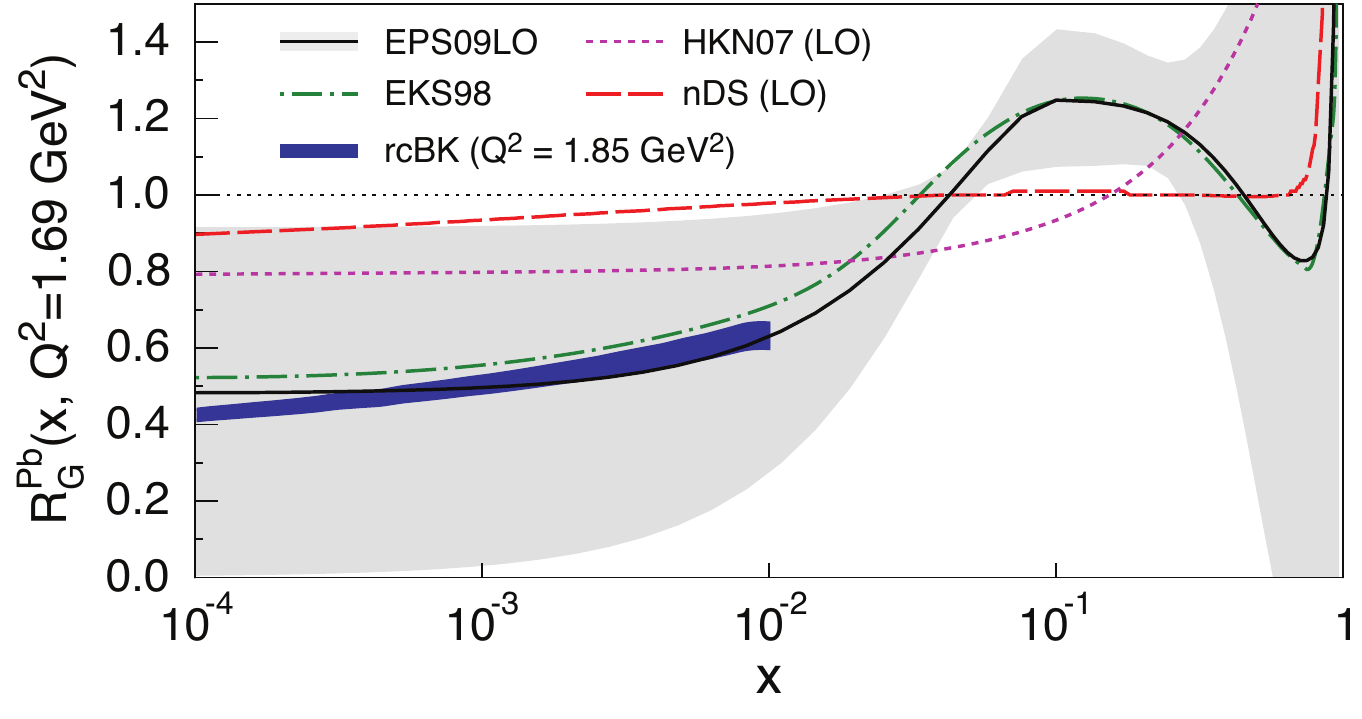}
\end{center}
\vspace{-8mm}
\caption{Theoretical predictions for $R_G (x, Q^2)$ plotted at $Q^2 =
  1.69$~GeV$^2$ for a $Pb$ nucleus: the models corresponding to
  different curves are explained in the plot legend. The models are:
  EPS09 \cite{Eskola:2009uj}, EKS 98 \cite{Eskola:1998df} (based on
  the leading-order (LO) global DGLAP analysis), HKN 07
  \cite{Hirai:2007sx}, nDS \cite{deFlorian:2003qf}
  (next-to-leading-order (NLO) DGLAP analysis), and rcBK
  \cite{Albacete:2009fh}, plotted for $Q^2 = 1.85$~GeV$^2$ (based on
  BK non-linear evolution with the running-coupling corrections (rcBK)
  \cite{Balitsky:2006wa,Kovchegov:2006vj,Gardi:2006rp,Albacete:2007yr},
  referred to as Model-II in Sec.~\ref{sec_map}). The light-gray
  shaded area depicts the uncertainty band of EPS09, while the blue
  shaded area indicates the uncertainty band of the rcBK
  approach.}
\label{fig:shadowing}
\end{figure*}

A sample of theoretical predictions for the ratio $R_G (x, Q^2)$ for
the gluon PDFs is plotted in \fig{fig:shadowing}, comprising several
DGLAP-based models along with the saturation-based prediction. Note
that the DGLAP equation, describing evolution in $Q^2$, can not predict
the $x$ dependence of distribution functions at low-$x$ without the
data at comparable values of $x$ and at lower $Q^2$: hence the
DGLAP-based ``predictions'' in \fig{fig:shadowing} strongly suffer
from the uncertainty in various {\sl ad hoc} parameterizations of the
initial conditions for DGLAP evolution. Conversely, the saturation
prediction is based on the BK equation \eqref{BK}, which is an
evolution equation in $x$, generating a very specific $x$-dependence
of the distribution functions that follows from QCD: this leads to a
narrow error band for the saturation prediction.

All existing approaches predict that the ratio $R_G$ would be below
one at small-$x$: this is the {\sl nuclear shadowing} phenomenon
\cite{Frankfurt:1988nt}, indicating that the number of small-$x$
gluons per nucleon in a nucleus is lower than that in a free
proton. In the DGLAP-based description of nuclear PDFs, shadowing
is included in the parameterizations of the initial conditions for
DGLAP evolution.  In the saturation/CGC approach, gluon mergers and
interactions dynamically lead to the decrease in the number of gluons
(and other partons) per nucleon as compared to that in a single
proton: this results in the shadowing of PDFs and reduction of
structure functions as well.

One can clearly see from \fig{fig:shadowing} that new data is
desperately needed to constrain the DGLAP-based prediction and/or to
test the prediction of saturation physics. It is also clear that
such data would eliminate some of the predictions shown in
\fig{fig:shadowing}, allowing us to get closer to finding the model
describing the correct physics. Still, as one can infer from
\fig{fig:shadowing}, due to the multitude of theoretical predictions,
the $R_G$ (or $R_L$) measurement alone may only rule out some of them,
leaving several predictions in agreement with the data within the
experimental error bars. As we detail further in Sec.~\ref{sec:strf},
one would need other measurements, like measurements of $R_2$,
$F_2^A$, $F_L^A$, along with those described below in Sec.~\ref{keyM},
to uniquely determine the physics involved in high-energy DIS on the
nucleus.

Nuclear effects in the structure functions can also be quantified
using their expansion in powers of $1/Q^2$
\cite{Bukhvostov:1985rn}. The standard linear perturbative QCD
approaches calculate the leading term in $1/Q^2$ expansion of
structure functions, the order-$1$ contribution, referred to as the
{\sl `leading twist'} term. The multiple re-scatterings of
Sec.~\ref{class_sec} along with the gluon mergers of
Sec.~\ref{nonlin_ev} contribute to all orders in the $1/Q^2$
expansion. Of particular interest is their contribution to the
non-leading powers of $1/Q^2$, known as {\sl `higher twists'}: 
the main parts of those corrections are enhanced by the nuclear
``oomph'' factor $A^{1/3}$ and by a power of $(1/x)^\lambda$, coming
in as
\begin{align}
  \label{eq:ht}
  \sim \, \frac{\Lambda_{QCD}^2 \, A^{1/3}}{Q^2} \, \left( \frac{1}{x}
  \right)^\lambda .
\end{align}
We see that the telltale sign of saturation physics are the higher
twist corrections, which are enhanced in DIS on a nucleus, and at
smaller-$x$.\footnote{In fact, equating the correction in \eq{eq:ht}
  to the leading-twist order-1 term gives the saturation scale of
  \eq{Qs} as the value of $Q^2$ at which the higher-twist corrections
  become important.}

\begin{figure*}[htb]
\begin{center}
\includegraphics[width=0.98\textwidth]{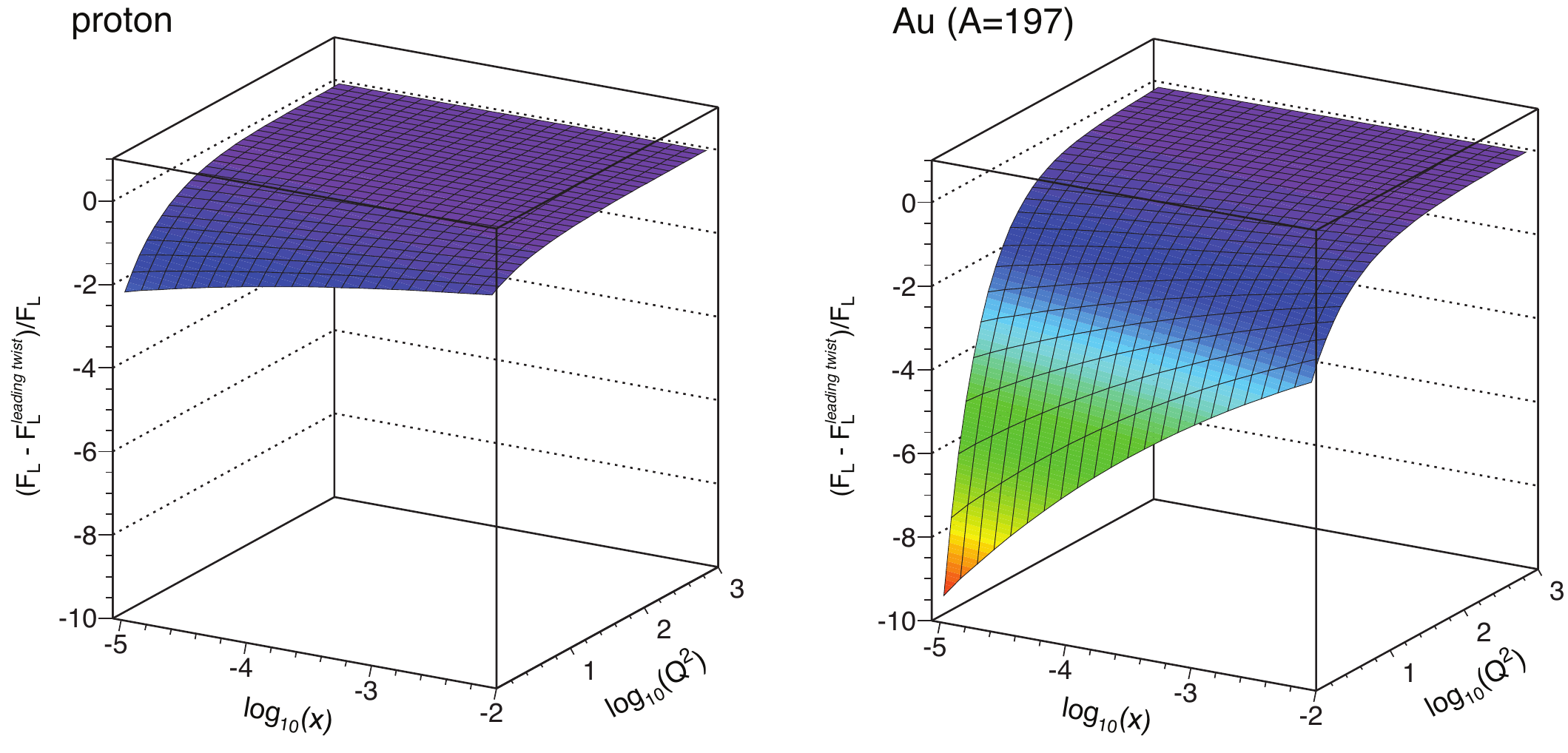}
\end{center}
\vskip -0.1in
\caption{Plots of the ratio from \eq{eq:bgbmratio} for $e$+$p$ and $e$+Au
  scattering from \cite{Bartels:2009tu}, demonstrating the sensitivity
  of nuclear structure function $F_L$ to the higher-twist effects. The
  plots go down to $x=10^{-5}$ as the smallest-$x$ reachable at an EIC
  (see \fig{fig:lA-experiments}).}
\label{fig:bgbm_combo}
\end{figure*}

To illustrate the effect of higher twist corrections on the nuclear
structure function we plot their relative contribution to $F_L$
defined by
\begin{align}
  \label{eq:bgbmratio}
  \frac{F_L - F_L (\mbox{leading twist})}{F_L}
\end{align}
in \fig{fig:bgbm_combo} as a function of $x$ and $Q^2$ as expected in
the framework of the saturation-inspired Golec-Biernat--Wusthoff (GBW)
model \cite{GolecBiernat:1998js,GolecBiernat:1999qd}, which has been
quite successful in describing the HERA $e$+$p$ data. The left panel of
\fig{fig:bgbm_combo} is for $e$+$p$ scattering, while the right one is
for $e$+Au. Note that the ratio is negative in both plots, indicating
that higher twists tend to decrease the structure function. It is also
clear from both plots that the effect of higher twists becomes
stronger at smaller-$x$, as expected from \eq{eq:ht}. Comparing the
two panels in \fig{fig:bgbm_combo}, we see that the higher twist
effects are also stronger in $e$+Au scattering due to nuclear
enhancement. \fig{fig:bgbm_combo} demonstrates that the structure
function $F_L$ is rather sensitive to parton
saturation. Experimentally, it is impossible to single out the
higher-twist contribution if the $Q^2$ of interest is too high, making
it difficult to plot the ratio from \eq{eq:bgbmratio} to verify the
prediction in \fig{fig:bgbm_combo}. At lower $Q^2$, experimental
separation of the leading twist contribution from the higher-twist
terms may also become a problem. Theoretical work is currently under
way to enable the separation of higher twist terms in $F_L$ (and
$F_2$), which is likely to make the ratio \eqref{eq:bgbmratio} an
observable which could be measured at an
EIC. 


\subsubsection{Diffractive Physics}
\label{diff_sec}

\begin{figure*}[htb]
\begin{center}
\includegraphics[width=0.95\textwidth]{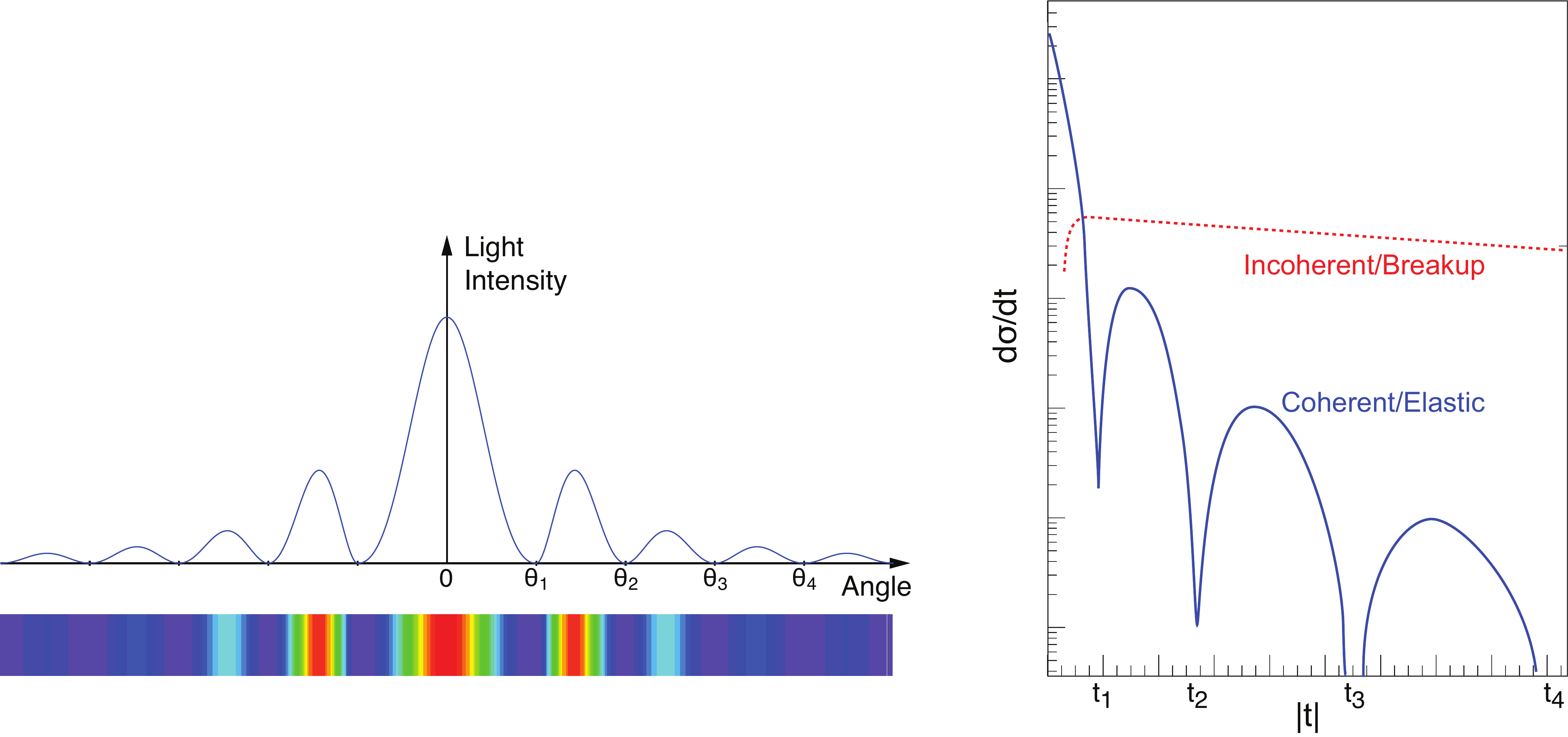}
\end{center}
\vskip -0.1in
\caption{Left panel: The diffractive pattern of light on a circular
  obstacle in wave optics. Right panel: The diffractive cross-section in
  high energy scattering. The elastic cross-section in the right panel
  is analogous to the diffractive pattern in the left panel if we
  identify $|t| \approx k^2 \, \theta^2$.}
\label{fig:diff_light}
\end{figure*}

The phenomenon of diffraction is familiar to us from many areas of
physics and is generally understood to arise from the constructive or
destructive interference of waves. Perhaps the best analogy of
diffraction in high-energy QCD comes from optics: imagine a standard
example of a plane monochromatic wave with the wave number $k$
incident on a circular screen of radius $R$ (an obstacle). The
diffractive pattern of the light intensity on a plane screen behind
the circular obstacle is shown in the left panel of
\fig{fig:diff_light} as a function of the deflection angle $\theta$,
and features the well-known diffractive maxima and minima. The
positions of the diffractive minima are related to the size of the
obstacle by $\theta_i \sim 1/(k \, R)$ for small-angle diffraction.

Elastic scattering in QCD has a similar structure.  Imagine a hadron (a
projectile) scattering on a target nucleus. If the scattering is
elastic, both the hadron and the nucleus will be intact after the
collision. The elastic process is described by the differential
scattering cross-section $d \sigma_{el}/dt$ with the Mandelstam
variable $t$ describing the momentum transfer between the target and
the projectile. A typical $d \sigma_{el}/dt$ is sketched by the solid
line in the right panel of \fig{fig:diff_light} as a function of
$t$. Identifying the projectile hadron with the incident plane wave in
the wave optics example, the target nucleus with the obstacle, and
writing $|t| \approx k^2 \, \theta^2$ valid for small angles, we can
see that the two panels of \fig{fig:diff_light} exhibit analogous
diffractive patterns and, therefore, describe very similar physics!
The minima (and maxima) of the cross-section $d \sigma_{el}/dt$ in the
right panel of \fig{fig:diff_light} are also related to the inverse
size of the target squared, $|t_i| \sim 1/R^2$. This is exactly the
same principle as employed for spatial imaging of the nucleons as
described in Sec.~\ref{sec:tmd}.

The essential difference between QCD and wave optics is summarized by
two facts: (i) The proton/nuclear target is not always an opaque
``black disk'' obstacle of geometric optics. A smaller projectile,
which interacts more weakly due to color-screening and asymptotic freedom,
is likely to produce a different diffractive pattern from the larger,
more strongly interacting, projectile. 
(ii) The scattering in QCD does not have to be completely elastic: the
projectile or target may break up. The event is still called
diffractive if there is a rapidity gap, as described in
the Sidebar on page~\pageref{sdbar:diffraction}. 
The cross-section for the target
breakup (leaving the projectile intact) is plotted by the dotted line
in the right panel of \fig{fig:diff_light}, and does not exhibit the
diffractive minima and maxima.

The property (i) is very important for diffraction in DIS in relation
to saturation/CGC physics. As we have seen above, owing to the
uncertainty principle, at higher $Q^2$, the virtual photon probes
shorter transverse distances, and is less sensitive to saturation
effects. Conversely, the virtual photon in DIS with the lower $Q^2$ is
likely to be more sensitive to saturation physics. Due to the presence
of a rapidity gap, the diffractive cross-section can be thought of as
arising from an exchange of several partons with zero net color between
the target and the projectile (see the Sidebar on page~\pageref{sdbar:diffraction}). 
In high-energy scattering, which is
dominated by gluons, this color neutral exchange (at the lowest order)
consists of at least two exchanged gluons. We see that compared to the
total DIS cross-section, which can be mediated by a single gluon or
quark exchange, the diffractive cross-section includes more
interactions, and, therefore, is likely to be more sensitive to
saturation phenomena, which, at least in the MV model, are dominated
by multiple re-scatterings. In fact, some diffractive processes are
related to the {\sl square} of the gluon distribution $x G$. We
conclude that the diffractive cross-section is likely to be a very
sensitive test of saturation physics.

Of particular interest is the process of elastic vector meson ($V$)
production, $e + A \to e + V + A$. The cross-section $d\sigma/dt$ for
such processes at lower $Q^2$ is sensitive to the effects of parton
saturation \cite{Munier:2001nr}, as we will explicitly demonstrate
below. For a vector meson with a sufficiently spread-out wave-function
(a large meson, like $\phi$ or $\rho$), varying $Q^2$ would allow one
to detect the onset of the saturation phenomenon \cite{Munier:2001nr}.

Diffraction can serve as a trigger of the onset of the black disk
limit of \eq{bd}. In that regime, the total diffractive cross-section
$\sigma_\mathrm{diff}$ (including all the events with rapidity gaps),
would constitute $50 \%$ of the total cross-section,
\begin{align}
  \label{eq:diff_fract}
  \frac{\sigma_\mathrm{diff}}{\sigma_{\mathrm{tot}}} = \frac{1}{2}. 
\end{align}
This may sound counter-intuitive: indeed, the naive expectation in QCD
is that events with gaps in rapidity are exponentially suppressed. It
was therefore surprising to see that a large fraction (approximately
$15 \%$) of all events reported by HERA experiments are rapidity gap
events \cite{Abramowicz:1998ii}. This corresponds to a situation where
the projectile electron slams into the proton at rest with an energy
50,000 times the proton rest energy and in about 1 in 7 such
scatterings, nothing happens to the proton. In the black disk regime
this ratio should increase to 1 in 2 events.
\end{multicols}

\subsection{Key Measurements}
\label{keyM}

\begin{figure*}[thb]
    \begin{center}
        \includegraphics[width=0.6\textwidth]{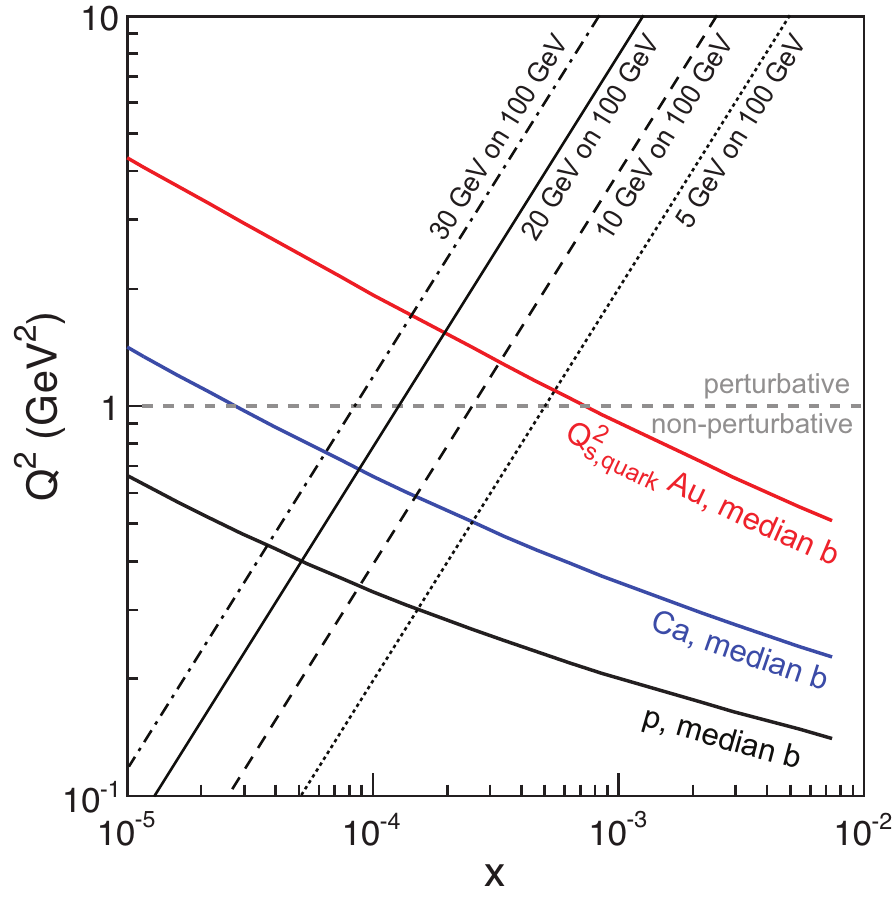}
    \end{center}
    \vskip -5mm
    \caption{The kinematic reach in $x$ and $Q^2$ of the EIC for different
      electron beam energies, given by the regions to the right of the
      diagonal black lines, compared with predictions of the
      saturation scale, $Q_s$, in $p$, Ca, and Au from Model-I
      (see Sec.~\ref{sec_map} and note that $x < 0.01$ in the figure). 
      }
    \label{fig:QsEICAcceptance}
\end{figure*}

\begin{multicols}{2}
The main goal of the $e$+A program at an EIC is to unveil the
collective behavior of densely packed gluons under conditions where
their self-interactions dominate, a regime where non-linear QCD
supersedes ``conventional'' linear QCD.  The plain fact that there is
no data from this realm of the nuclear wave-function available is a
already a compelling enough reason to build an EIC. It is truly {\em
  terra incognita}.  However, our goal is not only to observe the
onset of saturation, but to explore its properties and reveal its
dynamical behavior.  As explained above, the saturation scale squared
for nuclei includes an ``oomph'' factor of $A^{1/3}$ making it larger
than in the proton (cf. \eq{oomph}); \fig{fig:QsEICAcceptance}
demonstrates that. While at an EIC, a direct study of the saturation
region in the proton is impossible (while remaining in the
perturbative QCD region where the coupling $\as$ is small, i.e., above
the horizontal dashed line in the figure), this $A^{1/3}$ enhancement
may allow us to study the saturation region of large nuclei, such as
gold (Au). 
In \fig{fig:QsEICAcceptance}, the borders of the kinematic reach of 
the EIC are indicated by the diagonal black lines corresponding to
different combinations of electron and hadron beam energies; the actual
kinematic reach regions are to the right of the border lines.


A wide range of measurements with an EIC can distinguish between
predictions in the CGC, or other novel frameworks, and those following
from the established DGLAP evolution equations.  However, these
comparisons have to be made with care. Non-linear models are valid only
at or below the saturation scale, \Qss, while perturbative QCD (pQCD)
based on the linear DGLAP evolution equation is strictly only
applicable at large $Q^2$. In the range $Q^2 < \Qss$, solely non-linear
theories such as the CGC can provide quantitative calculations. It is
only in a small window of approximately $1 \lesssim Q^2 \lesssim 4$
GeV$^2$ where a comparison between the two approaches can be made (see
Fig.~\ref{fig:QsEICAcceptance}). 
Due to the complexity of high energy nuclear physics, at the end, the
final insight will come from the thorough comparison of models
calculations with a multitude of measurements, each investigating
different aspects of the low-$x$ regime. We will learn from varying
the ion species, A, from light to heavy nuclei, studying the $Q^2$,
$x$, and $t$ dependence of the cross-section in inclusive,
semi-inclusive, and exclusive measurements in DIS and diffractive
events.

In what follows we discuss a small set of key measurements whose
ability to extract novel physics is beyond question.  They serve
primarily to exemplify the very rich physics program available at an
EIC.  These ``golden'' measurements are summarized in
Tab.~\ref{tab:keyMeasurements} with two EIC energy options.
These measurements are discussed in further detail in the remainder of
this section.  It should be stressed that the low-$x$ physics program
will only reach its full potential when the beam energies are large
enough to reach sufficiently deep into the saturation
regime. Ultimately this will only be possible 
at an EIC where $x \sim 10^{-4}$ can be reached at $Q^2$ values of
1--2 GeV$^2$ as indicated in Fig.~\ref{fig:QsEICAcceptance}. Only the
highest energies will give us enough of a lever arm in $Q^2$ to study
the crossing into the saturation region allowing us to, at the same
time, make the comparison with DGLAP-based pQCD and CGC predictions.
The statistical error bars depicted in the figures described in this
section are derived by assuming an integrated luminosity of
$\int{\mathcal{L}} dt = 10\ \mathrm{fb}^{-1}/\mathrm{A}$ for each
species and include experimental cuts (acceptance and momentum).
Systematical uncertainties were estimated in a few cases based on
experience from HERA.  Ultimately they will depend on the details of
detectors and machine and hence cannot be fully addressed at this
time.
\end{multicols}

\begin{table}[ht]
\centering

\noindent\makebox[\textwidth]{%
\footnotesize
\begin{tabular}{|c|c|c|c|c|}
\hline
Deliverables & Observables & What we learn & Low energy option & High energy option \\
\hline
\hline
Integrated gluon  &   $F_{2}$, $F_L$, and $F_2^{c\bar{c}}$  & Nuclear wave       & Gluons at                      &  Exploration \\
momentum          &  {}         & function;          & $10^{-3}\lesssim x \lesssim 1$ &  of the saturation \\
distributions $G_A (x, Q^2)$     &  {}         & saturation  & {}     &  regime \\
\hline
$k_T$-dependent       &   Di-hadron        &  Non-linear QCD             &  Onset of     &  Non-linear  \\
gluons $f (x, k_T)$;               &   correlations     &  evolution/universality;     &  saturation;  &  small-$x$ \\
gluon correlations    &   {}               &  saturation scale $Q_s$                         &  $Q_s$ measurement       &  evolution \\
\hline
Spatial gluon      & Diffractive dissociation    & Non-linear small-$x$    &   saturation    & Spatial \\
distributions $f (x, b_T)$;     &  $\sigma_\mathrm{diff}/\sigma_\mathrm{tot}$              & evolution;              &           vs. non-saturation     & gluon \\
gluon correlations & vector mesons $\&$ DVCS              & saturation dynamics;     &     models         & distribution;  \\
& $d\sigma/dt$, $d\sigma/dQ^2$ & black disk limit &  & $Q_s$ vs centrality \\
\hline

\end{tabular}}

\caption{\small Key measurements in $e$+A collisions at an EIC with two energy options, as shown in \fig{fig:lA-experiments}, addressing the physics of high gluon densities.}
\label{tab:keyMeasurements}
\end{table}

\begin{multicols}{2}
\subsubsection{Structure Functions}
\label{sec:strf}

As we mentioned above in Sec.~\ref{sec_nsf}, the differential
unpolarized cross-section for DIS is fully described by a set of basic
kinematic variables and two structure functions, $F_2(x,Q^2)$ and
$F_L(x,Q^2)$, that encapsulate the rich structure of valence quarks,
sea quarks and anti-quarks ($F_2$) and gluons ($F_L$). The structure
function $F_L$ is directly proportional to the gluon distribution
function, $F_L(x,Q^2) \propto \alpha_s \, x G(x,Q^2)$, at low $x$ and
not-very-small $Q^2$ \cite{Dokshitzer:1977sg,Nikolaev:1994ce}.  A
precise knowledge of $F_L$ is mandatory for the study of gluons and
their dynamics in nucleons and nuclei (see the Sidebar on
page~\pageref{sdbar:crosssection}).

As demonstrated in Sec.~\ref{sec_nsf} and shown in
\fig{fig:shadowing}, various models have different predictions for the
gluon distribution ratio $R_G (x, Q^2)$. The same is true for the
ratios $R_2 (x, Q^2)$ and $R_L (x, Q^2)$, along with the nuclear
structure functions $F_2^A (x, Q^2)$ and $F_L^A (x, Q^2)$. These
observables can be measured at 
the EIC as functions of $x$, $Q^2$, and $A$. (For the $A$-dependence
one will need to perform machine runs with different types of nuclei,
while to extract $F_L$ one needs to vary the center-of-mass energy.)
The multitude of theoretical predictions should be counter-balanced by
the multitude of possible data points for the four observables in the
3-dimensional $(x, Q^2, A)$ parameter space. It is possible that the
abundance of data obtained with sufficient statistics would allow one
to rule out many models, hopefully pinpointing the one that best
describes all the data to be obtained.

\begin{figure*}[thb]
    \begin{center}
        \includegraphics[width=\textwidth]{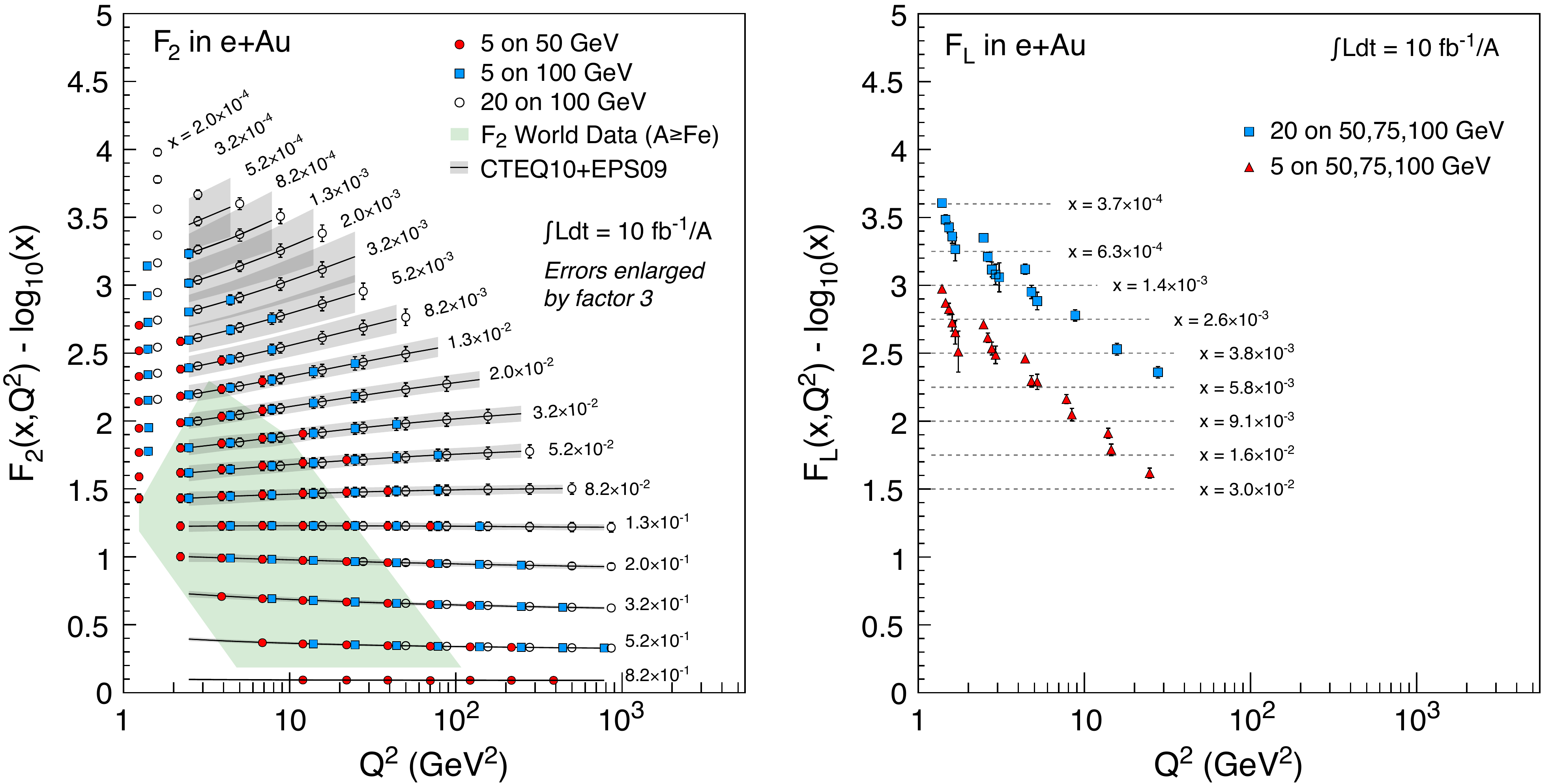}
    \end{center}
    \vskip -0.5cm
    \caption{The structure functions $F_2$ (left) and
      $F_L$ (right) as functions of $Q^2$ for various $x$-values in
      $e$+Au collisions at an EIC generated by using PYTHIA with EPS09
      nuclear PDFs \cite{Eskola:2009uj}. $F_2$ and $F_L$ are
      offset by $\log_{10}({x})$ for clarity.  Measurements
      and corresponding errors at different energies (indicated in the
      panels) are presented and illustrate the respective kinematic reach. Data
      points from different energies at the same $Q^2$ are slightly
      offset along the abscissa for visibility where necessary.
      Statistical errors for $F_2$ and $F_L$ are based on
      $10$~fb$^{-1}$/A integrated luminosity for the sum of all
      measurements at all indicated energies.  Both for $F_2$ and
      $F_L$ we assumed a 3$\%$ systematic uncertainty and added it to
      the statistical errors in quadrature; for $F_2$ the combined
      errors are scaled up by a factor of 3 to make them visible. 
      For $F_2$, we also depict the curves and respective uncertainty
      bands from the EPS09 parameterization of the nuclear parton
      distribution functions \cite{Eskola:2009uj,Martin:2009iq}. The
      green shaded area indicates the $(Q^2, x)$ range of existing
      measurements for nuclei larger than iron, demonstrating the
      kinematic reach of an EIC.}
    \label{fig:F2FL}
\end{figure*}
 
In order to verify the EIC's capability to measure the structure functions
$F_2$ and $F_L$, we conducted simulation of inclusive events in $e$+Au
collisions using PYTHIA with EPS09 nuclear parton distribution functions
\cite{Eskola:2009uj}. 
 Figure~\ref{fig:F2FL} shows the resulting structure
  functions $F_2$ (left) and $F_L$ (right) as functions of $Q^2$ with
  their respective $x$ values. 
  The curves and error bands for $F_2$ derived from the
  EPS09 distribution function in NLO
  ~\cite{Eskola:2009uj, Martin:2009iq} are overlaid. The comparison of the 
  current EPS09 uncertainty bands with the errors of the respective data
  points demonstrates that for $x \lesssim 0.01$, the EIC will have a
  substantial impact on reducing the uncertainty of leading-twist
  shadowing models.  
 
  Any measurement of $F_L$ requires data at a wide range of
  $\sqrt{s}$.  In our $F_L$ studies presented on the right in
  Fig.~\ref{fig:F2FL}, we varied the beam energies over the range
  indicated in the panel. 
  The final values for $F_L$ were extracted using the standard
  Rosenbluth method. This method is extremely sensitive to the quality
  of the absolute normalization achieved at the various
  energies. Since systematic uncertainties depend on the quality of
  the final detectors and on the accuracy of luminosity measurements,
  their ultimate magnitude is hard to estimate. In our studies we
  assumed systemic normalization uncertainties of 3$\%$ per energy,
  the same as the values that were achieved at HERA. The presented errors
  include both systematical and statistical
  contributions. 

  A comparison of $F_2$ and $F_L$ clearly shows the intricacy of the
  $F_L$ studies. While $F_L$ is of enormous importance for the study
  of gluons, its measurement is very difficult. In addition, the
  kinematic reach of $F_L$ measurements is much narrower than that of
  $F_2$.

\begin{figure*}[tbh]
    \begin{center}
        \includegraphics[width=\textwidth]{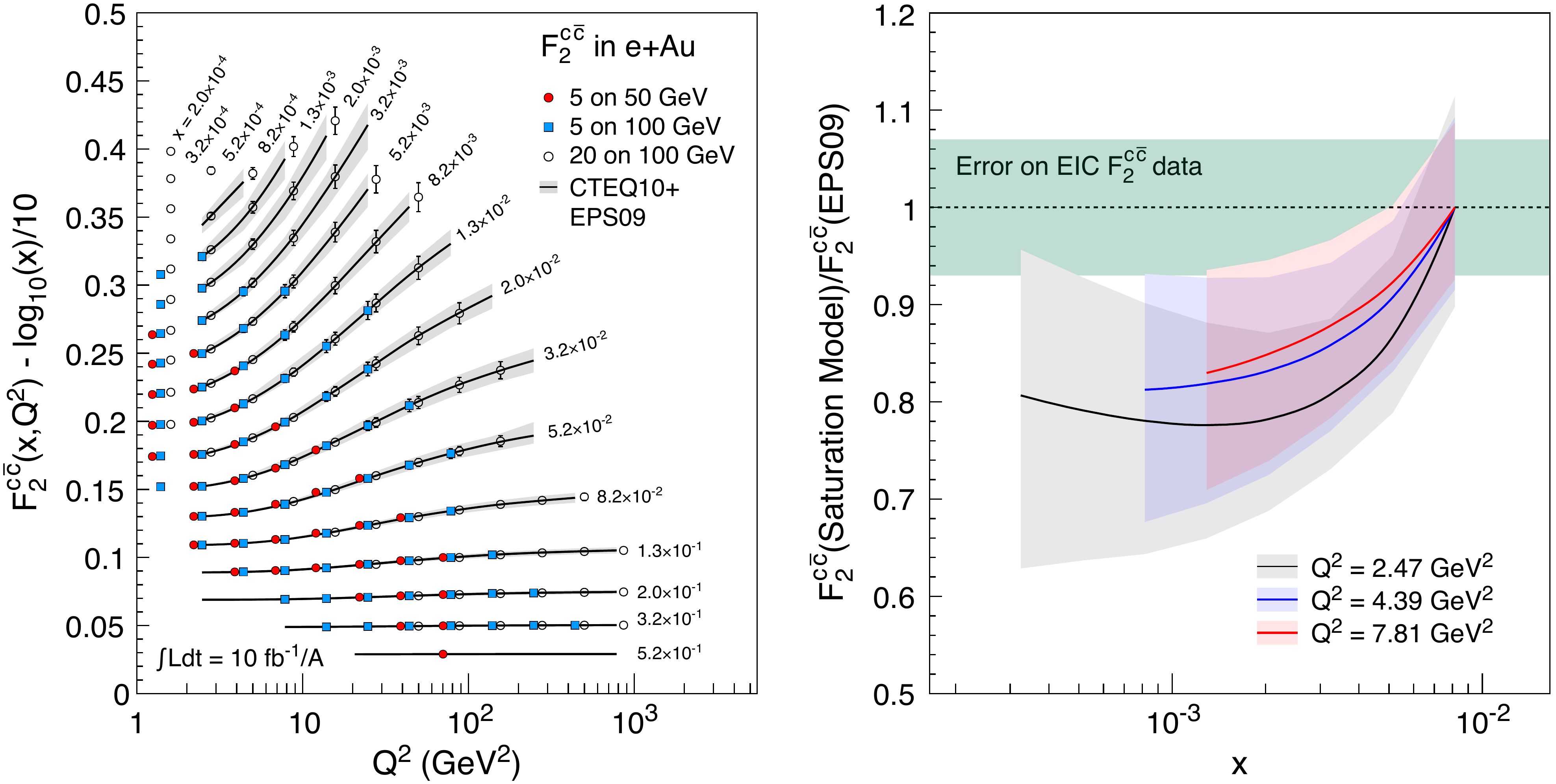}
    \end{center}
   \vskip -0.5cm
   \caption{Left panel: The charm structure function
     $F_2^{c\bar{c}}$ versus $Q^2$ for various $x$-values in $e$+Au
     collisions at an EIC generated by using PYTHIA with EPS09
     nuclear PDFs \cite{Eskola:2009uj}.  $F_2^{c\bar{c}}$
     values are offset by $\log_{10} ({x})/10$ for clarity.  
     Measurements and corresponding errors for three different
     energies are presented and illustrate the respective kinematic reach. Data
     points from different energies at the same $Q^2$ and $x$ are
     slightly offset along the abscissa for better visibility.
     Statistical errors are based on $10$~fb$^{-1}$/A integrated
     luminosity for the sum of all three energies. The depicted errors
     are derived from the statistical errors and a 7\% systematic
     uncertainty added in quadrature.  Also shown are curves and
     respective uncertainty bands from the EPS09 parameterization of
     the nuclear parton distribution functions
     \cite{Eskola:2009uj,Martin:2009iq}.  Right panel: Ratio of
     $F_2^{c\bar{c}}$ predictions from a saturation model (rcBK)
     \cite{Albacete:2009fh} and EPS09 for three different $Q^2$
     values. The uncertainty band for each $Q^2$ value reflects the
     combined uncertainties in both models. The green band depicts the
     approximate uncertainties of EIC measurements of $F_2^{c\bar{c}}$
     thus indicating in what kinematic range an EIC will be able to
     distinguish between traditional leading-twist shadowing and
     saturation models.  }
    \label{fig:F2charm}
\end{figure*}

An alternative and complementary method for studying the gluon density
is via the charm structure function $F_2^{c\bar{c}}$.  The left plot
in Fig.~\ref{fig:F2charm} shows $F_2^{c\bar{c}}$ versus $Q^2$ for
various $x$-values in $e$+Au collisions at an EIC. 
Also shown are curves and respective uncertainty bands resulting from the 
EPS09 parameterization of  nuclear parton distribution functions
 \cite{Eskola:2009uj,Martin:2009iq}.  
While an EIC will certainly
constrain these leading-twist shadowing models further for $x \lesssim
5\times10^{-3}$, it appears that the improvement would be rather
modest. Here, one has to keep in mind that through the charm structure
function, one probes the PDFs at a somewhat higher value of Bjorken
$x$, namely at $x^\prime \approx x (1 + (4 m_c^2)/Q^2)$, where the
PDFs are better constrained by the existing data. The fact that
$F_2^{c\bar{c}}$ is so surprisingly well-predicted in DGLAP-based
approaches compared to $F_L$ can be used to test for differences
between the traditional leading-twist shadowing models (such as EPS09)
and  models that involve non-linear dynamics.  The right plot in
Fig.~\ref{fig:F2charm} compares one such model, the rcBK model, to
EPS09 by depicting the ratio of these models predictions for
$F_2^{c\bar{c}}$ for three different $Q^2$ values as functions of $x$
where we expect these non-linear dynamics to be important.
rcBK is a saturation model in the CGC framework based on the BK
non-linear evolution with the running-coupling corrections
\cite{Balitsky:2006wa,Kovchegov:2006vj,Gardi:2006rp,Albacete:2009fh}:
we referred to it as Model-II in Sec.~\ref{sec_map}. As follows from the right plot in
Fig.~\ref{fig:F2charm}, it predicts a markedly different $x$-dependence
than NLO pQCD calculations based on EPS09: importantly, the difference
between the models (together with the combined uncertainty of both
models) exceeds the expected uncertainty of EIC measurements (the green
band).
It appears that with a sufficient experimental effort the EIC would be
able to distinguish between the saturation and leading-twist shadowing
predictions for $F_2^{c\bar{c}}$, providing us with another
measurement capable of identifying saturation dynamics.
\begin{figure*}[tbh]
    \begin{center}
        \includegraphics[width=\textwidth]{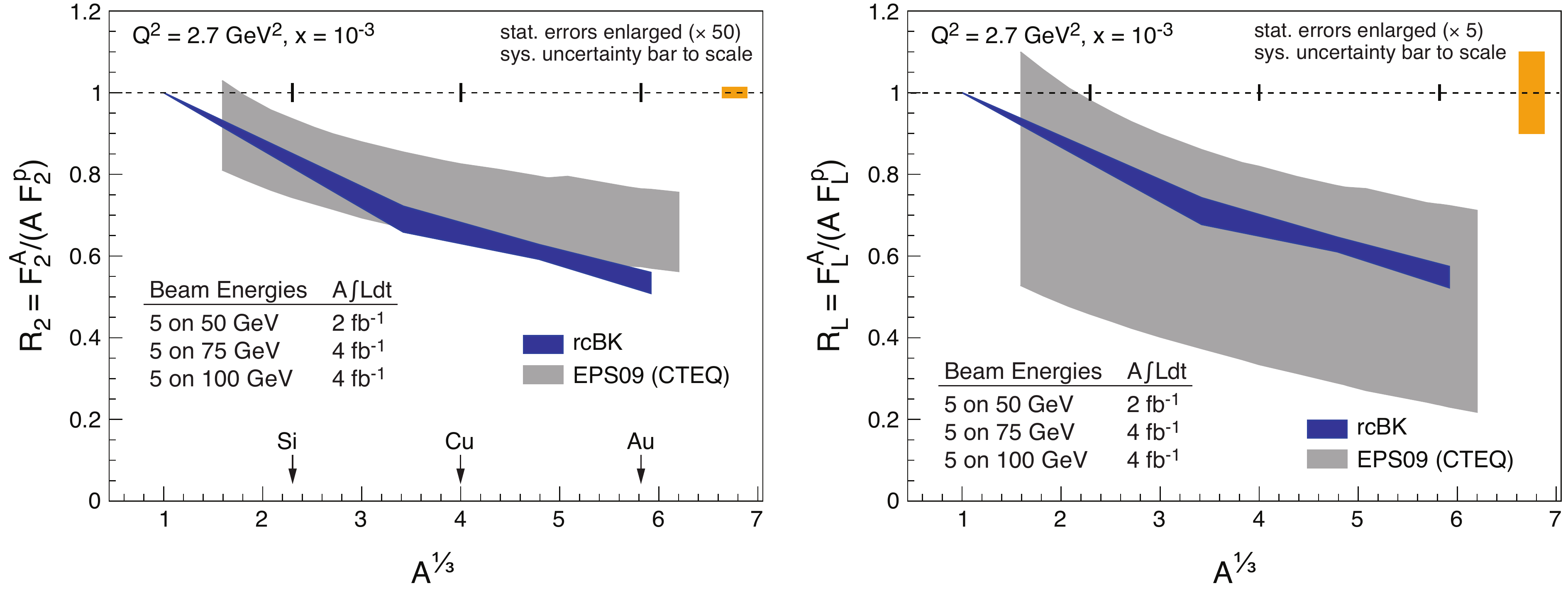}
    \end{center}
    \vskip -6mm
    \caption{Left: The ratio $R_2$ of the $F_2$ structure function in
      a nucleus over that of the proton scaled by mass number A as a
      function of A$^{1/3}$. The predictions from a CGC based
      calculation (rcBK) \cite{Albacete:2009fh} and from a linear
      evolution using the latest nuclear PDFs (EPS09) and CTEQ6 for
      the proton are shown \cite{Martin:2009iq,Eskola:2009uj}.  Right:
      The same for the longitudinal structure function $F_L$ (see text
      for details).}
    \label{fig:F2FLcombo}
\end{figure*}

Clearly, the EIC will reach into unexplored regions with unprecedented
precision and will be able to distinguish between traditional and
non-linear QCD models.  These measurements will have a profound impact
on our knowledge of nuclear structure functions and the underlying
evolution scheme, likely allowing to rule out many theoretical models
and to establish the correct underlying physics.  For a better
discrimination between models, especially involving non-linear
dynamics, several observables sensitive to the gluon distribution will
be essential: ({\it i}) scaling violation of $F_2$, ({\it ii}) the
direct measurement of $F_L$, and ({\it iii}) $F_2^{c\bar{c}}$.

Note that all three observables can be measured already at moderate
luminosities with good statistical precision. The final experimental
errors for the structure functions to be measured at EIC will be
dominated by systematic uncertainties.  High luminosities are not
required for the measurement of structure functions, while precise
knowledge of the actual luminosity is paramount.

In the context of model comparisons, it is important to note that
DGLAP-based models can not predict the $A$-dependence of PDFs and
structure functions without making additional data-driven assumptions:
this is the origin of the broad error bars of the EPS09 model
in~\fig{fig:shadowing}. However, this broad error band may also be
indicative of the ability of such models to indiscriminately describe
a broad range of $F_2$ and $F_L$ data: in such cases, further
experimental tests of DGLAP-based approaches can be carried out using
other observables described in the sections below.

To further illustrate this point, we show in Fig.~\ref{fig:F2FLcombo}
two theoretical predictions for the ratio $R_2$ ($R_L$), i.e., the
ratio of the $F_2$ ($F_L$) structure function in a nucleus over that
of the proton scaled by mass number $A$. The calculations are shown as
a function of A$^{1/3}$ at $Q^2$ = 2.7 GeV$^2$ and $x = 10^{-3}$.  In
the absence of any nuclear effects, both ratios $R_2$ and $R_L$ should
be unity.  Due to the lack of precise $e$+A data, the models are not
strongly constrained and we use error bands to indicate the range of
the referring predictions.  In Fig.~\ref{fig:F2FLcombo} we depict two
calculations for $R_2$ (left) and $R_L$ (right). The calculation shown
in blue is based on the CGC framework (rcBK) \cite{Albacete:2009fh}
which was already discussed earlier. It features an approximate
A$^{1/3}$ scaling of the saturation scale squared (see
Sec.~\ref{class_sec}), which allows us to make reasonably precise
predictions for $R_2$ and $R_L$; the second calculation (gray band)
uses the linear NLO DGLAP evolution in pQCD resulting in the nuclear
parton distribution set EPS09 \cite{Martin:2009iq,Eskola:2009uj}: it
exhibits a broader error band, similar to the case of $R_G$ in
\fig{fig:shadowing}. Even in linear DGLAP evolution, non-linear
effects may be absorbed into the non-perturbative initial conditions
for the nuclear PDFs, where the A-dependence is obtained through a fit
to available data, resulting in the ability of DGLAP-based approaches
to indiscriminately describe a broad range of nuclear data. This leads
to the wide error bands of EPS09, especially for $F_L$, clearly
demonstrating the lack of existing nuclear structure function data.
Due to these large theoretical error bars, the measurements of $R_2$
and $R_L$ as functions of A$^{1/3}$, while significantly extending our
knowledge of nuclear structure functions, may not allow one to
directly distinguish between a non-linear (saturation) and linear
(DGLAP) evolution approaches at an EIC with low collision energies.


Shown along the line at unity by vertical notches in
\fig{fig:F2FLcombo} are the statistical errors that were obtained from
the Rosenbluth separation technique using the range of energies
indicated in the figure.
The statistical error bars were generated from a total of 10~fb$^{-1}/$A
of Monte Carlo data, spread over three beam energies (see plot legend
for details).  The statistical error bars are scaled up by a factor of
50 for $R_2$ and a factor of 5 for $R_L$;  as the statistical
errors are clearly small, the experimental errors will be dominated by
the systematic uncertainties shown by the orange bars drawn to scale
in the two panels of \fig{fig:F2FLcombo}. 
This measurement, together with the ones described below, will
constrain models to such an extent that the ``true'' underlying
evolution scheme can be clearly identified.  It is also possible that
data from a lower-energy EIC would decrease the error band of
DGLAP-based predictions, allowing for the $R_2$ and $R_L$ measurement
at a higher energy EIC (smaller $x$) to discriminate between
saturation and DGLAP approaches.
However it is also possible that, on its own, the $R_2$ and $R_L$
measurements may turn out to be insufficient to uniquely differentiate
DGLAP-based models with nuclear ``shadowing'' in the initial
conditions from the saturation/CGC effects; in such a case, the
measurements presented below along with $F_2^{c\bar{c}}$ shown above
will be instrumental in making the distinction.


\subsubsection{Di-Hadron Correlations}
\label{sec:dihadrons}

One of the experimentally easiest and compelling measurement in $e$+A
is that of di-hadron azimuthal correlations in $e + \mathrm{A}
\rightarrow e^\prime + h_1 + h_2 + X$ processes.  These correlations are
not only sensitive to the transverse momentum dependence of the gluon
distribution, but also to that of gluon correlations for which first
principles CGC computations are only now becoming available. The
precise measurements of these di-hadron correlations at an EIC would
allow one to extract the spatial multi-gluon correlations and study
their non-linear evolution.
\begin{figure*}[tbh!]
    \begin{center}
\includegraphics[width=\textwidth]{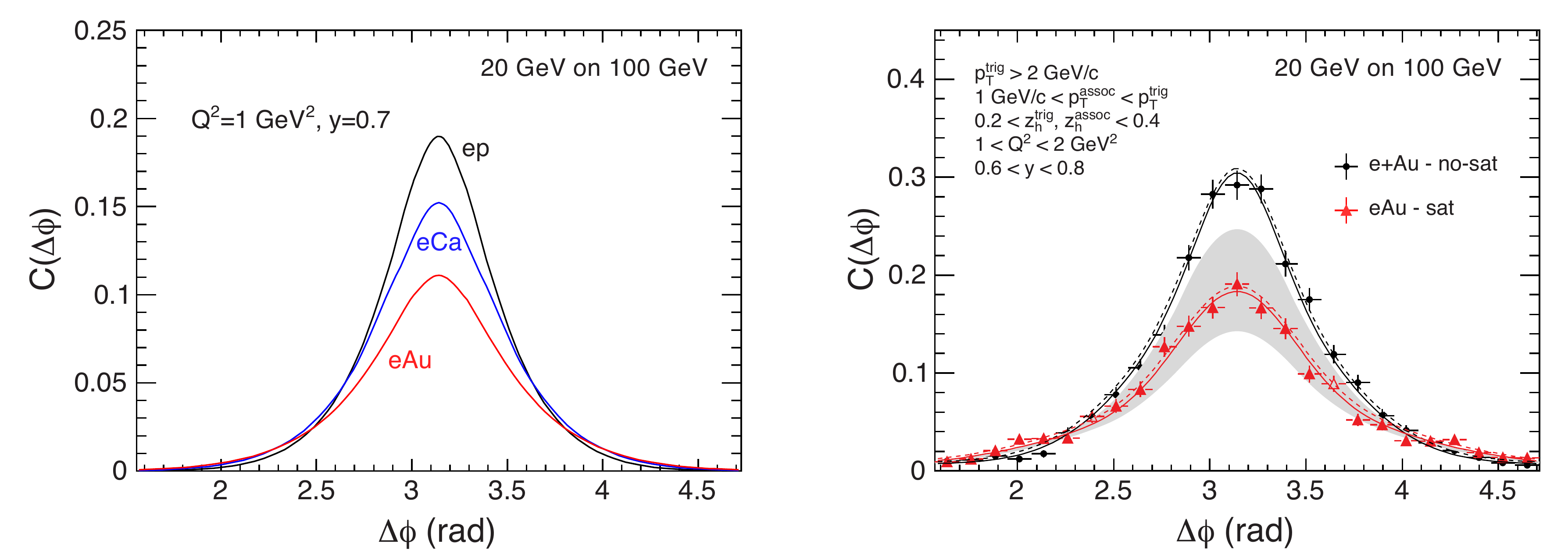}
    \end{center}
    \vskip -6mm
    \caption{Left: A saturation model prediction of the coincidence
      signal versus azimuthal angle difference $\Delta \varphi$
      between two hadrons in $e$+$p$, $e$+Ca, and $e$+A collisions
      \cite{Dominguez:2011wm,Dominguez:2010xd,Zheng:2014vka}. Right: A
      comparison of saturation model prediction for $e$+A collisions
      with calculations from conventional non-saturated model.
      Statistical error bars correspond to
      1~fb$^{-1}$/A integrated luminosity.}
    \label{fig:dihadronsdphi}
\end{figure*}
Saturation effects in this channel correspond to a progressive
disappearance of the back-to-back correlations of hadrons with
increasing atomic number A. These correlations are usually measured in
the plane transverse to the beam axis (the `transverse plane'), and
are plotted as a function of the azimuthal angle $\Delta \varphi$
between the momenta of the produced hadrons in that
plane. Back-to-back correlations are manifested by a peak at $\Delta
\varphi = \pi$ (see \fig{fig:dihadronsdphi}). In the conventional pQCD
picture, one expects from momentum conservation that the back-to-back
peak will persist as one goes from $e+p$ to $e+A$. In the saturation
framework, due to multiple re-scatterings and multiple gluon emissions,
the large transverse momentum of one hadron is balanced by the momenta
of several other hadrons (instead of just one back-to-back hadron),
effectively washing out the correlation at $\Delta \varphi = \pi$
\cite{Kharzeev:2004bw}. A comparison of the heights and widths of the
di-hadron azimuthal distributions in $e+A$ and $e+p$ collisions
respectively would clearly mark out experimentally such an effect.

An analogous phenomenon has already been observed for di-hadrons
produced at forward rapidity in comparing $d$+Au with $p$+$p$
collisions at RHIC (see Sec.~\ref{sec:pAConnections}). In that case,
di-hadron production is believed to proceed from valence quarks in the
deuteron (proton) scattering on small-x gluons in the target Au
nucleons (proton). Lacking direct experimental control over $x$, the
onset of the saturation regime is controlled by changing the
centrality of the collision, the di-hadron rapidity and the transverse
momenta of the produced particles. (Note that the gluon density and,
consequently, the saturation scale $Q_s$ depend on the impact
parameter and on rapidity/Bjorken-$x$.)  Experimentally, a striking
flattening of the $\Delta \varphi = \pi$ peak in $d$+Au collisions as
compared to $p$+$p$ collisions is observed in central collisions
\cite{Adare:2011sc,Braidot:2010ig}, but the peak re-appears in
peripheral collisions, in qualitative agreement with the CGC
predictions, since saturation effects are stronger in central
collisions.

There are several advantages to studying di-hadron correlations in
$e$+A collisions versus $d$+Au. Directly using a point-like electron
probe, as opposed to a quark bound in a proton or deuteron, is
extremely beneficial. It is experimentally much cleaner as there is no
``spectator'' background to subtract from the correlation function.
The access to the exact kinematics of the DIS process at an EIC would
allow for more accurate extraction of the physics than is possible at
RHIC or the LHC. Because there is such a clear correspondence between the
physics of this particular final state in $e$+A collisions to the same
in $p$+A collisions, this measurement is an excellent testing ground
for universality of multi-gluon correlations.

The left plot in Fig.~\ref{fig:dihadronsdphi} shows prediction in the
CGC framework for di-hadron $\Delta \varphi$ correlations in deep
inelastic $e$+$p$, $e$+Ca, and $e$+Au collisions 
\cite{Dominguez:2011wm,Dominguez:2010xd,Zheng:2014vka}. The
calculations are made for $Q^2 = 1$ GeV$^2$ and include a Sudakov form
factor to account for generated radiation through parton showers; only
$\pi^0$'s were used. The highest transverse momentum hadron in the
di-hadron correlation function is called the ``trigger'' hadron, while
the other hadron is referred to as the ``associate'' hadron.  The
``trigger'' hadrons have transverse momenta of $p_T^{trig} > 2 \,
\gevc$ and the ``associate'' hadrons were selected with $1 \, \gevc <
p_T^{assoc} < p_T^{trig}$. The CGC based calculations show a dramatic
``melting'' of the back-to-back correlation peak with increasing ion
mass.  The right plot in Fig.~\ref{fig:dihadronsdphi} compares the
prediction for $e$+A with a conventional non-saturated correlation
function. The latter was generated by a hybrid Monte Carlo generator,
consisting of PYTHIA-6 \cite{Sjostrand:2006za} for parton generation,
showering and fragmentation and DPMJet-III \cite{Roesler:2000he} for
the nuclear geometry, and a cold matter energy-loss afterburner
\cite{Salgado:2003gb}. The EPS09 \cite{Eskola:2009uj} nuclear parton
distributions were used to include leading twist shadowing.  The
resulting correlation function is shown in the right panel of
Fig.~\ref{fig:dihadronsdphi} by the black solid and dashed lines. The
solid black curve includes detector smearing effects, while the dashed
curve shows the result without taking into account any detector
response. The red curve in the right panel of
Fig.~\ref{fig:dihadronsdphi} represents the CGC predictions. While the
underlying model is identical to that shown in the left panel of
Fig.~\ref{fig:dihadronsdphi}, the simulations include all charged
hadrons as well as the quark channel contributions. The solid and
dashed red lines represent detector response effects switched on and
off, respectively. The shaded region reflects uncertainties in the CGC
predictions due to uncertainties in the knowledge of the saturation
scale, $Q_s$. This comparison nicely demonstrates the discrimination
power of these measurements. In fact, already with a fraction of the
statistics used here one will be able to exclude one of the scenarios
conclusively.

\begin{figure*}[tbh!]
    \begin{center}
\includegraphics[width=\textwidth]{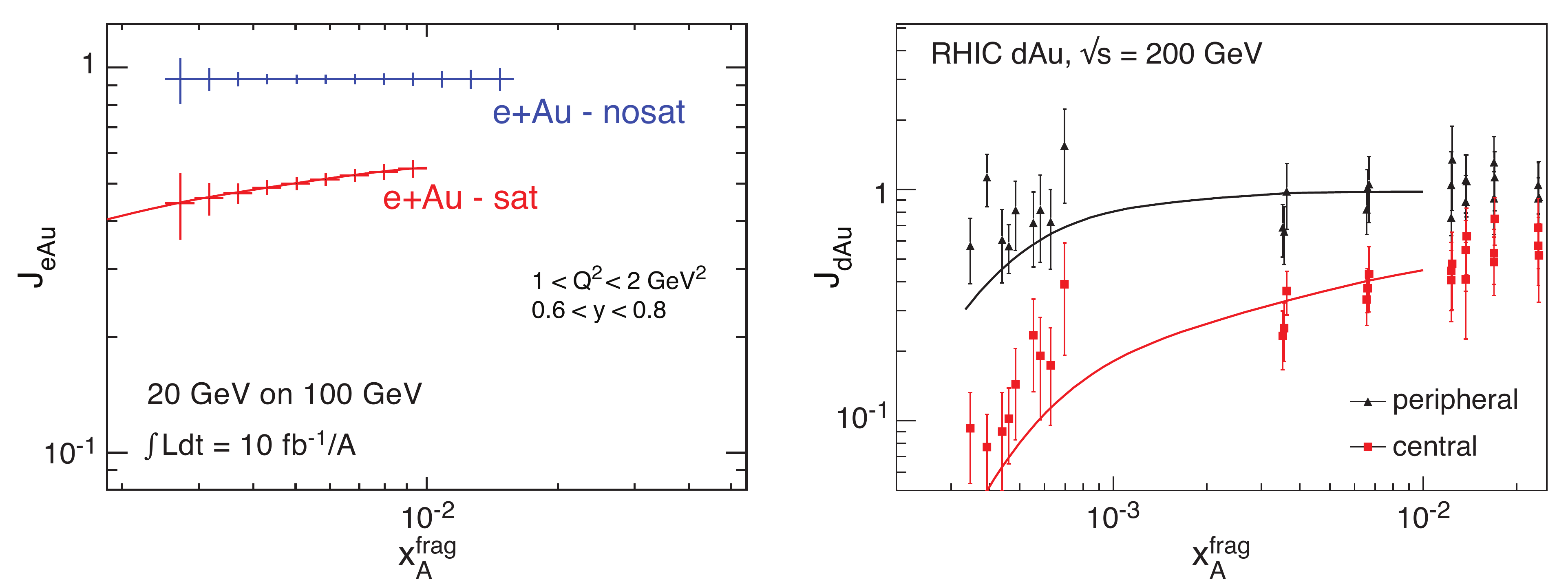}
    \end{center}
    \vskip -6mm
    \caption{Left: The relative yield of di-hadrons in $e$+Au compared
      to $e$+$p$ collisions, $J_{e\mathrm{Au}}$, plotted versus
      $x_\mathrm{A}^{frag}$, which is an approximation of the the
      longitudinal momentum fraction of the probed gluon
      $x_g$. Predictions for linear (nosat) and non-linear (sat) QCD
      models 
      are presented. The statistical
      error bars correspond to 10~fb$^{-1}$/A integrated
      luminosity. Right: The corresponding measurement in $\sqrt{s} =
      200$~GeV per nucleon $d$+Au collisions at RHIC
      \cite{Adare:2011sc}. 
      The curves in both panels
      depict calculations in the CGC framework
      \cite{Zheng:2014vka,Dominguez:2011wm,Dominguez:2010xd}.}
    \label{fig:dihadronJ}
\end{figure*}
The left panel of \fig{fig:dihadronJ} depicts the predicted
suppression through $J_{e\mathrm{Au}}$, the relative yield of
correlated back-to-back hadron pairs in $e$+Au collisions compared to
$e$+$p$ collisions scaled down by $A^{1/3}$ (the number of nucleons at a
fixed impact parameter)
\begin{align}
  \label{eq:JeA}
  J_{eA} = \frac{1}{A^{1/3}} \, \frac{\sigma^{pair}_{eA} /
    \sigma_{eA}}{\sigma^{pair}_{ep} / \sigma_{ep}}.
\end{align}
Here, $\sigma$ and $\sigma^{pair}$ are the total inelastic and the
di-hadron pair production cross-sections (or normalized yields). The
absence of collective nuclear effects in the pair production cross
section, $\sigma^{pair}_{eA}$, would correspond to $J_{eA}
=1$,\footnote{Without collective nuclear effects the hadron pairs are
  produced in independent electron--nucleon scatterings, such that
  $\sigma^{pair}_{eA} = A \, \sigma^{pair}_{ep}$. The cross-section
  for inelastic $e$+A collisions, $\sigma_{eA}$, is related to the
  probability for the incoming electron (or, more precisely, $\gamma^*
  \rightarrow q {\bar q}$) to get the first inelastic collision, which
  usually takes place on the nuclear surface: hence $\sigma_{eA} =
  A^{2/3} \, \sigma_{ep}$. Combining these results we get $J_{eA} =1$
  \cite{Kopeliovich:2003tz}.} while $J_{eA} < 1$ would signify
suppression of di-hadron correlations. In the left panel of
\fig{fig:dihadronJ}, $J_{e\mathrm{Au}}$ is plotted as a function of
$x_\mathrm{A}^{frag}$, which is an approximation of the the
longitudinal momentum fraction of the probed gluon $x_g$ derived from
the kinematics of the measured hadrons assuming they carry the full
parton energy. Compared to the measurement shown in the right panel of
Fig.~\ref{fig:dihadronsdphi} this study requires the additional
$e$+$p$ baseline measurement but has the advantage of several
experimental uncertainties canceling out. It is instructive to compare
this plot with the equivalent measurement in $d$+Au collisions at RHIC
\cite{Adare:2011sc} shown in the right panel of
Fig.~\ref{fig:dihadronJ}.  In $d$+Au collisions $J_{d\mathrm{Au}}$ is
defined by analogy to \eq{eq:JeA} with $A^{1/3}$ in the denominator
replaced by the number of the binary nucleon--nucleon collisions
$N_{coll}$ at a fixed impact parameter \cite{Adare:2011sc}.  
In both colliding systems, $e$+Au and $d$+Au, the exact momentum
fraction of the gluon $x_g$ cannot be directly measured experimentally
and has to be ultimately modeled. However, these calculations are much
better constrained in DIS where the key kinematic variables $x$ and
$Q^2$ are known precisely, allowing for tighter constraints on $x_g$.
The two curves in the right panel of
\fig{fig:dihadronJ} represent the CGC calculations from
\cite{Zheng:2014vka,Dominguez:2011wm,Dominguez:2010xd} but without the
Sudakov form-factor and appear to describe the data rather well. This
example nicely demonstrates on the one hand the correspondence between
the physics in $p$(d)+A and $e$+A collisions but on the other hand the
lack of precise control in $p$+A that is essential for precision
studies of saturation phenomena.


\subsubsection{Measurements of Diffractive Events}
\label{sec:diffmeasurements}

Diffractive interactions result when the electron probe in DIS
interacts with a proton or nucleus by exchanging several partons
with zero net color. This exchange, which in QCD may be visualized as a 
colorless combination of two or more gluons, is commonly referred to
as the ``Pomeron'' (see the Sidebar on page~\pageref{sdbar:diffraction}).

The HERA physics program of $e$+$p$ collisions surprisingly showed a
large fraction of diffractive events contributing about $15$\% to the
total DIS cross-section~\cite{Abramowicz:1998ii}.  One of the key
signatures of these events is an intact proton traveling at near-to
beam energies, together with a {\em gap} in rapidity before some
final-state particles are produced at mid-rapidity (i.e., at
$90^\circ$ angle to the beam axis).  While linear pQCD is able to
describe some aspects of diffraction, it fails to describe other major
features without introducing new types of structure functions, the
diffractive structure functions (see the Sidebar on page~\pageref{sdbar:diffraction}),
which describe the rapidity gap. A striking example is the fact that
the ratio of the diffractive to the total cross-section is constant
with energy, an observation not easily reconciled in a conventional
pQCD scenario without introducing the diffractive structure functions
\cite{Abramowicz:1998ii}.  As may therefore be anticipated, and as we
have argued above, the strongest hints for a manifestations of new,
non-linear effects in $e$+A collisions are likely to come from
diffractive measurements.

\begin{figure*}[thb]
    \begin{center}
  \includegraphics[width=\textwidth]{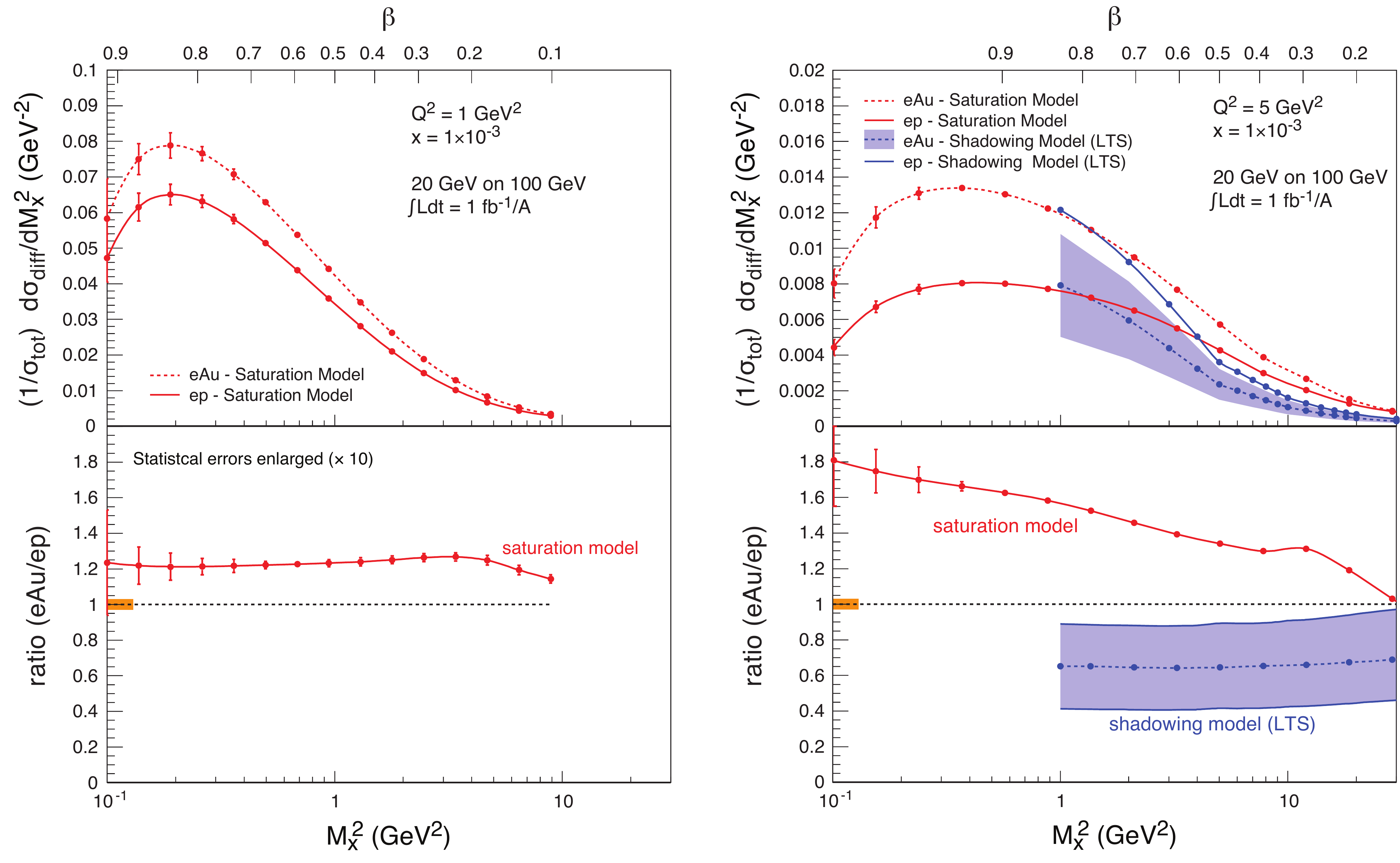}
    \end{center}
    \vskip -0.32in
    \caption{Top of each panel: the ratio of diffractive over total
      cross-sections, plotted as a function of the invariant mass of
      the produced particles, $M_X^2$. 
      The bottom of each panel contains the double ratio
      $[(d\sigma_{\mathrm{diff}}/dM_X^2)/\sigma_{\mathrm{tot}}]_{eA}/[(d\sigma_{\mathrm{diff}}/dM_X^2)/\sigma_{\mathrm{tot}}]_{ep}$
      plotted as a function of $M_X^2$ for the same kinematics as used
      at the top of each panel. The statistical error bars for the
      integrated luminosity of 1~fb$^{-1}/$A are too small to depict
      and are enlarged by a factor 10.  The non-monotonicity of the
      saturation curve in the lower panels is due to crossing the
      $c\bar{c}$ threshold; this threshold is not included in the LTS
      prediction.  }
    \label{fig:cyrilleDiffCombo}
    \vspace{-0.06in}
\end{figure*}

What makes the diffractive processes so interesting is that they are
most sensitive to the underlying gluon distribution, and that they are
the {\em only} known class of events that allows us to gain insight
into the spatial distribution of gluons in nuclei.  The reason for
this sensitivity is that the diffractive structure functions 
depend, in a wide kinematic range,
quadratically on the gluon momentum distribution and not linearly as
in DIS.  However, while the physics goals are golden, the technical
challenges are formidable but not insurmountable, and require careful
planning of the detector and interaction region. Diffractive events are
characterized by a rapidity gap, i.e. an angular region in the
direction of the scattered proton or nucleus without particle flow.
Detecting events with rapidity gaps requires a largely hermetic
detector.

As discussed earlier (see Sec.~\ref{diff_sec}) we distinguish two
kinds of diffractive events: coherent (nucleus stays intact) and
incoherent (nucleus excites and breaks up). Both contain a rich set of
information.  Coherent diffraction is sensitive to the space-time
distribution of the partons in the nucleus, while incoherent
diffraction (dominating at larger $t$ and thus small impact parameter
$b_T$) is most sensitive to high parton densities where saturation
effects are stronger.  In $e$+$p$ collisions, the scattered intact protons
can be detected in a forward spectrometer placed many meters down the
beam line.  This is not possible for nuclei which, due to their large
mass, stay too close to the ion beam. However, studies showed that the
nuclear breakup in incoherent diffraction can be detected with close
to 100\% efficiency by measuring the emitted neutrons in a zero degree
calorimeter placed after the first dipole magnet that bends the hadron
beam. This tagging scheme could be further improved by using a forward
spectrometer to detect charged nuclear fragments. A rapidity gap and
the absence of any break-up fragments was found sufficient to identify
coherent events with very high efficiency.

In the following, we present several measurements focusing on the
discrimination power between non-linear saturation models and a 
prediction from conventional linear QCD DGLAP evolution. Saturation
models incorporate the effects of linear small-$x$ evolution for $Q >
Q_s$ and saturation non-linear evolution effects for $Q < Q_s$.


\vspace{4mm}
\noindent
{\sl Ratio of diffractive and total cross-sections.}\hfill\\
Fig.~\ref{fig:cyrilleDiffCombo} depicts predictions for one of the
simplest inclusive measurements that can be performed with diffractive
events: the measurement of the ratio of the coherent diffractive
cross-section over the total cross-section in $e$+$p$ and $e$+A
collisions is shown at the top of each panel.  It is plotted here as a
function of the diffractive mass of the produced final state
particles, $M_X^2$ (see the Sidebar on
page~\pageref{sdbar:diffraction}), for $x = 10^{-3}$ and $Q^2 = 1$ and
$5$~GeV$^2$. For fixed $Q^2$ and $x$, $M_X^2$ can also be expressed in
terms of the fraction of the momentum of the pomeron that is carried
by the struck quark within the proton or nucleus, $\beta$, shown along
the alternative abscissa on the top of each plot where $\beta \approx
\frac{Q^2}{Q^2 + M_X^2}$, 
corresponding to a rapidity gap $\approx\ln(\beta/x)$. 
The red curves represent the predictions of
the saturation model
\cite{Kowalski:2008sa,Kowalski:2007rw,Toll:2012mb,Toll:2013gda} based
on Model-I of Sec.~\ref{sec_map} combined with the theoretical
developments of
\cite{GolecBiernat:1999qd,Buchmuller:1996xw,Kovchegov:1999kx}, while
the blue curves and bands in the right panel represent the
leading-twist shadowing (LTS) model
\cite{Frankfurt:2003gx,Frankfurt:2011cs}. The bottom part of each
panel depicts the double ratio
$[(d\sigma_{\mathrm{diff}}/dM_X^2)/\sigma_{\mathrm{tot}}]_{eA}/
[(d\sigma_{\mathrm{diff}}/dM_X^2)/\sigma_{\mathrm{tot}}]_{ep}$,
illustrating the fact that the fraction of diffractive over total
cross section is expected to be higher in $e$+A than in $e$+$p$ in the
saturation framework.  The curves in \fig{fig:cyrilleDiffCombo} are
plotted for the range of $x$ and $Q^2$ values which will be accessible
already at low to moderate EIC energies.
The $e$+$p$ curves in both approaches are in a
reasonable agreement with the available HERA data
\cite{Boer:2011fh,Kowalski:2008sa}.  The statistical error bars, shown
in the bottom parts of the panels in \fig{fig:cyrilleDiffCombo} are
very small, and had to be scaled up by a factor of 10 to become
visible.  We conclude that the errors of the actual measurement would
be dominated by the systematic uncertainties dependent on the quality
of the detector and on the luminosity measurements. The size of the
error bars shows that the two scenarios can be clearly distinguished
over a wide $x$ and $Q^2$ range, allowing for a clear 
early measurement aimed at finding evidence of parton saturation.

Note that in the saturation predictions plotted in
Fig.~\ref{fig:cyrilleDiffCombo}, the nuclear effects, responsible for the
difference between the $e$+Au and $e$+$p$ curves, are stronger at 
large $Q^2$: the effect of saturation is to weaken the $A$-dependence in the
$\sigma_\mathrm{diff}/\sigma_\mathrm{tot}$ ratio at low $Q^2$. Also,
in agreement with the expectation that diffraction would be a large
fraction of the total cross-section with the onset of the black disk
limit (see \eq{eq:diff_fract}), the ratio
$(d\sigma_{\mathrm{diff}}/dM_X^2)/\sigma_{\mathrm{tot}}$ plotted in
\fig{fig:cyrilleDiffCombo} both for $e$+$p$ and $e$+Au grows with
decreasing $Q^2$, getting larger as one enters the saturation
region. 

 \begin{figurehere}
    \begin{center}
        \includegraphics[width=\columnwidth]{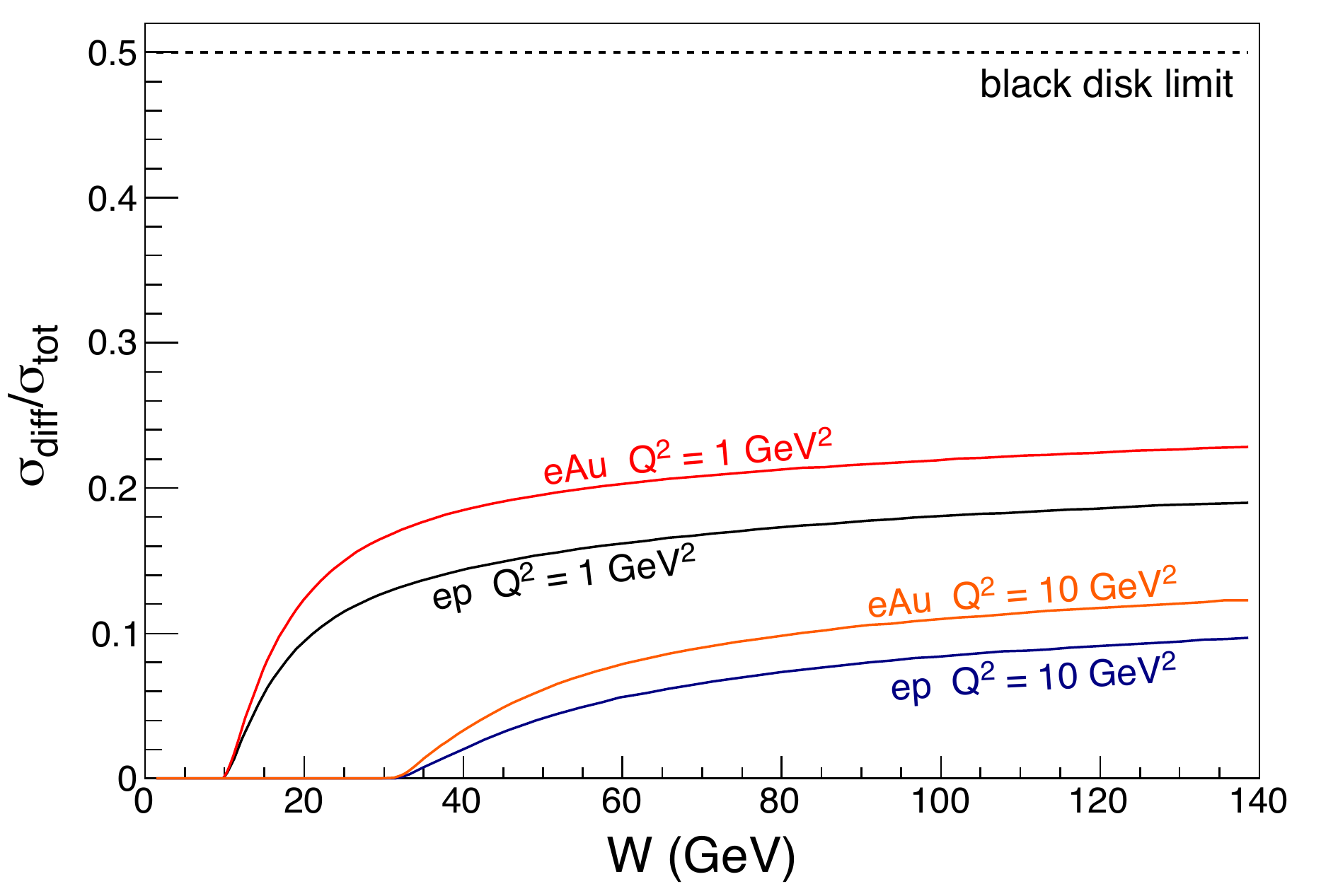}
    \end{center}
    \vskip -5mm
    \caption{The ratio of the diffractive to total cross sections as a
      function of the center-of-mass energy of the virtual
      photon--proton (nucleus) system $W$.}
          \label{fig:W_singleRatios}
  \end{figurehere}
 \vspace{5mm}


\begin{figure*}[tbh]
    \begin{center}
        \includegraphics[width=\textwidth]{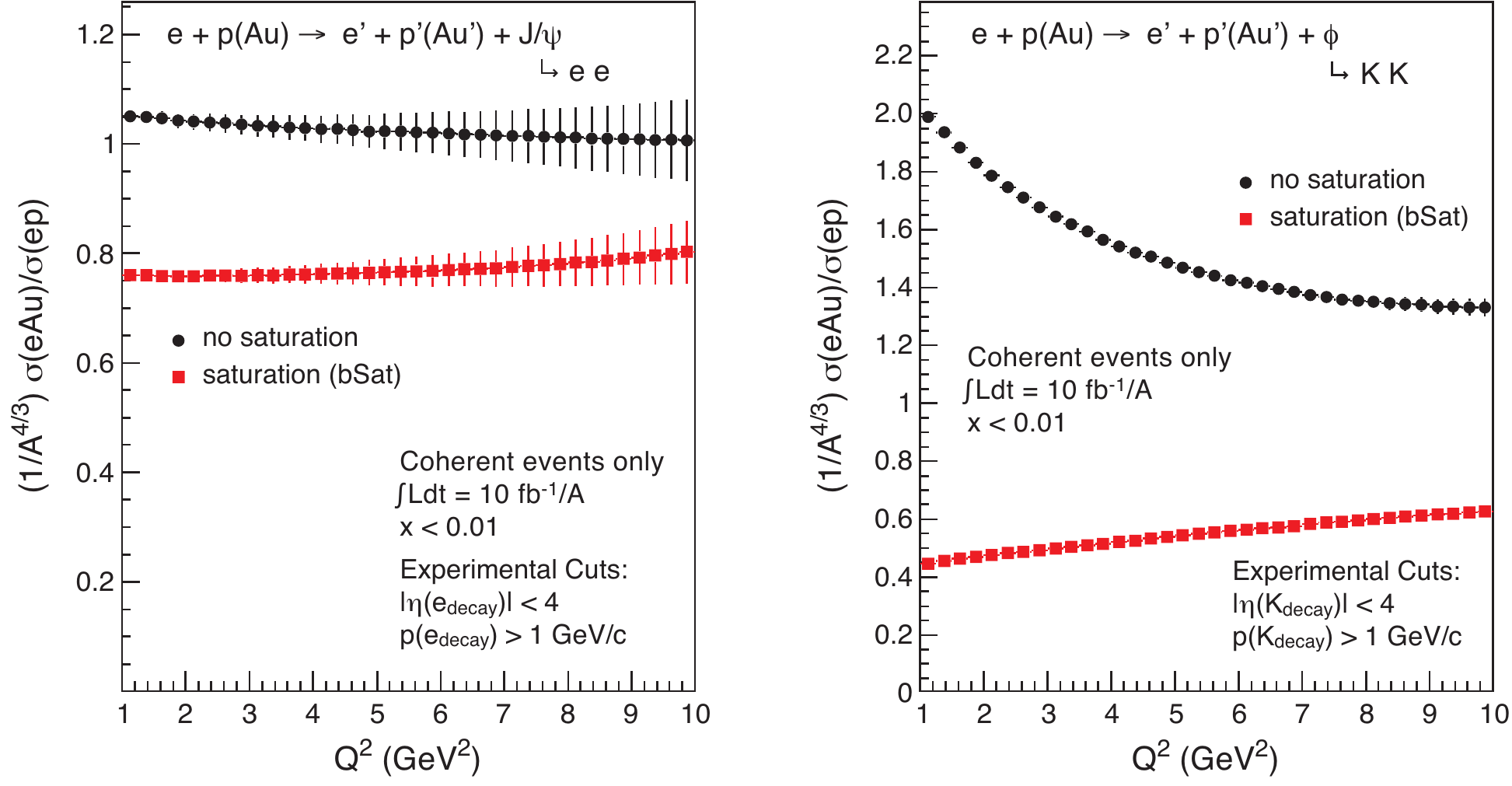}
    \end{center}
    \vskip -0.3in
    \caption{Ratios of the cross-sections for exclusive $J/\psi$ (left
      panel) and $\phi$ (right panel) meson production in coherent
      diffractive $e$+A and $e$+$p$ collisions as a function of
      $Q^2$ at an EIC with 20~GeV on 100~GeV beam energies. 
      Prediction for saturation and non-saturation
      models are presented. The ratios are scaled by $1/\mathrm{A}^{4/3}$.}
          \label{fig:diffQ2vm}
\end{figure*}

The ratio of the diffractive to total cross section,
$\sigma_{diff}/\sigma_{tot}$, evaluated in a CGC model
\cite{Toll:2012mb,Toll:2013gda}, is plotted in
Figure~\ref{fig:W_singleRatios} as a function of the center of mass
energy of the virtual photon--proton (nucleus) system $W$ (see the
Sidebar on page~\pageref{sdbar:DIS}) for $e$+$p$ and $e+$A scattering
with $Q^2 = 1$ and $10$~GeV$^2$. Again the diffractive to total cross
section ratio is higher in $e+$A than in $e+p$. Intriguingly, the
ratio becomes almost independent of energy $W$ for high enough $W$:
such behavior was already observed in $e+p$ scattering at HERA
\cite{Breitweg:1998gc}. (The ratio in Figure~\ref{fig:W_singleRatios}
is always much lower than its black-disk value of $1/2$ due to the
fact that even at very high energies saturation is not yet reached at
the edges of the proton or nucleus.) This energy-independence has a
particularly simple explanation in the saturation framework as being
due to the energy-dependent infrared cutoff $Q_s$
\cite{Kovchegov:1999kx}, suggesting that saturation effects may
possibly have been observed at HERA: it would be important to make
sure that this energy-independence of the diffractive to total cross
section ratio remains to be the case at EIC.

\vskip 4mm
\noindent
{\sl Diffractive vector meson production.} \hfill\\
The production of vector mesons
in diffractive processes, $e + \mathrm{A} \rightarrow e^\prime + \mathrm{A}^\prime +
V$ where $V = J/\psi$, $\phi$, $\rho$, or $\gamma$, is a unique
process, for it allows the measurement of the momentum transfer, $t$,
at the hadronic vertex even in $e$+A collisions where the 4-momentum of
the outgoing nuclei cannot be measured. Since only one new final state
particle is generated, the process is experimentally clean and can be
unambiguously identified by the presence of a rapidity gap. The study
of various vector mesons in the final state allows a systematic
exploration of the saturation regime \cite{Munier:2001nr}.  The $J/\psi$ is the
vector meson least sensitive to saturation effects due to the
small size of its wave-function. Larger mesons such as $\phi$ or
$\rho$ are considerably more sensitive to saturation effects
\cite{Kowalski:2006hc}.

The two panels in Fig.~\ref{fig:diffQ2vm} show the ratios
$[d\sigma(e\mathrm{Au})/dQ^2]/[d\sigma(ep)/dQ^2]$ (scaled down by
A$^{4/3}$) of the cross-sections $\sigma(e+\mathrm{Au})$ and
$\sigma(e+p)$ for exclusive $J/\psi$ (left panel) and $\phi$ (right
panel) production in coherent diffractive events for $e$+Au and $e$+$p$
collisions respectively. The ratios are plotted as functions of
$Q^2$ for saturation and non-saturation models. The parameters of both
models were tuned to describe the $e$+$p$ HERA data
\cite{ Kowalski:2003hm, Kowalski:2006hc}.  
All curves were generated with the Sartre event generator
\cite{Aaron:2012zz}, an $e$+A event generator specialized for
diffractive exclusive vector meson production based on the bSat
\cite{Kowalski:2006hc} dipole model.
We limit the calculation to $1 < Q^2 < 10$ GeV$^2$ and $x < 0.01$ to
stay within the validity range of saturation and non-saturation
models. The produced events were passed through an experimental filter
and scaled to reflect an integrated luminosity of 10 fb$^{-1}$/A. The
basic experimental cuts are listed in the legends of the panels in
Fig.~\ref{fig:diffQ2vm}. 
As expected, the difference between the saturation and non-saturation
curves is small for the smaller-sized $J/\psi$ ($< 20\%$), which is
less sensitive to saturation effects, but is substantial for the
larger $\phi$, which is more sensitive to the saturation region. In
both cases, the difference is larger than the statistical errors.  In
fact, the small errors for diffractive $\phi$ production indicate that
this measurement can already provide substantial insight into the
saturation mechanism after a few weeks of EIC running. 
Although this measurement could be already feasible at an EIC with low 
collision energies, the saturation effects would be less pronounced
due to the larger values of $x$.
For large $Q^2$, the two ratios
asymptotically approach unity.
\begin{figure*}[thb]
    \begin{center}
        \includegraphics[width=\textwidth]{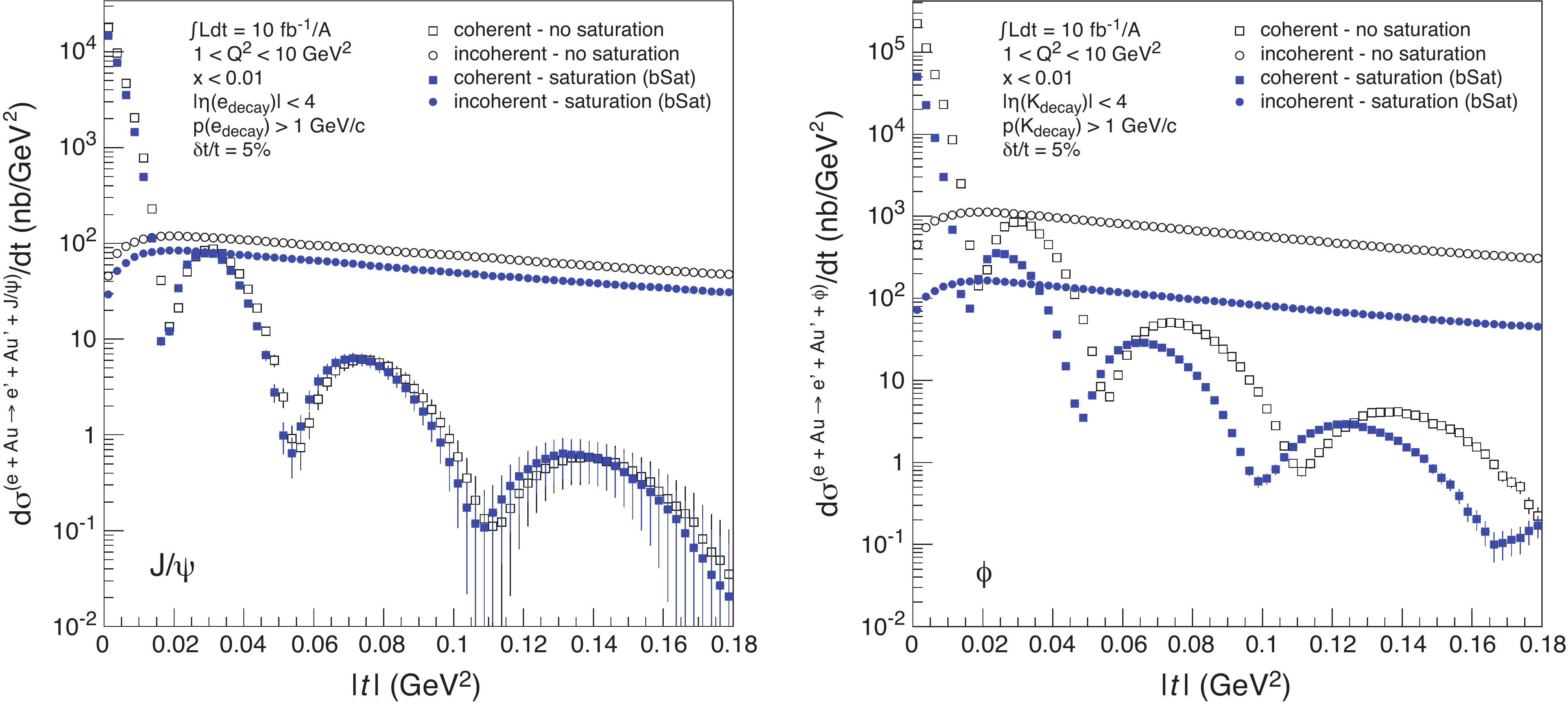}
    \end{center}
    \vskip -0.3in
    \caption{$d\sigma/dt$ distributions for exclusive $J/\psi$ (left)
      and $\phi$ (right) production in coherent and incoherent events
      in diffractive $e$+Au collisions. Predictions from saturation and
      non-saturation models are shown.}
    \label{fig:dsigdt}
\end{figure*}


As explained earlier in Sec.~\ref{diff_sec}, coherent diffractive
events allow one to learn about the shape and the degree of
``blackness'' of the black disk: this enables one to study the spatial
distribution of gluons in the nucleus. Exclusive vector meson
production in diffractive $e$+A collisions is the cleanest such process,
due to the low number of particles in the final state.
This would not only provide us with further insight into saturation
physics but also constitute a highly important contribution to
heavy-ion physics by providing a quantitative understanding of the
initial conditions of a heavy ion collision as described in
Sec.~\ref{AAconn_sec}. It might even shed some light on the role of
glue and thus QCD in the nuclear structure of light nuclei (see
Sec.~\ref{sec:nuclei}).  As described above, in diffractive DIS, the
virtual photon interacts with the nucleus via a color-neutral
exchange, which is dominated by two gluons at the lowest order.
It is precisely this two gluon exchange which yields a diffractive
measurement of the gluon density in a nucleus.

Experimentally the key to the spatial gluon distribution is the
measurement of the $d\sigma/dt$ distribution. As follows from the
optical analogy presented in Sec.~\ref{diff_sec}, the
Fourier-transform of (the square root of) this distribution is the
source distribution 
of the object probed, {\em i.e.}, the
dipole scattering amplitude $N (x, r_T, b_T)$ on the nucleus with
$r_T^2 \sim 1/(Q^2~+~M_V^2)$, where $M_V$ is the mass of the vector
meson \cite{Munier:2001nr} (see also the Sidebar on page~\pageref{sdbar:GPD}).
Note that related studies can be conducted in ultra-peripheral collisions of nuclei, albeit with a 
limited kinematic reach. This is discussed in section \ref{sec::upc-connections}.
\begin{figure*}[thb]
    \begin{center}
        \includegraphics[width=\textwidth]{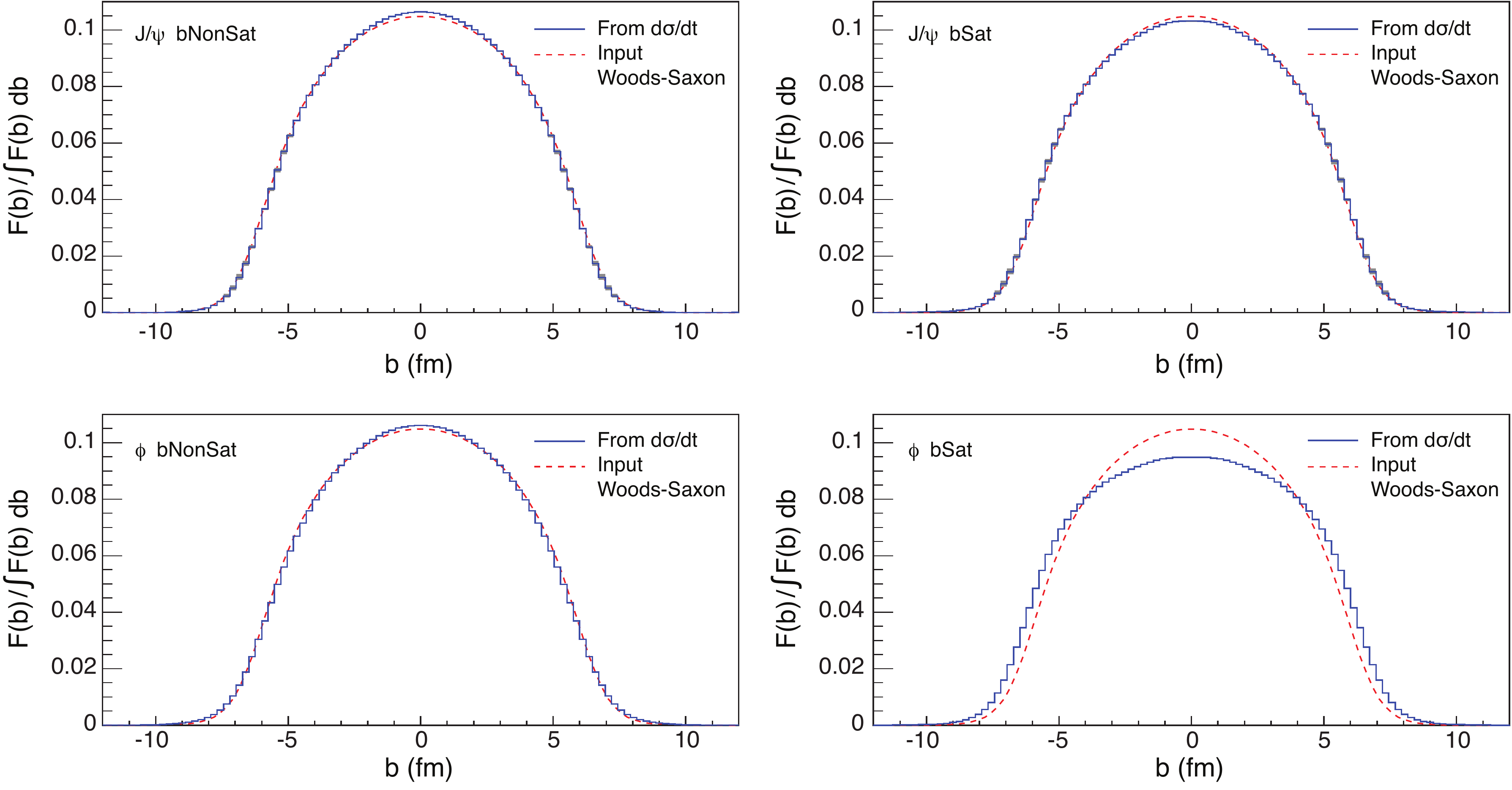}
    \end{center}
    \vskip -0.2in
    \caption{The Fourier transforms obtained in \cite{Toll:2012mb}
      from the distributions in Figure~\ref{fig:dsigdt} for
      $J/\psi$-mesons in the upper row and $\phi$-mesons in the lower
      row. The results from both saturation (right) and non-saturation
      (left) models are shown.  The used input Woods-Saxon
      distribution is shown as a reference in all four plots.}
    \label{fig:WoodsSaxon}
\end{figure*}

Figure \ref{fig:dsigdt} shows the $d\sigma/dt$ distribution for
$J/\psi$ on the left and $\phi$ mesons on the right. The {\em
  coherent} distribution depends on the shape of the source while the
{\em incoherent} distribution provides valuable information on the
fluctuations or ``lumpiness'' of the source \cite{Kowalski:2008sa}. As
discussed above, we are able to distinguish both by detecting the
neutrons emitted by the nuclear breakup in the incoherent case. Again,
we compare to predictions of saturation and non-saturation models. Just as for
the previous figures, the curves were generated with the Sartre event
generator and had to pass through an experimental filter. The
experimental cuts are listed in the figures.

As the $J/\psi$ is smaller than the $\phi$, one sees little difference
between the saturation and no-saturation scenarios for exclusive
$J/\psi$ production but a pronounced effect for the $\phi$, as
expected.  For the former, the statistical errors after the 3$^{rd}$
minimum become excessively large requiring substantially more than the
simulated integrated luminosity of 10 fb$^{-1}$/A. The situation is
more favorable for the $\phi$, where enough statistics up to the
4$^{th}$ minimum are available. The $\rho$ meson has even higher rates
and is also quite sensitive to saturation effects.  However, it
suffers currently from large theoretical uncertainties in the
knowledge of its wave-function, making calculations less reliable.

The coherent distributions in Figure~\ref{fig:dsigdt} can be used to
obtain information about the gluon distribution in impact parameter
space $F(b)$ through a two-dimensional Fourier transform of the square
root of the coherent elastic cross section
\cite{Munier:2001nr,Toll:2012mb}
\begin{align}
  F (b) = \int\limits_0^\infty \frac{d q \, q}{2 \pi} \, J_0 (q \, b)
  \, \sqrt{\frac{d \sigma_{coherent}}{dt}}
\end{align}
with $t = - q^2$. In Figure~\ref{fig:WoodsSaxon} we show the resulting
Fourier transforms of the coherent points in Figure~\ref{fig:dsigdt},
using the range $-t < 0.36$~GeV$^2$ which is achievable at the EIC
given enough statistics. As a reference, we show (dotted line) the
original input source distribution used in the generator, which is the
Woods-Saxon function integrated over the longitudinal direction. The
obtained distributions have been normalized to unity. The
uncertainties due to the statistical error are negligible, and are
barely visible in Figure~\ref{fig:WoodsSaxon}. Strictly-speaking, the
integral over $t$ in the Fourier transformation should be performed up
to $|t| \to \infty$.  We studied the effects of using the finite
$t$-range in the Fourier transform by varying the upper integration
limit and found fast convergence towards the input Woods-Saxon
distribution already for the upper limit of $|t| \sim 0.1$~GeV$^2$.

The non-saturation curves for $\phi$ and $J/\psi$-meson production
reproduce the shape of the input distribution perfectly. For the
saturation model, the shape of the $J/\psi$ curve also reproduces the
input distribution, while the $\phi$ curve does not. As explained
above, this is expected, as the size of the $J/\psi$ meson is much
smaller than that for $\phi$, making the latter more susceptible to
non-linear effects as already observed in Figures~\ref{fig:diffQ2vm}
and \ref{fig:dsigdt}. We conclude that the $J/\psi$ meson is better
suited for probing the transverse structure of the nucleus. However,
by measuring $F(b)$ with both $J/\psi$ and $\phi$ mesons, one can
obtain valuable information on how sensitive the measurement is to
non-linear effects. Thus, both measurements are important and
complementary to each other. The results in
Figure~\ref{fig:WoodsSaxon} provide a strong indication that EIC will
be able to obtain the nuclear spatial gluon distribution from the
measured coherent $t$-spectrum from exclusive $J/\psi$ and $\phi$
production in $e+$A, in a model-independent fashion.

\end{multicols}

\newpage

%% file: files_tex/eA-cold.tex
\newpage

\section{Quarks and Gluons in the Nucleus}
\label{sec:nuclei}

{\large\it Conveners:\ William Brooks and Jian-Wei Qiu} \vskip 0.15in

In this section we present a few key measurements that will allow us
to answer the fundamental questions from the beginning of this chapter
and to explore the properties of quarks and gluons and their
interactions in a nuclear environment.  In
Table~\ref{tab:keyMeasurements-large-x}, we list the key measurements
to be carried out at an EIC.  The measurement of nuclear structure
functions with various ion beams at intermediate-$x$ will enable
the first glimpses of collective nuclear effects at the partonic
level and the onset of the breakdown of DGLAP evolution.  The
semi-inclusive production of energetic hadrons will probe
nuclear matter's response to a fast moving color charge as well as the
mass of the particle carrying the charge.  
The multiple scattering of the fast moving color charge off the color field
inside the nucleus could modify the distribution of produced hadrons. 
The transverse momentum broadening of the produced hadrons in 
$e$+A collisions provides a sensitive probe to the characteristic time scale 
(or distance) of color neutralization, as well as the response of the nuclear 
medium to a fast moving color charge.  
It thus allows access to the transport coefficients
of the nuclear system and to medium induced energy loss mechanisms.
With the well-determined leptonic and hadronic scattering planes, and 
the azimuthal angle $\phi$ between the planes in semi-inclusive DIS, 
on an event-by-event basis, 
the nuclear modification to the angular $\phi$ modulation of the 
produced hadrons could be a sensitive probe of the fluctuation of 
spatial distributions of quarks and gluons inside a large nucleus
\cite{Brooks:2011nu}, which is very important for understanding the
initial condition of relativistic heavy ion collisions.

\begin{table}[h]
\centering

\noindent\makebox[\textwidth]{%
\footnotesize
\begin{tabular}{|c|c|c|c|c|}
\hline
Deliverables & Observables & What we learn \\
\hline
\hline
Collective           & Ratios  $R_2$          & $Q^2$ evolution:  onset of DGLAP violation, beyond DGLAP\\
nuclear effects   & from inclusive DIS    & $A$-dependence of shadowing and antishadowing \\
at intermediate-$x$ &  & Initial conditions for small-$x$ evolution\\
\hline
Transport  &  Production of light & Color neutralization: mass dependence of hadronization \\
coefficients in  &  and heavy hadrons,   & Multiple scattering and mass dependence of energy loss\\
 nuclear matter     &  and jets in SIDIS    & Medium effect of heavy quarkonium production \\
\hline
Nuclear density &  Hadron production     &  Transverse momentum broadening of produced hadrons \\
and its fluctuation    & in SIDIS  &  Azimuthal $\phi$-modulation of produced hadrons \\
\hline

\end{tabular}}

\caption{Key measurements in $e$+A collisions at an EIC to explore the dynamics of 
quarks and gluons in a nucleus in the non-saturation regime.}
\label{tab:keyMeasurements-large-x}
\end{table}

\subsection{Distributions of Quarks and Gluons in a Nucleus}
\label{subsec:sf}

\begin{multicols}{2}
The momentum distribution of quarks and gluons inside a fast moving
proton was best measured by lepton DIS on a proton beam at HERA.
Although the scattering could take place between the lepton and a
single quark (or gluon) state as well as a multiple quark-gluon state
of the proton, the large momentum transfer of the scattering, $Q$,
localizes the scattering, suppresses the contribution from multiple
scattering, and allows us to express the complex DIS cross-sections in
terms of a set of momentum distributions of quarks and gluons.  These
are probability density distributions to find a parton (quark,
anti-quark or gluon) to carry the momentum fraction $x$ of a fast
moving hadron.  Actually, it is a triumph of QCD that one set of
universal parton distributions, extracted from HERA data, plus
calculable scatterings between quarks and gluons, can successfully
interpret all existing data of high energy proton collisions with a
momentum transfer larger than 2 GeV (corresponding to hard scatterings
taking place at a distance less than one tenth of a femtometer).

\begin{figure*}[th!]
\begin{center}
\includegraphics[width=0.8\textwidth]{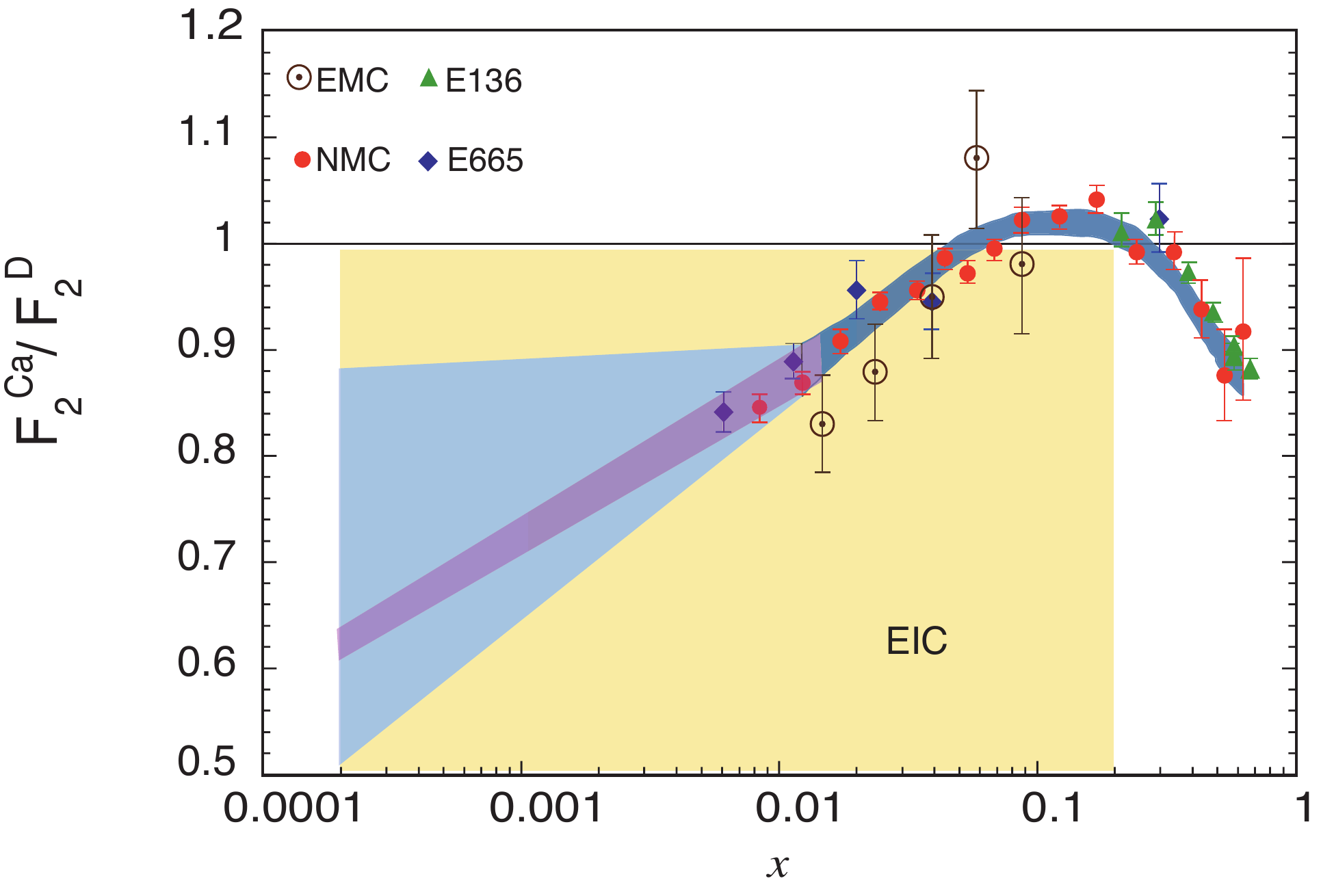}
\end{center} \vskip -0.2in
\caption{\label{fig:R2-x-A40} The ratio of nuclear over nucleon $F_2$
structure function, $R_2$, as a function of Bjorken $x$, with data
from existing fixed target DIS experiments at $Q^2> 1~$GeV$^2$, along
with the QCD global fit from EPS09 \cite{Eskola:2009uj}.  Also shown
is the expected kinematic coverage of the inclusive measurements at
the EIC.  The purple error band is the expected systematic uncertainty 
at the EIC assuming a $\pm $2\% (a total of 4\%) systematic error, 
while the statistical uncertainty is expected to be much smaller. }
\end{figure*}

Are the quarks and gluons in a nucleus confined within the individual
nucleons?  or does the nuclear environment significantly affect their
distributions?  The EMC experiment at CERN~\cite{Aubert:1983xm} and
experiments in the following two decades clearly revealed that the
momentum distribution of quarks in a fast moving nucleus is not a
simple superposition of their distributions within nucleons.  Instead,
the measured ratio of nuclear over nucleon structure functions, as
defined in Eq.~(\ref{eq:shadowing}), follows a non-trivial function of
Bjorken $x$, significantly different from unity, and shows the
suppression as $x$ decreases, as shown in Fig.~\ref{fig:R2-x-A40}.
The observed suppression at $x\sim 0.01$, which is often referred to
as the phenomenon of nuclear shadowing, is much stronger than what the
Fermi motion of nucleons inside a nucleus could account for. This
discovery sparked a worldwide effort to study the properties of quarks
and gluons and their dynamics in the nuclear environment both
experimentally and theoretically.

Using the same very successful QCD formulation at the leading power in
$Q$ for proton scattering, and using the DGLAP evolution for the scale
dependence of parton momentum distributions, several QCD global
analyses have been able to fit the observed non-trivial nuclear
dependence of existing data, attributing all observed nuclear
dependences --- including its $x$-dependence and nuclear atomic weight
$A$-dependence --- to a set of nucleus-dependent quark and gluon
distributions at an input scale $Q_0 \gtrsim 1$ GeV
\cite{Eskola:2009uj,deFlorian:2003qf,Hirai:2007sx}.  As an example,
the fitting result of Eskola {\it et al.}\ is plotted along with the
data on the ratio of the $F_2$ structure function of calcium divided
by that of deuterium in Fig.~\ref{fig:R2-x-A40}, where the dark blue
band indicates the uncertainty of the EPS09 fit \cite{Eskola:2009uj}.
The success of the QCD global analyses clearly indicates that the
response of the nuclear cross-section to the {\it variation} of the
probing momentum scale $Q \gtrsim Q_0$ is insensitive to the nuclear
structure, since the DGLAP evolution itself does not introduce any
nuclear dependence.  However, it does not answer the fundamental
questions: Why are the parton distributions in a nucleus so different
from those in a free nucleon at the probing scale $Q_0$?  How do the
nuclear structure and QCD dynamics determine the distributions of
quarks and gluons in a nucleus?

The nucleus is a ``molecule'' in QCD, made of nucleons --- which, in
turn, are bound states of quarks and gluons. Unlike the molecule in
QED, nucleons in the nucleus are packed next to each other, and there
are many soft gluons inside nucleons when probed at small $x$.  The
DIS probe has a high resolution in transverse size $\sim 1/Q$.  But
its resolution in the longitudinal direction, which is proportional to
$1/xp\sim 1/Q$, is not necessarily sharp in comparison with the
Lorentz contracted size of a light-speed nucleus, $\sim 2R_A(m/p)$,
with nuclear radius $R_A\propto A^{1/3}$ and the Lorentz contraction
factor $m/p$ and nucleon mass $m$.  That is, when $1/x p > 2R_A(m/p)$,
or at a small $x \sim 1/2mR_A \sim 0.01$, the DIS probe could interact
coherently with quarks and gluons of all nucleons at the same impact
parameter of the largest nucleus moving nearly at the speed of light,
$p\gg m$.  The destructive interference of the coherent multiple
scattering could lead to a reduction of the DIS cross-section
\cite{Gribov:1984tu,Frankfurt:2011cs}.  Such coherent multi-parton
interactions at small $x$ could take place non-perturbatively to
generate a nuclear dependence of the parton distributions at the input scale
$Q_0$, including shadowing \cite{Frankfurt:2011cs} and
anti-shadowing \cite{Brodsky:1989qz}, which could be systematically
extracted by using the DGLAP-based leading power QCD formalism.  In
addition, coherent multiple scattering could also take place at a
perturbative scale $Q>Q_0$, and its contribution to the inclusive DIS
cross-section could be systematically investigated in QCD in terms of
corrections to the DGLAP-based QCD formulation
\cite{Mueller:1985wy,Qiu:2003vd}.  Although such corrections are
suppressed by the small perturbative probing size, they can be enhanced by the 
number of nucleons at the same impact parameter in a nucleus and large
number of soft gluons in nucleons.  Coherent multiple scattering
naturally leads to the observed phenomena of nuclear shadowing: more
suppression when $x$ decreases, $Q$ decreases, and $A$ increases.
But, none of these dependences could have been predicted by the very
successful leading power DGLAP-based QCD formulation.

When the gluon density is so large at small-$x$ and the coherent
multi-parton interactions are so strong that their contributions are equally 
important as that from single-parton scattering, measurements of the 
DIS cross-section could probe a
new QCD phenomenon - the saturation of gluons discussed in the last
section.  In this new regime, which is referred to as a Color Glass
Condensate (CGC) \cite{McLerran:1993ni,Iancu:2000hn}, the standard
fixed order perturbative QCD approach to the coherent multiple
scattering would be completely ineffective.  The resummation of all powers
of coherent multi-parton interactions or new effective field theory
approaches are needed.  The RHIC data \cite{ Adare:2011sc,
Braidot:2010ig} on the correlation in deuteron-gold collisions
indicate that the saturation phenomena might take place at $x \lesssim
0.001$ \cite{Adare:2011sc, Braidot:2010ig}.  Therefore, the region of
$0.001 < x < 0.1$, at a sufficiently large probing scale $Q$, could be
the most interesting place to see the transition of a large nucleus
from a diluted partonic system --- whose response to the resolution of
the hard probe (the $Q^2$-dependence) follows linear DGLAP
evolution --- to matter composed of condensed and saturated gluons.

This very important transition region with Bjorken $x \in (0.001,
0.1)$ could be best explored by the EIC, as shown in
Fig.~\ref{fig:R2-x-A40}.  The EIC will not only explore
this transition region, but will also have a wide overlap with regions
that have been and will be measured by fixed target experiments, as
indicated by the yellow box in Fig.~\ref{fig:R2-x-A40}. 
At its full operation, the coverage of EIC in $x$ could be extended down 
to $10^{-4}$ while maintaining a sufficiently large $Q$.
The EIC will have ideal kinematic coverage for the systematic study of
QCD dynamics in this very rich transition region, as well as the new
regime of saturated gluons.

If the nuclear effect on the DIS cross-section, as shown in
Fig.~\ref{fig:R2-x-A40}, is mainly due to the abundance of nucleons at
the same impact parameter of the nucleus (proportional to $A^{1/3}$),
while the elementary scattering is still relatively weak, one would
expect the ratio of nuclear over nucleon structure functions to
saturate when $x $ goes below $0.01$, or equivalently, the nuclear
structure function to be proportional to the nucleon structure
functions, as shown, for example, by the upper line of the blue area
extrapolated from the current data in Fig.~\ref{fig:R2-x-A40}.  In
this case, there is no saturation in nuclear structure functions since
the proton structure function is not saturated at this
intermediate-$x$ region, and the ratio could have a second drop at a
smaller $x$ when nuclear structure functions enter the saturation
region.  On the other hand, if the soft gluons are a property of the
whole nucleus and the coherence is strong, one would expect the ratio
of the nuclear to nucleon structure function to fall continuously as
$x$ decreases, as sketched by the lower line of the blue band, and
eventually, reach a constant when both nuclear and nucleon structure
functions are in the saturation region.  From the size of the purple
error band in Fig.~\ref{fig:R2-x-A40}, which is the expected
systematic uncertainty at the EIC (while the statistical uncertainty
is expected to be much smaller), the EIC could easily distinguish
these two extreme possibilities to explore the nature of sea quarks
and soft gluons in a nuclear environment.

With the unprecedented energy and luminosity of lepton-nucleus
collisions at the EIC, the precision measurements of the $Q$
dependence of the nuclear structure functions could extract nuclear
gluon distributions at small $x$ that are effectively unknown now, 
and identify the momentum scale $Q_0$ below which the
DGLAP-based QCD formulation fails, to discover the onset of the new
regime of non-linear QCD dynamics.  With its variety of nuclear
species, and the precise measurements of the $x$ and $Q$ dependence in
this transition region, the EIC is an ideal machine to explore the
transition region and to provide immediate access to the first
glimpses of collective nuclear effects caused by coherent multi-parton
dynamics in QCD.  Inclusive DIS measurements at the EIC provide an
excellent and unique testing ground to study the transition to new and
novel saturation physics.
\end{multicols}

\subsection{Propagation of a Fast Moving Color Charge in QCD Matter}
\label{subsec:energyloss}

\begin{multicols}{2}
The discovery of the quark-gluon plasma (QGP) in the collision of two
heavy ions at the Relativistic Heavy Ion Collider (RHIC) at Brookhaven
National Laboratory made it possible to study in a laboratory the
properties of quark-gluon matter at extremely high temperatures
and densities, which were believed to exist only a few microseconds
after the Big Bang. One key piece of evidence of the discovery was the
strong suppression of fast moving hadrons produced in relativistic
heavy-ion collisions \cite{d'Enterria:2009am}, which is often referred
to as jet quenching \cite{Wang:1991xy}.  It was found that the
production rate of the fast moving hadrons in a central gold-gold
collision could be suppressed by as much as a factor of five compared to that of a 
proton-proton collision at the same energy, and the same
phenomenon was confirmed by the heavy ion program at the LHC.

Fast moving hadrons at RHIC are dominantly produced by the
fragmentation of colored fast moving quarks or gluons that are
produced during hard collisions at short distances.  Fragmentation
(or in general, Hadronization) -- the transition of a colored and
energetic parton to a colorless hadron -- is a rich and dynamical
process in QCD quantified by the fragmentation function $D_{{\rm
parton}\to{\rm hadron}} (z)$, with $z$ the momentum fraction of the
fast moving parton to be carried by the produced hadron in the DGLAP
based QCD formulation.  Although QCD calculations are consistent with
hadron production in high-energy collisions, knowledge about the
dynamics of the hadronization process remains limited and strongly model
dependent.  It is clear that color is ultimately confined in these
dynamical processes.  The color of an energetic quark or a gluon
produced in high-energy collisions has to be neutralized so that it
can transmute itself into hadrons.  Even the determination of a
characteristic time scale for the color neutralization would shed some
light on the properties of color confinement and help answer the
question of what governs the transitions of quarks and gluons to
hadrons.

\begin{figure*}[th!]
\begin{center}
\includegraphics[width=0.45\textwidth]{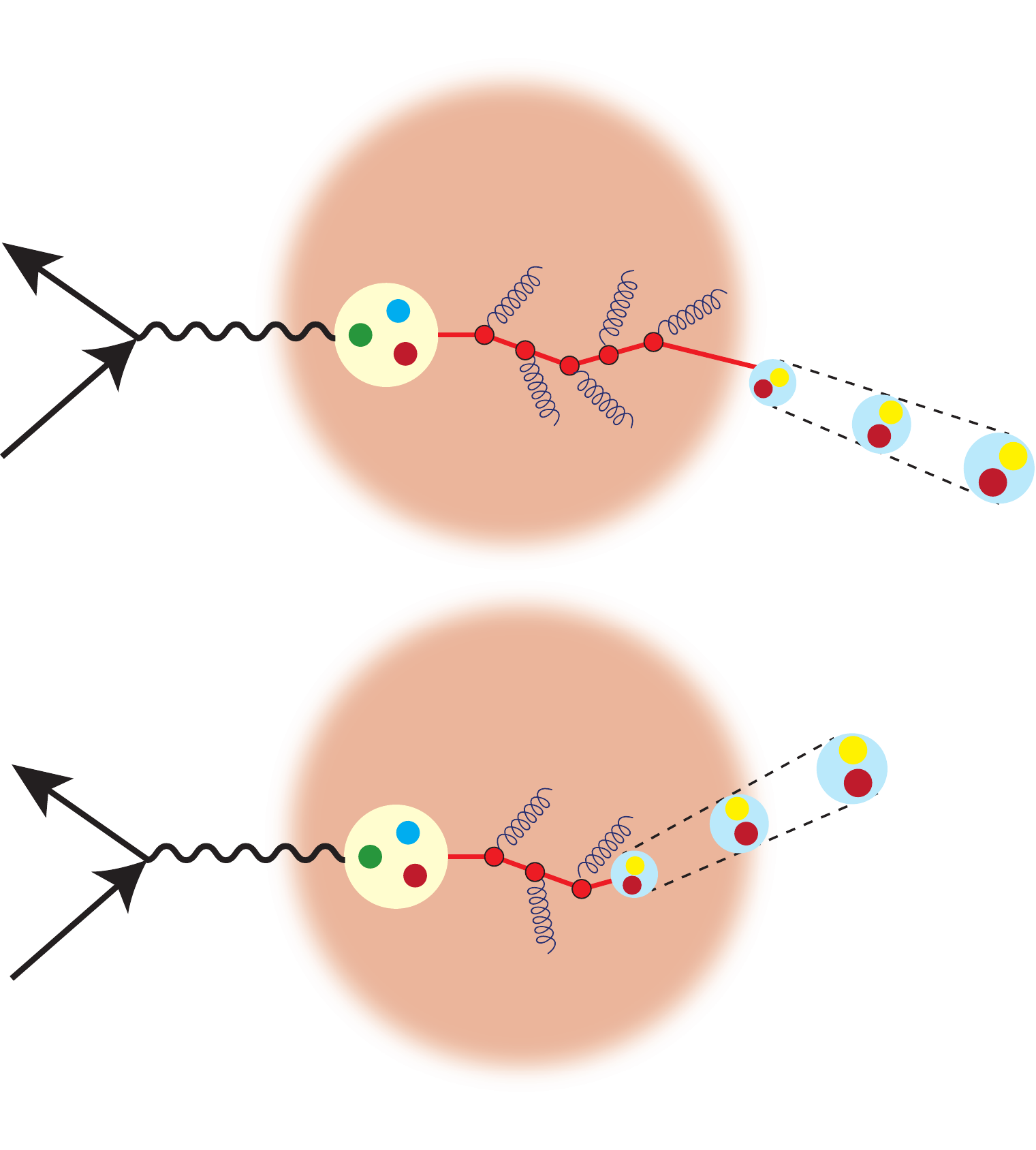}
\hskip 0.1\textwidth
\includegraphics[width=0.4\textwidth]{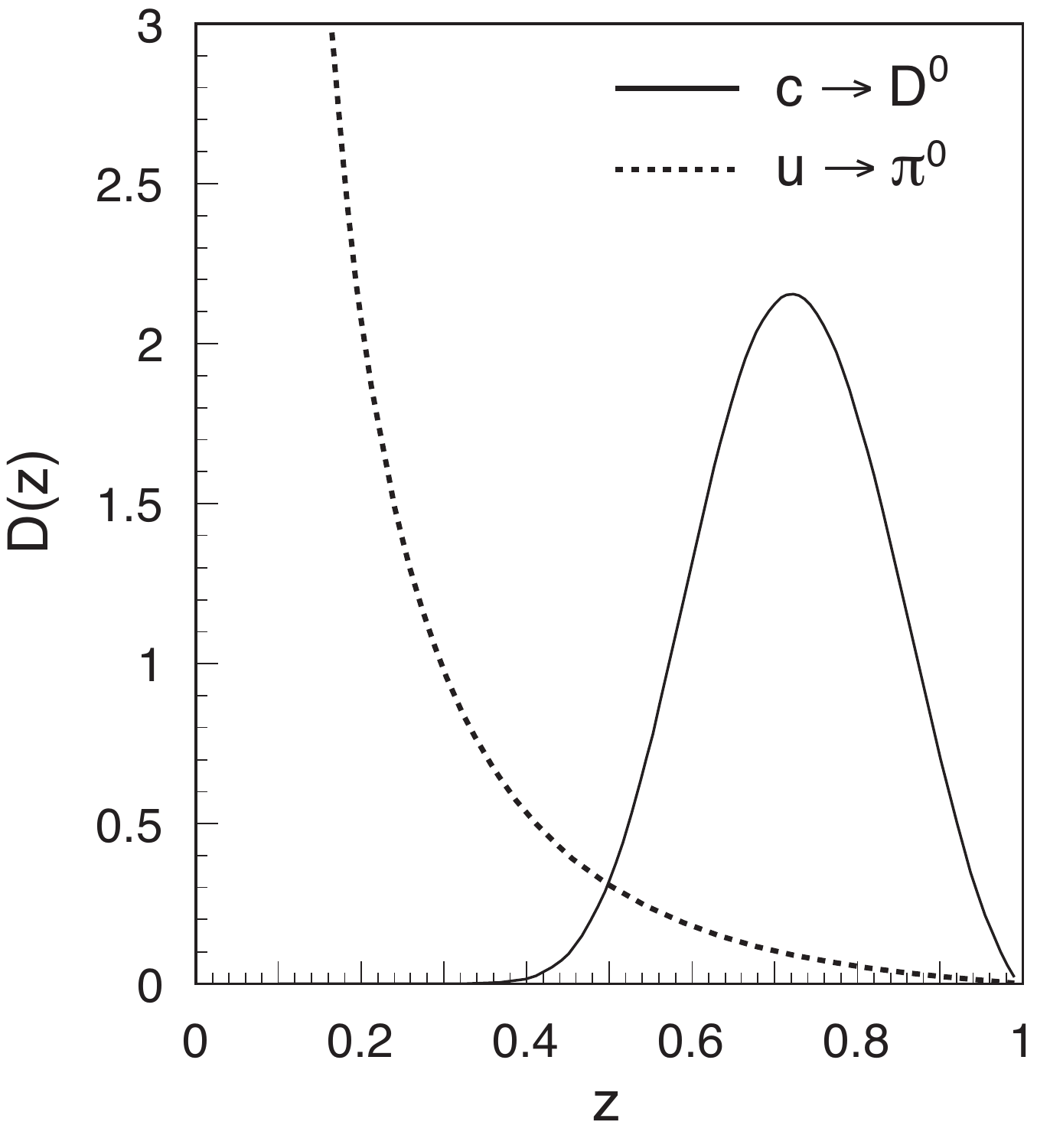}
\end{center} \vskip -0.1in
\caption{\label{fig:hadronization} {\bf Left:}\ A cartoon for the
interactions of the parton moving through cold nuclear matter when the
produced hadron is formed outside (upper) and inside (lower) the
nucleus.  {\bf Right:}\ Fragmentation functions as a function of $z$:
from the charm quark to the $D^0$ meson (solid) \cite{Kneesch:2007ey}
and from up quark to $\pi^0$ meson (dashed) \cite{Albino:2008fy}.  }
\end{figure*}

The collision of a fast moving parton within the QGP could induce gluon
radiation to reduce the parton's forward momentum and energy, while
the parton-to-hadron fragmentation functions might not be affected
since the energetic hadrons are likely to be formed outside the QGP
due to time dilation, as indicated by the cartoon in
Fig.~\ref{fig:hadronization} (Left - upper plot).  The energy loss of
the active parton would require a fragmentation function of a larger
$z$ in order to produce a hadron with the same observed momentum as
that produced in proton-proton collisions without energy loss
\cite{Guo:2000nz}.  However, it has been puzzling
\cite{Tarafdar:2012ef} that heavy meson production in the same
experiments at RHIC seems to be suppressed as much as the production
of light mesons, although a heavy quark is much less likely to lose
its energy via medium induced radiation.  It is critically important
to have new and clean measurements, as well as independent tests of
the energy-loss mechanisms, in order to have full confidence in
jet quenching as a hard probe of QGP properties.

Semi-inclusive DIS in $e$+A collisions provides a known and stable
nuclear medium (``cold QCD matter''), well-controlled kinematics of
hard scattering, and a final state particle with well-known
properties.  The time for the produced quark (or gluon) to neutralize
its color depends on its momentum and virtuality when it was produced.
The process could take place entirely inside the nuclear medium, or
outside the medium, or somewhere in-between, as indicated by the
cartoon in Fig.~\ref{fig:hadronization} (Left)
\cite{Kopeliovich:2003py,Accardi:2005jd}.  Cold QCD matter could be an
excellent femtometer-scale detector of the hadronization process from
its controllable interaction with the produced quark (or gluon).  By
facilitating studies on how struck partons propagate through cold
nuclear matter and evolve into hadrons, as sketched in
Fig.~\ref{fig:hadronization} (Left), the EIC would provide independent
and complementary information essential for understanding the response
of the nuclear medium to a colored fast moving (heavy or light) quark.
With its collider energies and thus the much larger range of $\nu$,
the energy of the exchanged virtual photon, the EIC is unique for
providing clean measurements of medium induced energy loss when the
hadrons are formed outside the nuclear medium, while it is also
capable of exploring the interplay between hadronization and
medium-induced energy loss when the hadronization takes place inside
the medium.  In the latter case, color transparency may also play a
role
\cite{Kopeliovich:2003py,Bertsch:1981py,Brodsky:1988xz,Frankfurt:1994hf},
and this is yet another important topic that can be independently
explored with various techniques and measurements at the EIC
\cite{Dutta:2012ii}.

\begin{figure*}[th!]
\begin{center} \vskip 0.1in
\includegraphics[scale=0.36]{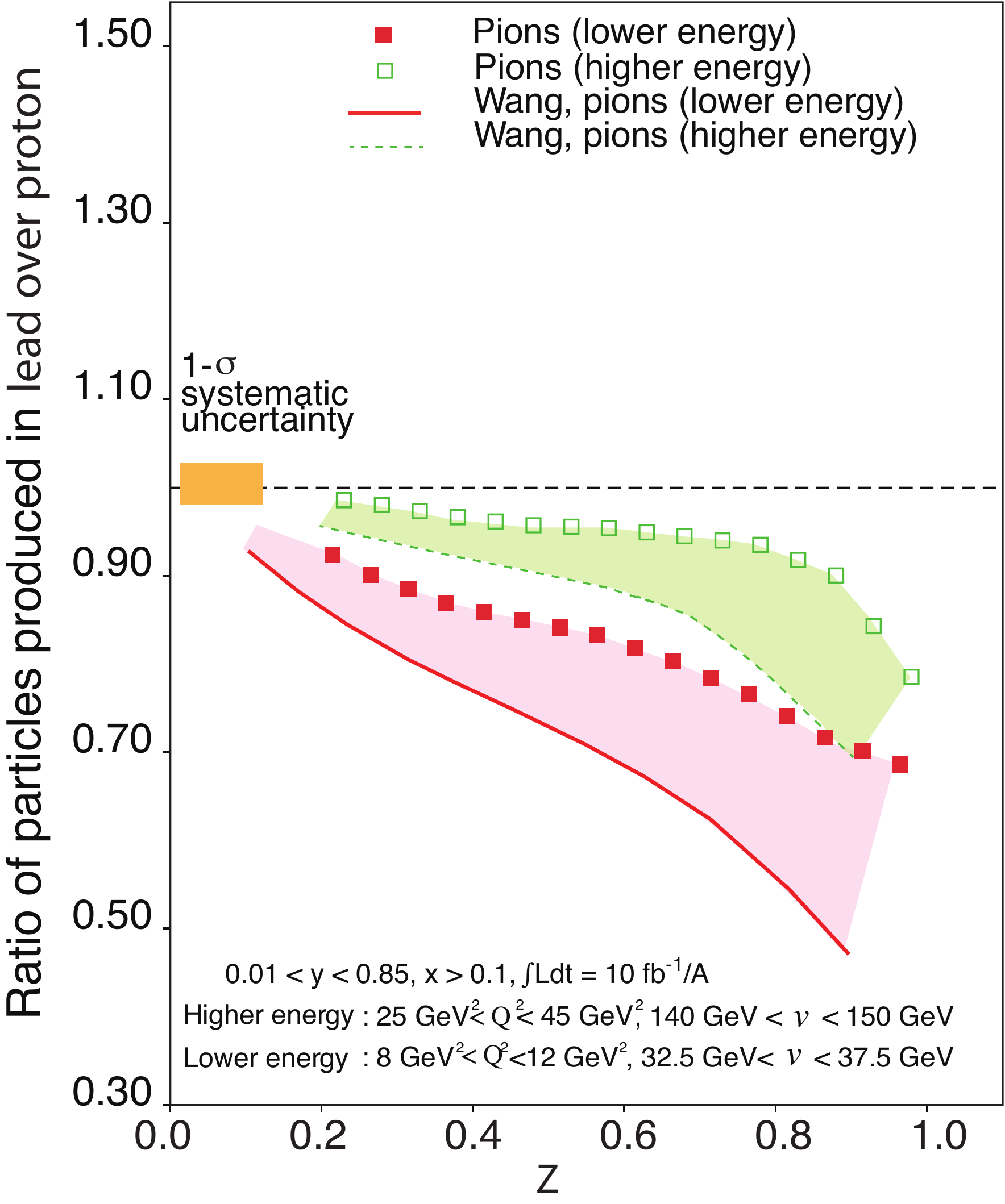}
\hskip 0.3in
\includegraphics[scale=0.36]{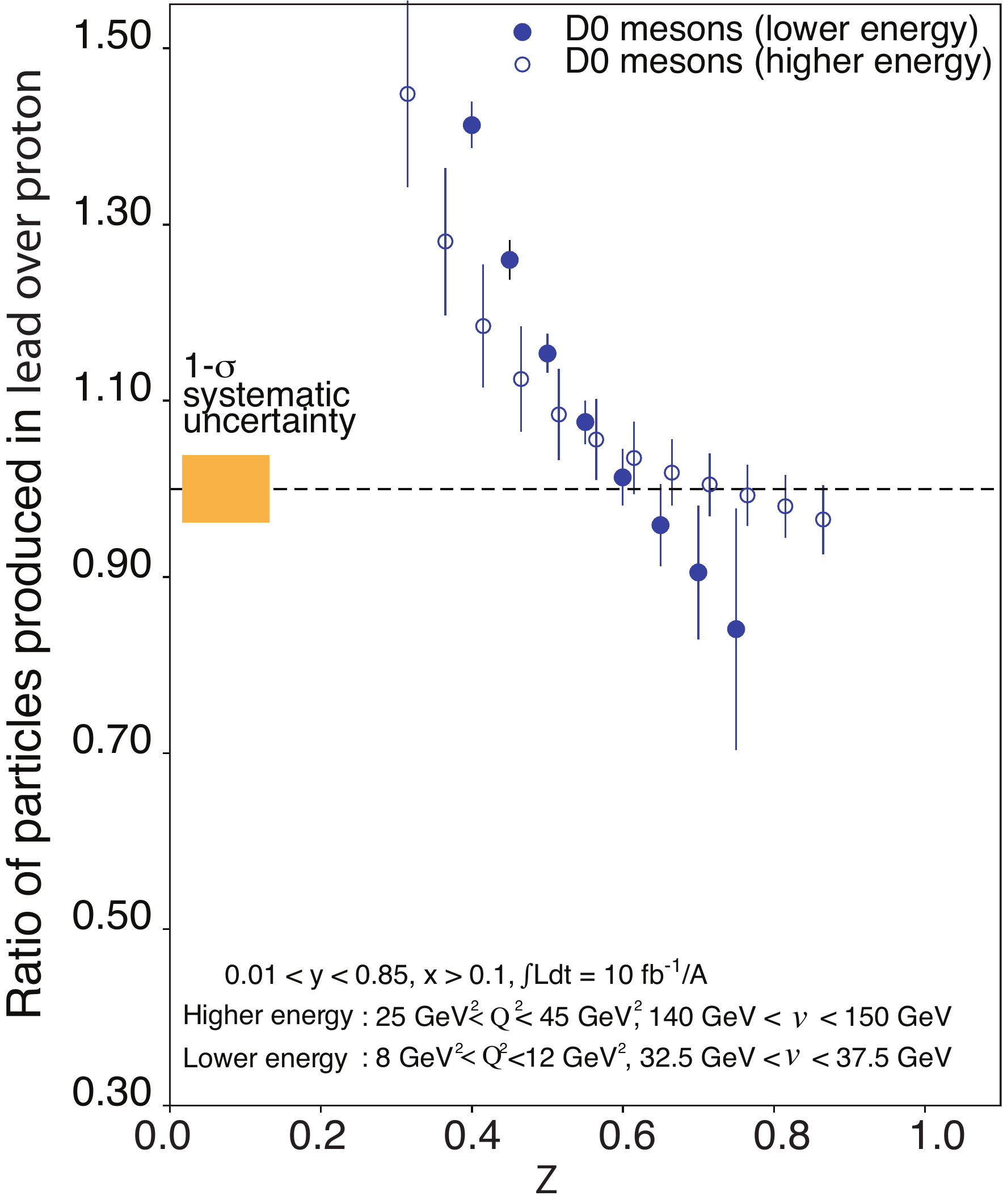}
\end{center} \vskip -0.2in
\caption{\label{fig:Rsidis-zh} The ratio of semi-inclusive cross-sections
for producing a single pion (Left) and a single $D^0$
(Right) in electron-lead collisions to the same produced in
electron-deuteron collisions as a function of $z$ at the EIC with two
different photon energies $\rm \nu=35~GeV$ at $\rm Q^2=10~GeV^2$
(solid symbols) and $\rm \nu=145~GeV$ at $\rm Q^2=35~GeV^2$ (open
symbols) ($p_T$ of the hadron is integrated).  The solid lines are
predictions of pure energy loss calculations for pion production (see
the text). }
\end{figure*}

The amount of the medium-induced energy loss and the functional form
of the fragmentation functions should be the most important cause for
the multiplicity ratio of hadrons produced in a large
nucleus compared to the same process on a proton, if the hadrons are formed
outside the nuclear medium.  It was evident from hadron production in
$e^-+e^+$ collisions that the fragmentation functions for
light mesons, such as pions, have a very different functional form
with $z$ from that of heavy mesons, such as $D$-mesons.  As shown in
Fig.~\ref{fig:hadronization} (Right), the heavy $D^0$-meson
fragmentation function has a peak while the pion fragmentation
function is a monotonically decreasing function of $z$.  The fact that
the energy loss matches the active parton to the fragmentation
function at a larger value of $z$ leads to two dramatically different
phenomena in the semi-inclusive production of light and heavy mesons
at the EIC, as shown in Fig.~\ref{fig:Rsidis-zh} \cite{Guo:2007ve}.
The ratio of light meson ($\pi)$ production in $e$+Pb
collisions over that in $e$+d collisions (red square
symbols) is always below unity, while the ratio of heavy meson ($D^0$)
production can be less than as well as larger than unity due to the
difference in hadronization.

In Fig.~\ref{fig:Rsidis-zh}, simulation results are plotted for
the multiplicity ratio of semi-inclusive DIS cross-sections for
producing a single pion (Left) and a single $D^0$ (Right) in
$e$+Pb collisions to the same produced in the $e$+d
as a function of $z$ at the EIC with two different photon energies: $\rm
\nu=35~GeV$ at $\rm Q^2=10~GeV^2$ (solid line and square symbols) and
$\rm \nu=145~GeV$ at $\rm Q^2=35~GeV^2$ (dashed line and open
symbols).  The $p_T$ of the observed hadrons is integrated.  The ratio
for pions (red square symbols) was taken from the calculation of
\cite{Kopeliovich:2003py}, extended to lower $\rm z$, and extrapolated
from a copper nucleus to a lead nucleus using the prescription of
\cite{Accardi:2005jd}.  In this model approach, pions are suppressed
in $e+A$ collisions due to a combination of the attenuation
of pre-hadrons as well as medium-induced energy loss.  In this figure,
the solid lines (red - $\nu=145$~GeV, and blue - $\nu=35$~GeV) are
predictions of pure energy loss calculations using the energy loss
parameters of \cite{Wang:2002ri}.  The large differences in the
suppression between the square symbols and solid lines are immediate
consequences of the characteristic time scale for the color
neutralization and the details of the attenuation of pre-hadrons, as
well as the model for energy loss.  With the size of the systematic
errors shown by the yellow bar on the left of the unity ratio, the
multiplicity ratio of pion production at the EIC will provide an
excellent and unique opportunity to study hadronization by using the
nucleus as a femtometer detector.

The dramatic difference between the multiplicity ratios of $D^0$ meson
production and that of pions, as shown in Fig.~\ref{fig:Rsidis-zh}, is
an immediate consequence of the difference in the fragmentation
functions shown in Fig.~\ref{fig:hadronization} (Right).  The
enhancement of the ratio is caused by the peak in the $D^0$'s
fragmentation function.  The slope of the enhancement is sensitive to
the amount of energy loss, or equivalently, the transport coefficient,
$\hat{q}$ of cold nuclear matter, and the shape of the fragmentation
function \cite{Guo:2007ve}.  The energy loss used in the simulation is
a factor of 0.35 less than that of light quarks as derived in
\cite{Zhang:2004qm} by taking into account the limited cone for
gluon radiation caused by the larger charm quark mass.  The solid
symbols are for $ x=0.1$ and $\rm Q^2=10~GeV^2$.  
 In the same figure we also show the same type of plot but for
$\rm \nu=145~GeV$ and $\rm Q^2=35~GeV^2$. The expected reduction in
the level of pion suppression relative to $\rm \nu=35~GeV$ is visible
and the shape of the $D^0$ data is quite different from that for $\rm
\nu=35~GeV$. In addition to the $D^0$ meson, similar studies could be
carried out with the $D_s^+$ and other heavy meson states, from which
more complete information on heavy quark energy loss could be 
extracted.  This strong sensitivity of the shape to the value of
$\nu$ will be a unique and powerful tool in the understanding of energy loss of
heavy quarks in cold nuclear systems.  The discovery of such a
dramatic difference in multiplicity ratios between light and heavy
meson production in Fig.~\ref{fig:Rsidis-zh} at the EIC would shed
light on the hadronization process and on what governs the transition
from quarks and gluons to hadrons.
\end{multicols}

\subsection{Spatial Fluctuation of Parton Density Inside a Large Nucleus}
\label{subsec:fluctuation}

\begin{multicols}{2}
The transverse flow of particles is a key piece of evidence for the
formation of a strongly interacting QGP in relativistic heavy-ion
collision.  It was recognized that fluctuations in the geometry of the
overlap zone of heavy-ion collisions lead to some\break 
\noindent unexpected 
azimuthal $\phi$ modulations $v_n$ of particle multiplicity 
with respect to the reaction plane.  In particular, $v_3$
leads to very interesting features of two particle correlations. 
The initial-state density
fluctuations seem to influence the formation and expansion of the QGP. 
An independent measurement of the spatial fluctuations of quark and 
gluon densities inside a large nucleus is hence critically important for 
understanding both, the formation of QGP in heavy-ion collisions and
nuclear structure in terms of quarks and gluons.
\end{multicols}

\begin{figure*}[h!]
\begin{center} \vskip -0.1in
\includegraphics[width=0.85\textwidth]{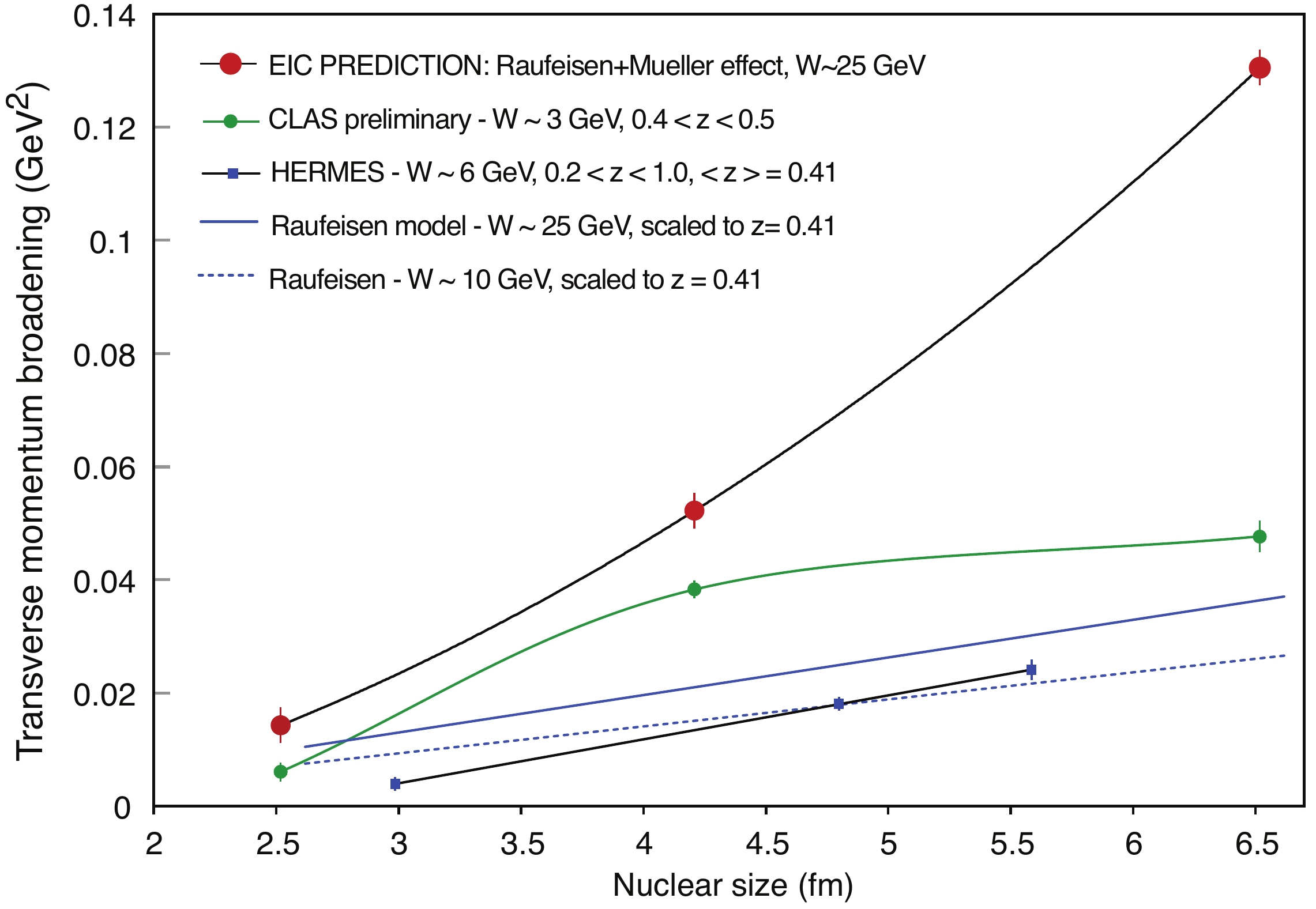}
\end{center} \vskip -0.2in
\caption{\label{fig:broadening}  
Transverse momentum broadening as a function of nuclear size in $e$+A collisions 
as defined in Eq.~(\ref{eq:broadening}).  See the text for the details.}
\end{figure*}

\vspace{-0.2in}
\begin{multicols}{2}
Multiple scattering between the produced parton and the nuclear medium 
in semi-inclusive $e$+A collisions can broaden the transverse momentum spectrum 
of the produced hadron in comparison with that in corresponding $e$+$p$ collisions.  
The nuclear modification to the transverse momentum spectrum could be 
quantified by defining the transverse momentum broadening in terms of 
the azimuthal angle dependent  broadening,
\end{multicols}
\begin{equation} 
\Delta\langle p_T^2 \rangle_{AN} \equiv
\int_0^{2\pi} d\phi\,  \Delta\langle p_T^2(\phi) \rangle_{AN} 
\equiv
\int_0^{2\pi} d\phi\, \langle p_T^2(\phi) \rangle_A 
- \int_0^{2\pi} d\phi\, \langle p_T^2(\phi) \rangle_N
\label{eq:broadening}
\end{equation} 
with the averaged transverse momentum squared at a given $\phi$,
\begin{equation} \langle p_T^2(\phi) \rangle_A = \left.  \int
dp_T^2 \, p_T^2 \frac{d\sigma_{eA}}{dx_BdQ^2 dp_T^2 d\phi} \right/
\frac{d\sigma_{eA}}{dx_BdQ^2}\, .
\end{equation} 
\begin{multicols}{2}
\noindent
The azimuthal angle $\phi$ is defined as the angle between the leptonic 
and hadronic scattering planes in semi-inclusive DIS, 
as shown in the Sidebar on page~\pageref{sdbar:tmd}.
The measurement of the transverse momentum broadening in Eq.~(\ref{eq:broadening})
provides important information on the strength and distribution of the color fields 
inside the colliding nucleus and the color neutralization of the fast moving parton, 
since the color clearly affects the interaction between the fragmenting parton and 
the nucleus, and hence the amount of the broadening.  In addition,
the transverse momentum broadening also depends on the underlying QCD mechanism
of multiple scatterings, as well as on $Q_s$, 
the typical virtuality of the scattering partons inside the nucleus.   
The larger $Q_s$ is, the broader the transverse momentum distribution gets.
However, our understanding of the fundamental QCD mechanism controlling color 
propagation and interaction inside a nuclear medium is still at an early stage. 
With its high energy and luminosity, better detector(s), and precise 
measurements of the transverse broadening, the EIC will 
enable rapid advance in our knowledge of color neutralization 
and multiple scattering of colored partons.

Figure~\ref{fig:broadening} shows the broadening of the transverse momentum spectrum 
of positive pions as a function of nuclear radius for various nuclei. 
Existing measurements from HERMES and CLAS are shown, as well as a calculation 
from Raufeisen~\cite{Raufeisen:2003zk} who has compiled and compared the results 
of various theoretical approaches to the transverse momentum broadening. 
In these approaches a linear dependence on the nuclear radius is obtained. 
As shown in Fig.~\ref{fig:broadening}, the HERMES data exhibit a linear dependence, 
while the CLAS data (for which $3.7<\nu <4.3$~GeV; $1.8\leq Q^2<4.2$~GeV$^2$; and 
$0.4<Z_h<0.5$) show a saturation of the broadening at large nuclear radii, 
which is likely related to the reduced lifetime of the colored virtual quark 
at the lower energies where the rescattering becomes weaker 
once the color of the fragmenting quark is neutralized. 
However, at the EIC with a much higher energy, more phase space opens up for radiation,
and a qualitatively different behavior is expected \cite{Liou:2013qya}.
As shown in Fig.~\ref{fig:broadening}, the points, labelled for EIC from \cite{Liou:2013qya},
predict a nonlinear increase of the broadening caused 
by a logarithmic enhancement of the medium induced radiation, which
contributes substantially to the broadening beyond the contribution from the elastic rescattering. 
In Fig.~\ref{fig:broadening}, error bars of the EIC data points result from a PYTHIA simulation 
for which $x>0.1$, $Q^2> 1$~GeV, $\langle z\rangle=0.41$ (matching the HERMES data), 
and $\langle W\rangle = 25$~GeV. The size of the scattering centers was taken to be that 
of a constituent quark for the purpose of this plot; scattering from a smaller sized object 
will logarithmically enhance the size of the effect.  
An integrated luminosity of $10~{\rm fb}^{-1}{\hskip -0.03in}/$A has been assumed, 
and systematic uncertainties similar to the statistical uncertainties have been employed; 
the two uncertainties are combined in quadrature. 

In semi-inclusive DIS, the uniquely determined leptonic plane 
plays the role of the reaction plane in relativistic heavy ion collisions, 
and helps define the azimuthal angle $\phi$ distribution of produced hadrons.  
Therefore, like the relativistic heavy-ion (A+A) collisions, 
the $\phi$-modulation of produced hadrons in SIDIS, or $v_n$,
is well-defined on an event-by-event basis.  
The non-uniform spatial distribution of the scattering centers (the parton densities)
inside a large nucleus could naturally generate a $\phi$-dependence 
of the transverse momentum broadening of the observed hadron,
which was observed by the CLAS collaboration at Jefferson
Lab \cite{Brooks:2011nu}.

With a large enough $Q$ to localize the production of a fast moving fragmenting
parton at the EIC, the strength of exotic  $\phi$-modulation of hadrons could shed 
light on the spatial fluctuation of parton densities inside a large colliding nucleus.
Within the one-photon-exchange approximation for unpolarized semi-inclusive DIS, 
the produced hadrons naturally have the $\cos(n\phi)$ modulation with $n=1,2$, 
due to the interference of two scattering amplitudes with the virtual photon
in different spin states.  Non-vanishing exotic modulation of transverse momentum broadening 
on an event-by-event basis, $\Delta \langle p_T^2(\phi)  \rangle_{AN} \propto \cos(n\phi)$ 
with $n$ other than 1 and 2, 
is a direct and clean evidence of spatial fluctuation of parton densities
in a large colliding nucleus \cite{PitonyakQiu:2014}.
In addition, the A-dependence of $\cos(\phi)$ and $\cos(2\phi)$ could also
shed light on the spatial fluctuation of parton densities in the nucleus.  
The EIC could provide an independent verification and study of spatial fluctuations of 
parton densities inside the colliding nuclei observed in relativistic heavy ion collisions,
and help us understand the initial condition of the collision to produce the QGP. 
\end{multicols}

%% file: files_tex/eA-connection.tex
\newpage
\setlength{\parskip}{0mm}
\section{Connections to $p$+A,  A+A  and Cosmic Ray  Physics}
\label{sec::connections}

{\large\it Conveners:\ Yuri Kovchegov and Thomas Ullrich}
\vskip 0.1in

\begin{figure}[tbh]
    \begin{center}
    \includegraphics[width=0.6\textwidth]{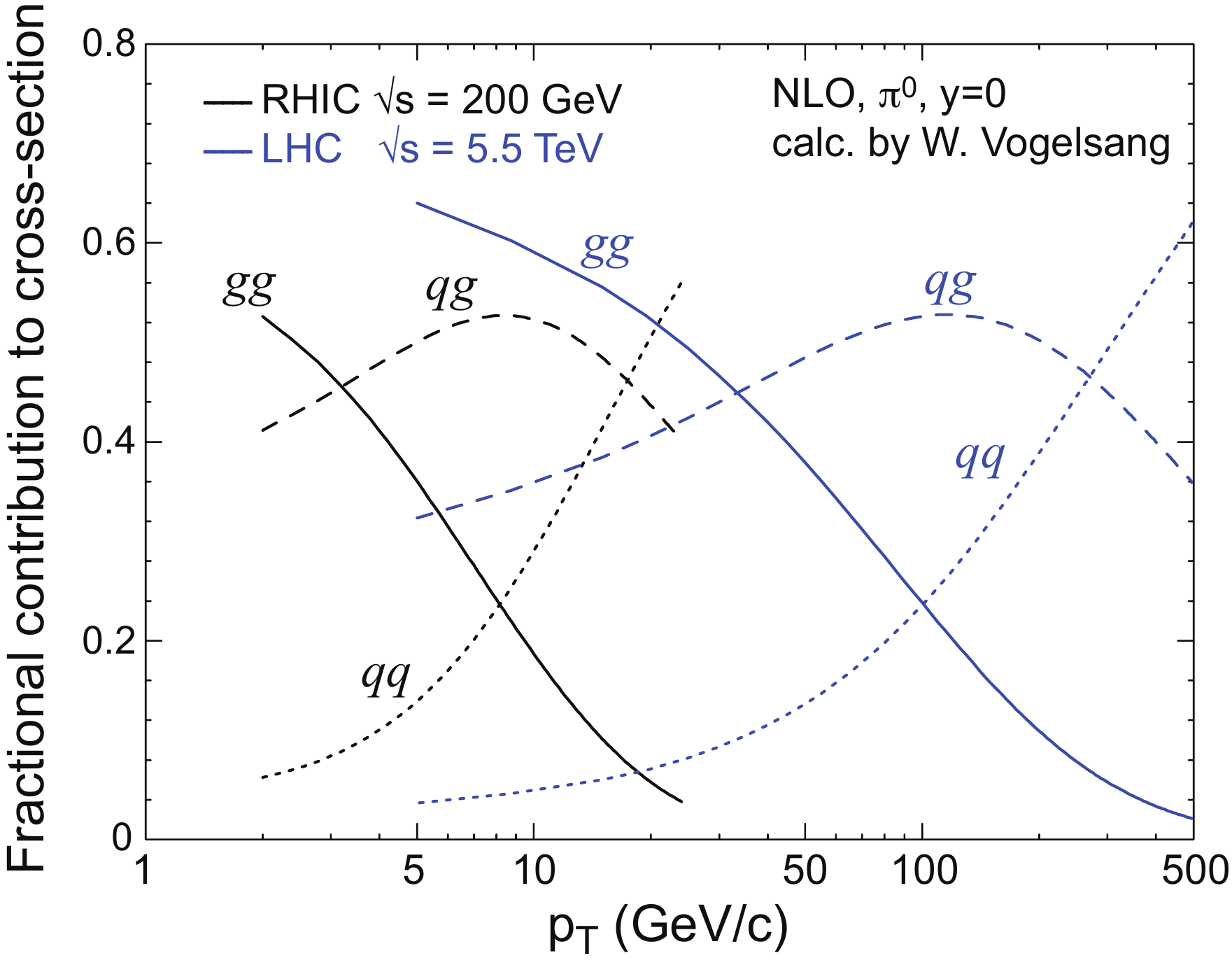}
    \end{center}
   \vskip -5mm
     \caption{Fractional contributions from $gg$, $qg$, and $qq$
      scattering processes to pion production at mid-rapidity $p$+$p$
      collisions at RHIC (black) and LHC (blue).}
    \label{fig::jetComposition}
\end{figure}

\subsection{Connections to $p$+A Physics}
\label{sec:pAConnections}

\begin{multicols}{2}
Both $p$+A and $e$+A collisions can provide excellent information on the
properties of gluons in the nuclear wave-functions. It is therefore
only logical to compare the strengths and weaknesses of the two
different programs in exploring the saturation regime.

In the beginning of the RHIC era, the $d$+Au program was perceived as
merely a useful baseline reference for the heavy-ion program. It very
soon turned out that, due to a wise choice of colliding energy, RHIC
probes the transition region to a new QCD regime of gluon
saturation. While only marginal hints of non-linear effects were
observed in DIS experiments at HERA \cite{Caola:2010cy}, it is fair to
say that very tantalizing hints for gluon saturation were observed in
$d$+Au collisions at RHIC
\cite{Adare:2011sc,Braidot:2010ig,Arsene:2004ux,Adler:2006wg,Adams:2006uz}.
In the $p$+A program at the LHC, these effects should be even more
pronounced as the data from forward rapidities become available. While
$p$+A and $p$+$p$ colliders provide superior access to the low-$x$
region, they also have some severe disadvantages that impede
systematic studies of the saturation phenomena that we will describe
below.

As shown in Fig.~\ref{fig::jetComposition}, in $p$+$p$ collisions
at mid-rapidity at RHIC and the LHC, the bulk of particles produced
originate from processes involving gluons.   This is a simple manifestation of
the dominance of gluons at low-$x$ in hadrons (see
Fig.~\ref{fig:hera_10GeV2}).  While it is unlikely that saturation
phenomena are observed at RHIC energies in $p$+$p$ collisions due to the
small values of $Q_s$ even at the lowest accessible $x$, the amplified
$Q_s$ scale in $p$+A collisions opens the experimentally accessible range
where saturation effects become detectable.
The relation between rapidity $y$ and transverse momentum $p_T$ of the
final state partons/particles with mass $m$ and their fractional
longitudinal momenta $x_{1,2}$ is $x_{1,2} = e^{\pm y} \, \sqrt{(p_T^2
  + m^2)/s}$. Hence, at mid-rapidity ($y=0$) at RHIC, only particle
production with very small \pT\ will be sensitive to the saturation
region in parton densities while at the LHC the region of transverse
momenta will be much larger. At RHIC, saturation effects are largely
absent at central rapidities but become measurable at large forward
rapidities (that is, for particles coming out close to the incoming
proton or deuteron direction with $y = 2 - 4$ corresponding to small
$x_2$).

\begin{figurehere}
    \begin{center}
    \vskip 0.4cm
    \includegraphics[width=0.46\textwidth]{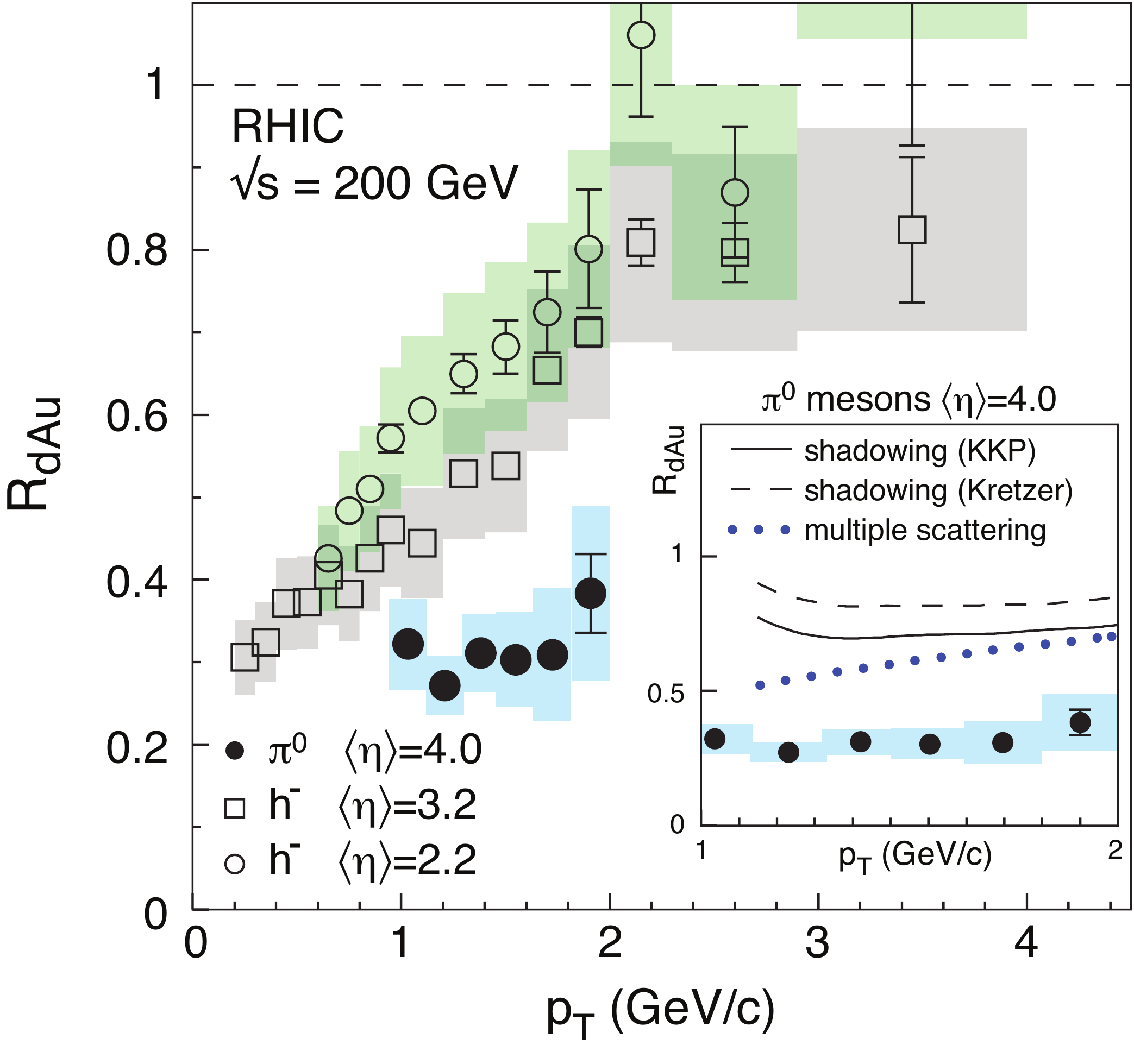}
    \end{center}
    \vskip -6mm
    \caption{The nuclear modification factor ($R_{dAu}$) versus \pT\ for
      minimum bias $d$+Au collisions measured at RHIC. The solid
      circles are for $\pi^0$ mesons \cite{Adams:2006uz}, the open
      circles and boxes are for negative hadrons
      \cite{Arsene:2004ux}. The error bars are statistical, the shaded
      boxes are point-to-point systematic errors. (Inset) $R_{dAu}$
      for $\pi^0$ mesons compared with pQCD calculations based on
      collinear factorization.  Note that none of the curves can describe the data.}
    \label{fig:hforwardRHIC}
\end{figurehere}

\vspace{0.5cm} First hints for the onset of saturation in $d$+Au
collisions at RHIC have been observed by studying the rapidity
dependence of the nuclear modification factor, $R_{dAu}$,
as a function of \pT\ for charged hadrons \cite{Arsene:2004ux} and
$\pi^0$ mesons \cite{Adams:2006uz}, and more recently through
forward-forward hadron-hadron correlations
\cite{Braidot:2010ig,Adare:2011sc}. 

\begin{figurehere}
    \begin{center}
   \includegraphics[width=0.48\textwidth]{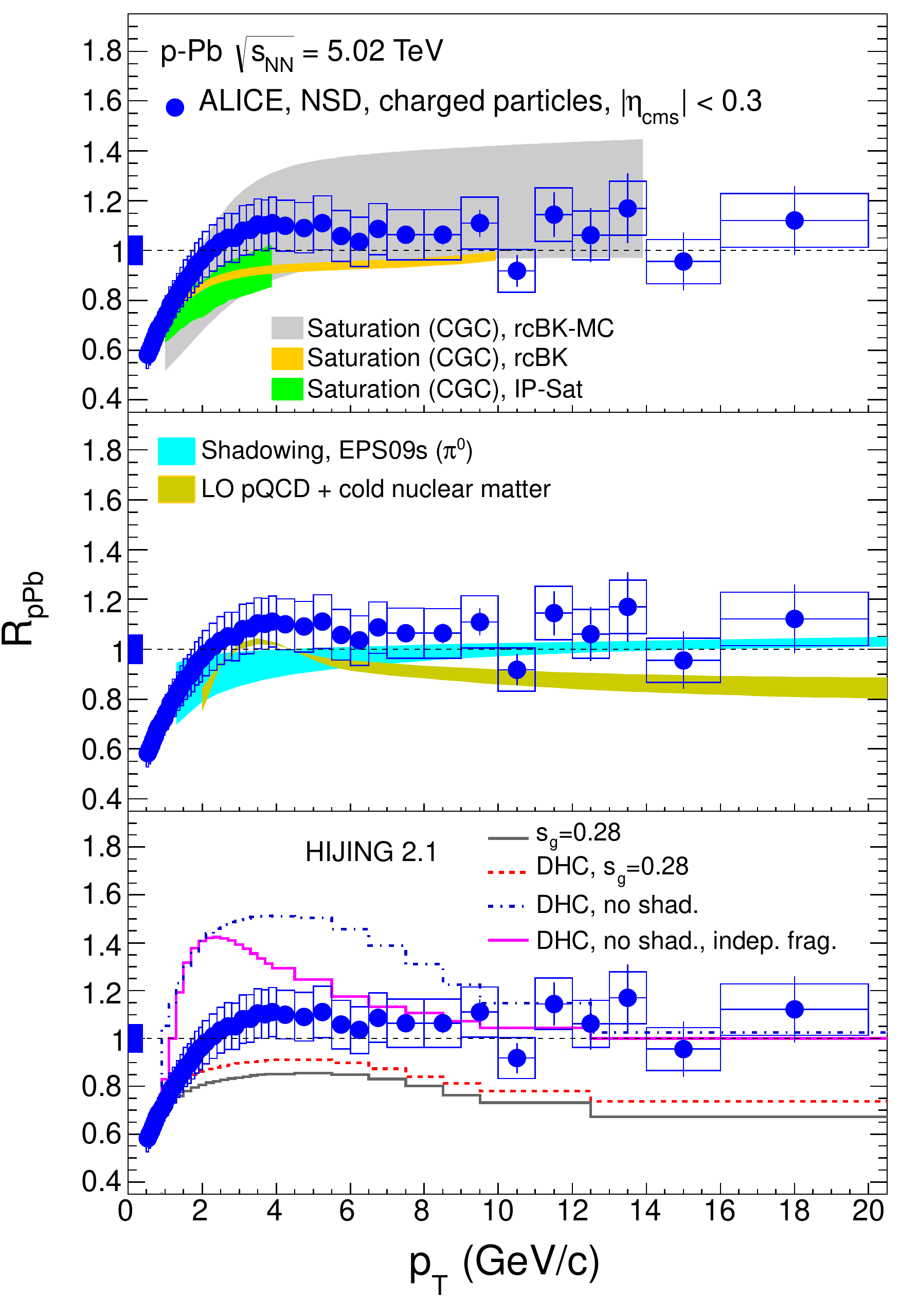}
    \end{center}
    \vskip -0.8cm
    \caption{The nuclear modification factor ($R_{pPb}$) versus \pT\
      for charged particles produced in $p$+Pb collisions at LHC
      \cite{ALICE:2012mj} compared to various theoretical models.}
    \label{fig:RpPb_LHC}
\end{figurehere}
\vspace{5mm}

\begin{figure*}[!bth]
\begin{center}
  \includegraphics[width=\textwidth]{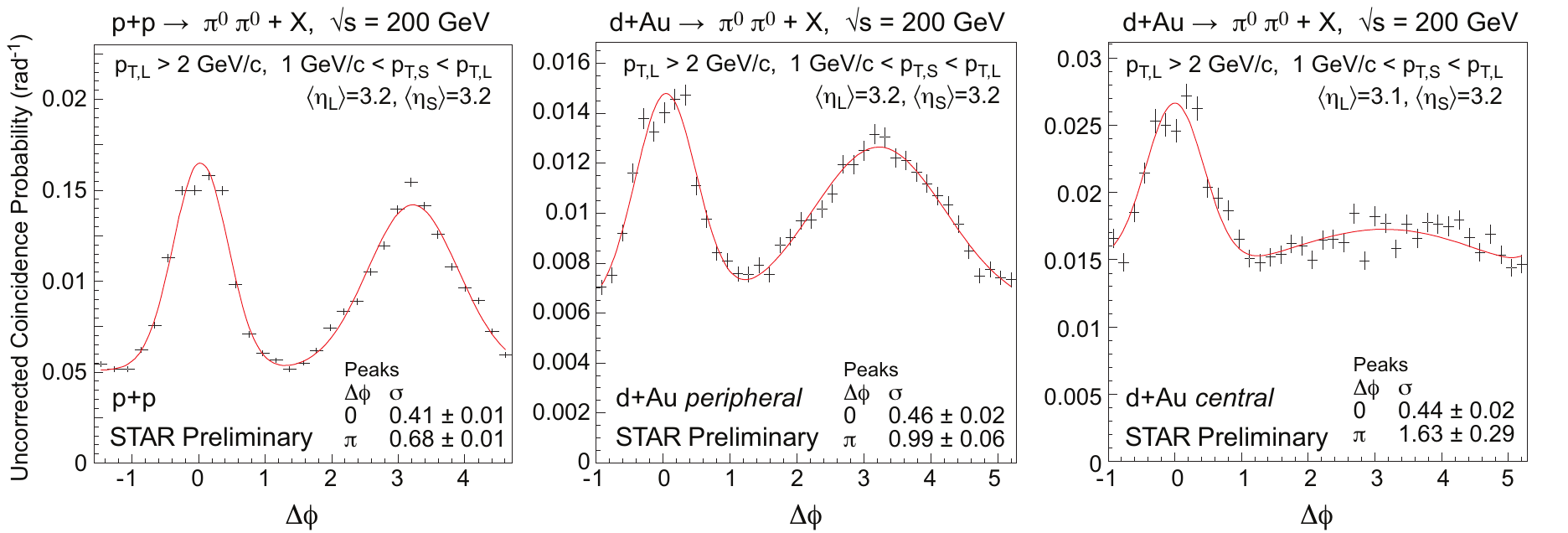}
\end{center}
\vskip -7mm
  \caption{Di-hadron correlations measured at forward rapidities at RHIC: 
        the uncorrected coincidence signal plotted versus the azimuthal angle
        difference between two forward neutral pions in $p$+$p$
        collisions (left) compared to peripheral (center) and
        central $d$+Au collisions (right) \cite{Braidot:2010ig}. Data
        are shown with statistical errors and fit with a constant plus
        two Gaussian functions (in red).}
  \label{fig:pAhhRHIC}
\end{figure*}

The nuclear modification factor for a $p$+A\ collision is defined by
\begin{align}
  \label{eq:RpA}
  R_{pA} = \frac{1}{N_{coll}} \, \frac{d N_{pA}/d^2 p_T \, dy}{d
    N_{pp}/d^2 p_T \, dy},
\end{align}
where $d N/d^2 p_T \, dy$ is the produced hadron multiplicity in a
given region of phase space while $N_{coll}$ is the number of 
binary nucleon--nucleon collisions. The nuclear modification factor
$R_{pA}$ is equal to $1$ in the absence of collective nuclear effects.
Figure \ref{fig:hforwardRHIC} shows $R_{dAu}$ versus
\pT\ for minimum bias $d$+Au collisions for charged hadrons measured by
the BRAHMS experiment \cite{Arsene:2004ux} and $\pi^0$ mesons by STAR
\cite{Adams:2006uz}.  While the inclusive yields of hadrons ($\pi^0$
mesons) at \sqrts=200 GeV in $p$+$p$ collisions generally agree with
pQCD calculations based on DGLAP evolution and collinear
factorization, in $d$+Au collisions, the yield per binary collision is
suppressed with increasing $\eta$, decreasing to $\sim$30\% of the
$p$+$p$ yield at $\langle \eta \rangle = 4$, well below shadowing and
multiple scattering expectations. The \pT\ dependence of the $d$+Au
yield is found to be consistent with the gluon saturation picture of
the Au nucleus (e.g., CGC model calculations \cite{Dumitru:2005gt,
  Albacete:2010rh,Kharzeev:2002pc,wiedemann,Kharzeev:2003wz,Kharzeev:2004yx})
although other interpretations cannot be ruled out based on this
observable alone \cite{Qiu:2003vd,Guzey:2004zp,Kopeliovich:2005ym}.

Recent result from the $p$+Pb scattering experiments at the LHC appear
to confirm this picture. Figure~\ref{fig:RpPb_LHC} depicts the data
for $R_{pPb}$ of charged particles reported by ALICE collaboration as
compared to different theoretical models. The data is the same in all
three sub-panels. Saturation models, whose predictions are depicted in
the top panel of Figure~\ref{fig:RpPb_LHC}, do a good job in
describing the data, though other models' predictions, most notably
that of EPS09 shown in the middle panel, also describe the data well.

A more powerful technique than single inclusive measurements is the use of 
two-particle azimuthal correlations, as discussed in Section
\ref{sec:dihadrons}.  In collinear factorization-based pQCD at leading
order, particle production in high-energy hadronic interactions
results from the elastic scattering of two partons ($2 \rightarrow 2$
scattering) leading to back-to-back jets. When high-\pT\ hadrons are
used as jet surrogates, we expect the azimuthal correlations of hadron
pairs to show a peak at $\Delta \phi = 0$, and a `back-to-back' peak
at $\pi$. When the gluon density increases, the basic dynamics for the
particle production is expected to change. Instead of elastic $2
\rightarrow 2$ scattering, particle production can proceed by the
interaction of a probe parton from the proton (deuteron) beam with
multiple gluons from the heavy-ion beam. At sufficiently high gluon
densities, the transverse momentum from the fragments of the probing
parton may be compensated by several gluons with lower \pT.  Two
particle azimuthal correlations are expected to show a broadening of
the back-to-back peak (loss of correlation: $2 \rightarrow$ many
processes) and eventually to disappear. In the CGC framework, the
hadronic wave-function is saturated as a consequence of gluon
recombination. At very low values of the $x$ of the probed gluons, 
the occupation numbers become large
and the probe scatters {\em coherently} off the dense gluon field of
the target, which recoils collectively, leading to a modification in
$\Delta \phi$ \cite{Kharzeev:2004bw}.

Figure \ref{fig:pAhhRHIC} shows the (efficiency uncorrected)
probability to find an associated $\pi^0$ given a trigger $\pi^0$,
both in the forward region measured by the STAR detector.  
The coincidence signal versus azimuthal angle difference between the
two pions in $p$+$p$ collisions (left) compared to peripheral (middle)
and central $d$+Au collisions (right) is shown~\cite{Braidot:2010ig} (see also
\fig{fig:dihadronJ} for a similar measurement and the related discussion, along with
\cite{Adare:2011sc}).  All the distributions present two signal
components, surmounting a constant background representing the
underlying event contribution (larger in $d$+Au). The near-side peak
represents the contribution from pairs of pions belonging to the same
jet. It is not expected to be affected by saturation effects. The
away-side peak represents the back-to-back contribution to the
coincidence probability, which should disappear in going from $p$+$p$ to
$d$+Au if saturation sets in \cite{Kharzeev:2004bw}. The data show that
the width of the near-side peak remains nearly unchanged from $p$+$p$ to
$d$+Au, and particularly from peripheral to central $d$+Au
collisions. Central $d$+Au collisions show a substantially reduced away
side peak that is significantly broadened.  Again, pQCD calculations
based on linear DGLAP evolution without coherent multiple scattering
fail to describe this observation, while those including non-linear
effects describe the data considerably well \cite{Albacete:2010rh, Marquet:2007vb,
  Albacete:2010pg}.  This measurement represents the 
strongest hint yet for saturation phenomena and also indicates
that the kinematic range of the EIC is well suited to explore
saturation physics with great precision.

One of the most important results from the LHC $p$+Pb program is the
observation of the 'ridge' correlation in high-multiplicity $p$+Pb
collisions. 
The 'ridge' is a di-hadron correlation which is very broad in rapidity
($\Delta \eta$) and very narrow in the azimuthal angle ($\Delta
\phi$). The near-side 'ridge' ($\Delta \phi \approx 0$) was originally
discovered in heavy ion collisions at RHIC
\cite{Adams:2005ph,Adare:2008ae,Alver:2009id,Abelev:2009af}. With the
advent of the LHC experimental program, it was also seen in
high-multiplicity $p$+$p$ \cite{Khachatryan:2010gv} and $p$+Pb
\cite{CMS:2012qk,Abelev:2012ola,Chatrchyan:2013nka} collisions. A
simple causality argument \cite{Dumitru:2008wn} indicates that the
long-range rapidity correlation in the 'ridge' is due to dynamics in
the early stages of the collisions, and hence may possibly be due to
the saturation effects. While saturation effects may also explain the
narrow azimuthal structure of the 'ridge' correlation
\cite{Dumitru:2008wn}, in heavy ion collisions the azimuthal shape of
the correlation is likely to be strongly affected by the final-state
QGP effects.

\begin{figurehere}
    \begin{center}
    \vskip 0.4cm
    \includegraphics[width=0.48\textwidth]{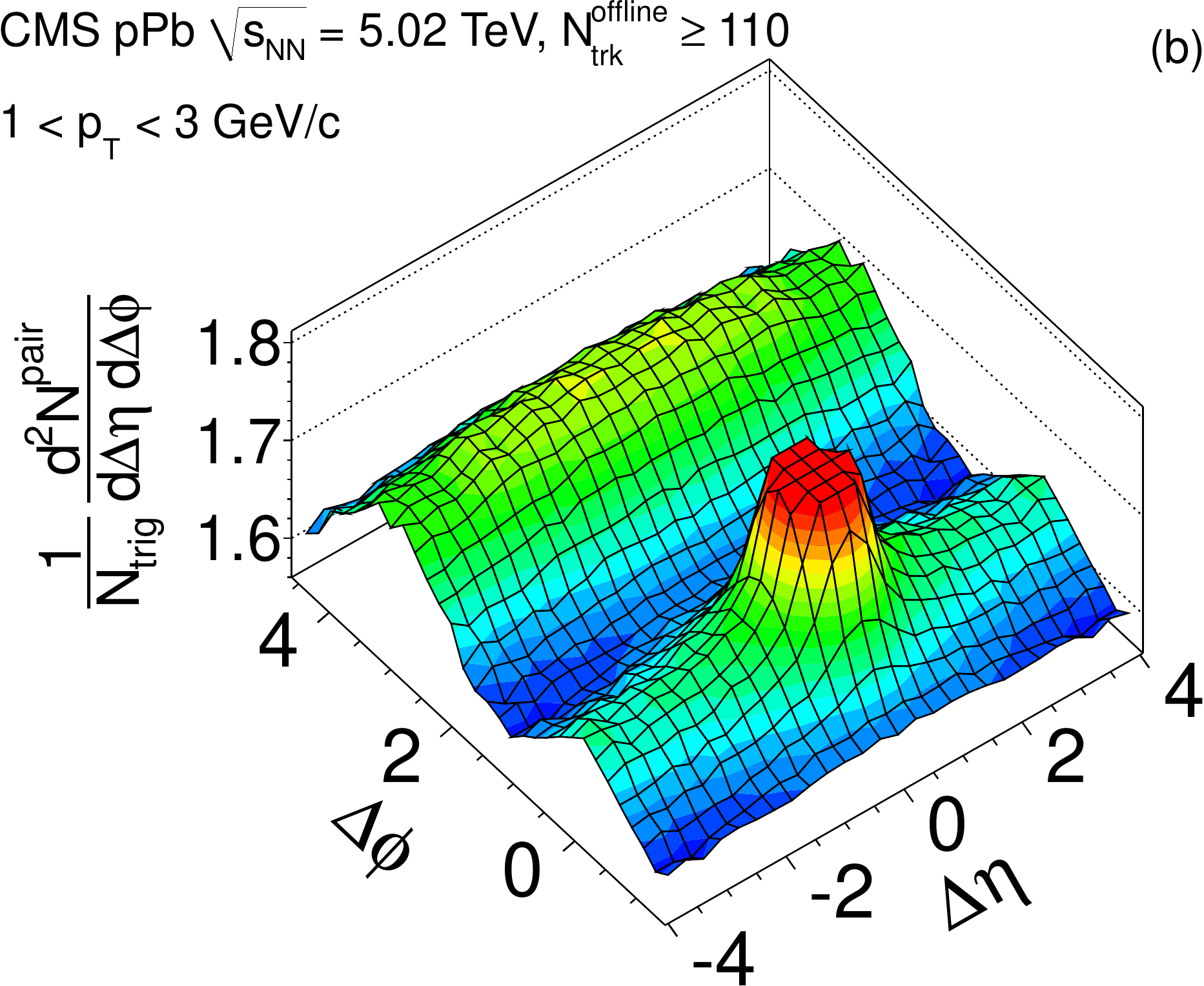}
    \end{center}
    \vskip -0.4cm
    \caption{The two-particle correlation function in
      high-multiplicity $p$+Pb collisions as a function of $\Delta
      \phi$ and $\Delta \eta$ reported by the CMS collaboration
      \cite{CMS:2012qk}. The 'ridge' structure is seen as a
      correlation near $\Delta \phi = 0$ stretching over many units of
      rapidity $\Delta \eta$.}
    \label{fig:ridge_pPb_CMS}
\end{figurehere}

~\vspace{0.5cm} 

The dynamical origin of the azimuthal shape of the 'ridge' in $p$+$p$
and $p$+Pb collisions is less clear, since collective QGP effects are
usually not expected in such systems. A quadrupole azimuthal
anisotropy with symmetric peaks at $\Delta\phi \approx 0$ and
$\Delta\phi \approx \pi$ was predicted in
\cite{Dumitru:2008wn,Dumitru:2010iy,Dusling:2012wy,Kovner:2012jm,Kovchegov:2012nd}
based on CGC physics and was experimentally confirmed in
\cite{Abelev:2012ola}, suggesting a saturation origin of the
correlation. A quantitative comparison of CGC theory to data was made
in \cite{Dusling:2013oia}. However, at the moment all saturation-based
explanations of the 'ridge' appear to predict the correlation function
which is expandable into a Fourier series over even cosine harmonics,
$\cos (2 \, n \, \Delta \phi)$, whereas the data presents clear
evidence of odd harmonics as well, even in high-multiplicity $p$+Pb
collisions \cite{Chatrchyan:2013nka}, similar to those generated by
hydrodynamic expansion of QGP in $A$+$A$ collisions on top of the
event-by-event fluctuations in the initial conditions (see
Sec.~\ref{AAconn_sec} below). At the same time, preliminary
measurements of the system size \cite{Abelev:2014pja} appear to be in
contradiction with the hydrodynamic interpretation. Clearly the jury
is still out regarding the origin of the 'ridge' in high-multiplicity
$p$+$p$ and $p$+Pb collisions: it is possible that both CGC and
hydrodynamic effects are at play. While LHC can reach down to very low
values of $x$, it is possible that the ability to study small-$x$
physics and saturation at LHC $p$+Pb experiments is somewhat blunted
by the final-state interactions. The final-state effects should not be
present at an $e$+A collider, which should allow for a cleaner probe
of low-$x$ dynamics.


Although the results of the $d$+Au program at RHIC and the $p$+Pb program
at LHC show tantalizing evidence of saturation phenomena, alternative
explanations for each of the {\it individual} observations exist.  The
unambiguous ultimate proof of existence of saturation can only come
from an $e$+A collider.  While in $e$+A collisions the probe (the
electron) is point-like and structureless, in $p$+A collisions, one
has to deal with a probe whose structure is almost as complex as that
of the target nucleus to be studied.  The EIC's usefulness as a gluon
``microscope'' is somewhat counterintuitive since electrons do not
directly interact with gluons.  However, the presence and dynamics of
the gluons in the ion will modify the precisely understood
electromagnetic interaction of the electron with quarks in ways that
allow us to infer the gluon properties. Deeply inelastic $e$+A
collisions are dominated by one photon exchange (see the Sidebar on
page~\pageref{sdbar:DIS}).
The photon could interact with one parton to probe parton
distributions, as well as multiple partons coherently to probe
multi-parton quantum correlations.  
One	of the major advantages of DIS
is that it allows for the direct, model-independent, determination of
the momentum fraction $x$ carried by the struck parton before the
scattering and $Q^2$, the momentum transferred to the parton in the
scattering process. Only the control of these variables ultimately
will allow us a precise mapping of the gluon distributions and their
dynamics.


One may wonder whether physics similar to what one can probe at an
EIC could be studied in the Drell-Yan process in a $p$+A collider. Due
to crossing symmetry, the Drell-Yan process can be related to DIS
\cite{Kopeliovich:1998nw} with the invariant mass of the di-lepton
pair $M^2$ playing the role of $Q^2$. Owing to the very broad reach in
$x$ and $M^2$, $p$+A collisions at RHIC and even more so at the LHC
clearly have significant discovery potential for the physics of
strong color fields in QCD. However, the di-lepton signal in $p$+A is
contaminated by the leptons resulting from decays of heavy-flavor
hadrons, such as $J/\psi$, up to a rather large invariant masses of
$M^2 = 16$~GeV$^2$ and even beyond \cite{Accardi:2004be}. This
contamination does not allow one to cleanly probe the saturation
region of $M^2 < 16$~GeV$^2$. To avoid hadronic decay background one
may study large values of the net transverse momentum $p_T$ of the
pair.  However, this would also push one away from the lower-$p_T$
saturation region.

Ultimately it will be the combination of strong $p$+A and $e$+A
programs, each providing complementary measurements, that will answer the
questions raised above in full.
\end{multicols}


\newpage

\subsection{Connections to Ultra-relativistic Heavy-Ion Physics}
\label{AAconn_sec}

\begin{multicols}{2}
Measurements over the last decade in heavy-ion collision
experiments at RHIC indicate the formation of a strongly coupled
plasma of quarks and gluons (sQGP). Striking results include: (i) the
strong collective flow of all mesons and baryons, and especially that 
of heavy charm quarks, and (ii) the opaqueness of the hot and dense 
medium to hadron jets up to $p_\perp \sim 20$ GeV.  

This sQGP appears to behave like a ``near-perfect fluid" with a ratio
of the shear viscosity to entropy density, $\eta/s$, approaching zero
\cite{Adcox:2004mh,Adams:2005dq,Back:2004je,Arsene:2004fa,Gyulassy:2004zy}.
Recent experiments at the LHC, with substantially higher energies and
thus a hotter and longer lived plasma phase, confirm this picture
\cite{Tserruya:2011dy}.

Despite the significant insight that the QGP is a strongly correlated
nearly perfect fluid, little is understood about how the QGP is
created and what its properties are.
Qualitative questions that the heavy ion community would like to answer include how the
dynamics of gluons in the nuclear wave-functions generates entropy
after the collision, what the properties and dynamics of the
pre-equilibrium state are, why the thermalization of the system occurs
rapidly, and whether the system is fully or only partially thermalized
during its evolution.
Furthermore, though it is widely accepted that the QGP medium is a
strongly correlated one, it is less clear whether the coupling is weak
or strong. In the weak-coupling scenario, the strongly correlated
dynamics are generated by the scales that characterize the electric and
magnetic sectors of the hot fluid.  In the strongly-coupled
scenario, progress has been made by exploiting the Anti-de Sitter
space/Conformal Field Theory (AdS/CFT) correspondence
\cite{Maldacena:1997re,Aharony:1999ti} of weakly coupled gravity
(which is calculable) to strongly coupled supersymmetric Yang-Mills
theory with many features in common with QCD.

Quantitative questions the heavy-ion community would like to answer
include determining the shear viscosity of the medium averaged over
its evolution, measuring the values of other transport coefficients
such as the bulk viscosity and the heavy quark diffusion coefficient,
and perhaps most importantly, identifying the equation of state of
finite-temperature QCD medium. Some of these questions can be
addressed in numerical lattice QCD computations. It is still not
entirely clear how these results can be cross-checked and improved
upon in the environment of a rapidly evolving and incredibly complex
heavy ion event.

Despite the significant progress achieved in the qualitative understanding of
several aspects of this matter, there is still no comprehensive
quantitative framework to understand all the stages in the creation
and expansion of the hot and dense QGP medium.  In the following we
outline how an EIC can contribute to a better understanding of the
dynamics of heavy ion collisions, from the initial formation of
bulk partonic matter to jet quenching and hadronization that probe
the properties of the sQGP.

\subsubsection{Initial Conditions in A+A collisions}

Understanding the dynamical mechanisms that generate the large flow in
heavy ion collisions is one of the outstanding issues in the RHIC
program.
Hydrodynamic modeling of RHIC data is consistent with the system
rapidly thermalizing at times of around 1 -- 2 \fmc\ after the initial
impact of the two nuclei
\cite{Kolb:2003dz,Teaney:2000cw,Luzum:2008cw}.  These hydrodynamic
models are very sensitive to the initial pre-equilibrium properties of
the matter formed immediately after the collision of the two nuclei.

Our current understanding, based on the CGC framework, suggests that the
wave-functions of the nuclei, due to their large occupancy, can be
described as classical fields, as was explained above.  Therefore, 
at the leading order, the collision can be approximated by the collision of
``shock waves" of classical gluon fields
~\cite{Kovner:1995ja,Krasnitz:1998ns} resulting in the production of
non-equilibrium gluonic matter.  It is generally believed that the
instability and consequent exponential growth of these intense gluon
fields would be the origin of early thermalization
\cite{Mrowczynski:1993qm,Arnold:2003rq}, though the exact mechanism
for the process is not completely understood. Alternatively, within
the strong coupling paradigm, thermalization in heavy ion collisions
is achieved very rapidly, and there has been considerable recent work
in this direction~\cite{Kovchegov:2011ub,Chesler:2010bi}.

The properties of the nuclear wave-functions will be studied in great
detail in $e$+A collisions. They promise a better understanding of the
initial state and its evolution into the sQGP.  Specifically, the
saturation scale $Q_s$, which can be independently extracted in $e$+A
collisions, sets the scale for the formation and thermalization of
strong gluon fields. Saturation effects of these low-$x$ gluon fields
affect the early evolution of the pre-QGP system in heavy-ion
collisions. Their spatial distribution governs the eccentricity of the
collision volume and this affects our understanding of collective flow
and its interpretation profoundly. However, the features of these
gluon fields --- their momentum and spatial distributions at energies
relevant for RHIC --- are only vaguely known. More detailed information of
relevance to the properties of the initial state (such as the spatial
distributions of gluons and sea quarks) and thereby improved
quantitative comparisons to heavy ion data, can be attained with an
EIC.

The high-energy wave-functions of nuclei can be viewed as coherent
superpositions of quantum states that are ``frozen" configurations of
large numbers of primarily gluons. How these states decohere, produce
entropy and subsequently interact is clearly essential to a deep
understanding of high-energy heavy ion collisions. Remarkably, models
based on the CGC framework manage to describe particle production in
A+A collisions over a broad range of energies and centralities
extraordinary well.  These models are constrained by HERA inclusive
and diffractive DIS data on \ep\ collisions, and the limited fixed
target $e$+A DIS data available. One such model is the IP-Sat
model~\cite{Bartels:2002cj,Kowalski:2003hm} (Model-I from
Sec.~\ref{sec_map})
Another saturation model (Model-II from Sec.~\ref{sec_map} ) is based
on BK nonlinear evolution including the running-coupling corrections
(rcBK)
\cite{Balitsky:2006wa,Kovchegov:2006vj,Gardi:2006rp,Albacete:2007yr}
and the impact parameter independence \cite{ALbacete:2010ad}. 

\begin{figurehere}
    \centerline{\includegraphics[width=0.4\textwidth]{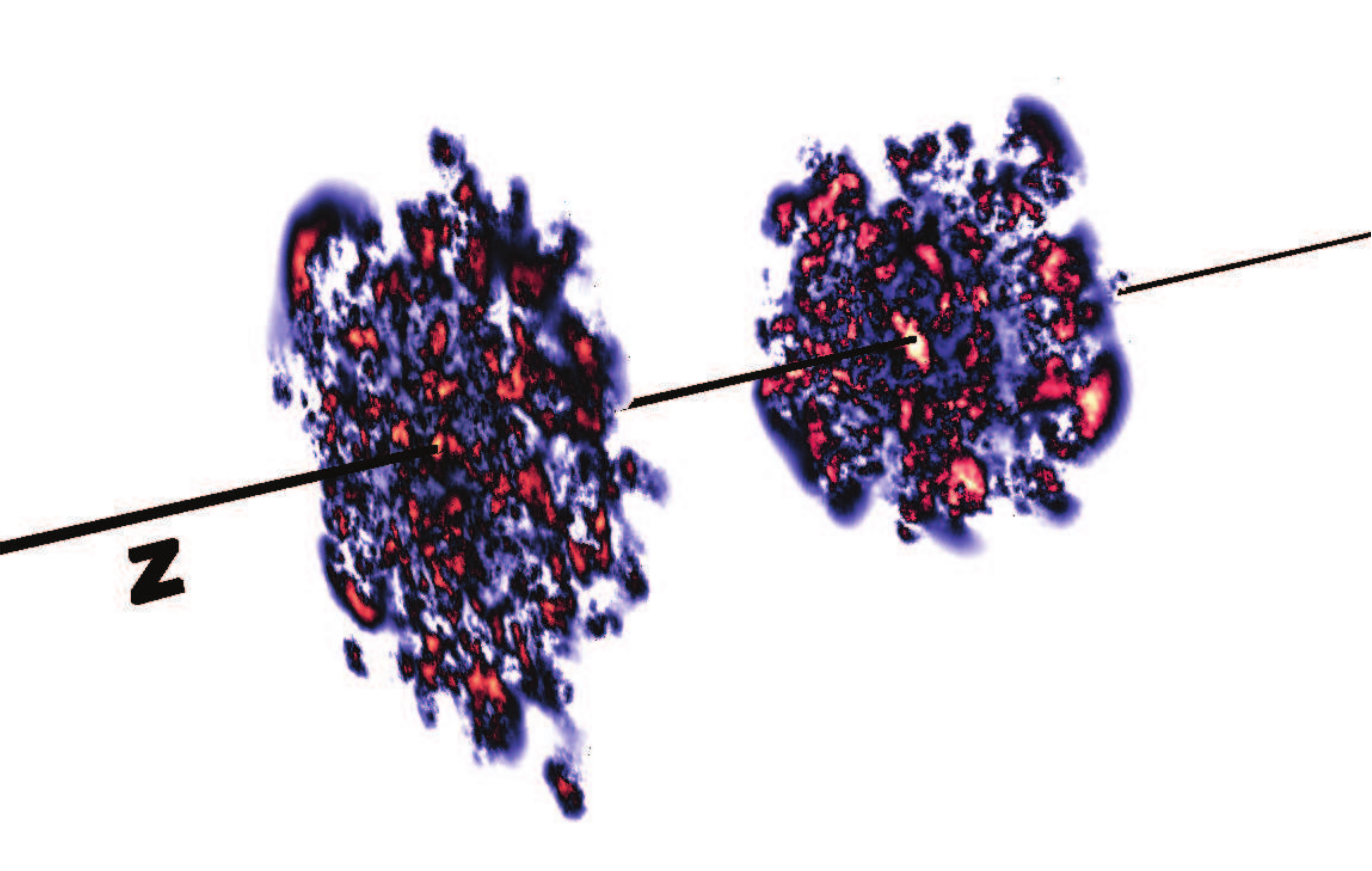}}
        \vskip -0.4cm
    \caption{The spatial distribution of gluon fields of the incoming
      nuclei for a collision of lead ions at $\sqrt{s}=2760$ GeV. The
      colors -- from blue to red -- denote increasing strength of
      gluon correlations.}
    \label{fig:VdagVLHC}
\end{figurehere}
\vspace{0.5cm}

\begin{figure*}[htb]
    \centerline{\includegraphics[width=\textwidth]{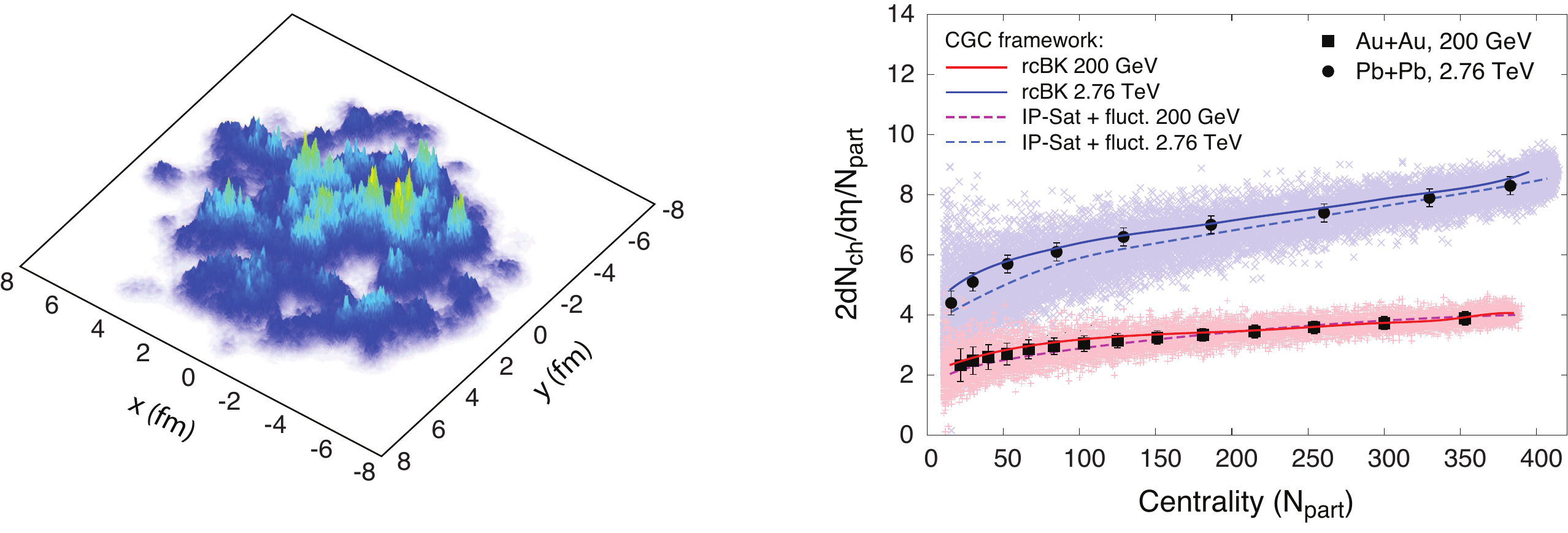}}
    \caption{Left panel: The spatial variation of the energy density in a
      single heavy ion event (based on IP-Sat model with
      fluctuations). The variations occur on distance scales $1/Q_s$,
      much smaller than the nucleon size.  Right panel: The centrality
      dependence of the multiplicity at $\sqrt{s_{_{NN}}} = 200 \gev$
      and $2760 \gev$.  (\AuAu\ \cite{Back:2002uc} data from RHIC,
      \PbPb\ data \cite{Aamodt:2010pb} from LHC.). The experimental
      data are compared to results from two model realizations in the
      CGC framework.  Solid curves represent the results from
      $k_T$-factorization with running-coupling BK unintegrated gluon
      distributions~\cite{ALbacete:2010ad} (Model-II from before)
      while dashed curves represent the result in the IP-Sat model
      with fluctuations~\cite{Schenke:2012hg} (Model-I).  The pale
      blue (LHC) and pink bands (RHIC) denote the referring range of
      event-by-event values of the single inclusive multiplicity.}
    \label{fig:rcBKAAmult}
\end{figure*}

The IP-Sat model can be used to construct nucleon color charge
distributions event-by-event.  Convoluting this with Woods-Saxon
distributions of nucleons enables one to construct Lorentz contracted
two dimensional nuclear color charge distributions of the incoming
nuclei event-by-event. Such a nuclear color charge density profile is
shown in Fig.~\ref{fig:VdagVLHC} for a heavy ion collision. 
The scale of transverse event-by-event fluctuations in
Fig.~\ref{fig:VdagVLHC} is $1/Q_s$, not the nucleon size.  The
resulting model~\cite{Schenke:2012wb,Schenke:2012hg} employs the
fluctuating gluon fields generated by the IP-Sat model to study the
event-by-event evolution of gluon fields.  Here, the corresponding
energy density distributions vary on the scale $1/Q_s$ and are
therefore highly localized (as shown in the left panel of
Fig.~\ref{fig:rcBKAAmult}).
 
The right panel in Fig.~\ref{fig:rcBKAAmult} shows data for the
centrality dependence of charged particle production for heavy-ion
collisions at $\sqrt{s_{_{NN}}} = 200 \ \gev$ and $2760 \ \gev$
compared to both Model-I (IP-Sat with fluctuations) and Model-II
(based on rcBK evolution).  Both models do an excellent job of
describing the data (note that Model-II is a prediction). The pale
bands shown in this figure are the event-by-event fluctuations of the
multiplicity in the Model-I.  
The successful descriptions of the
energy and centrality dependence of multiplicity distributions at RHIC
and the LHC are strong indications that
the CGC provides the right framework for entropy production.
Therefore, a fuller understanding of the small $x$ formalism promises
to enable us to separate these initial state effects from final state
entropy production during the thermalization process and thereby
constrain mechanisms (by their centrality and energy dependence) that
accomplish this.

\begin{figure*}[htb]
\centering
\includegraphics[width=0.8\textwidth]{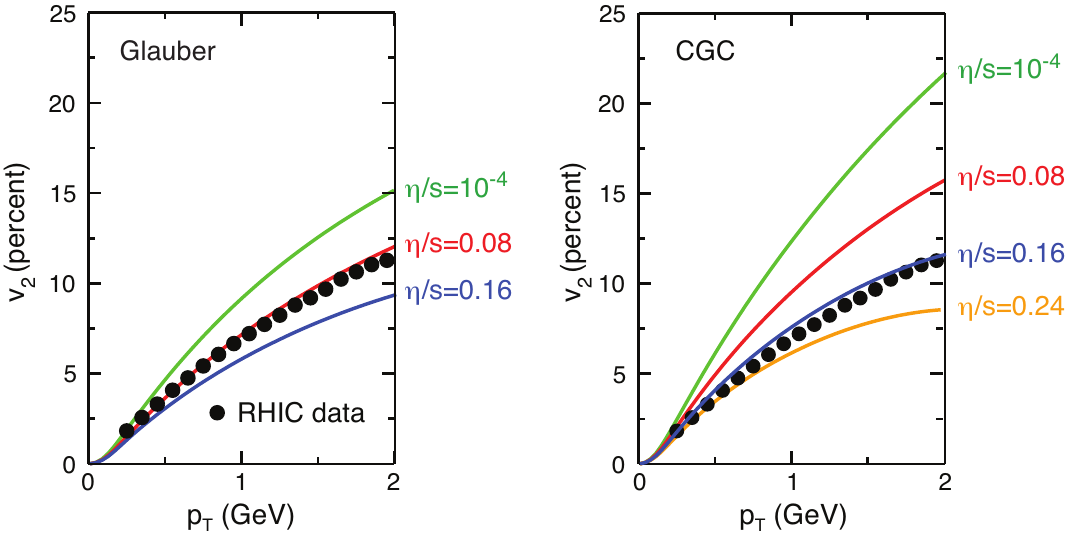}
    \vskip -0.4cm
\caption{ A comparison of data and theoretical predictions using viscous
  relativistic hydrodynamics for 
  $v_2^h(p_T)$ (right) with Glauber-like initial conditions (left) or a simplified
  implementation of CGC physics (KLN) model (right). Figures adapted from \cite{Luzum:2008cw}.}
\label{fig:v2Viscosity}
\end{figure*}

The other bulk quantity very sensitive to the properties of the
initial state is the collective flow generated in heavy ion
collisions. A useful way to characterize flow~\cite{Ollitrault:1992bk}
is through measured harmonic flow coefficients $v_n$, defined through
the expansion of the azimuthal particle distribution as
\begin{equation}
  \frac{dN}{d\phi} = \frac{N}{2\pi} \left(1 + \sum_n  2 \, v_n \, 
    \cos(n {\tilde \phi})\right)\,,
\end{equation}
where $v_n(p_T) = \langle cos(n {\tilde \phi})\rangle$, with
$\langle\cdots\rangle$ denoting an average over particles in a given
$p_T$ window and over events in a given centrality class, and ${\tilde
  \phi}= \phi-\psi_n$ with the event plane angle $\psi_n =
\frac{1}{n}\arctan\frac{\langle \sin (n\phi)\rangle}{\langle
  \cos(n\phi)\rangle}$.  Spatial eccentricities, extant at the instant
a hydrodynamic flow description becomes applicable, are defined {\it
  e.g.} for the second harmonic as
$\varepsilon_2 = \langle y^{\,2}{-}x^{\,2}\rangle/ \langle
y^{\,2}{+}x^{\,2}\rangle$, where now $\langle \cdots \rangle$ is the
energy density-weighted average in the transverse $x$-$y$ plane.
These are in turn converted to momentum space anisotropies by
hydrodynamic flow.  How efficiently this is done is a measure of the
transport properties of the strongly coupled QCD matter such as the
shear and bulk viscosities. Early flow studies focused on the second
flow harmonic coefficient $v_2$, which is very large at RHIC and the LHC,
and particularly sensitive to the ratio of the shear viscosity to
entropy density ratio $\eta/s$.  In Fig.~\ref{fig:v2Viscosity}, we
show $v_2$ for a Glauber model used in hydrodynamic simulations (left)
and the Kharzeev-Levin-Nardi CGC (KLN-CGC) model
\cite{Kharzeev:2002ei} (right).  The eccentricity $\varepsilon_2$ in
the Glauber model has a weaker dependence on collision centrality
relative to the KLN-CGC model, and therefore requires a lower $\eta/s$
to fit the data. The value of $\eta/s$ for the Glauber eccentricity in
this model study is equal to $1/4\pi$ in natural
units~\cite{Policastro:2001yc,Kovtun:2004de} conjectured to be a
universal bound for strongly-coupled liquids based on applications of
the Anti-de Sitter space/Conformal Field Theory (AdS/CFT)
correspondence~\cite{Maldacena:1997re,Aharony:1999ti}.  The KLN-CGC
value on the other hand gives a number that's twice as large as this
prediction.

Experimental and theoretical developments can help settle what is the
true value of $\eta/s$, and in particular, potentially provide essential
information on its temperature dependence.
Interestingly, the effect of $\eta/s$ on each of the $v_n$ harmonics
is different. This is shown strikingly in comparisons of results for
the $v_n$ moments from event-by-event viscous hydrodynamic simulations
relative to equivalent ideal hydrodynamic simulations in
Fig.~\ref{fig:flowHarmonics}. 

The figure shows the ratio of viscous to
ideal moments~\cite{Schenke:2011bn} for $n=2,\cdots, 5$ for the
previously discussed values of $\eta/s$. The damping of the higher
moments $v_n$ is quite dramatic with increasing $\eta/s$. 

\begin{figurehere}
    \vskip 0.4cm 
   \centering
    \includegraphics[width=0.48\textwidth]{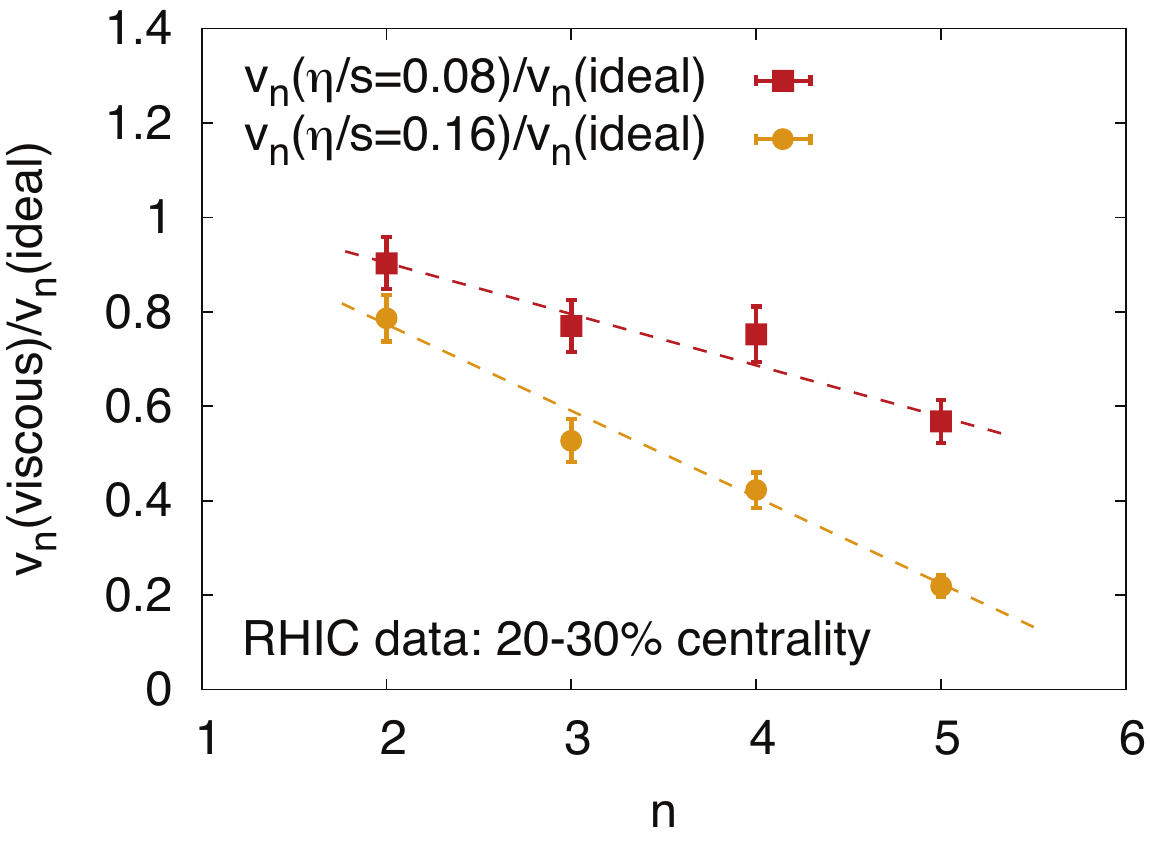}
    \vskip -0.4cm 
    \caption{The ratio of charged hadron flow $v_n$ harmonics from viscous
      hydrodynamic simulations and from ideal hydrodynamics
      \cite{Schenke:2011bn} (lines only to guide the eye).  The ratio
      is shown for two different values of $\eta/s$. Higher harmonics
      are substantially more affected by shear viscosity than
      $v_2$.}
      \label{fig:flowHarmonics}
\end{figurehere}
\vspace{0.5cm}
\noindent
The results
shown are for the Glauber model, and the values for $v_{2,3,4}$ are in
good agreement with available RHIC data~\cite{Adare:2011tg} for
$\eta/s=0.08$. In contrast, the CGC-KLN model, which we saw fits data
for the larger $\eta/s=0.16$, does poorly with $v_3$ because of the
damping effect noted. The poor agreement of the CGC-KLN model with the
higher $v_n$ moments can be traced primarily to the absence of
event-by-event color charge fluctuations we discussed previously. The
IP-Sat model with fluctuations~\cite{Schenke:2012hg,Schenke:2012wb}
includes these and the results for the moments are closer to the
Glauber model that is tuned to fit the data.

It is here that we see that the EIC might have significant additional
impact on bulk observables in A+A collisions.  This is, of course, in addition
to the absolutely crucial input of establishing the saturation
paradigm of entropy production, extracting information on the energy
and centrality dependence of $Q_s$ and providing information on gluon
correlations that may influence thermalization and long range rapidity
correlations.  The measurement of $d\sigma/dt$ in diffractive $e$+A
collisions (see Sec.~\ref{sec:diffmeasurements} and
Fig.~\ref{fig:dsigdt}) allows for a clean determination of the spatial
gluon density on average and will help constrain the scale and
magnitude of event-by-event fluctuations of color charge
densities. Inelastic vector meson production can further constrain
the spatial extent of these event-by-event fluctuations
\cite{Marquet:2010cf}. Hadronic multiplicity fluctuations, along with
the di-hadron correlation, would also allow one to pinpoint the
dynamical origin of the energy density fluctuations in heavy ion
collisions. Such direct access to spatial information is unique to
$e$+A collisions (in contrast to $p$+A collisions) and therefore can
only be provided by an EIC.

\subsubsection{Energy Loss and Hadronization} 

The dramatic suppression of high transverse momentum (high-$p_T$)
hadrons discovered at RHIC is important evidence for the production of
a dense medium in nuclear collisions.  It is commonly accepted that
partonic energy loss could be the main cause of the observed
suppression of hadrons at a sufficiently high $p_T$ (much larger than
the hadron mass) assuming that the hadronization of a high-$p_T$
hadron is taking place outside the medium.  However, if color
neutralization of the hadronization process starts inside the medium,
partonic energy loss might not be the only mechanism contributing to
the observed suppression, and additional forms of suppression could be
relevant.  Furthermore, the experimental evidence for suppression of
hadrons composed of heavy quarks is quite complex. So far, the
observed suppression could not be explained with the pure partonic
energy loss treatment, even though a description of heavy quarks
should be a straightforward extension of the approach for light
quarks, taking into account the more important role of collisional
losses.  These observations strongly imply that there are, at the very
least, missing elements in our understanding of what needs to be
included in describing the observed suppression of high-$p_T$ hadrons,
and require us to better understand the partonic energy loss and
time-evolution of hadronization, and to explore other independent
measurements which can test the suppression mechanism.

\begin{figure*}[thb]
\begin{center}
\includegraphics[width = 0.6 \textwidth]{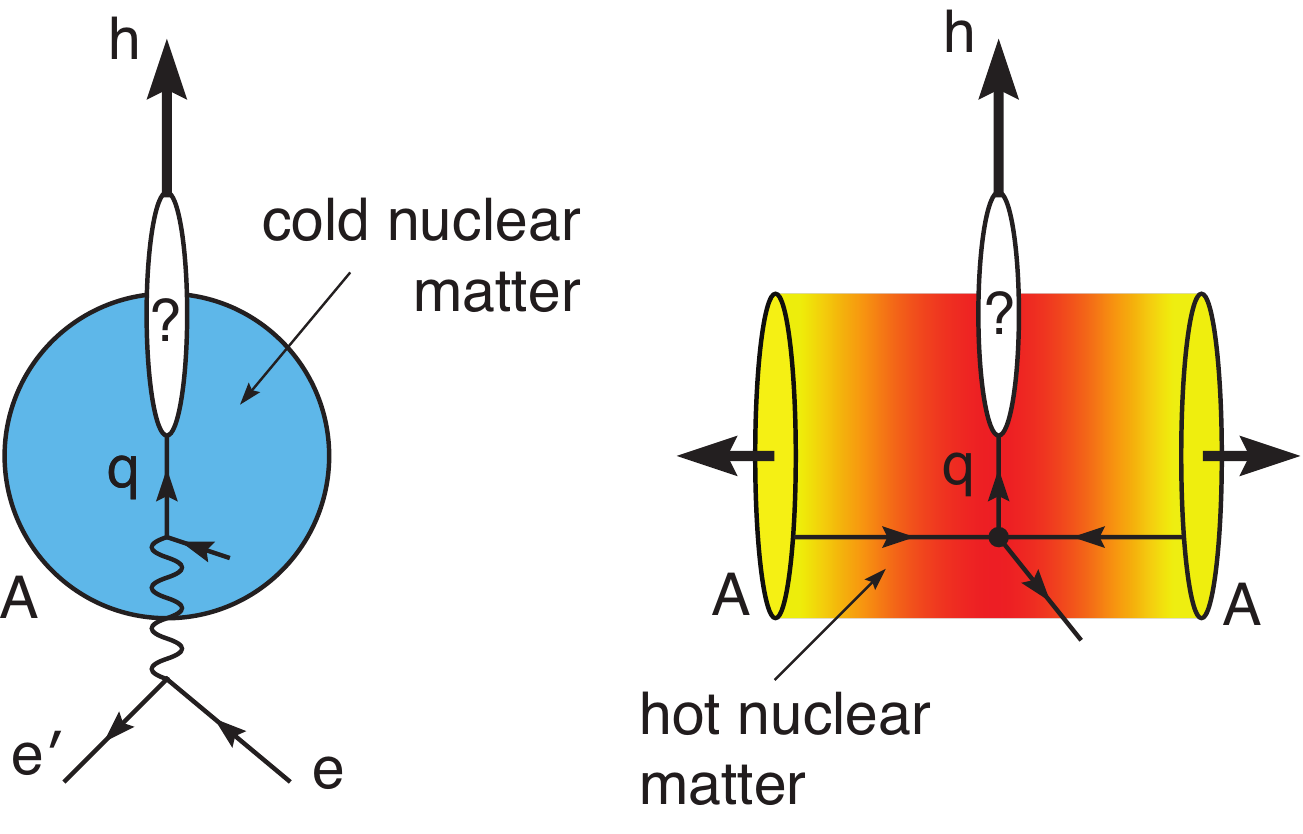}
\end{center}
    \vskip -0.4cm 
\caption{ \label{fig:DISvsAA} Cartoon showing a similarity of the
  kinematics and geometry in production of hadrons in a hot matter
  created in A+A collisions and in SIDIS on nuclei.}
 \end{figure*}

 If energy loss of a colored fast moving parton is the sole
 suppression mechanism, the inclusive hadron suppression at high $p_T$
 in A+A collisions could be represented by a single parameter -- the
 jet transport coefficient $\hat{q}$ of the medium, defined as the
 average transverse momentum squared acquired by a parton traversing a
 medium per unit distance traveled.  However, the extraction of
 $\hat{q}$ from the A+A data hardly provides a clean test of the
 energy loss mechanism because of the complexity of the created
 medium's dynamics, which includes, in particular, the first 1-2 fm/c
 after the collision where the medium is in a non-equilibrium stage
 proceeding towards thermalization.  Furthermore, because of the
 complexity of partonic kinematics in hadronic and nuclear collisions,
 the suppression of inclusive hadron production does not provide a
 simple connection of the momentum fractions $x_1$ and $x_2$ of
 colliding partons and the fraction $z_h$ of the fragmenting parton
 momentum carried by the produced hadron; these parameters are crucial
 in determining the medium-induced partonic energy loss.

 High energy hadron production in electron-ion collisions could offer
 an alternative and cleaner way to study the mechanisms of energy loss
 and in-medium hadronization of energetic virtual partons moving
 through nuclear medium.  Semi-inclusive deep-inelastic processes
 (SIDIS, see Chap. 2) can be used as a testing ground for the
 suppression mechanism of high-$p_T$ hadron production seen in the
 nuclear collisions, as illustrated in Fig.~\ref{fig:DISvsAA}, where
 the similarity of the kinematics and geometry in hadron production in
 SIDIS and A+A collisions was presented.

 The transverse momentum~$p_T$ of the detected hadrons in A+A
 collisions varies up to $10$~GeV and higher at RHIC and up to about
 $100\,$GeV at the LHC.  Available data from SIDIS in fixed target
 experiments, such as HERMES \cite{Airapetian:2007vu} and CLAS
 \cite{Brooks:2009xg}, cover only a small part of the hadron momentum
 range observed at RHIC and the LHC.  As demonstrated in
 Sec.~\ref{subsec:energyloss}, the coverage could be significantly
 extended by SIDIS measurements at a future EIC.  The path lengths in
 the cold nuclear matter and hot medium are similar, of the order of
 the nuclear radius. However, SIDIS on nuclear targets allows to test
 suppression models in much more specific and controlled conditions.
 The nuclear density does not vary with time, its value and spatial
 distribution are well known, while the probe is characterized by the
 virtual photon's energy $\nu$ and the photon's four-momentum squared
 $Q^2$ are also uniquely determined (see the Sidebar on
 page~\pageref{sdbar:DIS}).  At the leading order of the strong
 interaction, the momentum of the hadronizing quark, as well as the
 fractional energy $z_h$ of the detected hadron, are effectively
 measured.

 Accurate measurements of different observables, like the magnitude of
 suppression and broadening at different $\nu$, $z_h$ and $Q^2$ with
 different nuclei should provide a stringent test for the models of
 energy loss and in-medium hadronization. If the suppression is
 dominated by the partonic energy loss, these measurements would help
 constrain the value of the jet quenching parameter ${\hat q}$ of a
 known medium.  This parameter is central to the energy loss studies
 in A+A collisions: its value for hot nuclear matter in the early
 stages of the collision is presently narrowed down to the range of
 $1$~GeV$^2$/fm to $10$~GeV$^2$/fm. The cold nuclear matter
 experiments at an EIC would help further pinpoint the value of this
 important parameter.

Furthermore, at an EIC, for the first time one will be able to study
open charm and open bottom meson production in $e$+A collisions, as
well as the in-medium propagation of the associated heavy quarks:
these measurements would allow one to fundamentally test high-energy
QCD predictions for partonic energy loss, and confront puzzling
measurements of heavy flavor suppression in the QGP at RHIC (see
Sec.~\ref{subsec:energyloss}).

With a wide energy coverage, the EIC could be an excellent machine to
study the space-time development of hadronization by varying the
energy and virtuality of the probe -- the exchanging virtual photon in
SIDIS.  As discussed in Sec.~\ref{subsec:energyloss}, the color
neutralization of the fragmenting quark could take place inside the
nuclear medium to form the so-called ``pre-hadron'' state, which is a
name for a state of partons with zero net color but with the same
quantum numbers of a hadron that the state is about to transmute into.
The ``pre-hadron'' state represents an intermediate stage of the
hadronization process from an energetic single parton produced in a
hard collision to the hadron observed in the detector. This stage is
expected to exist from general theoretical considerations, but it is
likely non-perturbative.  If it does exist, the interaction of the
``pre-hadron'' state with nuclear medium should certainly be different
from that of a colored and fast moving single parton.  As indicated in
Fig.~\ref{fig:Rsidis-zh}, the EIC is capable to distinguish the
suppression caused by a purely partonic energy loss from that
involving a ``pre-hadron'' stage.


\subsubsection{Ultra-Peripheral Collisions} 
\label{sec::upc-connections}

Ultra-peripheral collisions (UPC) are defined as interactions among
two nuclei that are separated by impact parameters larger than the sum
of their radii. The ions do not interact directly with each other and
move essentially undisturbed along the beam direction. Due to the
coherent action of all the protons in the nuclei, the electromagnetic
fields are very strong and the resulting flux of equivalent photons is
large ($\propto Z^2$); the otherwise dominant hadronic interactions
are strongly suppressed.  The only possible interaction is
electromagnetic involving a long-range photon exchange. A photon
stemming from the electromagnetic field of one of the two colliding
nuclei can penetrate into the other nucleus and interact with one or
more of its hadrons, giving rise to photon-nucleus collisions
\cite{Bertulani:2005ru}.  Ultra-peripheral heavy-ion collisions are an
important, albeit kinematically limited alternative, which is being
used to study QCD dynamics until an EIC becomes a reality.

 \begin{figurehere}
 \begin{center}
 \includegraphics[width = \columnwidth]{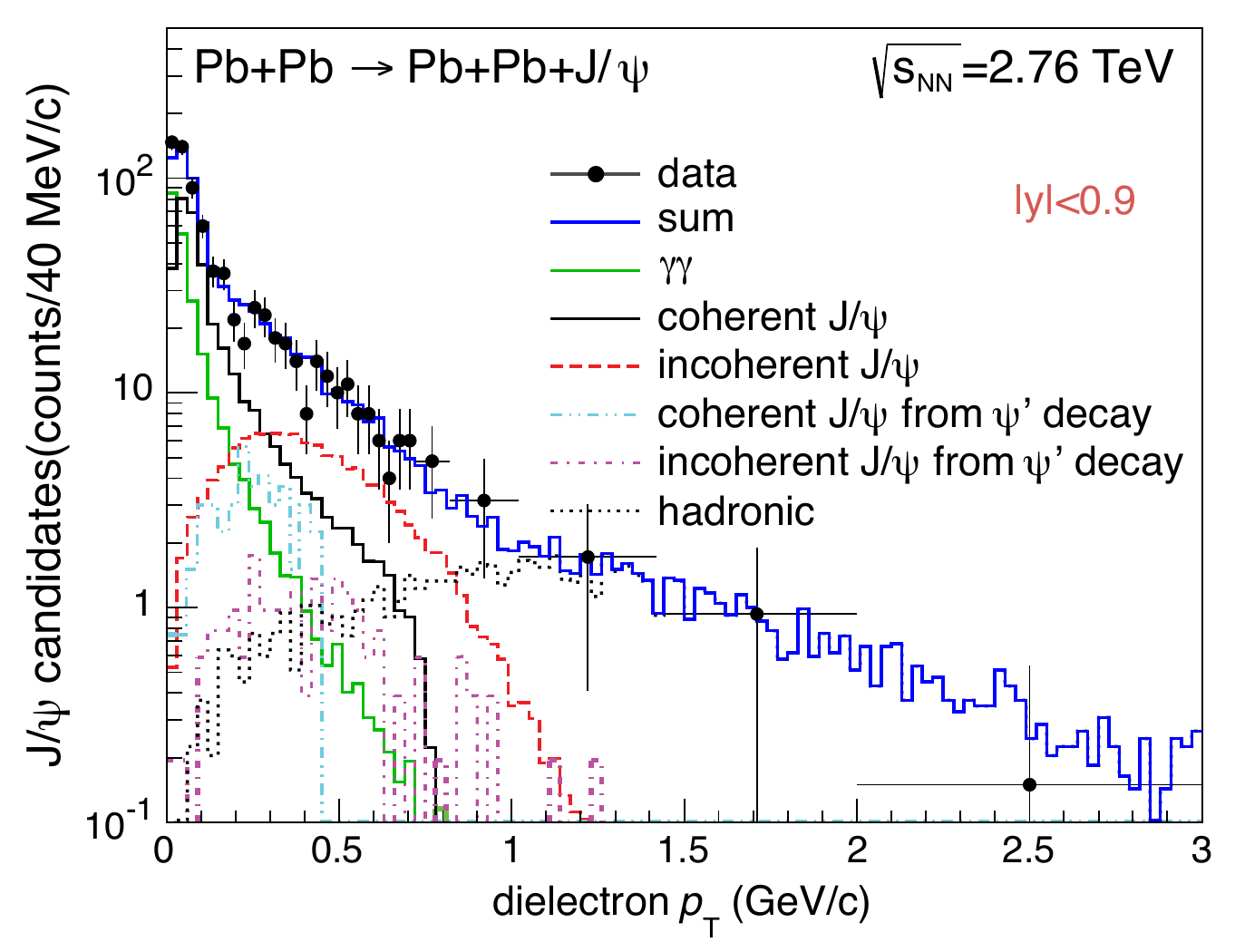}
 \end{center}
 \vskip -0.8cm
 \caption{ \label{fig:upc-alice} Plot of the $p_T$-distribution of
   $J/\psi$ mesons produced in ultra-peripheral Pb+Pb collisions at
   $\sqrt{s_{_{NN}}}$=2.76 TeV as measured by  ALICE.
   \cite{Abbas:2013oua}. Overlaid are coherent and incoherent
   components derived from Monte Carlo simulations.}
\end{figurehere}
  \vspace{0.5cm}

  A special focus of the UPC program at RHIC and LHC are photo-nuclear
  reactions involving the exclusive production of heavy vector mesons
  such as $J/\psi$ and $\psi$(2S) \cite{Goncalves:2005sn}. They
  provide a good tool for evaluating the behavior of the gluon
  distribution functions at low $x$, since the photo-production cross
  section scales at leading order as the square of the gluon
  distribution $G(x,\mu^2)$; the scale $\mu$ is typically approximated
  by $M_V/2$, where $M_V$ is the mass of the vector meson. However, as
  pointed out in section \ref{sec:diffmeasurements}, heavy vector
  mesons are less sensitive to gluon saturation than lighter vector
  mesons such as $\phi$ and $\rho$, whose production cross sections
  are much harder to measure in UPCs.

  The exclusive photo-production can be either coherent, where the
  photon couples coherently to almost all the nucleons, or incoherent,
  where the photon couples to a single nucleon. Coherent production is
  characterized by low transverse momentum of vector mesons ($\langle
  p_T \rangle \approx$ 60 MeV/c at LHC) where the nucleus normally
  does not break up. Incoherent production, corresponding to
  quasi-elastic scattering off a single nucleon, is characterized by a
  somewhat higher transverse momentum ($\langle p_T \rangle \approx$
  500 MeV/c at LHC). Unlike the $e$+A collisions with their wide
  coverage in $Q^2$, vector meson production in UPCs is limited to
  vanishingly small $Q^2$, much smaller than most of the other
  momentum scales in the problem.  The value of $Q^2$ is inversely
  proportional to the square of the impact parameter between the
  nuclei, such that the photon is quasi-real.
The momentum transfer $t$ can be approximated by
$t \approx -p_T^2$.  

Figure \ref{fig:upc-alice} shows as an example the
$p_T$-distribution  of  $J/\psi$ mesons produced in ultra-peripheral Pb+Pb collisions at
$\sqrt{s_{_{NN}}}$=2.76 TeV measured by the ALICE collaboration \cite{Abbas:2013oua}.
The clear peak at low $p_T$ is mainly due to coherent interactions,
while the tail extending out to 1 GeV/c comes from incoherent
production.

In general, measurements of ultra-peripheral collisions at RHIC and
LHC energies can provide useful insights into gluon densities at
low-$x$. They lack, however, the wide kinematic coverage in $Q^2$ and
the precise knowledge of the relevant kinematic variables that will be
available at an EIC.  

\end{multicols}


\subsection{Connections to Cosmic Ray Physics}
\begin{multicols}{2}

\begin{figurehere}
    \begin{center}
        \includegraphics[width = 0.48\textwidth]{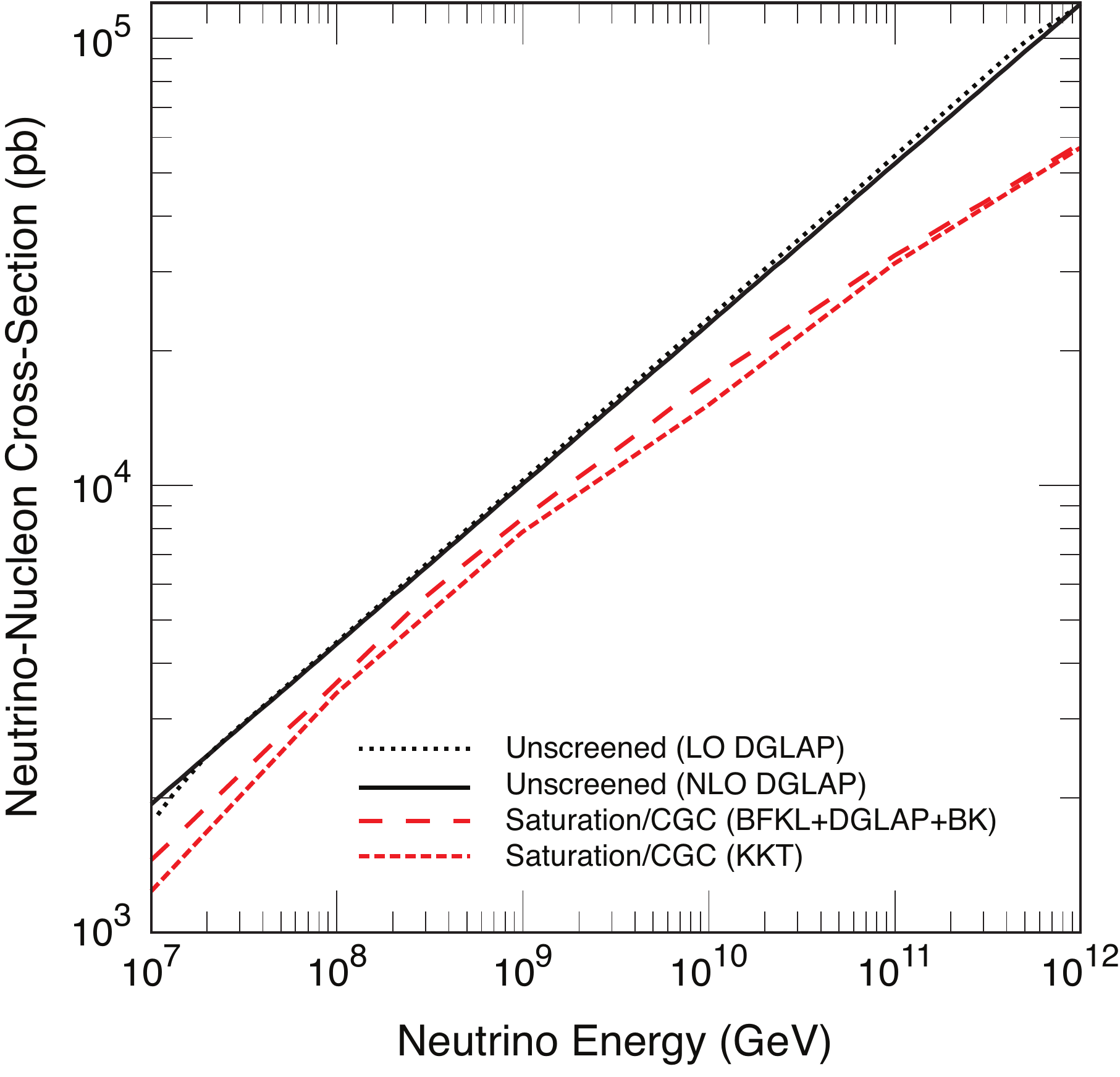}
    \end{center}
    \vskip -0.6cm
    \caption{\label{fig:cray}
    Predictions of several models with (red) and without (black) parton saturation (CGC) physics for the cross-sections for neutrino--nucleon scattering at high energies as calculated in \cite{Anchordoqui:2006ta,Henley:2005ms,Kutak:2003bd}. The KKT saturation model is defined in \cite{Kharzeev:2004yx}. Saturation effects appear to lower the neutrino--nucleon cross-section at very high energies in agreement with general expectations of saturation taming the growth of the gluon numbers.}
\end{figurehere}
\vspace{0.5cm}

Decisive evidence in favor of parton saturation, which could be
uncovered at an EIC, would also have a profound impact on the physics of
Cosmic Rays. 

The sources of the observed ultra-high energy cosmic rays must also
generate ultra--high energy neutrinos. Deep inelastic scattering of
these neutrinos with nucleons on Earth is very sensitive to the strong
interaction dynamics. This is shown in \fig{fig:cray} for the cross
sections for neutrino--nucleon scattering plotted as a function of the
incident neutrino energy for several models.
As we argued above, the experiments at an EIC would be able to rule
out many of the models of high energy strong interactions, resulting
in a more precise prediction for the neutrino--nucleon cross-section,
thus significantly improving the precision of the theoretical
predictions for the cosmic ray interactions.  The improved precision
in our understanding of strong interactions will enhance the ability
of the cosmic ray experiments to interpret their measurements
accurately and will thus allow them to uncover new physics beyond the
Standard Model of particle physics.

\begin{figurehere}
    \begin{center}
        \includegraphics[width = 0.48 \textwidth]{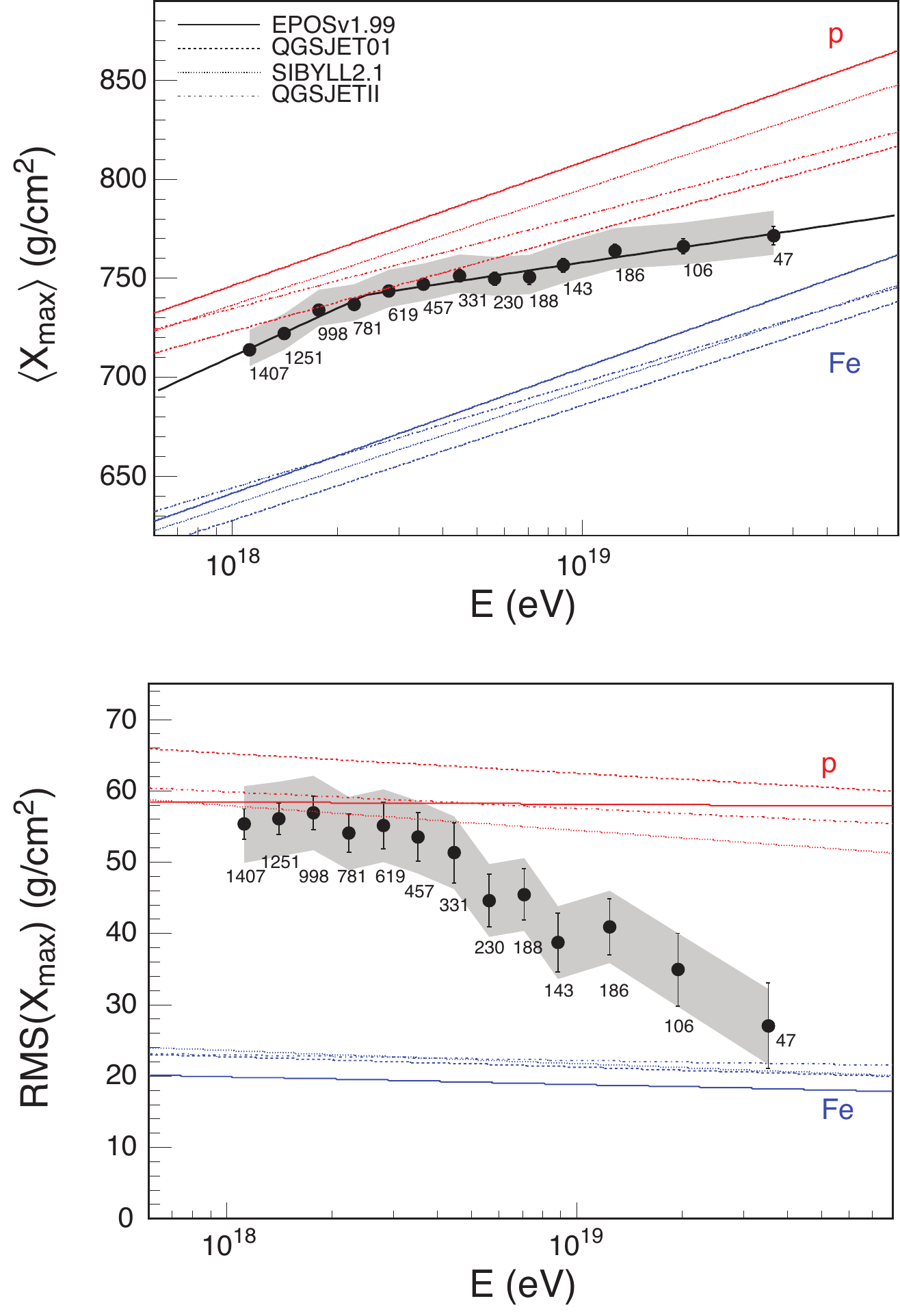}
    \end{center}
    \vskip -0.4cm
    \caption{ \label{fig:AugerPlot} The data on the atmospheric depth
      of the cosmic ray shower maximum $X_{max}$ (upper panel) and on
      its RMS (lower panel) as a function of the cosmic ray energy $E$
      reported by the Pierre Auger Observatory
      \cite{PierreAuger:2011aa}. The error bars reflect the
      statistical uncertainty, while the bands depict the systematic
      uncertainty. The numbers next to the data points indicate the
      number of events in each data bin. The solid lines represent
      predictions of various Monte-Carlo simulations for the cosmic
      ray being a proton ($p)$ and an iron nucleus ($Fe$). }
\end{figurehere}
\vspace{0.5cm}

Saturation physics that is likely to be discovered and studied at an
EIC has other important connections with cosmic-ray physics.  One key
question concerns the nuclear composition of ultra-high energy cosmic
rays: are they made out of protons or out of heavier nuclei?  At
energies above about $10^{16}$ eV, the low cosmic ray flux forces us
to rely on indirect measurements of the composition. These indirect
measurements necessarily depend on the modeling of the hadronic
showers that the cosmic-rays produce.  Variables such as the depth of
the shower maximum ($X_{max}$) in the atmosphere and the muon content
of the showers depend strongly on the hadronic modeling.

The Pierre Auger collaboration has measured the depth of shower
maxima in air showers with energies above $10^{18}$ eV
\cite{PierreAuger:2011aa} shown here in \fig{fig:AugerPlot}.  At
energies below $10^{18.4}$~eV, they see a composition with a constant
elongation rate (the slope of $X_{max}$ plotted versus the cosmic ray
energy $E_{cr}$, $dX_{max}/dE_{cr}$) at a position that is consistent
with a composition that is largely protons.  
However, as one can see
from the upper panel of \fig{fig:AugerPlot}, at higher energies, there
is a significant shift in elongation rate, and, by an energy of
$10^{19.4}$~eV, the depth $X_{max}$ is more consistent with an
all-iron composition \cite{PierreAuger:2011aa}.  At the same time, the
root mean square (RMS) variation in the position of $X_{max}$ (plotted
in the lower panel of \fig{fig:AugerPlot}) drops by a factor of two,
also consistent with a change in composition.  This is a rather abrupt
change of composition in one decade of energy; an alternate
possibility is that there is a shift in hadronic physics, such as the
onset of saturation. The EIC could shed light on this possibility.

At somewhat lower energies, the IceCube collaboration has measured the
production of high-energy ($\approx 1$~TeV) and high-$p_T$ (roughly
$p_T> 2 $~GeV/c) muons in cosmic-ray air showers
\cite{IceCube:2012ij}, and needs to interpret the data using modern
pQCD, again with a view to probing the cosmic ray composition.  These
forward muons come from the collision of a high-$x$ parton in the
incident cosmic ray with a low-$x$ parton in the nitrogen/oxygen
target in the atmosphere.  Saturation will alter the distribution of
low-x partons in the target, and so must be considered in the
calculations.  EIC data is needed to pin down this possible saturation
effect.
\end{multicols}

%% file: files_tex/electroweak.tex
\vspace*{-0.2in}
{\large\it Conveners:\ Krishna Kumar and Michael Ramsey-Musolf}

\section{Introduction}
\label{sec:bsm}

\vspace*{-0.15in}
\begin{multicols}{2}
It is natural to ask whether the envisioned machine parameters of the
EIC could enable new discoveries in the broad subfield of Fundamental
Symmetries (FS), which addresses one of the overarching goals of
nuclear physics, namely, the exploration of the origin and evolution of
visible matter in the early universe.  The theoretical and
experimental studies in this subfield are complementary to those of
particle physics and cosmology. Indeed, a broader categorization of the
full range of initiatives that encompass the FS goals falls under the
titles ``Energy Frontier", ``Cosmic Frontier" and
``Intensity/Precision Frontier".
 
The FS subfield of nuclear physics is part of the intensity/precision
frontier, the specific primary goal of which is the study of
electroweak interactions of leptons and hadrons with progressively
higher sensitivity. By comparing the measured interaction amplitudes
with theoretical predictions within the framework of the Standard
Model (SM) of strong, weak and electromagnetic interactions, insights
are gained into the symmetries and interactions of matter in the
universe at its earliest moments of existence, indirectly accessing
energy scales similar to, and sometimes beyond the reach of, the highest
energy accelerators.

The EIC offers a unique new combination of experimental probes given
the high center-of-mass energy, high luminosity and the ability to
polarize the electron and hadron beams. Electron-hadron collisions
would be analyzed by a state-of-the-art hermetic detector package with
high efficiency and resolution.  In this section, we explore new FS
measurements that become possible with these capabilities, the physics
impact of potential measurements, and the experimental requirements to
enable the measurements.

Electroweak interaction studies at the EIC can also be used to probe
novel aspects of nucleon structure via measurements of spin
observables constructed from weak interaction amplitudes mediated by
the W and Z bosons. Indeed, some parity-violating observables become
accessible that have never before been measured. These measurements
are considered in detail in Chapter~\ref{sec:helicity} along with
other fundamental observables that probe the longitudinal spin
structure of the nucleon.
\end{multicols}

\section{Specific Opportunities in Electroweak Physics}

\subsection{Charged Lepton Flavor Violation}

\begin{multicols}{2}
With the discovery of neutrino oscillations, we now know that lepton
flavor is not a conserved quantity in fundamental interactions. It is
natural to ask whether lepton flavor non-conservation can be observed
in {\it charged} lepton interactions. In addition, the implication
that neutrinos have mass leads to the fundamental question of whether
neutrinos are their own anti-particles (Majorana neutrinos) which
could have profound implications for the origin of the
matter-antimatter asymmetry in the universe. Speculative new theories
of the early universe that predict Majorana neutrinos often also
predict observable rates of charged lepton flavor violation
(CLFV). Searches for CLFV are thus one of the most sensitive
accelerator-based low-energy probes of the dynamics of the early
universe and the physics of the smallest length scales, in a manner
complementary to searches for new physics at the energy frontier at
the Large Hadron Collider.

The most sensitive CLFV searches to date have come from searches for
the neutrinoless conversion of stopped muons to electrons in nuclei,
searches for the rare decay of a free muon to an electron and photon,
and searches for the rare decay of a kaon to an electron and muon. The
limits from these processes, though extremely sensitive, all involve
the $e\leftrightarrow\mu$ transition.  Speculative CLFV theories can
predict enhanced rates for $e\leftrightarrow\tau$
transitions. Existing limits for the $e\leftrightarrow\tau$ transition
come from searches for rare $\tau$ decays at the high luminosity
$e^+e^-$ colliders at a center of mass energy of 5 to 10 GeV, the
so-called B-factories.

In lepton-hadron interactions, one could search for the rare cases
where an electron converts to a muon or tau lepton, or a muon converts
to a tau lepton. However, this is impossible to observe due to large
and irreducible background in fixed target experiments.  The only
successful such searches for $e\rightarrow\tau$ transitions have been
carried out at the HERA electron-hadron collider experiments ZEUS and
H1. In a collider environment, the event topology for rare signal
events can be differentiated from conventional electroweak deep
inelastic scattering (DIS) events~
\cite{Adloff:1999tp,Chekanov:2005au, Aktas:2007ji}

\begin{figure*}
\begin{center}
  \includegraphics[width=0.95\textwidth]{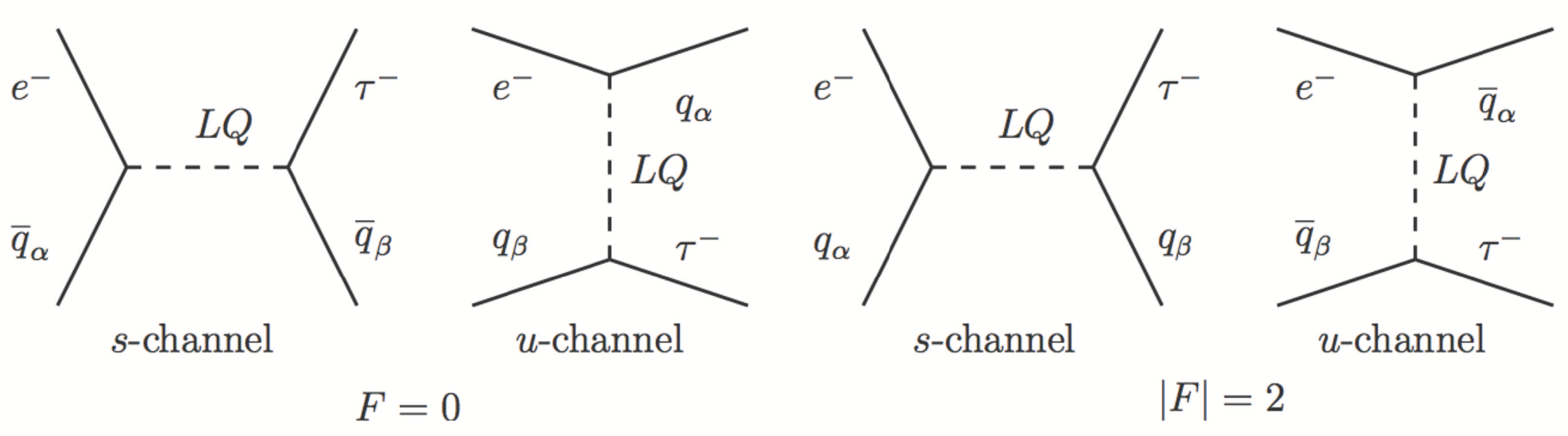}
\end{center}
	\caption{\label{etaufeyn}Feynman diagrams for
$e\rightarrow\tau$ scattering processes via leptoquarks, which carry
fermion number $F=3B+L$ equal to 0 or $\pm
2$~\cite{Gonderinger:2010yn}}
\end{figure*}

The CLFV process could be mediated by a hypothesized new heavy boson
known as a leptoquark, which carries both lepton and baryon quantum
numbers and appears naturally in many SM extensions such as Grand
Unified Theories, supersymmetry, and compositeness and technicolor
models (for a concise review, see~\cite{Albright:2008ke}).
Figure~\ref{etaufeyn} shows the Feynman diagrams that could be
responsible for the CLFV transition that might be observed at an EIC.
The most recent published search by H1 finds no evidence for CLFV
$e\rightarrow\tau$ transitions~\cite{Aaron:2012zz}, which can in turn
be converted to a limit on the mass and the couplings of leptoquarks
in specific SM extensions~\cite{Buchmuller:1986zs}.

A high energy, high luminosity EIC, with 100 to 1000 times the
accumulated luminosity of HERA experiments would allow a large
increase in sensitivity. A recent study has shown that an EIC, with 90
GeV center-of-mass energy, could surpass the current limits with an
integrated luminosity of $10\fb^{-1}$~\cite{Gonderinger:2010yn}.  The
study also showed that the EIC could compete or surpass the updated
leptoquark limits from rare CLFV tau decays for a subset of quark
flavor-diagonal couplings.  A follow-up study beyond this, including
knowledge of inefficiencies from the H1 and ZEUS collaborations for
$\tau$ reconstruction, indicates that these estimates
are too optimistic by a factors of 10-20, thus requiring
$100-200\fb^{-1}$ luminosity integrated over the EIC lifetime~\cite{Boer:2011fh}. At the
highest possible luminosities envisioned for the EIC, these
luminosities are deemed achievable.  Over the lifetime of the EIC, the
$e\rightarrow\tau$ reach would thus be comparable to the reach of rare
$\tau$ decays at future high-luminosity super-B factories.

It must be emphasized that the unambiguous observation of a CLFV
process would be a paradigm-shifting discovery in subatomic physics,
with wide-ranging implications for nuclear physics, particle physics
and cosmology. It is quite possible that future potential discoveries
at the energy and cosmic frontiers could make CLFV searches at the EIC
even more compelling.
\end{multicols}

\subsection{Precision Measurements of Weak Neutral Current Couplings}

\begin{multicols}{2}
A comprehensive strategy to indirectly probe for new high energy
dynamics via sensitive tests of electroweak interactions at the
intensity frontier must also include precision measurements of
flavor-diagonal weak neutral current interactions mediated by the Z
boson. For electron-hadron interactions at $Q^2\ll M_Z^2$, weak
neutral current amplitudes are accessed via parity violation since
pseudoscalar observables, sensitive to weak-electromagnetic
interference terms, can be constructed from the product of vector and
axial-vector electron and quark electroweak currents.  The
parity-violating part of the electron-hadron interaction at $Q^2\ll
M_Z^2$ can be given in terms of phenomenological couplings $C_{ij}$
\end{multicols}
\[\mathcal{L}^{PV}=\frac{G_F}{\sqrt{2}} [\overline
e\gamma^\mu\gamma_5e(C_{1u}\overline u\gamma_\mu u+C_{1d}\overline
d\gamma_\mu d) +\overline e\gamma^\mu e(C_{2u}\overline
u\gamma_\mu\gamma_5 u+C_{2d}\overline d\gamma_\mu\gamma_5 d)]\]
\begin{multicols}{2}
\noindent
with additional terms as required for the heavy quarks.  Here $C_{1j}$
($C_{2j}$) gives the vector (axial-vector) coupling to the $j^{th}$
quark.

Within the SM context, each coupling constant is precisely predicted
since they are all functions of the weak mixing angle
$\sin^2\theta_W$.  Under the assumption that the recently discovered 
scalar resonance at the Large Hadron Collider~\cite{atlas,cms} is the SM Higgs boson, 
the value of the weak mixing angle is now known to better than 0.03\%.
Over the past two decades, the $C_{1i}$ couplings have been
measured with steadily improving precision in tabletop atomic parity
violation experiments and in fixed target parity-violating electron
scattering experiments, most recently at Jefferson Laboratory (JLab)~\cite{qweakfirst}.
Comparing these measurements to SM predictions has produced strong
constraints on new high energy dynamics, such as limits on TeV-scale
heavy Z' bosons and certain classes of interactions in supersymmetric
theories, in a manner complementary to direct searches at
colliders~\cite{mikeprc, young}. This an active field with new experimental
tools under development, as described in recent reviews~\cite{sujens,mikevincenzo,kkreview}.
\begin{figure*}[t!]
\begin{center}
    \includegraphics[width=0.75\textwidth]{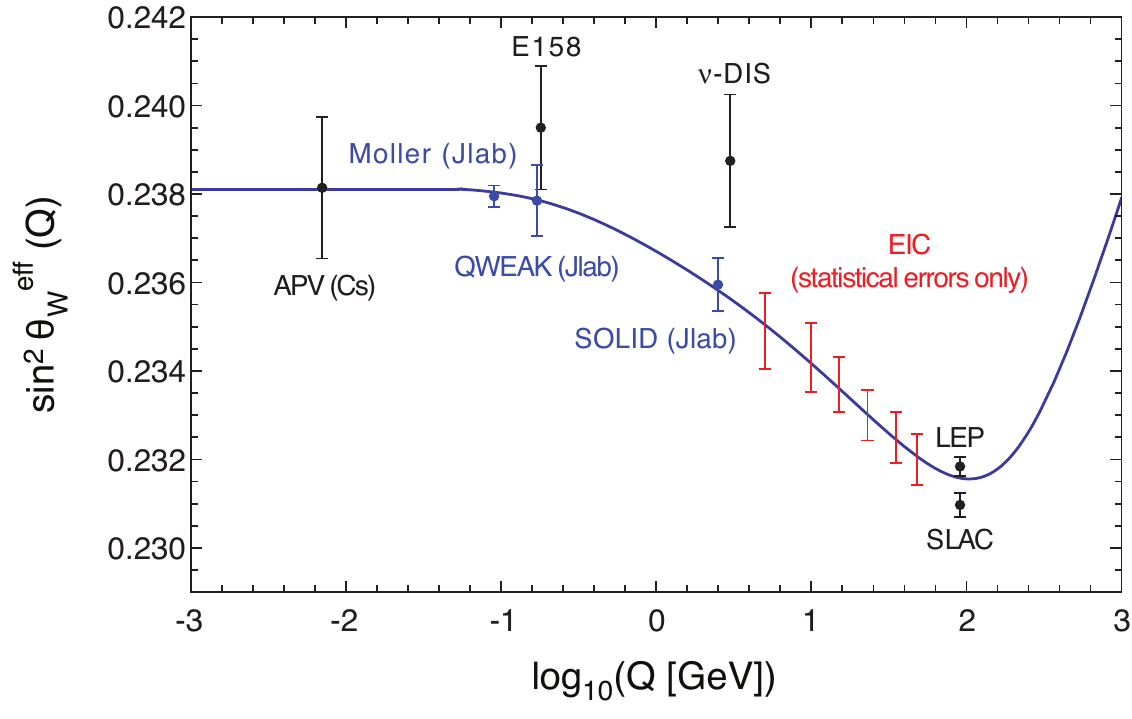}
\end{center}
\vskip -0.5cm
\caption{\label{sinthetarunning} Projected statistical uncertainties
on the $\sin^2\theta_W$ in a series of $Q^2$ bins ($\sqrt{s} = 140$
GeV, 200 fb$^{-1}$.)  The black points are published results while the
blue points are projections from the JLab program.}
\end{figure*}

At the EIC, the availability of high luminosity collisions of
polarized electrons with polarized $^1$H and $^2$H would allow the
construction of parity-violating observables that are sensitive to all
four semi-leptonic coupling constants introduced above. The observable
with the best sensitivity to cleanly measure coupling constants
without significant theoretical uncertainty is $A_{PV}$ in $e-^2$H
collisions. $A_{PV}$ is constructed by averaging over the hadron
polarization and measuring the fractional difference in the deep
inelastic scattering (DIS) rate for right-handed vs left-handed
electron bunches.

The collider environment and the hermetic detector package at high
luminosity will allow precision measurements of $A_{PV}$ over a wide
kinematic range.  In particular, the EIC will provide the opportunity
to make highly precise measurements of $A_{PV}$ at high values of the
4-momentum transfer $Q^2$, and in the range $0.2\lsim x \lsim 0.5$ for
the fraction of the nucleon momentum carried by the struck quark, such that hadronic uncertainties from limited knowledge
of parton distribution functions and higher-twist effects are expected
to be negligible.

By mapping $A_{PV}$ as a function of $Q^2$ and the inelasticity of the
scattered electron $y$ (something that is very challenging to do in
fixed target experiments), a clean separation of two linear
combination of couplings namely $2C_{1u}-C_{1d}$ and $2C_{2u}-C_{2d}$
will become feasible as a function of $Q^2$.  Thus, at the highest
luminosities and center-of-mass energies envisioned at the EIC, very
precise measurements of these combinations can be achieved at a series
of $Q^2$ values, providing an important and complementary validation
of the electroweak theory at the quantum loop level.
Figure~\ref{sinthetarunning} shows a first estimate of projected
uncertainties on the weak mixing angle extracted from such a
dataset~\cite{Boer:2011fh}, for a center-of-mass energy of 140 GeV and
an integrated luminosity of 200 fb$^{-1}$. The effects of radiative
corrections and detector effects need to be considered in the future
to further refine this study.

A unique feature of DIS $A_{PV}$ measurements is the sensitivity to
the $C_{2i}$ coupling constants that involve the amplitudes with
axial-vector quark currents.  While the couplings are kinematically
accessible at large scattering angle measurements in fixed target
elastic electron scattering, axial-hadronic radiative correction
uncertainties cloud the interpretation of the measurements in terms of
fundamental electroweak physics.  Parity-violating DIS using $^2$H is
the only practical way to measure one combination accurately, namely
$2C_{2u}-C_{2d}$.  A recent measurement at 6 GeV at JLab made the
first non-zero measurement of this combination~\cite{pvdisnature}, and a new experiment has 
been proposed at 11 GeV to
constrain this combination to better than 10\%. At the highest
envisioned luminosities, the EIC would offer the opportunity to
further improve on this constraint by a further factor of 2 to 3.
\begin{figurehere}
\begin{center}
  \includegraphics[height=1.5in]{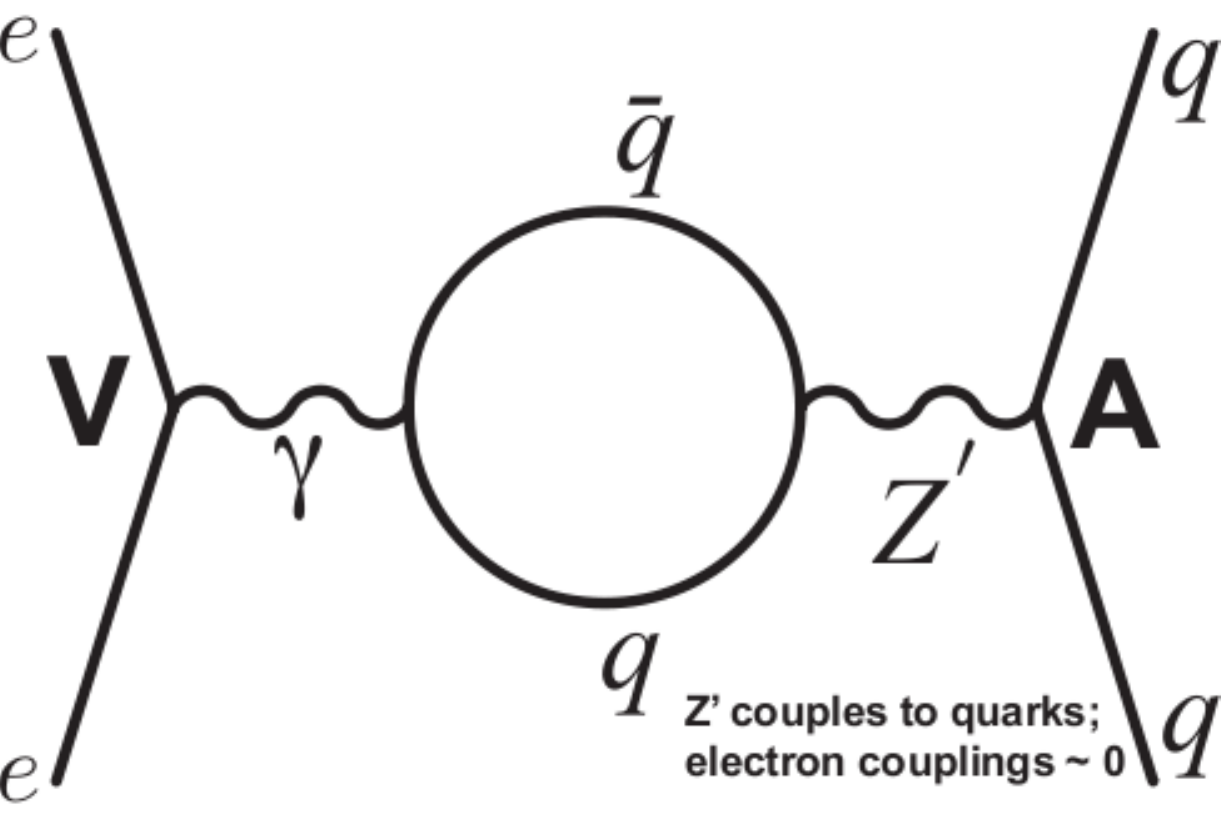}
\end{center}
\vskip -0.2cm
\caption{\label{leptophobiczprime} A Feynman diagram for an amplitude
with a vector electron current and axial-vector hadron current which
would be sensitive to a heavy new vector boson that couples to quarks
and has no couplings to leptons.~\cite{Buckley:2012tc}}
\end{figurehere}

\vspace{0.5cm}

One example of the importance of achieving sensitive constraints on
the $C_{2i}$ couplings is depicted in Fig.~\ref{leptophobiczprime},
which shows how a heavy Z$^\prime$ boson (predicted in many SM
extensions) could introduce an additional amplitude and induce a
deviation in the measured $C_{2i}$ couplings~\cite{Buckley:2012tc}.  A
remarkable feature of this amplitude is the fact it is sensitive to
the Z$^\prime$ boson even in the case that it might not couple to
leptons (so-called lepto-phobic Z$^\prime$).  The limits on the
existence of such bosons from other precision weak neutral current
measurements as well as from colliders is very weak because all
signatures require non-zero lepton-Z$^\prime$ couplings.  Note that
this amplitude cannot contribute to any tree-level amplitudes nor
amplitudes involving the $C_{1i}$ couplings at the quantum loop
level. The projected uncertainty from the JLab measurements will be
sensitive to a lepto-phobic Z$^\prime$ with a mass $\lsim 150$ GeV,
significantly better than the current limit from indirect searches
when there is no significant Z-Z$^\prime$ mixing.

The JLab extraction will rely on a simultaneous fit of electroweak
couplings, higher-twist effects and violation of charge symmetry to a
series of $A_{PV}$ measurements in narrow $x$ and $Q^2$ bins. It is
highly motivated to find ways to improve the sensitivity to the
$C_{2i}$ couplings further, given its unique sensitivity for TeV-scale
dynamics such as the aforementioned Z$^\prime$ bosons.  The
kinematical range for the $A_{PV}$ measurement at the EIC would enable
a significantly improved statistical sensitivity in the extraction of
the $C_{2i}$ couplings.  Apart from statistical reach, the EIC
measurements will have the added advantage of being at significantly
higher $Q^2$ so that higher-twist effects should be totally
negligible.

A study of the statistical reach shows that an EIC measurement can
match the statistical sensitivity of the 12 GeV JLab measurement with
$\sim 75\fb^{-1}$.  It is also worth noting that the EIC measurements
will be statistics-limited, unlike the JLab measurement. The need for
precision polarimetry, the limiting factor in fixed target
measurements, will be significantly less important at the
corresponding EIC measurement because $2C_{2u}-C_{2d}$ would be
extracted by studying the variation of $A_{PV}$ as a function of the fractional energy loss parameter, $y$. Thus, with
an integrated luminosity of several 100 $\fb^{-1}$ in Stage II of the
EIC, the precision could be improved by a further factor of 2 to
3. Depending on the discoveries at the LHC over the next decade, it is
quite possible that such sensitivity to $C_{2i}$ couplings, which is
quite unique, would prove to be critical to unravel the nature of
TeV-scale dynamics.
\end{multicols}

\newpage
\section{EIC Requirements for Electroweak Physics Measurements}

\begin{multicols}{2}
For the CLFV $e\rightarrow\tau$ transition search, it was pointed out
that the collider environment facilitates separating potential signal
events from conventional DIS events, as demonstrated by successful
searches carried out at modest integrated luminosity at HERA. This is
because the lepton in the final state tends to be isolated at low
$Q^2$ from the hadron jet. The detector will have to be suitably
designed so as to allow high-energy electron identification at high
$Q^{2}$ where it might be buried in the jet fragment.

In addition, compared to HERA, it is reasonable to expect that the EIC
detector will have significant technological enhancements that will
allow increased sensitivity, and improved background rejection.  The
momentum resolution for tracks and the granularity of the calorimeter
will be improved. Detector coverage will extend down to much smaller
angles.  Most importantly, we envision a vertex detector that will
greatly improve the robustness of the search. Since the lifetime of
the $\tau$ lepton is 290 fs, for the typical energies expected for
signal events, the decay length will be between a few 100 $\mu$m to
several mm, which will allow displaced vertices to be easily
identified.

For the flavor-diagonal precision electroweak measurements, the
apparatus being designed will be adequate to select the events
required to make the precision asymmetry measurements. The challenge
will be in controlling normalization errors, particularly the electron
beam polarization. For the anticipated precision of the $A_{PV}$
measurements, the electron beam polarization must be monitored to
significantly better than 1\%. At the completion of the JLab12
program, it is expected that techniques will be developed to monitor
the beam polarization at the level of $~0.5$\%. It will be necessary
to transfer this technology to the collider environment.
\end{multicols}

%% file: files_tex/accelerator.tex

\hyphenation{eRHIC}

{\large\it Conveners:\ Andrew Hutton and Thomas Roser}

\section{eRHIC}
\label{sec:eRHIC}

\vspace*{-0.25cm}
\begin{multicols}{2} 
eRHIC is a future Electron-Ion Collider (EIC) based on the existing
Relativistic Heavy Ion Collider (RHIC) hadron facility with its two
intersecting superconducting rings, each 3.8 km in circumference.  The
replacement cost of the RHIC facility is about two billion US dollars,
and eRHIC will take full advantage of, and build on, this investment.

A polarized electron beam with an energy up to 21 GeV would collide
with a number of ion species accelerated in the existing RHIC
accelerator complex, from polarized protons with a top energy of 250
GeV to fully-stripped uranium ions with energies up to 100 GeV/u
covering a center-of-mass energy range from 30 to 145 GeV for
polarized $e$+$p$, and from 20 to 90 GeV for
$e$+A (for large A). Using the present significant margin of
the RHIC superconducting magnets, the maximum beam energy could be
increased by 10 or more percent.

\begin{figure*}[t!]
\begin{center}
\hskip 0.45in
\includegraphics[width=0.99\textwidth]{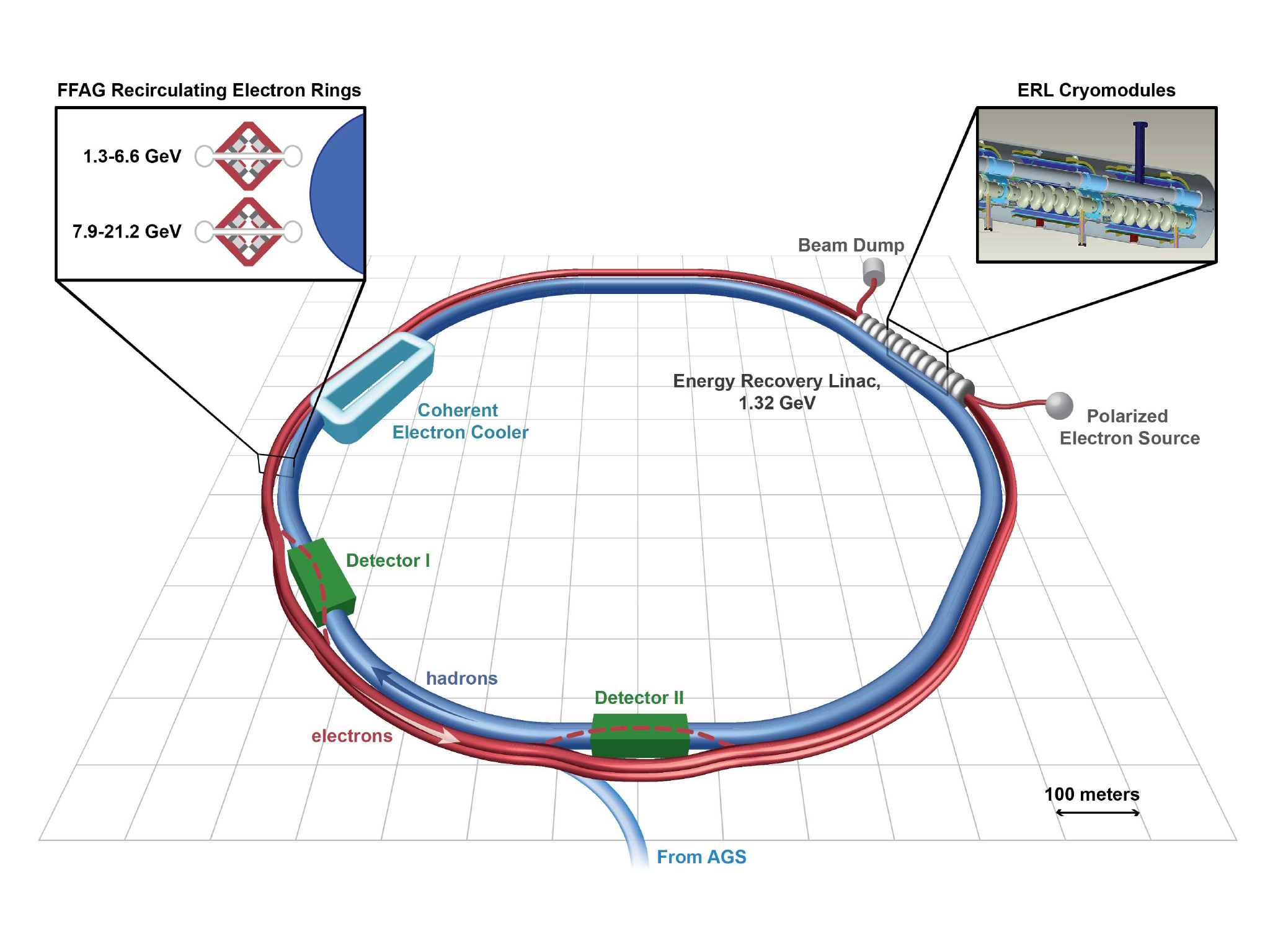}
\end{center}
\vskip -0.1in
\caption{\label{fig:roser:eRHIC_layout}
The layout of the ERL-based, 21 GeV x 250 GeV high-energy high-luminosity eRHIC.}
\end{figure*}

The eRHIC design is based on using one of the two RHIC hadron rings
and a multi-pass Energy Recovery Linac (ERL). Using an ERL as the
electron accelerator assures high luminosity in the $10^{33} -
10^{34}$ cm$^{-2}$ s$^{-1}$ range
(Fig.~\ref{fig:roser:eRHIC_layout}). Most of the electron accelerator components,
including the injector, the ERL and the recirculation passes, are located inside the RHIC
tunnel.  eRHIC will be able to provide electron-hadron collisions in up
to three interaction regions.

eRHIC employs a cost effective way to provide multiple electron beam recirlculations
by using Fixed-Field Alternating Gradient optics with very high momentum acceptance.
It allows for up to 16 recirculations in only two vertically stacked beamlines. 
Additional savings are expected from the use of low-cost permanent magnets.

To reach the required performance, eRHIC will employ several novel
technologies such as a polarized electron gun delivering a current of
50 mA, strong hadron beam cooling using Coherent electron Cooling
(CeC), a high current multi-pass Energy Recovery Linac (ERL), and
acceleration of polarized He-3 to high energy. BNL, in collaboration
with JLab and MIT, is pursuing a vigorous R\&D program to address
these technical challenges. Projected performance values for eRHIC 
are shown in Tab.~\ref{roser:eRHIC_parameters}.
\end{multicols}
\vspace*{0.1cm}

\begin{table}
\begin{center}
\begin{tabular}{|l|c|c|c|} \hline &electron&proton&Au\\ \hline
Max. beam energy [GeV/n] &15.9&250&100 \\ Bunch frequency
[MHz]&9.38&9.38&9.38\\ Bunch intensity (nucleons/electrons)
[$10^{11}$]&0.33&0.3 (3) &0.6 (2.2)\\ Beam current [mA]&50&42 (420) &33 (120)\\ Polarization
[\%]&80&70&\\ RMS bunch length [mm]&4&50 (84) &50 (84) \\ RMS norm. emittance
(e-p/e-Au) [$\mu$m]&32/58&0.3&0.2\\ $\beta^*$ [cm]&5&5&5\\ Luminosity
[$10^{33}$ cm$^{-2}$s$^{-1}$]&&1.7 (13) &1.7 (5.2) \\ \hline
\end{tabular}
\end{center}
\caption{\label{roser:eRHIC_parameters} Projected eRHIC parameters and
luminosities. The parameters  in parentheses correspond to possible future upgrade, HL-eRHIC.}
\end{table}

\subsection{eRHIC Design}

\begin{figure*}
\begin{center}
\includegraphics[width=.47\textwidth]{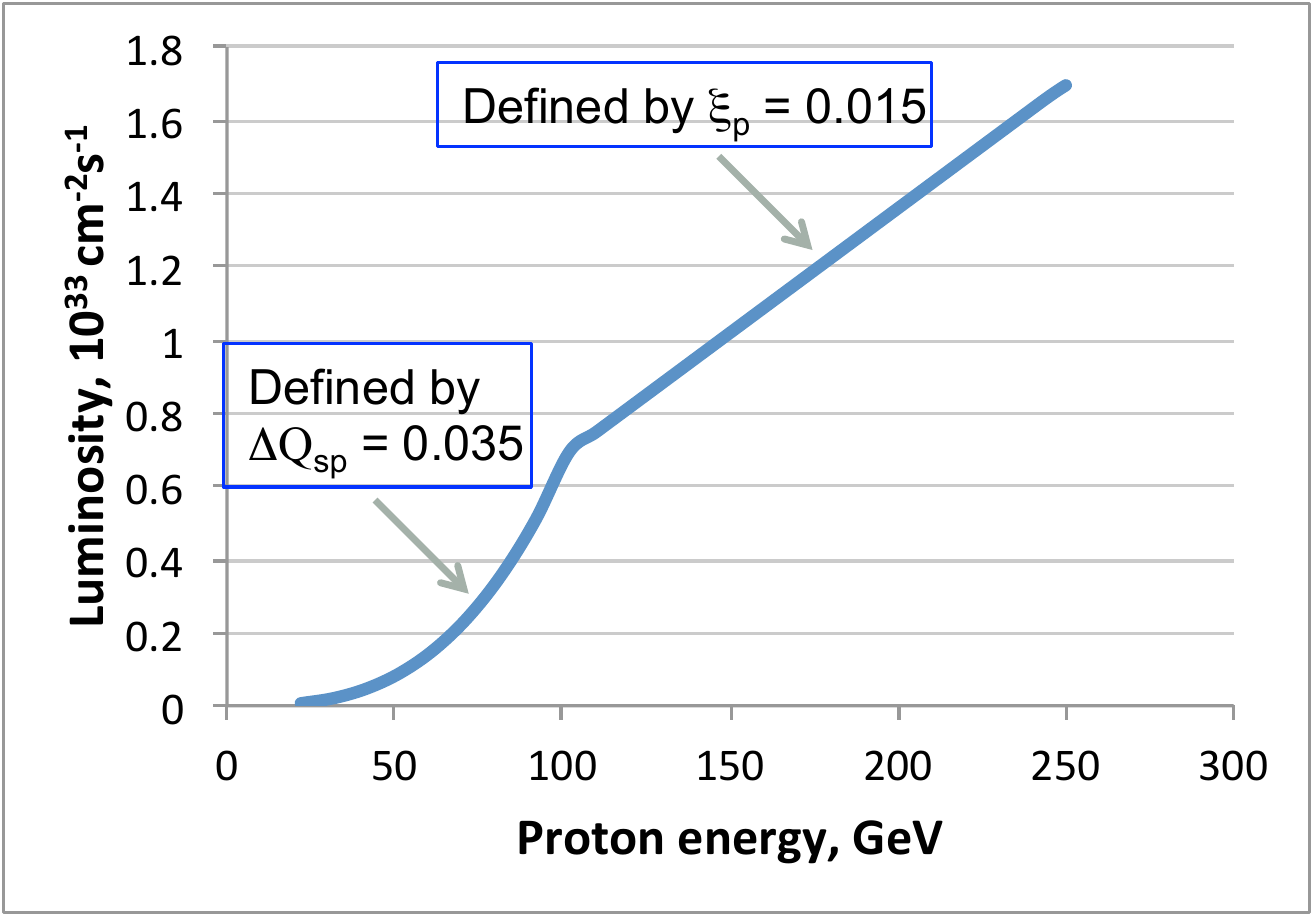}
\includegraphics[width=.51\textwidth]{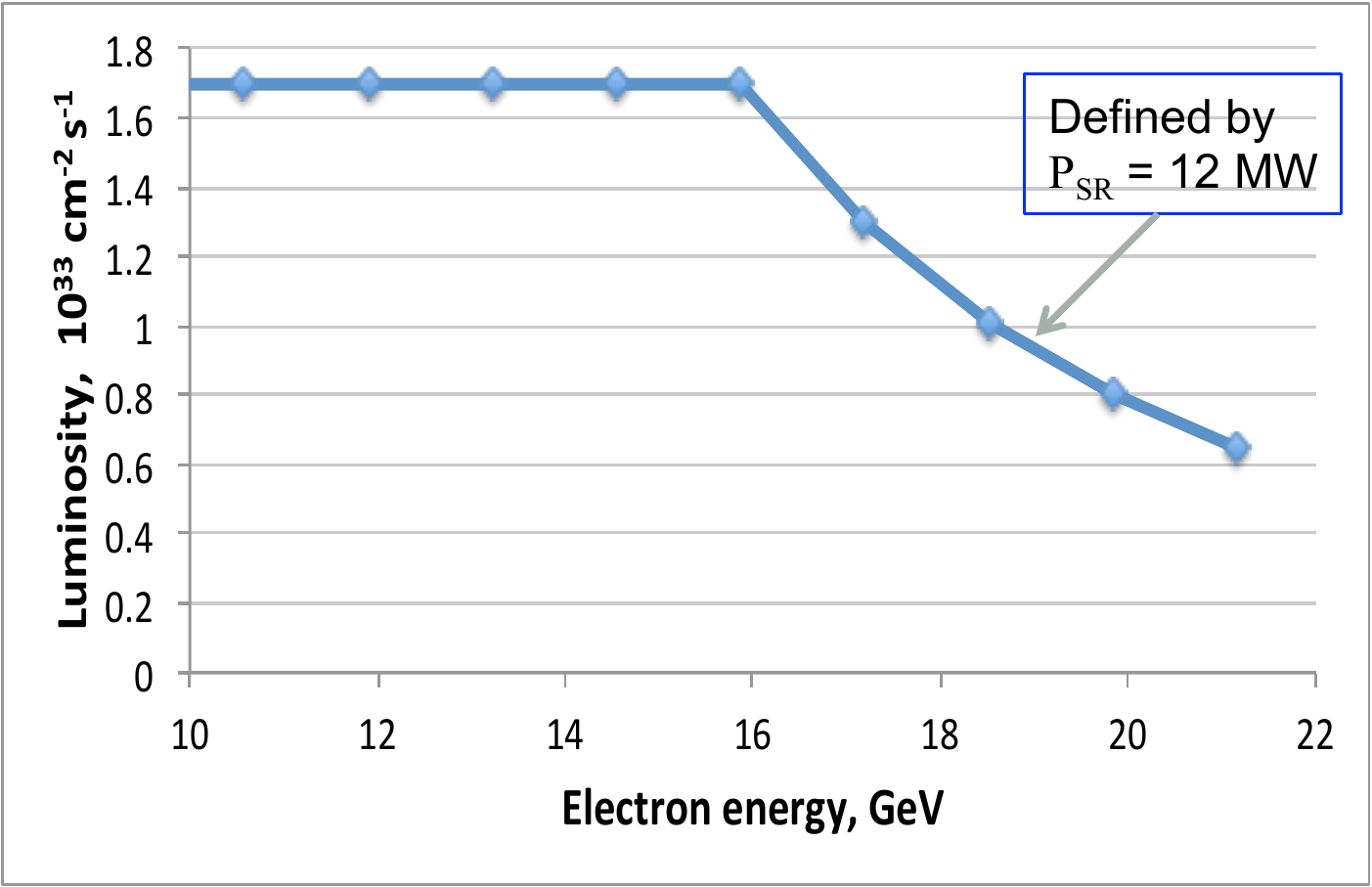}
\end{center} 
\vskip -0.2in
\caption{\label{fig:roser:eRHIC_lumi_plot}eRHIC luminosity as a
function of electron and proton beam energy. Left plot shows the luminosity dependence
on proton energy at $E_e = 15.9 \gev$. Right plot presents the luminosity 
as a function of electron energy at $E_p = 250 \gev$.}
\end{figure*}

\begin{multicols}{2}
The eRHIC design was guided by beam dynamics limitations
experimentally observed at existing colliders such as  beam-beam tune spread of less than
0.015 and accelerator technology limits such as the focusing required
to reach $\beta^*$ = 5 cm for hadron beams. The incoherent space
charge tune spread is limited to about 0.035 to support an adequate beam lifetime. 
 For practical and cost considerations, we limited the
maximum electron beam power loss due to synchrotron radiation to about
12 MW, which corresponds to a 50 mA electron beam current at 16 GeV and
about 18 mA at 21 GeV. This means that the luminosity of eRHIC
operating with 21 GeV electrons will be about 35\% of the luminosity
at 16 GeV or lower electron energy. The luminosity reachable with eRHIC is shown
in Fig.~\ref{fig:roser:eRHIC_lumi_plot} as a function of electron and
proton beam energy.

\begin{figure*}[th!]
\begin{center}
\includegraphics[width=.49\textwidth]{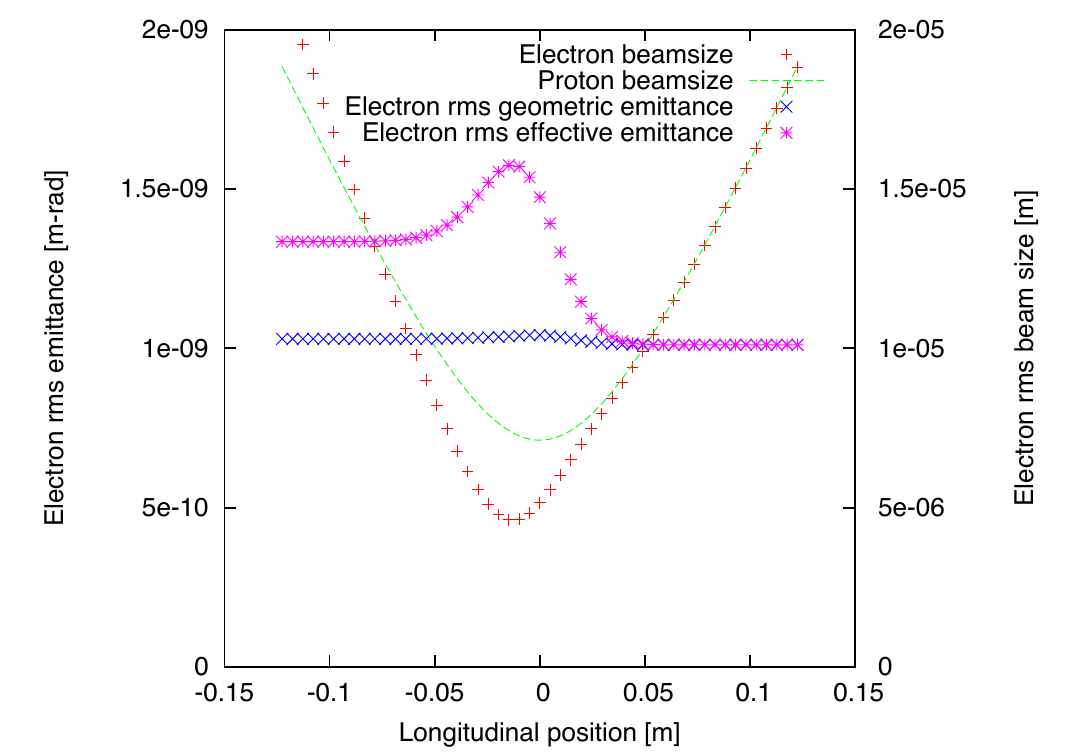}
\includegraphics[width=.49\textwidth]{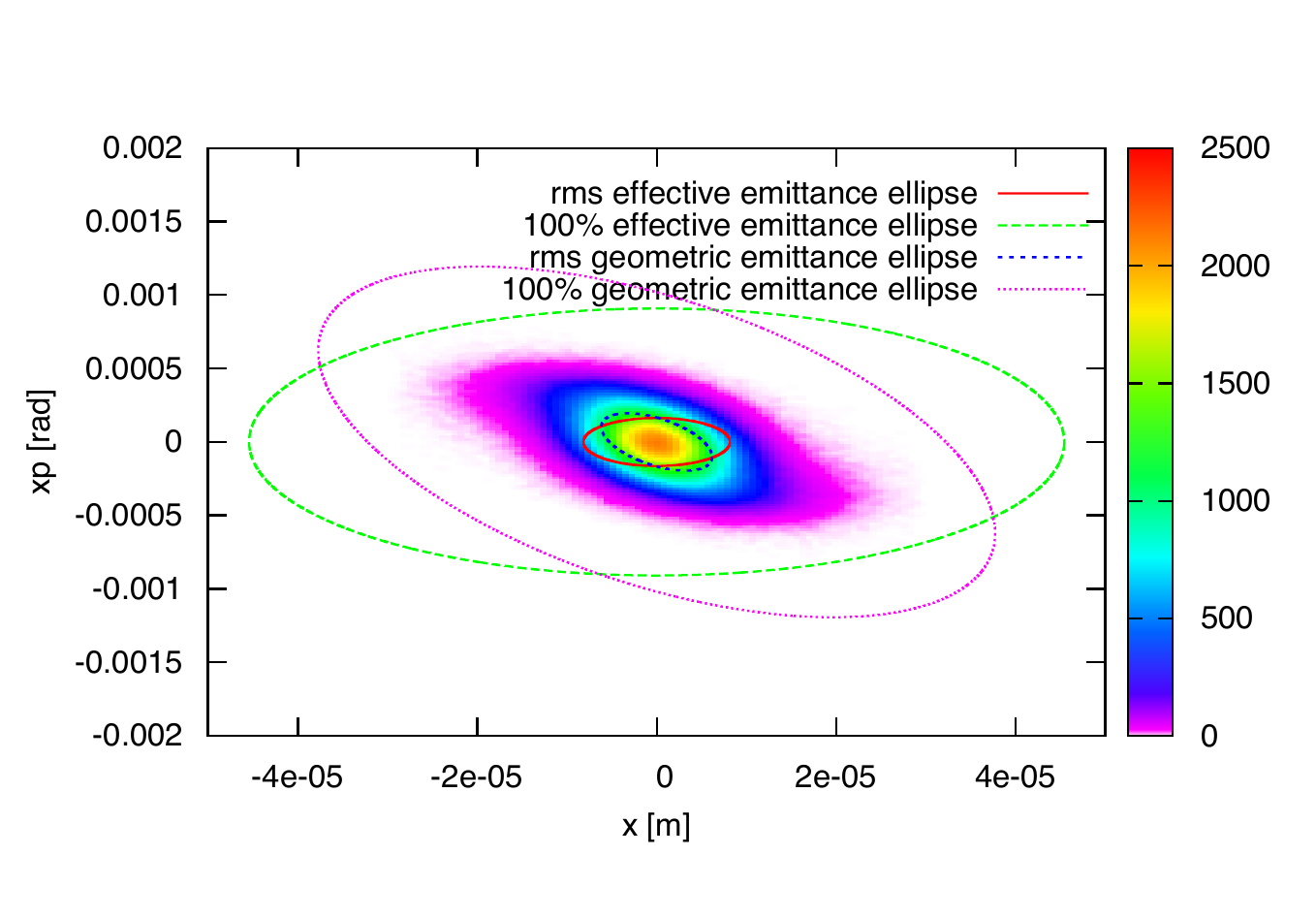}
\end{center} 
\vskip -0.2in
\caption{\label{fig:roser:e-beam_collision}The evolution of size and
emittance of the e-beam during the collision (left) and distribution of
electrons after collision with the hadron beam in eRHIC (right).}
\end{figure*}

Since the ERL provides fresh electron bunches for every collision, the
electron beam can be strongly distorted during the collision with the
much stiffer hadron beam. This allows for greatly exceeding the
beam-beam interaction limit that would apply for an electron beam in a
storage ring. The electrons are strongly focused during the collision
with the hadron beam (pinch effect), and the electron beam emittance
grows by about 30\% during the collision as shown in
Fig.~\ref{fig:roser:e-beam_collision}. This increased beam emittance
can still be easily accommodated by the beam transport during
deceleration in the ERL. The only known effect of concern is the
so-called kink instability. However, the ways of suppressing this
instability within the range of parameters accessible by eRHIC are
well understood.
Small transverse and longitudinal beam emittances of the hadron beam in eRHIC 
are of critical importance, both for the attainment of high luminosity as well as for 
separating and detecting collision products scattered at small angles
from the core of the hadron beam.  For instance, the transverse emittance should be 
 about ten times smaller than presently available in
hadron machines. This requires a level of beam cooling that can only
be achieved using Coherent electron Cooling (CeC), a novel form of
beam cooling that promises to cool ion and proton beams by a factor of
10, both transversely and longitudinally, in less than 30
minutes. Traditional stochastic or electron cooling techniques could
not satisfy this demand. CeC will be tested in a proof-of-principle
experiment at RHIC by a collaboration of scientists from BNL, JLab,
and TechX.

Unlike ring-ring colliders, the ERL allows for easy synchronization of
the electron beam with the hadron beam in RHIC over a wide energy
range from 20 to 250 GeV/n by using various sub-harmonics of the ERL
RF frequency for the electron bunches, plus tuning a warm magnet delay
 line in a straight section of the hadron ring by up to 15 cm. The ERL concept also
allows for full electron beam polarization with longitudinal direction
at the interaction point (IP) over the whole energy range. The
electron polarization from the polarized electron gun, with the sign
selectable for each bunch, is precessing in the horizontal plane
during acceleration over multiple recirculation passes in the ERL. In order to  
preserve the polarization at 80\% level special RF cavities operating at 5th harmonic
of main ERL cavities are applied to reduce beam energy spread and related spin decoherence.
The choice of the ERL energy, 1.322 GeV, warranties the same polarization orientation at all experimental IPs.

The eRHIC hadron bunch intensity is significantly smaller than the one used during
 present RHIC operation.  It leaves a straightforward path for a future luminosity 
 upgrade, HL-eRHIC, by increasing the proton intensity  to $3 \cdot 10^{11}$ p/bunch. 
 Since the bunch length of the eRHIC hadron beam is small due to cooling, 
 hadron ring upgrades will be required to allow for this intensity increase:
  copper coating of the beam pipe, an upgrade of the beam instrumentation 
  and a new RF system. With these relatively modest upgrades the luminosity of 
  the collider can be improved by one order of magnitude, exceeding $10^{34} \lumi$ (Tab.~\ref{roser:eRHIC_parameters}). 

\end{multicols}

\subsection{eRHIC Interaction Region}

\begin{figure*}[h!]
\begin{center}
\includegraphics[width=0.9\textwidth]{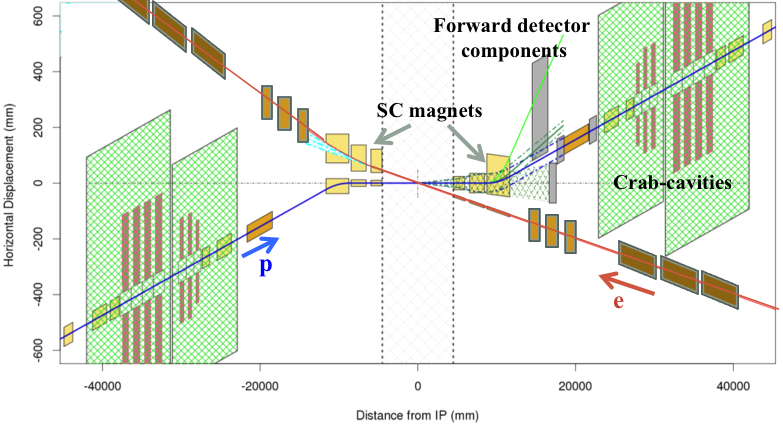}
\end{center}
\vskip -0.2in
\caption{\label{fig:roser:eRHIC_interaction_region} The layout of the eRHIC interaction region. }
\end{figure*}

\begin{multicols}{2}
In the eRHIC IR design, the hadron and electron beam trajectories cross inside the
detector at a 10 mrad horizontal angle,  as shown in Fig.~\ref{fig:roser:eRHIC_interaction_region}. The
hadron beam is focused to $\beta^*$ = 5 cm using both strong focusing  
by superconducting quadrupoles and an artificially excited $\beta$-function wave (ATS technique).
The quadrupoles closest to the detector have a field gradient as high as 170~T/m,
necessitating the application of superconducting magnet technology. 
Additional detector components are placed 
downstream of the hadron beam trajectory after a 16~mrad bending magnet  that
separates the beam and collision products of interest. The chromaticity 
correction is arranged with arc sextupole families, taking advantage of the 90 degree 
lattice of the hadron ring arcs.

Head-on collisions of the electron and hadron bunches are restored
with crab cavities located on either side of the interaction
region. With a hadron ring lattice that provides large beta functions
at the location of the crab cavities, an integral transverse RF field
of 16 MV on either side will provide the required 5 mrad bunch
rotation. The crab cavities for the electron ring are much more modest
requiring only about 2 MV transverse RF field.

The design of hadron superconducting magnets includes a free-field pass for the
electron beam which is arranged for some magnets through the low-field area 
between the superconducting coils and for other magnets through their iron yoke. This
configuration guarantees the absence of harmful high-energy X-ray
synchrotron radiation in the vicinity of the detector. Furthermore, the electron beam 
is brought into the collision via a 130-meter long merging system, of which the last
60 meters use only soft bends with a magnet strengths of less than 10
mT and less than 3 mT for the final bend. Only 1.9 W of soft radiation
from these magnets would propagate through the detector.

\end{multicols}
\vskip -0.5cm

\subsection{eRHIC R\&D}
\begin{figure*}
\begin{center}
\includegraphics[width=0.90\textwidth,height=2.0in]{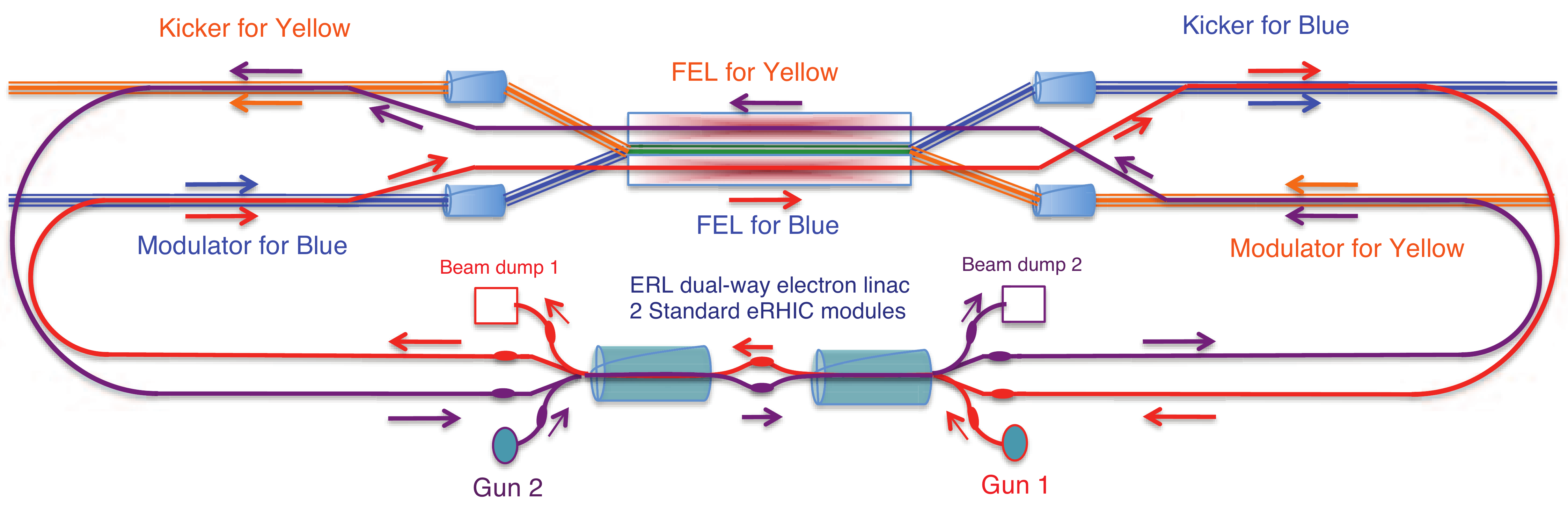}
\end{center}
\vskip -0.1in
\caption{\label{fig:roser:RHIC_CeC_layout} A possible layout of RHIC CeC
system cooling for both the yellow and blue beams.}
\end{figure*}

\begin{multicols}{2}
R\&D for eRHIC is focusing on three main areas. To study the behavior
of an ERL at very high beam intensity, an R\&D ERL that can accommodate
up to 500 mA electron current is being assembled at BNL using a
specially optimized 5-cell 704 MHz SRF cavity with design features that are similar 
to cavities planned at the eRHIC ERL.
The second project is the demonstration of Coherent
electron Cooling (CeC) in RHIC using a 20 MeV high brightness electron
bunch to cool a 40 GeV/n gold
bunch. Figure~\ref{fig:roser:RHIC_CeC_layout} shows a possible layout
of CeC for RHIC. Finally, two efforts are underway to demonstrate the
feasibility of producing a 50 mA polarized electron beam. One is based
on a single large GaAs cathode and the other employes multiple GaAs
cathodes that are used one at a time and the electron bunches are then
combined with a rotating dipole field into a continuous electron beam.
\end{multicols}


\vskip 1.0cm

\section{MEIC }

\subsection{Jefferson Lab Staged Approach}

\begin{multicols}{2} 
The JLab response to U.S. user demand is to
propose MEIC ~\cite{MEICDesignReport} based on the 12 GeV CEBAF recirculating SRF linac. This
first stage of the JLab EIC program aims to cover a medium CM energy
range up to 65 GeV while meeting all other facility requirements.
This approach achieves an optimized balance among the science program,
technology R\&D and project cost. The MEIC design maintains capability
for future upgrades with maximum flexibility for changes in science
goals and for cost-saving facility equipment reuse. Presently, MEIC
is designed to collide 3 to 12 GeV electrons with 25 to 100 GeV protons
or up to 40 GeV/u light to heavy ions, reach luminosities above $10^{34}$
cm$^{-2}$s$^{-1}$ per interaction point (IP), and deliver 80\% polarization
for both electron and light ion beams. An envisioned upgrade would
provide full coverage of the CM energy range up to 140 GeV or above,
and boost the peak luminosity close to $10^{35}$ cm$^{-2}$s$^{-1}$
per IP.

\begin{figure*}[h!]
\begin{center}
\vspace{-0.1in}
\includegraphics[width=0.70\textwidth]{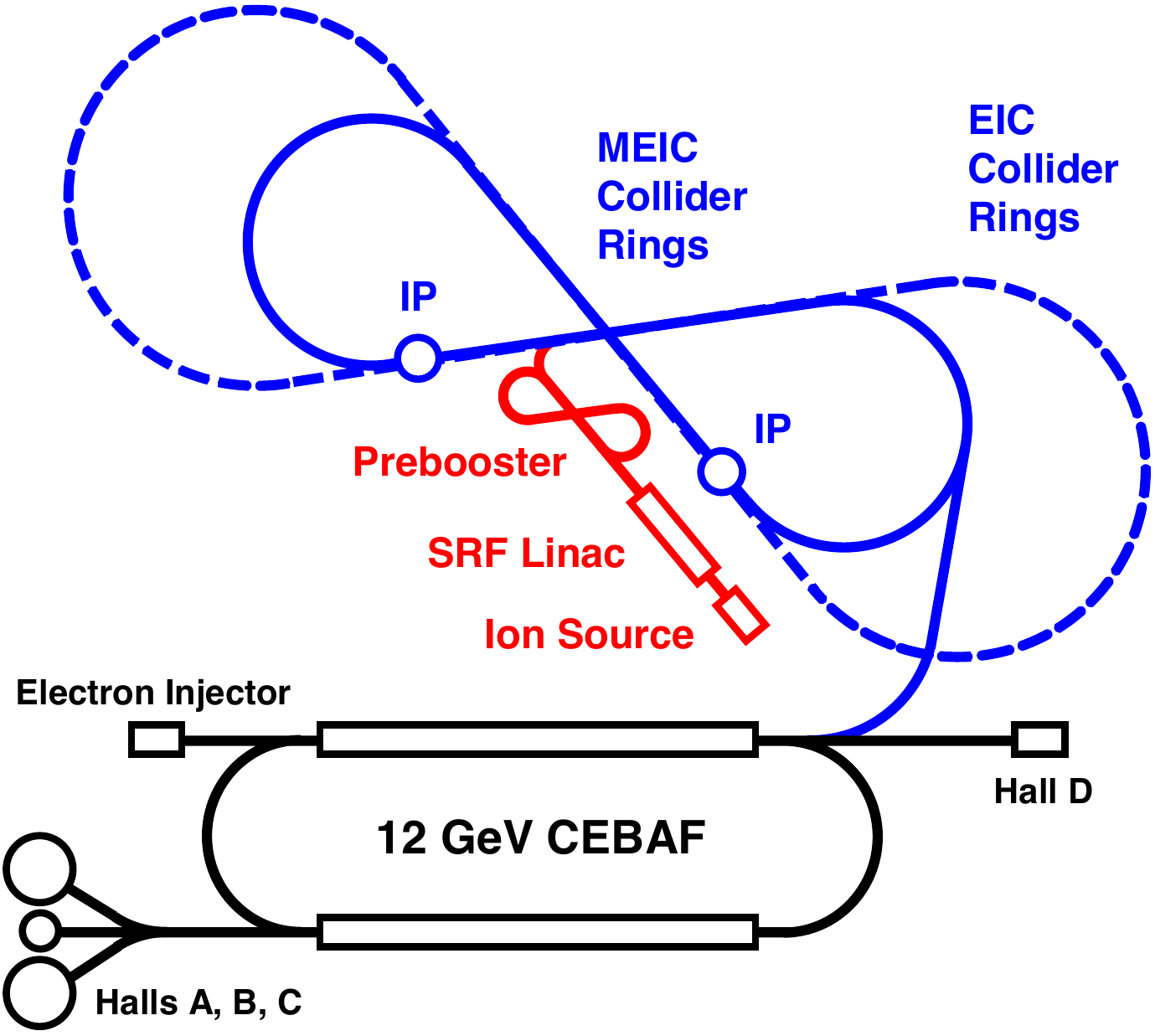}
\end{center}
\vskip -0.2in
\caption{\label{fig:MEIC_layout} Schematic layout of MEIC}
\end{figure*}

MEIC, designed as a traditional ring-ring collider and shown in Figure
~\ref{fig:MEIC_layout}, takes advantage of several unique machine design features for delivering
high performance. It utilizes a high repetition rate CW electron
beam from the CEBAF and matched ion beams from a new ion facility.
This enables MEIC to adopt a luminosity concept ~\cite{Derbenev2010HB} which is based on
high bunch repetition rate CW colliding beams and has been successfully
proven in several lepton-lepton colliders for achieving an ultra high
luminosity. A multi-phased cooling scheme ~\cite{Derbenev2009Electron} provides strong cooling
of ion beams not only at their formation stage but also during collisions.
The MEIC collider rings and ion boosters are in a figure-8 shape which
is a revolutionary solution ~\cite{Derbenev1996UM} for preserving and controlling the beam
polarization during acceleration and storage in a synchrotron. This
design feature can deliver superior polarization of ion beams for
experiments and is also the only practical way for accelerating and
storing a medium or high energy polarized deuteron beam. Furthermore,
the interaction regions are designed to provide ultra high to essentially
full detector acceptance capability. \end{multicols}

\begin{figure*}[h!]
\begin{center}
\includegraphics[width=0.80\textwidth]{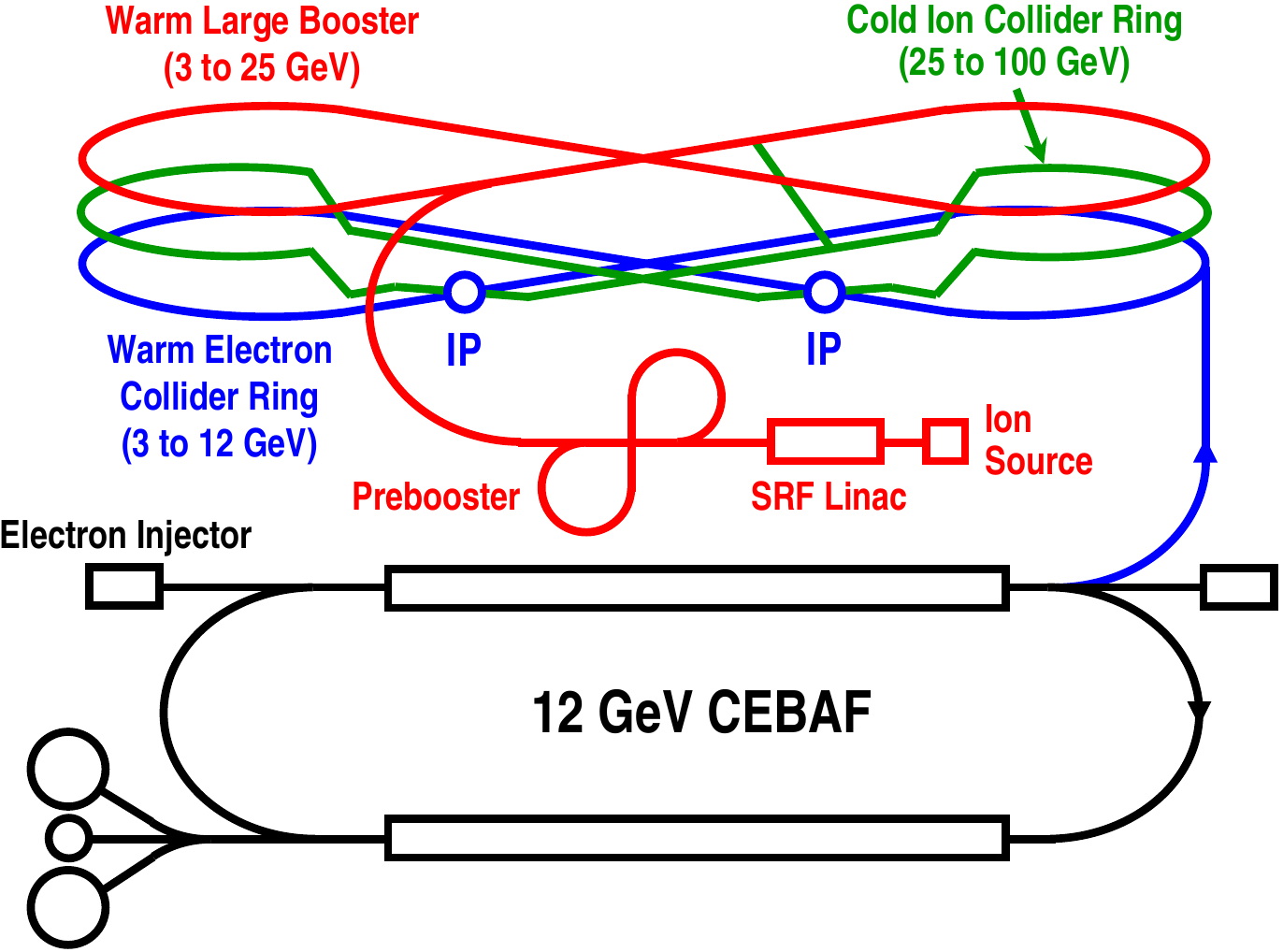}
\end{center}
\vskip -0.18in
\caption{\label{fig:MEIC_illustr} An illustration of MEIC, the large booster and collider rings are vertically stacked in a common tunnel. }
\end{figure*}

\subsection{Baseline Design}

\begin{multicols}{2} 
The two MEIC collider rings are stacked vertically
in a tunnel as illustrated in Figure~\ref{fig:MEIC_illustr}. The electron ring stores 3
to 12 GeV polarized electrons injected at full energy from the CEBAF,
while the SC ion collider ring stores 25 to 100 GeV protons or fully
stripped light to heavy ions with energies of the same magnetic rigidity.
The ions execute a vertical excursion to the plane of the electron
ring for collision at two IPs. An optional third detector may be placed
at another IP. There is a third ring, the ion large booster with energy
from 3 to 25 GeV made of normal conducting magnets, in the same tunnel
and stacked above the collider rings. The large figure-8 shaped ring
(the dashed line) in Fig.~\ref{fig:MEIC_layout} represents a future energy
upgrade for reaching up to 20 GeV electrons and 250 GeV protons or
100 GeV/u ions. 

Table~\ref{firsttable} summarizes the MEIC parameters at a design
point of 5 GeV electrons colliding with 60 GeV protons ~\cite{MEICDesignReport}. The luminosity
reaches $5.6\times10^{33}\mathrm{\ cm^{-2}s^{-1}}$ for a full-acceptance
detector. To reach such an acceptance, the machine-element-free detector
space must be 7 m for ion beams on the down-stream side; however,
it can be shortened to 3.5 m on the up-stream side. For the second
detector which is optimized for higher luminosities while still maintaining
a large detector acceptance, the detector space for ion beams can
be reduced to 4.5 m  so that the luminosity is increased to above
$10^{34}$ cm$^{-2}$s$^{-1}$.

MEIC achieves high luminosities through a design choice characterized
by a very high bunch repetition rate for both colliding beams ~\cite{Derbenev2010HB}. It
is about one to two orders of magnitude higher than that of the typical
hadron colliders, however, it is similar to the bunch repetition rate
of $e^{+}e^{-}$ colliders. The bunch intensities of such beams are
extremely small (usually by one to two orders of magnitude) even
though the average current is several Amperes. This opens up the possibility
of very short bunch lengths for the ion beams, thus enabling a drastic
reduction of the final focusing beta-star (to several cm or even less).
As a result, the collider can reach a high luminosity. 

Table~\ref{secondtable} shows
the luminosities of e-ion collisions for several different ion species.

To derive this parameter set, certain limits were imposed on several
machine or beam parameters in order to improve robustness of the design
and to reduce accelerator R\&D. These limits are largely based on
previous experiences of lepton and hadron colliders and the present
state of the art of accelerator technologies. They include 
\begin{itemize}
\itemsep -0.3em
\item Ion SC magnet field is up to 6 T.
\item The stored beam currents are up to 0.5 A for protons or ions and 3
A for electrons.
\item The electron synchrotron radiation power should not exceed 20 kW/m.
\item Maximum betatron function near an IP is 2.5 km. 
\end{itemize}
\end{multicols}

\begin{table}[htb]
\begin{centering}
\begin{tabular}{lccc}
\hline 
 &  & Proton  & Electron \tabularnewline
Beam energy  & GeV  & 60  & 5 \tabularnewline
Collision frequency  & MHz  & \multicolumn{2}{c}{748.5}\tabularnewline
Beam current/Particles per bunch  & A/$10^{10}$  & 0.5/0.416  & 3/2.5 \tabularnewline
Polarization  &  & $\sim$80\% & $\sim$80\%\tabularnewline
RMS bunch length  & mm  & 10  & 7.5 \tabularnewline
Normalized emit. ($\varepsilon_{x}$ / $\varepsilon_{y}$)  & $\mathrm{\mu m}$ & 0.35/0.07  & 53.5/10.7 \tabularnewline
Horizontal and vertical $\beta^{*}$  & cm  & \multicolumn{2}{c}{10/2 (4/0.8)}\tabularnewline
Vertical beam-beam tune shift  &   & 0.015 & 0.03\tabularnewline
Laslett tune-shift  &  & 0.06  & Small \tabularnewline
Detector space  & m  & $\pm$7 (4.5)  & $\pm$3.5 \tabularnewline
Luminosity per IP  & $10^{33}$cm$^{-2}$s$^{-1}$  & \multicolumn{2}{c}{5.6 (14.2)}\tabularnewline
\hline 
\end{tabular}
\par\end{centering}
\begin{centering}
\protect\caption{\label{firsttable} MEIC parameters at an example design point of particle energies for a full-acceptance detector (values for a high-luminosity detector are given in parentheses)}
\par\end{centering}
\end{table}

\begin{table}[htb]
\small
\center
\begin{tabular}{| l | c | c | c | c | c | c | c |}
\hline
&       & e      & P	    & D	       & ${}^3$He$^{++}$ & ${}^{40}$Ca$^{20+}$&${}^{208}$Pb$^{82+}$ \\ \hline
Energy & GeV/u & 6      & 100    & 50       & 66.7   & 50       & 40   \\ \hline
Current&	A  & 3 & \multicolumn{5}{|c|}{0.5} \\ \hline
Particles per bunch&	$10^9$& 25 &	4.2&	4.2&	2.1&	0.2&	0.05 \\ \hline
\multirow{2}{*}{$\beta^{*}$ (x/y)} & \multirow{2}{*}{cm }&1.6 to 2.8& \multicolumn{5}{|c|}{\multirow{2}{*}{6/2 (2.4/0.8)}}\\
 & & (0.61 to 1.1) & \multicolumn{5}{|c|}{}\\ \hline
Beam-beam tune  & \multirow{2}{*}{ }& 0.023 &  \multirow{2}{*}{0.014}	&\multirow{2}{*}{0.008}	&\multirow{2}{*}{0.01}&	\multirow{2}{*}{0.008}&	\multirow{2}{*}{0.006} \\ 
shift (vertical)& & to 0.029 & & & & & \\ \hline
\multirow{2}{*}{Luminosity/IP } & \multirow{2}{*}{$10^{34}$cm$^{-2}$s$^{-1}$} & &0.8 & 1.1 & 1.1 & 1.1 & 1.1 \\
& & &(2.1) & (2.8) & (3.7) & (2.8) & (2.8) \\ \hline
\end{tabular}
\caption{\label{secondtable}
MEIC luminosities for different ion species (values for a high-luminosity detector with a 4.5 m ion detector space are given in parentheses.)}
\end{table}

\subsection{Ion Complex}
\begin{figure*}[th!]
\begin{center}
\includegraphics[width=0.90\textwidth]{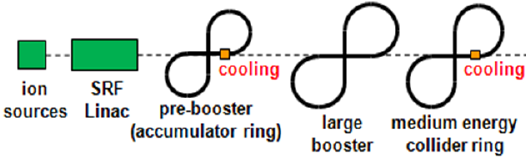}
\end{center}
\vskip -0.1in
\caption{\label{fig:MEIC_ion} Schematic layout of the ion complex. }
\end{figure*}

\begin{multicols}{2}
Figure~\ref{fig:MEIC_ion} illustrates the schematic
layout of the MEIC ion complex. The ions from polarized or un-polarized
sources are accelerated step-by-step to the colliding energies in
the following major machine components: a 285 MeV pulsed SRF linac,
a 3 GeV pre-booster synchrotron, a 25 GeV large booster synchrotron,
and finally a collider ring of 25 to 100 GeV. 

\noindent \textbf{Ion Sources: } 
The MEIC ion sources will rely on
existing technologies: an Atomic Beam Polarized Ion Source (ABPIS) ~\cite{Clegg1995ANew}
with Resonant Charge Exchange Ionization for producing polarized light
ions $\mathrm{H^{-}/D^{+}}$ and $\mathrm{^{3}He^{++}}$, and an Electron Beam
Ion Source (EBIS) similar to the one currently in operation at BNL ~\cite{Alessi2006Status}
for producing un-polarized light to heavy ions. 

\noindent\textbf{Ion Linac: }
The technical design of a pulsed SRF ion linac ~\cite{Mustapha2007ADriver},
originally developed at Argonne National Laboratory as a heavy-ion
driver linac for FRIB, has been adopted for the MEIC proposal. This
linac is very effective in accelerating a wide variety of ions from
H$^{-}$ to $^{208}$Pb$^{30+}$.

\noindent \textbf{Pre-Booster/Accumulator Ring: } 
The pre-booster synchrotron
accepts linac pulses of any ion species, after accumulation and acceleration,
transfers them to the large booster for further acceleration. It utilizes
either the painting technique for H$^{-}$/D$^{-}$ or the DC electron
cooling for lead or other heavy ions during multi-turn injections
from the linac. 

\noindent \textbf{Large Booster: } 
This booster synchrotron is responsible for
accelerating protons to 25 GeV or ions to 12.5 GeV/u before transporting
them to the collider ring. Its circumference is four times that of
the pre-booster.

A key design requirement for both booster synchrotrons is
sufficiently high transition gamma such that the ions never cross
the transition energy during acceleration in order to prevent particle
loss associated with such a crossing. 
\end{multicols}

\subsection{Collider Rings}
\begin{multicols}{2} 
The two MEIC collider rings have nearly identical footprints with
a circumference of approximately 1470 m. The figure-8 crossing angle
is 60$^{\circ}$. The two rings intersect at two symmetric points
in two long straights for medium energy collisions. A third crossing
point can be arranged for an extra detector. The long straights also
accommodate necessary utility components such as injection, ejection,
RF systems and electron cooling. One universal spin rotator consisting
of two SC solenoids and two sets of arc dipoles is placed at each
end of the two electron arcs. In the ion collider ring, a transition
from a low bunch frequency to 750 MHz repetition rate takes place.
\end{multicols}

\subsection{Interaction Regions}
\begin{multicols}{2} 
The primary detector of MEIC is unique in its ability to provide essentially full acceptance to all fragments from collisions. The interaction region (IR) design ~\cite{Morozov2012IPAC} is optimized to support this detector acceptance. It relies on several features including a relatively large 50 mrad crab crossing angle, large-aperture final focusing (FF) quadrupoles and spectrometer dipoles as well as a large (7 m) machine-element-free detection space downstream of the ion beam.

\setlength{\parskip}{0pt}
The large crab crossing angle of the MEIC design not only allows quick separation of the two colliding beams near an IP for avoiding parasitic collisions and makes sufficient space for placement of IR magnets but also moves the spot of poor resolution along the solenoid axis into the periphery and minimizes the shadow of the electron FF quadrupoles. Crab cavities will be utilized for restoring head-on collisions.

The IR design takes special care to minimize radiation in the detectors and maintain good background. Bending of electrons in the straight sections is reduced to the minimum, thus the ions are arranged to travel to the plane of the electron ring for collisions. The electron beam line is parallel to the detector solenoid axis for avoiding extra bending by the solenoid. The detectors are placed far from the electron arc exits to minimize the synchrotron radiation background and close to the ion arc exits to minimize the hadronic background due to the ion beam scattering on the residual gas.

\begin{figure*}[hb!]
\begin{center}
\includegraphics[angle=0,width=0.9\textwidth]{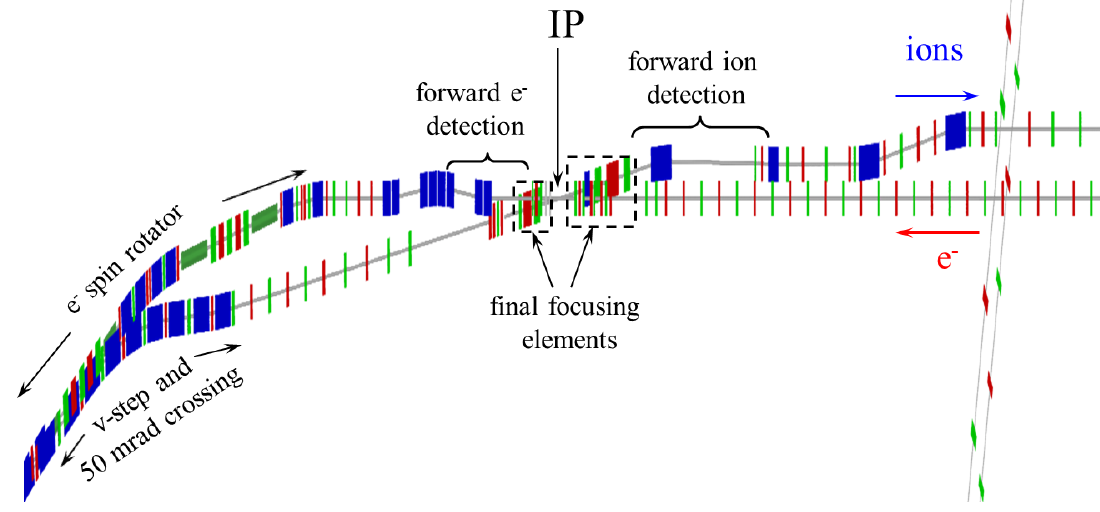}
\end{center}
\vskip -0.2in
\caption[figure]{\label{fig:MEIC_IP} The layout of the interaction region associated with a full acceptance detector.}
\end{figure*}

\setlength{\parskip}{0pt}
Figure~\ref{fig:MEIC_IP} shows the layout of an IR. The end section of the ion arc upstream of an IP is shaped to produce a net 50 mrad horizontal angle between the ion and electron beams while the ion beam line segment downstream of the IP is designed to make a 2 m transverse separation between the ion and electron beams. 
 
\begin{figure*}[ht!]
\begin{center}
\includegraphics[angle=0,width=0.55\textwidth]{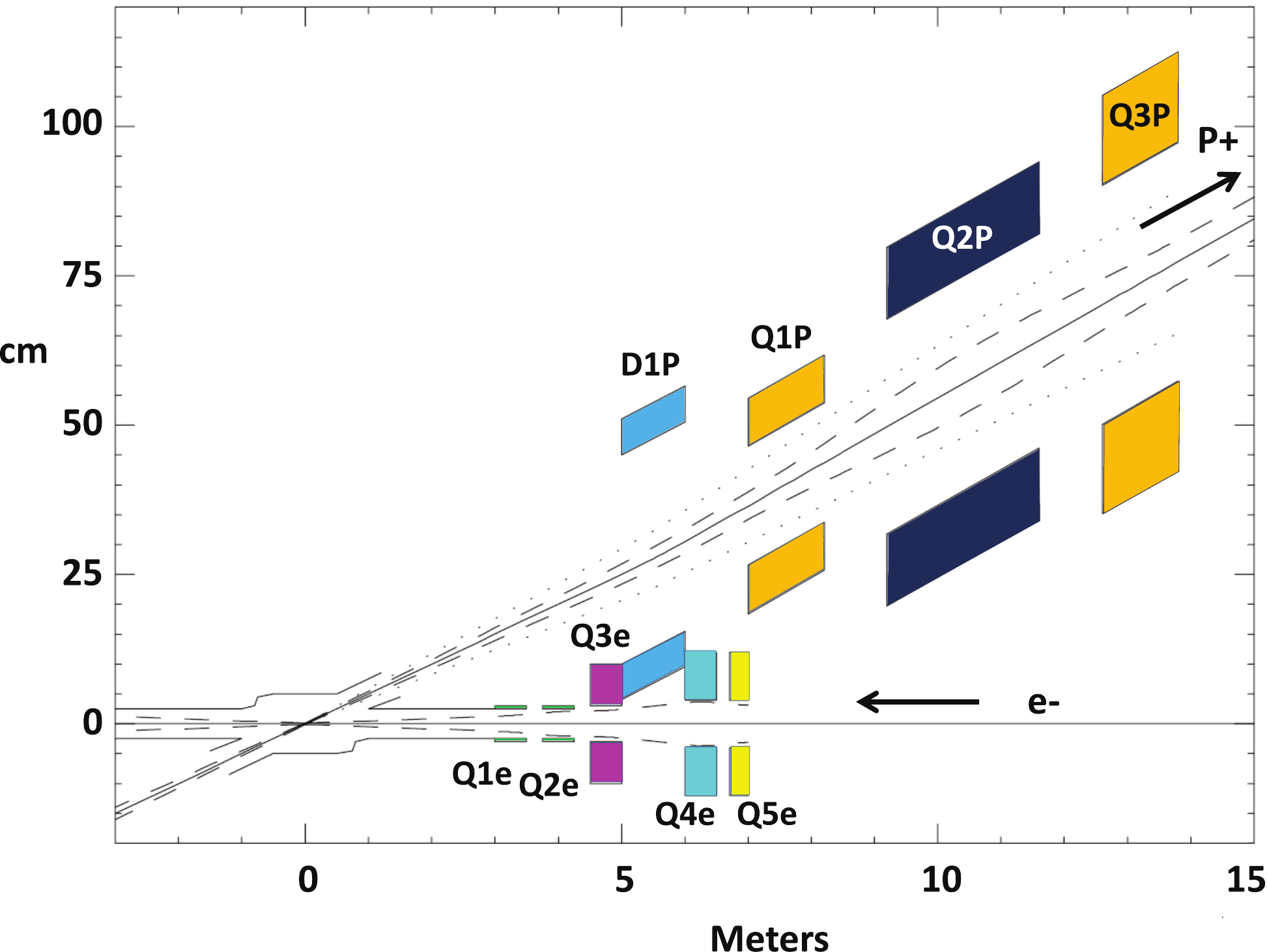}
\end{center} 
\vskip -0.2in
\caption[figure]{\label{fig:MEIC_IP2} Layout of the IR in the forward ion direction.}
\end{figure*}

Due to kinematic considerations, more detector space is needed downstream of the IP than upstream along the ion beam direction. Consequently, the upstream ion final focusing block (FFB) is placed closer to the IP (at a distance of 3.5 m) than the downstream one (at a distance of 7 m), yielding an asymmetric detector region. Each ion FFB is a quadrupole triplet allowing for a more flexible control of the beta functions. Electron FFBs are also based on quadrupole triplets but include additional permanent-magnet quadrupoles placed at the front of the FFBs. The permanent-magnet quadrupoles have a small size and can be placed closer to the IP. Change of their focusing strength with energy is compensated by adjusting the regular electromagnetic FFB quadrupoles. The electron FFBs are placed 3 m away from the IP. The downstream ion and electron FF quadrupoles are designed with large apertures for forward detection and are followed by spectrometer dipoles. Additionally, there is a weak spectrometer dipole in front of the downstream ion FFB. Such a design shown in Figure~\ref{fig:MEIC_IP2} satisfies the detector requirements while minimizing the chromatic contribution of both the ion and electron FFBs.

Sufficient machine-element-free space is reserved beyond the downstream FFBs and spectrometer dipoles for detection purposes. Both the ion and electron beams are focused again towards the end of this element-free space to allow closer placement of the detectors, which, in combination with relatively large dispersion at those points, enhances the forward detector's momentum resolution. The dispersion generated by the spectrometer dipoles is suppressed on the ion side by a specially designed section, which also controls the beam line geometry, while on the electron side the dispersion suppression is done by a simple dipole chicane whose parameters are chosen to avoid a significant impact on the electron equilibrium emittance. 

Due to the strong beam focusing at the IPs, the chromatic effect of the FFBs in both the ion and electron collider rings is very significant and requires proper compensation. MEIC employs a local compensation approach where dedicated chromaticity compensation blocks cancel the chromatic kick of the FFBs. Initial simulations using this concept yielded encouraging results ~\cite{Morozov2013Symmetric}. Detailed studies and optimization of the non-linear dynamics are underway.
\end{multicols} 

\newpage
\subsection{Ion Polarization}
\begin{multicols}{2} 
 The MEIC is designed to preserve and control
high polarization of proton, light ion (deuteron, helium-3 and possibly
lithium), and electron beams as required by the nuclear physics program. 

A figure-8 shape is adopted for all ion booster synchrotrons and both
collider rings to preserve and control beam polarization during acceleration
and storage. The complete cancellation of the spin procession in the
two halves of a figure-8 ring leads to an energy-independent zero
spin tune and the lack of a preferred periodic spin direction so that
the polarization can be effectively controlled by small magnetic fields.
In particular, a figure-8 design is the only practical way for accelerating
polarized medium energy deuterons due to their small anomalous magnetic
moment. The polarization can be stabilized by weak solenoid fields
lower than 3 T$\cdot$m in all ion rings with polarization directions
matched at the beam injection and extraction locations. In the ion
collider ring, either longitudinal or transverse polarization can
be obtained at IPs using weak radial-field dipoles ($<$0.25 T$\cdot$m
each) for protons or weak solenoids ($<$1.5 T$\cdot$m each) for deuterons.
The required spin flipping can be implemented by changing the source
polarization, manipulating the polarization direction in the collider
ring using weak fields, or using RF magnetic fields to flip the polarization
of a stored beam. A polarization of up to 85\% for ion beams is expected ~\cite{Derbenev2013PSTP}. 

In the ion pre-booster and large booster, one small solenoid placed
in a straight section is sufficient to attain longitudinal
polarization for both deuterons and protons. The maximum integral
of longitudinal solenoid field is about 0.3 T$\cdot$m for deuterons
and 1.5 T$\cdot$m for protons. The spin tune induced by the solenoid
field is much greater than the strength of the zero-harmonic spin
resonance. 

In the collider ring, deuteron polarization can be efficiently controlled
by small solenoids. A symmetric scheme has been developed for deuteron
polarization control with two solenoids on both sides of the experimental
straight. The maximum field integral in a single solenoid at the maximum
energy does not exceed 1.5 T$\cdot$m. The proton polarization in
the collider ring can be controlled using the schemes for deuterons
as well. However, at higher proton energies, it is more efficient
to use radial fields that can be significantly lower than the longitudinal
fields.\end{multicols}

\subsection{Electron Polarization}
\begin{figure*}[h!]
\begin{center}
\includegraphics[angle=0,width=0.52\textwidth]{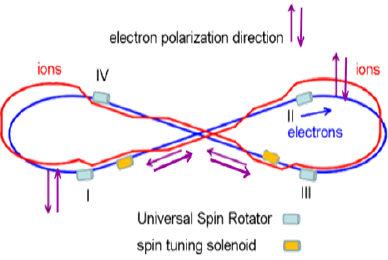}
\end{center} 
\vskip -0.2in
\caption[figure]{\label{fig:MEIC_Polarization} Polarization configuration in the MEIC electron collider ring}
\end{figure*}

\begin{figure*}[h!]
\begin{center}
\includegraphics[angle=0,width=0.52\textwidth]{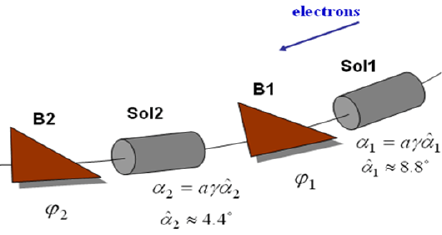}
\end{center} 
\vskip -0.1in
\caption[figure]{\label{fig:MEIC_SpinRotator} Schematic drawing of a universal spin rotator consisting of two sets of arc dipoles (B1 and B2) and two solenoids (Sol1 and Sol2).}
\end{figure*}

\begin{multicols}{2}
 A highly polarized electron beam is injected
from the CEBAF into the electron collider ring at full energy. As
shown in Figure~\ref{fig:MEIC_Polarization}, the electron polarization is designed to be vertical
in the arcs to minimize spin diffusion (i.e. depolarization) and longitudinal
at IPs for experiments. This is achieved by means of a universal spin
rotator ~\cite{Chevtsov2010JLabTN} illustrated in Figure~\ref{fig:MEIC_SpinRotator}. Four such spin rotators, located
at the ends of the two arcs, rotate the polarization in the whole
energy range, leaving the design orbit intact. 

Desired spin flipping can be attained by alternating the helicity
of the photo-injector driver laser to provide two long opposite polarization
bunch trains. The polarization configuration is chosen to have the
same polarization direction (either up or down) in the two arcs by
setting opposite solenoid fields in the two spin rotators at the both
ends of the same experimental straight ~\cite{Lin2013Polirized}. Such a configuration, with
a figure-8 shape, removes the spin tune energy dependence, therefore
significantly reducing the quantum depolarization. The spin tune can
be easily controlled by weak solenoid(s) in the experimental straights,
where the polarization is along the longitudinal direction. 

The polarization lifetime is estimated to be reasonably long (a few
hours) at low energies however it drops to tens to a few minutes at
higher energies (9 GeV and above). To obtain a high polarization in the whole energy range, continuous injection
(top-off) of highly polarized electrons from the CEBAF is used to
assist preservation of the stored beam\textquoteright s polarization,
especially at higher energies. An equilibrium polarization of up to
80\% in the whole energy range can be achieved ~\cite{Lin2013Polirized}. \end{multicols}

\vskip -0.2in

\subsection{Electron Cooling}
\begin{multicols}{2}
 Cooling of ion beams is essential to achieve
high luminosities over a broad CM energy range in MEIC. The design
relies on the traditional electron cooling method and adopts a concept
of multi-phase cooling of bunched ion beams of medium energies. Electron
cooling is first utilized for assisting accumulation of ions in the
pre-booster. It then provides initial cooling at the ejection energy
of the pre-booster, taking advantage of high cooling efficiency at
low energies. In the ion collider ring, electron cooling is used at
the injection energy and also after acceleration to the collision
energy. Most importantly, electron cooling will be operated continuously
during collisions to suppress IBS-induced beam emittance growth. Shortening
of the bunch length (1 cm or less) that results from electron cooling
of the ion beam captured in a high-voltage SRF field is critical for
high luminosity in the MEIC since it facilitates the strong
focusing of the colliding beams and also implementation of crab crossing
at the IPs for achieving an ultra-high bunch collision rate.

\begin{figure*}[h!]
\begin{center}
\includegraphics[angle=0,width=0.8\textwidth]{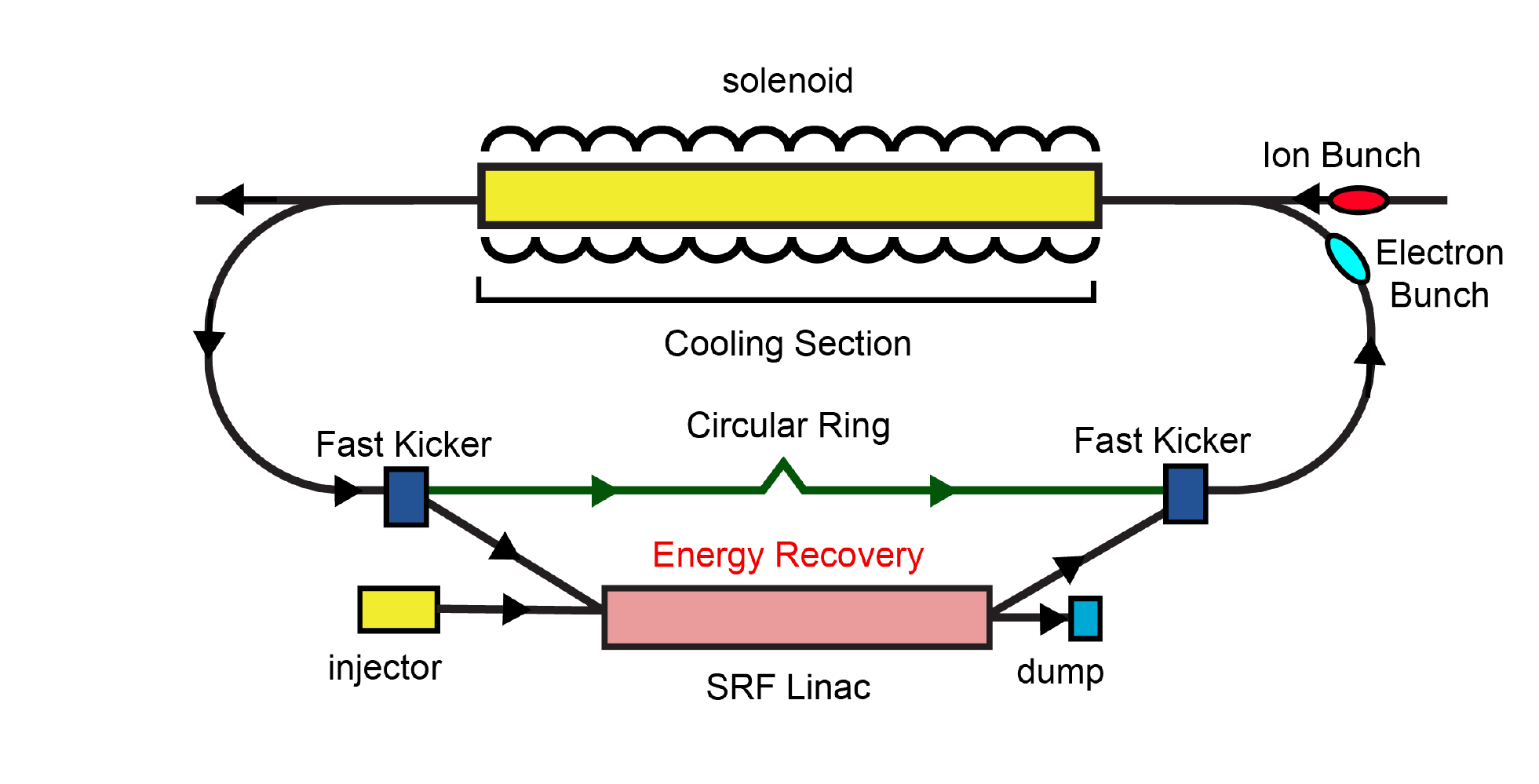}
\end{center} 
\vskip -0.2in
\caption[figure]{\label{fig:MEIC_Cooling} Schematic of electron cooling for the MEIC}
\end{figure*}

Two electron coolers are required to implement the MEIC cooling scheme.
In the pre-booster, a DC cooler with an up to 2 MeV electron beam
energy is needed and is within the state-of-art. In the ion collider
ring, an energy-recovery-linac based electron cooler ~\cite{Derbenev2009Electron} illustrated
by a schematic drawing in Figure~\ref{fig:MEIC_Cooling} will be responsible for cooling
the medium energy ions. Two accelerator technologies\textemdash an
ERL and a circulator ring\textemdash play critical roles in the success
of this facility by providing perfect solutions to the two challenging aspects
of the facility: the high current and high power of the cooling electron
beam. For example, a 1.5 A 50 MeV cooling beam (75 MW of power) can
effectively be provided by a 15 mA (30 kW of active beam
power) from the injector/ERL if the cooling beam makes 100 turns in
the circulator cooler ring. 

The MEIC will reach its ultimate full luminosity at the $10^{34} \mathrm{\ cm^{-2}s^{-1}}$
scale with envisioned electron cooling scheme utilizing a circulator cooling ring. Nonetheless,
to reduce an dependence on this scheme, the MEIC electron cooling can be implemented in various 
stages. Utilizing DC cooling at pre-booster energies - with similar requirements as the established
DC cooling at FNAL ~\cite{Nagaitsev2006Experimental} and FZ-J\"{u}elich ~\cite{Dietrich2010Status} - will already allow a peak luminosity above $3\times 10^{33} \mathrm{\ cm^{-2}s^{-1}}$
if only projecting a single-turn ERL cooler without a circulator ring. 
\end{multicols}
\vskip -0.1in


%% file: files_tex/detector.tex

{\large\it Conveners:\ Elke C. Aschenauer and Tanja Horn}
\label{sec:detectors}

\section{Introduction}
\label{sec:det-intro}

The physics program of an EIC imposes several challenges on the design
of a detector, and more globally the extended interaction region, as
it spans a wide range in center-of-mass energy, different combinations
of both beam energy and particle species, and several distinct physics
processes. The various physics processes encompass inclusive
measurements ($ep/A \rightarrow e'+X$), which require the detection of
the scattered lepton and/or the hadrons of the full scattered hadronic
debris for which $E-p_z^{had}$ is different from zero; semi-inclusive
processes ($ep/A \rightarrow e'+h+X$), which require detection in
coincidence with the scattered lepton of at least one (current or
target region) hadron; and exclusive processes ($ep/A \rightarrow
e'+N'/A'+\gamma/m$), which require the detection of all particles in the
reaction with high precision. The figures in Sec.~\ref{sec:kinmat}
demonstrate the differences in particle kinematics of some
representative examples of these reaction types, as well as differing
beam energy combinations. The directions of the beams are defined as
for HERA at DESY: the hadron beam is in the positive z direction
(0$^o$) and the lepton beam is in the negative z-direction (180$^o$).

\section{Kinematic Coverage}
\label{sec:kinmat}
\subsection{$y$ Coverage}

\begin{figure*}[htbp]
\begin{center}
\begin{tabular}{cc}
\includegraphics[width=0.46\textwidth]{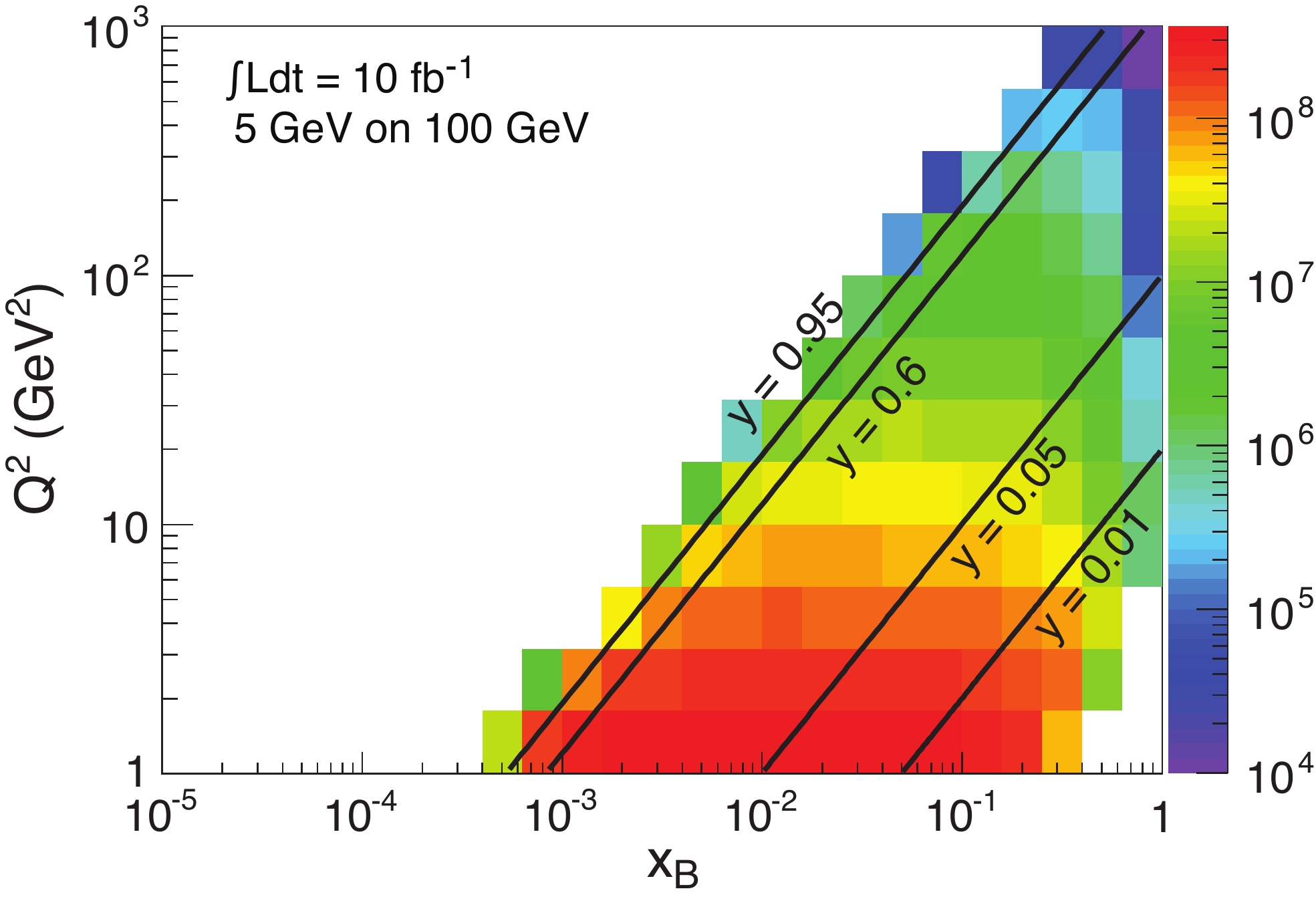}
&
\includegraphics[width=0.46\textwidth]{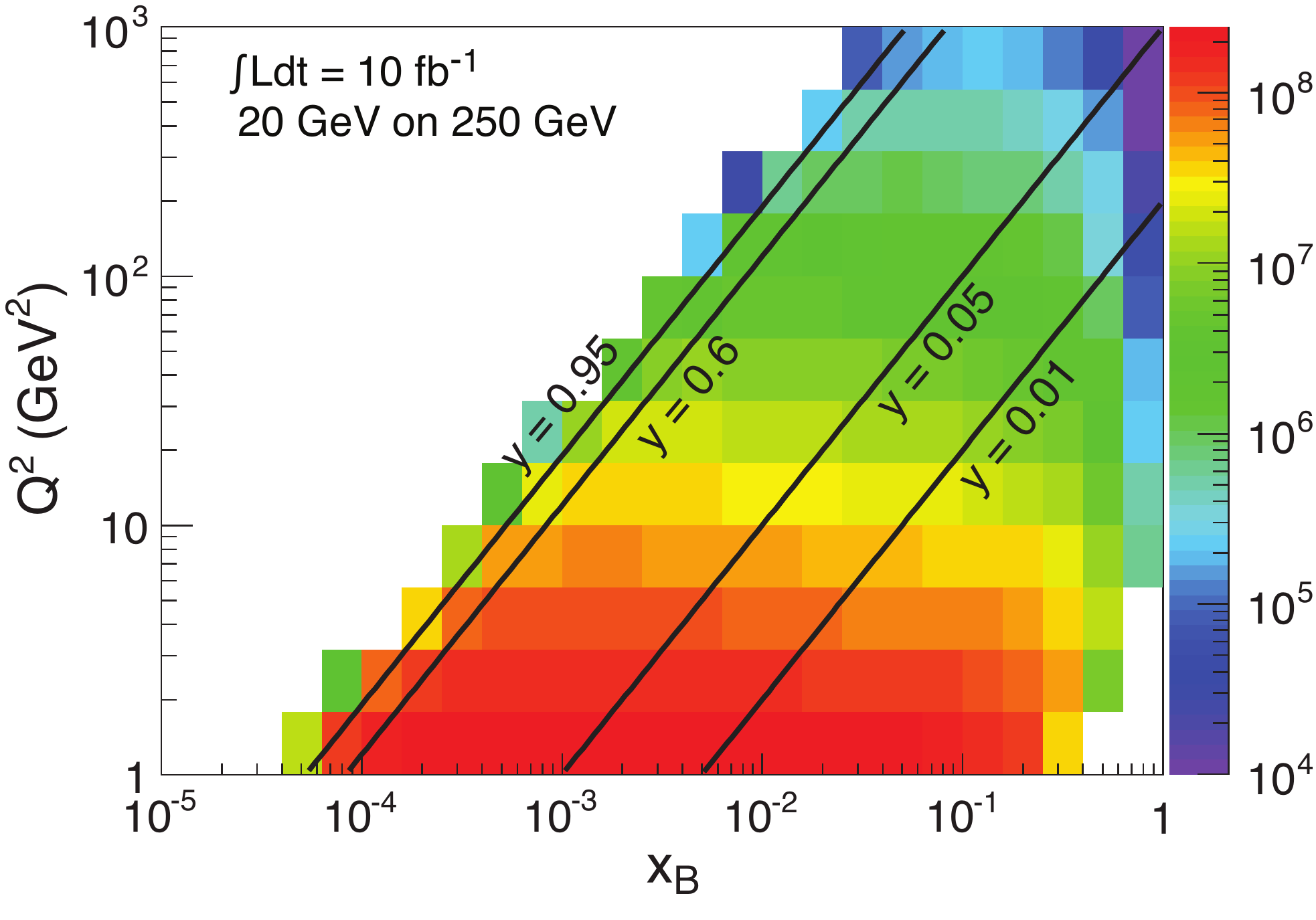}\\
\end{tabular}
\end{center} \vskip -0.2in
\caption{\label{fig:xQ2y} The $x-Q^2$ plane for center-of-mass energy
45 GeV (left) and 140 GeV (right). The black lines indicate different
$y$-cuts placed on the scattered lepton kinematics.}
\end{figure*}

\begin{multicols}{2}
Figure~\ref{fig:xQ2y} shows the x-Q$^2$ plane for two different
center-of-mass energies.  In general, the correlation between $x$ and
$Q^2$ for a collider environment is weaker than for fixed target
experiments. However, an important consideration is the extreme range
of values of the inelasticity $y$. At large $y$, radiative corrections
become large, as illustrated in Fig. 7.25 in Ref.~\cite{Boer:2011fh}.
There are two ways to address this: one is to calculate radiative
corrections and correct for them; the other is utilize the hadronic
activity in the detector together with cuts on the invariant mass of
the hadronic final state.

The x-Q$^2$ correlations become stronger for small scattering angles
or correspondingly small inelasticity. Here, radiative corrections
are small, but the momentum and scattering angle resolution for the
scattered lepton deteriorates.  
This problem is addressed by reconstructing the lepton kinematics 
from the hadronic final state using the Jacquet-Blondel
method~\cite{Jacquet:1979jb,Bassler:1994uq}. At HERA, this method was
successfully used down to $y$ of 0.005.  The main reason this hadronic
method renders better resolution at low $y$ follows from the equation
$y_{JB} = E-P_z^{had} / 2 E_e$, where $E-P_z^{had}$ is the sum over
the energy minus the longitudinal momentum of all hadronic final-state
particles and $E_e$ is the electron beam energy. This quantity has no
degradation of resolution for $y<0.1$ as compared to the electron
method, where $y_e = 1 - (1-cos \theta{_e})E'_e / 2 E_e$.

Typically, one can obtain for a given center-of-mass energy squared,
roughly a decade of $Q^2$ reach at fixed $x$ when using only an
electron method to determine lepton kinematics, and roughly two
decades when including the hadronic method. If only using the electron
method, one can increase the range in accessible $Q^2$ by lowering the
center-of-mass energy, as can be seen from comparing the two panels of
Fig.~\ref{fig:xQ2y}. This is
relevant for some semi-inclusive
and exclusive processes. The coverage of each setting is given by the
product of $y \times s$. With a low $y_{min}$ cut, one thus needs fewer
settings in $s$. However, this is an important consideration for any
measurement, which needs to separate the cross-section components due
to longitudinal and transverse photon polarization, 
i.e. the measurement of $F_L$ where one
needs to have full $y$-coverage at all energies. The advantages and
disadvantages of this solution are discussed in the two
machine-specific detector sections of this chapter.
\end{multicols}

\subsection{Angle and Momentum Distributions}
\begin{figure*}[t]
\begin{center}
\includegraphics[width=0.85\textwidth]{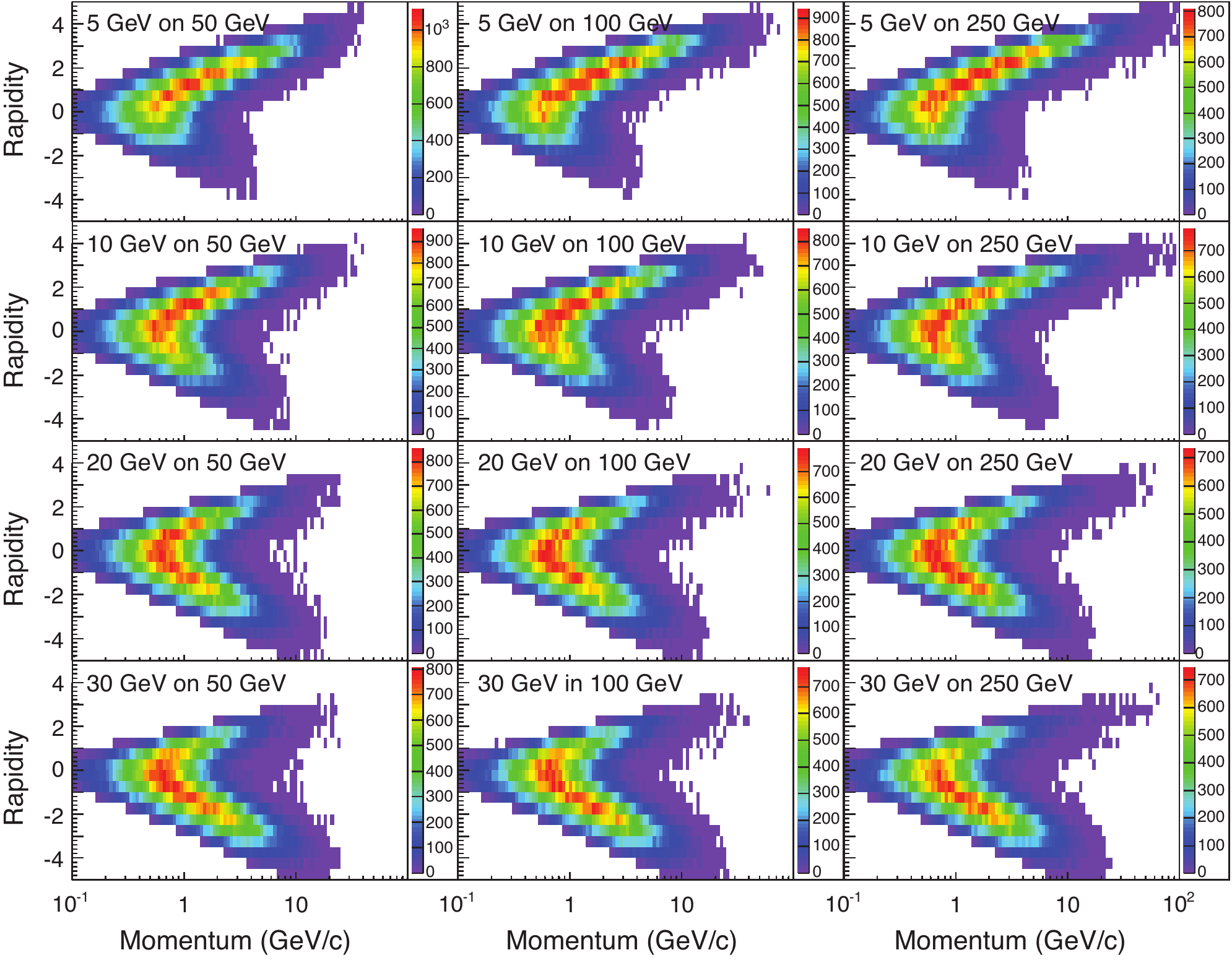}
\end{center}
\caption{\label{fig:pion.kinematic} Momentum vs. rapidity in the
laboratory frame for pions from non-exclusive reactions.  The
following cuts have been applied: Q$^2 > 1$ GeV$^2$, 0.01 $<$ y $<$
0.95, 0.1 $<$ z and -5 $<$ rapidity $<$ 5}
\end{figure*}

\begin{multicols}{2}
Figure~\ref{fig:pion.kinematic} shows the momentum versus rapidity
distributions in the laboratory frame for pions originating from
semi-inclusive reactions for different lepton and proton beam energy
combinations. For lower lepton energies, pions are scattered more in
the forward (ion) direction.  With increasing lepton beam energy, the
hadrons increasingly populate the central region of the detector. At
the highest lepton energies, hadrons are even largely produced going
backward (i.e. in the lepton beam direction). The kinematic
distributions for kaons and additional protons/anti-protons 
are essentially identical to
those of the pions. The distributions for semi-inclusive events in
electron-nucleus collisions may be slightly altered due to nuclear
modification effects, but the global features will remain.

Figure~\ref{fig:pion.kinematic} also indicates the momentum range of
pions in the central detector region (-1 $<$ rapidity $<$ 1) of
typically 0.3 GeV/c to 4 GeV/c with a maximum of about 10 GeV/c.  A
combination of high resolution time-of-flight (ToF) detectors (with
timing resolutions $\delta$t $\sim$ 10ps), a DIRC or a proximity
focusing Aerogel RICH may be considered for particle identification in
this region.  
Hadrons with higher momenta go typically in the
forward (ion) direction for low lepton beam energies, and in the
backward direction for higher lepton beam energies. The most viable
detector technology for this region of the detector is a Ring-Imaging
Cherenkov (RICH) detector with dual-radiators.

\begin{figure*}[t]
\begin{center}
\includegraphics[width=0.85\textwidth]{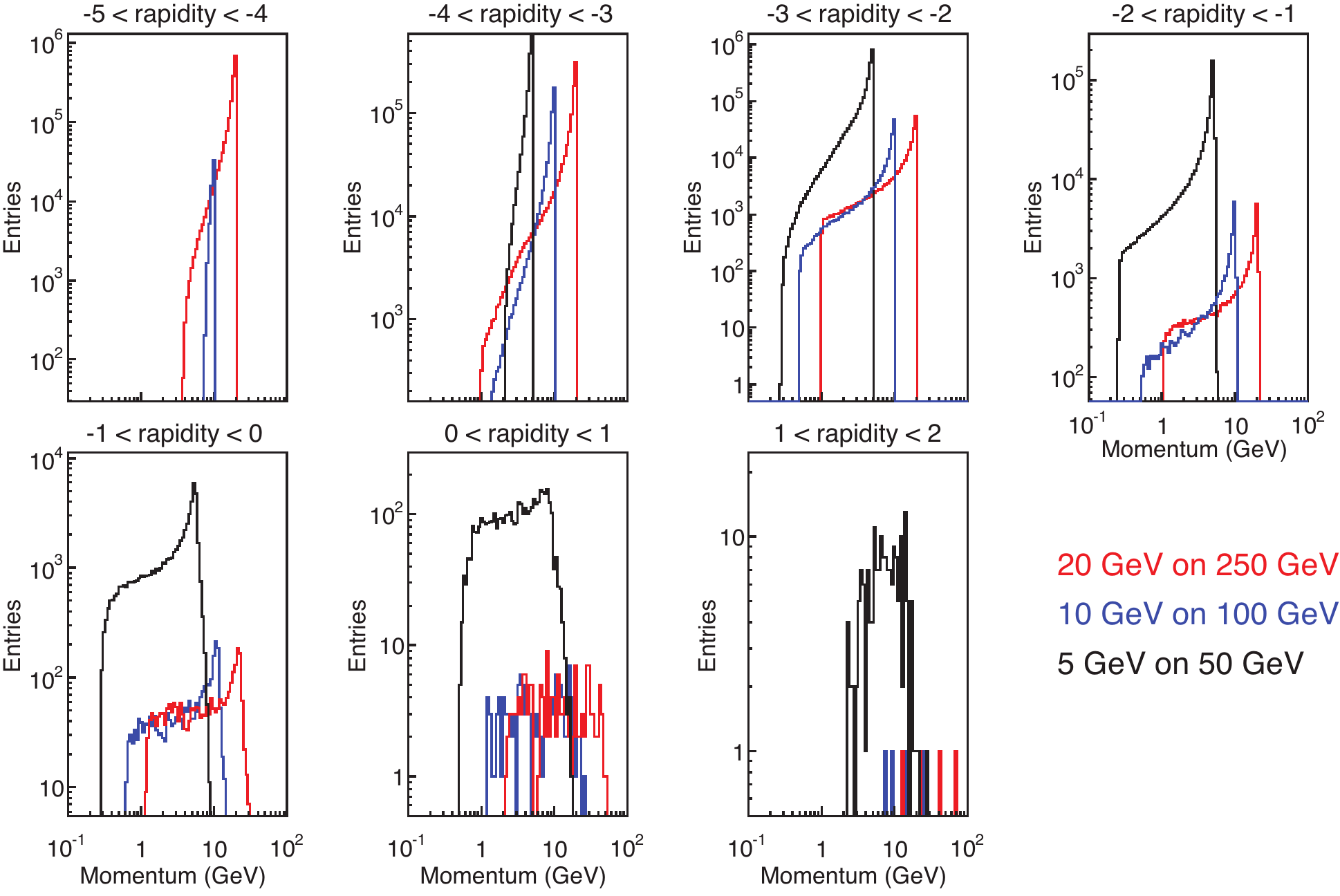}
\end{center} \vskip -0.1in
\caption{\label{fig:scat.lepton.kinematic} The momentum distribution for
the scattered lepton for different center-of-mass energies and
different rapidity bins in the laboratory frame.  The following cuts
have been applied: Q$^2 > 0.1$ GeV$^2$, 0.01 $<$ y $<$ 0.95 and -5 $<$
rapidity $<$ 5}
\end{figure*}

Figure~\ref{fig:scat.lepton.kinematic} shows the momentum distribution
for the scattered lepton for different rapidity bins and three
different lepton-proton beam energy combinations. The $Q^2 < 10$
GeV$^{2}$ events typically correspond to negative rapidities ($ \eta <
-3$) and $Q^2 > 10$ GeV$^{2}$ correspond to rapidities $\eta > -2$ for
5 GeV x 50 GeV and $\eta > -3$ 30 GeV x 50 GeV.  Depending on the
center-of-mass energy the rapidity distributions for hadrons (both
charged and neutral) and the scattered lepton overlap and need to be
disentangled. The kinematic region in rapidity over which hadrons and
photons need to be suppressed with respect to electrons depends on the
center-of-mass energy.  
For lower center-of-mass energies,
electron, photon and charged hadron rates are roughly comparable at 1
GeV/c total momentum and rapidity = -3. For the higher center-of-mass
energy, electron rates are a factor of 10-100 smaller than photon and
charged hadron rates, and comparable again at a 10 GeV/c total
momentum (see Fig.~7.18 in Ref.~\cite{Boer:2011fh}).  This adds
another requirement to the detector: good electron identification. The
kinematic region in rapidity over which hadrons and also photons need
to be suppressed, typically by a factor of 10 - 100, shifts to more
negative rapidity with increasing center-of-mass energy.

Measuring the ratio of the energy and momentum of the scattered
lepton, typically gives a reduction factor of $\sim$100 for
hadrons. This requires the availability of both tracking detectors (to
determine momentum) and electromagnetic calorimetry (to determine
energy) over the same rapidity coverage. By combining information from these two
detectors, one also immediately suppresses the
misidentification of photons in the lepton sample by requiring that a
track must point to the electromagnetic cluster. Having good tracking
detectors over similar coverage as electromagnetic calorimetry
similarly aids in $y$ resolution at low y from a lepton method only
(as explained earlier), as the angular as well as the momentum
resolution for trackers are much better than for electromagnetic
calorimeters.  The hadron suppression can be further improved by
adding a Cherenkov detector to the electromagnetic calorimetry or
having tracking detectors, (e.g., a Time Projection Chamber) to
provide good $dE/dx$. Combining the responses from the electromagnetic
calorimeter and the Cherenkov detectors or $dE/dx$ may especially help
in the region of low-momentum scattered leptons, about 1 GeV/c. Other
detector technologies, such as transition radiation detectors, may
provide hadron rejection by a factor of 100 for leptons with 
$\gamma > 1000$ ($\gamma=1/\sqrt{(1-v^{2}/c^{2}})$).

There is specific interest in extracting structure functions with
heavy quarks from semi-inclusive reactions for mesons, which contain
charm or bottom quarks. To measure such structure functions as
F$_2^C$, F$_L^C$, and F$_2^B$, it is sufficient to tag the charm and
the bottom quark content via the detection of additional leptons
(electrons, positrons, muons) in addition to the scattered (beam)
lepton. The leptons from charmed mesons can be identified via a
displaced vertex of the second lepton ($<\tau> \sim 150 \mu$m). This
can be achieved by integrating a high-resolution vertex detector into
the detector design. For measurements of the charmed (bottom)
fragmentation functions, or to study medium modifications of heavy
quarks in the nuclear environment, at least one of the charmed
(bottom) mesons must be completely reconstructed to have access to the
kinematics of the parton.  This requires, in addition to measuring the
displaced vertex, good particle identification to reconstruct the
meson via its hadronic decay products, e.g. $D^0 \rightarrow
K^{\pm}+\pi^{\mp}$.

\begin{figure*}[htbp]
\begin{center} \hskip 0.5in
\includegraphics[width=0.85\textwidth]{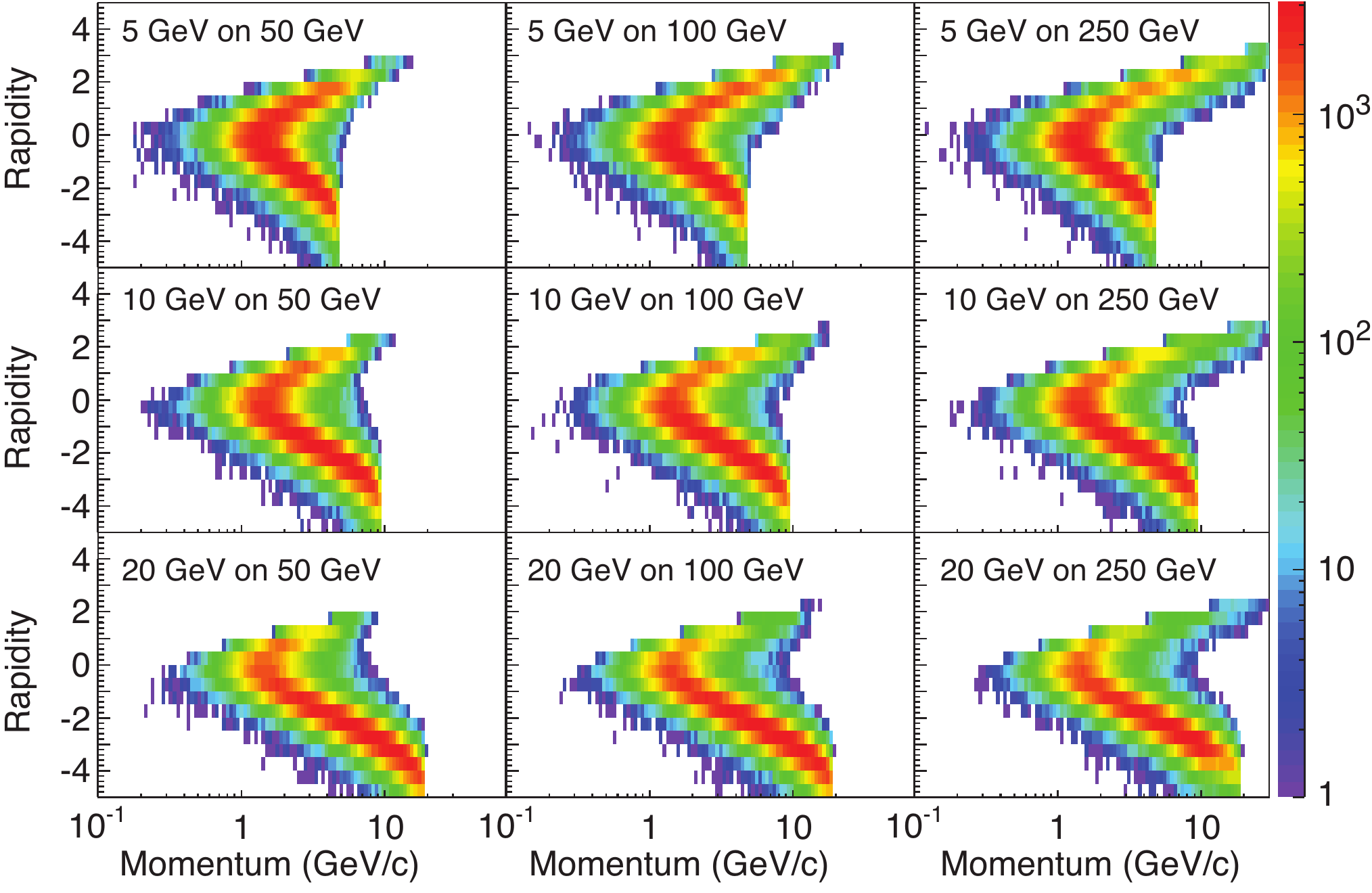}\\
\vskip 0.2in
\includegraphics[width=0.85\textwidth]{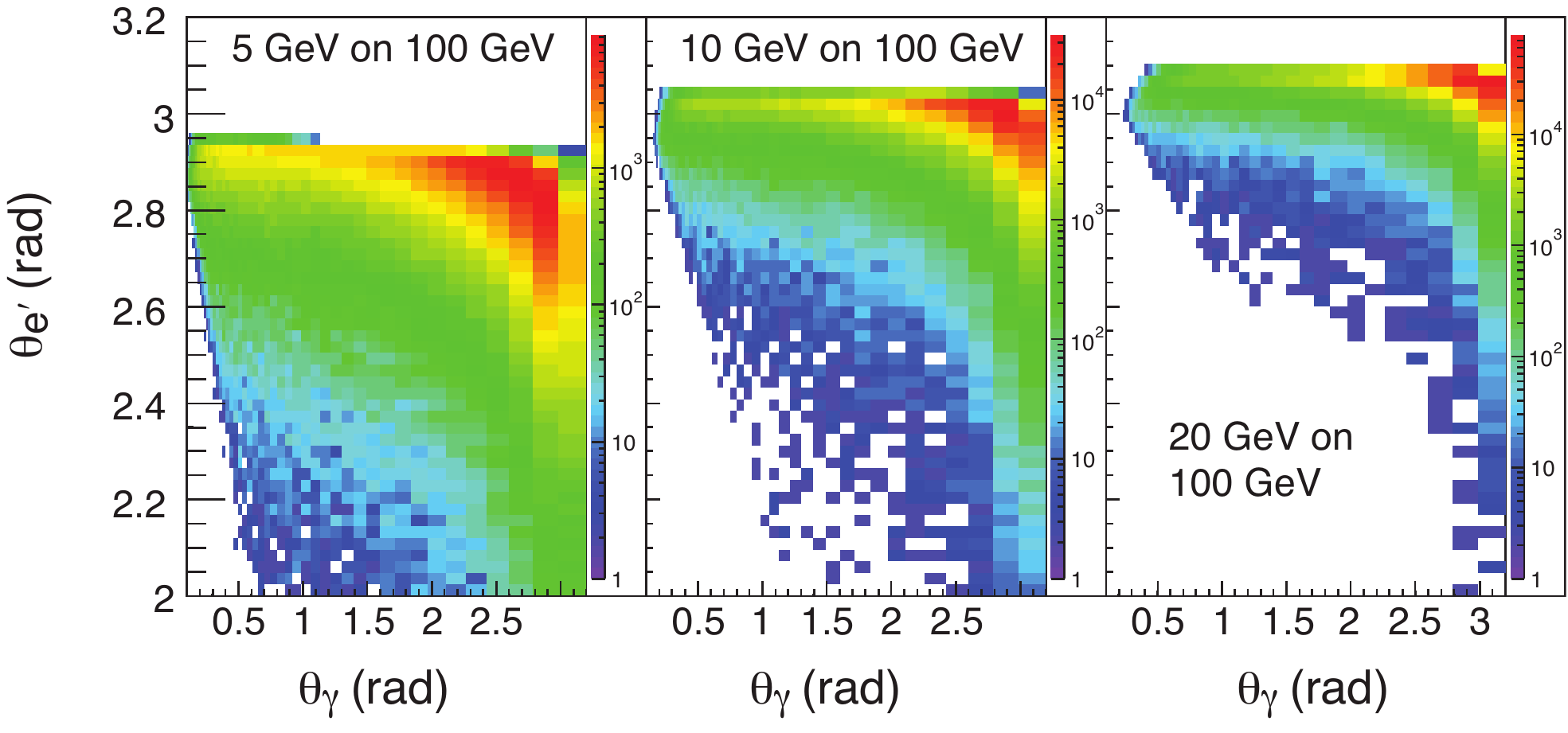}
\end{center} \vskip -0.2in
\caption{\label{fig:DVCS_photon_kinematics} The energy vs. rapidity in
the laboratory frame for photons from DVCS for different
center-of-mass energies (top) and the correlation between the
scattering angle of the DVCS photon and the scattered lepton for three
different center-of-mass energies.  The following cuts have been
applied: Q$^2 > 1.0$ GeV$^2$, 0.01 $<$ y $<$ 0.95, E$_{\gamma} > 1$
GeV and -5 $<$ rapidity $<$ 5.}
\end{figure*}

Figure~\ref{fig:DVCS_photon_kinematics} shows the energy vs. rapidity
distributions for photons from deeply virtual Compton scattering
(DVCS), and the correlation between the scattering angle of the DVCS
photon and the scattered lepton in the laboratory frame for different
beam energy combinations. The general patterns are as in
Fig.~\ref{fig:pion.kinematic}, but even at the low lepton beam
energies the DVCS photons go more into the backward
direction. However, for imaging studies through exclusive reactions
involving light mesons, a $Q^2$ cut must be applied for a valid
partonic interpretation.  Since exclusive low-$Q^2$ hadrons are
produced in the forward direction, a $Q^2 >$ 10 GeV$^2$ cut changes
the kinematic patterns from Fig.~\ref{fig:pion.kinematic}.

The most challenging constraints on the detector design for exclusive
reactions compared to semi-inclusive reactions is, however, not given
by the final state particle ($\pi$, K, $\rho, \phi$, J/$\psi,
\gamma$), but to ensure the exclusivity of the event.
\end{multicols}

\subsection{Recoil Baryon Angles and $t$ Resolution}
\label{sec:recoil}
\begin{multicols}{2}
For exclusive reactions
on the nucleon or coherent nuclear processes
, it is extremely important to ensure that the
nucleon (or the nucleus) remains intact during the scattering
process. Hence, one has to ensure exclusivity by measuring all
products. In general, for exclusive reactions, one wishes to map the
four-momentum transfer (or Mandelstam variable) $t$ of the hadronic
system, and then obtain an image by a Fourier transform, for $t$ close
to its kinematic limit $t_{min}$ up to about 1-2 GeV (for details see
Chap. 3.6 in Ref.~\cite{Boer:2011fh}).

\begin{figure*}[h!]
\begin{center}
\includegraphics[width=0.80\textwidth]{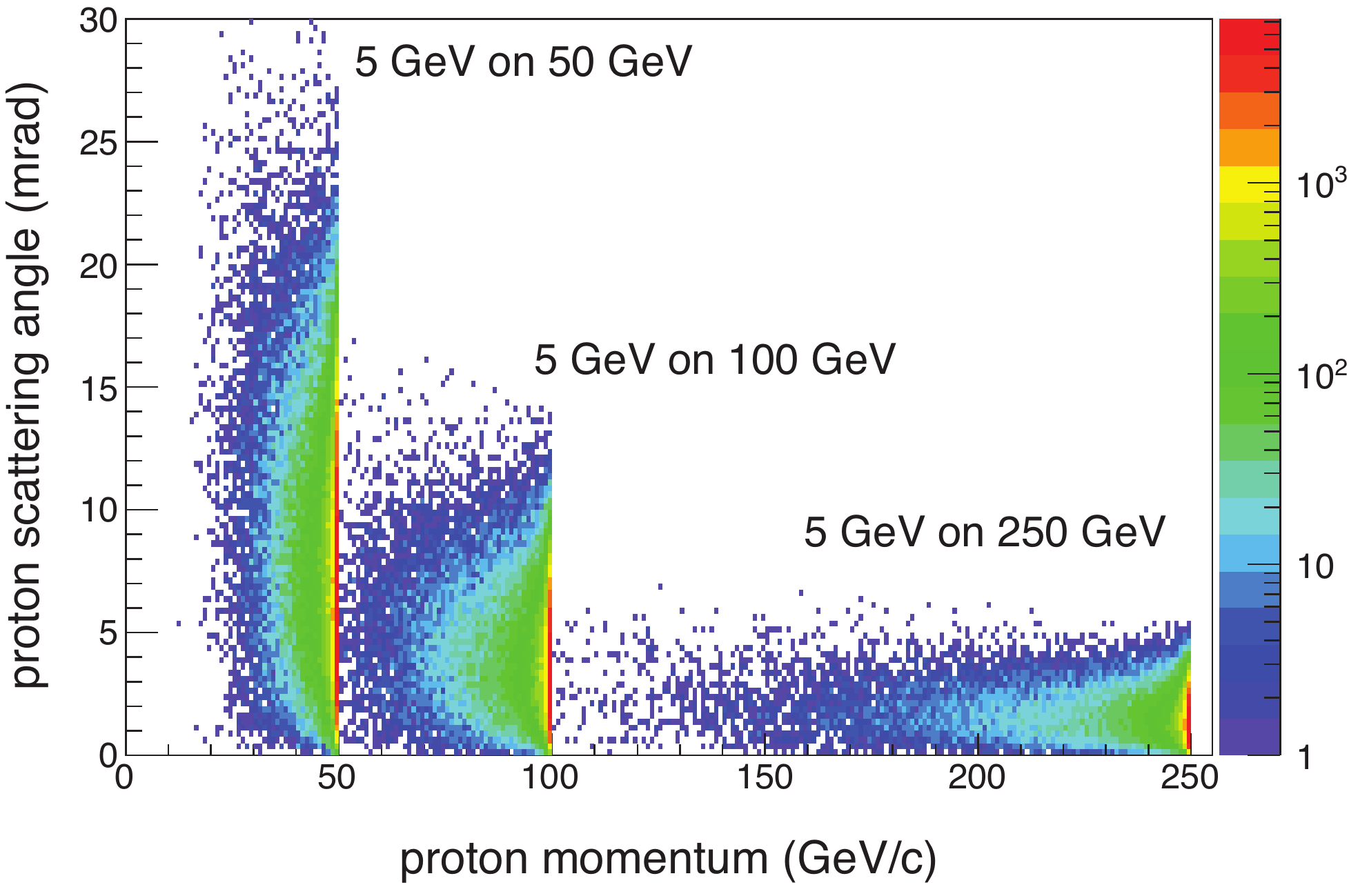}
\end{center} 
\caption{\label{fig:DVCS_baryon_kinematics} The scattered proton momentum
vs. scattering angle in the laboratory frames for DVCS events with
different beam energy combinations. The following cuts have been
applied: 1 GeV$^2 <$ Q$^2 <$ 100 GeV$^2$, $0.00001 < x <$ 0.7 and 0 $<
t <$ 2 GeV$^2$.  The angle of the recoiling hadronic system is
directly and inversely correlated with the proton energy.  It thus
decreases with increasing proton energy. }
\end{figure*}

Figure~\ref{fig:DVCS_baryon_kinematics} shows one of the most
challenging constraints on the detector and interaction region design
from exclusive reactions, the need to detect the full hadronic final
state. The figure shows the correlation between proton scattering
angle and its momentum, and illustrates that the remaining baryonic
states go in the very forward ion direction.  Even at a proton energy
of 50 GeV, the proton scattering angles only range to about
2$^\circ$. At proton energies of 250 GeV, this number is reduced to
one/fifth. In all cases, the scattering angles are small. Because of
this, the detection of these protons, or more general recoil baryons,
is extremely dependent on the exact interaction region design and will
therefore be discussed in more detail in the machine-dependent part of
this chapter.

In the case of nucleus breakup as in, e.g., measurements of the 
quasi-free reaction on the nucleon in the nucleus, detection of the 
nuclear spectators and fragments is required. Unlike the recoil baryons 
from, e.g., DVCS the ion fragments have rigidities different from the beam. 
Examples of these processes are spectator tagging with polarized ion 
beams requiring a resolution in the transverse momentum better than 
the Fermi momentum and detection of the final state in heavy-ion collisions. 
\end{multicols}

\subsection{Luminosity Measurement}
\label{sec:luminosity}
\begin{multicols}{2}
The Bethe-Heitler bremsstrahlung process $ep \longrightarrow ep\gamma$
was successfully used to measure luminosity by the experiments at the
HERA $e$+$p$ collider. It has a large cross-section, allowing rapid
measurements with negligible statistical uncertainty. The cross
section of this process can be calculated entirely within QED, and is
known to a precision of $\sim$ 0.2\%. The luminosity measurement was
typically carried out by detecting the final state photons; the final
state electron was also measured in some cases for experimental cross
checks. Limitations in determining the geometric acceptance of the
very-forward photons resulted in a systematic uncertainty of 1-2\% on
the HERA luminosity measurements.  For a polarized $e$+$p$ collider, the
bremsstrahlung cross-section has a dependence on the beam
polarizations, which may be expressed as $\sigma =
\sigma_0(1+aP_eP_p)$.  Preliminary estimates indicate that the
coefficient $a$ is small, but detailed studies are currently underway
to understand the size of $a$ relative to the magnitude of the double
spin asymmetries $A_{LL}$ at small $x_B$. The theoretical uncertainty
on $a$, and the experimental uncertainties on the measured beam
polarizations $P_e$ and $P_p$, will limit the precision of the
absolute and relative luminosity measurements.
\end{multicols}

\newpage
\subsection{Hadron and Lepton Polarimetry}
\label{sec:polarimetry}
\begin{multicols}{2}
Compton back-scattering is the established method to measure lepton
beam polarization in $e$+$p$ colliders.  At HERA, there were two Compton
back-scattering polarimeters \cite{Hera:Polarimeters}: one measuring
the transverse polarization (TPOL) of the beam through a position
asymmetry and one measuring the longitudinal polarization (LPOL) of
the beam through an energy asymmetry in Compton back-scattered
photons. The TPOL and LPOL systematic uncertainties of RUN-I were
3.5\% and 1.6\% and of Run-II 1.9\% and 2.0\%, respectively.  In spite
of the expected high luminosity at the EIC, these systematic
uncertainties could be reduced to $\sim$1\% if special care is taken
to reduce the impact of beam orbit instabilities and laser light
polarization on the measurement. The detection of the lepton and the
Compton photon in coincidence will provide an energy self-calibration
of the polarimeter. 

To measure the hadron beam polarization is very difficult as, contrary
to the lepton case, there is no process that can be calculated from
first principles. Therefore, a two tier measurement is needed: one
providing the absolute polarization, which has low statistical power
and a high statistical power measurement, which measures the relative
polarization. At RHIC \cite{RHIC:Polarimeters}, the single spin
asymmetry $A_N$ of the elastically scattered polarized proton beam on
a polarized hydrogen jet is used to determine the absolute
polarization. This measurement provides the average polarization per
fill and beam with a statistical uncertainty on the order of $\sim$
5\% and a systematic uncertainty of 3.2\%. High-statistics
bunch-by-bunch relative polarization measurements are provided,
measuring the single spin asymmetry $A_N$ for scattering the polarized
proton beam of a carbon fiber target. To obtain absolute measurements,
the pC-measurements are cross normalized to the absolute polarization
measurements from the hydrogen-jet polarimeter. The pC-measurements
provide the polarization lifetime and the polarization profile per
fill with high statistical precision.  The achieved total systematic
uncertainty for single spin asymmetries is 3.4\%.  The systematic
uncertainties could be further reduced by monitoring continuously the
molecular hydrogen contamination in the jet, improving the operational
stability of the carbon fiber targets, and by developing methods to
monitor the silicon detector energy calibration at the recoil carbon
energy.  All are under development for the polarized $p$+$p$ program at
RHIC.

To have minimal impact from potential
bunch-to-bunch polarization fluctuations on the luminosity measurement,
it is important to have both hadron and lepton beam polarimeters that 
can provide high statistics polarization information for individual bunches.
\end{multicols}


\section{Detector and Interaction Region (IR) Layout}
\label{sec:layout}

\subsection{eRHIC Detectors \& IR Considerations and Technologies}

Three studies on a possible implementation for an eRHIC detector have been performed. 
Two studies are built on the existing RHIC detectors. Both the PHENIX and STAR 
collaborations have studied how the sPHENIX \cite{Aidala:2012nz} and STAR \cite{STAR:NIM} 
detectors would have to be 
upgraded/modified to fulfill the performance requirements as laid out by the eRHIC physics 
program \cite{Adare:2014aaa,eSTAR}. The third study is based on a “green field” design for an 
eRHIC detector, 
which is completely optimized to the physics requirements and the change in particle kinematics 
resulting from varying the center of mass energies from 55 GeV to 140 GeV. In the following 
mainly details about the model detector will be described.

\begin{figure*}[h]
\begin{center}
\includegraphics[width=0.90\textwidth]{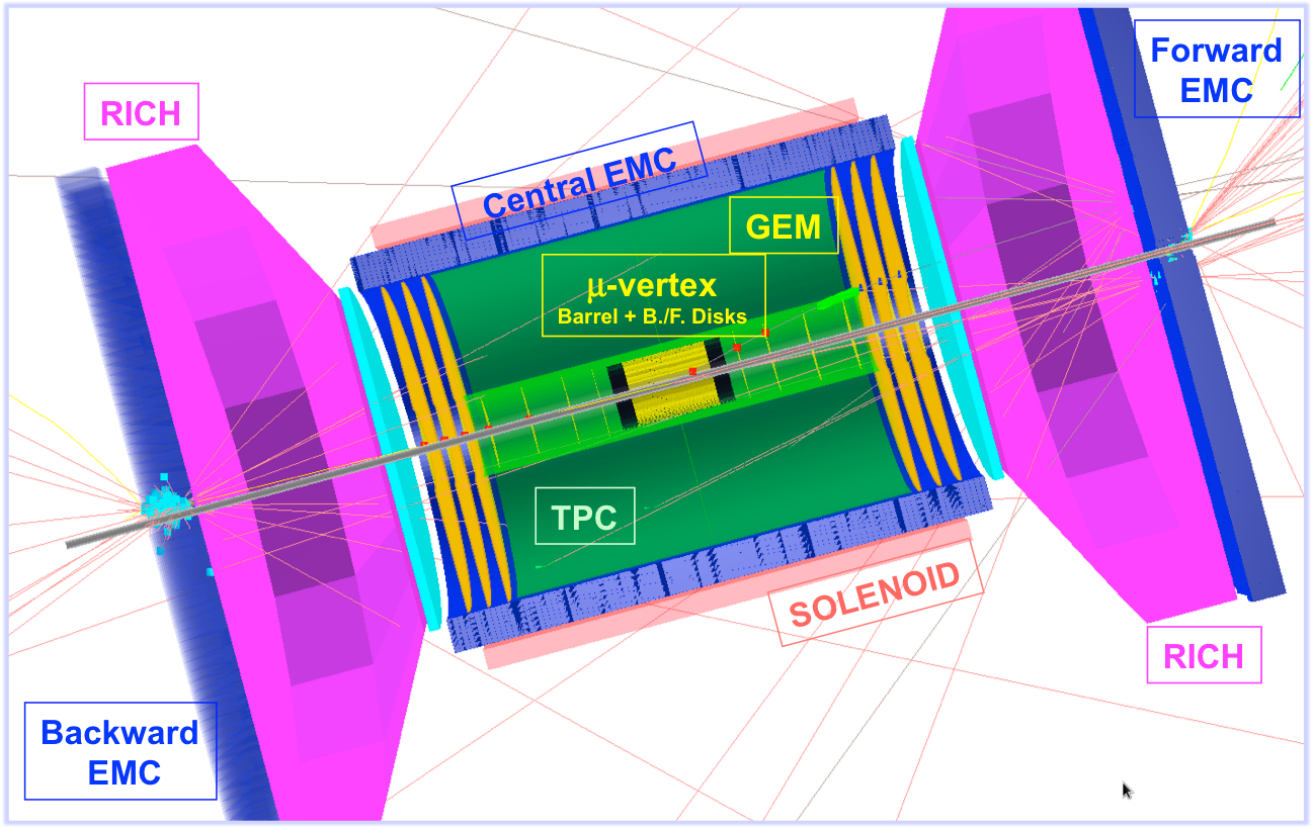}
\end{center}
\caption{\label{fig:eRHIC-det} The eRHIC model detector implementation (BeAST = Brookhaven eA 
Solenoidal Tracker) with tracker and calorimeter components implemented in the EicRoot GEANT 
simulation framework \protect\cite{EicRoot}}
\end{figure*}

\begin{multicols}{2}
Combining all the requirements described in Sec.~\ref{sec:kinmat} and
in the preceding physics chapters, a schematic view of the emerging
dedicated eRHIC detector is shown in Fig.~\ref{fig:eRHIC-det}.

The compact tracker, located symmetrically with respect to the IP, consists of: a MAPS 
silicon barrel vertex detector and a set of forward/backward disks; a 2m long TPC with 
a gas volume outer radius of 0.8m and several GEM stations, all placed into a $\sim$3T solenoid 
field. The TPC is specifically chosen as the main tracking element because of its small overall 
material budget, minimizing the rate of photon conversions on detector components, which is 
required in particular for the DVCS measurements. Besides this, the TPC should provide good 
charged PID in the momentum range up to a few GeV/c at central rapidities. Other detector options 
for the main tracker, such as a set of cylindrical micromegas planes are considered as well \cite{MicroMega}.
Significant progress in the last decade in the development of
Monolithic Active Pixel Sensors (MAPS) in which the active detector,
analog signal shaping, and digital conversion take place in a single
silicon chip (i.e., on a single substrate; see \cite{Gaycken:2006se}
and references therein) provides for a unique opportunity for a
$\mu$-vertex detector for an eRHIC detector.  As a result, CMOS pixel
detectors can be built with high segmentation, limited primarily by
the space required for additional shaping and digital conversion
elements.  The key advantage of CMOS MAPS detectors is the reduced
material required for the detector and the (on substrate) on-detector
electronics. Such detectors have been fabricated and extensively
tested (see e.g.~\cite{HuGuo:2009zz}) with thicknesses of about 50
$\mu$m, corresponding to 0.05\% of a radiation length.
The vertex detector, covering the central rapidity range $-1 < \eta < 1$, 
is strongly inspired by the STAR HFT tracker design \cite{Kapitan:2008ae} a similar 
design is now considered by the ALICE experiment at LHC. 
The projected rates for a
luminosity in the 10$^{34}$ cm$^{-2}$ s$^{-1}$ range, depending on the
center-of-mass energy, between 300 and 600 kHz, with an average of 6
to 8 charged tracks per event. These numbers do not impose strong
constraints on any of the above technology for tracking detectors.

To have equal rapidity coverage for tracking and
electromagnetic calorimetry will provide good electron
identification and give better momentum and angular resolutions at low
inelasticity, $y$, than with an electromagnetic calorimeter alone.
Therefore the detector will be equipped with a set of electromagnetic calorimeters, 
hermetically covering a pseudorapidity range of at least $-4 < \eta < 4$. The calorimeter 
technology choice is driven by the fact that a moderately high-energy resolution, on order of 
$\sim$2-3\%/$\sqrt{E}$ , is needed only at backward (electron-going) rapidities. 
Therefore in the present design the backward endcap calorimeter for the $-4 < \eta < -1$ range is 
composed of PWO crystals at room temperature, with the basic performance parameters taken from the 
very extensive PANDA R\&D studies \cite{PANDA:2008uqa}. The calorimeter is located 
$\sim$  500 mm away from the IP. 
The crystal length corresponds to $\sim$22.5 $\chi_0$, and both the crystal shape and grouping follow the ideas 
of the PANDA and CMS \cite{Musienko:2002ji}  calorimeter designs. 
For the barrel and forward endcap electromagnetic calorimeters, covering a pseudo-rapidity range of 
$-1 < \eta < 4$, a noticeably worse energy resolution suffices. In order to save costs, at present 
it is planned to use the STAR upgrade R\&D building blocks of tungsten powder scintillating fiber sampling 
calorimeter towers, with a design goal of $\sim$12\%/$\sqrt{E}$ energy resolution . 
The forward endcap calorimeter will be located at $\sim$2500 mm from the IP in hadron-going 
direction. The barrel calorimeter will have an average installation radius of $\sim$900 mm and be 
composed of slightly tapered towers, in order to avoid gaps in the azimuthal direction. 
Both calorimeter types will have a non-projective geometry and tower length corresponding 
to $\sim$23 $\chi_0$. 
\end{multicols}
\begin{figure*}[h]
\begin{center}
\begin{tabular}{cc}
Electron Method & Jacquet-Blondel Method \\ 
\includegraphics[width=0.50\textwidth]{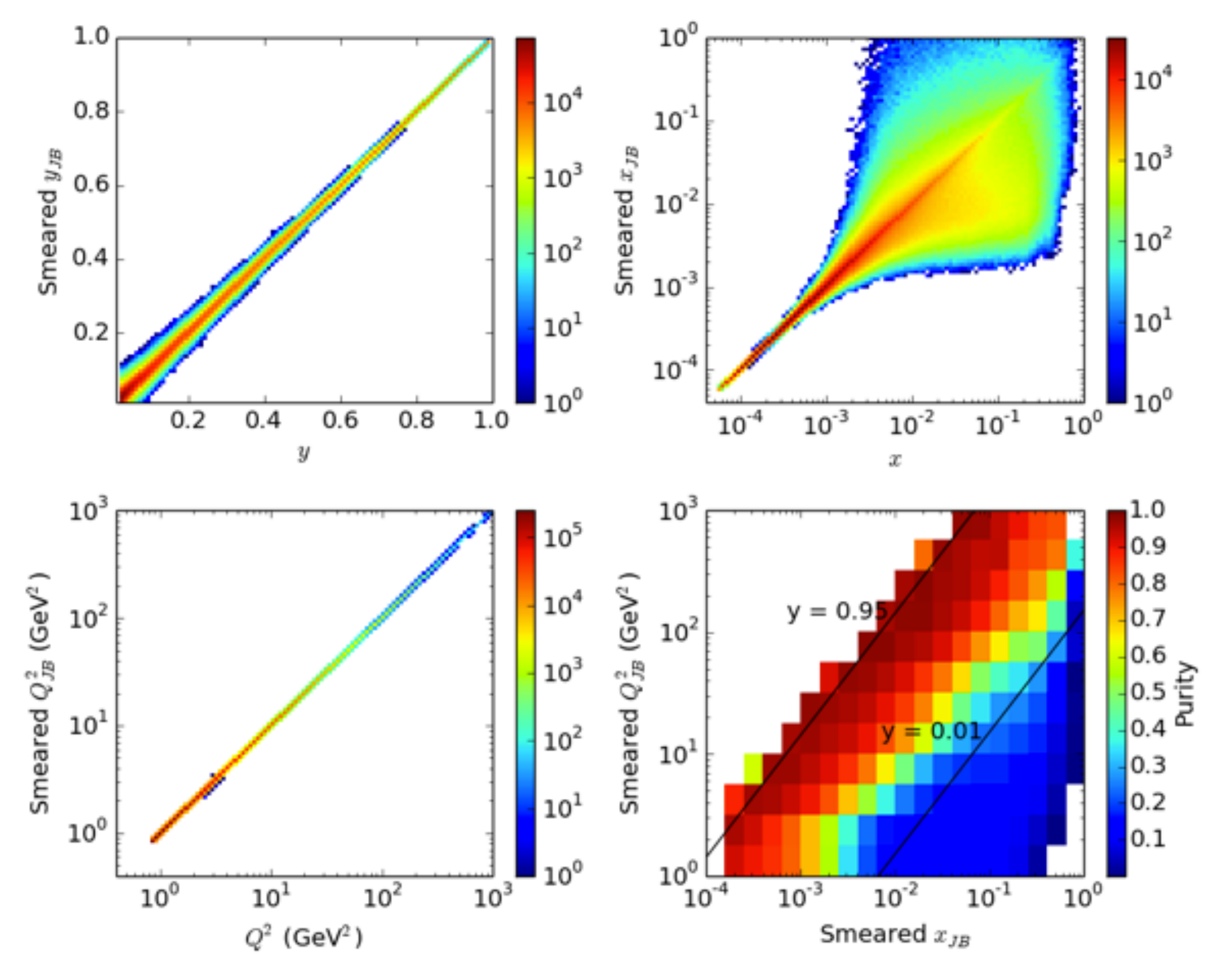} &
\includegraphics[width=0.50\textwidth]{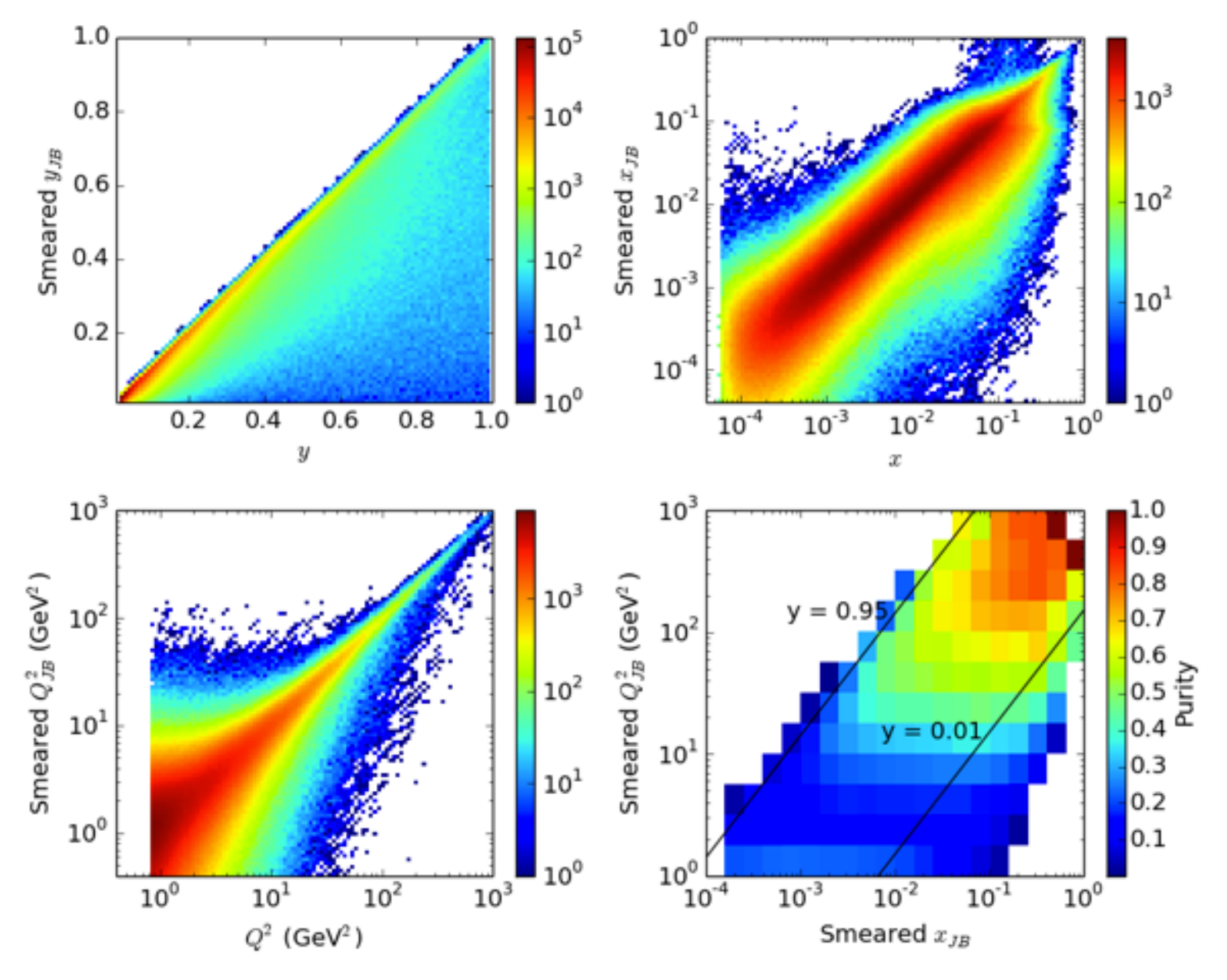} \\
\end{tabular}
\end{center}
\caption{\label{fig:eRHICDetResolution} 
The correlation between smeared and true $y, x$ and $Q^2$ (top to bottom left), and the resulting 
bin-by-bin event purity in the $x-Q^2$ plane (bottom right), reconstructed using the electron method. 
Purity is defined as $(N_{gen} - N_{out}) / (N_{gen} - N_{out} + N_{in})$, where $N_{gen}, out, in$ 
are the number of events generated in a bin, smeared out of it, and smeared into it from other bins, 
respectively. Both the electron (left) and Jacquet-Blondel (right) method are shown.}
\end{figure*}

\begin{multicols}{2}
To quantify the performance of the model detector to reconstruct the event kinematics PYTHIA events,
generated for a 15 GeV electron beam colliding with a 250 GeV proton beam, were passed through
the detector simulation. Figure~\ref{fig:eRHICDetResolution} shows as example
the results of detector smearing on event kinematics calculated using the electron method (left) and
the Jacquet-Blondel method (right), crucial for charge current (CC) events.
As expected, due to the excellent resolution in both momentum and electron energy, $y, x$ and $Q^2$
are exceedingly well reconstructed. Event purity is excellent at moderate-to-large $y$
(typically $>$  90\%) even with a relatively fine $x-Q^2$ binning of five bins per decade in $x$
and four per decade in $Q^2$. The Jacquet-Blondel (JB) method is a purely hadronic method of
kinematic calculation, meaning it can be used in the absence of a measured scattered lepton.
A drawback of this method is that it suffers from very poor resolution at low $Q^2$.
Fortunately, as the majority of the CC cross-section resides at large $Q^2 >$ 100, the JB
method can be very successfully applied to the analysis of these events \cite{Aschenauer:2013iia}.

To increase the separation of
photons and $\pi^0$s to high momenta and to improve the matching of
charged tracks to the electromagnetic cluster, it would be an
advantage to add, in front of all calorimetry, a high-resolution
To have at least one pre-shower layer with 1--2 radiation lengths of
tungsten and silicon strip layers (possibly with two spatial
projections) would allow to separate single photons from $\pi^0$ to up
$p_T \approx 50$\,GeV, as well as enhanced electron-identification.  A
straw-man design could have silicon strips with $\Delta \eta = 0.0005$
and $\Delta \phi = 0.1$.  

Due to the momentum range to be covered the only solution for PID in
the forward direction is a dual radiator RICH, combining either
Aerogel with a gas radiator like C$_4$F$_{10}$ or C$_4$F$_8$O if
C$_4$F$_{10}$ is no longer available, or combining the gas radiator
with a liquid radiator like C$_6$F$_{14}$.
In the barrel part of the detector, several solutions are possible as
the momenta of the majority of the hadrons to be identified are
between 0.5 GeV and 5 GeV.  The technologies available in this
momentum range are high resolution ToF detectors (t $\sim$ 10ps), a
DIRC or a proximity focusing Aerogel RICH.

To achieve the physics program as described in earlier sections, it is
extremely important to integrate the detector design into the
interaction region design of the collider.  Particularly challenging
is the detection of forward-going scattered protons from exclusive
reactions, as well as of decay neutrons from the breakup of heavy ions
in non-diffractive reactions.  The eRHIC design features a 10 mrad
crossing angle between the protons or heavy ions during collisions
with electrons. This choice removes potential problems for the
detector induced by synchrotron radiation. To obtain luminosities
higher than 10$^{34}$ cm$^{-2}$ s$^{-1}$, very strong focusing close
to the IR is required to have the smallest beam sizes at the
interaction point. A small beam size is only possible if the beam
emittance is also very small. The focusing triplets are symmetrically 
around the interaction point (IP) starting at 4.5 meters

While the above accomplishes a small-emittance electron beam, the ions
and protons need to be cooled by coherent electron cooling to have
small emittance. The eRHIC interaction region design relies on the
existence of small emittance beams with a longitudinal RMS of $\sim$5
cm, resulting in a $\beta^*$ = 5 cm. Strong focusing is obtained by
three high-gradient quadrupole magnets using recent results from the
LHC quadrupole magnet upgrade program (reaching gradients of 200 T/m
at 120 mm aperture).  To ensure the previously described requirements
from physics are met, four major requirements need to be fulfilled:
high luminosity ($>$ 100 times that of HERA); the ability to detect
neutrons; measurement of the scattered proton from exclusive reactions
(i.e. DVCS); and the detection of spectator protons
from deuterium and He-3 breakup. The eRHIC IR design
fulfills all these requirements. 
The apertures of the interaction region magnets allow detection of neutrons with a solid
angle of $\pm$ 4 mrad, as well as the scattered proton from exclusive
reactions, i.e. DVCS, up to a solid angle of $\sim$ 9 mrad. The
detection of the scattered proton from exclusive reactions is realized
by integrating several ``Roman Pot" stations into the warm section of the IR. 
The electrons are transported to the interaction point through the heavy-ion/proton
triplets, seeing zero magnetic field.

Figure~\ref{fig:roser:eRHIC_interaction_region} 
shows the current eRHIC interaction region design in the direction of
the outgoing hadron beam. The other side of the IR is mirror symmetric
for the incoming hadron beam. A low scattering-angle lepton tagger for
events with $Q^2 < 0.1$ GeV$^2$ is currently integrated in the outgoing lepton beam
line design.
An extensive R\&D program has been started to design and integrate the lepton beam polarimeter and the
luminosity monitor into the interaction region \cite{eRHICLumiePOL}.
\end{multicols}

\begin{multicols}{2}
\subsubsection{ePHENIX}
\begin{figure*}[h]
\begin{center}
\includegraphics[width=0.90\textwidth]{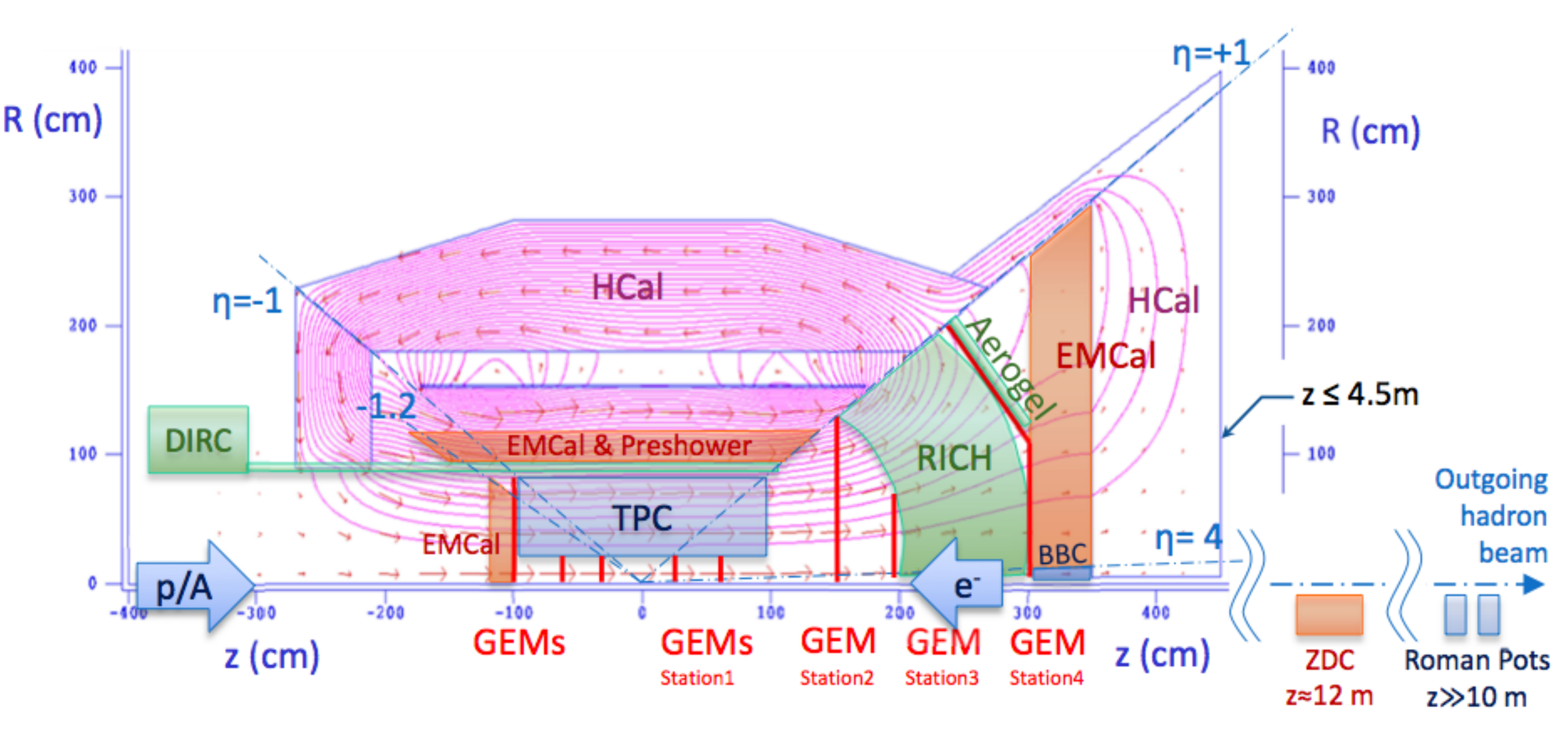}
\end{center} \vskip -0.2in
\caption{\label{fig:ePHENIX-det}A cross section through the top-half of the ePHENIX detector
concept, showing the location of the superconducting solenoid, the barrel calorimeter system,
the EMCal in the electron-going direction and the system of tracking, particle identification
detectors and calorimeters in the hadron-going direction. Forward detectors are also shown
along the outgoing hadron beamline. The magenta curves are contour lines of magnetic field
potential as determined using the 2D magnetic field solver, POISSON.}
\end{figure*}

The PHENIX Collaboration has proposed to build an eRHIC detector, here referred to as ePHENIX, 
upon sPHENIX \cite{Aidala:2012nz}, which is designed to further advance the study of 
cold and hot nuclear matter in nuclear collisions, with its main emphasis on jet measurements.
In addition to fully utilizing the sPHENIX superconducting solenoid and barrel calorimetry, 
ePHENIX adds new detectors in the barrel, electron-going and hadron-going directions 
\cite{Adare:2014aaa}, see Figure~\ref{fig:ePHENIX-det} .  
In the electron-going direction, a crystal calorimeter is 
added for electron identification and precision resolution. A compact time projection chamber, 
augmented by additional forward and backward angle GEM detectors, provides full tracking coverage.  
In the hadron-going direction, behind the tracking is electromagnetic and hadronic calorimetry.  
Critical particle identification capabilities are incorporated via a barrel DIRC, and in 
the hadron-going direction, a gas RICH and an aerogel RICH. 
\end{multicols}

\begin{multicols}{2}
\subsubsection{eSTAR}
\begin{figure*}[h]
\begin{center}
\includegraphics[width=1.00\textwidth]{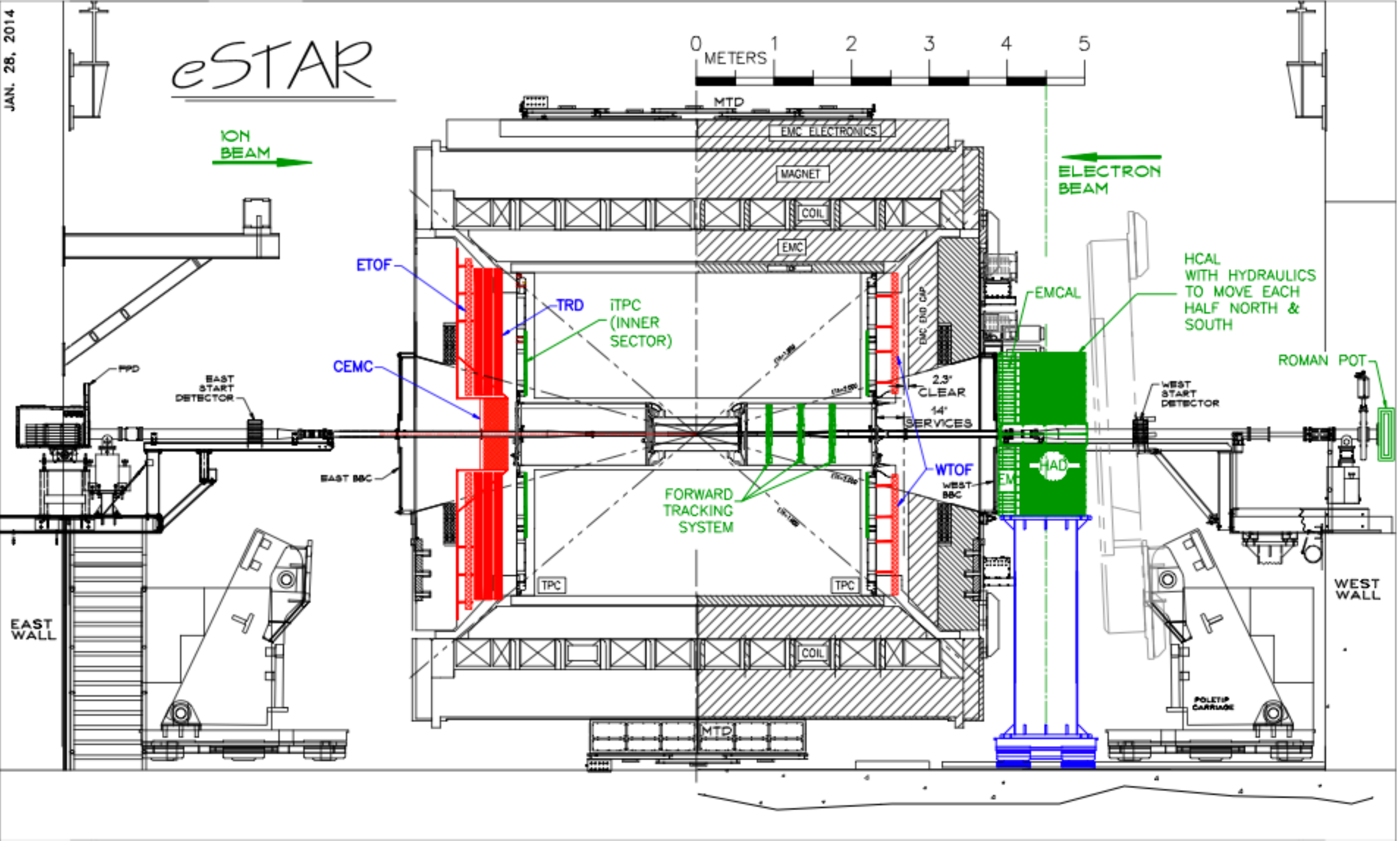}
\end{center} \vskip -0.2in
\caption{\label{fig:eSTAR-det} The eSTAR layout with the proposed upgrades of iTPC, 
Forward Calorimetry System (FCS), the Forward Tracking System (FTS), Endcap TOF (E/W TOF), BSO 
Crystal Calorimeter (CEMC), GEM based TRD. In this configuration, the electron beam is 
from right to left (eastward) while hadron beam from left to right (westward).
}
\end{figure*}

The STAR collaboration has proposed a path to evolve the existing STAR detector
\cite{Aidala:2012nz} to an initial-stage eRHIC detector, eSTAR.
In this plan an optimized suite of detector upgrades will maintain and extend
the existing low-mass mid-central rapidity tracking and particle-identification capabilities
towards more forward rapidities in both the electron and hadron going beam directions.
This plan is described in \cite{Adare:2014aaa}, which contains also a capability assessment
for key measurements of the eRHIC science program.
Figure~\ref{fig:eSTAR-det} shows a side-view of the baseline eSTAR detector layout. This
baseline plan consists of three essential upgrade projects, namely endcap Time-of-Flight
walls located between the TPC and the magnet pole-tips on the East and West sides of the
interaction region (ETOF and WTOF, covering the regions $1 < |\eta| < 2$ in pseudo-rapidity),
a GEM-based Transition Radiation Detector (GTRD) between the TPC and ETOF in the forward
electron direction, covering $-2 < \eta < -1$, and a Crystal ElectroMagnetic Calorimeter with
preshower (CEMC, covering  $-4 < \eta < -2$).  Furthermore, eSTAR will rely
on a replacement upgrade of the Inner Sectors of the existing Time-Projection-Chamber prior
to a completion of the RHIC Beam-Energy Scan program with A + A collisions and
on a subsequent upgrade in the form of a new Forward Calorimeter System (FCS) with associated
Forward Tracking System (FTS) on the West side of STAR.
\end{multicols}

\newpage
\subsection{Detector Design for MEIC/ELIC}
\begin{figure*}[h!]
\begin{center}
\includegraphics[width=0.96\textwidth]{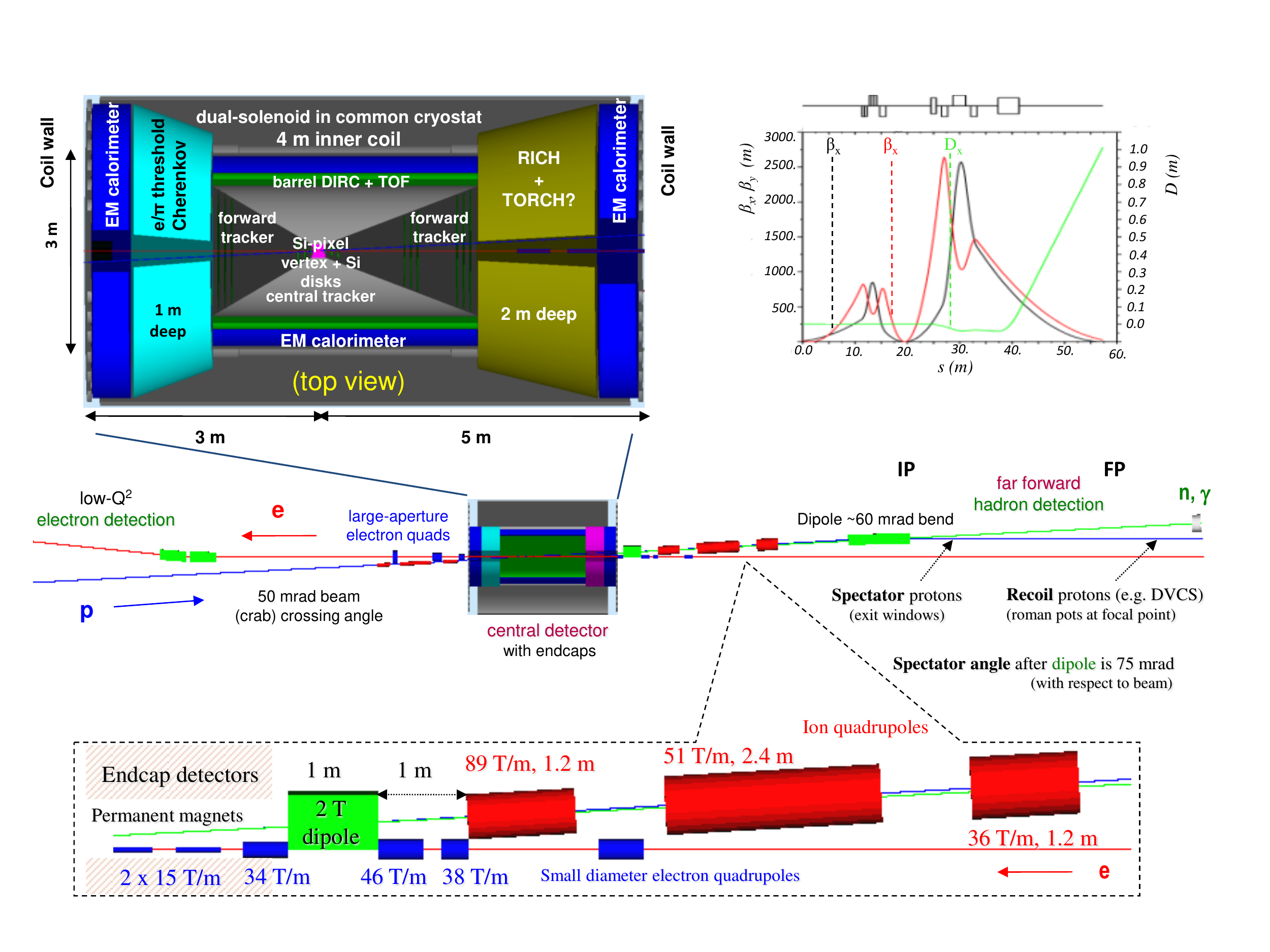}
\end{center} \vskip -0.1in
\caption{\label{fig:ir_and_central_detector_layout} The interaction region
and central detector layout, and its placement in the general
integrated detector and interaction region. The central detector
includes endcaps in both the electron and ion direction.}
\end{figure*}

\begin{multicols}{2}
A global outline of the fully integrated MEIC detector and interaction
region (IR) is given in Fig.~\ref{fig:ir_and_central_detector_layout}. 
A detailed description of the central detector 
as well as the extended interaction region strategy for achieving a 
full-acceptance detector can be found in Ref.  \cite{Boer:2011fh}. 
Since the publication of this article the central detector design has 
been optimized by, e.g., using innovative design features to relax 
specifications and/or to improve its performance. All basic requirements 
and technologies/solutions are understood. Furthermore, new 
opportunities for small-angle hadron and electron detection have 
been identified, and inter-lab and university collaborations on general 
detector R\&D have been formed. The subsequent sections will focus 
on the main aspects of one detector compatible with the full-acceptance 
interaction region and optimized for the physics goals of SIDIS and 
exclusive reactions (see Chapter 2) while keeping in mind that 
accelerator integration is the highest priority since it allows the storage 
ring to be designed around the detector needs. Since a ring-ring 
collider configuration can support multiple detectors without time 
sharing, the full-acceptance detector could be complemented by, 
for instance, a high-luminosity detector at another interaction point. 
Such a second detector could use Time Projection Chambers and 
focus on hadron calorimetry (jets).

To achieve full-acceptance, small-angle detection is required on either 
side of the central detector. 
The low-$Q^2$ electron detection required for heavy flavor 
photoproduction processes is relatively simple to incorporate, including 
a dipole chicane for tagger electrons, which would also be used for 
a Compton polarimeter. The latter would have a laser in the middle of the 
chicane, where the polarization would be identical to that at the IP. In 
addition to the photons, the Compton electrons would also be detected.
The space on the side of the low-$Q^2$ tagger will also be instrumented 
for luminosity monitoring. Measuring forward and ultra-forward going 
hadronic or nuclear fragments along the ion direction 
is more challenging and we make critical use of various ingredients of the 
MEIC detector/interaction region design: 
i) the 50 mrad crossing angle, which moves the spot of poor resolution 
along the solenoid axis into the periphery and minimizes the shadow from
the electron magnets (see, e.g., section 5.2.5); 
ii) the range of proton energies (see, e.g., section 5.2.2); 
iii) a small 2 Tm dipole magnet before the ion final focusing quadrupole 
magnets (FFQs) to allow high-resolution tracking of particles that do not 
enter the FFQs; 
iv) Low-gradient FFQs with apertures sufficient for particles scattered at 
initial angles of 10-15 mrad in each direction for all ion fragment rigidities; and 
v) a 20 Tm large-acceptance dipole magnet a few meters
downstream of the FFQs to peel off spectator particles and allow for
very small-angle detection with high resolution (essentially only
limited by the intrinsic momentum spread of the beam).

As illustrated in Fig.~\ref{fig:ir_and_central_detector_layout} detectors 
will be placed in front of the FFQs, between the FFQs and the 20 Tm 
dipole, and/or in an extended, magnet-free drift space downstream of the 
latter providing far forward hadron detection. In particular, the FFQ 
acceptance for neutral particles will depend on the choice of the peak 
field (6 T baseline) but is
generally in the $\pm$ 10-15 mrad range, centered close to zero.
The neutrons (and boosted nuclear photons) will be detected in a
zero-degree calorimeter (ZDC) on the outside of the ring. 
In this configuration, any desired angular resolution can be achieved simply
by adjusting the distance of the ZDC (as well as its size). This then
results in an essentially 100\% full-acceptance detector.
\end{multicols}

\begin{multicols}{2}
To minimize synchrotron radiation and improve the small-angle hadron 
acceptance and resolution, the electron beam travels along the center 
of the central solenoid, while the proton/ion beam traverses it at the 
crab crossing angle.

To fulfill the requirement of hermeticity, the central detector will
be built around a solenoid magnet (with a coil length of about 4 m). Due to
the asymmetric beam energies, the interaction point (IP) will be
slightly offset towards the electron side (1.5 m + 2.5 m). This will allow
more distance for the tracking of high-momentum hadrons produced at
small angles, and a larger bore angle for efficient detection of the
scattered beam leptons.
Existing superconducting detector solenoid magnets like those from 
CLEO or Babar~\footnote{both are 4m long, have a 3 m diameter, a 
1.5 T field, and an iron yoke} would be suitable for use in the MEIC at 
either IP. Like many detector solenoids these employ an iron yoke for 
the flux return, which encapsulates the detector and the endcaps. An 
interesting alternative is a dual solenoid where the inner and outer 
solenoid have opposite polarity thus providing an iron-free flux return 
in the space between. This design was proposed for the 4th detector 
concept for the ILC. The main advantages of the dual solenoid include 
light weight, high field capability (3 T), improved endcap acceptance, 
compact endcaps (coils instead of iron), easy detector access, low 
external field, and precise internal field map (no hysteresis). These 
features are ideal for a detector optimized for SIDIS, e.g., partonic 
fragmentation, and exclusive processes with recoils (see section~\ref{sec:recoil}). 
The initial magnetic design for the MEIC dual-solenoid- based detector 
has been completed. 

Figure~\ref{fig:ir_and_central_detector_layout} shows the dual-solenoid-based 
MEIC detector with three layers of forward- and central trackers including 
a vertex detector.
The current tracker layout is compatible with both a dual solenoid and 
the CLEO magnet.
Particle identification in the central detector would be provided by a TOF, and a
radially compact detector providing $e$/$\pi$, $\pi$/$K$, and $K$/$p$
identification. The current baseline design includes a DIRC
whose performance at an EIC compared with state of the art (BaBar) is the 
topic of an R\&D proposal. 
Optimizations and alternatives to the global baseline design are discussed 
in more detail in 
Ref.~\cite{Boer:2011fh}.

Small-angle tracking in the central detector could be an extension of
the vertex tracker, using semiconductor detectors, while larger angles
could be covered by planar micro-pattern detectors (GEMs/micromegas). On the
electron side, where the particle momenta are generally lower, one
could even consider drift chambers with a small cell size, in
particular for a final tracking region that could be added outside of
the solenoid itself. 
Lepton identification in the end-cap will be
performed using an electromagnetic calorimeter and a High-Threshold
Cherenkov Counter (HTCC) with CF$_4$ gas or equivalent. The details of
hadron identification in the electron end cap can be found in
Ref.~\cite{Boer:2011fh}.

The ion-side end-cap would have to deal with hadrons with a wide range
of momenta, some approaching that of the ion beam. While the
small-angle tracking resolution on this side is greatly enhanced by
the 50 mrad crossing angle~\footnote{particles scattered at zero degrees are
not moving parallel to the B-field} and dipole in front of the FFQs,
the forward tracking would nevertheless greatly benefit from good
position resolution, making this a priority. To identify particles of
various species over the full momentum range, one would ideally want
to use a RICH with several radiators, such as aerogel, C$_4$F$_{10}$,
and CF$_4$. Possible implementations are detailed in
Ref.~\cite{Boer:2011fh}.
\begin{figure*}[t]
\begin{center}
\includegraphics[trim=0cm 6cm 0cm 5cm, clip=true, width=0.96\textwidth]{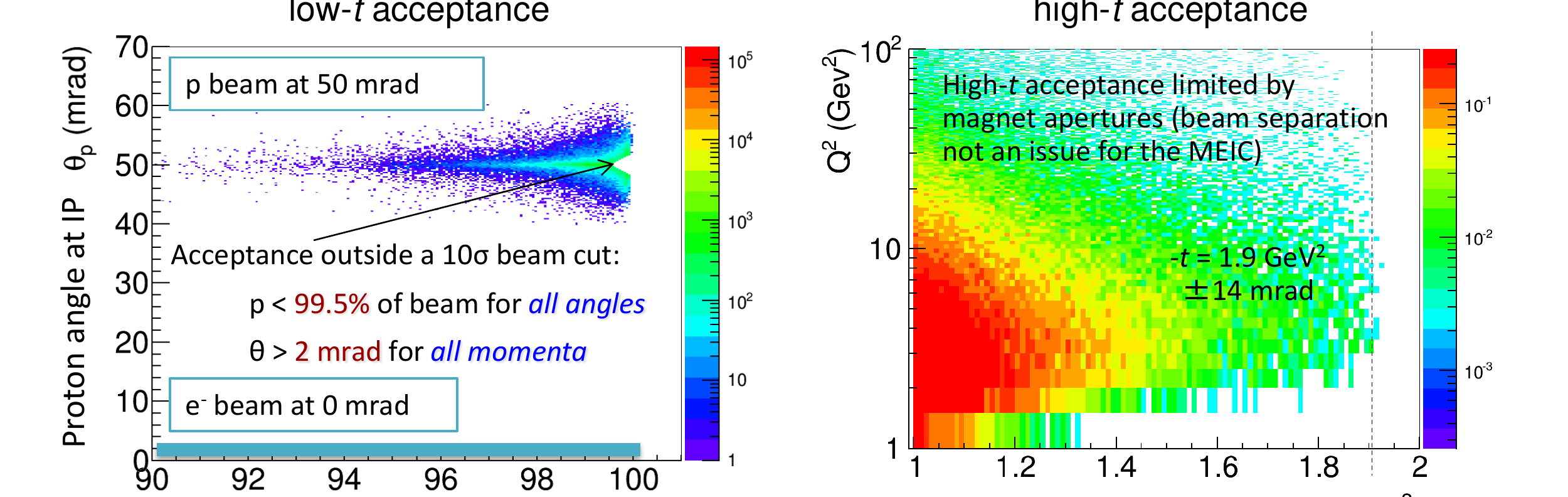}
\end{center}
\caption{\label{fig:ir_performance} The projected recoil baryon acceptance outside a 10~$\sigma$ beam cut for the MEIC for 5 GeV electrons colliding with 100 GeV protons at a crossing angle of 50 mrad. The following cuts have been applied: $Q^2 >$ 1 GeV$^2$, $x <$ 0.1, and $E^\prime_e >$ 1 GeV. 
}
\end{figure*}

On the ion side, the detection will be performed in three stages. The
first stage is the endcap, which will cover all angles down to the
acceptance of the forward spectrometer (several degrees around the ion
beam line). This in turn has two stages: one upstream of the ion FFQs, 
and one downstream of them. 
The acceptance of all stages is
matched so that there are minimal gaps in the coverage. 
The last stage
will cover angles up to 10-15 mrad on either side of the beam (more
vertically) for all ion fragments with different charge-to-mass ratios
and fractions of the beam momentum, with modest requirements on
magnet peak fields ~\footnote{Good performance can be achieved with 
peak fields for two magnets at 6 T and one at 5 T.}

The intermediate stage will use a 2 Tm dipole to augment the solenoid
at small angles where the tracking resolution otherwise would be
poor. The magnet will be about 1 m long and its aperture will cover
the distance to the electron beam (corresponding to the horizontal
crossing angle of 50 mrad), while the acceptance in the other three
directions is not restricted and can be larger. An important feature
of the magnet design is to ensure that the electron beam line stays
field free. The dipole will have trackers at the entrance and exit,
followed by a calorimeter covering the ring-shaped area in front of
the first ion FFQ. The intermediate stage is essential for providing a
wide coverage in $-t$ also for the lowest beam energies, and 
to investigate target fragmentation.

The last, small-angle stage provides the ultra-forward detection that
is crucial for detecting recoil baryons and tagging of spectator
protons in deuterium, as well as other nuclear fragments. The design
is heavily integrated with the accelerator, and the 4 m long, 20 Tm
downstream dipole serves not only as a spectrometer, but also
``corrects'' the 50 mrad crossing angle, and allows the neutrons to
escape on a tangent to the ring, separating cleanly from the beam area
before detection. This makes the electron and ion beam lines parallel
in the $\sim$15 m long drift space after the dipole, with separation
of more than 1 m, providing ample space for detectors. 
To optimize the low-$t$ coverage, it is essential that the 
10$\sigma$~\footnote{the proton beam size at 60 GeV (the 
mid-range point for the 20-100 GeV MEIC coverage).} beam size is as 
small as possible. This is achieved through cooling (which also reduces 
the angular spread), and by introducing a weaker secondary focus 16 m 
downstream of exit from the large 20 Tm analyzing dipole. Even the 
preliminary optics give full angular acceptance for charged particles 
with rigidities (momenta) of up to 99.5\% of the beam momentum 
(or more than 100.5\%) down to zero degrees, and full momentum 
acceptance for particles scattered at more than about 2-3 mrad with 
respect to the central beam. 
As shown in Fig.~\ref{fig:ir_performance}, the high $-t$ recoil baryon 
acceptance is only limited by the magnet apertures, while the low $-t$ 
acceptance requiring small beam size at the detection point and large 
dispersion after the IP to move the recoils away from the beam is limited 
by the beam itself. 
The dipole aperture can also be made sufficiently large to accept all off-angle
and off-momentum particles that exit the FFQs with the exception of
some ``spectator'' protons from deuterium scattered at very large
angles. These can, however, easily be detected in between the FFQs and
the dipole. 
Tracking studies show that the momentum resolution for
particles up to the beam momentum will only be limited by the
intrinsic momentum spread of the beam 
(longitudinal 4$\times 10^{-4}$), and
the angular resolution will also be excellent (0.2 mrad for all $\phi$). 
This is very important since the four-momentum transfer of the hadronic system is
proportional to $t \sim \theta^2_p E^2_p$, and the $t$-resolution for
instance determines the quality of the 3-D imaging that can be
achieved (see Sec.~\ref{sec:recoil}).
\end{multicols}


%% file: files_tex/acknowledgment.tex
\label{sec:acknowledgement}

\newpage
\noindent{\Large\bf Acknowledgments}

\bigskip\bigskip

This white paper is a result of a community wide effort through the many EIC 
workshops organized by the physics communities associated with both BNL and 
JLab, which culminated in the INT Report~\cite{Boer:2011fh}.  
We thank the following colleagues who made valuable comments
to the draft of this document:
\medskip

Christine Aidala (Univ.~of Michigan), 
Yasuyuki Akiba (RIKEN, Japan)
Kieran Boyle (RIKEN BNL Research Center at BNL),
Jian-Ping Chen (Jefferson Laboratory)
Leonard Gamberg (Penn State University)
Yuji Goto (RIKEN, Japan), 
John Harris (Yale University),
Thomas Hemmick (Stony Brook University),
Barbara Jacak (Stony Brook University),
Peter Jacobs (Lawrence Berkeley Laboratory),
Zhongbo Kang (Los Alamos National Laboratory),
David Kaplan (Institute of Nuclear Theory \& U. of Washington)
Dmitry Kharzeev (BNL \& Stony Brook University),
Sebastian Kuhn (Old Dominion University), 
Paul Newman (Univ.~ of Liverpool, United Kingdom),
Joakim Nystrand (Univ.~of Bergen, Norway),
Kent Paschke (Univ.~of Virginia),
Krishna Rajagopal (Massachusetts Institute of Technology),
Seamus Riordan (Stony Brook University)
Oscar Rondon (Univ.~of Virginia), 
Patrizia Rossi (Jefferson Laboratory), 
Berndt Surrow (Temple University)
Michael Tannenbaum (BNL), 
Mikhail Tokarev (JINR, Russia), 
Xin-Nian Wang (Lawrence Berkeley National Lab \& Central China Normal Univ.).
\medskip

This work was supported in part by the U.S. Department of Energy, contract numbers
DE-AC02-06CH11357 (R.H.), 
DE-AC02-98CH10886 (E.C.A., T.B., S.F., Y.H., M.A.C.L, J.-H.L., Y.L., V.L., T.W.L., B.M.,
V.P., J.-W.Q., Th.R., M.S., T.T., D.T., Th.U., R.V., S.V., L.Z.), 
DE-AC02-05CH11231 (E.S., F.Y.),
DE-AC02-76SF00515 (M.Su.),
DE-AC05-06OR23177 (A.A., R.E., V.G., A.H., F.-L.L., R.M., V.S.M, P.N.-T., A.P., C.W., Y.-H.Z.), 
and grant numbers 
DE-FG02-05ER41372 (A.De.), 
DE-FG02-09ER41620 (A.Du.), 
DE-FG02-03ER41231 (H.G.), 
DE-SC0004286 (Yu.K.), 
DE-FG02-88ER40415 (K.K.), 
DE-FG02-94ER40844 (Z.-E.M), 
DE-FG02-92ER40699 (A.H.M.), 
DE-FG02-08ER41531 (M.R.-M.), 
by the National Science Foundation, grant number 
PHY-1019521 (T.H.), 
by The ERC grant HotLHC ERC-2001-StG-279579; 
     Ministerio de Ciencia e Innovacion of Spain grants FPA2008-01177, 
     FPA2009-06867-E and Consolider-Ingenio 2010 CPAN CSD2007-00042; 
     Xunta de Galicia grant PGIDIT10PXIB206017PR; and FEDER (N.A.), 
by Chilean CONICYT: grant number FB0821, Anillo grant ACT-119 and Fondecyt grant 1120953 (W.B.), 
by City University of NY PSC-CUNY Research Award Program Grant 65041-0043 (A.Du), 
by Chilean Fondecyt Grant 11121448 (H.H.), 
by Fondecyt (Chile) Grant No. 1090291 (B.K.), 
by Croatian Ministry of Science, contract no. 119-0982930-1016 (K.Ku.),
and by CEA-Saclay and GDR 3034 PH-QCD (F.S.).